\documentclass[12pt]{iopart}
\pdfoutput=1 

\usepackage{iopams}
\usepackage{amstext} 
\usepackage{amssymb} 
\usepackage{bbold} 
\usepackage{bbm} 
\usepackage{amsthm} 
\usepackage{stmaryrd} 
\usepackage{stix}
\usepackage{etoolbox} 
\usepackage{xparse}
\usepackage{pgffor}

\usepackage[utf8]{inputenc} 
\usepackage{textgreek} 
\usepackage{relsize} 
\usepackage{microtype} 
\usepackage{ulem} 
\usepackage[compress]{cite} 
\usepackage{csquotes} 
\usepackage{setspace} 
\usepackage{scalerel}

\usepackage{graphicx} 
\usepackage[usenames,dvipsnames]{xcolor} 
\usepackage{tikzit} 
\usepackage{stackengine}
\usepackage{circuitikz} 
\usetikzlibrary{
    arrows,
    shapes,
    decorations,
    intersections,
    backgrounds,
    positioning,
    circuits.ee.IEC
    }

\tikzstyle{inline text}=[text height=1.5ex, text depth=0.25ex, yshift=0.5mm]
\tikzstyle{upground}=[circuit ee IEC, thick, ground, rotate=90, scale=2]
\tikzstyle{downground}=[circuit ee IEC, thick, ground, rotate=-90, scale=1.5]
\tikzstyle{point}=[regular polygon, regular polygon sides=3, draw, scale=0.75, inner sep=-0.5pt, minimum width=9mm, fill=white, regular polygon rotate=180, tikzit fill={rgb,255: red,242; green,255; blue,92}]
\tikzstyle{wide copoint}=[fill=white, draw, shape=isosceles triangle, shape border rotate=90, isosceles triangle stretches=true, inner sep=0pt, minimum width=1.5cm, minimum height=6.12mm]
\tikzstyle{wide point}=[fill=white, draw, shape=isosceles triangle, shape border rotate=-90, isosceles triangle stretches=true, inner sep=0pt, minimum width=1.5cm, minimum height=6.12mm, yshift=-0.0mm]
\tikzstyle{wide dpoint}=[wide point, doubled]
\tikzstyle{copoint}=[regular polygon, regular polygon sides=3, draw, scale=0.75, inner sep=-0.5pt, minimum width=9mm, fill=white, tikzit fill={rgb,255: red,255; green,128; blue,0}, tikzit draw={rgb,255: red,255; green,128; blue,0}]
\tikzstyle{dot}=[inner sep=0mm, minimum width=2mm, minimum height=2mm, draw, shape=circle]
\tikzstyle{black dot}=[dot, fill={gray!30}, text depth=-0.2mm]
\tikzstyle{white dot}=[dot, fill=white, text depth=-0.2mm]
\tikzstyle{small box}=[rectangle, inline text, fill=white, draw, minimum height=5mm, yshift=-0.5mm, minimum width=5mm, font={\small}]
\tikzstyle{small gray box}=[small box, fill={gray!30}]
\tikzstyle{medium box}=[rectangle, inline text, fill=white, draw, minimum height=5mm, yshift=-0.5mm, minimum width=10mm, font={\small}]
\tikzstyle{square box}=[small box]
\tikzstyle{medium gray box}=[small box, fill={gray!30}]
\tikzstyle{semilarge box}=[rectangle, inline text, fill=white, draw, minimum height=5mm, yshift=-0.5mm, minimum width=12.5mm, font={\small}]
\tikzstyle{large box}=[rectangle, inline text, fill=white, draw, minimum height=5mm, yshift=-0.5mm, minimum width=15mm, font={\small}]
\tikzstyle{large gray box}=[small box, fill={gray!30}]
\tikzstyle{dpoint}=[point, doubled]
\tikzstyle{dcopoint}=[copoint, doubled]
\tikzstyle{boldedge}=[doubled, shorten <=-0.17mm, shorten >=-0.17mm]
\tikzstyle{normal}=[line width=0.9pt]
\tikzstyle{doubled}=[line width=1pt]
\tikzstyle{boldedge}=[doubled, shorten <=-0.17mm, shorten >=-0.17mm]
\tikzstyle{small dbox}=[small box, doubled]
\tikzstyle{white ddot}=[white dot, doubled]
\tikzstyle{black ddot}=[black dot, doubled, tikzit fill=black]
\tikzstyle{map}=[draw, shape=NEbox, inner sep=2pt, minimum height=6mm, fill=white]
\tikzstyle{box}=[draw, shape=rectangle, inner sep=2pt, minimum height=6mm, minimum width=6mm, fill=white]
\tikzstyle{dbox}=[draw, doubled, shape=rectangle, inner sep=2pt, minimum height=6mm, minimum width=6mm, fill=white]
\tikzstyle{dmap}=[draw, doubled, shape=NEbox, inner sep=2pt, minimum height=6mm, fill=white]
\tikzstyle{dmapdag}=[draw, doubled, shape=SEbox, inner sep=2pt, minimum height=6mm, fill=white]
\tikzstyle{dmapadj}=[draw, doubled, shape=SEbox, inner sep=2pt, minimum height=6mm, fill=white]
\tikzstyle{dmaptrans}=[draw, doubled, shape=SWbox, inner sep=2pt, minimum height=6mm, fill=white]
\tikzstyle{dmapconj}=[draw, doubled, shape=NWbox, inner sep=2pt, minimum height=6mm, fill=white]
\tikzstyle{map}=[draw, shape=NEbox, inner sep=2pt, minimum height=6mm, fill=white]
\tikzstyle{dashedmap}=[draw, dashed, shape=NEbox, inner sep=2pt, minimum height=6mm, fill=white]
\tikzstyle{mapdag}=[draw, shape=SEbox, inner sep=2pt, minimum height=6mm, fill=white]
\tikzstyle{mapadj}=[draw, shape=SEbox, inner sep=2pt, minimum height=6mm, fill=white]
\tikzstyle{maptrans}=[draw, shape=SWbox, inner sep=2pt, minimum height=6mm, fill=white]
\tikzstyle{mapconj}=[draw, shape=NWbox, inner sep=2pt, minimum height=6mm, fill=white]
\tikzstyle{semilarge map}=[draw, shape=NEbox, inner sep=2pt, minimum height=6mm, fill=white, minimum width=9.5mm]
\tikzstyle{semilarge dmap}=[draw, doubled, shape=NEbox, inner sep=2pt, minimum height=6mm, fill=white, minimum width=9.5mm]
\tikzstyle{kpointdag}=[kpoint adjoint]
\tikzstyle{kpointadj}=[kpoint adjoint]
\tikzstyle{kpointconj}=[kpoint conjugate]
\tikzstyle{kpointtrans}=[kpoint transpose]
\tikzstyle{kpoint common}=[draw, fill=white, inner sep=1pt, minimum height=4mm]
\tikzstyle{kpoint sc}=[shape=cornerpoint, kpoint common]
\tikzstyle{kpoint adjoint sc}=[shape=cornercopoint, kpoint common]
\tikzstyle{kpoint}=[shape=cornerpoint, shorten left=5pt, kpoint common, tikzit fill={rgb,255: red,255; green,128; blue,0}]
\tikzstyle{kpoint adjoint}=[shape=cornercopoint, shorten left=5pt, kpoint common, tikzit fill={rgb,255: red,255; green,128; blue,0}]
\tikzstyle{kpoint conjugate}=[shape=cornerpoint, shorten right=5pt, kpoint common]
\tikzstyle{kpoint transpose}=[shape=cornercopoint, shorten right=5pt, kpoint common]
\tikzstyle{kpoint symm}=[shape=cornerpoint, shorten left=5pt, shorten right=5pt, kpoint common]
\tikzstyle{wide kpoint}=[kpoint, minimum width=1 cm, inner sep=2pt]
\tikzstyle{wide kpointdag}=[kpointdag, minimum width=1 cm, inner sep=2pt]
\tikzstyle{wide kpointconj}=[kpointconj, minimum width=1 cm, inner sep=2pt]
\tikzstyle{wide kpointtrans}=[kpointtrans, minimum width=1 cm, inner sep=2pt]
\tikzstyle{wider kpoint}=[kpoint, minimum width=1.25 cm, inner sep=2pt]
\tikzstyle{wider kpointdag}=[kpointdag, minimum width=1.25 cm, inner sep=2pt]
\tikzstyle{wider kpointconj}=[kpointconj, minimum width=1.25 cm, inner sep=2pt]
\tikzstyle{wider kpointtrans}=[kpointtrans, minimum width=1.25 cm, inner sep=2pt]
\tikzstyle{dkpoint}=[kpoint, doubled, tikzit fill={rgb,255: red,255; green,85; blue,210}]
\tikzstyle{wide dkpoint}=[wide kpoint, doubled, tikzit fill={rgb,255: red,68; green,255; blue,0}]
\tikzstyle{dkpointdag}=[kpoint adjoint, doubled]
\tikzstyle{wide dkpointdag}=[wide kpointdag, doubled]
\tikzstyle{label}=[fill=white, draw=white, shape=circle, tikzit draw={rgb,255: red,10; green,26; blue,255}, tikzit fill={rgb,255: red,0; green,12; blue,255}, font={\small}]
\tikzstyle{squarelabel}=[fill=white, draw=white, shape=rectangle, tikzit draw=black]
\tikzstyle{eslabel}=[tikzit draw={rgb,255: red,255; green,191; blue,191}, tikzit fill={rgb,255: red,255; green,191; blue,191}, font={\tiny}]
\tikzstyle{large dmap}=[draw, doubled, shape=NEbox, inner sep=2pt, minimum height=6mm, fill=white, minimum width=12mm]
\tikzstyle{gray point}=[point, fill={gray!40!white}]
\tikzstyle{gray dpoint}=[gray point, doubled, tikzit draw={rgb,255: red,128; green,128; blue,128}, tikzit fill={rgb,255: red,128; green,128; blue,128}]
\tikzstyle{gray copoint}=[copoint, fill={gray!40!white}, tikzit fill={rgb,255: red,128; green,128; blue,128}]
\tikzstyle{gray dcopoint}=[gray copoint, doubled, tikzit fill={rgb,255: red,128; green,128; blue,128}]
\tikzstyle{circlenew}=[draw=black, shape=circle, inner sep=1pt]
\tikzstyle{blue label}=[text=NavyBlue, tikzit draw={rgb,255: red,0; green,96; blue,167}, tikzit fill={rgb,255: red,35; green,68; blue,255}]
\tikzstyle{big dot}=[fill=white, draw=black, shape=circle, minimum width=6mm, minimum height=6mm]
\tikzstyle{3d box}=[fill=white, draw=black, shape=trapezium, trapezium left angle=-70, trapezium right angle=70, rotate=10]
\tikzstyle{slant red box}=[fill={rgb,255: red,191; green,0; blue,64}, draw={rgb,255: red,191; green,0; blue,64}, shape=rectangle, xslant=0.5, font={\tiny}, text={rgb,255: red,191; green,0; blue,64}, fill opacity=0.5, line width=1pt]
\tikzstyle{slant point}=[regular polygon, regular polygon sides=3, draw, scale=0.75, inner sep=-0.5pt, minimum width=9mm, fill white, regular polygon rotate=180, yslant=-0.3]
\tikzstyle{tiny orange label}=[font={\tiny}, text={rgb,255: red,255; green,128; blue,0}, tikzit draw={rgb,255: red,255; green,128; blue,0}]
\tikzstyle{tiny red label}=[font={\tiny}, text={rgb,255: red,191; green,0; blue,64}, tikzit draw={rgb,255: red,191; green,0; blue,64}, draw=none]
\tikzstyle{red label}=[text={rgb,255: red,191; green,0; blue,64}, tikzit draw={rgb,255: red,191; green,0; blue,64}]
\tikzstyle{slant label black}=[font={\tiny}, xslant=0.5, tikzit draw=black]
\tikzstyle{slant label red}=[font={\tiny}, xslant=0.5, text={rgb,255: red,191; green,0; blue,64}, tikzit draw={rgb,255: red,191; green,0; blue,64}]
\tikzstyle{slant label orange}=[font={\tiny}, xslant=0.5, text={rgb,255: red,255; green,128; blue,0}, tikzit draw={rgb,255: red,255; green,128; blue,0}]
\tikzstyle{slanted point}=[fill={rgb,255: red,191; green,0; blue,64}, draw={rgb,255: red,191; green,0; blue,64}, shape=triangle, regular polygon, regular polygon sides=3, scale=0.75, inner sep=-0.5pt, minimum width=5mm, regular polygon rotate=90, xslant=0.5, fill opacity=0.5, font={\tiny}, line width=1pt, text={rgb,255: red,191; green,0; blue,64}]
\tikzstyle{slanted point black}=[draw=black, shape=triangle, regular polygon, regular polygon sides=3, scale=0.75, inner sep=-0.5pt, minimum width=5mm, regular polygon rotate=90, xslant=0.5, font={\tiny}, line width=0.2pt, text=black, fill=white, tikzit fill=white]
\tikzstyle{red dot}=[fill={rgb,255: red,191; green,0; blue,64}, draw={rgb,255: red,191; green,0; blue,64}, shape=circle, inner sep=0, minimum width=1.5mm, minimum height=1.5mm]
\tikzstyle{black dot}=[fill=black, draw=black, shape=circle, inner sep=0, minimum width=1.5mm, minimum height=1.5mm]
\tikzstyle{orange dot}=[fill={rgb,255: red,255; green,128; blue,0}, draw={rgb,255: red,255; green,128; blue,0}, shape=circle, inner sep=0, minimum width=1.5mm, minimum height=1.5mm]
\tikzstyle{blue dot}=[fill={rgb,255: red,0; green,0; blue,228}, draw={rgb,255: red,0; green,0; blue,228}, shape=circle, inner sep=0, minimum width=1.5mm, minimum height=1.5mm]
\tikzstyle{slant white}=[fill=white, draw=black, shape=rectangle, xslant=0.5, font={\tiny}, line width=1pt]
\tikzstyle{slant small map}=[fill=white, draw=black, xslant=0.5, shape=rectangle, font={\tiny}, line width=1pt, inner sep=0.6mm]
\tikzstyle{slanted copoint black}=[draw=black, shape=triangle, regular polygon, regular polygon sides=3, scale=0.75, inner sep=-0.5pt, minimum width=5mm, regular polygon rotate=-90, xslant=0.5, font={\tiny}, line width=0.2pt, text=black, fill=white, tikzit fill=white]
\tikzstyle{purple dot}=[fill={rgb,255: red,128; green,0; blue,128}, draw={rgb,255: red,128; green,0; blue,128}, shape=circle, inner sep=0, minimum width=1.5mm, minimum height=1.5mm]
\tikzstyle{white dot 2}=[fill=white, draw=black, shape=circle]
\tikzstyle{horizontal point}=[style=point, rotate=-90, tikzit shape=rectangle, tikzit fill={rgb,255: red,191; green,128; blue,64}]
\tikzstyle{pslant orange}=[style=slanted point black, fill={rgb,255: red,255; green,128; blue,0}, draw={rgb,255: red,255; green,128; blue,0}, tikzit fill={rgb,255: red,255; green,128; blue,0}, tikzit draw={rgb,255: red,255; green,128; blue,0}]
\tikzstyle{upground horizontal}=[style=upground, rotate=-90]
\tikzstyle{double horizontal point}=[style=horizontal point, line width=1pt]
\tikzstyle{double point}=[style=point, line width=1pt]
\tikzstyle{double copoint}=[style=copoint, line width=1pt]
\tikzstyle{horizontal copoint}=[style=double copoint, rotate=-90]
\tikzstyle{slant label purple}=[style=slant label black, tikzit draw={rgb,255: red,128; green,0; blue,128}, text={rgb,255: red,128; green,0; blue,128}]
\tikzstyle{orange copoint}=[style=pslant orange, rotate=-180, tikzit fill={rgb,255: red,255; green,128; blue,0}]
\tikzstyle{new style 0}=[style=slant white, draw={rgb,255: red,0; green,0; blue,228}, fill={rgb,255: red,0; green,0; blue,228}, fill opacity=0.5, shape=rectangle]
\tikzstyle{wide slanted point}=[style=wide point, xslant=0.5, fill=white, rotate=-90, minimum width=0.8cm, fill={rgb,255: red,128; green,128; blue,128}, fill opacity=0.5, line width=1pt]
\tikzstyle{black dot white}=[style=black dot, text=white, draw=none, tikzit draw={rgb,255: red,191; green,255; blue,0}, shape=circle]

\tikzstyle{new edge style 1}=[-, line width=1pt, shorten <=-0.17mm, shorten >=-0.17mm, tikzit draw={rgb,255: red,204; green,0; blue,3}]
\tikzstyle{diredge}=[-, postaction=decorate, decoration={markings, mark=at position 0.55 with \edgearrow}]
\tikzstyle{bold diredge}=[-, diredge, line width=1pt, tikzit draw={rgb,255: red,128; green,0; blue,128}]
\tikzstyle{grey}=[-, draw={rgb,255: red,188; green,188; blue,188}]
\tikzstyle{classical}=[-, dashed, tikzit draw={rgb,255: red,255; green,128; blue,0}]
\tikzstyle{reddashed}=[-, dashed, draw={rgb,255: red,0; green,128; blue,128}, postaction=decorate, decoration={markings, mark=at position 0.55 with \edgearrow}]
\tikzstyle{reddahednoarrow}=[-, dashed, draw={rgb,255: red,179; green,40; blue,40}]
\tikzstyle{arrow edge}=[-, ->, draw={rgb,255: red,191; green,191; blue,191}, tikzit draw={rgb,255: red,191; green,191; blue,191}, ultra thick]
\tikzstyle{tarrow edge}=[-, ->, draw={rgb,255: red,191; green,191; blue,191}, tikzit draw={rgb,255: red,191; green,191; blue,191}]
\tikzstyle{gray edge}=[-, draw={rgb,255: red,191; green,191; blue,191}, tikzit draw={rgb,255: red,191; green,191; blue,191}, ultra thick]
\tikzstyle{lightgrayedge}=[-, draw={rgb,255: red,207; green,207; blue,207}]
\tikzstyle{green edge}=[-, tikzit draw={rgb,255: red,128; green,128; blue,0}, draw={rgb,255: red,128; green,128; blue,0}]
\tikzstyle{red edge}=[-, draw={rgb,255: red,191; green,0; blue,64}, tikzit draw={rgb,255: red,191; green,0; blue,64}]
\tikzstyle{arrow edge black}=[-, ->]
\tikzstyle{solid blue}=[-, draw={rgb,255: red,0; green,96; blue,167}, tikzit draw={rgb,255: red,0; green,96; blue,167}]
\tikzstyle{classical blue}=[-, draw={rgb,255: red,0; green,96; blue,167}, tikzit draw={rgb,255: red,0; green,96; blue,167}, dashed]
\tikzstyle{fill gray}=[-, fill=gray]
\tikzstyle{bold gray}=[-, line width=1pt, tikzit draw={rgb,255: red,128; green,128; blue,128}]
\tikzstyle{fill pink}=[-, fill={rgb,255: red,193; green,100; blue,94}, fill opacity=0.5, draw={rgb,255: red,134; green,68; blue,65}, line width=1pt, tikzit draw={rgb,255: red,134; green,68; blue,65}, tikzit fill={rgb,255: red,193; green,100; blue,94}]
\tikzstyle{fill carta da zucchero}=[-, fill={rgb,255: red,129; green,158; blue,219}, fill opacity=0.5, line width=0.4mm]
\tikzstyle{fill white}=[-, fill=white]
\tikzstyle{fill purple}=[-, fill={rgb,255: red,113; green,69; blue,128}, fill opacity=0.5, draw={rgb,255: red,79; green,48; blue,90}, tikzit fill={rgb,255: red,113; green,69; blue,128}, tikzit draw={rgb,255: red,79; green,48; blue,90}, line width=1pt]
\tikzstyle{fill green}=[-, fill={rgb,255: red,62; green,128; blue,120}, fill opacity=0.5, draw={rgb,255: red,33; green,68; blue,63}, tikzit fill={rgb,255: red,62; green,128; blue,120}, tikzit draw={rgb,255: red,33; green,68; blue,63}, line width=1pt]
\tikzstyle{bold orange}=[-, draw={rgb,255: red,255; green,128; blue,0}, fill=none, line width=1pt]
\tikzstyle{bold black}=[-, line width=1pt, draw=black, fill=none, tikzit draw=black]
\tikzstyle{bold red}=[-, draw={rgb,255: red,191; green,0; blue,64}, fill=none, line width=1pt]
\tikzstyle{fill light green}=[-, fill={rgb,255: red,166; green,166; blue,112}, fill opacity=0.5, draw={rgb,255: red,121; green,121; blue,81}, line width=1pt]
\tikzstyle{new edge style 0}=[-, fill=yellow, fill opacity=0.5, draw={rgb,255: red,146; green,146; blue,0}, tikzit fill=yellow, tikzit draw={rgb,255: red,146; green,146; blue,0}]
\tikzstyle{bold dashed red}=[-, draw={rgb,255: red,191; green,0; blue,64}, fill=none, line width=1pt, dashed]
\tikzstyle{bold dashed orange}=[-, draw={rgb,255: red,255; green,128; blue,0}, dashed, line width=1pt]
\tikzstyle{bold blue}=[-, draw={rgb,255: red,0; green,0; blue,228}, line width=1pt]
\tikzstyle{arrow red}=[draw={rgb,255: red,191; green,0; blue,64}, ->, line width=1pt]
\tikzstyle{new edge style 2}=[-, draw={rgb,255: red,191; green,0; blue,64}, line width=1pt]
\tikzstyle{boldish}=[-, line width=0.6mm, fill=cyan]
\tikzstyle{white edge}=[-, draw=white]
\tikzstyle{purple edge}=[-, draw={rgb,255: red,128; green,0; blue,128}, line width=1pt]
\tikzstyle{light gray}=[-, fill={rgb,255: red,191; green,191; blue,191}, draw={rgb,255: red,191; green,191; blue,191}, tikzit fill={rgb,255: red,191; green,191; blue,191}, tikzit draw={rgb,255: red,191; green,191; blue,191}, fill opacity=0.3]
\tikzstyle{invisible edge}=[-, fill opacity=0, fill=none]
\tikzstyle{carta da zucchero thin}=[-, style=fill carta da zucchero, line width=0.1pt, fill={rgb,255: red,129; green,158; blue,219}, tikzit fill={rgb,255: red,129; green,158; blue,219}]
\tikzstyle{pink thin}=[-, style=fill pink, line width=0.1pt, fill={rgb,255: red,193; green,100; blue,94}]
\tikzstyle{fill green thin edge}=[-, style=fill green, tikzit fill={rgb,255: red,62; green,128; blue,120}, line width=0.1pt]

\definecolor{evred}{rgb}{0.996, 0.403, 0.537}
\definecolor{evgreen}{rgb}{0.501, 1.0, 0.505}
\definecolor{evblue}{rgb}{0.2, 0.588, 1.0}

\usepackage[frozencache]{minted} 

\setminted[python]{
bgcolor=lightgray!10,
xleftmargin=2mm,
fontsize=\footnotesize,
frame=lines,
linenos,
python3=true}
\setminted[text]{
bgcolor=lightgray!10,
xleftmargin=2mm,
fontsize=\footnotesize}

\theoremstyle{definition}
\newtheorem{definition}{Definition}[section]

\usepackage{xifthen}

\newcommand{\opapp}[2]{\ensuremath{#1\left(#2\right)}} 
\newcommand{\opapptxt}[2]{\ensuremath{\text{#1}\left(#2\right)}} 

\newcommand{\tsuchthat}[2]{\ensuremath{\left\{#1\middle|#2\right\}}} 
\newcommand{\suchthat}[2]{\tsuchthat{\,#1\,}{\,#2\,}} 

\newcommand{\downset}[1]{\ensuremath{#1\!\downarrow}}
\newcommand{\upset}[1]{\ensuremath{#1\!\uparrow}}
\newcommand{\domSym}{\text{dom}}
\newcommand{\dom}[1]{\opapp{\domSym}{#1}}

\newcommand{\restrict}[2]{#1|_{#2}}
\newcommand{\PFun}[1]{\opapptxt{PFun}{#1}} 

\newcommand{\ev}[1]{\text{#1}} 
\newcommand{\discrete}[1]{\opapptxt{discrete}{#1}} 
\newcommand{\total}[1]{\opapptxt{total}{#1}} 
\newcommand{\seqcomposeSym}{\rightsquigarrow}
\newcommand{\causeqcls}[1]{\ensuremath{\left[#1\right]_{\simeq}}}
\newcommand{\LsetsSym}{\Lambda} 
\newcommand{\Lsets}[1]{\opapp{\LsetsSym}{#1}} 
\newcommand{\Hist}[1]{\opapptxt{Hist}{#1}} 
\newcommand{\ExtHist}[1]{\opapptxt{ExtHist}{#1}} 
\newcommand{\Ext}[1]{\opapptxt{Ext}{#1}} 
\newcommand{\Prime}[1]{\opapptxt{Prime}{#1}} 
\newcommand{\allJoinsSym}{\;\dot{\vee}\;}
\newcommand{\tips}[2]{\opapp{\text{tips}_{#1}}{#2}} 
\newcommand{\tip}[2]{\opapp{\text{tip}_{#1}}{#2}} 
\newcommand{\CausCompl}[1]{\opapptxt{CausCompl}{#1}} 
\newcommand{\Events}[1]{{E}^{#1}} 
\newcommand{\Inputs}[1]{{I}^{#1}} 
\newcommand{\AllSpaces}{\ensuremath{\text{Spaces}}} 
\newcommand{\Spaces}[1]{\opapptxt{Spaces}{#1}} 
\newcommand{\SpacesFC}[1]{\opapp{\text{Spaces}_{\text{FC}}}{#1}} 
\newcommand{\CCSpaces}[1]{\opapptxt{CCSpaces}{#1}} 
\newcommand{\CSwitchSpaces}[1]{\opapptxt{CSwitchSpaces}{#1}} 

\newcommand{\hist}[1]{
    \ensuremath{
        \left\{
            \foreach \i\j [count=\idx] in {#1}{%
                \ifnum\idx=1%
                    \ev{\i}\!:\!\j%
                \else%
                    ,\,\ev{\i}\!:\!\j%
                \fi%
            }
        \right\}
    }
}
\newcommand{\evset}[1]{
    \ensuremath{
        \left\{
            \foreach \i [count=\idx] in {#1}{%
                \ifnum\idx=1%
                    \ev{\i}%
                \else%
                    ,\ev{\i}%
                \fi%
            }
        \right\}
    }
}

\usetikzlibrary{external}
\begin{document}

\title{Classification of causally complete spaces on 3 events with binary inputs}

\author{Stefano Gogioso$^{1,2}$ and Nicola Pinzani$^{1,3}$}

\address{$^1$Hashberg Ltd, London, UK}
\address{$^2$Department of Computer Science, University of Oxford, Oxford, UK}
\address{$^3$QuIC, Universit\'{e} Libre de Bruxelles, Brussels, BE}
\ead{$^1$stefano.gogioso@cs.ox.ac.uk, $^2$nicola.pinzani@ulb.be}
\vspace{10pt}

\begin{abstract}
    In this work, we present an exhaustive classification of the 2644 causally complete spaces on 3 events with binary inputs, together with the algorithm used for the classification.
    This paper forms the supplementary material for a trilogy of works: spaces of input histories, our dynamical generalisation of causal orders, were introduced in ``The Combinatorics of Causality''; the sheaf-theoretic treatment of causal distributions was detailed in ``The Topology of Causality''; the polytopes formed by the associated empirical models were studied in ``The Geometry of Causality''.
\end{abstract}

\maketitle










\section{Causal orders}
\label{section:causal-orders}

In this Section we recap basic notions about causal orders that form the basis for spaces of input histories, as described in the next Section.
For a longer discussion of causal orders and their operations, we refer the reader to Section 2 of the companion work ``The Combinatorics of Causality'' \cite{gogioso2022combinatorics}.

\subsection{Causal Orders and Hasse Diagrams}
\label{subsection:causal-orders-intro}

\begin{definition}
A \emph{causal order} $\Omega$ is a preorder: a set $|\Omega|$ of events---finite, in this work---equipped with a symmetric transitive relation $\leq$, which we refer to as the \emph{causal relation}.
In cases where multiple cause orders are involved, we might also use the more explicit notation $\leq_{\Omega}$, to indicate that the relation is order-dependent.
\end{definition}

\begin{definition}
There are four possible ways in which two distinct events $\omega, \xi \in \Omega$ can relate to each other causally:
\begin{itemize}
\item $\omega$ \emph{causally precedes} $\xi$ if $\omega \leq \xi$ and $\xi \not \leq \omega$, which we write succinctly as $\omega \prec \xi$ (to distinguish it from $\omega < \xi$, meaning instead that $\omega \leq \xi$ and $\omega \neq \xi$)
\item $\omega$ \emph{causally succeeds} $\xi$ if $\xi \leq \omega$ and $\omega \not \leq \xi$, which we write succinctly as $\omega \succ \xi$ (to distinguish it from $\omega > \xi$, meaning instead that $\omega \geq \xi$ and $\omega \neq \xi$)
\item $\omega$ and $\xi$ are \emph{causally unrelated} if $\omega \not\leq \xi$ and $\xi \not \leq \omega$
\item $\omega$ and $\xi$ are in \emph{indefinite causal order} if $\omega \neq \xi$, $\omega \leq \xi$ and $\xi \leq \omega$, which we write succinctly as $\omega \simeq \xi$
\end{itemize}
We say that a causal order is \emph{definite} when the last case cannot occur, i.e. when $\leq$ is anti-symmetric ($\omega \leq \xi$ and $\omega \geq \xi$ together imply $\omega = \xi$); otherwise, we say that it is \emph{indefinite}.
A definite causal order is thus a \emph{partial order}, or \emph{poset}: in this case, $\omega\prec\xi$ is the same as $\omega<\xi$, and $\omega\succ\xi$ is the same as $\omega > \xi$.
\end{definition}

\begin{definition}
We say that two events $\omega, \xi$ are \emph{causally related} if they are not causally unrelated, i.e. if at least one of $\omega \leq \xi$ or $\omega \geq \xi$ holds.
We also define the \emph{causal past} $\downset{\omega}$ and \emph{causal future} $\upset{\omega}$ of an event $\omega \in \Omega$, as well as its \emph{causal equivalence class} $\causeqcls{\omega}$:
\begin{eqnarray}
    \downset{\omega} & := \suchthat{\xi \in \Omega}{\xi \leq \omega} \\
    \upset{\omega} & := \suchthat{\xi \in \Omega}{\xi \geq \omega} \\
    \causeqcls{\omega} & :=  \suchthat{\xi \in \Omega}{\xi \simeq \omega} = \downset{\omega} \cap\; \upset{\omega}
\end{eqnarray}
Note that the $\omega$ always lies in both its own causal future and its own causal past, but also that their intersection can comprise more events (if the order is indefinite).
\end{definition}

Our interpretation of causality is a "negative" one, as "no-signalling from the future": when $\omega$ causally precedes $\xi$, for example, we are not so much interested in the "possibility" of causal influence from $\omega$ to $\xi$ (because $\omega \leq \xi$) as we are in the "impossibility" of causal influence from $\xi$ to $\omega$ (because $\xi \not \leq \omega$).
This generalises the "spatial" no-signalling case, where one is interested in the statements $\omega \not\leq \xi$ and $\xi \not \leq \omega$.
Far from being merely an interpretation, such no-signalling approach to causality permeates the entirety of this work.

Definite causal orders have an equivalent presentation as directed acyclic graphs (DAGs), known as \emph{Hasse diagrams}: vertices in the graph correspond to events $\omega \in \Omega$, while edges $x \rightarrow y$ correspond to those causally related pairs $\omega \leq \xi$ with no intermediate event (i.e. where there is no $\zeta \in \Omega$ such that $\omega < \zeta < \xi$).
For example, below are the Hasse diagrams for three definite causal orders on three events \ev{A}, \ev{B} and \ev{C}.
\begin{center}
    \includegraphics[height=2.5cm]{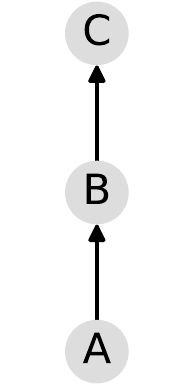}
    \hspace{1.5cm}
    \includegraphics[height=2cm]{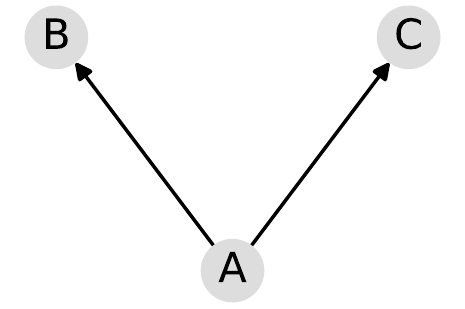}
    \hspace{1.5cm}
    \includegraphics[height=2cm]{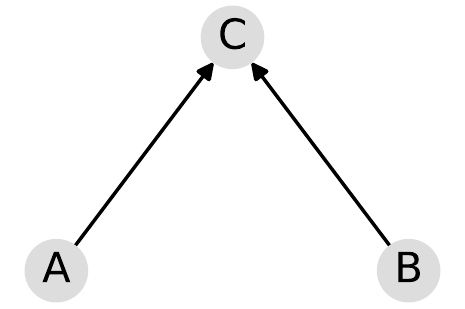}
\end{center}
Causal orders are naturally ordered by inclusion: $\Omega \leq \Xi$ if $|\Omega| \subseteq |\Xi|$ as sets and $\leq_{\Omega} \subseteq \leq_{\Xi}$ as relations (i.e. as subsets $\suchthat{(\omega, \omega')}{\omega \leq_\Omega \omega'} \subseteq |\Omega|^2$ and $\suchthat{(\xi, \xi')}{\xi \leq_\Xi \xi'} \subseteq |\Xi|^2$).
The requirement that $\leq_{\Omega} \subseteq \leq_{\Xi}$ explicitly means that for all $\omega, \omega' \in \Omega$ the constraint $\omega \not \leq_\Xi \omega'$ in $\Xi$ implies the constraint $\omega' \not \leq_\Omega \omega'$.
Put in different words:
\begin{itemize}
    \item If $\omega$ and $\omega'$ are causally unrelated in $\Xi$ , then they are causally unrelated in $\Omega$.
    \item If $\omega$ causally precedes $\omega'$ in $\Xi$, then it can either causally precede $\omega'$ in $\Omega$ or it can be causally unrelated to $\omega'$ in $\Omega$; it cannot causally succeed $\omega'$ or be in indefinite causal order with it.
    \item If $\omega$ and $\omega'$ are in indefinite causal order in $\Xi$, then their causal relationship in $\Omega$ is unconstrained: $\omega$ can causally precede $\omega'$, causally succeed it, be causally unrelated to it or be in indefinite causal order with it.
\end{itemize}
From a causal standpoint, $\Omega \leq \Xi$ means that $\Omega$ imposes on its own events at least the same causal constraints as $\Xi$, and possibly more.
In particular, if $\Xi$ is definite (no two events in indefinite causal order) then so is $\Omega$; conversely, if $\Omega$ is indefinite, then so is $\Xi$.

\begin{figure}[h]
    \centering
    \includegraphics[height=10cm]{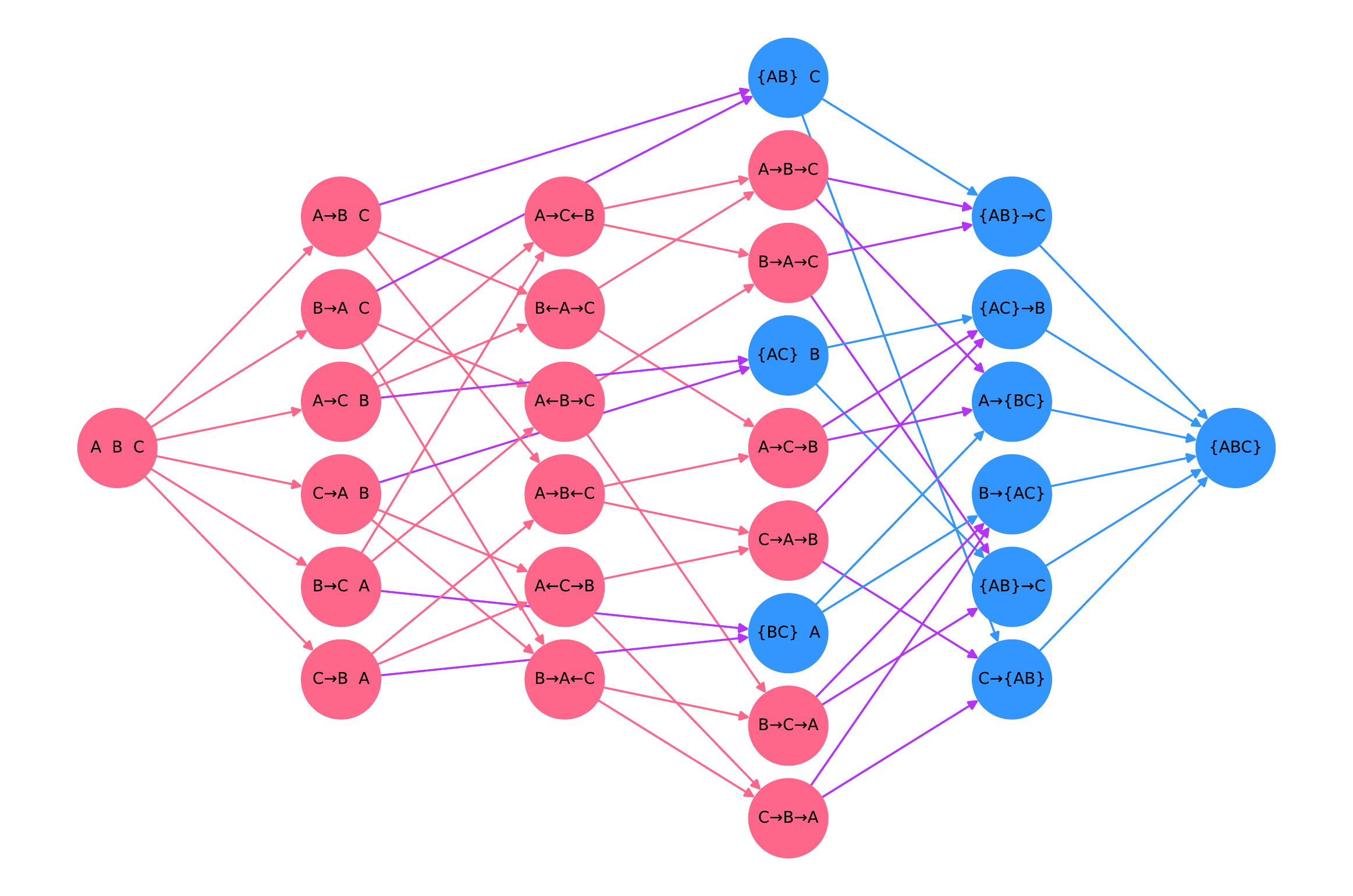}
    \caption{
    Hasse diagram for the hierarchy of causal orders on three events $\{\ev{A},\ev{B},\ev{C}\}$, left-to-right in inclusion order.
    See Figure 2 p.13 of ``The Combinatorics of Causality'' for full description.
    }
\label{fig:hierarchy-orders-3}
\end{figure}

Causal orders on a given set of events form a finite lattice, which we refer to as the \emph{hierarchy of causal orders}.
The join and meet operations on this lattice are those described in the previous subsection, the indiscrete order is the unique maximum (all events in indefinite causal order, i.e. no causal constraints), while the discrete order is the unique minimum (all events are causally unrelated).
The hierarchy of causal orders on three events $\{\ev{A},\ev{B},\ev{C}\}$ is displayed by Figure \ref{fig:hierarchy-orders-3} (p.\pageref{fig:hierarchy-orders-3}), with definite causal order coloured red and indefinite ones coloured blue.

\subsection{Lattice of Lowersets}
\label{subsection:causal-orders-lowersets}

As discussed in detail by Section \ref{section:spaces}, this work is concerned with a certain class of operational scenarios: blackbox devices are operated locally at events in spacetime, determining a probability distribution on their joint outputs conditional to their (freely chosen) joint inputs.
In such scenarios, causality constraints essentially state that the output at any subset of events cannot depend on inputs at events which causally succeed them or are causally unrelated to them.
Furthermore, the output at any event is only well-defined conditional to inputs for all events in its past: we are not interested in all sub-sets of events of a causal order, but rather in its lowersets.

The discussion above indicates that the object we seek to understand is not the causal order $\Omega$ itself, but rather its \emph{lattice of lowersets} $\Lsets{\Omega}$.
This is the subsets of events closed in the past, ordered by inclusion:
\begin{equation*}
    \Lsets{\Omega}
    :=
    \suchthat{U \subseteq \Omega}{\forall \omega \in U.\,\downset{\omega} \subseteq U}
\end{equation*}
In this case, being a lattice means that lowersets are closed under both intersection and union; we always omit the empty set from our Hasse diagrams, for clarity.

Inclusions between lowersets determine the causality constraints for the causal order: if $U, V \in \Lsets{\Omega}$ are such that $U \subseteq V$, then the output at events in $U$ cannot depend on the inputs at events in $V \backslash U$.
Consider the total order $\ev{A}\rightarrow\ev{B}\rightarrow\ev{C}$, and its associated lattice of lowersets: the inclusion $\{\ev{A},\ev{B}\} \subseteq \{\ev{A},\ev{B},\ev{C}\}$, for example, tells us that the outputs at events \ev{A} and \ev{B} cannot depend on the input at event \ev{C}; the inclusion $\{\ev{A}\} \subseteq \{\ev{A},\ev{B}\}$, additionally, tells us that the outputs at event \ev{A} cannot depend on the input at event \ev{B}.
\begin{center}
    \raisebox{1.40cm}{$\LsetsSym$}
    \raisebox{1.40cm}{$\left(\rule{0cm}{1.35cm}\right.$}
    \hspace{0.0cm}
    \raisebox{0.25cm}{
        \includegraphics[height=2.5cm]{svg-inkscape/total-ABC_svg-tex.pdf}
    }
    \hspace{0.0cm}
    \raisebox{1.40cm}{$\left.\rule{0cm}{1.35cm}\right)$}
    \hspace{0.75cm}
    \raisebox{1.40cm}{$=$}
    \hspace{0.5cm}
    \includegraphics[height=3cm]{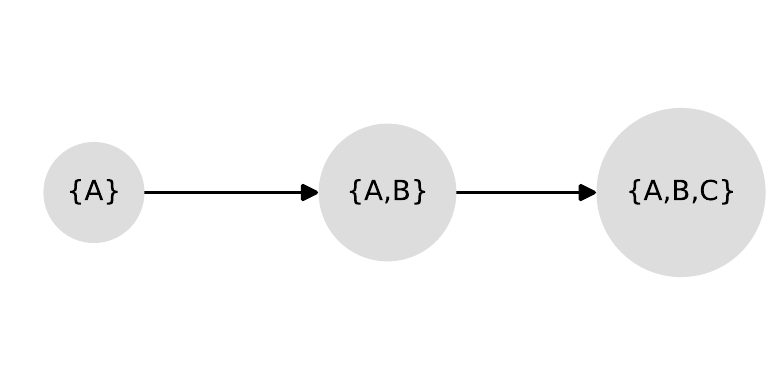}
\end{center}
Below is a more complicated example, for the diamond order: the inclusion $\{\ev{A},\ev{B}\} \subseteq \{\ev{A},\ev{B},\ev{C},\ev{D}\}$, for example, tells us that the outputs at events \ev{A} and \ev{B} cannot depend on the input at events \ev{C} and \ev{D}.
\begin{center}
    \raisebox{2.4cm}{$\LsetsSym$}
    \raisebox{2.4cm}{$\left(\rule{0cm}{1.35cm}\right.$}
    \hspace{0.0cm}
    \raisebox{1.25cm}{
        \includegraphics[height=2.5cm]{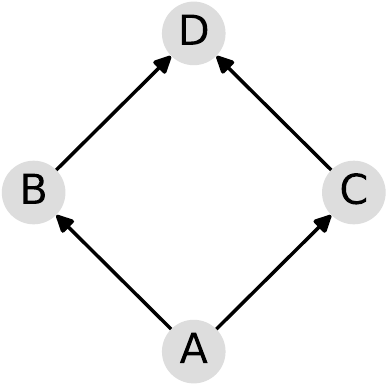}
    }
    \hspace{0.0cm}
    \raisebox{2.4cm}{$\left.\rule{0cm}{1.35cm}\right)$}
    \hspace{0.75cm}
    \raisebox{2.4cm}{$=$}
    \hspace{0.5cm}
    \includegraphics[height=5cm]{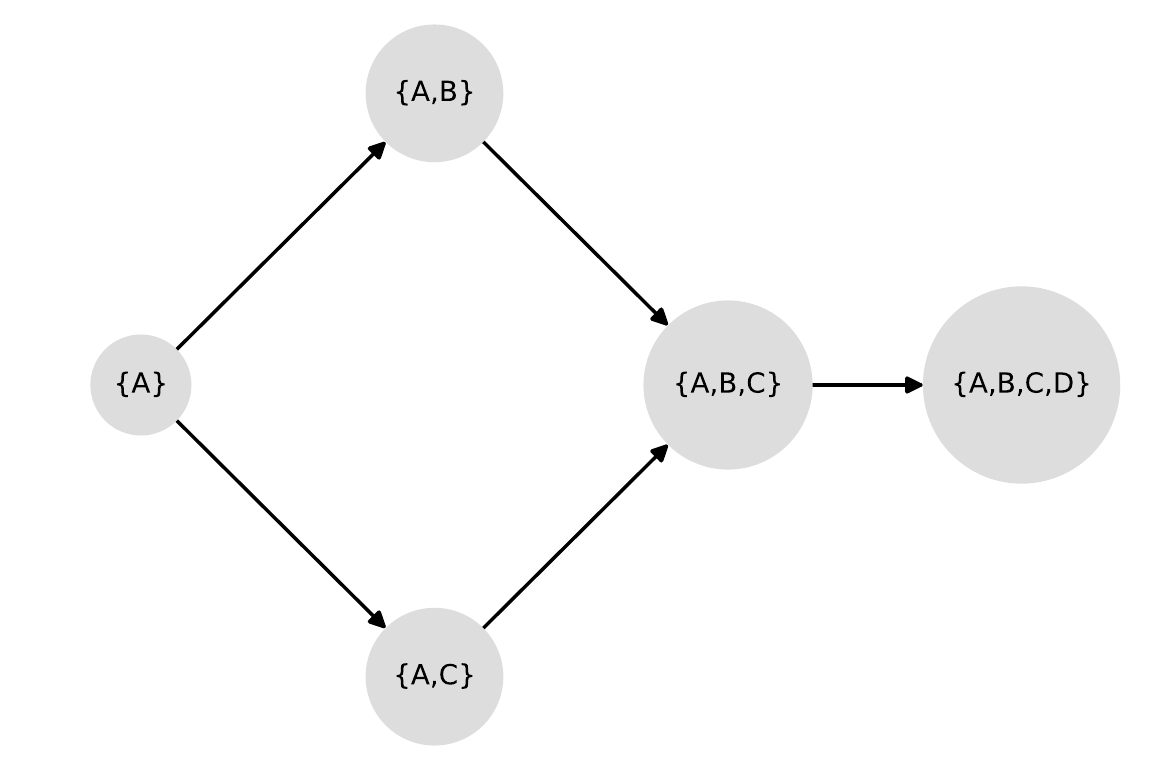}
\end{center}
Here, we note for the first time how lowersets are more general than downsets: we have $\downset{\ev{A}}=\{\ev{A}\}$, $\downset{\ev{B}}=\{\ev{A},\ev{B}\}$, $\downset{\ev{C}}=\{\ev{A},\ev{C}\}$ and $\downset{\ev{D}}=\{\ev{A},\ev{B},\ev{C},\ev{D}\}$, but lowerset $\{\ev{A},\ev{B},\ev{C}\}$ does not originate from any individual event.
Hence, lowersets strictly generalise the notion of causal past from individual events to arbitrary subsets of events:
\[
\{\ev{A},\ev{B},\ev{C}\}
=
\downset{\ev{B}} \cup \downset{\ev{C}}
=
\downset{\{\ev{B},\ev{C}\}}
\]
When the causal order is indefinite, lowersets cannot split causal equivalence classes: either no event from the class is in the lowerset, or all events are.
We can see this in the lattice of lowersets for the indefinite causal order $\ev{A}\rightarrow\{\ev{B},\ev{C}\}\rightarrow\ev{D}$, where events $\{\ev{B},\ev{C}\}$ form a causal equivalence class.
\begin{center}
    \raisebox{1.40cm}{$\LsetsSym$}
    \raisebox{1.40cm}{$\left(\rule{0cm}{1.35cm}\right.$}
    \hspace{0.0cm}
    \raisebox{0.25cm}{
        \includegraphics[height=2.5cm]{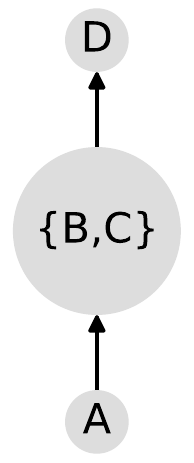}
    }
    \hspace{0.0cm}
    \raisebox{1.40cm}{$\left.\rule{0cm}{1.35cm}\right)$}
    \hspace{0.75cm}
    \raisebox{1.40cm}{$=$}
    \hspace{0.5cm}
    \includegraphics[height=3cm]{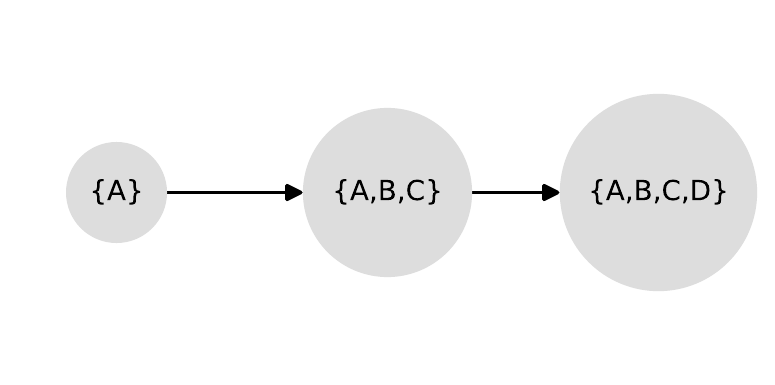}
\end{center}
The hierarchy of causal orders is contravariantly related to the hierarchy formed by the corresponding lowersets under inclusion:
\begin{equation*}
    \Omega \leq \Omega'
    \;\Leftrightarrow\;
    \Lsets{\Omega}\supseteq\Lsets{\Omega'}
\end{equation*}
The lowersets of a join of causal orders are exactly the lowersets common to both, while the lowersets of a meet of causal orders contain at least the lowersets of the two orders:
\begin{equation*}
\begin{array}{rcl}
    \Lsets{\Omega}\cap\Lsets{\Omega'} &=& \Lsets{\Omega \vee \Omega'}\\
    \Lsets{\Omega}\cup\Lsets{\Omega'} &\subseteq& \Lsets{\Omega \wedge \Omega'}
\end{array}
\end{equation*}
The inclusion above for the meet of causal orders cannot be strengthened to an equality: in general, the union of lowersets on the left hand side is not a lattice.

\section{Spaces of input histories}
\label{section:spaces}

In this Section we recap notions about spaces of input histories that form the basis for the topological framework presented in the next Section.
For a full discussion, including proofs of all results, we refer the reader to Section 3 of the companion work ``The Combinatorics of Causality'' \cite{gogioso2022combinatorics}.
For applications of spaces of input histories, we refer the reader to the comapnion works ``The Topology of Causality'' \cite{gogioso2022topology} and ``The Geometry of Causality'' \cite{gogioso2022geometry}

This work is concerned with the causal structure of a certain class of experiments or protocols, where events correspond to the local operation of black-box devices.
At each event, an input to the device is freely chosen from a finite input set, in response to which the device produces (probabilistically) an output in a finite output set.
The ensuing probability distribution on joint outputs for all devices, conditional on joint inputs for all devices, forms the basis of our causal analysis.
In the most general case, no causal constraints are given on the events.

When we say that the devices are operated locally at each event, we mean that no information about the other events is explicitly used in the operation: every dependence on the inputs and outputs at other events must be entirely mediated by the causal structure.
If event \ev{A} causally succeeds event \ev{B}, for example, then the output at \ev{A} is allowed depend on the input and output at \ev{B}: the devices being operated at the two events are black-box, and it is causally possible for one of them to signal the other.
However, the input at \ev{A} is still freely chosen, regardless of what happened at \ev{B}, and the input/output sets for the device at \ev{A} are fixed beforehand.
In the absence of causal constraints, it is therefore possible for the output at each event to arbitrarily depend on inputs at all events, and for the outputs at any set of events to be correlated.
As a consequence, the only conditional probability distribution that is well-defined in general is one on joint outputs for all events, conditional on joint inputs for all events.

\subsection{Partial Functions}
\label{subsection:spaces-pfuns}

Given a family $\underline{Y} = (Y_x)_{x \in X}$ of sets, the \emph{partial functions} $\PFun{\underline{Y}}$ on $\underline{Y}$ are defined to be all possible functions $f$ having subsets $D \subseteq X$ as their domain $\dom{f} := D$ and such that $f(x) \in Y_x$ for all $x \in D$.
\begin{equation*}
    \PFun{\underline{Y}}
    :=
    \bigcup_{D \subseteq X}
    \prod_{x \in D}
    Y_x
\end{equation*}
Partial functions are partially ordered by restriction:
\begin{equation*}
    f \leq g
    \hspace{2mm}\stackrel{def}{\Leftrightarrow}\hspace{2mm}
    \dom{f} \subseteq \dom{g}
    \text{ and }
    \restrict{g}{\dom{f}} = f
\end{equation*}
We observe that, under their restriction order, partial functions form a lower semilattice, with the empty function $\emptyset$ as its minimum and meets given by:
\begin{equation*}
\begin{array}{rcl}
    \dom{f \wedge g}
    &=&
    \suchthat{x \in \dom{f}\cap\dom{g}}{f(x) = g(x)}
    \\
    f \wedge g
    &=&
    \restrict{f}{\dom{f \wedge g}}
    =
    \restrict{g}{\dom{f \wedge g}}
\end{array}
\end{equation*}
We say that $f$ and $g$ are \emph{compatible} when the inclusion above is an equality:
\begin{equation*}
    \text{$f$ and $g$ compatible}
    \hspace{2mm} \Leftrightarrow \hspace{2mm}
    \dom{f \wedge g}
    =
    \dom{f} \cap \dom{g}
\end{equation*}
More generally, we say that a set $\mathcal{F} \subseteq \PFun{\underline{Y}}$ of partial functions is \emph{compatible} if $f$ and $g$ are compatible for all $f, g \in \mathcal{F}$.
The \emph{join} of a set $\mathcal{F}$ of partial functions exists exactly when the set is compatible, in which case it is given by:
\begin{equation*}
\label{eq:definition:join}
\begin{array}{rcl}
    \dom{\bigvee \mathcal{F}}
    &=&
    \bigcup\limits_{f \in \mathcal{F}} \dom{f}
    \\
    \bigvee \mathcal{F}
    &=&
    x \mapsto f(x) \text{ for any $f$ such that } x \in \dom{f}
\end{array}
\end{equation*}
The \emph{compatible joins} in a set $\mathcal{F}'$ of partial functions are all possible joins $\bigvee\mathcal{F}$ of compatible subsets $\mathcal{F} \subseteq \mathcal{F}'$.

\subsection{Input Histories for Causal Orders}
\label{subsection:spaces-order-induced}

Consider a causal order $\Omega$ and an associated family of input sets $\underline{I}$.
Because of causality, the output at an event $\xi$ can only depend on choices of inputs for the events $\omega$ in $\downset{\xi}$, the causal past of $\xi$. This observation motivates the following definition.

\begin{definition}
The \emph{input histories} for a given choice of order $\Omega$ and inputs $\underline{I} = (I_\omega)_{\omega \in \Omega}$ are defined to be the partial functions in the following set:
\begin{equation*}
    \Hist{\Omega, \underline{I}}
    \;:=\;
    \bigcup_{\xi \in \Omega}
    \prod_{\omega \in \downset{\xi}}
    I_\omega
    \;\;\subseteq\;\;
    \PFun{\underline{I}}    
\end{equation*}
Input histories inherit the restriction order of partial functions, and we refer to the partially ordered set $\Hist{\Omega, \underline{I}}$ as a \emph{space of input histories}.
\end{definition}

As a simple concrete example, let $\Omega$ be the total order on 3 events and consider its associated lattice of lowersets $\Lsets{\Omega}$.
In the lattice, the causal pasts of individual events have been colour-coded.
\begin{center}
    \raisebox{1.15cm}{$\LsetsSym$}
    \raisebox{1.15cm}{$\left(\rule{0cm}{1.35cm}\right.$}
    \hspace{-0.3cm}
    \raisebox{0cm}{
        \includegraphics[height=2.5cm]{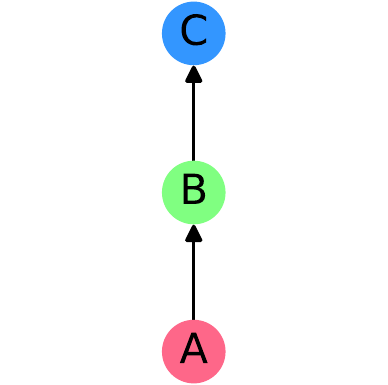}
    }
    \hspace{-0.3cm}
    \raisebox{1.15cm}{$\left.\rule{0cm}{1.35cm}\right)$}
    \hspace{0.5cm}
    \raisebox{1.15cm}{$=$}
    \hspace{0.25cm}
    \includegraphics[height=2.5cm]{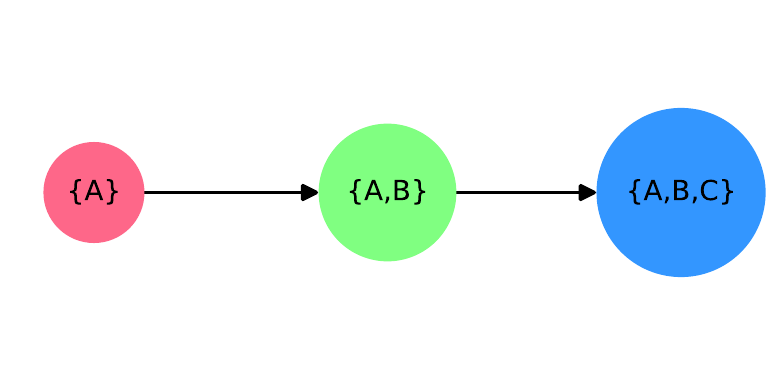}
\end{center}
Below is the Hasse diagram for the space of input histories, consisting of all binary functions on $\evset{A}$, $\evset{A,B}$ and $\evset{A,B,C}$.
\begin{center}
    \includegraphics[height=3cm]{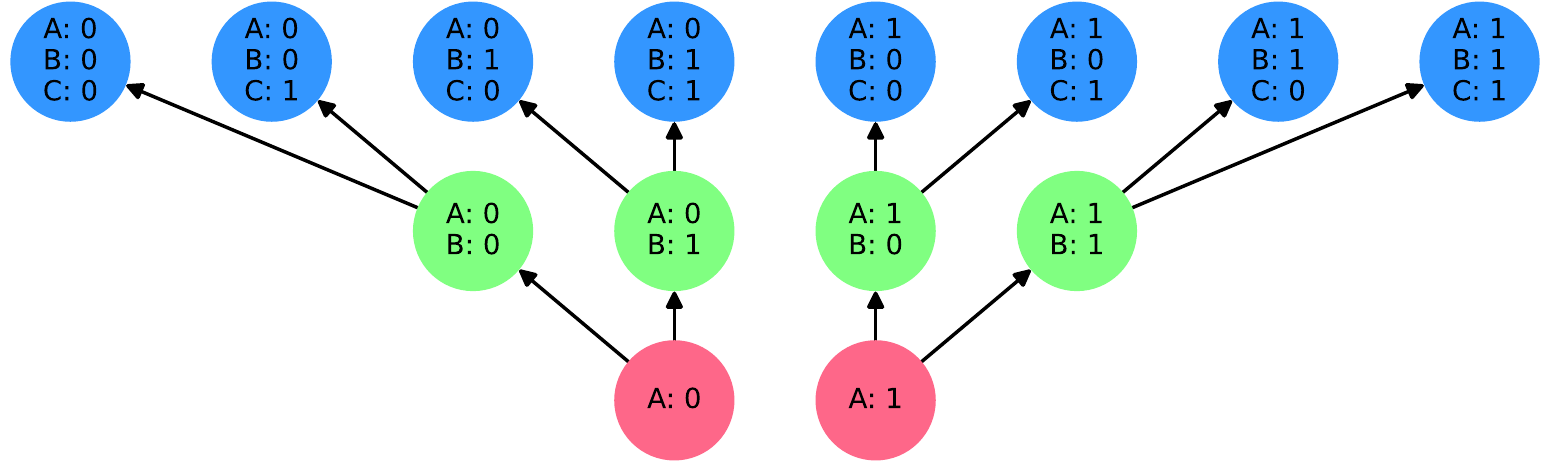}
\end{center}

Input histories are not generally closed under meets, and the only subsets closed under joins are chains (in which case the join is the maximum).
When talking about the meet of two (or more) input histories, we will always mean the meet in $\PFun{\underline{I}}$; similarly, when saying that a subset $\mathcal{F} \subseteq \Hist{\Omega, \underline{I}}$ of input histories is compatible, we will always mean that it is compatible in $\PFun{\underline{I}}$.

\begin{definition}
The \emph{extended input histories} for a given choice of order $\Omega$ and inputs $\underline{I} = (I_\omega)_{\omega \in \Omega}$ are defined to be the partial functions in the following set:
\begin{equation*}
    \ExtHist{\Omega, \underline{I}}
    \;:=\;
    \bigcup_{U \in \Lsets{\Omega}}
    \prod_{\omega \in U}
    I_\omega
    \;\;\subseteq\;\;
    \PFun{\underline{I}}    
\end{equation*}
Extended input histories inherit the restriction order of partial functions, and we refer to the partially ordered set $\ExtHist{\Omega, \underline{I}}$ as a \emph{space of extended input histories}.
\end{definition}

The spaces of (extended) input histories derived from causal orders work quite well when the orders are definite, but they do not quite capture the full desired gamut of possibilities for indefinite causal orders.
Indeed, consider the following indefinite causal order $\Omega$ on 3 events, and its associated lattice of lowersets $\Lsets{\Omega}$.
\begin{center}
    \raisebox{1.15cm}{$\LsetsSym$}
    \raisebox{1.15cm}{$\left(\rule{0cm}{1.35cm}\right.$}
    \hspace{-0.4cm}
    \raisebox{0cm}{
        \includegraphics[height=2.5cm]{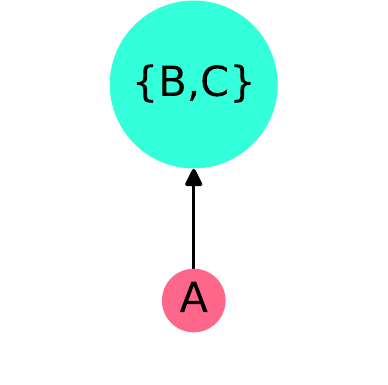}
    }
    \hspace{-0.5cm}
    \raisebox{1.15cm}{$\left.\rule{0cm}{1.35cm}\right)$}
    \hspace{0.5cm}
    \raisebox{1.15cm}{$=$}
    \hspace{0.25cm}
    \raisebox{0.0cm}{
        \includegraphics[height=2.5cm]{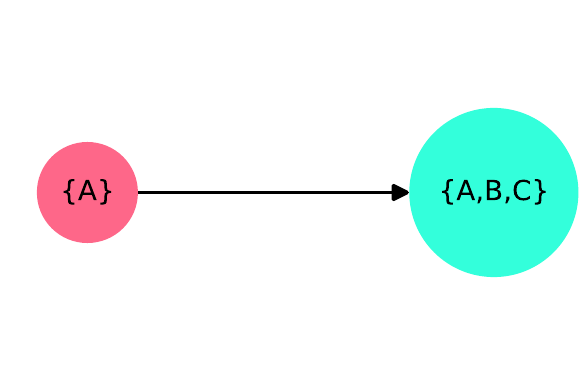}
    }
\end{center}
Because events \ev{B} and \ev{C} are in indefinite causal order, they have the same causal past, and hence they are never separated by input histories.
\begin{center}
    \includegraphics[height=2.25cm]{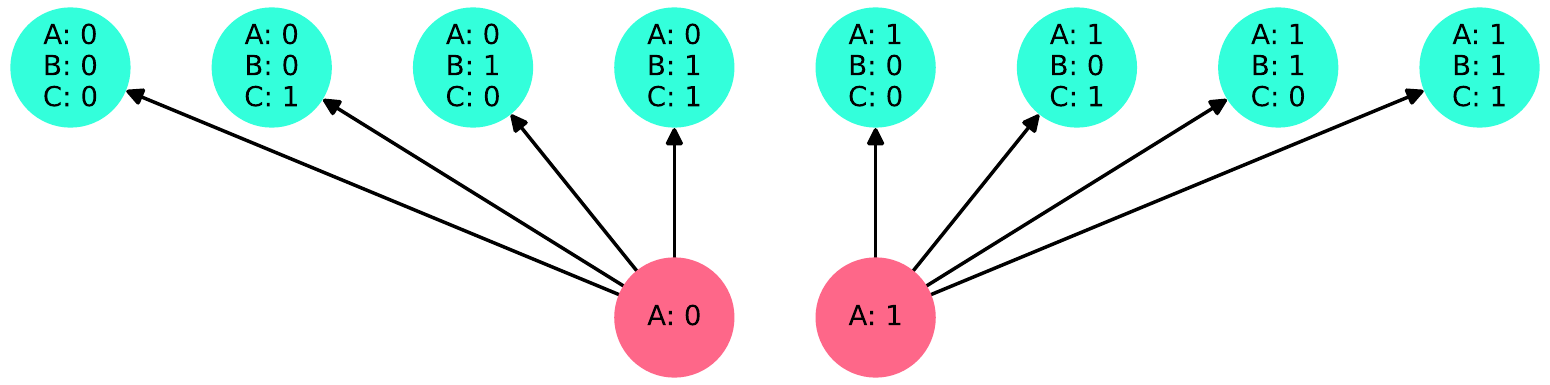}
\end{center}

\subsection{Spaces of Input Histories}
\label{subsection:spaces-definition}

\begin{definition}
A subset $\Theta \subseteq \PFun{\underline{I}}$ is said to be \emph{$\vee$-prime} (read ``join-prime'') if no $h \in \Theta$ can be written as the compatible join $h = \bigvee \mathcal{F}$ of a subset $\mathcal{F} \subseteq \Theta$ such that $h \notin \mathcal{F}$:
\[
\left(
    \mathcal{F} \subseteq \Theta \text{ compatible and }
    \bigvee \mathcal{F} \in \Theta
\right)
\Rightarrow \bigvee \mathcal{F} \in \mathcal{F}
\]
Dually, a subset $W \subseteq \PFun{\underline{I}}$ is said to be \emph{$\vee$-closed} (read ``join-closed'') if for every pair of compatible $h, k \in W$ the join $h \vee k$ is itself in $W$.
This implies that, more generally:
\[
\mathcal{F} \subseteq \Theta \text{ compatible }
\Rightarrow
\bigvee \mathcal{F} \in \Theta
\]
\end{definition}

\begin{definition}
A \emph{space of input histories} is a finite set $\Theta$ of partial functions which is $\vee$-prime.
The associated \emph{event set} $\Events{\Theta}$ and family of \emph{input sets} $\underline{\Inputs{\Theta}} = (\Inputs{\Theta}_\omega)_{\omega \in \Events{\Theta}}$ are defined as follows:
\begin{equation*}
    \begin{array}{lcl}
        \Events{\Theta}
        &:=&
        \bigcup_{h \in \Theta} \dom{h}
        \\
        \Inputs{\Theta}_\omega
        &:=&
        \suchthat{h_\omega}{h \in \Theta, \omega \in \dom{h}}
    \end{array}
\end{equation*}
We have $\Theta \subseteq \PFun{\underline{\Inputs{\Theta}}}$ and the space $\Theta$ is equipped with the partial order inherited from $\PFun{\underline{\Inputs{\Theta}}}$.
The \emph{space of extended input histories} $\Ext{\Theta}$ associated to $\Theta$ is its $\vee$-closure:
\begin{equation*}
    \Ext{\Theta}
    :=
    \suchthat{\bigvee \mathcal{F}}{\mathcal{F} \subseteq \Theta \text{ compatible}}
\end{equation*}
We have $\Ext{\Theta} \subseteq \PFun{\underline{\Inputs{\Theta}}}$.
The space $\Ext{\Theta}$ is equipped with the partial order inherited from $\PFun{\underline{\Inputs{\Theta}}}$.
\end{definition}

We observe that, given any subset $W \subseteq \PFun{\underline{I}}$, we can obtain a space of input histories by taking its $\vee$-prime elements:
\begin{equation*}
    \Prime{W}
    :=
    \suchthat{w \in W}{
    \forall \mathcal{F} \subseteq W \text{ compatible}.\,
    w = \bigvee\mathcal{F} \Rightarrow w \in \mathcal{F}
    }
\end{equation*}
Below is an example of a space of input histories $\Theta$ (on the left) together with its corresponding space of extended input histories $\Ext{\Theta}$ (on the right).
Input histories have been colour-coded by the events whose output they refer to (more on this later), in both diagrams: grey coloured extended input histories on the right are those which are not input histories (i.e. they arise by join).
This space is a variation on the total order $\total{\ev{A}, \ev{B}, \ev{C}}$, where input 0 at event $\ev{B}$ causally disconnects $\ev{B}$ from $\ev{A}$ and input 0 at event $\ev{C}$ causally disconnects $\ev{C}$ from $\ev{B}$ and $\ev{A}$.
\begin{center}
    \begin{tabular}{cc}
    \includegraphics[height=3.5cm]{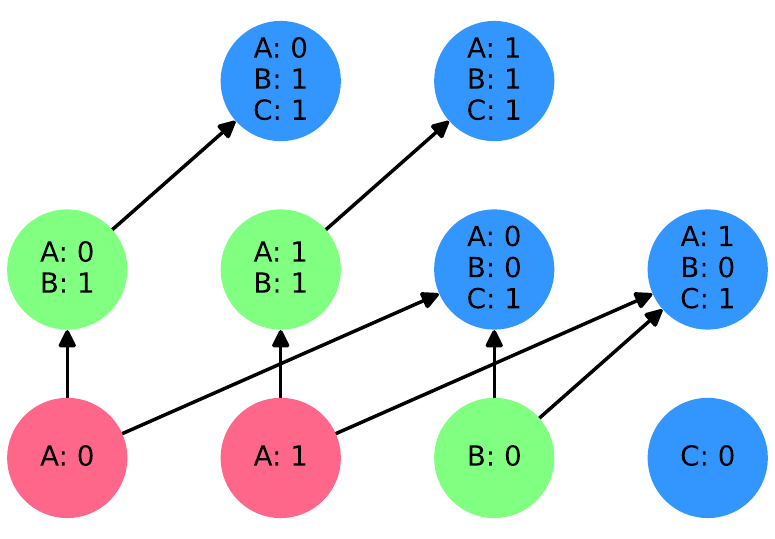}
    &
    \includegraphics[height=3.5cm]{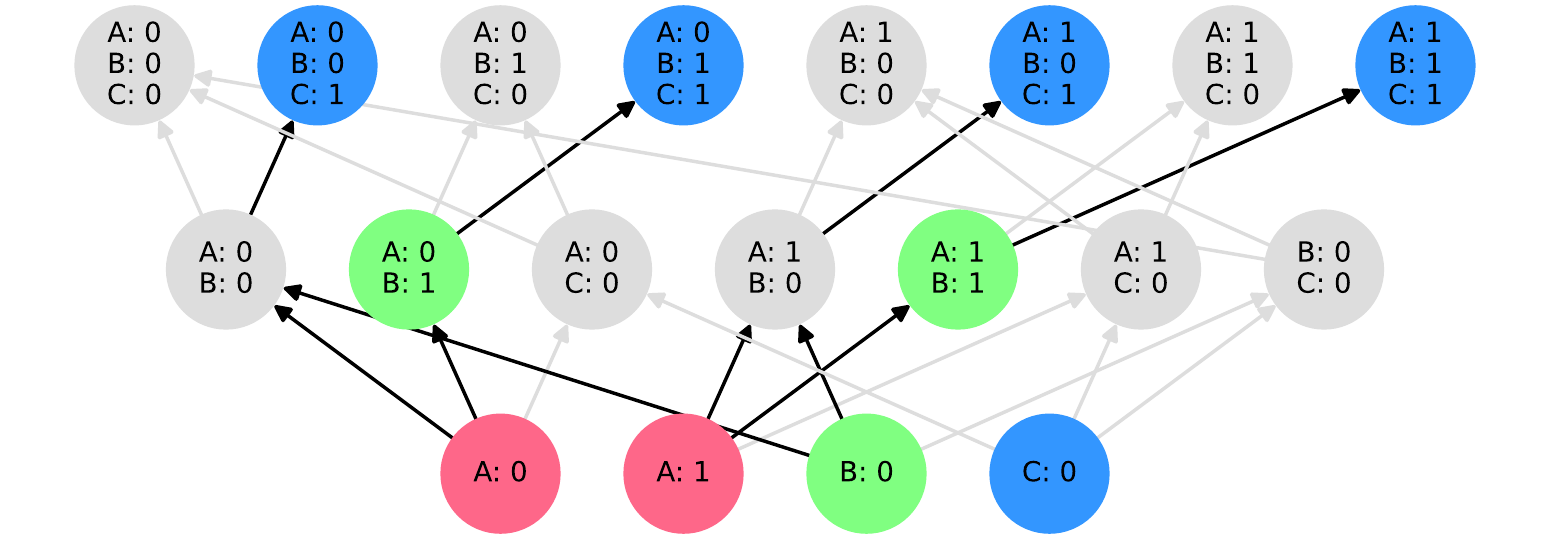}
    \\
    $\Theta$
    &
    $\Ext{\Theta}$
    \end{tabular}
\end{center}

In the introduction to this Section we stated that inputs at events are ``freely chosen'', i.e. without any local or global constraint.
By stitching together input histories, it must therefore be possible to obtain all possible combinations of joint input values over all events.

\begin{definition}
A space of input histories $\Theta$ is said to satisfy the \emph{free-choice condition} if:
\[
    \max\Ext{\Theta} = \prod_{\omega \in \Events{\Theta}} \Inputs{\Theta}_\omega
\]
In spaces satisfying the free-choice condition, we refer to the histories in $\prod_{\omega \in \Events{\Theta}} \Inputs{\Theta}_\omega$ as the \emph{maximal extended input histories}.
\end{definition}

Recall now that causal orders form a hierarchy (a lattice) when ordered by inclusion.
We would like this hierarchy to generalise from causal orders $\Omega$ to their spaces of input histories $\Hist{\Omega, \underline{I}}$, and then to all spaces of input histories, including ones that don't arise from orders.
Unfortunately, this is not as simple as ordering the spaces themselves by inclusion: $\Omega \leq \Xi$ does not in general imply an inclusion relationship between $\Hist{\Omega, \underline{I}}$ and $\Hist{\Xi, \underline{I}}$.
However, a suitable statement of inclusion holds for the corresponding spaces of extended input histories (cf. Proposition 3.8 p.28 \cite{gogioso2022combinatorics}):
\begin{equation*}
    \Omega \leq \Xi
    \;\;\Leftrightarrow\;\;
    \ExtHist{\Omega, \underline{I}} \supseteq \ExtHist{\Xi, \underline{I}}    
\end{equation*}

\begin{definition}
\label{definition:spaces-order}
We define the following partial order on spaces of input histories:
\begin{equation*}
    \Theta' \leq \Theta
    \stackrel{def}{\Leftrightarrow}
    \Ext{\Theta'} \supseteq \Ext{\Theta}
\end{equation*}
We say that $\Theta'$ is a \emph{causal refinement} of $\Theta$ (more causal constraints), or that $\Theta$ is a \emph{causal coarsening} of $\Theta'$ (fewer causal constraints).
Equivalently, sometimes we say that $\Theta'$ is a \emph{sub-space} of $\Theta$ or that $\Theta'$ is a \emph{super-space} of $\Theta$.
\end{definition}

Note that Definition \ref{definition:spaces-order} allows us to compare spaces of input histories with different underlying sets of events and inputs.
We refer to the partial order, or ``hierarchy'', of all spaces of input histories simply as $\AllSpaces$.
Spaces $\Theta$ with $\underline{\Inputs{\Theta}} = \underline{I}$ for a specific choice $\underline{I} = (I_\omega)_{\omega \in E}$ form a sub-hierarchy:
\[
    \Spaces{\underline{I}}
    \hookrightarrow
    \AllSpaces
\]
Spaces satisfying the free-choice condition form a further sub-hierarchy $\SpacesFC{\underline{I}}$:
\[
    \SpacesFC{\underline{I}}
    \hookrightarrow\Spaces{\underline{I}}
    \hookrightarrow \AllSpaces
\]
It is a fact (cf. Proposition 3.10 p.30 \cite{gogioso2022combinatorics}) the three ``hierarchies'' above are lattices, sharing the same notion of join and meet:
\begin{equation*}
\begin{array}{rcl}
    \Theta \vee \Theta' &=& \Prime{\Ext{\Theta} \cap \Ext{\Theta'}}\\
    \Theta \wedge \Theta' &=& \Prime{\Ext{\Theta} \cup \Ext{\Theta'}}
\end{array}
\end{equation*}
We refer to the join of two spaces as their \emph{closest common coarsening}, and to their meet as their \emph{closest common refinement}.

\begin{definition}
Let $\Theta, \Theta'$ be spaces of input histories with $\Events{\Theta} \cap \Events{\Theta'} = \emptyset$.
The \emph{parallel composition} of $\Theta$ and $\Theta'$ is defined to be their union as sets:
\begin{equation*}
    \Theta \cup \Theta'
\end{equation*}
The \emph{sequential composition} of $\Theta$ before $\Theta'$ is defined as follows:
\begin{equation*}
    \Theta \seqcomposeSym \Theta'
    :=
    \Theta \cup \left(\max{\Ext{\Theta}}\allJoinsSym\Theta'\right)
\end{equation*}
where we adopted the symbol $\allJoinsSym$ to indicate all possible compatible joins between two sets (or families) of partial functions:
\begin{equation*}
    \max{\Ext{\Theta}}\allJoinsSym\Theta'
    :=
    \suchthat{k\vee h'}{k \in \max{\Ext{\Theta}}, h' \in \Theta'}
\end{equation*}
Note: because $\Events{\Theta} \cap \Events{\Theta'} = \emptyset$, all joins above are necessarily compatible.
\end{definition}

As simple examples of sequential and parallel composition of spaces, we consider the spaces $\Theta$ and $\Theta'$ induced by the discrete order $\discrete{A,B}$ and total order $\total{C,D}$ respectively.
The two spaces are depicted below.
\begin{center}
    \begin{tabular}{cc}
    \includegraphics[height=2.5cm]{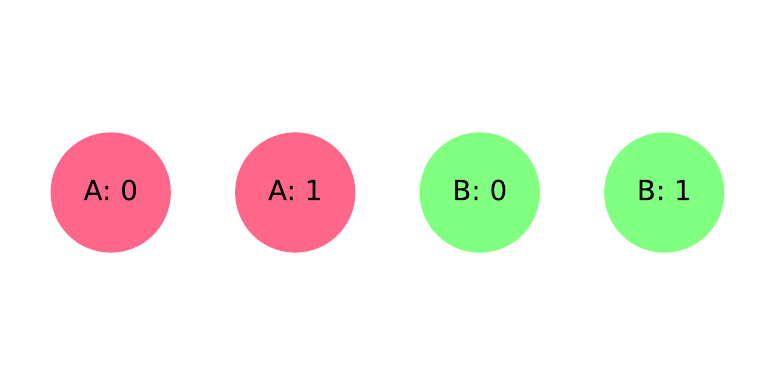}
    &
    \includegraphics[height=2.5cm]{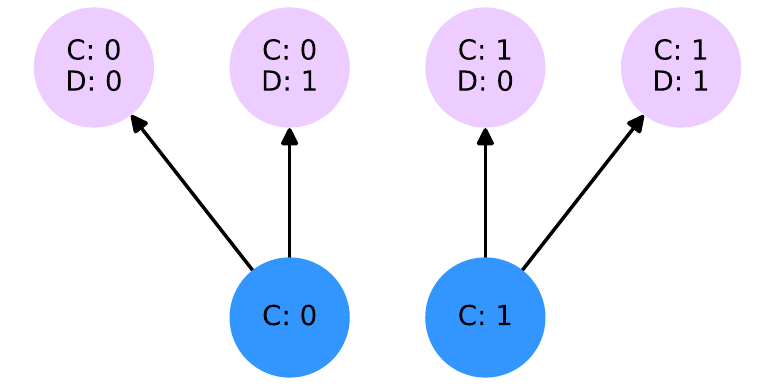}
    \\
    $\Theta$
    &
    $\Theta'$
    \end{tabular}
\end{center}
The parallel composition $\Theta \cup \Theta'$ of the two spaces is simply the disjoint union of their histories, with no additional causal relationship between them. 
\begin{center}
    \begin{tabular}{c}
    \includegraphics[height=2.5cm]{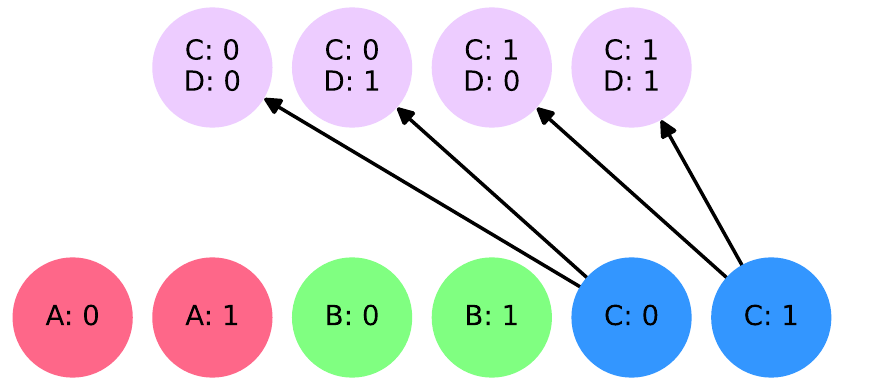}
    \\
    $\Theta \cup \Theta'$
    \end{tabular}
\end{center}
The sequential composition $\Theta \seqcomposeSym \Theta'$ of the two spaces consists of a copy of $\Theta'$ appearing after each maximal extended input history of $\Ext{\Theta}$, for a total of four copies.
\begin{center}
    \begin{tabular}{c}
    \includegraphics[height=5cm]{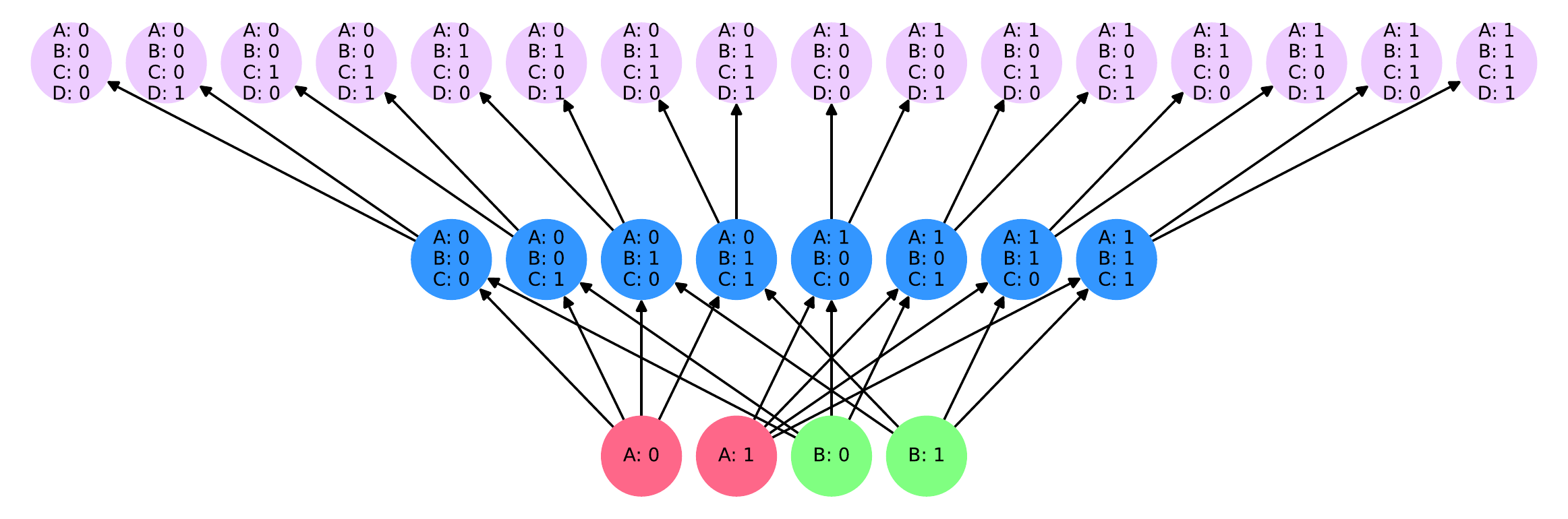}
    \\
    $\Theta \seqcomposeSym \Theta'$
    \end{tabular}
\end{center}

\begin{definition}
Let $\Theta$ be a space of input histories and let $\underline{\Theta'} := (\Theta'_k)_{k \in \max{\Ext{\Theta}}}$ be a family of spaces of input histories such that $\Events{\Theta} \cap \Events{\Theta'_k} = \emptyset$ for all $k \in \max{\Ext{\Theta}}$.
The \emph{conditional sequential composition} of $\Theta$ and $\underline{\Theta'}$ is defined as follows:
\begin{equation*}
    \Theta \seqcomposeSym \underline{\Theta'}
    :=
    \Theta \cup \suchthat{k\vee h'}{k \in \max{\Ext{\Theta}}, h' \in \Theta'_k}
\end{equation*}
Sequential composition $\Theta \seqcomposeSym \Theta'$ arises as the special case of conditional sequential composition where $\Theta'_k = \Theta'$ for all $k \in \max\Ext{\Theta}$.
\end{definition}

As a simple example of conditional sequential composition, we compose the space induced by the discrete order on one event $\ev{A}$ (having \hist{A/0} and \hist{A/1} as its maximal extended input histories) with the spaces induced by the two total orders on two events $\ev{B}$ and $\ev{C}$:
\begin{center}
    \begin{tabular}{ccc}
    \includegraphics[height=2.5cm]{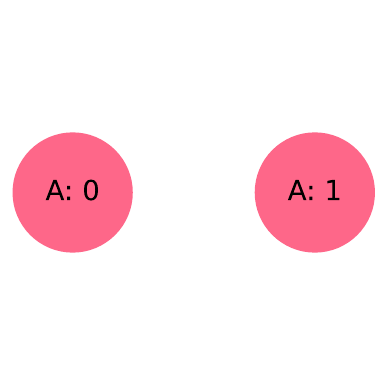}
    &
    \includegraphics[height=2.5cm]{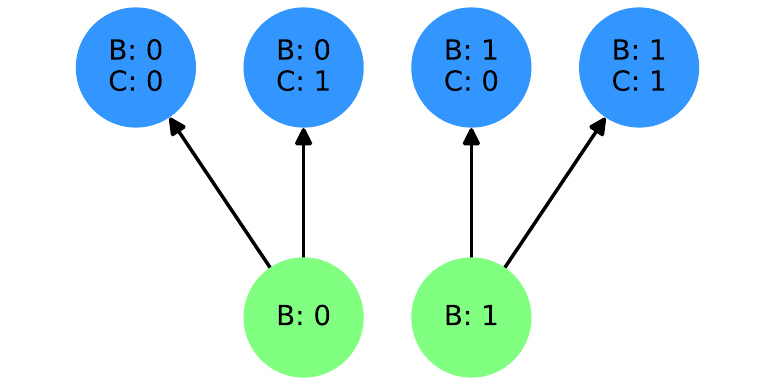}
    &
    \includegraphics[height=2.5cm]{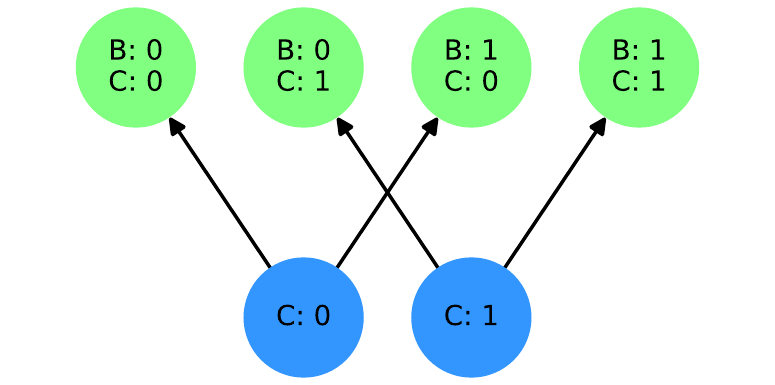}
    \\
    $\Theta$
    &
    $\Theta'_{\hist{A/0}}$
    &
    $\Theta'_{\hist{A/1}}$
    \end{tabular}
\end{center}
The result of this conditional sequential composition is a ``3-party causal switch'' space, in which event \ev{A} controls the order of events \ev{B} and \ev{C} with its input, e.g. by setting $\ev{B} < \ev{C}$ when the input is 0 and $\ev{C} < \ev{B}$ when the input is 1.
In such a setting, the output at \ev{B} is fully determined by the inputs at events \ev{A} and \ev{B} when the input at \ev{A} is 0, but the input at event \ev{C} is also needed---in the general case---when the input at \ev{A} is 1.
\begin{center}
    \begin{tabular}{c}
    \includegraphics[height=3.5cm]{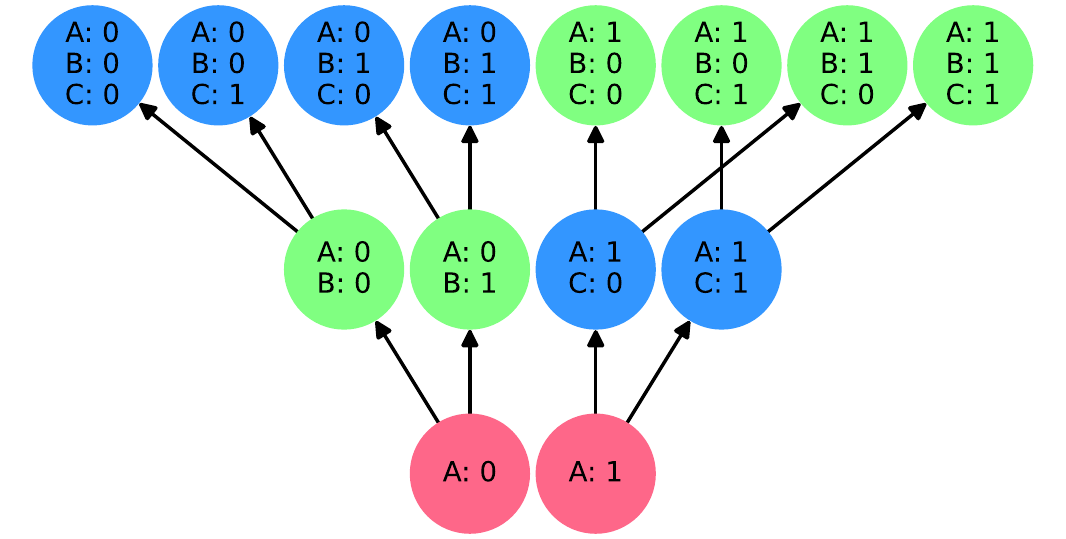}
    \\
    $\Theta \seqcomposeSym \underline{\Theta'}$
    \end{tabular}
\end{center}

\subsection{Causally Complete Spaces}
\label{subsection:spaces-cc}

In our operational interpretation, input histories are the data upon which the output values at individual events are allowed to depend.
When the causal order is given, it is always clear which histories refer to which outputs: the output at event $\omega$ is determined by the input histories $h$ with domain $\dom{h} = \downset{\omega}$.
In the more general setting of spaces of input histories, however, a causal order might not be given: in the absence of a $\downset{\omega}$, how do we associate events to the input histories that determine their outputs?
To get ourselves started, we consider the example of the causal diamond $\Omega$.
\begin{center}
    \includegraphics[height=2.5cm]{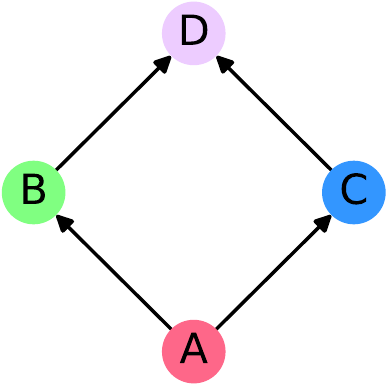}
\end{center}
Looking at the space of input histories $\Hist{\Omega,\{0,1\}}$---a shorthand by which we mean $\Hist{\Omega,(\{0,1\})_{\omega \in \Omega}}$---we observe that an association between input histories and events can be made from the order of histories alone.
Indeed, if $h$ is a history with $\dom{h} = \downset{\omega}$, then we can look at all input histories $k < h$ strictly below it and recover $\omega$ as the only event in $\dom{h}\backslash\bigcup_{k < h}\dom{k}$: this is the only event not covered by the domains of the histories strictly below $h$, which we will refer to as a ``tip event''.
In the Hasse diagram below, we have color-coded input histories according to the tip event associated to them by this procedure.
\begin{center}
    \includegraphics[height=4cm]{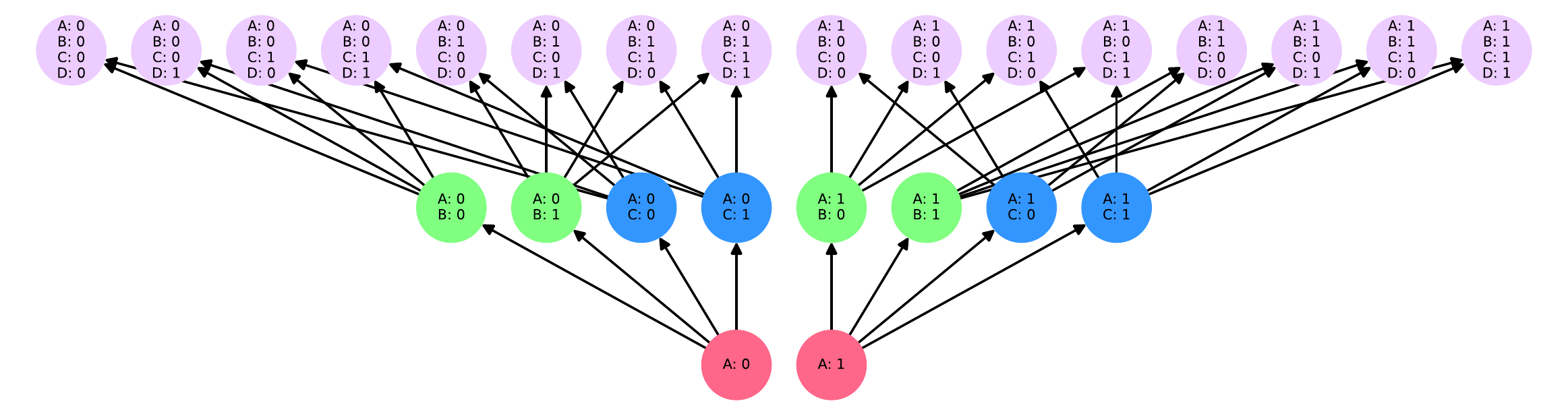}
\end{center}
The procedure works well for definite causal orders, but something goes wrong for indefinite ones: if two or more events are in indefinite causal order, they will appear together at the tip of histories.
Indeed, consider the following indefinite version of the diamond order above: the space $\total{\ev{A}, \evset{B,C}, \ev{D}}$, where the events \ev{B} and \ev{C} are in indefinite causal order rather than causally unrelated.
\begin{center}
    \includegraphics[height=3cm]{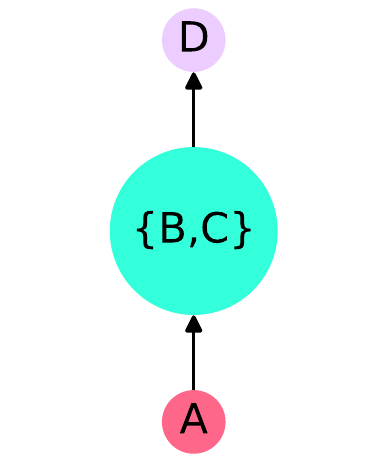}
\end{center}
Because \ev{B} and \ev{C} cannot be distinguished by input histories in the space, the histories in the middle layer now have two ``tip events'' instead of one.
\begin{center}
    \includegraphics[height=4cm]{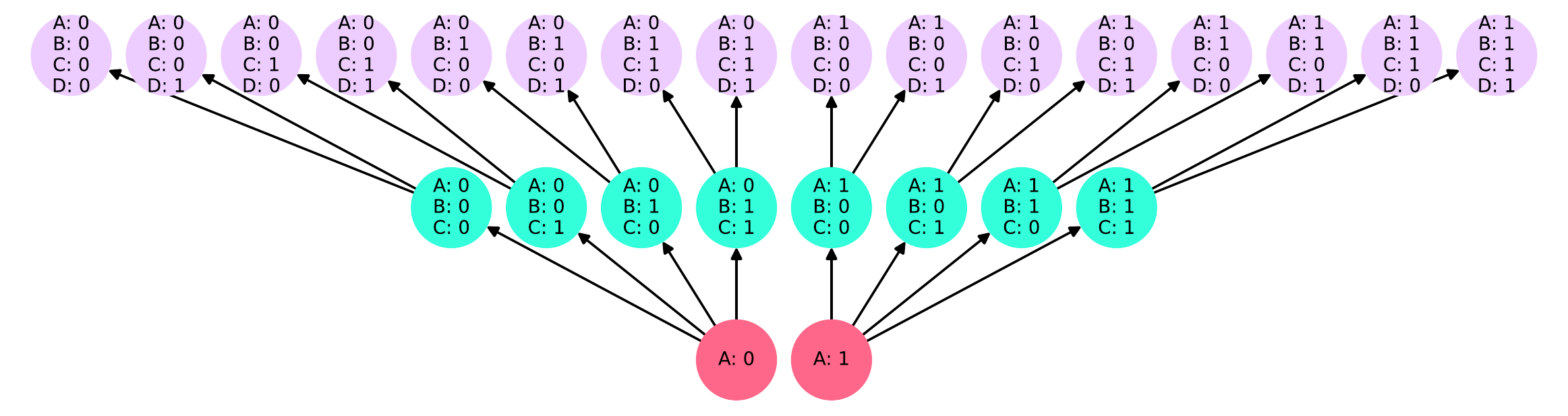}
\end{center}

The operational interpretation of multiple tip events is challenging: in a naive sense, it means that the output value at two events in indefinite causal order is to be produced ``simultaneously'', using the input values at both events.
This is problematic, because indefinite causal order should not trivialise to causal collapse: under our operational interpretation, distinct events should retain their independent local nature.
It should not, for example, be possible to perform the ``swap'' function $(b, c) \mapsto (c, b)$ on two events \ev{B} and \ev{C} in indefinite causal order: the devices would have to wait for both inputs to be given before producing their outputs, with the effect of delocalising the events.

However, there is an alternative way to look at the presence of multiple tip events, as a form of ``causal incompleteness''.
Rather than interpreting such spaces as allowing event delocalisation, we think of such spaces as not providing sufficient information for causal inference to be performed.
As such, we will focus our efforts on ``causally complete'' spaces, studying the incomplete spaces through the lens of their ``causal completions''.

\begin{definition}
Let $\Theta$ be a space of input histories.
Given an extended input history $h \in \Ext{\Theta}$, we define the \emph{tip events} of $h$ in $\Theta$ as the events which are in the domain of $h$ but not in the domain of any history strictly below it:
\begin{equation*}
    \begin{array}{rcl}
    \tips{\Theta}{h}
    &:=&
    \dom{h}\backslash\bigcup_{k < h}\dom{k}
    \\
    &=&
    \suchthat{\omega \in \dom{h}}{\forall k < h.\, \omega \notin \dom{k}}
    \end{array}
\end{equation*}
\end{definition}

\begin{definition}
Let $\Theta$ be a space of input histories satisfying the free-choice condition.
We say that $\Theta$ is \emph{causally complete} if all input histories $h \in \Theta$ have exactly one tip event, and that it is \emph{causally incomplete} otherwise.
If $\Theta$ is causally complete and $h \in \Theta$, we define the \emph{tip event} of $h$ in $\Theta$ to be the unique event in $\tips{\Theta}{h}$:
\begin{equation*}
    \Theta \text{ causally complete }
    \Leftrightarrow
    \forall h \in \Theta.\,
    \tips{\Theta}{h} = \{\tip{\Theta}{h}\}
\end{equation*}
\end{definition}

It is a fact (cf. Proposition 3.21 p.35 \cite{gogioso2022combinatorics}) that a space of input histories induced by a causal order is causally complete if and only if the causal order is causally definite.

\begin{definition}
Let $\Theta$ be a space of input histories satisfying the free-choice condition.
The \emph{causal completions} of $\Theta$ are the closest refinements of $\Theta$ which are causally complete, i.e. the maxima of the set of causally complete spaces which are causal refinements of $\Theta$:
\begin{equation*}
    \CausCompl{\Theta}
    :=
    \max
    \suchthat{\Theta' \leq \Theta}{\Theta' \text{ causally complete}}
\end{equation*}
Since the discrete space $\Hist{\discrete{E^\Theta}, \underline{I}^\Theta}$ is always causally complete, the set of causal completions of $\Theta$ is never empty. If $\Theta$ is itself causally complete, then $\CausCompl{\Theta} = \{\Theta\}$.
\end{definition}

As an example of causal completion, we refer back to the indefinite causal order $\total{\ev{A}, \evset{B,C}}$.
The associated space of input histories is causally incomplete, because \ev{B} and \ev{C} always appear together as tip events (coloured aquamarine, at the top).
\begin{center}
    \includegraphics[height=2.5cm]{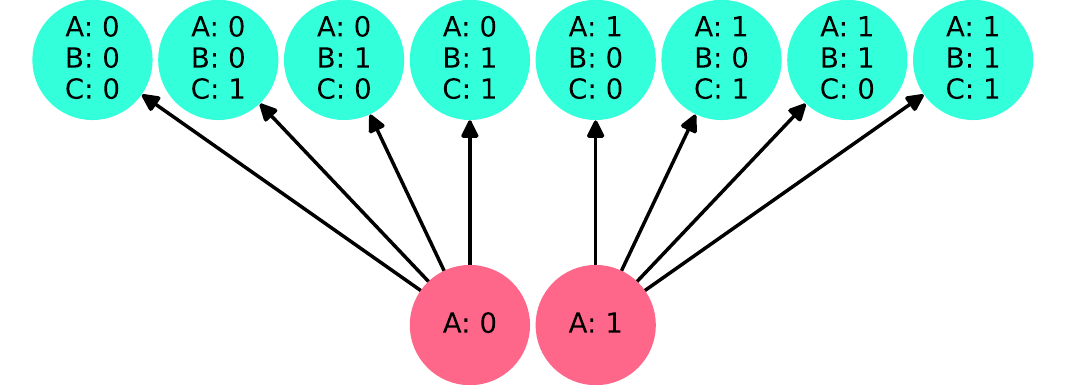}
\end{center}
There are four possible causal completions for this space.
Two of the causal completions are obtained by imposing a fixed order on events \ev{B} and \ev{C}: either \ev{B} causally precedes \ev{C} (left below) or \ev{B} causally succeeds \ev{C} (right below).
\begin{center}
\begin{tabular}{cc}
    \includegraphics[height=2.5cm]{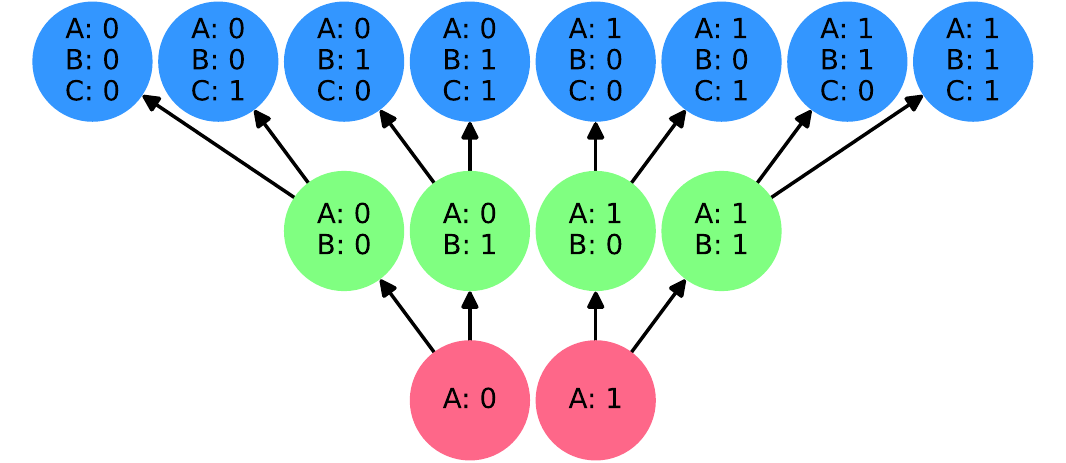}
    &
    \includegraphics[height=2.5cm]{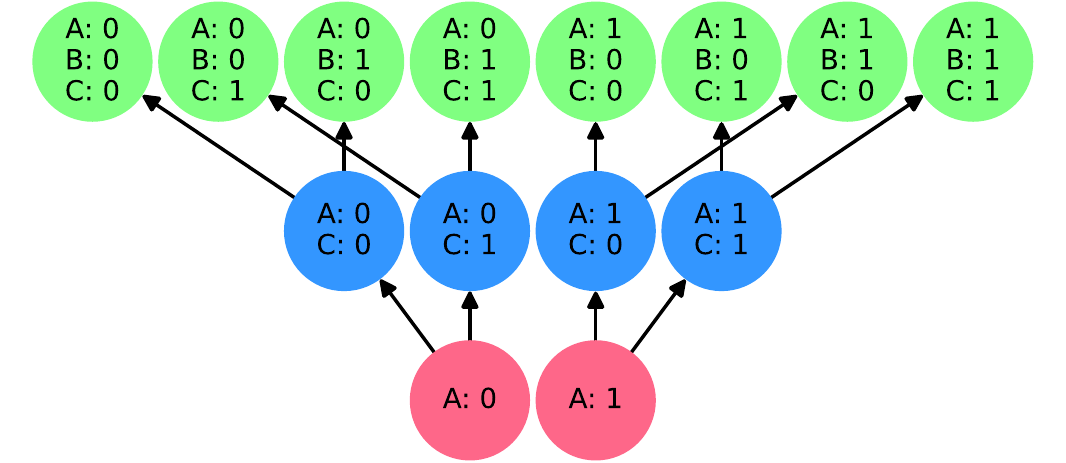}
\end{tabular}
\end{center}
The remaining two causal completions are obtained by imposing an order on events \ev{B} and \ev{C} that depends on the input at event \ev{A}: either \ev{B} causally precedes \ev{C} when the input at \ev{A} is 0 and causally succeeds \ev{C} when the input at \ev{A} is 1 (left below), or \ev{B} causally succeeds \ev{C} when the input at \ev{A} is 0 and causally precedes \ev{C} when the input at \ev{A} is 1 (right below). 
\begin{center}
\begin{tabular}{cc}
    \includegraphics[height=2.5cm]{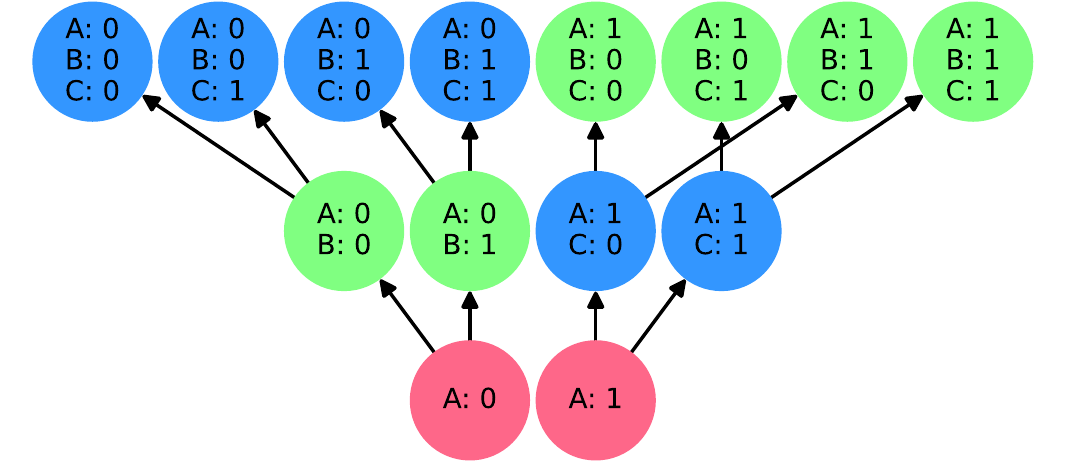}
    &
    \includegraphics[height=2.5cm]{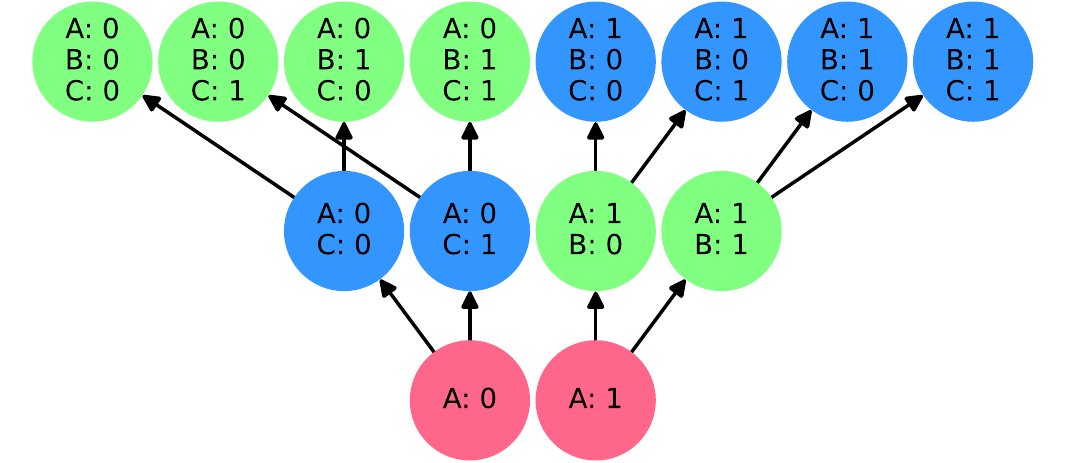}
\end{tabular}
\end{center}

It is a fact (cf. Theorem 3.26, Theorem 3.27 and Corollary 3.28 p.38 \cite{gogioso2022combinatorics}) that causal composition, sequential composition and conditional sequential composition respect causal completeness.
The parallel composition $\Theta \cup \Theta'$ of two spaces of input histories satisfying the free-choice condition has the following causal completions:
\begin{equation*}
    \CausCompl{\Theta \cup \Theta'}
    =
    \suchthat{\hat{\Theta} \cup \hat{\Theta}'\;\;}{
        \begin{array}{l}
        \hat{\Theta} \in \CausCompl{\Theta},\\
        \hat{\Theta}' \in \CausCompl{\Theta'}
        \end{array}
    }
\end{equation*}
The sequential composition $\Theta \seqcomposeSym \Theta'$ of two spaces of input histories satisfying the free-choice condition has the following causal completions:
\begin{equation*}
    \CausCompl{\Theta \seqcomposeSym \Theta'}
    =
    \suchthat{\hat{\Theta} \seqcomposeSym \hat{\Theta}'\;\;}{
        \begin{array}{l}
        \hat{\Theta} \in \CausCompl{\Theta},\\
        \hat{\Theta}' \in \CausCompl{\Theta'}
        \end{array}
    }
\end{equation*}
The conditional sequential composition $\Theta \seqcomposeSym \underline{\Theta'}$ where all spaces satisfy teh free-choice condition has the following causal completions, as long as the families of input sets $\underline{\Inputs{\Theta'_k}}$ are identical for all $k \in \max\Ext{\Theta}$:
\begin{equation*}
    \CausCompl{\Theta \seqcomposeSym \underline{\Theta'}}
    =
    \suchthat{\hat{\Theta} \seqcomposeSym \underline{\hat{\Theta}'}\;\;}{
        \begin{array}{l}
        \hat{\Theta} \in \CausCompl{\Theta},\\
        \forall k.\; \hat{\Theta}'_k \in \CausCompl{\Theta'_k}
        \end{array}
    }
\end{equation*}

\subsection{The Hierarchy of Causally Complete Spaces}
\label{subsection:spaces-hierarchy}

Causally complete spaces are the main focus of this work: for given inputs $\underline{\Inputs{\Theta}} = \underline{I}$, we denote them by $\CCSpaces{\underline{I}}$.
It is a fact that they are closed under meet but not generally under join (cf. Proposition 3.28 p.39 \cite{gogioso2022combinatorics}): we refer to the $\wedge$-semilattice $\CCSpaces{\underline{I}}$ as the \emph{hierarchy of causally complete spaces} for $\underline{I}$.

\begin{figure}[h]
    \centering
    \includegraphics[width=\textwidth]{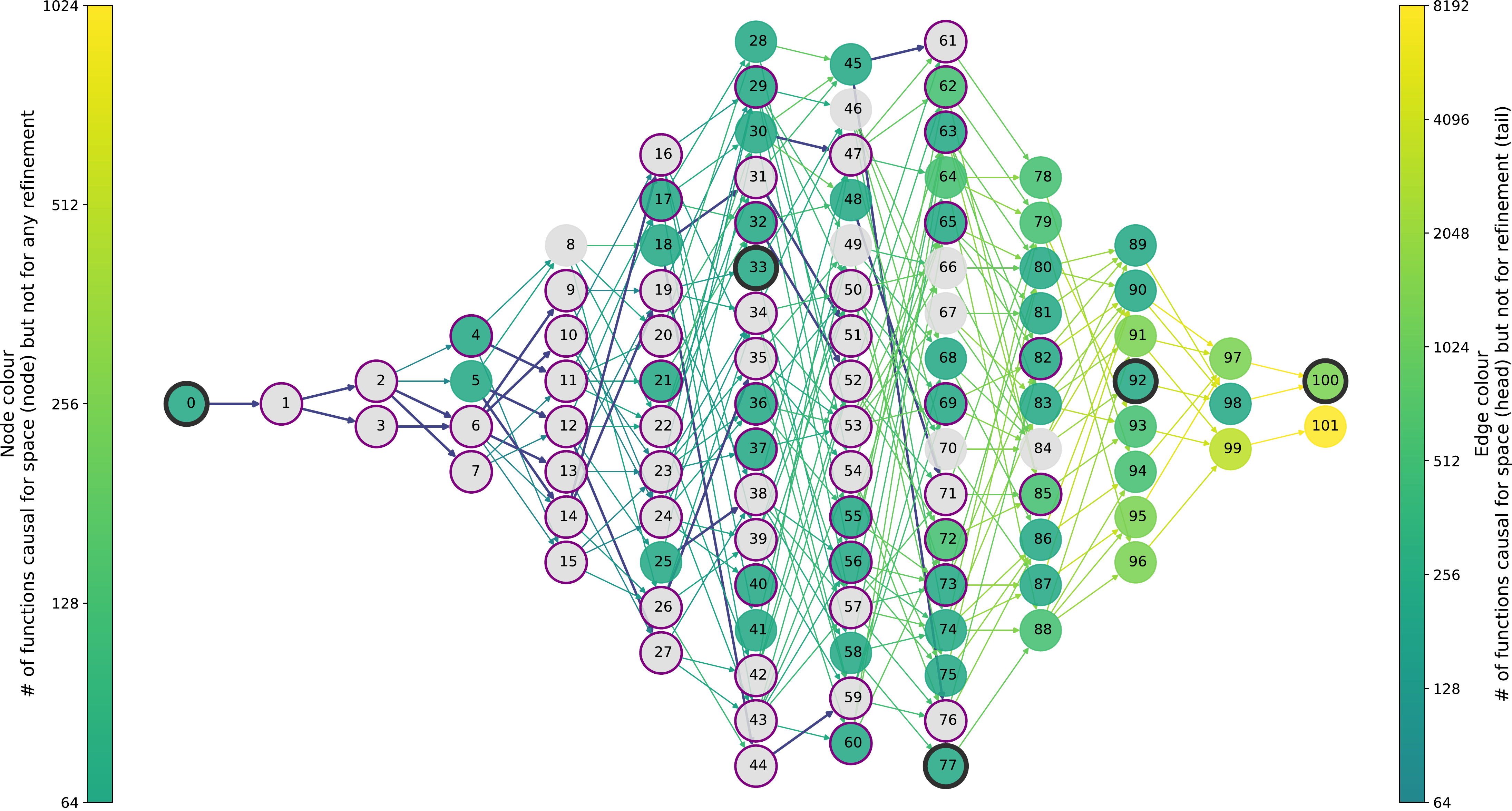}
    \caption{
    The hierarchy of causally complete spaces on 3 events with binary inputs, grouped into 102 equivalence classes under event-input permutation symmetry.
    An edge $i \rightarrow j$ indicates that some space---and hence every space---in equivalence class $i$ is a closest refinement for some space in equivalence class $j$.

    Node colour indicates the number of causal functions for a space which are not causal for any of its subspaces, while edge colour indicates the number of causal functions for the head space that are not causal for the tail space.
    Grey nodes (e.g. eq. class 1) indicate spaces where every causal function is also causal for some subspace, while thicker dark blue edges (e.g. edge $0 \rightarrow 1$) indicate that all causal functions for the head space are also causal a single tail space.

    Thin purple borders for nodes indicate eq. classes of non-tight spaces (e.g. eq. class 1).
    Thick black borders for nodes indicate the eq. classes of spaces induced by causal orders.
    }
\label{fig:hierarchy-spaces-ABC}
\end{figure}

To gain some intuition about this hierarchy, we look at several examples from the hierarchy $\CCSpaces{\left(\{0,1\}\right)_{\omega \in \evset{A,B,C}}}$ on three events.
This hierarchy has 2644 spaces, forming 102 equivalence classes under event-input permutation symmetry (cf. pp.40-41 of \cite{gogioso2022combinatorics}).
Figure \ref{fig:hierarchy-spaces-ABC} (p.\pageref{fig:hierarchy-spaces-ABC}) depicts condensed hierarchy formed by the 102 equivalence classes: in this condensed graph, an edge $i \rightarrow j$ indicates that some space---and hence every space---in equivalence class $i$ is a closest refinement of some space of equivalence class $j$.

At the bottom of the hierarchy we find the discrete space, induced by the discrete order $\discrete{\ev{A},\ev{B},\ev{C}}$, sitting alone in equivalence class 0.
This the \emph{no-signalling space}, where the output at each event depends only on the input at that event.
The corresponding space of extended input histories contains all 26 binary-valued partial functions on the 3 events: histories supported by more than one event are not $\vee$-prime in this space.
\begin{center}
    \begin{tabular}{cc}
    \includegraphics[height=3.5cm]{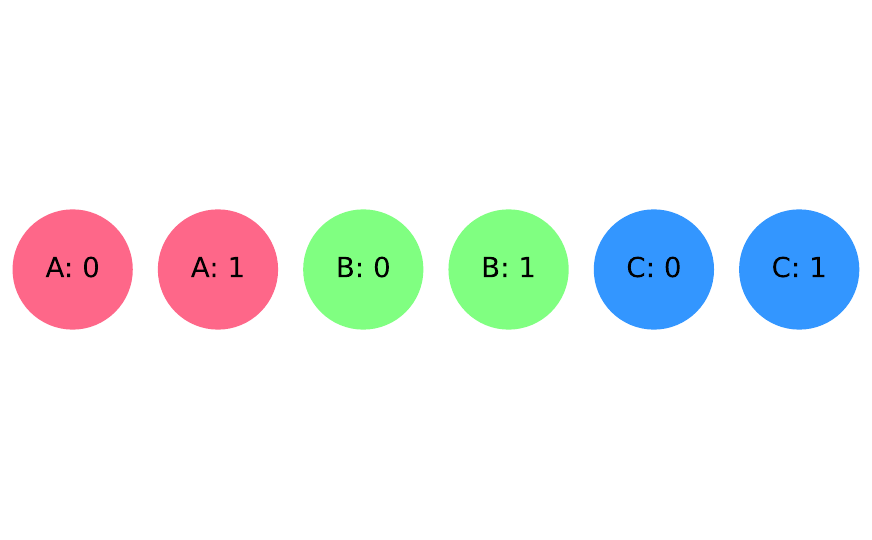}
    &
    \includegraphics[height=3.5cm]{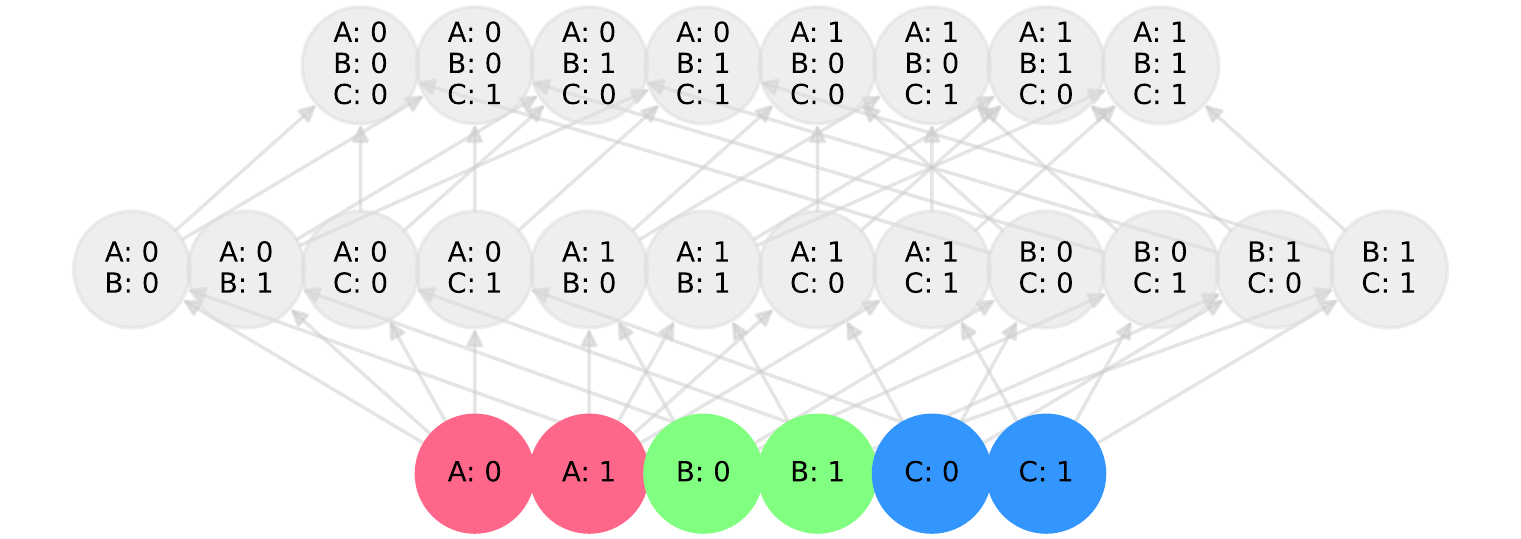}
    \\
    $\Theta_{0}$
    &
    $\Ext{\Theta_{0}}$
    \end{tabular}
\end{center}

At the top of the hierarchy we find two equivalence classes of spaces, labelled 100 and 101.
Equivalence class 100 contains the 6 spaces induced by total order: below is the space induced by $\total{\ev{A},\ev{B},\ev{C}}$.
This space coincides with its own space of extended input histories.
\begin{center}
    \begin{tabular}{cc}
    \includegraphics[height=3.5cm]{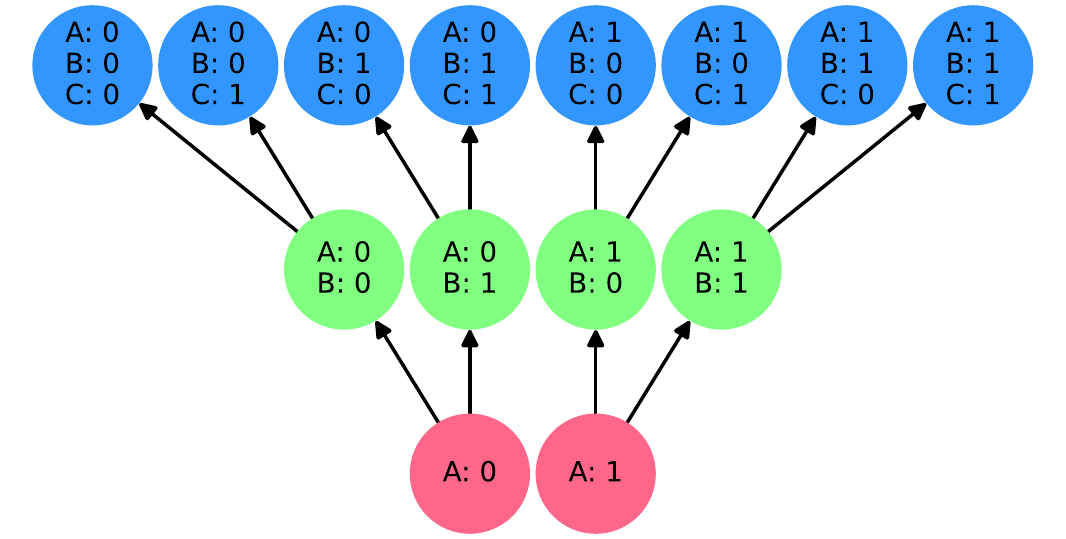}
    &
    \includegraphics[height=3.5cm]{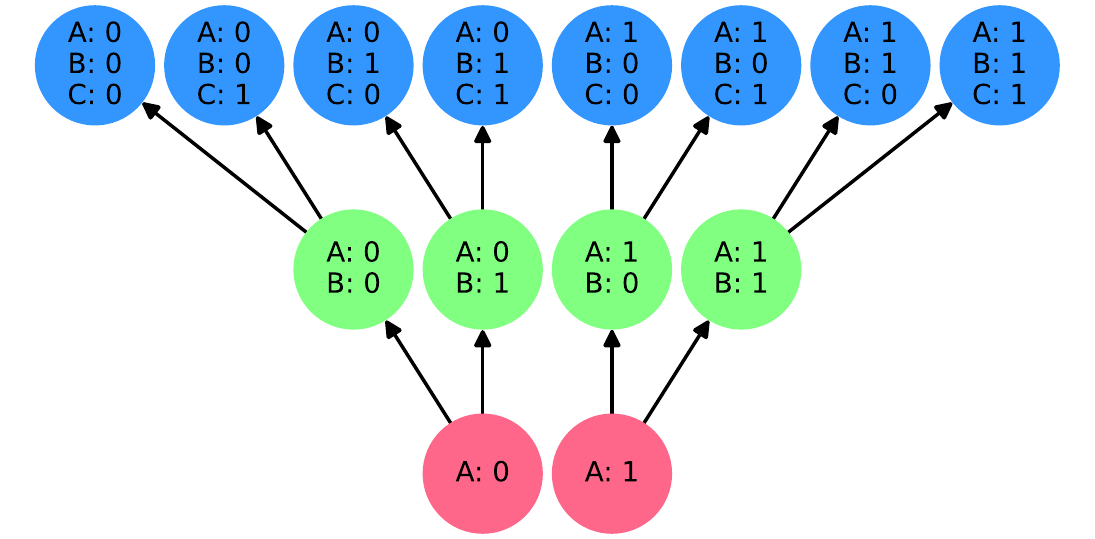}
    \\
    $\Theta_{100}$
    &
    $\Ext{\Theta_{100}}$
    \end{tabular}
\end{center}
Equivalence class 101 contains the 6 spaces for a 3-party causal switch: below is the space where the input of \ev{A} determines the total order between \ev{B} and \ev{C}, with input 0 at \ev{A} setting $\ev{B} < \ev{C}$ and input 1 at \ev{A} setting $\ev{C} < \ev{B}$.
This space coincides with its own space of extended input histories.
\begin{center}
    \begin{tabular}{cc}
    \includegraphics[height=3.5cm]{svg-inkscape/space-ABC-unique-tight-101-highlighted_svg-tex.pdf}
    &
    \includegraphics[height=3.5cm]{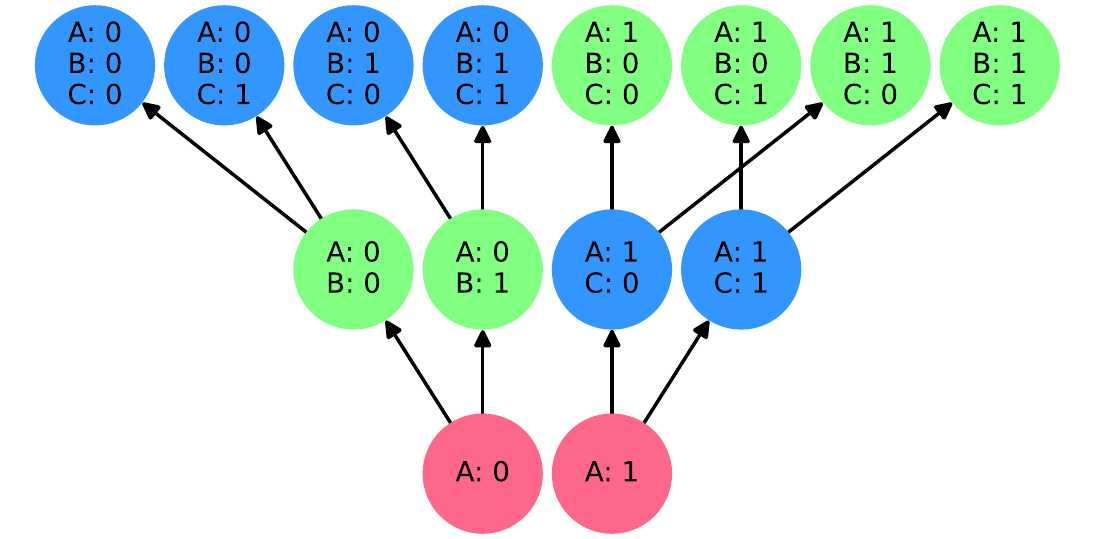}
    \\
    $\Theta_{101}$
    &
    $\Ext{\Theta_{101}}$
    \end{tabular}
\end{center}
The spaces in equivalence class 101 are examples of causally complete spaces not admitting a fixed definite causal order: they are not refinements of $\Hist{\Omega, \{0,1\}}$ for any definite causal order $\Omega$ on \ev{A}, \ev{B} and \ev{C}.
There are 13 equivalence classes consisting of spaces that don't admit a fixed definite causal order (cf. Figure 6 p.44 \cite{gogioso2022combinatorics}).

The 5 equivalence classes of spaces induced by total orders are marked by a thick black border in Figure \ref{fig:hierarchy-spaces-ABC} (p.\pageref{fig:hierarchy-spaces-ABC}).
We have already seen equivalence class 0 (for the discrete order) and equivalence class 100 (for total orders): we now look at the remaining three.
Equivalence class 92 contains the 3 spaces induced by wedge orders: below is the space induced by order $\total{\ev{A},\ev{C}}\vee\total{\ev{B},\ev{C}}$.
The extended input histories supported by $\evset{A,B}$ are not $\vee$-prime in this space.
\begin{center}
    \begin{tabular}{cc}
    \includegraphics[height=3.5cm]{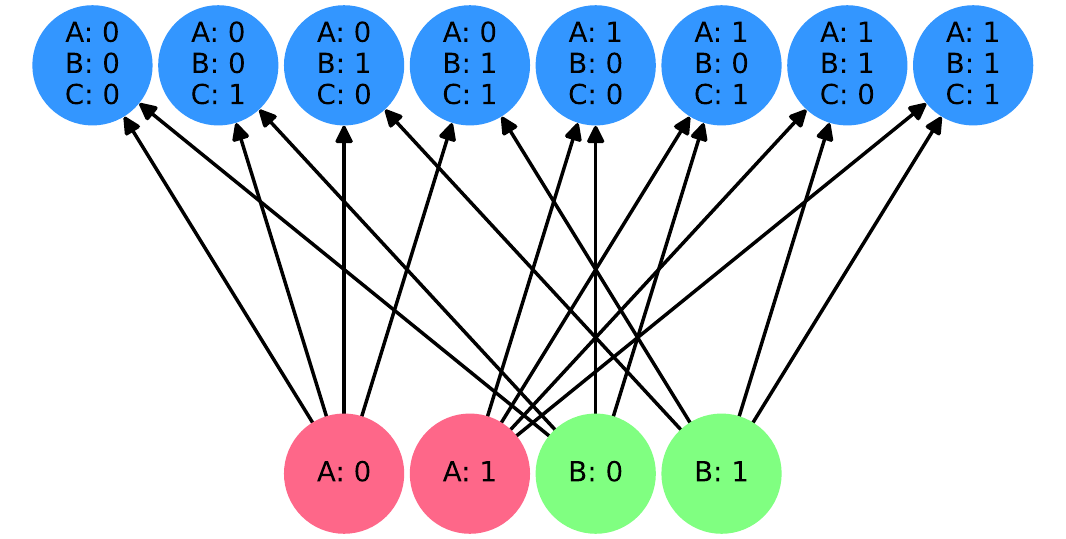}
    &
    \includegraphics[height=3.5cm]{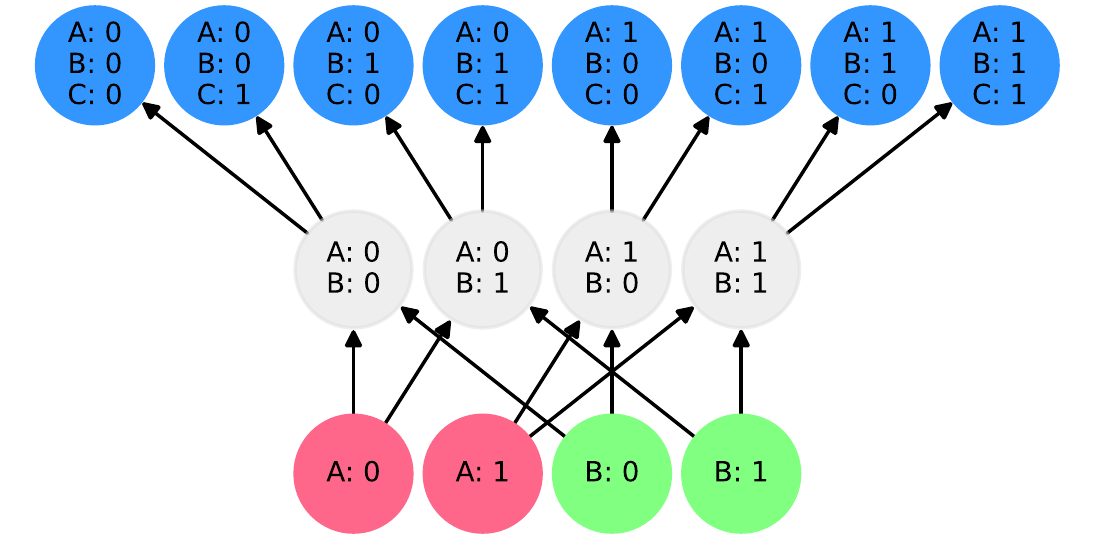}
    \\
    $\Theta_{92}$
    &
    $\Ext{\Theta_{92}}$
    \end{tabular}
\end{center}
Equivalence class 77 contains the 3 spaces induced by fork orders: below is the space induced by order $\total{\ev{A},\ev{B}}\vee\total{\ev{A},\ev{C}}$.
The extended input histories supported by all three events are not $\vee$-prime in this space.
\begin{center}
    \begin{tabular}{cc}
    \includegraphics[height=3.5cm]{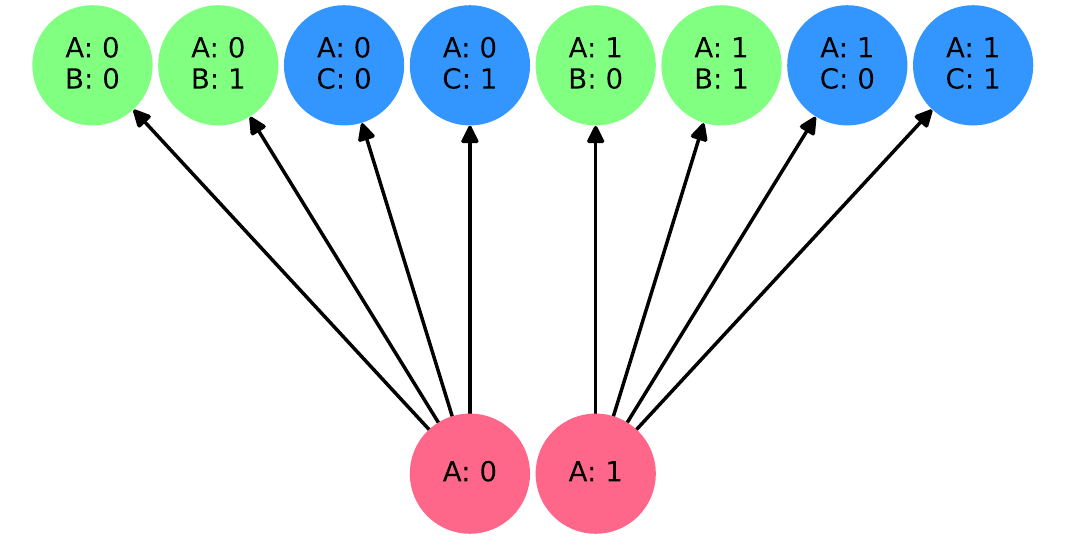}
    &
    \includegraphics[height=3.5cm]{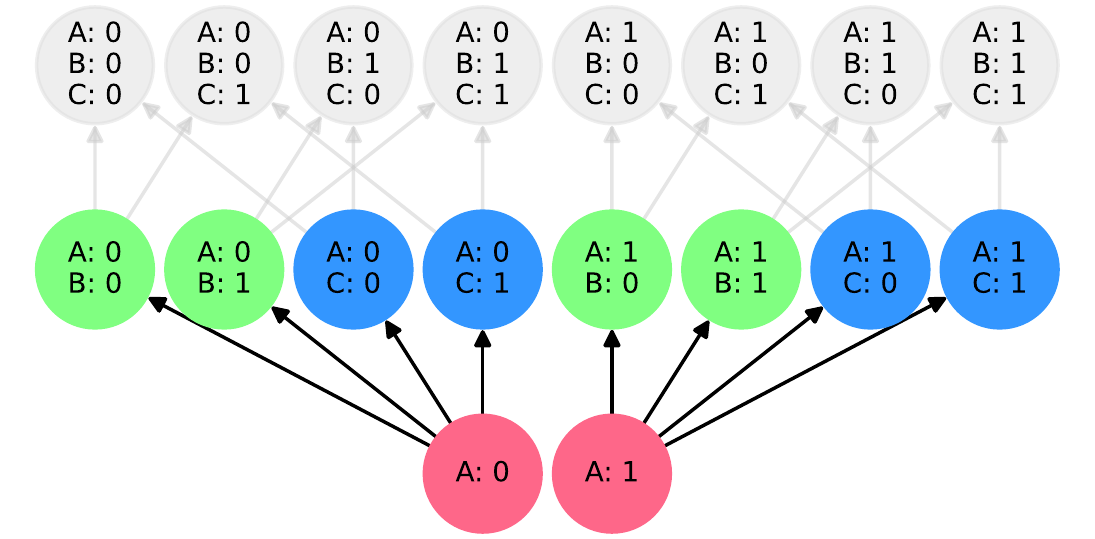}
    \\
    $\Theta_{77}$
    &
    $\Ext{\Theta_{77}}$
    \end{tabular}
\end{center}
Equivalence class 33 contains the 6 spaces induced by the disjoint join of a total order on two events with a discrete third event: below is the space induced by order $\total{\ev{A},\ev{B}}\vee\discrete{\ev{C}}$.
The extended input histories supported by either $\evset{A,C}$ or by all three events are not $\vee$-prime in this space.
\begin{center}
    \begin{tabular}{cc}
    \includegraphics[height=3.5cm]{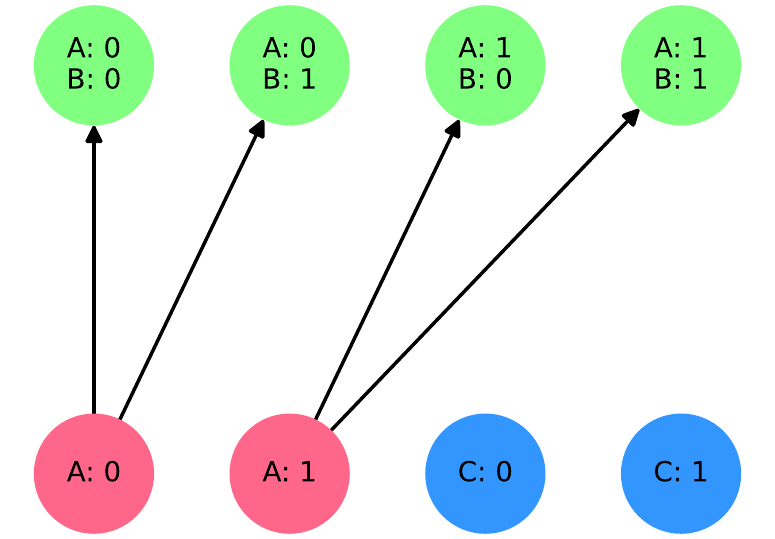}
    &
    \includegraphics[height=3.5cm]{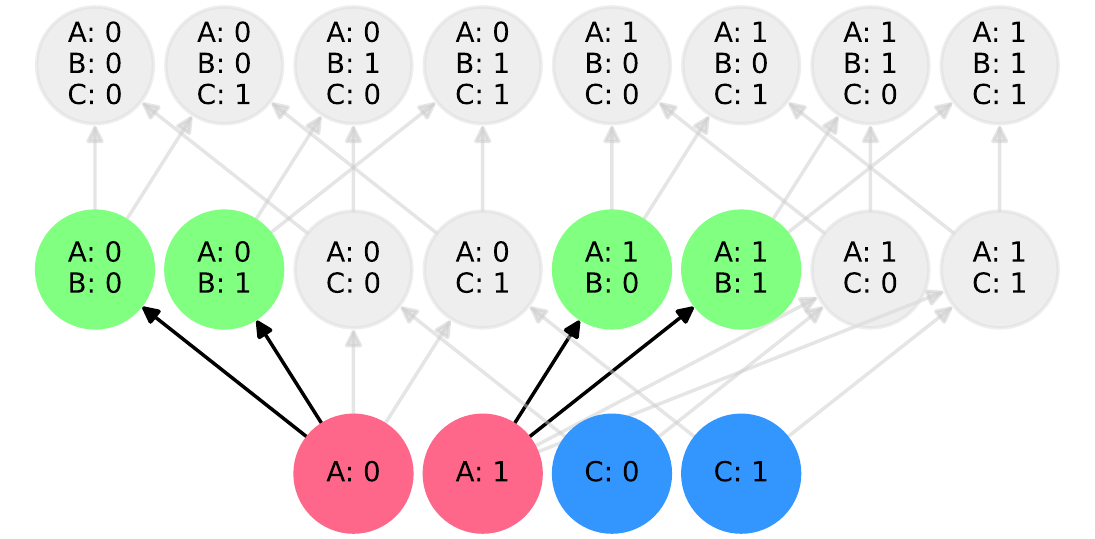}
    \\
    $\Theta_{33}$
    &
    $\Ext{\Theta_{33}}$
    \end{tabular}
\end{center}

Spaces not induced by causal orders can all be understood as introducing input-dependent causal constraints.
As a simple example, consider space $\Theta_{98}$ below, a representative from equivalence class 98 which is a closest refinement of $\Hist{\total{\ev{A},\ev{B},\ev{C}}, \{0,1\}}$.
\begin{center}
    \begin{tabular}{cc}
    \includegraphics[height=3.5cm]{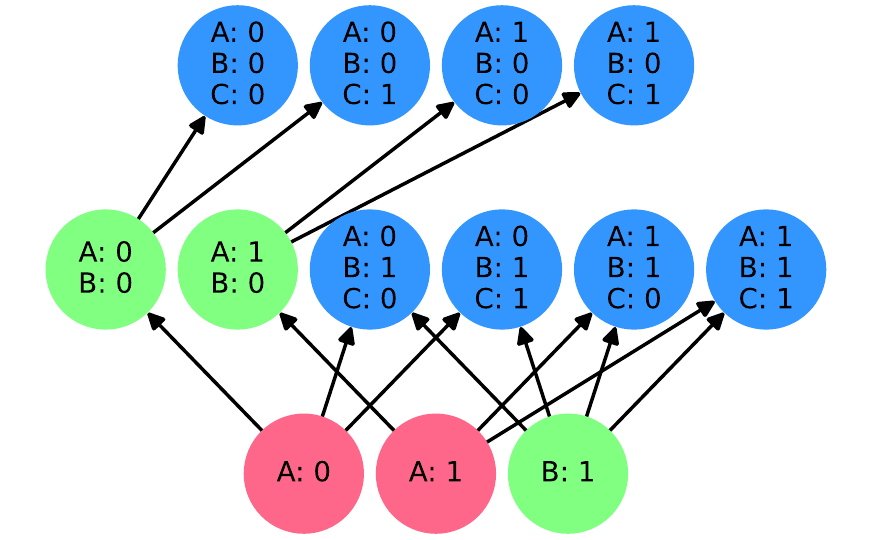}
    &
    \includegraphics[height=3.5cm]{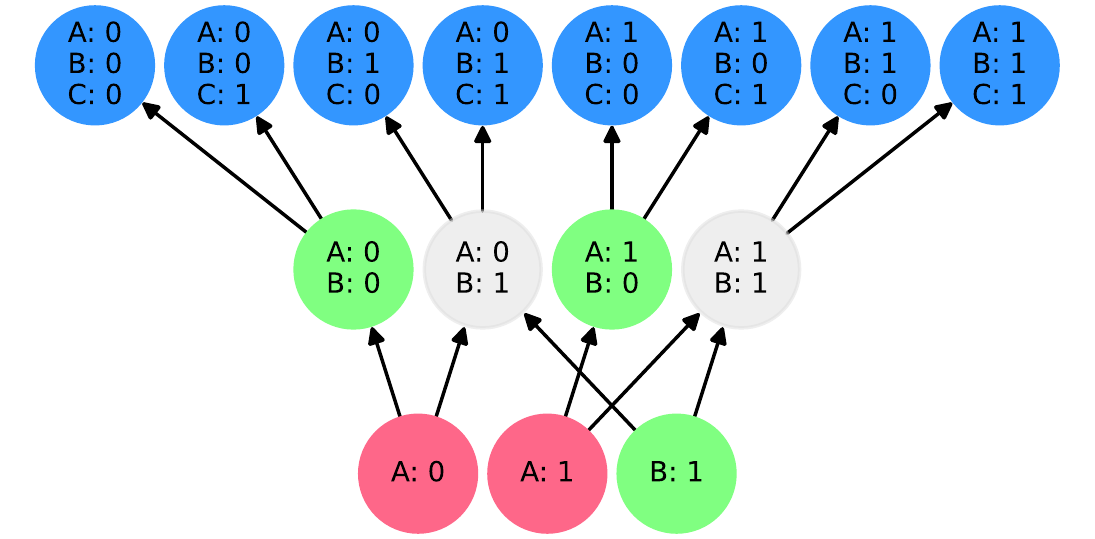}
    \\
    $\Theta_{98}$
    &
    $\Ext{\Theta_{98}}$
    \end{tabular}
\end{center}
The only history in space $\Theta_{98}$ above additional to  $\Hist{\total{\ev{A},\ev{B},\ev{C}}, \{0,1\}}$ is $\hist{B/1}$, imposing the following constraint: when the input at \ev{B} is 1, the output at \ev{B} is independent of the input at event \ev{A}.
However, we mentioned the additional causal constraints need not be truly input dependent, as witnessed by the meet of order-induced spaces for causal orders $\Omega = \total{\ev{A},\ev{B}}\vee\discrete{\ev{C}}$ and $\Omega' = \discrete{A}\vee\total{\ev{C},\ev{B}}$.
\begin{center}
    \begin{tabular}{ccc}
    \includegraphics[height=2.5cm]{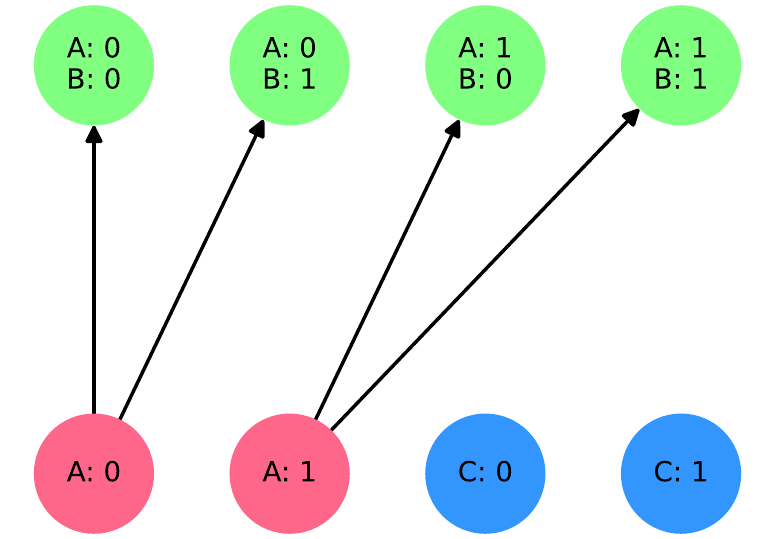}
    &
    \includegraphics[height=2.5cm]{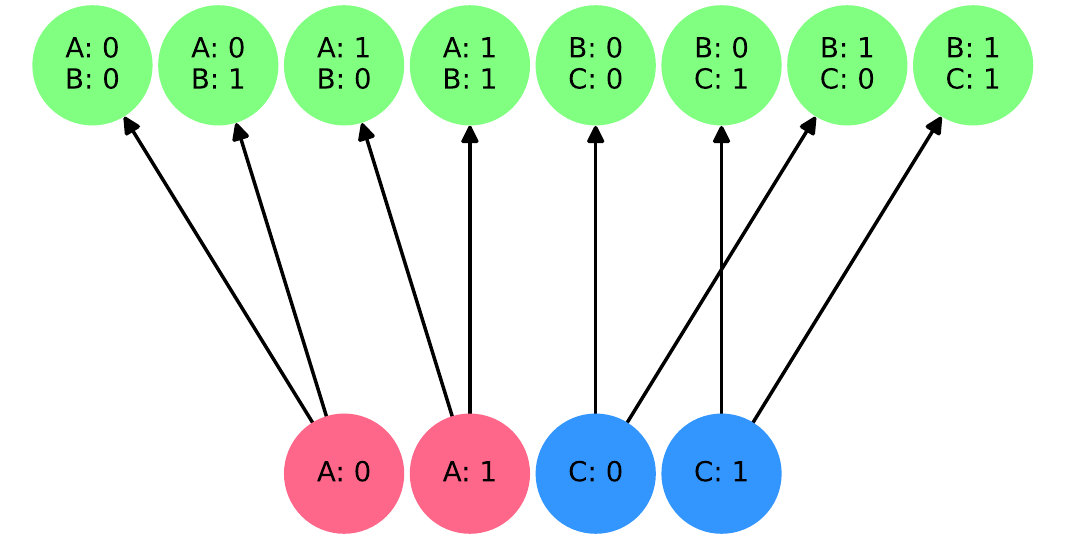}
    &
    \includegraphics[height=2.5cm]{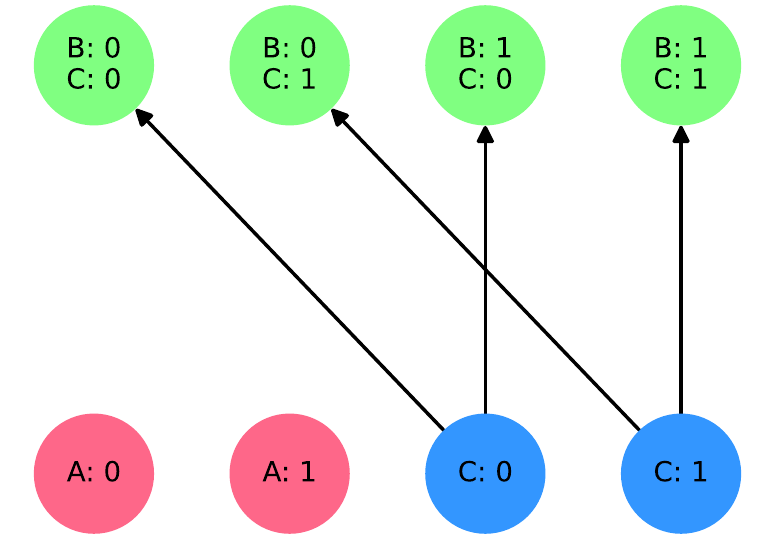}
    \\
    $\Theta_{33} = \Hist{\Omega,\{0,1\}}$
    &
    \hspace{2mm}
    $\Theta_3 = \Hist{\Omega',\{0,1\}}\wedge\Hist{\Omega',\{0,1\}}$
    \hspace{2mm}
    &
    $\Hist{\Omega',\{0,1\}}$
    \end{tabular}
\end{center}
Indeed, the spaces in equivalence class 3 are exactly the meets of 3 pairs of spaces from equivalence class 33 (the other 15 non-trivial meets of pairs in equivalence class 33 all yield the discrete space, in equivalence class 0).
For space $\Theta_{3}$, specifically, we get the following additional constraints:
\begin{itemize}
    \item as a coarsening of $\Hist{\total{\ev{A},\ev{B}}\vee\discrete{\ev{C}},\{0,1\}}$, the additional constraints come from the 4 histories with domain $\evset{B,C}$: they state that the outputs on \evset{B,C} are independent of the input on \ev{A} for all possible choices of inputs on $\evset{B,C}$.
    \item as a coarsening of $\Hist{\discrete{\ev{A}}\vee\total{\ev{C},\ev{B}},\{0,1\}}$, the additional constraints come from the 4 histories with domain $\evset{A,B}$: they state that the outputs on \evset{A,B} are independent of the input on \ev{C} for all possible choices of inputs on $\evset{A,B}$.
\end{itemize}
Because the additional constraints appear for all possible choices of inputs on their common support, they are not truly input-dependent in this case.
\begin{center}
    \begin{tabular}{cc}
    \includegraphics[height=3cm]{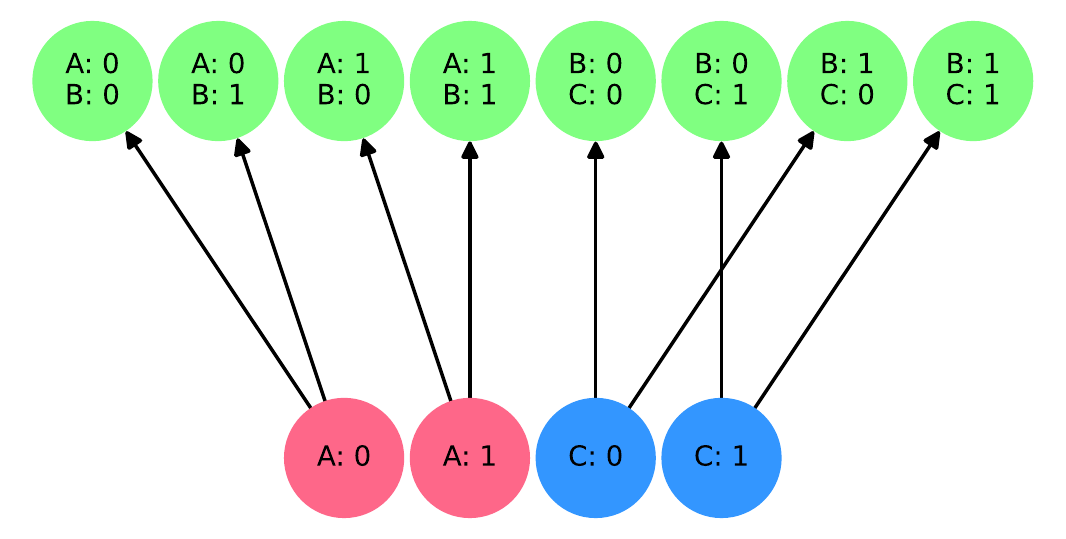}
    &
    \includegraphics[height=3cm]{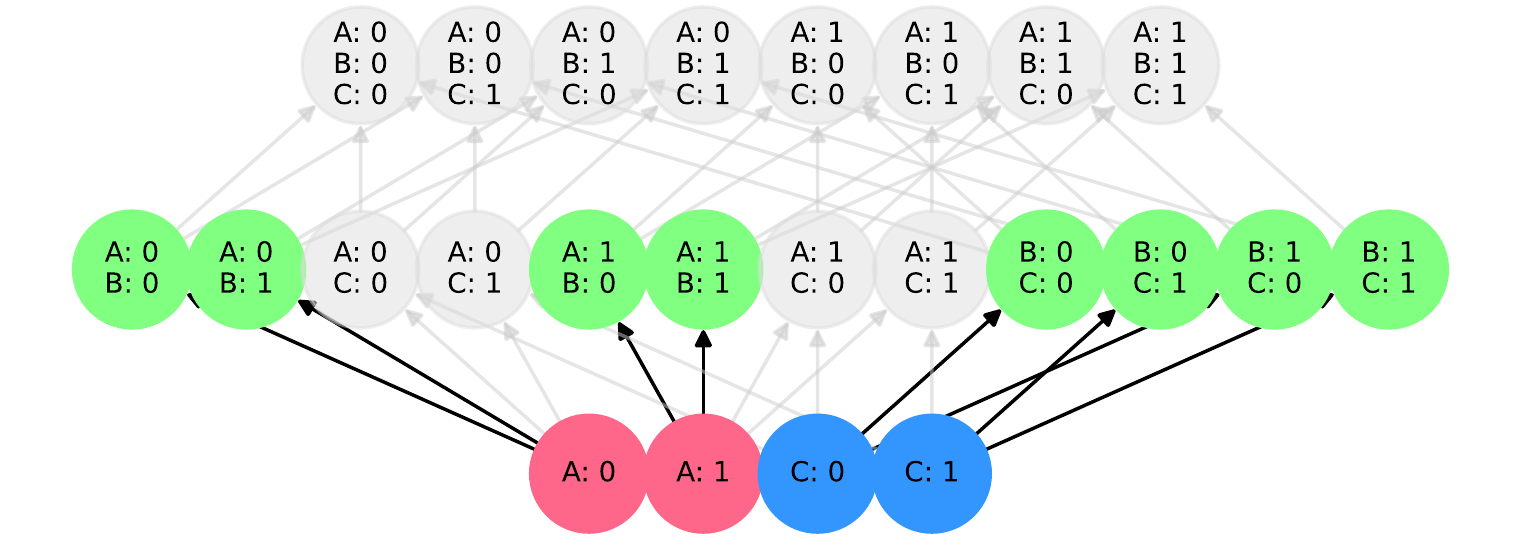}
    \\
    $\Theta_{3}$
    &
    $\Ext{\Theta_{3}}$
    \end{tabular}
\end{center}

The description of the constraints for space $\Theta_3$ is a bit confusing: one would certainly be forgiven for thinking that these constraints should be equivalent to the no-signalling ones, generated by the discrete space.
And, in a sense, they are: $\Theta_3$ has the same causal functions as the discrete space (cf. ``The Topology of Causality'' \cite{gogioso2022topology}) and the same causal distributions as the discrete space when non-locality is concerned (cf. ``The Geometry of Causality'' \cite{gogioso2022geometry}), but it admits strictly more causal distributions for more general notions of contextuality.

Space $\Theta_3$ is an example of a ``non-tight'' space, one where the events in some histories are constrained by multiple causal orders.
Lack of tightness is a peculiar pathology: in some cases, it implies a form of contextuality where deterministic functions defined compatibly on certain subsets of input histories cannot always be glued together into functions defined on all histories.

\begin{definition}
\label{definition:tight-space}
Let $\Theta$ be a space of input histories.
We say that $\Theta$ is \emph{tight} if for every (maximal) extended input history $k \in \Ext{\Theta}$ and every event $\omega \in \dom{k}$ there is a unique input history $h \in \Theta$ such that $h \leq k$ and $\omega \in \tips{\Theta}{h}$.
We say that $\Theta$ is \emph{non-tight} otherwise.
\end{definition}

Non-tight spaces are indicated in Figure \ref{fig:hierarchy-spaces-ABC} (p.\pageref{fig:hierarchy-spaces-ABC}) by a thin violet border, and they constitute the majority of examples: out of 102 equivalence classes, 58 consist of non-tight spaces and 44 consist of tight spaces.
To understand what lack of tightness means concretely, let's consider space $\Theta_{17}$ below.
In the input histories below extended input history $\hist{A/1,B/1,C/2}$ (circled in blue), the event $\ev{C}$ appears as a tip event in two separate histories, namely $\hist{A/1, C/1}$ and $\hist{B/1, C/1}$; edges from the latter input histories to the former extended input histories have also been highlighted blue, for clarity.
\begin{center}
    \begin{tabular}{cc}
    \includegraphics[height=3.5cm]{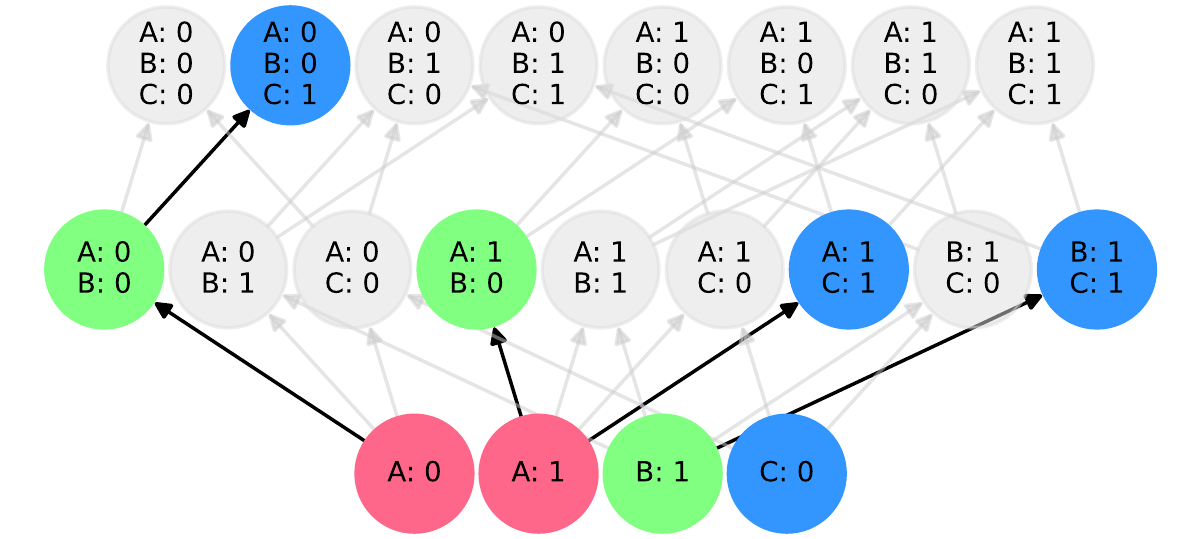}
    &
    \includegraphics[height=3.5cm]{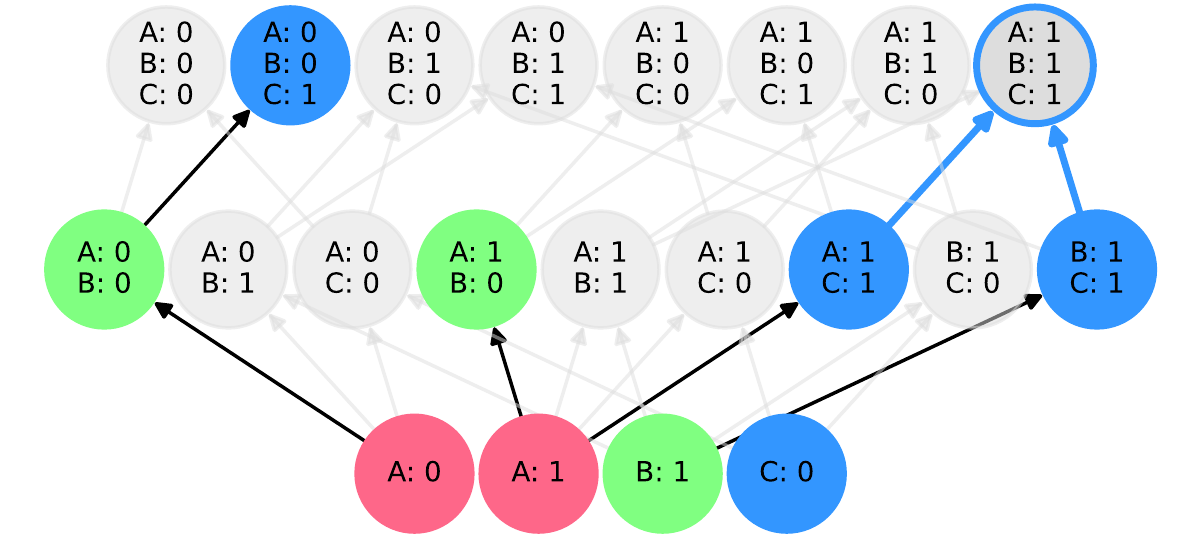}
    \\
    $\Ext{\Theta_{17}}$
    &
    $\Ext{\Theta_{17}}$ with highlights
    \end{tabular}
\end{center}
The effect of this multiple appearance of $\ev{C}$ as a tip event is that causal functions on space $\Theta_{17}$ must yield identical output values at event $\ev{C}$ for both input histories $\hist{A/1, C/1}$ and $\hist{B/1, C/1}$, which would have otherwise been unrelated.
Put in other words, in history $\hist{A/1,B/1,C/2}$ the output at event $\ev{C}$ must satisfy the constraints of two different causal orders: \total{\ev{A}, \ev{C}, \ev{B}} (from $\hist{A/1} \rightarrow \hist{A/1, C/1} \rightarrow \hist{A/1, B/1, C/1}$) and \total{\ev{B}, \ev{C}, \ev{A}} (from $\hist{B/1} \rightarrow \hist{B/1, C/1} \rightarrow \hist{A/1, B/1, C/1}$).

The hierarchy $\CSwitchSpaces{\underline{I}}$ of causally complete spaces is full of complicated examples.
Its ``canopy'', however, is significantly more tranquil: it is a fact (cf. Theorem 3.36 p.51 \cite{gogioso2022combinatorics}) that the maxima of $\CCSpaces{\underline{I}}$ are exactly the causal switch spaces $\CSwitchSpaces{\underline{I}}$ defined below.
This is what Figure \ref{fig:hierarchy-spaces-ABC} (p.\pageref{fig:hierarchy-spaces-ABC}) shows for the 3-event case and it is consistent with the approach taken by previous literature on indefinite causality.
It is furthermore a fact (cf. Theorem 3.34 and Corollary 3.35 p.51 \cite{gogioso2022combinatorics}) that causal switch spaces are exactly the causally complete spaces $\Theta$ such that $\Theta = \Ext{\Theta}$.

\begin{definition}
Let $E$ be a set of events and $\underline{I} = (I_\omega)_{\omega \in E}$ be a family of non-empty input sets.
The \emph{causal switch spaces} $\CSwitchSpaces{\underline{I}}$ are defined as follows.
If $E = \emptyset$, then $\CSwitchSpaces{\underline{I}} = \emptyset$.
Otherwise, for each $\omega_1 \in E$ we can consider:
\[
\begin{array}{rcl}
\restrict{\underline{I}}{\{\omega_1\}} &=& (I_\omega)_{\omega \in \{\omega_1\}}\\
\restrict{\underline{I}}{E\backslash\{\omega_1\}} &=& (I_\omega)_{\omega \in E\backslash\{\omega_1\}}
\end{array}
\]
Then the set $\CSwitchSpaces{\underline{I}}$ is defined inductively as follows:
\begin{equation*}
    \bigcup\limits_{\omega_1 \in E}
    \suchthat{
    \Hist{\{\omega_1\}, \restrict{\underline{I}}{\{\omega_1\}}}
    \seqcomposeSym
    \underline{\Theta}
    }
    {
    \underline{\Theta}
    \in
    \CSwitchSpaces{\restrict{\underline{I}}{E\backslash\{\omega_1\}}}^{I_{\omega_1}}
    }    
\end{equation*}
\end{definition}


\newpage

\section{Causally Complete Spaces on 3 Events with Binary Inputs}
\label{appendix:all-spaces-ABC}

\begin{figure}[h]
    \centering
    \includegraphics[width=\textwidth]{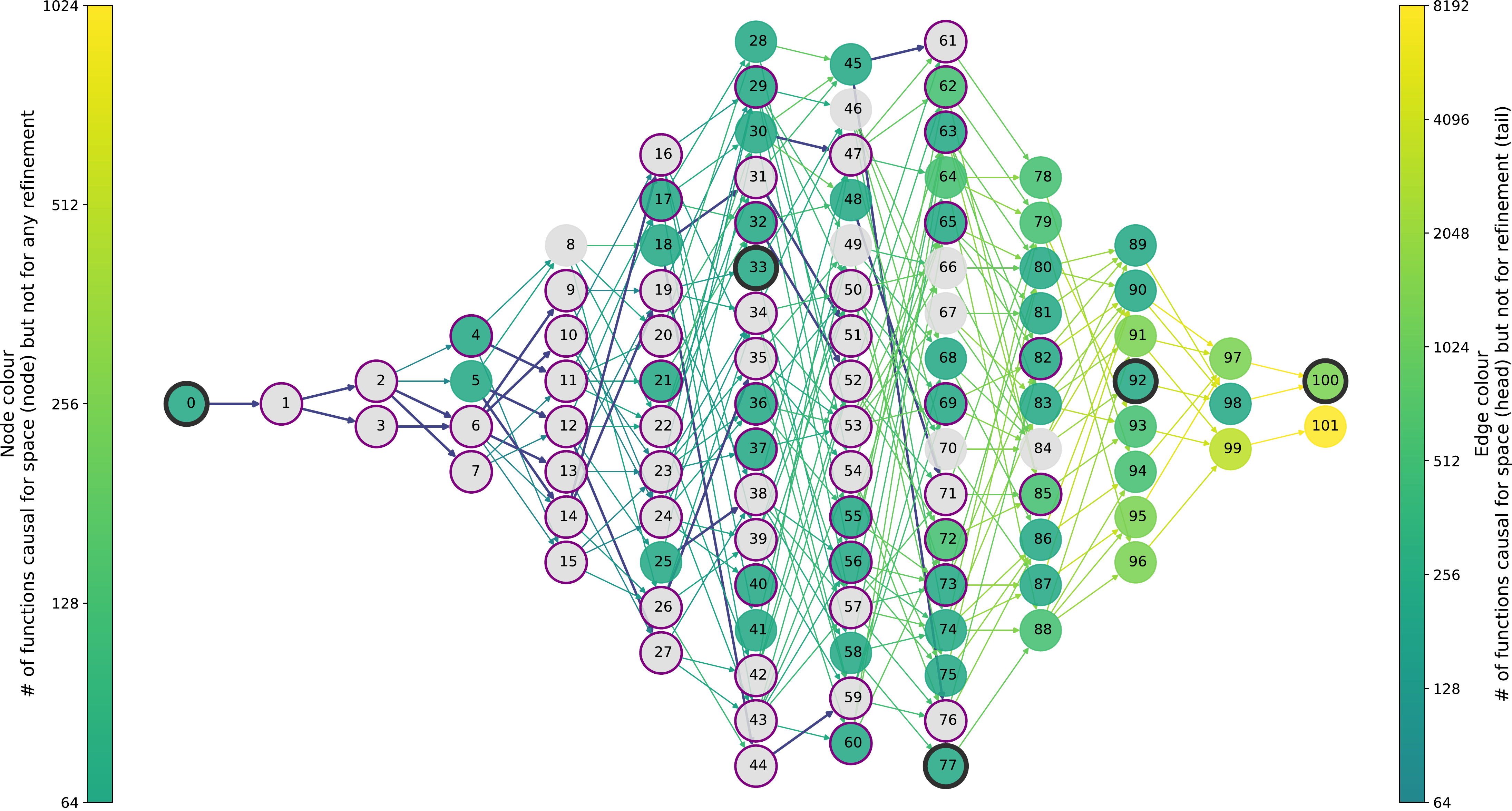}
    \caption{
    The hierarchy of causally complete spaces on 3 events with binary inputs, grouped into 102 equivalence classes under event-input permutation symmetry.
    See Figure \ref{fig:hierarchy-spaces-ABC} (p.\pageref{fig:hierarchy-spaces-ABC}) for the full description.
    }
\label{appendix-fig:hierarchy-spaces-ABC}
\end{figure}

\begin{figure}[h]
    \centering
    \includegraphics[width=\textwidth]{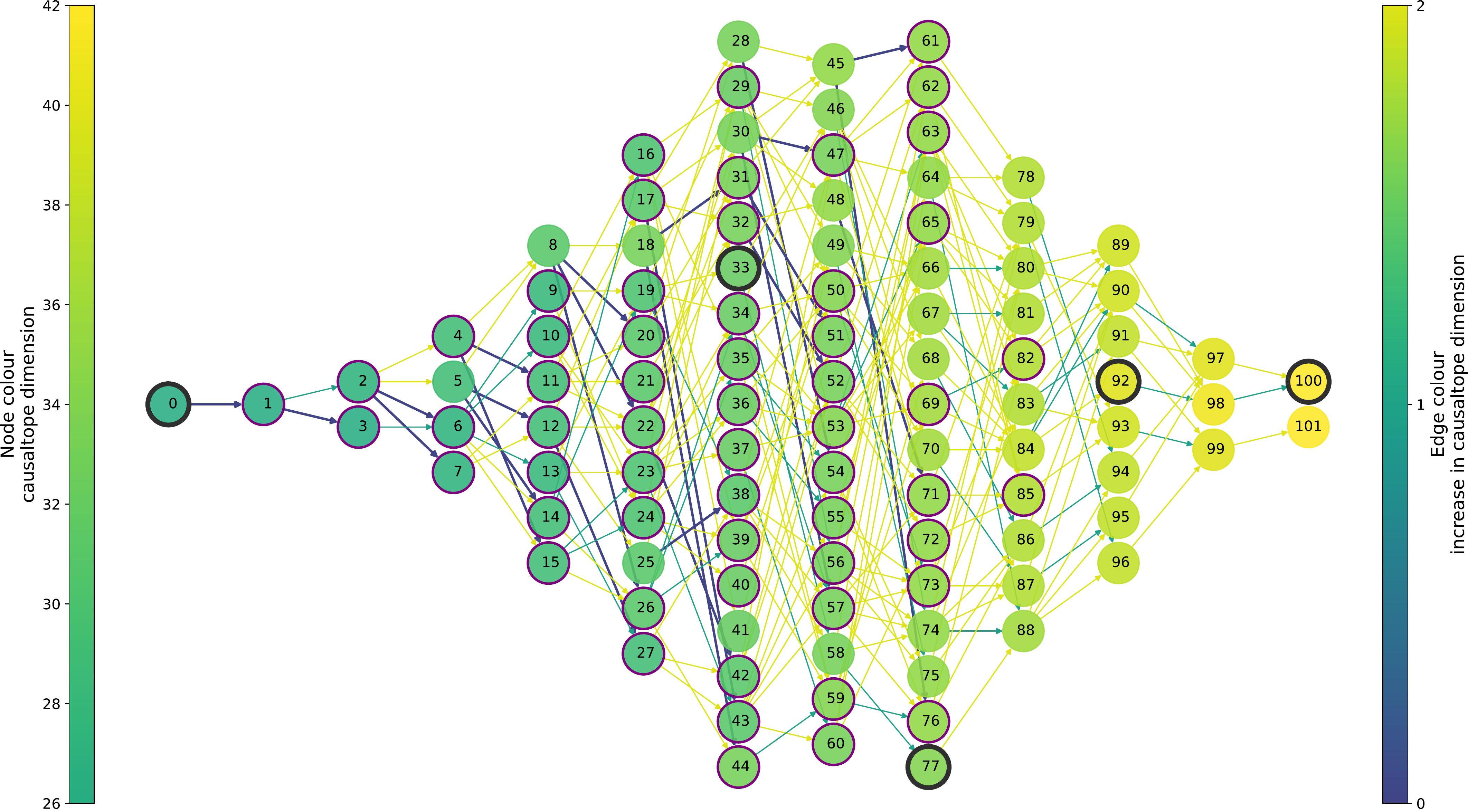}
    \caption{
    The hierarchy of causaltopes for standard empirical models on causally complete spaces with 3 events and binary inputs.
    See Figure 6 (p.88) of ``The Geometry of Causality'' \cite{gogioso2022geometry} for the full description.
    }
\label{appendix-fig:hierarchy-spaces-ABC-causaltope-dim}
\end{figure}

\newpage
\newpage
\subsection*{Space 0}

Space 0 is induced by the definite causal order $\discrete{\ev{A},\ev{B},\ev{C}}$.
Its equivalence class under event-input permutation symmetry contains 1 space.

\noindent Below are the histories and extended histories for space 0: 
\begin{center}
    \begin{tabular}{cc}
    \includegraphics[height=3.5cm]{svg-inkscape/space-ABC-unique-tight-0-highlighted_svg-tex.pdf}
    &
    \includegraphics[height=3.5cm]{svg-inkscape/space-ABC-unique-tight-0-ext-highlighted_svg-tex.pdf}
    \\
    $\Theta_{0}$
    &
    $\Ext{\Theta_{0}}$
    \end{tabular}
\end{center}

\noindent The standard causaltope for Space 0 has dimension 26.
Below is a plot of the homogeneous linear system of causality and quasi-normalisation equations for the standard causaltope, put in reduced row echelon form:

\begin{center}
    \includegraphics[width=11cm]{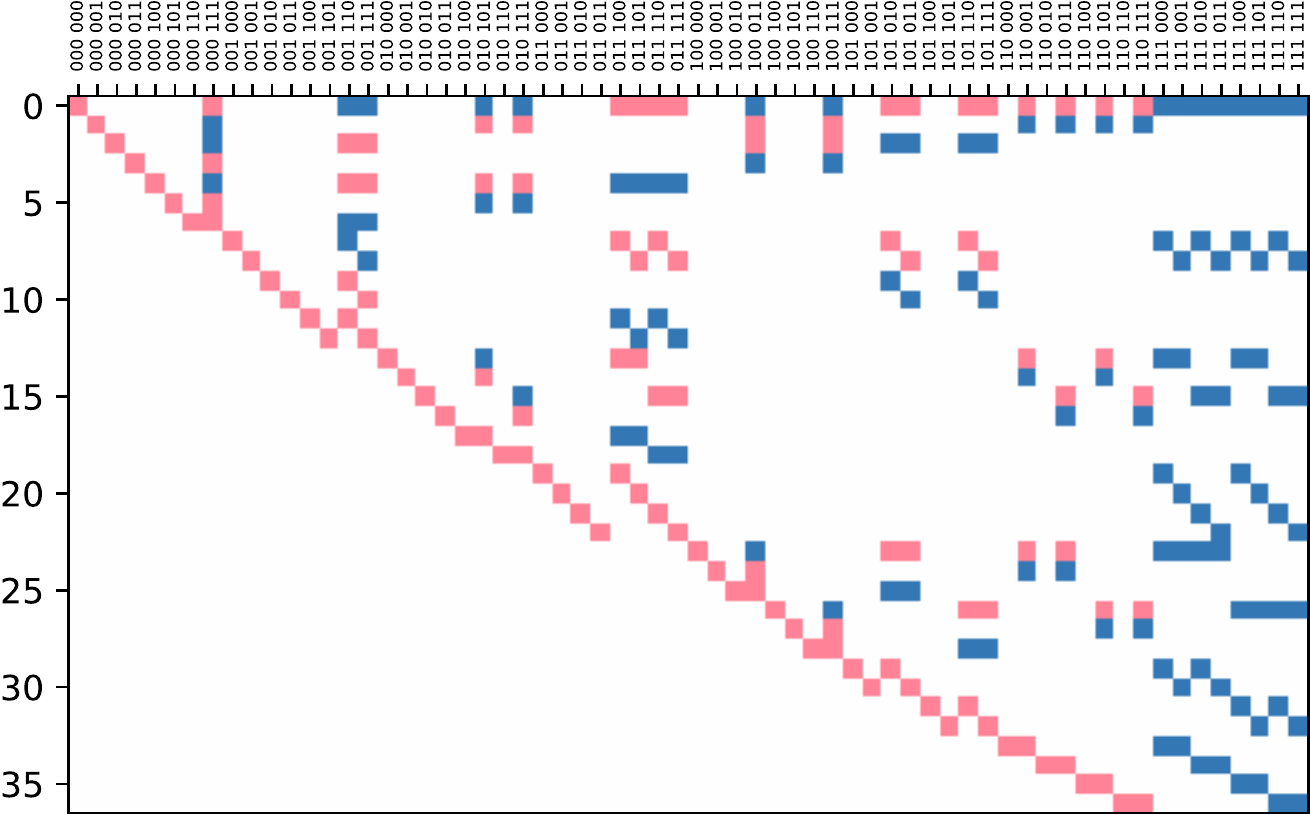}
\end{center}

\noindent Rows correspond to the 37 independent linear equations.
Columns in the plot correspond to entries of empirical models, indexed as $i_A i_B i_C$ $o_A o_B o_C$.
Coefficients in the equations are color-coded as white=0, red=+1 and blue=-1.

Space 0 is the global minimum of the hierarchy, with no refinements.
It has closest coarsenings in equivalence class 1; 
it is the meet of its (closest) coarsenings.
It has 64 causal functions.
It is a tight space.

The standard causaltope for Space 0 is the meet of the standard causaltopes for its closest coarsenings.
For completeness, below is a plot of the full homogeneous linear system of causality and quasi-normalisation equations for the standard causaltope:

\begin{center}
    \includegraphics[width=12cm]{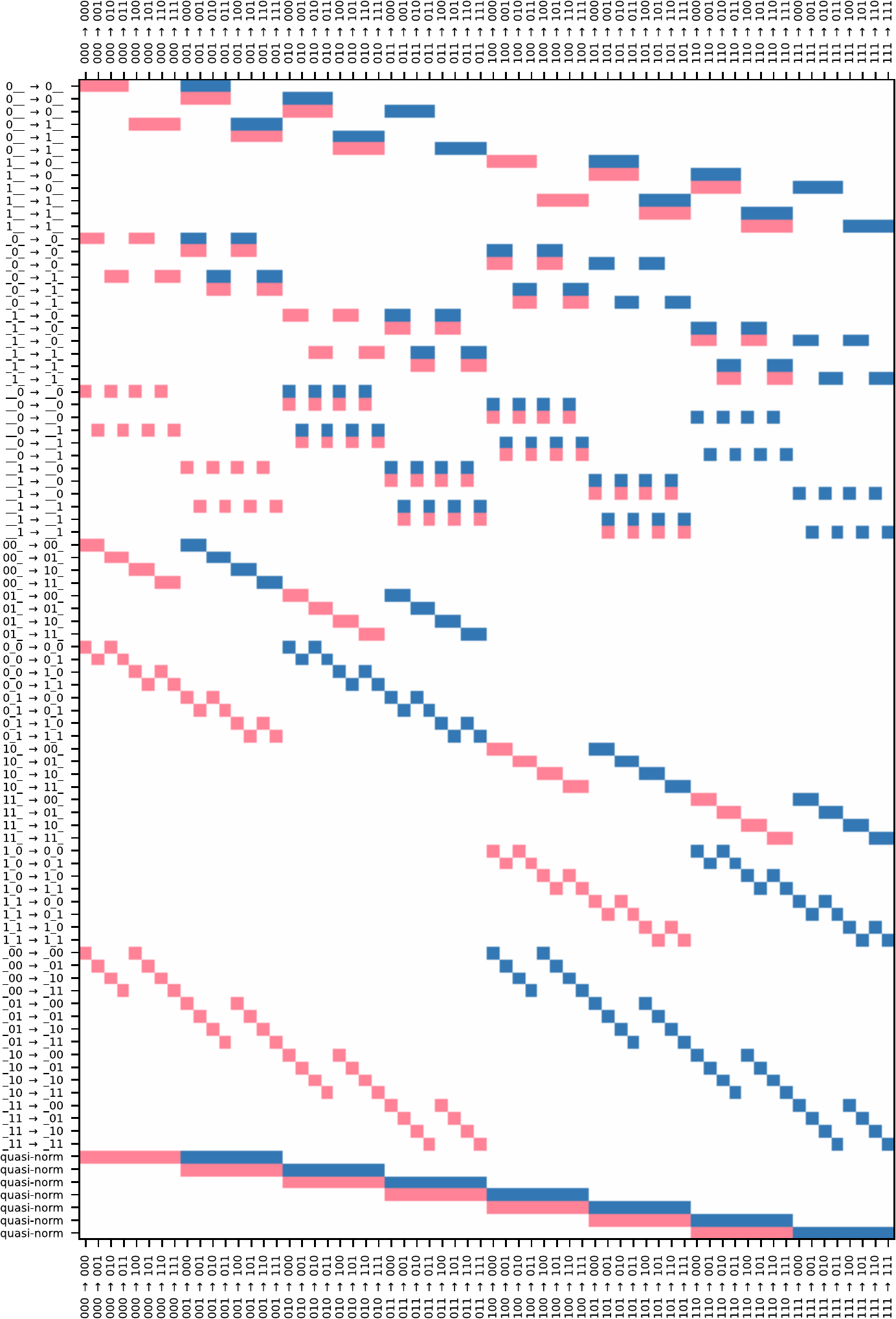}
\end{center}

\noindent Rows correspond to the 91 linear equations, of which 37 are independent.

\newpage
\subsection*{Space 1}

Space 1 is not induced by a causal order, but it is a refinement of the space in equivalence class 33 induced by the definite causal order $\total{\ev{B},\ev{A}}\vee\discrete{\ev{C}}$ (note that the space induced by the order is not the same as space 33).
Its equivalence class under event-input permutation symmetry contains 6 spaces.
Space 1 differs as follows from the space induced by causal order $\total{\ev{B},\ev{A}}\vee\discrete{\ev{C}}$:
\begin{itemize}
  \item The outputs at events \evset{\ev{A}, \ev{C}} are independent of the input at event \ev{B} when the inputs at events \evset{A, C} are given by \hist{A/0,C/1}, \hist{A/0,C/0}, \hist{A/1,C/0} and \hist{A/1,C/1}.
  \item The output at event \ev{A} is independent of the input at event \ev{B} when the input at event A is given by \hist{A/0}.
\end{itemize}

\noindent Below are the histories and extended histories for space 1: 
\begin{center}
    \begin{tabular}{cc}
    \includegraphics[height=3.5cm]{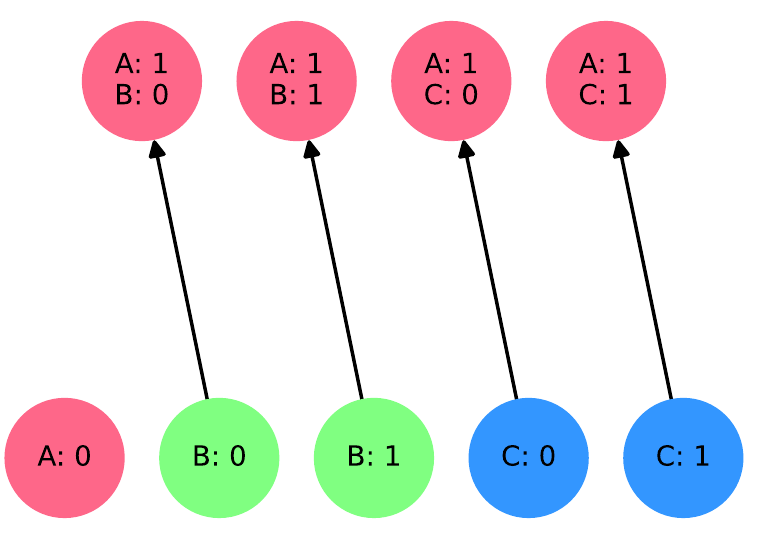}
    &
    \includegraphics[height=3.5cm]{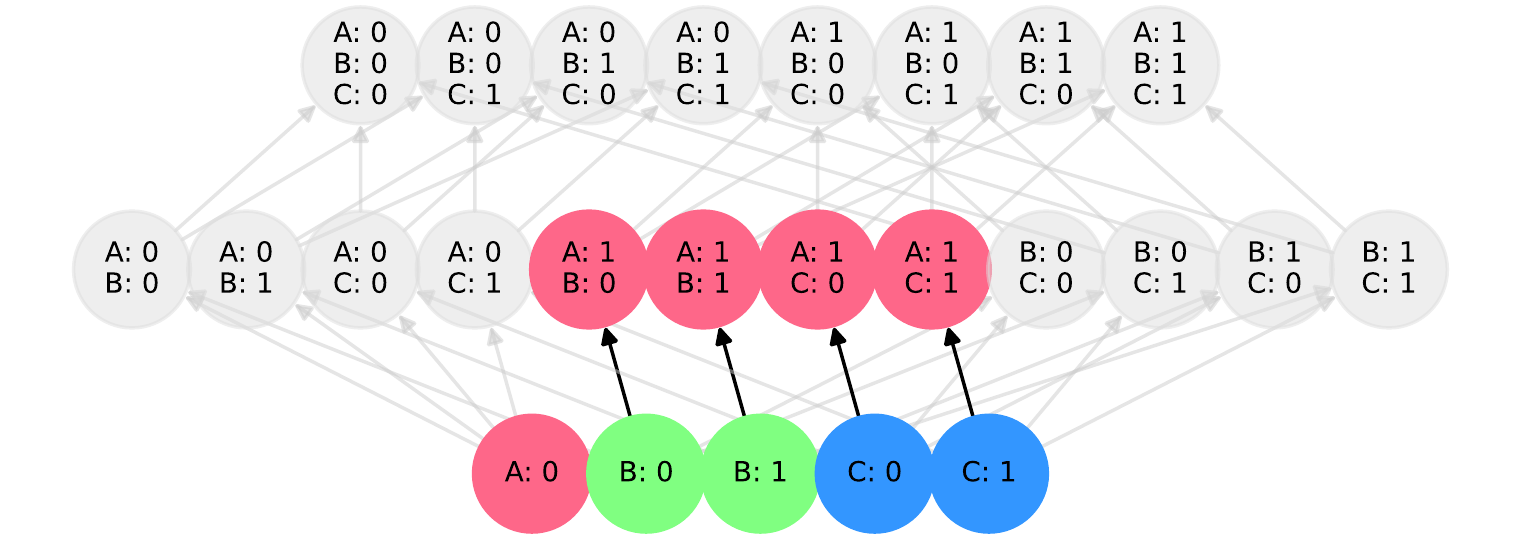}
    \\
    $\Theta_{1}$
    &
    $\Ext{\Theta_{1}}$
    \end{tabular}
\end{center}

\noindent The standard causaltope for Space 1 has dimension 26.
Below is a plot of the homogeneous linear system of causality and quasi-normalisation equations for the standard causaltope, put in reduced row echelon form:

\begin{center}
    \includegraphics[width=11cm]{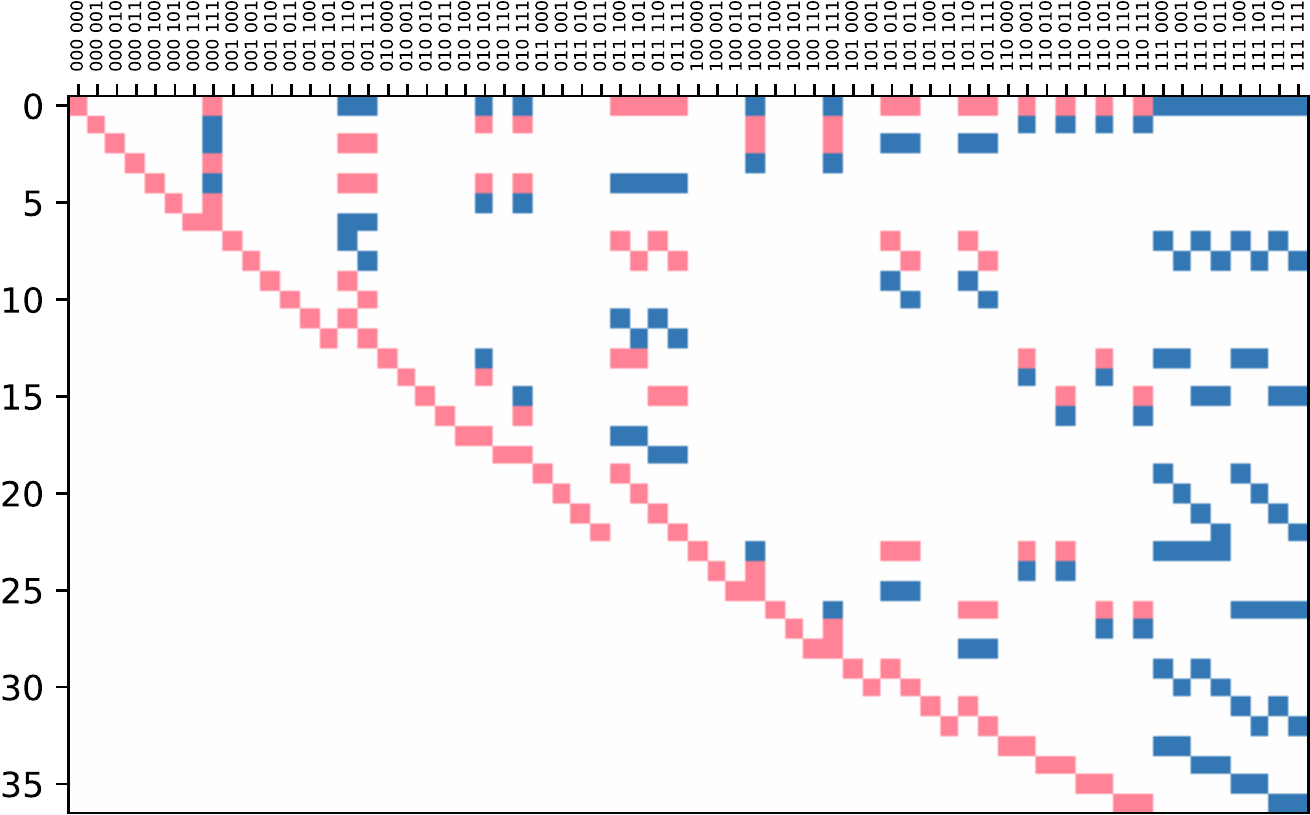}
\end{center}

\noindent Rows correspond to the 37 independent linear equations.
Columns in the plot correspond to entries of empirical models, indexed as $i_A i_B i_C$ $o_A o_B o_C$.
Coefficients in the equations are color-coded as white=0, red=+1 and blue=-1.

Space 1 has closest refinements in equivalence class 0; 
it does not arise as a nontrivial join in the hierarchy.
It has closest coarsenings in equivalence classes 2 and 3; 
it is the meet of its (closest) coarsenings.
It has 64 causal functions, all of which are causal for at least one of its refinements.
It is not a tight space: for event \ev{A}, a causal function must yield identical output values on input histories \hist{A/1,B/0}, \hist{A/1,B/1}, \hist{A/1,C/0} and \hist{A/1,C/1}.

The standard causaltope for Space 1 coincides with that of its subspace in equivalence class 0.
The standard causaltope for Space 1 is the meet of the standard causaltopes for its closest coarsenings.
For completeness, below is a plot of the full homogeneous linear system of causality and quasi-normalisation equations for the standard causaltope:

\begin{center}
    \includegraphics[width=12cm]{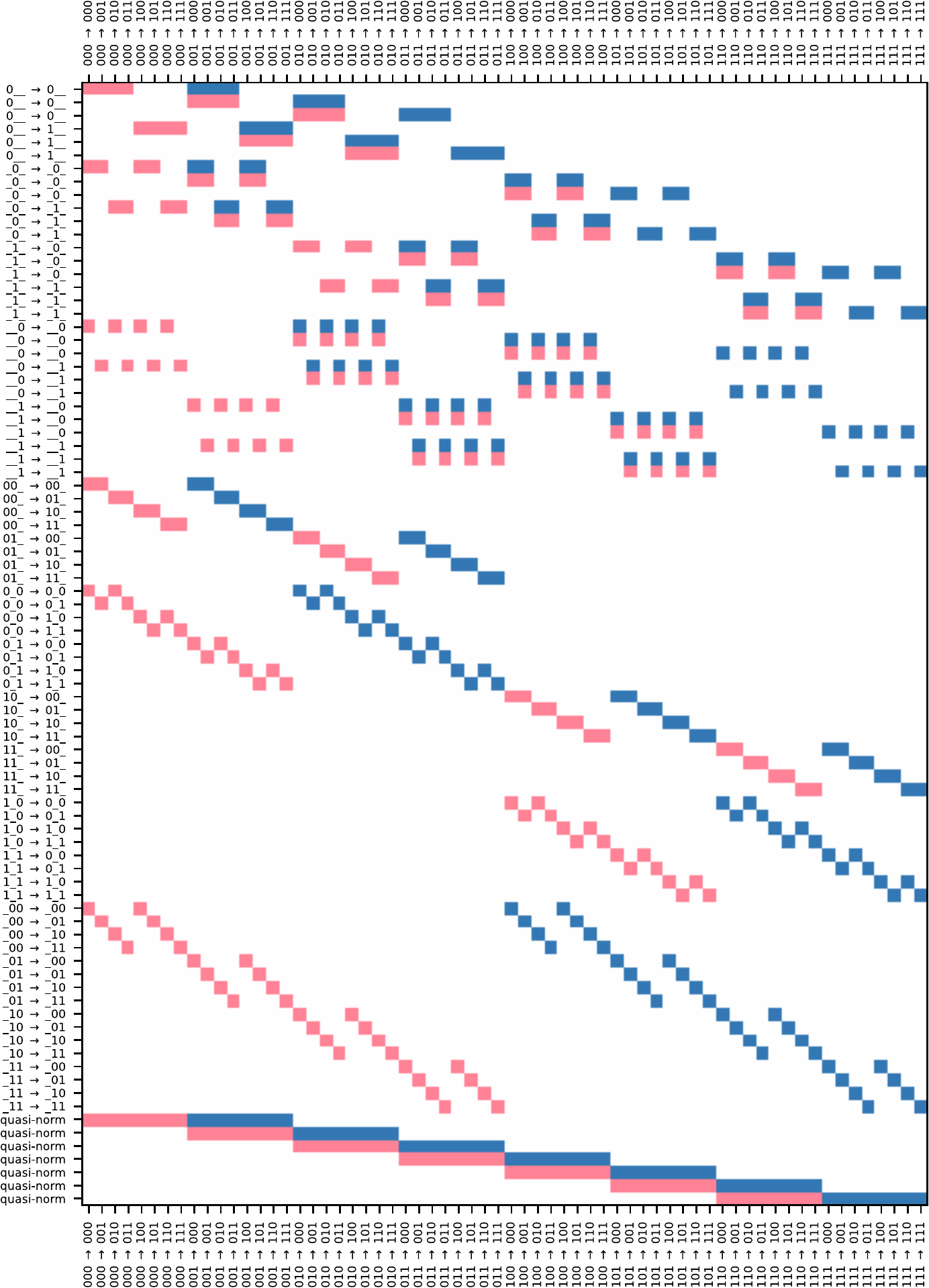}
\end{center}

\noindent Rows correspond to the 85 linear equations, of which 37 are independent.

\newpage
\subsection*{Space 2}

Space 2 is not induced by a causal order, but it is a refinement of the space in equivalence class 33 induced by the definite causal order $\total{\ev{C},\ev{B}}\vee\discrete{\ev{A}}$ (note that the space induced by the order is not the same as space 33).
Its equivalence class under event-input permutation symmetry contains 24 spaces.
Space 2 differs as follows from the space induced by causal order $\total{\ev{C},\ev{B}}\vee\discrete{\ev{A}}$:
\begin{itemize}
  \item The outputs at events \evset{\ev{A}, \ev{B}} are independent of the input at event \ev{C} when the inputs at events \evset{A, B} are given by \hist{A/0,B/0}, \hist{A/0,B/1} and \hist{A/1,B/0}.
  \item The output at event \ev{B} is independent of the input at event \ev{C} when the input at event B is given by \hist{B/0}.
\end{itemize}

\noindent Below are the histories and extended histories for space 2: 
\begin{center}
    \begin{tabular}{cc}
    \includegraphics[height=3.5cm]{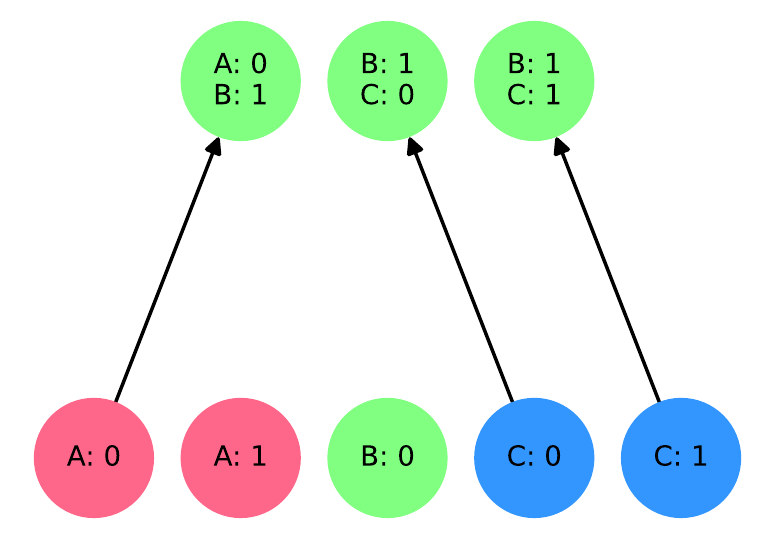}
    &
    \includegraphics[height=3.5cm]{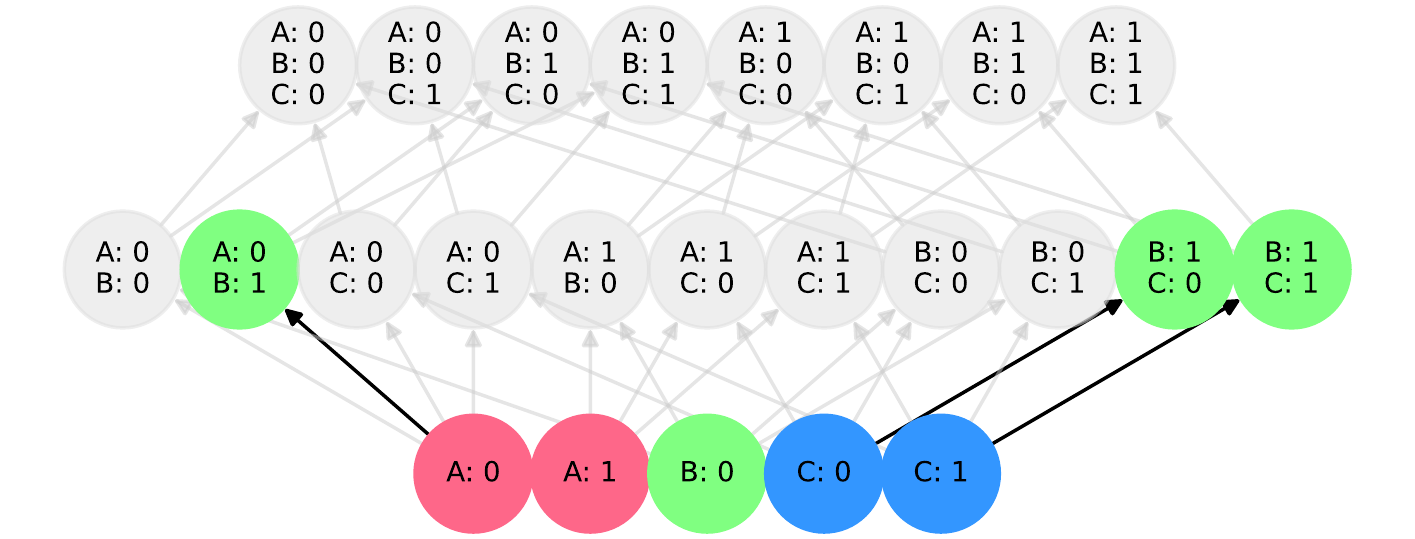}
    \\
    $\Theta_{2}$
    &
    $\Ext{\Theta_{2}}$
    \end{tabular}
\end{center}

\noindent The standard causaltope for Space 2 has dimension 27.
Below is a plot of the homogeneous linear system of causality and quasi-normalisation equations for the standard causaltope, put in reduced row echelon form:

\begin{center}
    \includegraphics[width=11cm]{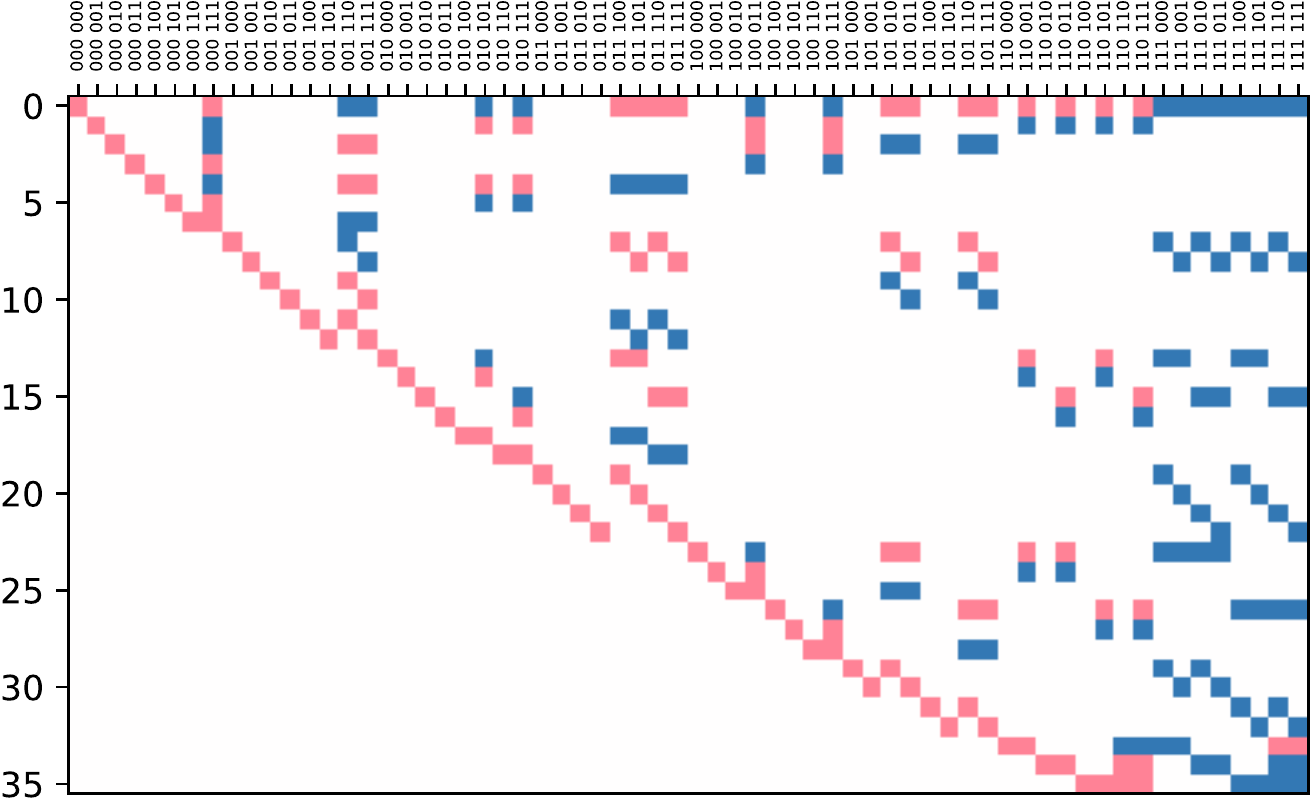}
\end{center}

\noindent Rows correspond to the 36 independent linear equations.
Columns in the plot correspond to entries of empirical models, indexed as $i_A i_B i_C$ $o_A o_B o_C$.
Coefficients in the equations are color-coded as white=0, red=+1 and blue=-1.

Space 2 has closest refinements in equivalence class 1; 
it does not arise as a nontrivial join in the hierarchy.
It has closest coarsenings in equivalence classes 4, 5, 6 and 7; 
it is the meet of its (closest) coarsenings.
It has 64 causal functions, all of which are causal for at least one of its refinements.
It is not a tight space: for event \ev{B}, a causal function must yield identical output values on input histories \hist{A/0,B/1}, \hist{B/1,C/0} and \hist{B/1,C/1}.

The standard causaltope for Space 2 has 1 more dimension than that of its subspace in equivalence class 1.
The standard causaltope for Space 2 is the meet of the standard causaltopes for its closest coarsenings.
For completeness, below is a plot of the full homogeneous linear system of causality and quasi-normalisation equations for the standard causaltope:

\begin{center}
    \includegraphics[width=12cm]{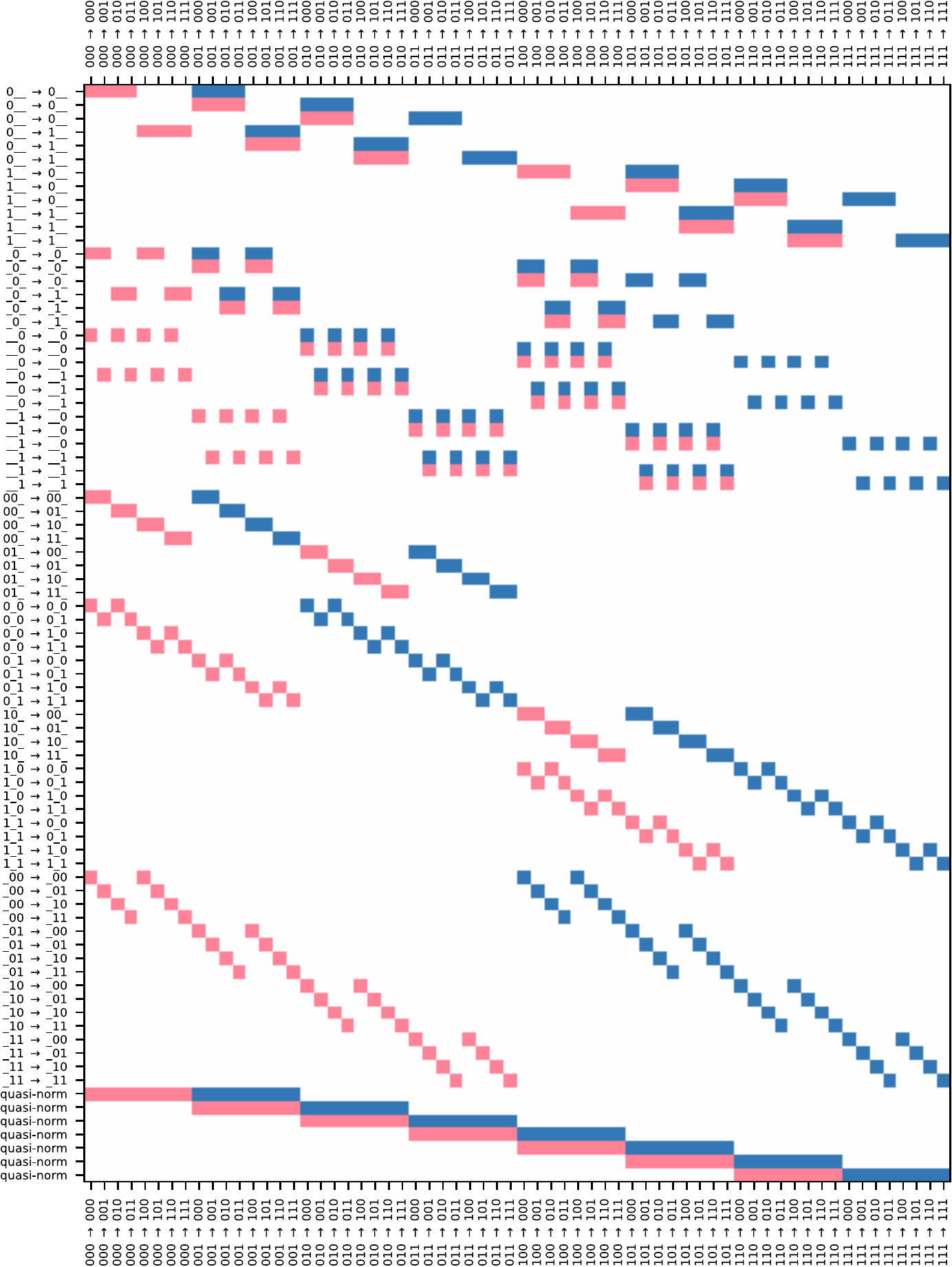}
\end{center}

\noindent Rows correspond to the 81 linear equations, of which 36 are independent.

\newpage
\subsection*{Space 3}

Space 3 is not induced by a causal order, but it is a refinement of the space 33 induced by the definite causal order $\total{\ev{A},\ev{B}}\vee\discrete{\ev{C}}$.
Its equivalence class under event-input permutation symmetry contains 3 spaces.
Space 3 differs as follows from the space induced by causal order $\total{\ev{A},\ev{B}}\vee\discrete{\ev{C}}$:
\begin{itemize}
  \item The outputs at events \evset{\ev{B}, \ev{C}} are independent of the input at event \ev{A} when the inputs at events \evset{B, C} are given by \hist{B/1,C/0}, \hist{B/1,C/1}, \hist{B/0,C/0} and \hist{B/0,C/1}.
\end{itemize}

\noindent Below are the histories and extended histories for space 3: 
\begin{center}
    \begin{tabular}{cc}
    \includegraphics[height=3.5cm]{svg-inkscape/space-ABC-unique-untight-3-highlighted_svg-tex.pdf}
    &
    \includegraphics[height=3.5cm]{svg-inkscape/space-ABC-unique-untight-3-ext-highlighted_svg-tex.pdf}
    \\
    $\Theta_{3}$
    &
    $\Ext{\Theta_{3}}$
    \end{tabular}
\end{center}

\noindent The standard causaltope for Space 3 has dimension 26.
Below is a plot of the homogeneous linear system of causality and quasi-normalisation equations for the standard causaltope, put in reduced row echelon form:

\begin{center}
    \includegraphics[width=11cm]{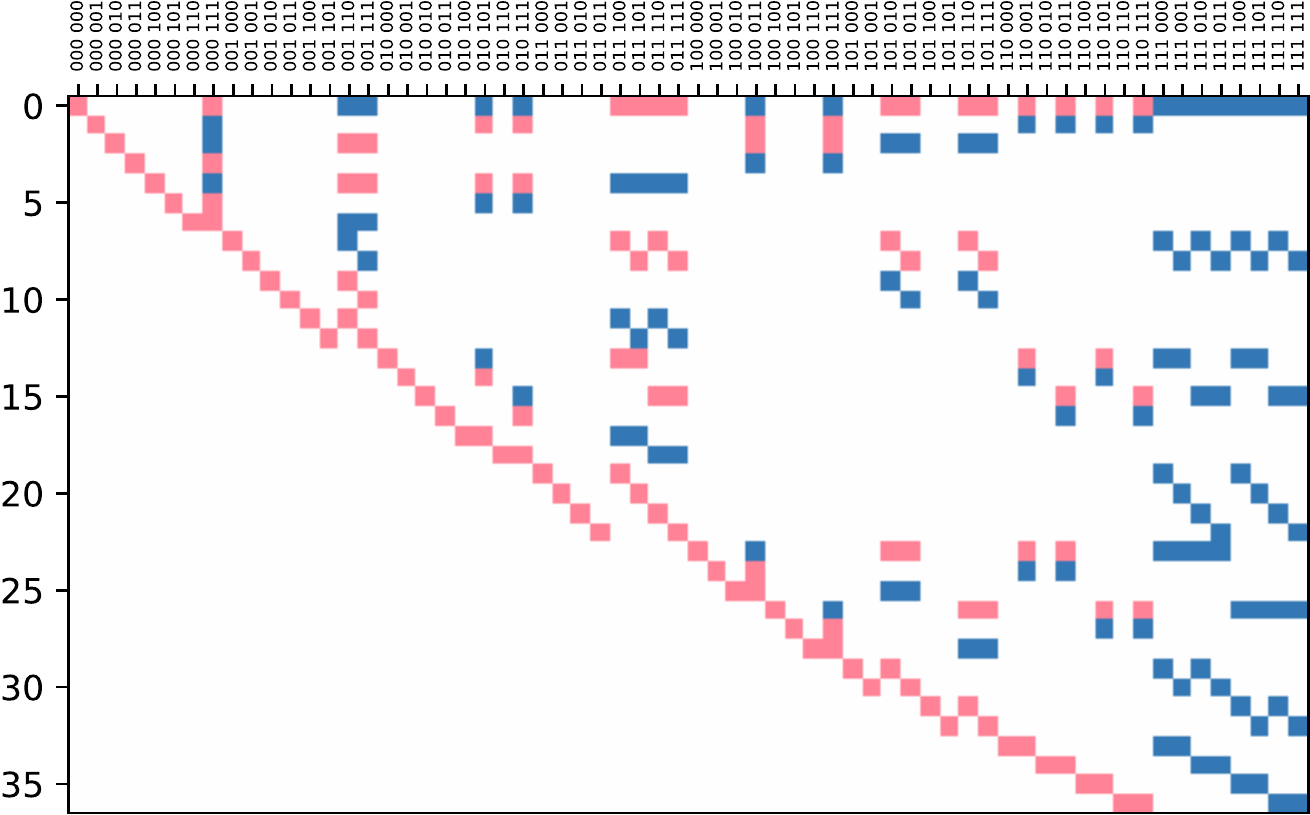}
\end{center}

\noindent Rows correspond to the 37 independent linear equations.
Columns in the plot correspond to entries of empirical models, indexed as $i_A i_B i_C$ $o_A o_B o_C$.
Coefficients in the equations are color-coded as white=0, red=+1 and blue=-1.

Space 3 has closest refinements in equivalence class 1; 
it is the join of its (closest) refinements.
It has closest coarsenings in equivalence class 6; 
it is the meet of its (closest) coarsenings.
It has 64 causal functions, all of which are causal for at least one of its refinements.
It is not a tight space: for event \ev{B}, a causal function must yield identical output values on input histories \hist{A/0,B/0}, \hist{A/1,B/0}, \hist{B/0,C/0} and \hist{B/0,C/1}, and it must also yield identical output values on input histories \hist{A/0,B/1}, \hist{A/1,B/1}, \hist{B/1,C/0} and \hist{B/1,C/1}.

The standard causaltope for Space 3 coincides with that of its 2 subspaces in equivalence class 1.
The standard causaltope for Space 3 is the meet of the standard causaltopes for its closest coarsenings.
For completeness, below is a plot of the full homogeneous linear system of causality and quasi-normalisation equations for the standard causaltope:

\begin{center}
    \includegraphics[width=12cm]{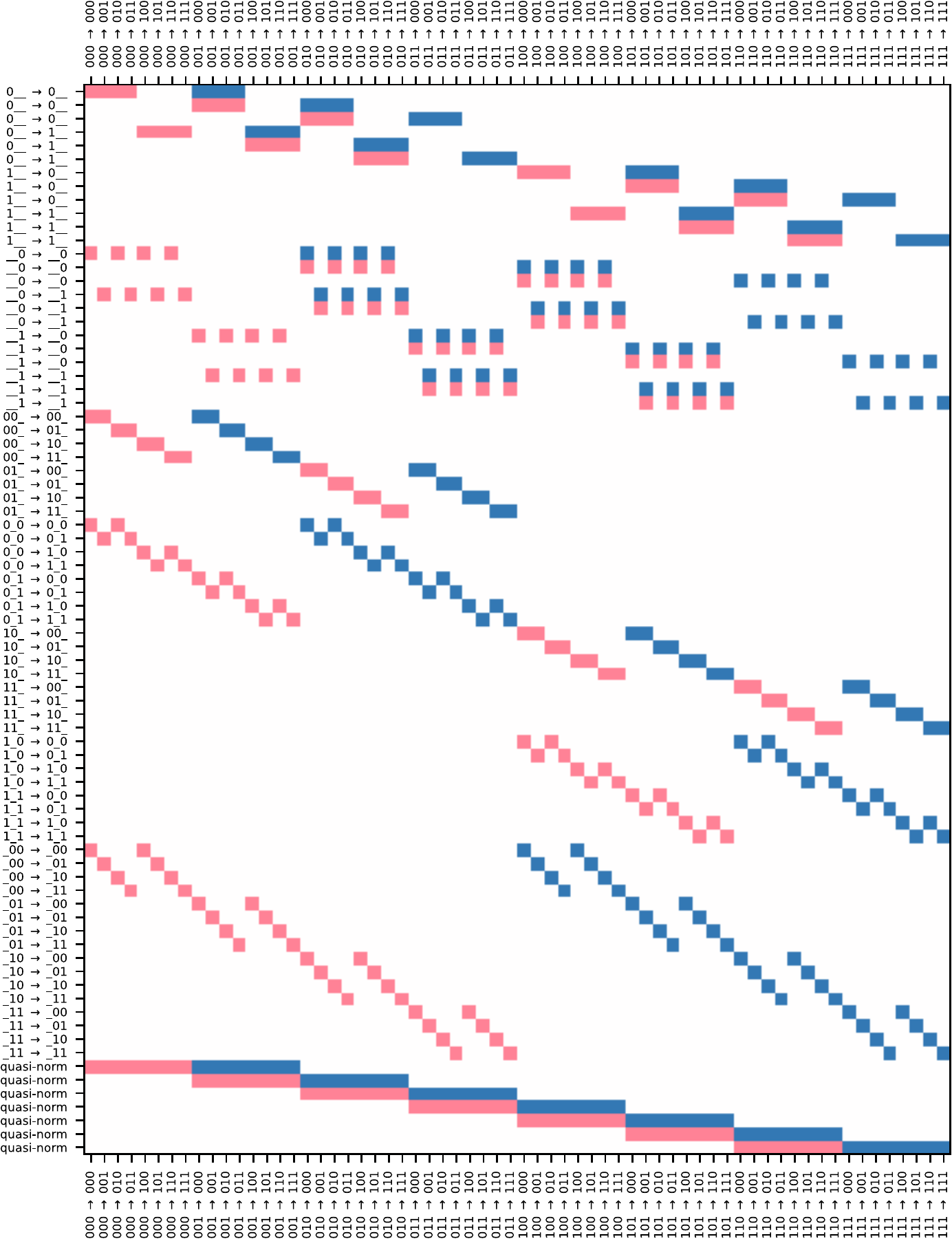}
\end{center}

\noindent Rows correspond to the 79 linear equations, of which 37 are independent.

\newpage
\subsection*{Space 4}

Space 4 is not induced by a causal order, but it is a refinement of the space in equivalence class 92 induced by the definite causal order $\total{\ev{A},\ev{B}}\vee\total{\ev{C},\ev{B}}$ (note that the space induced by the order is not the same as space 92).
Its equivalence class under event-input permutation symmetry contains 24 spaces.
Space 4 differs as follows from the space induced by causal order $\total{\ev{A},\ev{B}}\vee\total{\ev{C},\ev{B}}$:
\begin{itemize}
  \item The outputs at events \evset{\ev{A}, \ev{B}} are independent of the input at event \ev{C} when the inputs at events \evset{A, B} are given by \hist{A/0,B/0}, \hist{A/0,B/1} and \hist{A/1,B/0}.
  \item The outputs at events \evset{\ev{B}, \ev{C}} are independent of the input at event \ev{A} when the inputs at events \evset{B, C} are given by \hist{B/1,C/1}, \hist{B/0,C/0} and \hist{B/0,C/1}.
  \item The output at event \ev{B} is independent of the inputs at events \evset{\ev{A}, \ev{C}} when the input at event B is given by \hist{B/0}.
\end{itemize}

\noindent Below are the histories and extended histories for space 4: 
\begin{center}
    \begin{tabular}{cc}
    \includegraphics[height=3.5cm]{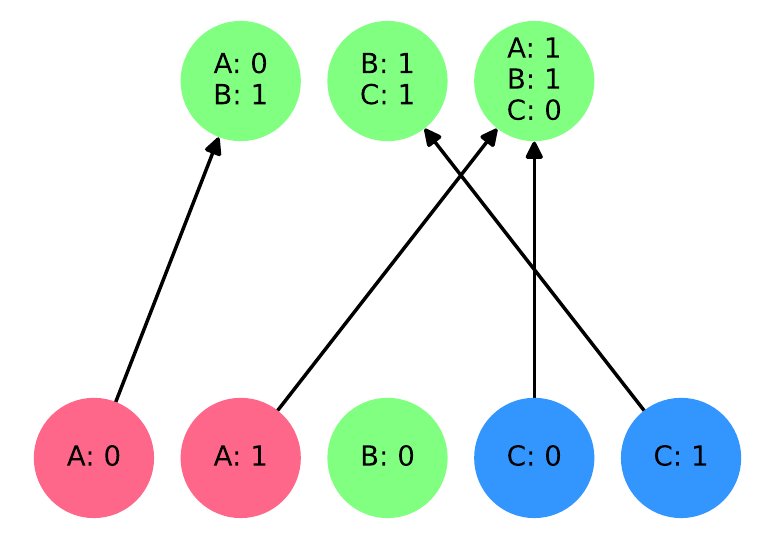}
    &
    \includegraphics[height=3.5cm]{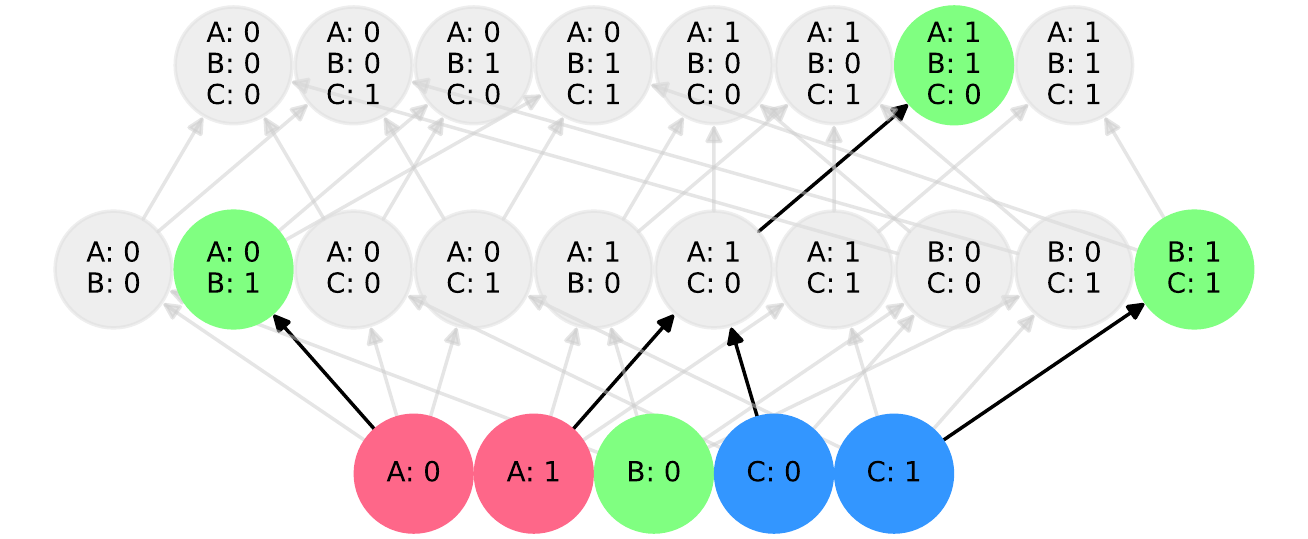}
    \\
    $\Theta_{4}$
    &
    $\Ext{\Theta_{4}}$
    \end{tabular}
\end{center}

\noindent The standard causaltope for Space 4 has dimension 29.
Below is a plot of the homogeneous linear system of causality and quasi-normalisation equations for the standard causaltope, put in reduced row echelon form:

\begin{center}
    \includegraphics[width=11cm]{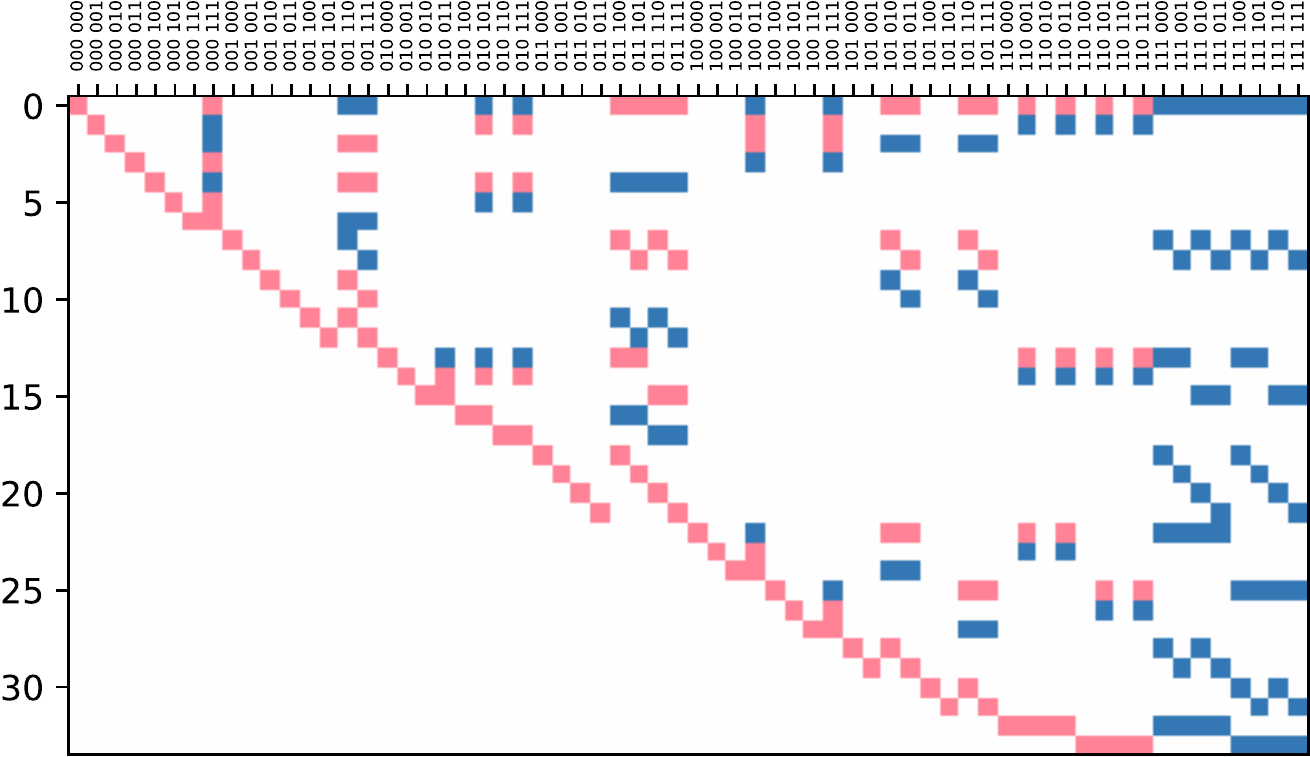}
\end{center}

\noindent Rows correspond to the 34 independent linear equations.
Columns in the plot correspond to entries of empirical models, indexed as $i_A i_B i_C$ $o_A o_B o_C$.
Coefficients in the equations are color-coded as white=0, red=+1 and blue=-1.

Space 4 has closest refinements in equivalence class 2; 
it is the join of its (closest) refinements.
It has closest coarsenings in equivalence classes 8, 11 and 15; 
it is the meet of its (closest) coarsenings.
It has 128 causal functions, 64 of which are not causal for any of its refinements.
It is not a tight space: for event \ev{B}, a causal function must yield identical output values on input histories \hist{A/0,B/1} and \hist{B/1,C/1}.

The standard causaltope for Space 4 has 2 more dimensions than those of its 2 subspaces in equivalence class 2.
The standard causaltope for Space 4 is the meet of the standard causaltopes for its closest coarsenings.
For completeness, below is a plot of the full homogeneous linear system of causality and quasi-normalisation equations for the standard causaltope:

\begin{center}
    \includegraphics[width=12cm]{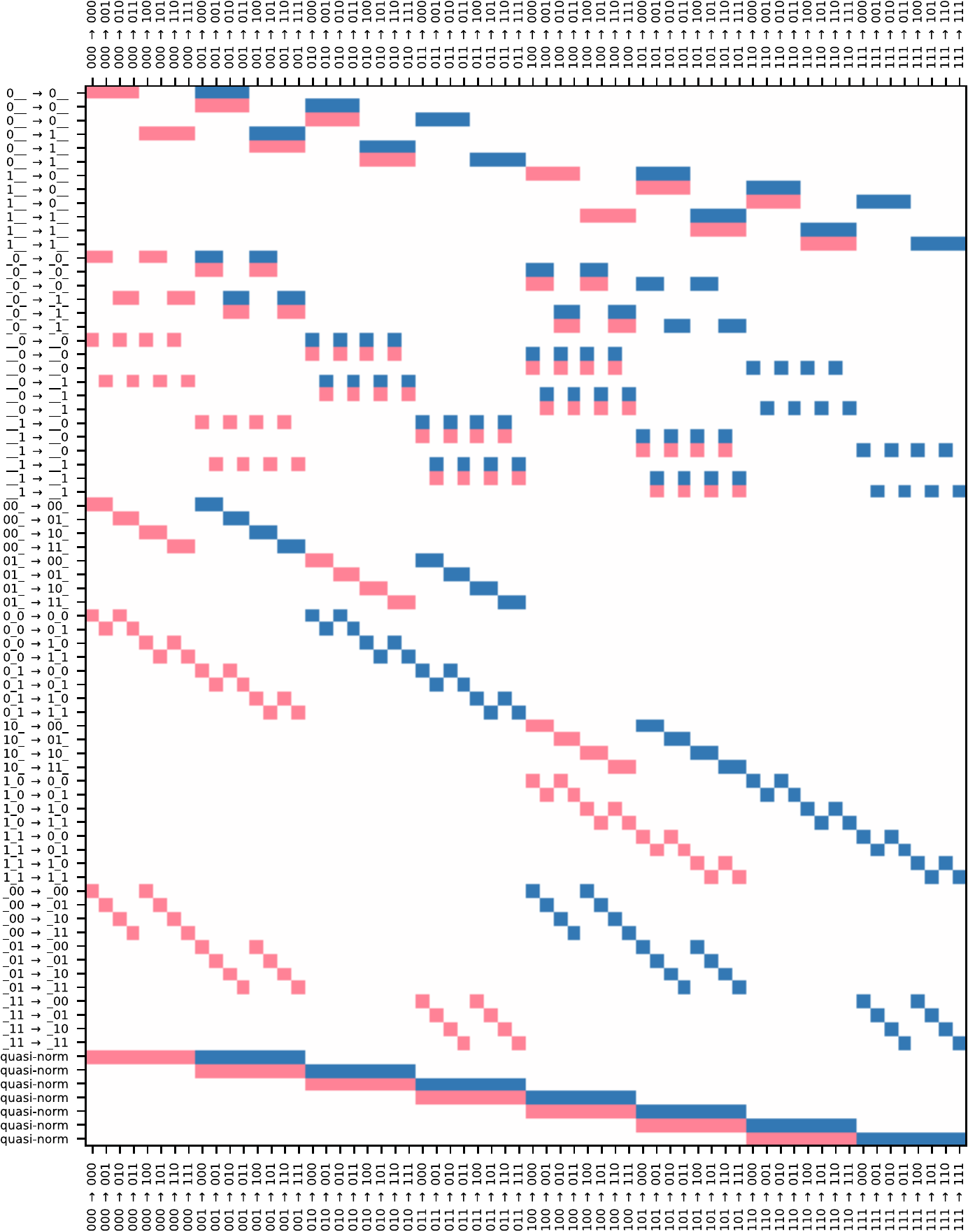}
\end{center}

\noindent Rows correspond to the 77 linear equations, of which 34 are independent.

\newpage
\subsection*{Space 5}

Space 5 is not induced by a causal order, but it is a refinement of the space 33 induced by the definite causal order $\total{\ev{A},\ev{B}}\vee\discrete{\ev{C}}$.
Its equivalence class under event-input permutation symmetry contains 12 spaces.
Space 5 differs as follows from the space induced by causal order $\total{\ev{A},\ev{B}}\vee\discrete{\ev{C}}$:
\begin{itemize}
  \item The outputs at events \evset{\ev{B}, \ev{C}} are independent of the input at event \ev{A} when the inputs at events \evset{B, C} are given by \hist{B/1,C/0} and \hist{B/1,C/1}.
  \item The output at event \ev{B} is independent of the input at event \ev{A} when the input at event B is given by \hist{B/1}.
\end{itemize}

\noindent Below are the histories and extended histories for space 5: 
\begin{center}
    \begin{tabular}{cc}
    \includegraphics[height=3.5cm]{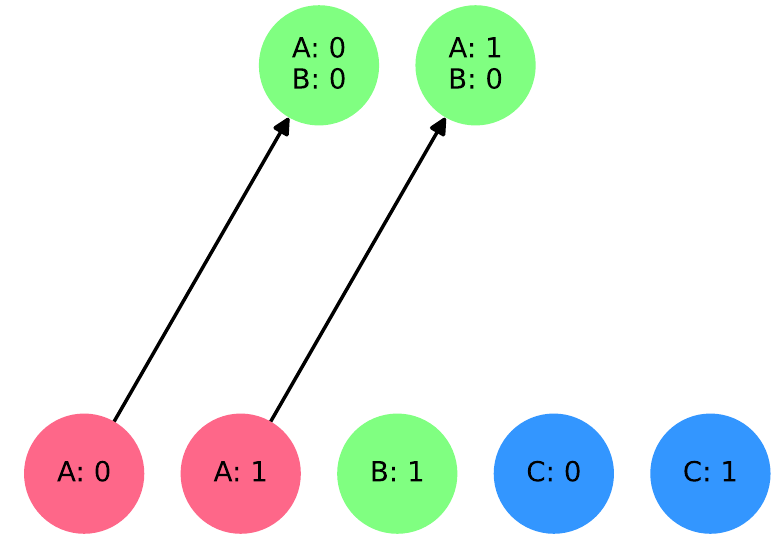}
    &
    \includegraphics[height=3.5cm]{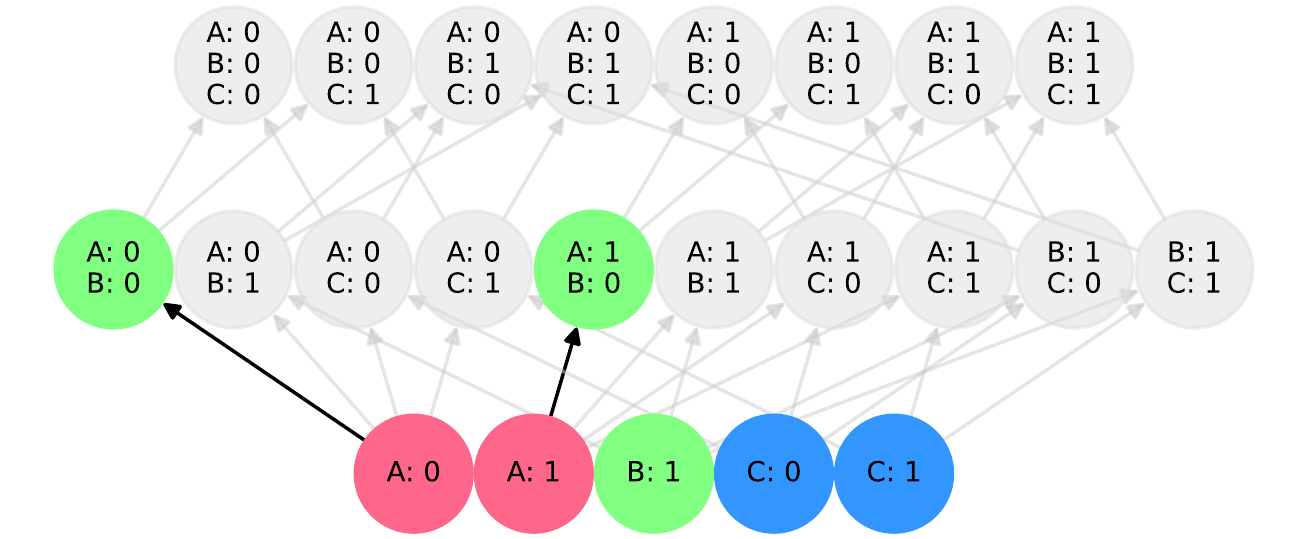}
    \\
    $\Theta_{5}$
    &
    $\Ext{\Theta_{5}}$
    \end{tabular}
\end{center}

\noindent The standard causaltope for Space 5 has dimension 29.
Below is a plot of the homogeneous linear system of causality and quasi-normalisation equations for the standard causaltope, put in reduced row echelon form:

\begin{center}
    \includegraphics[width=11cm]{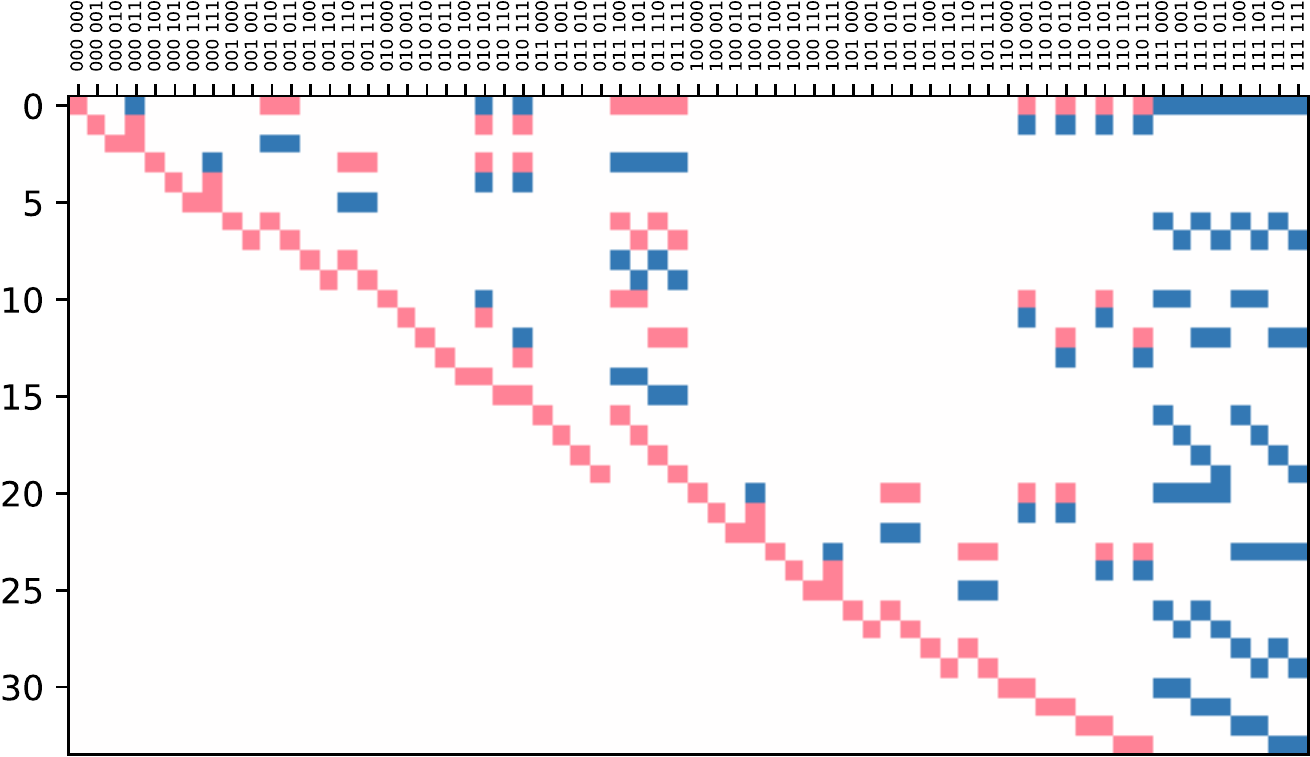}
\end{center}

\noindent Rows correspond to the 34 independent linear equations.
Columns in the plot correspond to entries of empirical models, indexed as $i_A i_B i_C$ $o_A o_B o_C$.
Coefficients in the equations are color-coded as white=0, red=+1 and blue=-1.

Space 5 has closest refinements in equivalence class 2; 
it is the join of its (closest) refinements.
It has closest coarsenings in equivalence classes 8, 12 and 14; 
it is the meet of its (closest) coarsenings.
It has 128 causal functions, 64 of which are not causal for any of its refinements.
It is a tight space.

The standard causaltope for Space 5 has 2 more dimensions than those of its 2 subspaces in equivalence class 2.
The standard causaltope for Space 5 is the meet of the standard causaltopes for its closest coarsenings.
For completeness, below is a plot of the full homogeneous linear system of causality and quasi-normalisation equations for the standard causaltope:

\begin{center}
    \includegraphics[width=12cm]{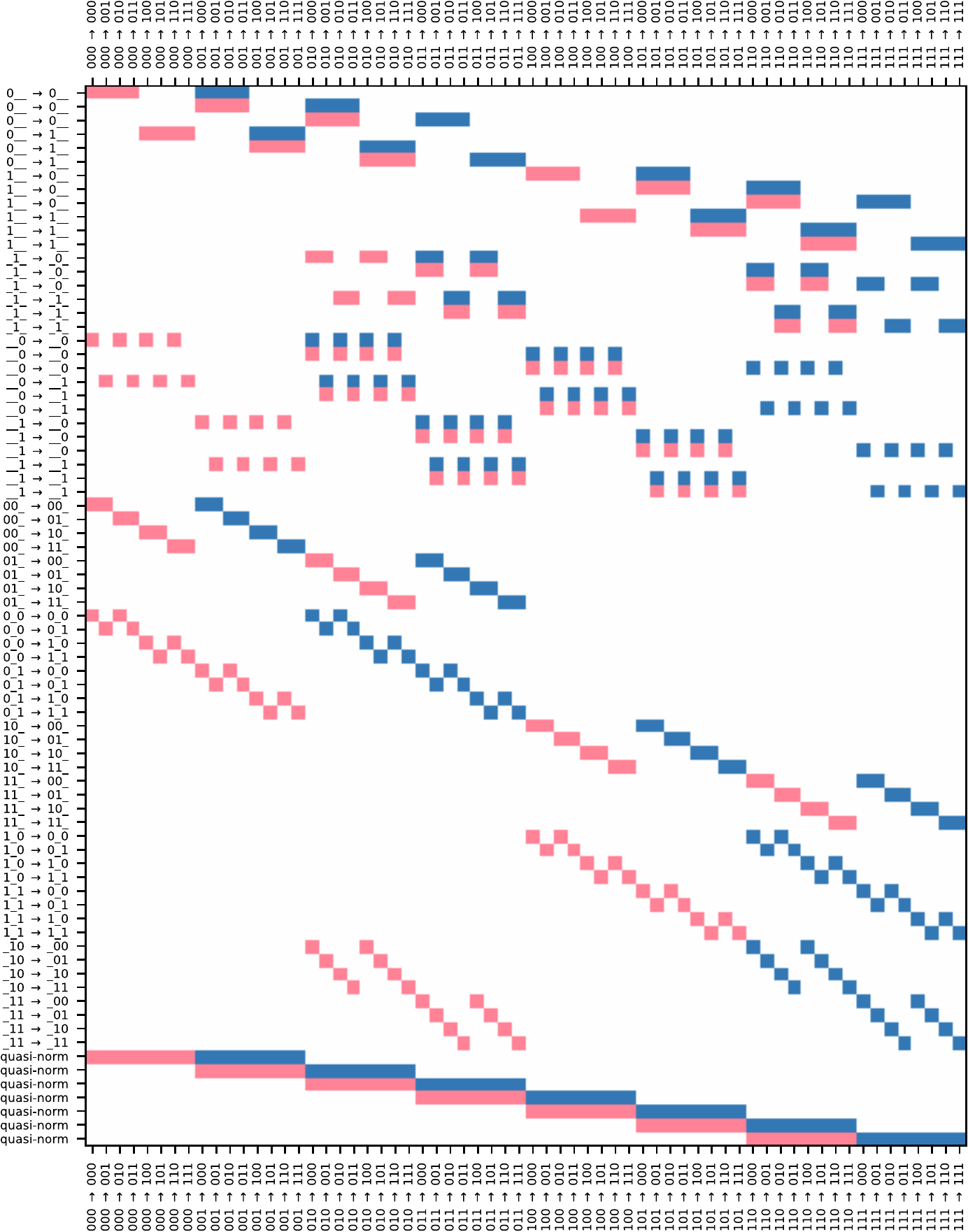}
\end{center}

\noindent Rows correspond to the 77 linear equations, of which 34 are independent.

\newpage
\subsection*{Space 6}

Space 6 is not induced by a causal order, but it is a refinement of the space 33 induced by the definite causal order $\total{\ev{A},\ev{B}}\vee\discrete{\ev{C}}$.
Its equivalence class under event-input permutation symmetry contains 24 spaces.
Space 6 differs as follows from the space induced by causal order $\total{\ev{A},\ev{B}}\vee\discrete{\ev{C}}$:
\begin{itemize}
  \item The outputs at events \evset{\ev{B}, \ev{C}} are independent of the input at event \ev{A} when the inputs at events \evset{B, C} are given by \hist{B/1,C/0}, \hist{B/1,C/1} and \hist{B/0,C/1}.
\end{itemize}

\noindent Below are the histories and extended histories for space 6: 
\begin{center}
    \begin{tabular}{cc}
    \includegraphics[height=3.5cm]{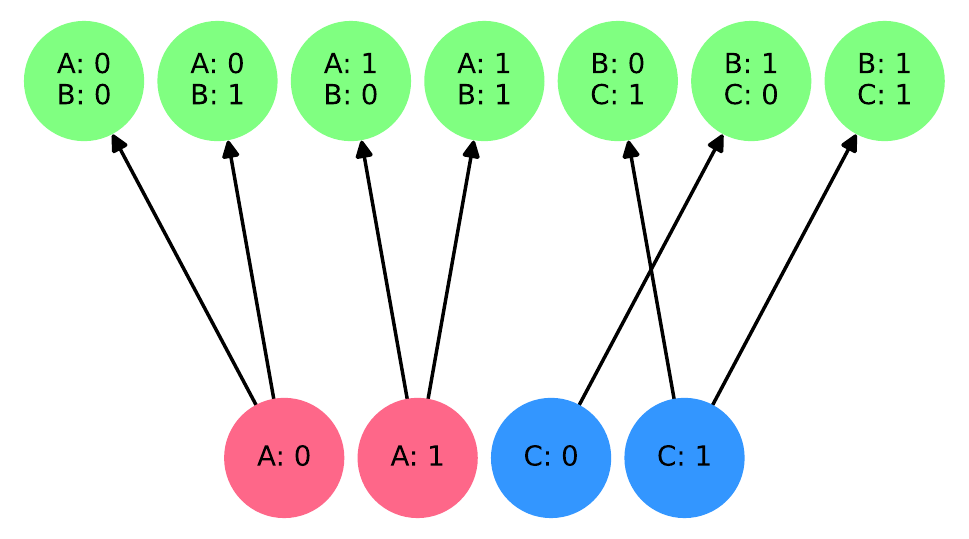}
    &
    \includegraphics[height=3.5cm]{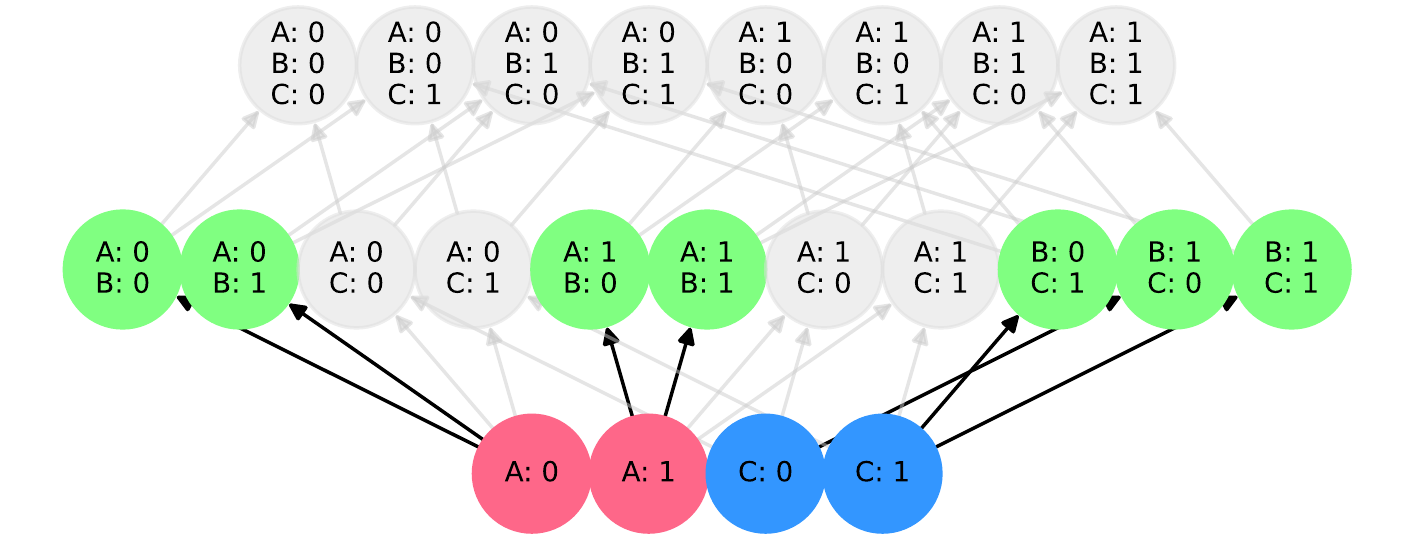}
    \\
    $\Theta_{6}$
    &
    $\Ext{\Theta_{6}}$
    \end{tabular}
\end{center}

\noindent The standard causaltope for Space 6 has dimension 27.
Below is a plot of the homogeneous linear system of causality and quasi-normalisation equations for the standard causaltope, put in reduced row echelon form:

\begin{center}
    \includegraphics[width=11cm]{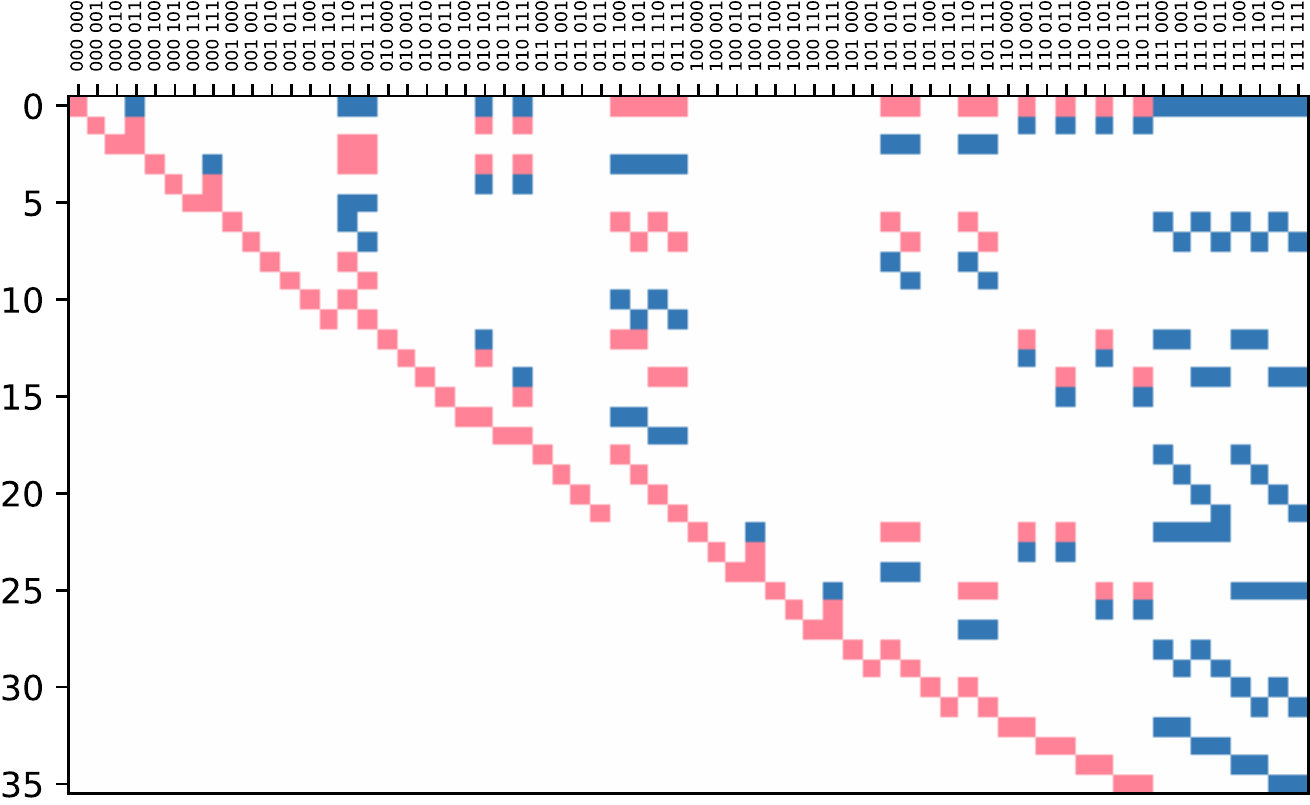}
\end{center}

\noindent Rows correspond to the 36 independent linear equations.
Columns in the plot correspond to entries of empirical models, indexed as $i_A i_B i_C$ $o_A o_B o_C$.
Coefficients in the equations are color-coded as white=0, red=+1 and blue=-1.

Space 6 has closest refinements in equivalence classes 2 and 3; 
it is the join of its (closest) refinements.
It has closest coarsenings in equivalence classes 9, 10, 13, 14 and 15; 
it is the meet of its (closest) coarsenings.
It has 64 causal functions, all of which are causal for at least one of its refinements.
It is not a tight space: for event \ev{B}, a causal function must yield identical output values on input histories \hist{A/0,B/0}, \hist{A/1,B/0} and \hist{B/0,C/1}, and it must also yield identical output values on input histories \hist{A/0,B/1}, \hist{A/1,B/1}, \hist{B/1,C/0} and \hist{B/1,C/1}.

The standard causaltope for Space 6 coincides with that of its subspace in equivalence class 2.
The standard causaltope for Space 6 is the meet of the standard causaltopes for its closest coarsenings.
For completeness, below is a plot of the full homogeneous linear system of causality and quasi-normalisation equations for the standard causaltope:

\begin{center}
    \includegraphics[width=12cm]{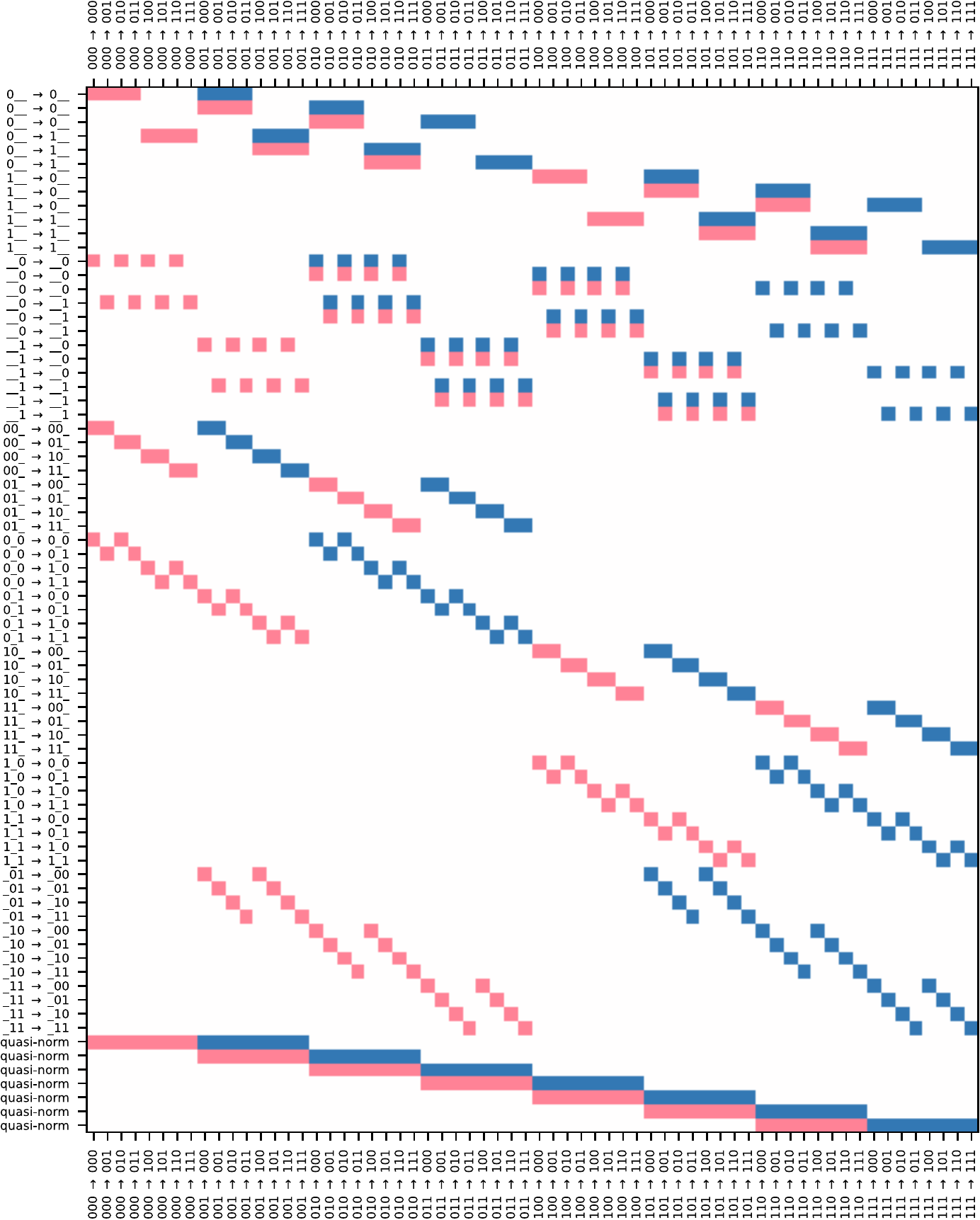}
\end{center}

\noindent Rows correspond to the 75 linear equations, of which 36 are independent.

\newpage
\subsection*{Space 7}

Space 7 is not induced by a causal order, but it is a refinement of the space in equivalence class 77 induced by the definite causal order $\total{\ev{C},\ev{A}}\vee\total{\ev{C},\ev{B}}$ (note that the space induced by the order is not the same as space 77).
Its equivalence class under event-input permutation symmetry contains 12 spaces.
Space 7 differs as follows from the space induced by causal order $\total{\ev{C},\ev{A}}\vee\total{\ev{C},\ev{B}}$:
\begin{itemize}
  \item The outputs at events \evset{\ev{A}, \ev{B}} are independent of the input at event \ev{C} when the inputs at events \evset{A, B} are given by \hist{A/0,B/0}, \hist{A/0,B/1} and \hist{A/1,B/0}.
  \item The output at event \ev{B} is independent of the input at event \ev{C} when the input at event B is given by \hist{B/0}.
  \item The output at event \ev{A} is independent of the input at event \ev{C} when the input at event A is given by \hist{A/0}.
\end{itemize}

\noindent Below are the histories and extended histories for space 7: 
\begin{center}
    \begin{tabular}{cc}
    \includegraphics[height=3.5cm]{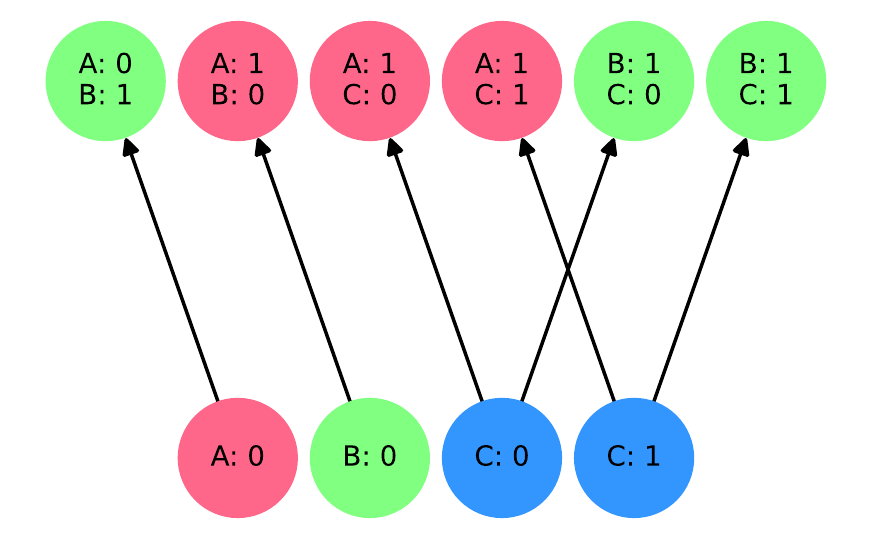}
    &
    \includegraphics[height=3.5cm]{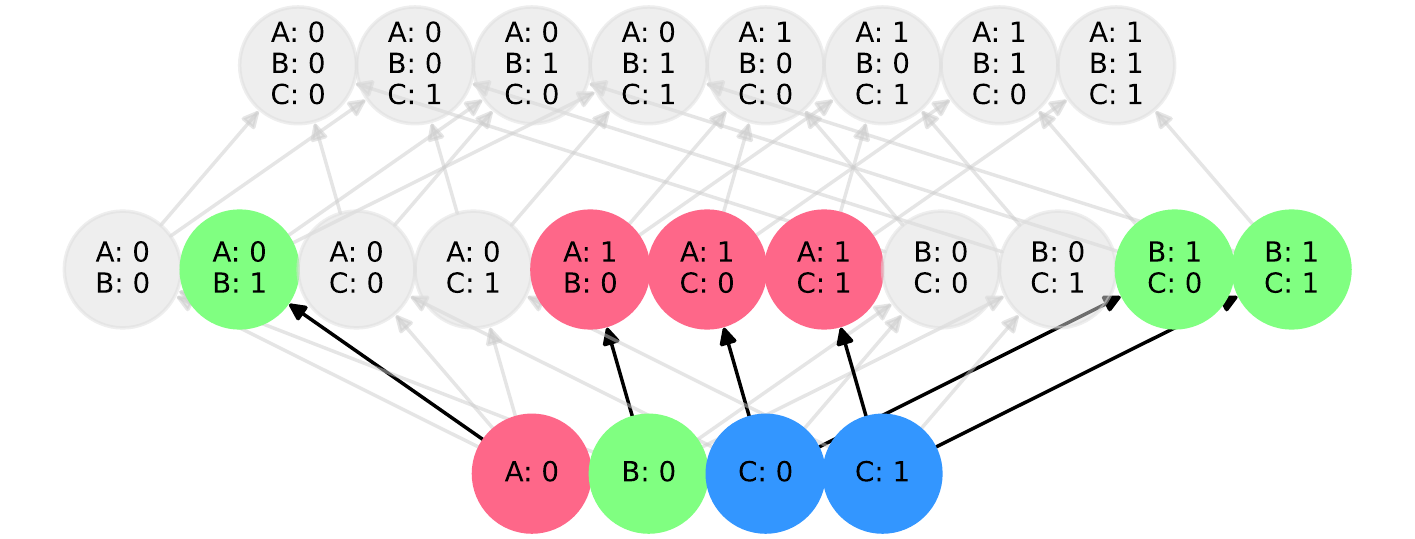}
    \\
    $\Theta_{7}$
    &
    $\Ext{\Theta_{7}}$
    \end{tabular}
\end{center}

\noindent The standard causaltope for Space 7 has dimension 27.
Below is a plot of the homogeneous linear system of causality and quasi-normalisation equations for the standard causaltope, put in reduced row echelon form:

\begin{center}
    \includegraphics[width=11cm]{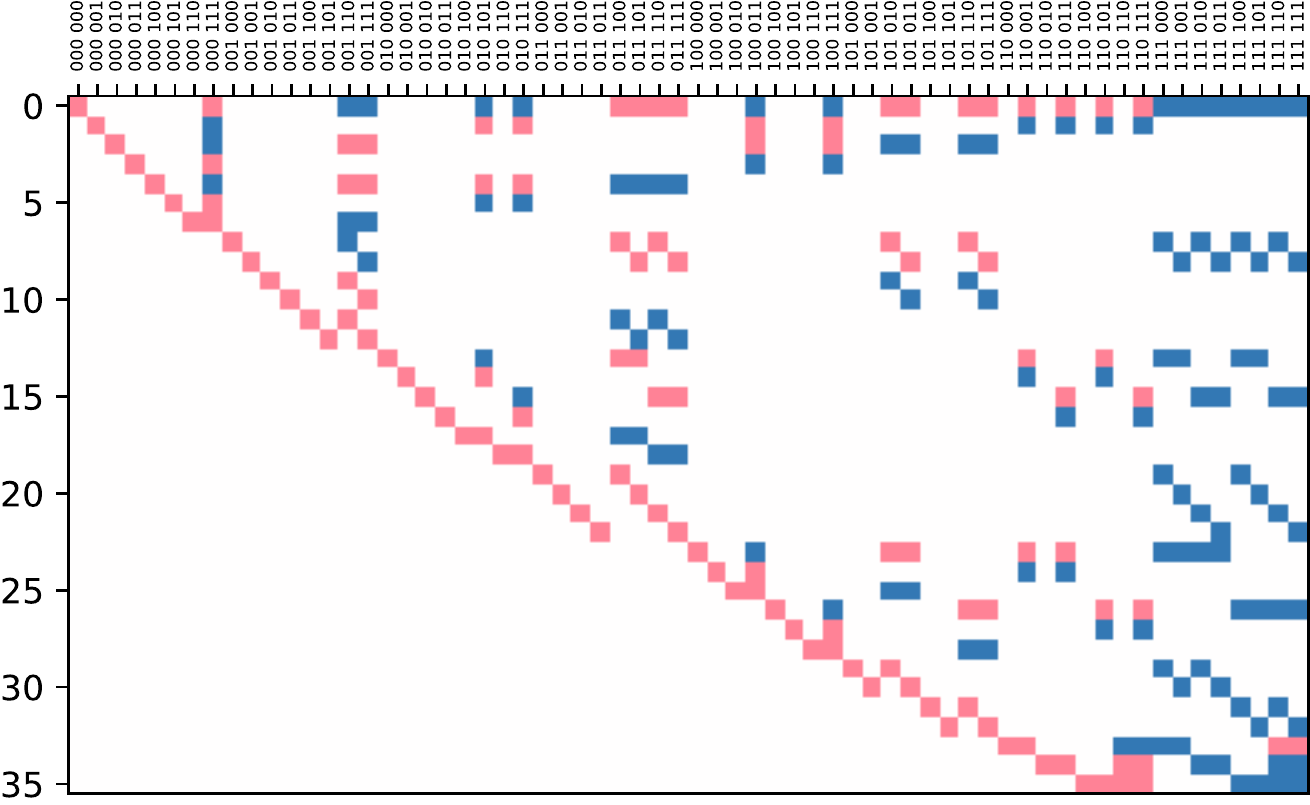}
\end{center}

\noindent Rows correspond to the 36 independent linear equations.
Columns in the plot correspond to entries of empirical models, indexed as $i_A i_B i_C$ $o_A o_B o_C$.
Coefficients in the equations are color-coded as white=0, red=+1 and blue=-1.

Space 7 has closest refinements in equivalence class 2; 
it is the join of its (closest) refinements.
It has closest coarsenings in equivalence classes 11 and 12; 
it is the meet of its (closest) coarsenings.
It has 64 causal functions, all of which are causal for at least one of its refinements.
It is not a tight space: for event \ev{A}, a causal function must yield identical output values on input histories \hist{A/1,B/0}, \hist{A/1,C/0} and \hist{A/1,C/1}; for event \ev{B}, a causal function must yield identical output values on input histories \hist{A/0,B/1}, \hist{B/1,C/0} and \hist{B/1,C/1}.

The standard causaltope for Space 7 coincides with that of its 2 subspaces in equivalence class 2.
The standard causaltope for Space 7 is the meet of the standard causaltopes for its closest coarsenings.
For completeness, below is a plot of the full homogeneous linear system of causality and quasi-normalisation equations for the standard causaltope:

\begin{center}
    \includegraphics[width=12cm]{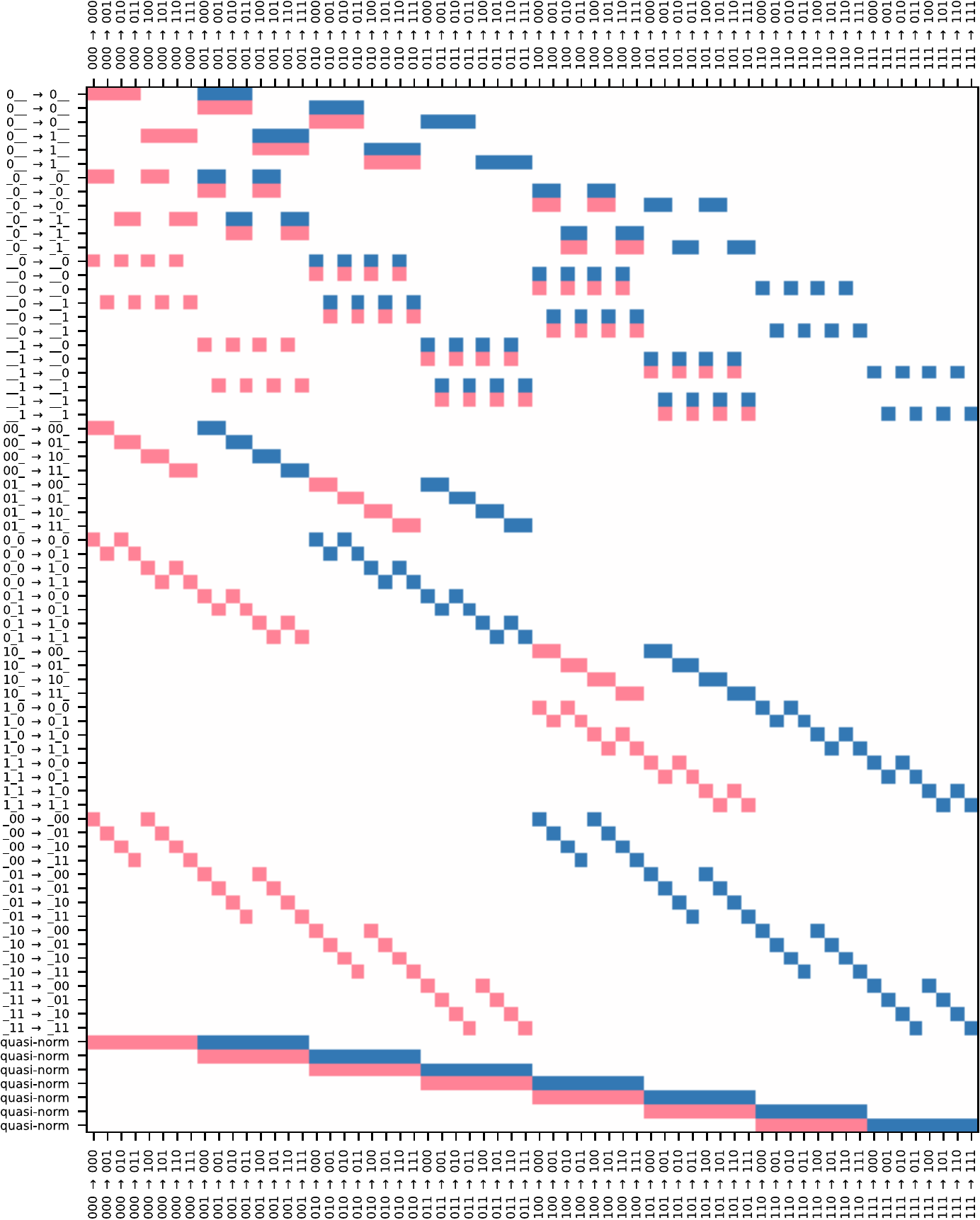}
\end{center}

\noindent Rows correspond to the 75 linear equations, of which 36 are independent.

\newpage
\subsection*{Space 8}

Space 8 is not induced by a causal order, but it is a refinement of the space in equivalence class 92 induced by the definite causal order $\total{\ev{A},\ev{B}}\vee\total{\ev{C},\ev{B}}$ (note that the space induced by the order is not the same as space 92).
Its equivalence class under event-input permutation symmetry contains 24 spaces.
Space 8 differs as follows from the space induced by causal order $\total{\ev{A},\ev{B}}\vee\total{\ev{C},\ev{B}}$:
\begin{itemize}
  \item The outputs at events \evset{\ev{A}, \ev{B}} are independent of the input at event \ev{C} when the inputs at events \evset{A, B} are given by \hist{A/0,B/0}, \hist{A/0,B/1} and \hist{A/1,B/0}.
  \item The outputs at events \evset{\ev{B}, \ev{C}} are independent of the input at event \ev{A} when the inputs at events \evset{B, C} are given by \hist{B/0,C/0} and \hist{B/0,C/1}.
  \item The output at event \ev{B} is independent of the inputs at events \evset{\ev{A}, \ev{C}} when the input at event B is given by \hist{B/0}.
\end{itemize}

\noindent Below are the histories and extended histories for space 8: 
\begin{center}
    \begin{tabular}{cc}
    \includegraphics[height=3.5cm]{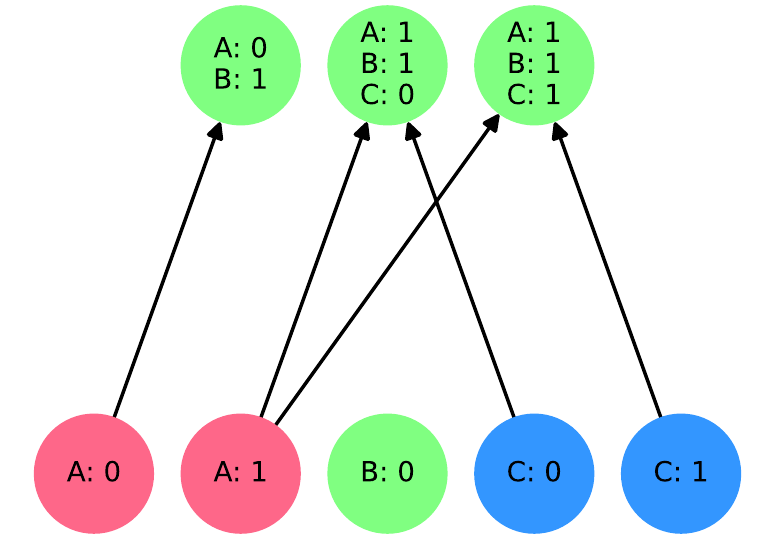}
    &
    \includegraphics[height=3.5cm]{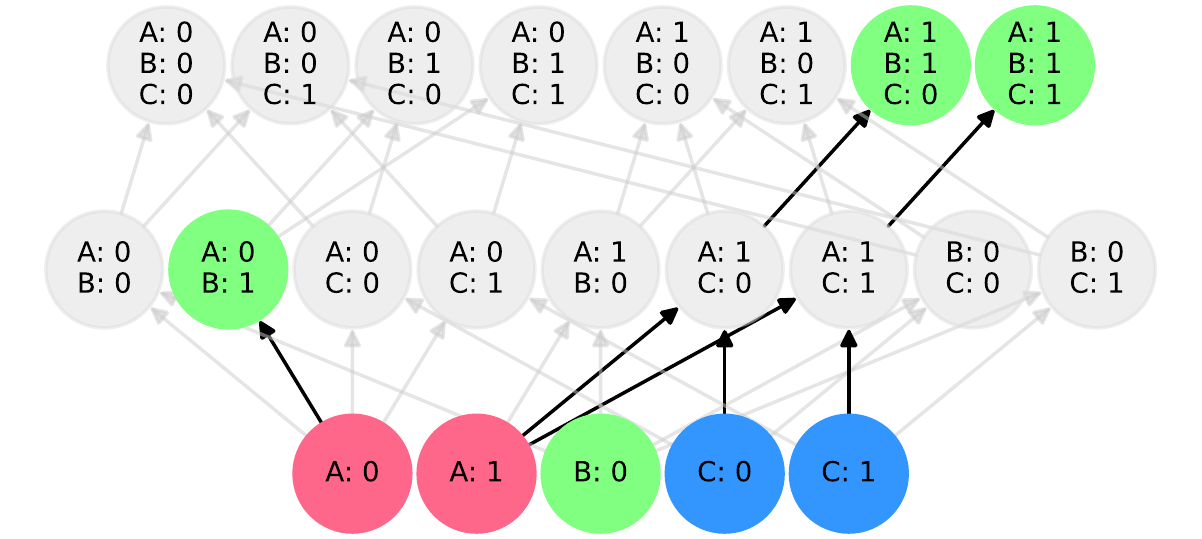}
    \\
    $\Theta_{8}$
    &
    $\Ext{\Theta_{8}}$
    \end{tabular}
\end{center}

\noindent The standard causaltope for Space 8 has dimension 31.
Below is a plot of the homogeneous linear system of causality and quasi-normalisation equations for the standard causaltope, put in reduced row echelon form:

\begin{center}
    \includegraphics[width=11cm]{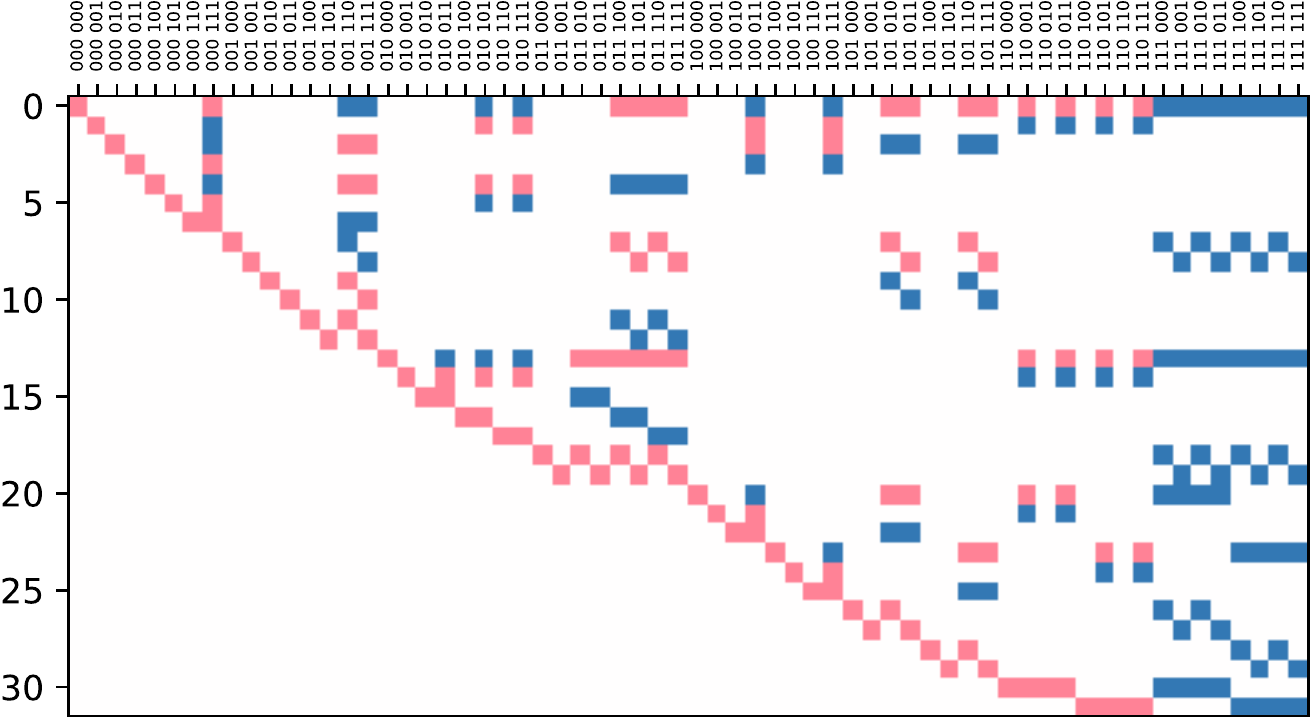}
\end{center}

\noindent Rows correspond to the 32 independent linear equations.
Columns in the plot correspond to entries of empirical models, indexed as $i_A i_B i_C$ $o_A o_B o_C$.
Coefficients in the equations are color-coded as white=0, red=+1 and blue=-1.

Space 8 has closest refinements in equivalence classes 4 and 5; 
it is the join of its (closest) refinements.
It has closest coarsenings in equivalence classes 18, 20, 22 and 26; 
it is the meet of its (closest) coarsenings.
It has 256 causal functions, 128 of which are not causal for any of its refinements.
It is a tight space.

The standard causaltope for Space 8 has 2 more dimensions than those of its 3 subspaces in equivalence classes 4 and 5.
The standard causaltope for Space 8 is the meet of the standard causaltopes for its closest coarsenings.
For completeness, below is a plot of the full homogeneous linear system of causality and quasi-normalisation equations for the standard causaltope:

\begin{center}
    \includegraphics[width=12cm]{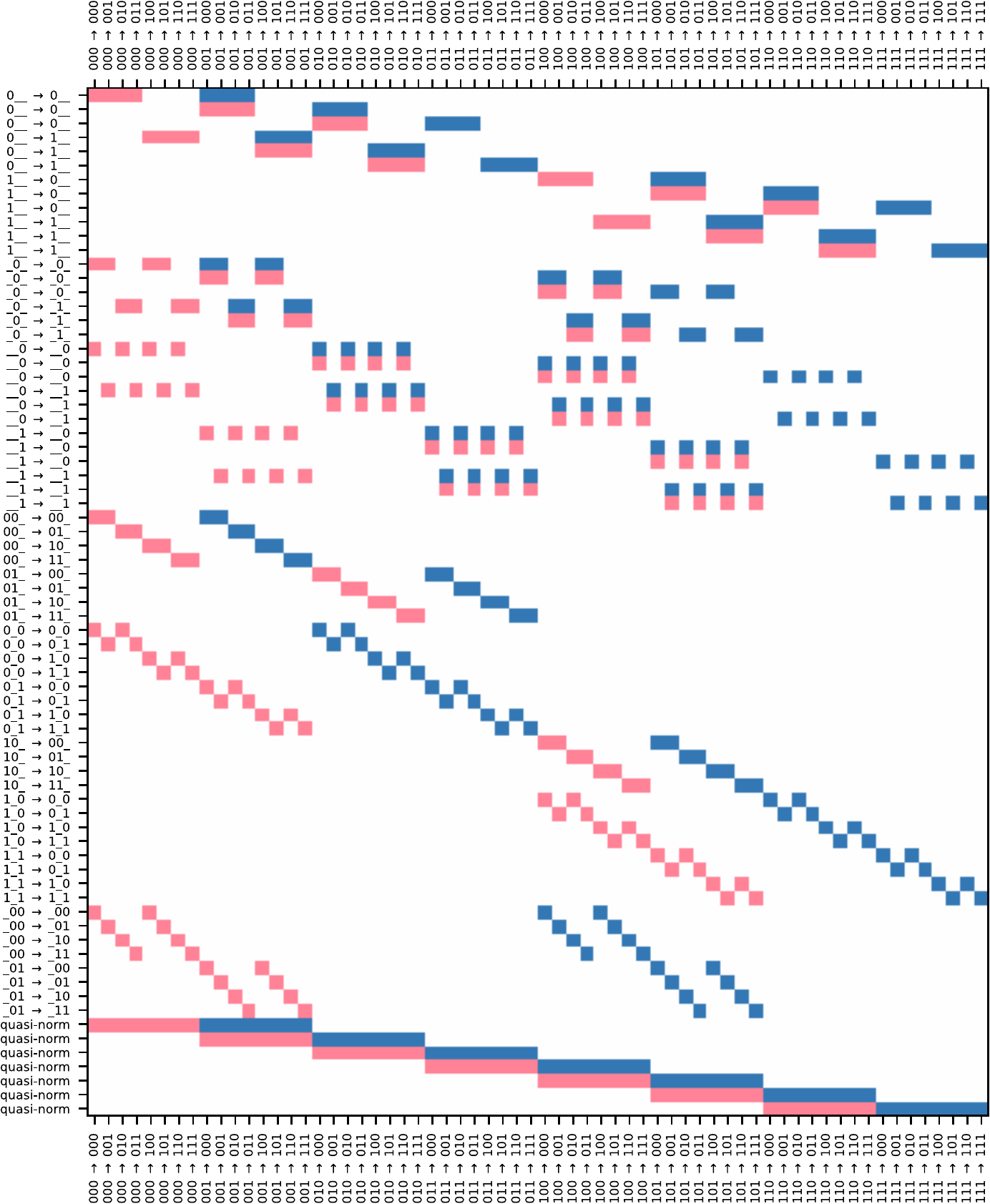}
\end{center}

\noindent Rows correspond to the 73 linear equations, of which 32 are independent.

\newpage
\subsection*{Space 9}

Space 9 is not induced by a causal order, but it is a refinement of the space 33 induced by the definite causal order $\total{\ev{A},\ev{B}}\vee\discrete{\ev{C}}$.
Its equivalence class under event-input permutation symmetry contains 12 spaces.
Space 9 differs as follows from the space induced by causal order $\total{\ev{A},\ev{B}}\vee\discrete{\ev{C}}$:
\begin{itemize}
  \item The outputs at events \evset{\ev{B}, \ev{C}} are independent of the input at event \ev{A} when the inputs at events \evset{B, C} are given by \hist{B/1,C/0} and \hist{B/0,C/1}.
\end{itemize}

\noindent Below are the histories and extended histories for space 9: 
\begin{center}
    \begin{tabular}{cc}
    \includegraphics[height=3.5cm]{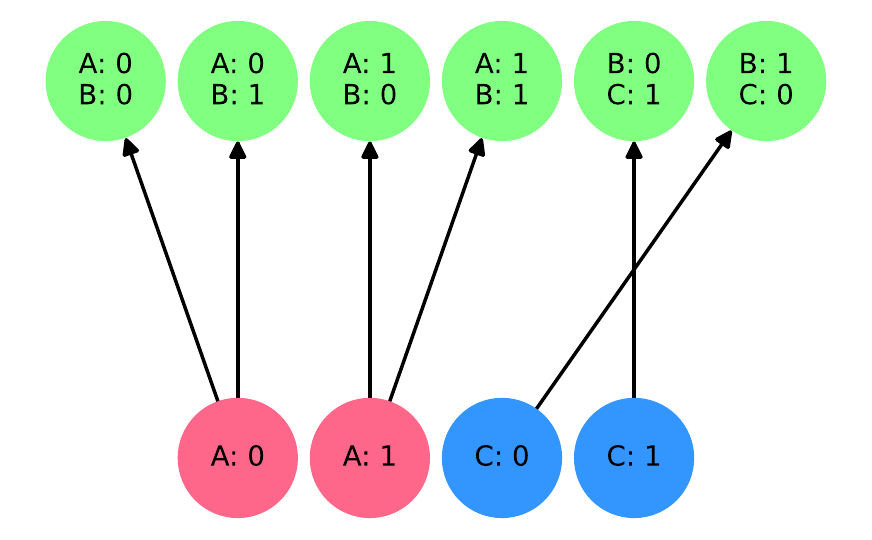}
    &
    \includegraphics[height=3.5cm]{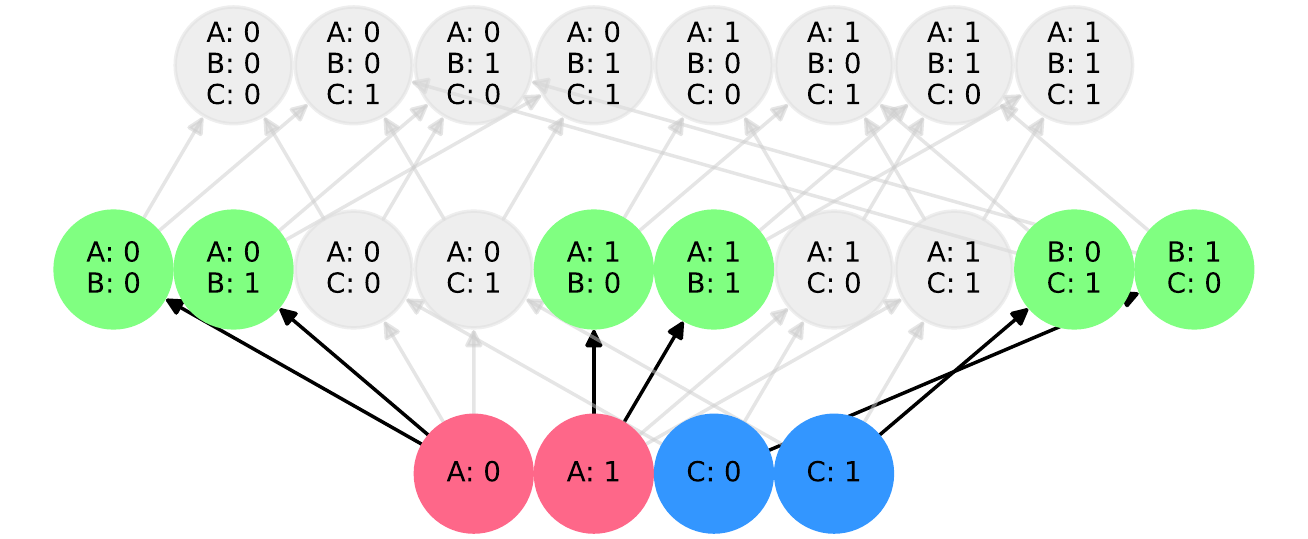}
    \\
    $\Theta_{9}$
    &
    $\Ext{\Theta_{9}}$
    \end{tabular}
\end{center}

\noindent The standard causaltope for Space 9 has dimension 28.
Below is a plot of the homogeneous linear system of causality and quasi-normalisation equations for the standard causaltope, put in reduced row echelon form:

\begin{center}
    \includegraphics[width=11cm]{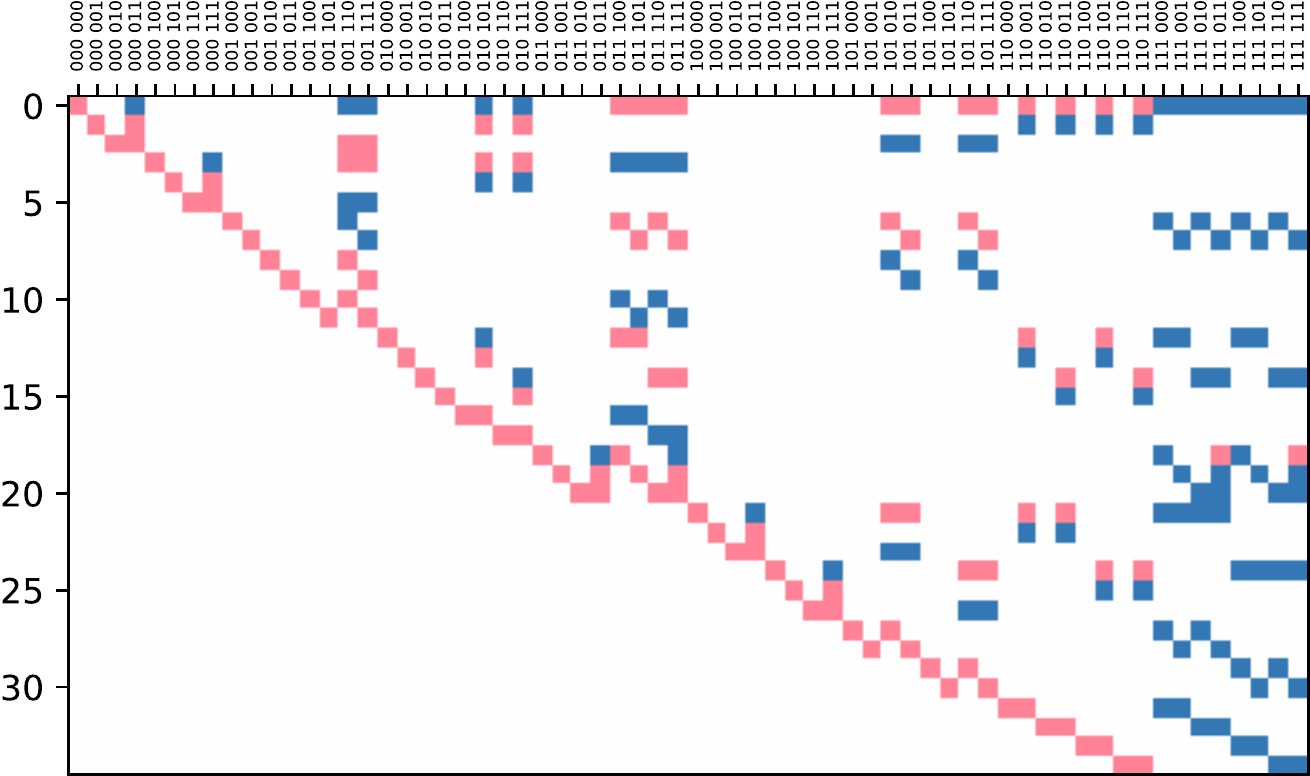}
\end{center}

\noindent Rows correspond to the 35 independent linear equations.
Columns in the plot correspond to entries of empirical models, indexed as $i_A i_B i_C$ $o_A o_B o_C$.
Coefficients in the equations are color-coded as white=0, red=+1 and blue=-1.

Space 9 has closest refinements in equivalence class 6; 
it is the join of its (closest) refinements.
It has closest coarsenings in equivalence classes 19 and 24; 
it is the meet of its (closest) coarsenings.
It has 64 causal functions, all of which are causal for at least one of its refinements.
It is not a tight space: for event \ev{B}, a causal function must yield identical output values on input histories \hist{A/0,B/0}, \hist{A/1,B/0} and \hist{B/0,C/1}, and it must also yield identical output values on input histories \hist{A/0,B/1}, \hist{A/1,B/1} and \hist{B/1,C/0}.

The standard causaltope for Space 9 has 1 more dimension than those of its 2 subspaces in equivalence class 6.
The standard causaltope for Space 9 is the meet of the standard causaltopes for its closest coarsenings.
For completeness, below is a plot of the full homogeneous linear system of causality and quasi-normalisation equations for the standard causaltope:

\begin{center}
    \includegraphics[width=12cm]{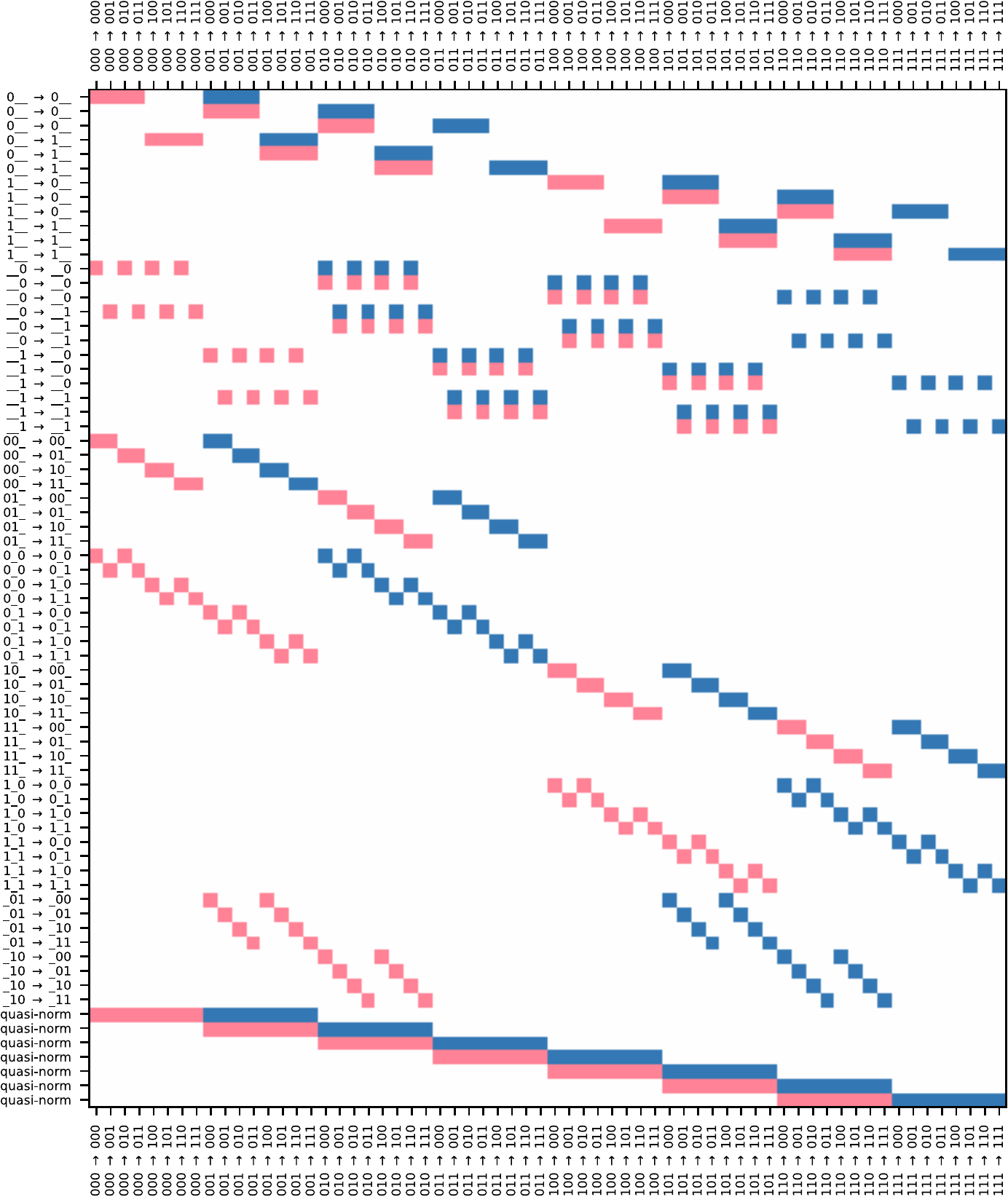}
\end{center}

\noindent Rows correspond to the 71 linear equations, of which 35 are independent.

\newpage
\subsection*{Space 10}

Space 10 is not induced by a causal order, but it is a refinement of the space in equivalence class 92 induced by the definite causal order $\total{\ev{A},\ev{B}}\vee\total{\ev{C},\ev{B}}$ (note that the space induced by the order is not the same as space 92).
Its equivalence class under event-input permutation symmetry contains 24 spaces.
Space 10 differs as follows from the space induced by causal order $\total{\ev{A},\ev{B}}\vee\total{\ev{C},\ev{B}}$:
\begin{itemize}
  \item The outputs at events \evset{\ev{A}, \ev{B}} are independent of the input at event \ev{C} when the inputs at events \evset{A, B} are given by \hist{A/0,B/0}, \hist{A/0,B/1} and \hist{A/1,B/0}.
  \item The outputs at events \evset{\ev{B}, \ev{C}} are independent of the input at event \ev{A} when the inputs at events \evset{B, C} are given by \hist{B/1,C/0}, \hist{B/1,C/1} and \hist{B/0,C/1}.
\end{itemize}

\noindent Below are the histories and extended histories for space 10: 
\begin{center}
    \begin{tabular}{cc}
    \includegraphics[height=3.5cm]{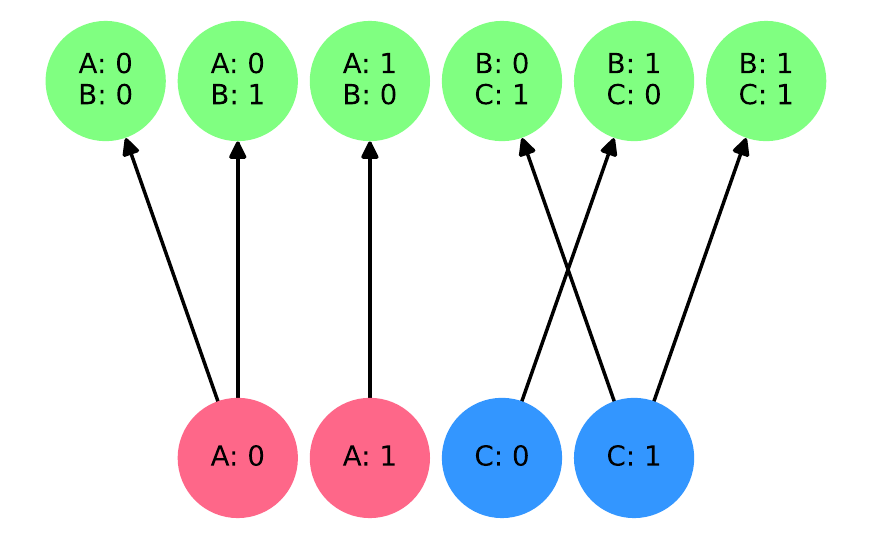}
    &
    \includegraphics[height=3.5cm]{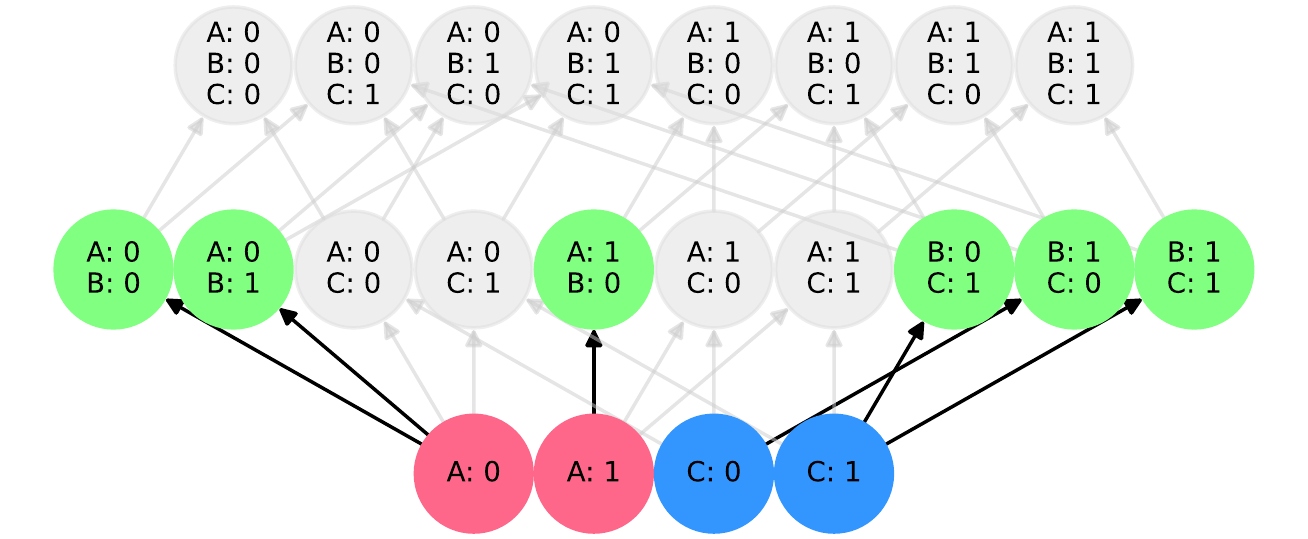}
    \\
    $\Theta_{10}$
    &
    $\Ext{\Theta_{10}}$
    \end{tabular}
\end{center}

\noindent The standard causaltope for Space 10 has dimension 28.
Below is a plot of the homogeneous linear system of causality and quasi-normalisation equations for the standard causaltope, put in reduced row echelon form:

\begin{center}
    \includegraphics[width=11cm]{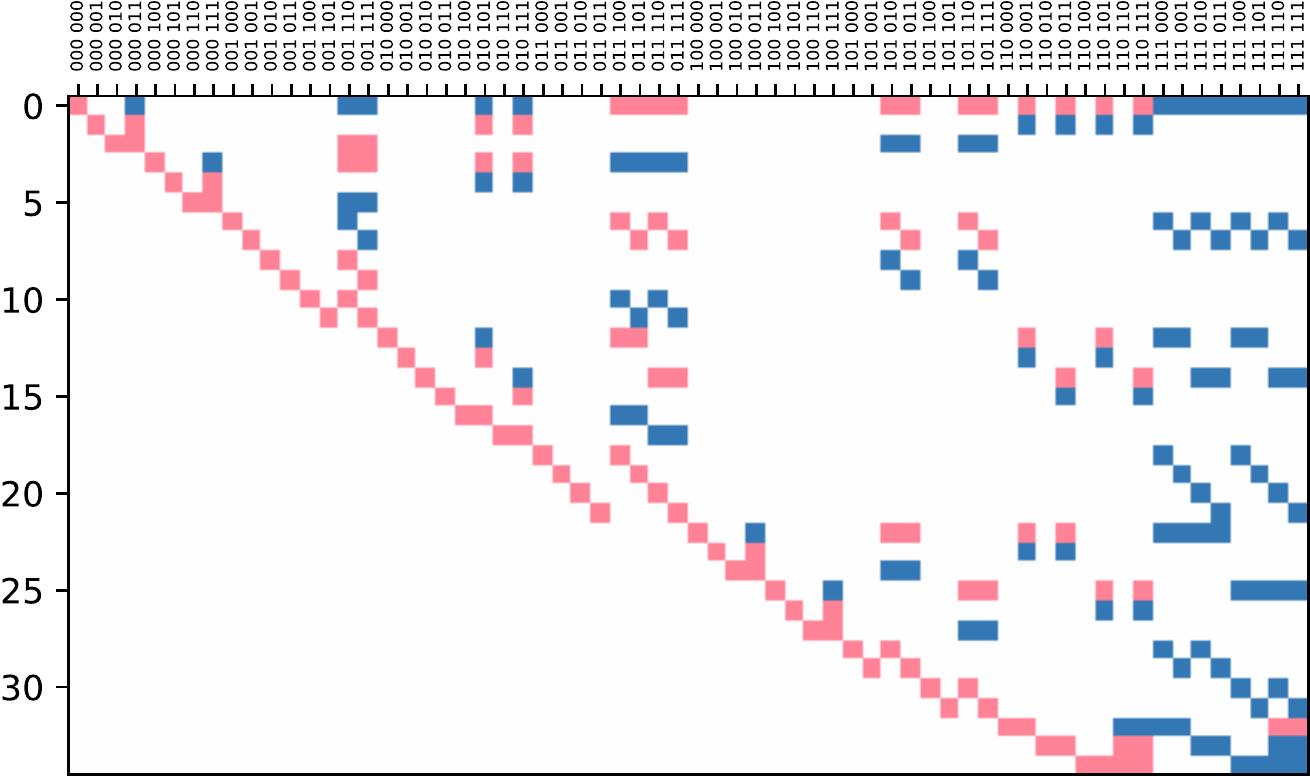}
\end{center}

\noindent Rows correspond to the 35 independent linear equations.
Columns in the plot correspond to entries of empirical models, indexed as $i_A i_B i_C$ $o_A o_B o_C$.
Coefficients in the equations are color-coded as white=0, red=+1 and blue=-1.

Space 10 has closest refinements in equivalence class 6; 
it is the join of its (closest) refinements.
It has closest coarsenings in equivalence classes 16, 23 and 24; 
it is the meet of its (closest) coarsenings.
It has 64 causal functions, all of which are causal for at least one of its refinements.
It is not a tight space: for event \ev{B}, a causal function must yield identical output values on input histories \hist{A/0,B/0}, \hist{A/1,B/0} and \hist{B/0,C/1}, and it must also yield identical output values on input histories \hist{A/0,B/1}, \hist{B/1,C/0} and \hist{B/1,C/1}.

The standard causaltope for Space 10 has 1 more dimension than those of its 2 subspaces in equivalence class 6.
The standard causaltope for Space 10 is the meet of the standard causaltopes for its closest coarsenings.
For completeness, below is a plot of the full homogeneous linear system of causality and quasi-normalisation equations for the standard causaltope:

\begin{center}
    \includegraphics[width=12cm]{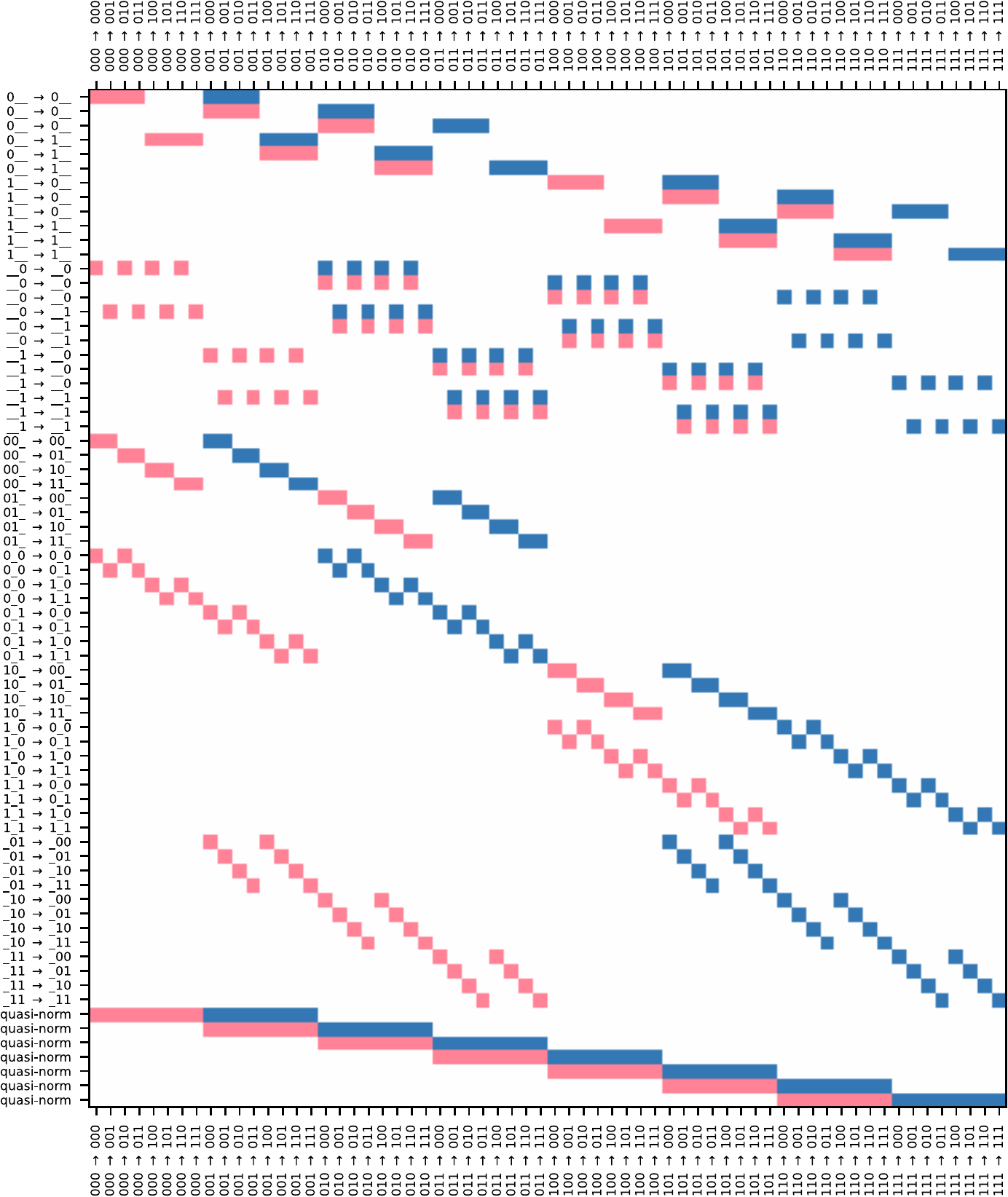}
\end{center}

\noindent Rows correspond to the 71 linear equations, of which 35 are independent.

\newpage
\subsection*{Space 11}

Space 11 is not induced by a causal order, but it is a refinement of the space in equivalence class 100 induced by the definite causal order $\total{\ev{C},\ev{A},\ev{B}}$ (note that the space induced by the order is not the same as space 100).
Its equivalence class under event-input permutation symmetry contains 48 spaces.
Space 11 differs as follows from the space induced by causal order $\total{\ev{C},\ev{A},\ev{B}}$:
\begin{itemize}
  \item The outputs at events \evset{\ev{A}, \ev{B}} are independent of the input at event \ev{C} when the inputs at events \evset{A, B} are given by \hist{A/0,B/0}, \hist{A/0,B/1} and \hist{A/1,B/0}.
  \item The outputs at events \evset{\ev{B}, \ev{C}} are independent of the input at event \ev{A} when the inputs at events \evset{B, C} are given by \hist{B/1,C/1}, \hist{B/0,C/0} and \hist{B/0,C/1}.
  \item The output at event \ev{B} is independent of the inputs at events \evset{\ev{A}, \ev{C}} when the input at event B is given by \hist{B/0}.
  \item The output at event \ev{A} is independent of the input at event \ev{C} when the input at event A is given by \hist{A/0}.
\end{itemize}

\noindent Below are the histories and extended histories for space 11: 
\begin{center}
    \begin{tabular}{cc}
    \includegraphics[height=3.5cm]{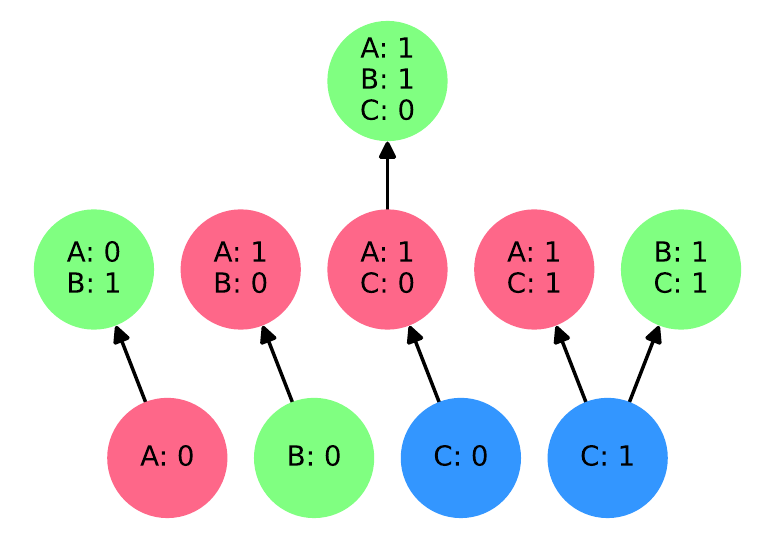}
    &
    \includegraphics[height=3.5cm]{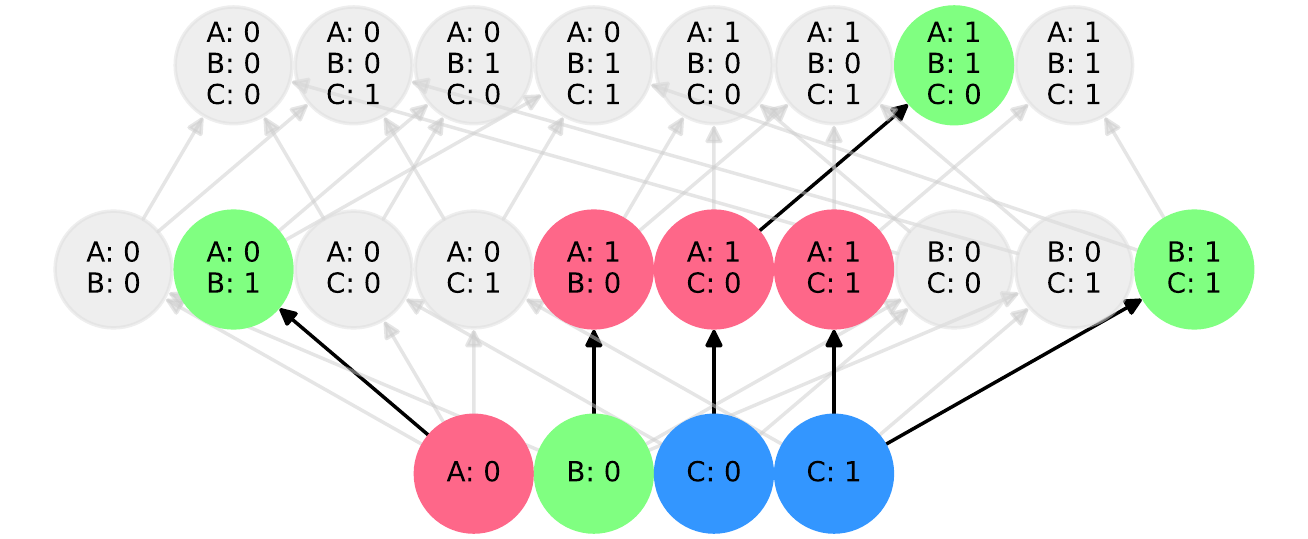}
    \\
    $\Theta_{11}$
    &
    $\Ext{\Theta_{11}}$
    \end{tabular}
\end{center}

\noindent The standard causaltope for Space 11 has dimension 29.
Below is a plot of the homogeneous linear system of causality and quasi-normalisation equations for the standard causaltope, put in reduced row echelon form:

\begin{center}
    \includegraphics[width=11cm]{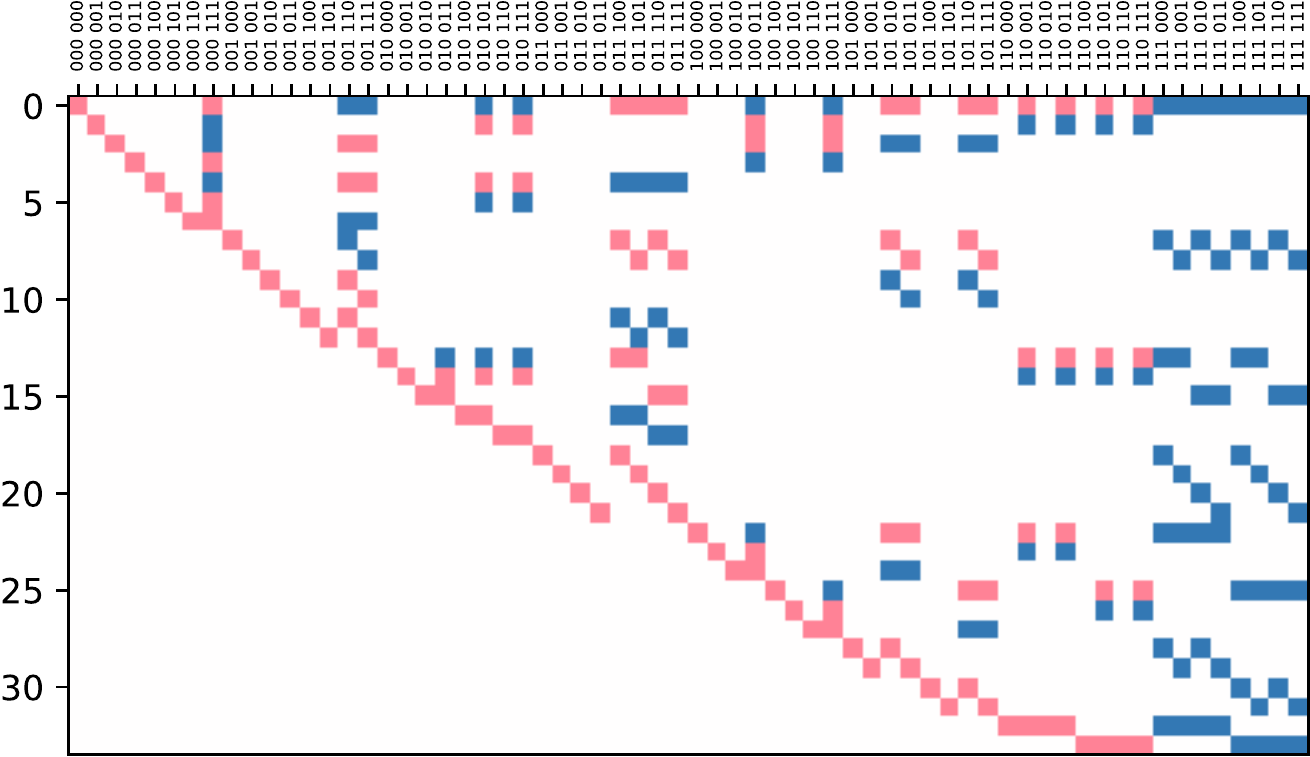}
\end{center}

\noindent Rows correspond to the 34 independent linear equations.
Columns in the plot correspond to entries of empirical models, indexed as $i_A i_B i_C$ $o_A o_B o_C$.
Coefficients in the equations are color-coded as white=0, red=+1 and blue=-1.

Space 11 has closest refinements in equivalence classes 4 and 7; 
it is the join of its (closest) refinements.
It has closest coarsenings in equivalence classes 17, 20, 21 and 22; 
it is the meet of its (closest) coarsenings.
It has 128 causal functions, all of which are causal for at least one of its refinements.
It is not a tight space: for event \ev{A}, a causal function must yield identical output values on input histories \hist{A/1,B/0}, \hist{A/1,C/0} and \hist{A/1,C/1}; for event \ev{B}, a causal function must yield identical output values on input histories \hist{A/0,B/1} and \hist{B/1,C/1}.

The standard causaltope for Space 11 coincides with that of its subspace in equivalence class 4.
The standard causaltope for Space 11 is the meet of the standard causaltopes for its closest coarsenings.
For completeness, below is a plot of the full homogeneous linear system of causality and quasi-normalisation equations for the standard causaltope:

\begin{center}
    \includegraphics[width=12cm]{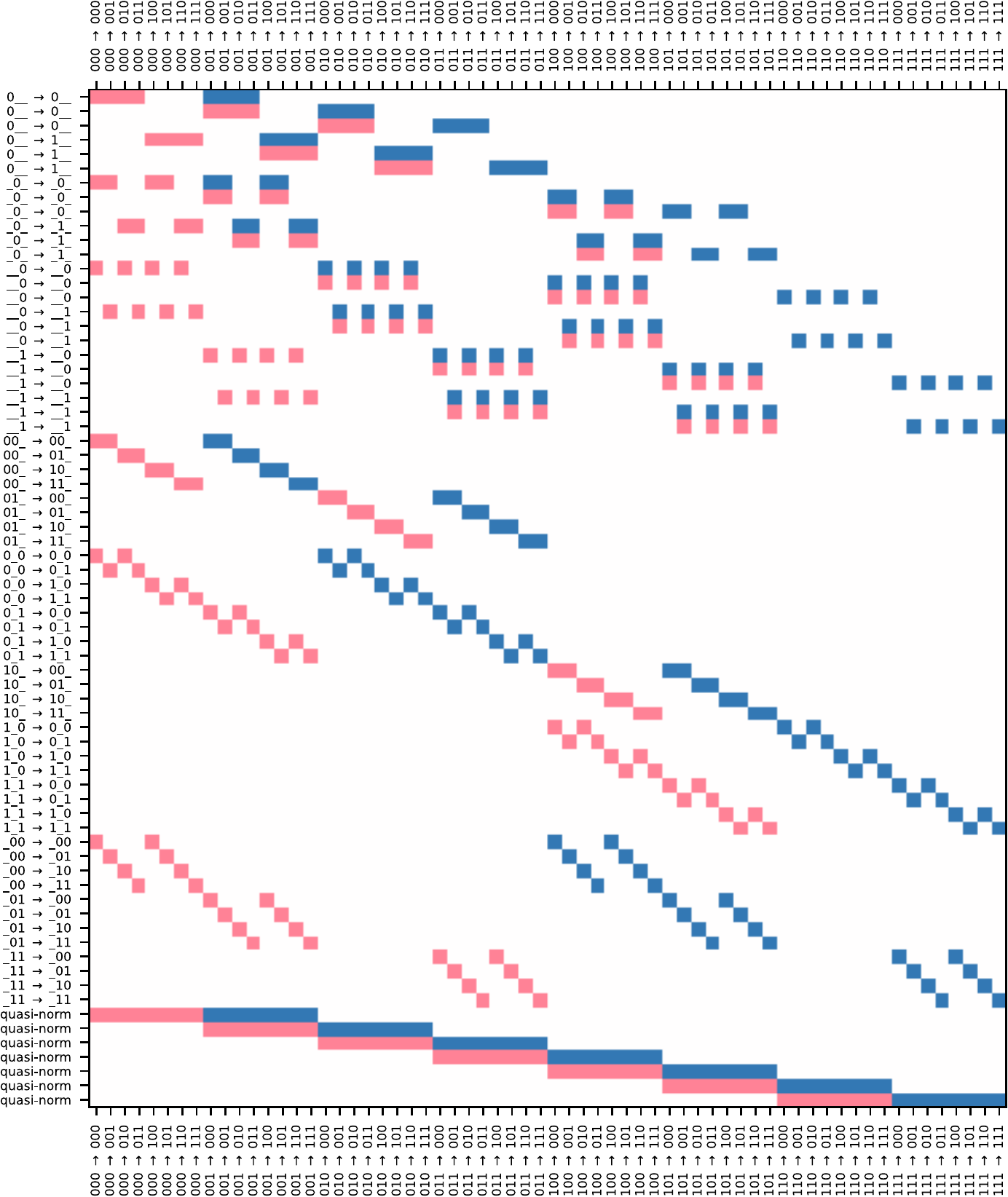}
\end{center}

\noindent Rows correspond to the 71 linear equations, of which 34 are independent.

\newpage
\subsection*{Space 12}

Space 12 is not induced by a causal order, but it is a refinement of the space 77 induced by the definite causal order $\total{\ev{A},\ev{B}}\vee\total{\ev{A},\ev{C}}$.
Its equivalence class under event-input permutation symmetry contains 24 spaces.
Space 12 differs as follows from the space induced by causal order $\total{\ev{A},\ev{B}}\vee\total{\ev{A},\ev{C}}$:
\begin{itemize}
  \item The outputs at events \evset{\ev{B}, \ev{C}} are independent of the input at event \ev{A} when the inputs at events \evset{B, C} are given by \hist{B/1,C/0} and \hist{B/1,C/1}.
  \item The output at event \ev{C} is independent of the input at event \ev{A} when the input at event C is given by \hist{C/0}.
  \item The output at event \ev{B} is independent of the input at event \ev{A} when the input at event B is given by \hist{B/1}.
\end{itemize}

\noindent Below are the histories and extended histories for space 12: 
\begin{center}
    \begin{tabular}{cc}
    \includegraphics[height=3.5cm]{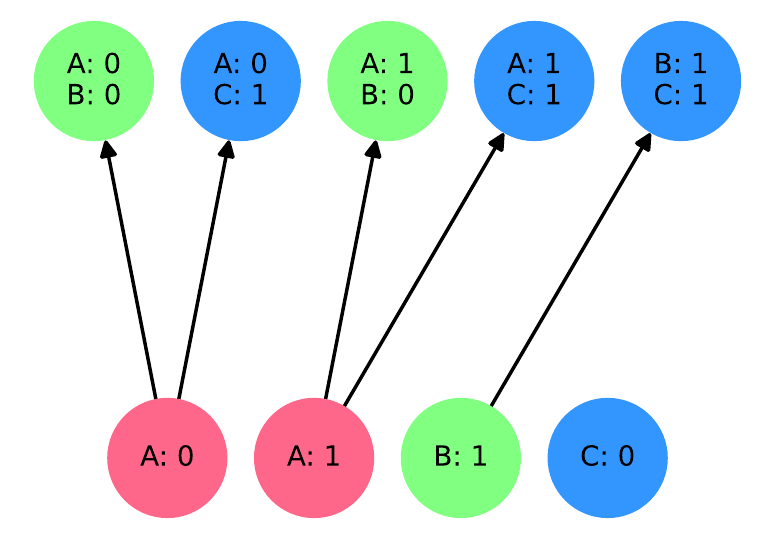}
    &
    \includegraphics[height=3.5cm]{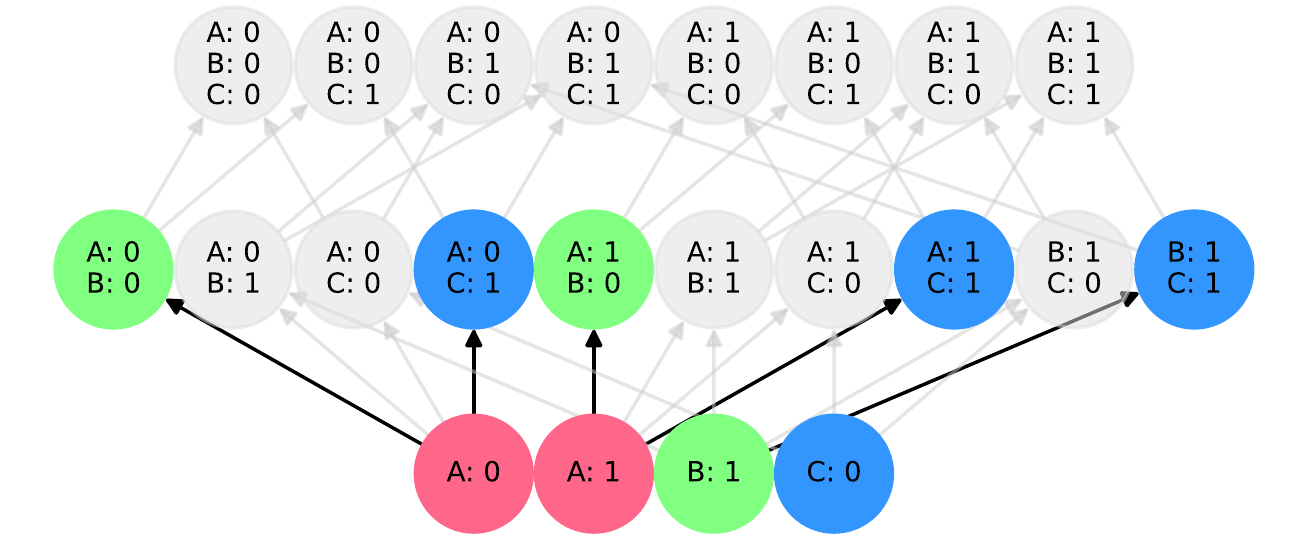}
    \\
    $\Theta_{12}$
    &
    $\Ext{\Theta_{12}}$
    \end{tabular}
\end{center}

\noindent The standard causaltope for Space 12 has dimension 29.
Below is a plot of the homogeneous linear system of causality and quasi-normalisation equations for the standard causaltope, put in reduced row echelon form:

\begin{center}
    \includegraphics[width=11cm]{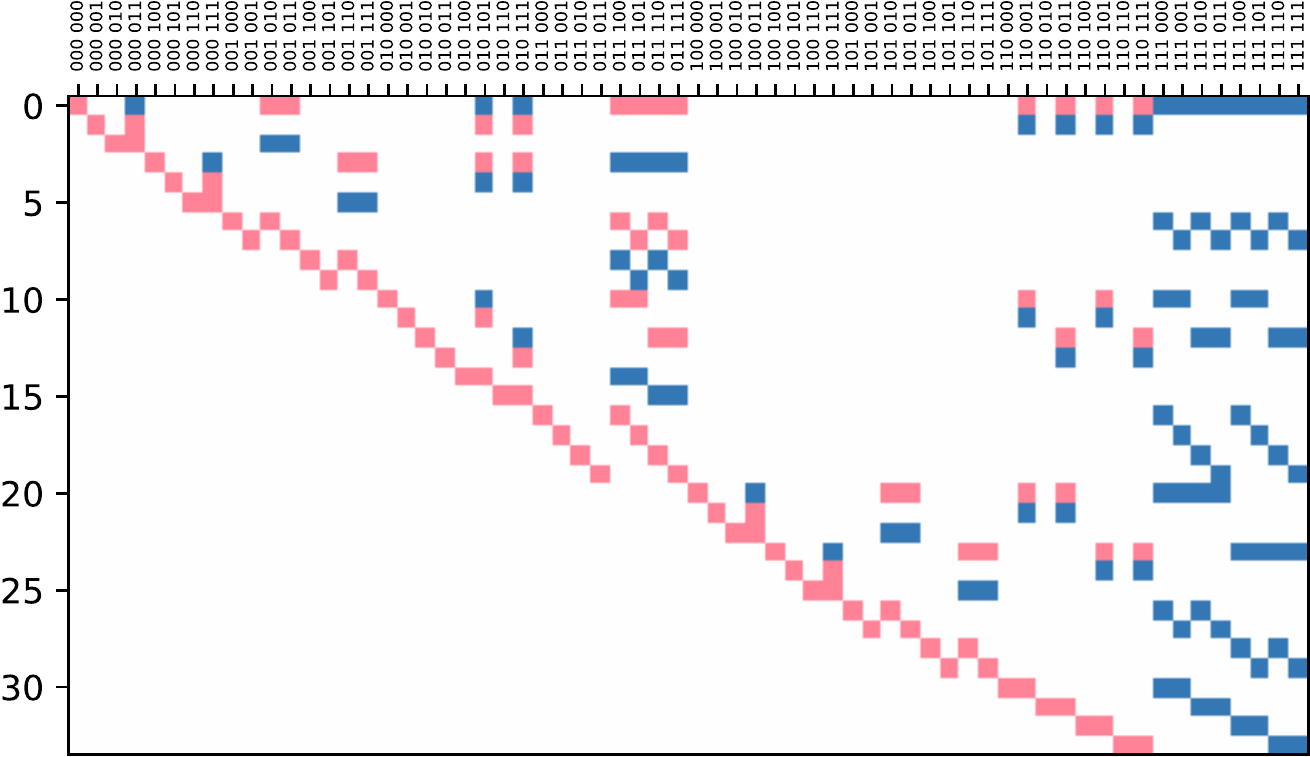}
\end{center}

\noindent Rows correspond to the 34 independent linear equations.
Columns in the plot correspond to entries of empirical models, indexed as $i_A i_B i_C$ $o_A o_B o_C$.
Coefficients in the equations are color-coded as white=0, red=+1 and blue=-1.

Space 12 has closest refinements in equivalence classes 5 and 7; 
it is the join of its (closest) refinements.
It has closest coarsenings in equivalence classes 17, 22, 25 and 27; 
it is the meet of its (closest) coarsenings.
It has 128 causal functions, all of which are causal for at least one of its refinements.
It is not a tight space: for event \ev{C}, a causal function must yield identical output values on input histories \hist{A/0,C/1}, \hist{A/1,C/1} and \hist{B/1,C/1}.

The standard causaltope for Space 12 coincides with that of its subspace in equivalence class 5.
The standard causaltope for Space 12 is the meet of the standard causaltopes for its closest coarsenings.
For completeness, below is a plot of the full homogeneous linear system of causality and quasi-normalisation equations for the standard causaltope:

\begin{center}
    \includegraphics[width=12cm]{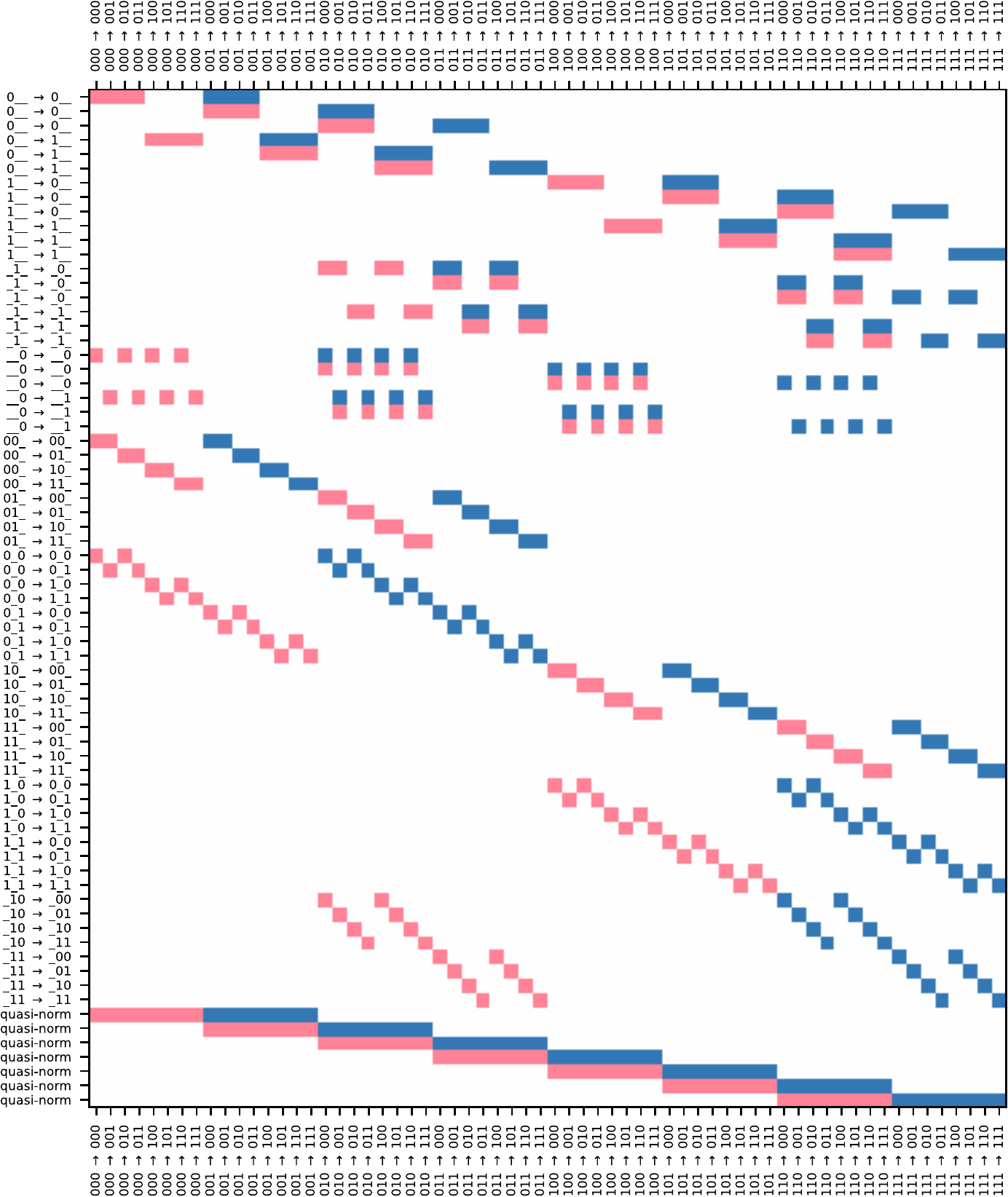}
\end{center}

\noindent Rows correspond to the 71 linear equations, of which 34 are independent.

\newpage
\subsection*{Space 13}

Space 13 is not induced by a causal order, but it is a refinement of the space 33 induced by the definite causal order $\total{\ev{A},\ev{B}}\vee\discrete{\ev{C}}$.
Its equivalence class under event-input permutation symmetry contains 12 spaces.
Space 13 differs as follows from the space induced by causal order $\total{\ev{A},\ev{B}}\vee\discrete{\ev{C}}$:
\begin{itemize}
  \item The outputs at events \evset{\ev{B}, \ev{C}} are independent of the input at event \ev{A} when the inputs at events \evset{B, C} are given by \hist{B/1,C/0} and \hist{B/0,C/0}.
\end{itemize}

\noindent Below are the histories and extended histories for space 13: 
\begin{center}
    \begin{tabular}{cc}
    \includegraphics[height=3.5cm]{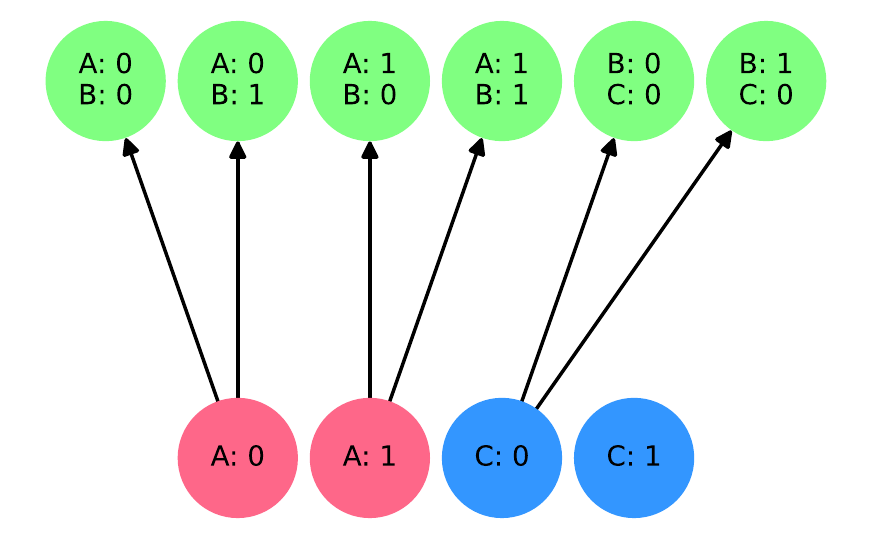}
    &
    \includegraphics[height=3.5cm]{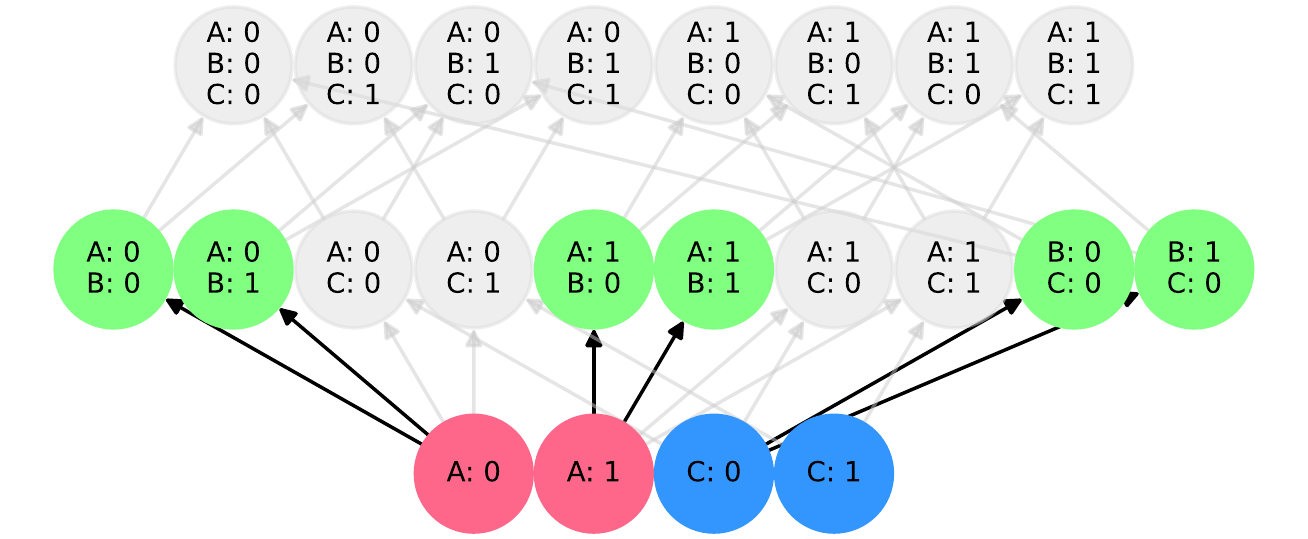}
    \\
    $\Theta_{13}$
    &
    $\Ext{\Theta_{13}}$
    \end{tabular}
\end{center}

\noindent The standard causaltope for Space 13 has dimension 28.
Below is a plot of the homogeneous linear system of causality and quasi-normalisation equations for the standard causaltope, put in reduced row echelon form:

\begin{center}
    \includegraphics[width=11cm]{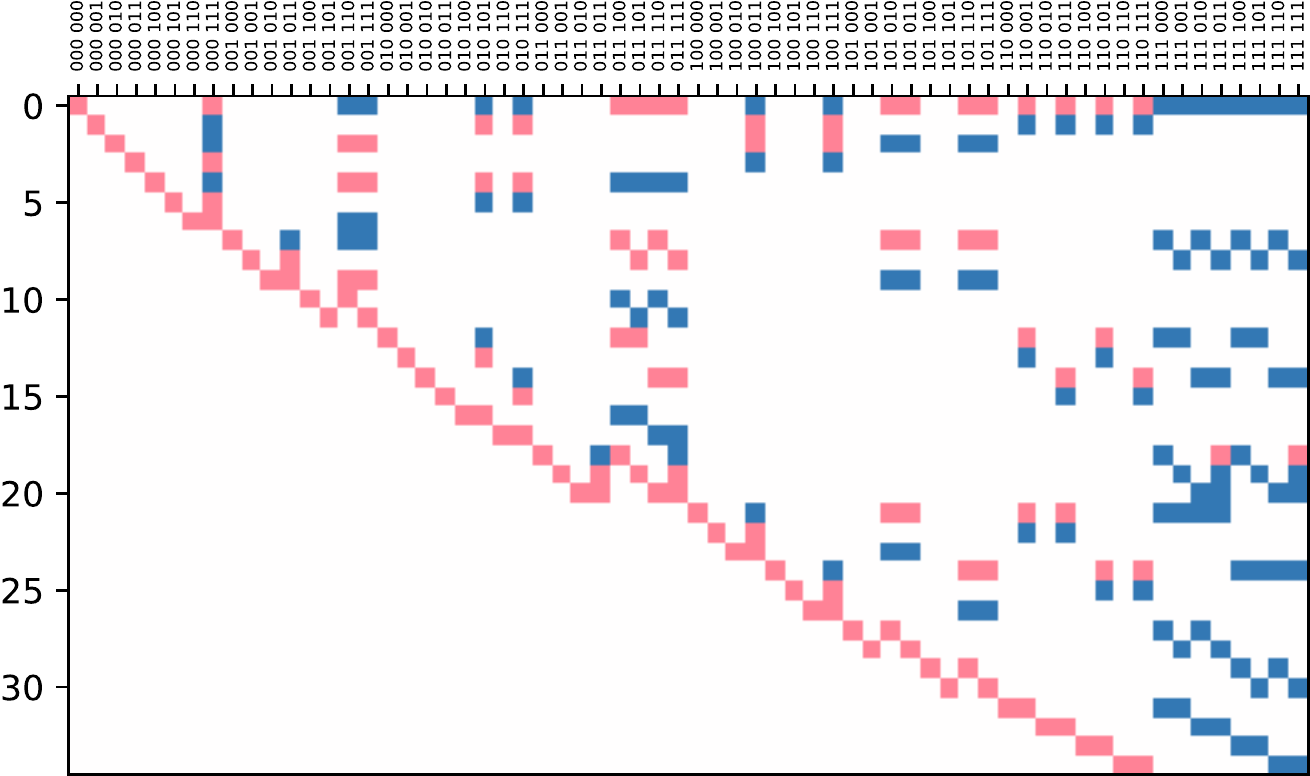}
\end{center}

\noindent Rows correspond to the 35 independent linear equations.
Columns in the plot correspond to entries of empirical models, indexed as $i_A i_B i_C$ $o_A o_B o_C$.
Coefficients in the equations are color-coded as white=0, red=+1 and blue=-1.

Space 13 has closest refinements in equivalence class 6; 
it is the join of its (closest) refinements.
It has closest coarsenings in equivalence classes 19, 23 and 27; 
it is the meet of its (closest) coarsenings.
It has 64 causal functions, all of which are causal for at least one of its refinements.
It is not a tight space: for event \ev{B}, a causal function must yield identical output values on input histories \hist{A/0,B/0}, \hist{A/1,B/0} and \hist{B/0,C/0}, and it must also yield identical output values on input histories \hist{A/0,B/1}, \hist{A/1,B/1} and \hist{B/1,C/0}.

The standard causaltope for Space 13 has 1 more dimension than those of its 2 subspaces in equivalence class 6.
The standard causaltope for Space 13 is the meet of the standard causaltopes for its closest coarsenings.
For completeness, below is a plot of the full homogeneous linear system of causality and quasi-normalisation equations for the standard causaltope:

\begin{center}
    \includegraphics[width=12cm]{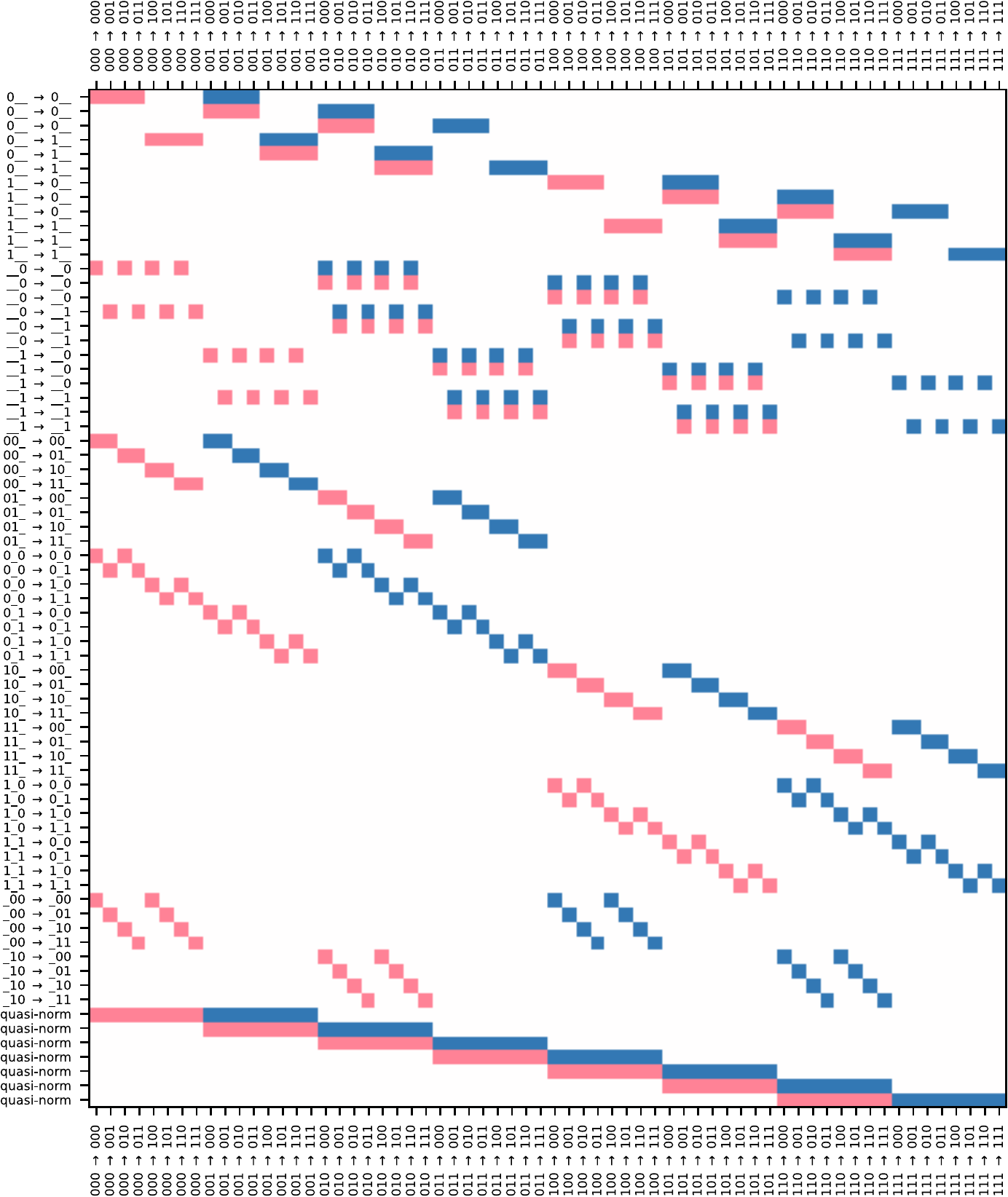}
\end{center}

\noindent Rows correspond to the 71 linear equations, of which 35 are independent.

\newpage
\subsection*{Space 14}

Space 14 is not induced by a causal order, but it is a refinement of the space 33 induced by the definite causal order $\total{\ev{A},\ev{B}}\vee\discrete{\ev{C}}$.
Its equivalence class under event-input permutation symmetry contains 12 spaces.
Space 14 differs as follows from the space induced by causal order $\total{\ev{A},\ev{B}}\vee\discrete{\ev{C}}$:
\begin{itemize}
  \item The outputs at events \evset{\ev{B}, \ev{C}} are independent of the input at event \ev{A} when the inputs at events \evset{B, C} are given by \hist{B/1,C/0} and \hist{B/1,C/1}.
\end{itemize}

\noindent Below are the histories and extended histories for space 14: 
\begin{center}
    \begin{tabular}{cc}
    \includegraphics[height=3.5cm]{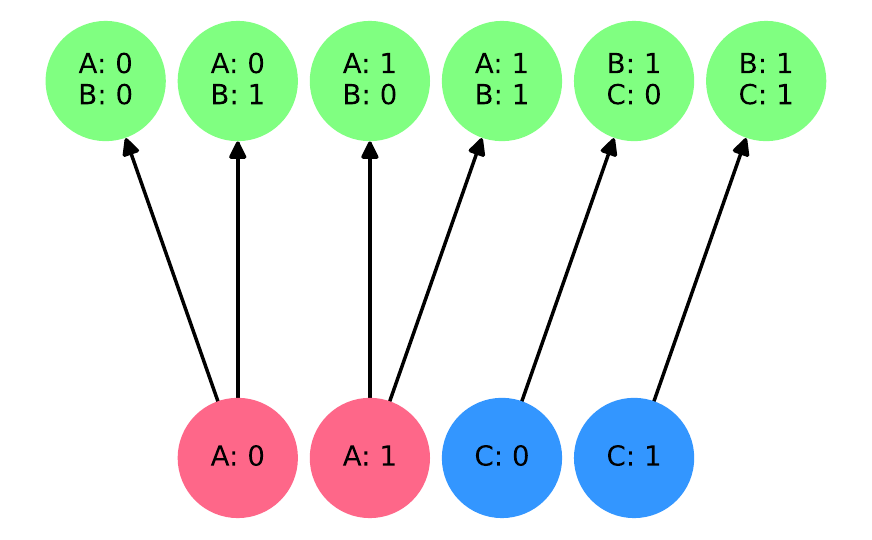}
    &
    \includegraphics[height=3.5cm]{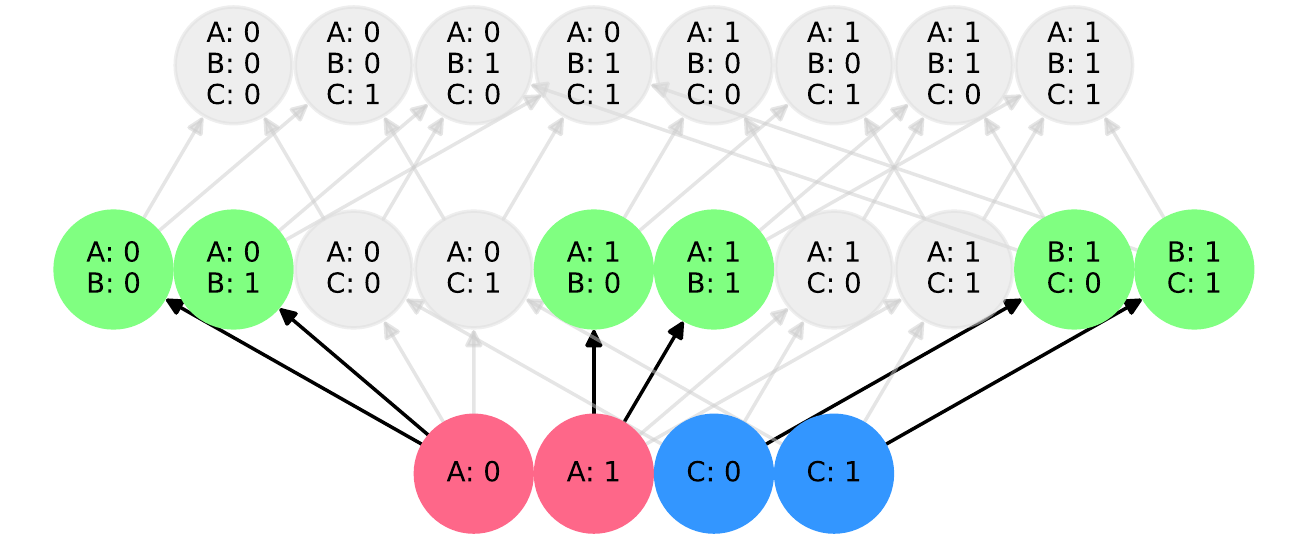}
    \\
    $\Theta_{14}$
    &
    $\Ext{\Theta_{14}}$
    \end{tabular}
\end{center}

\noindent The standard causaltope for Space 14 has dimension 29.
Below is a plot of the homogeneous linear system of causality and quasi-normalisation equations for the standard causaltope, put in reduced row echelon form:

\begin{center}
    \includegraphics[width=11cm]{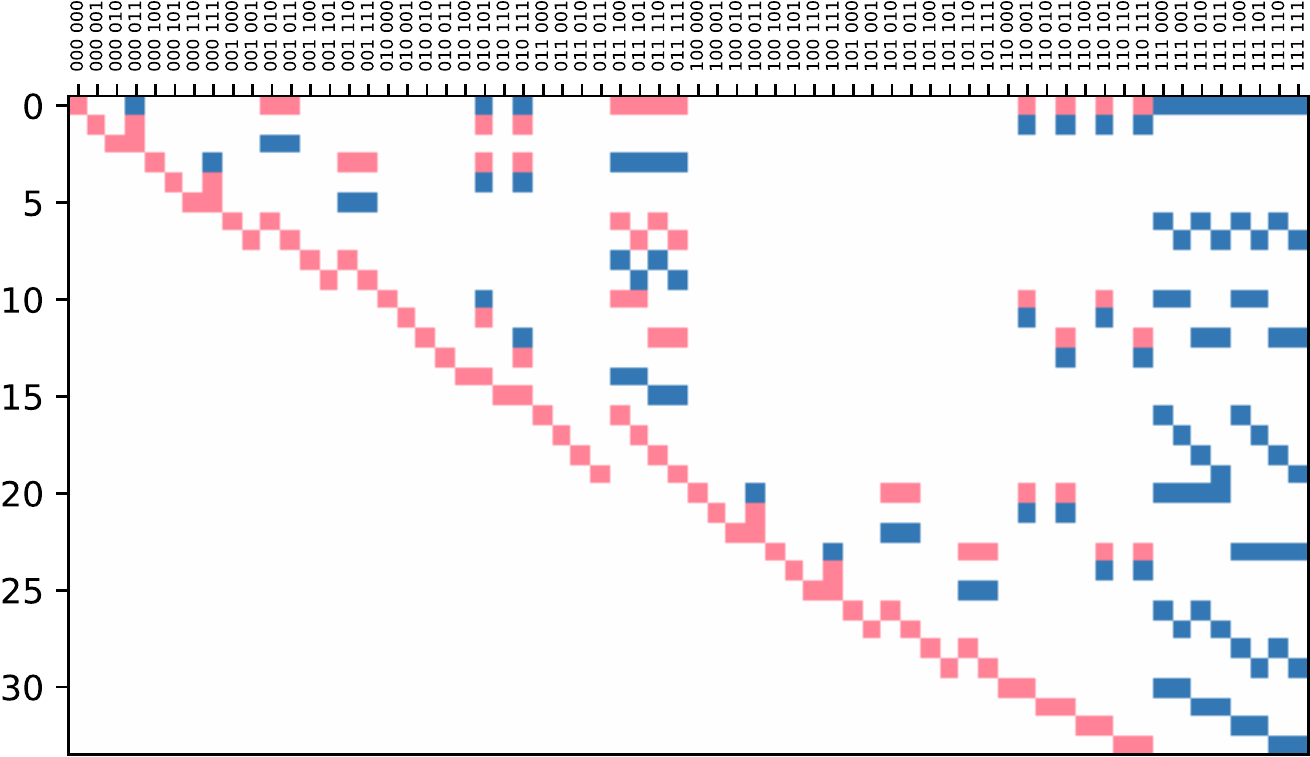}
\end{center}

\noindent Rows correspond to the 34 independent linear equations.
Columns in the plot correspond to entries of empirical models, indexed as $i_A i_B i_C$ $o_A o_B o_C$.
Coefficients in the equations are color-coded as white=0, red=+1 and blue=-1.

Space 14 has closest refinements in equivalence classes 5 and 6; 
it is the join of its (closest) refinements.
It has closest coarsenings in equivalence classes 16, 19 and 26; 
it is the meet of its (closest) coarsenings.
It has 128 causal functions, all of which are causal for at least one of its refinements.
It is not a tight space: for event \ev{B}, a causal function must yield identical output values on input histories \hist{A/0,B/1}, \hist{A/1,B/1}, \hist{B/1,C/0} and \hist{B/1,C/1}.

The standard causaltope for Space 14 coincides with that of its subspace in equivalence class 5.
The standard causaltope for Space 14 is the meet of the standard causaltopes for its closest coarsenings.
For completeness, below is a plot of the full homogeneous linear system of causality and quasi-normalisation equations for the standard causaltope:

\begin{center}
    \includegraphics[width=12cm]{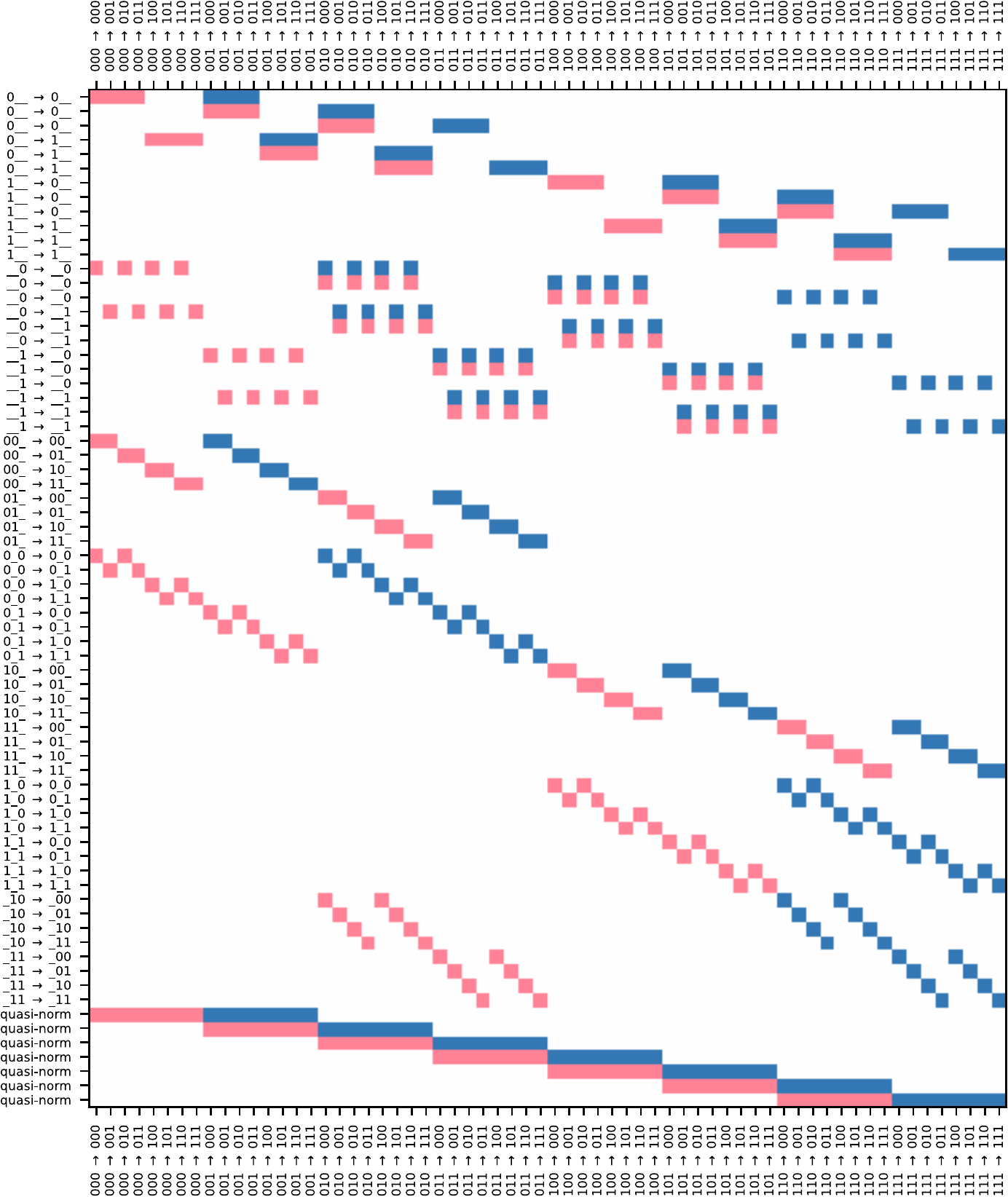}
\end{center}

\noindent Rows correspond to the 71 linear equations, of which 34 are independent.

\newpage
\subsection*{Space 15}

Space 15 is not induced by a causal order, but it is a refinement of the space in equivalence class 92 induced by the definite causal order $\total{\ev{A},\ev{B}}\vee\total{\ev{C},\ev{B}}$ (note that the space induced by the order is not the same as space 92).
Its equivalence class under event-input permutation symmetry contains 24 spaces.
Space 15 differs as follows from the space induced by causal order $\total{\ev{A},\ev{B}}\vee\total{\ev{C},\ev{B}}$:
\begin{itemize}
  \item The outputs at events \evset{\ev{A}, \ev{B}} are independent of the input at event \ev{C} when the inputs at events \evset{A, B} are given by \hist{A/0,B/0}, \hist{A/0,B/1} and \hist{A/1,B/1}.
  \item The outputs at events \evset{\ev{B}, \ev{C}} are independent of the input at event \ev{A} when the inputs at events \evset{B, C} are given by \hist{B/1,C/0}, \hist{B/1,C/1} and \hist{B/0,C/1}.
\end{itemize}

\noindent Below are the histories and extended histories for space 15: 
\begin{center}
    \begin{tabular}{cc}
    \includegraphics[height=3.5cm]{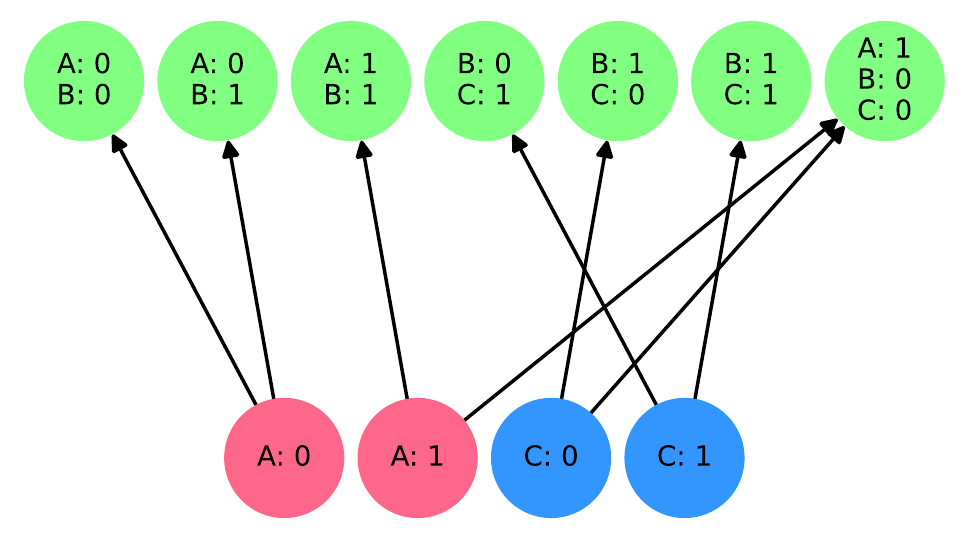}
    &
    \includegraphics[height=3.5cm]{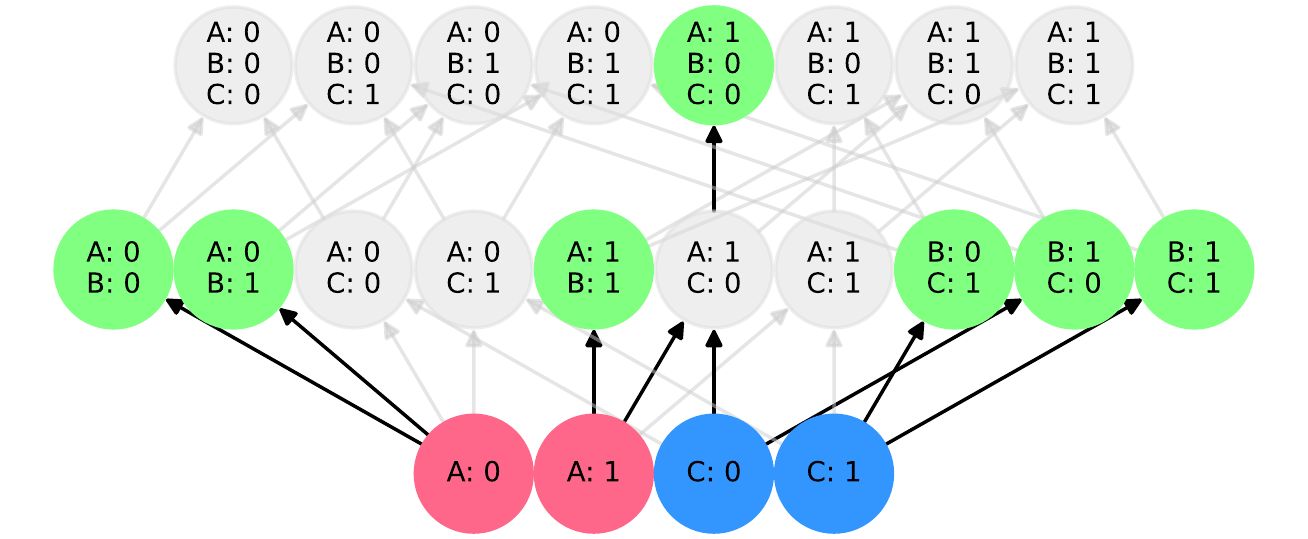}
    \\
    $\Theta_{15}$
    &
    $\Ext{\Theta_{15}}$
    \end{tabular}
\end{center}

\noindent The standard causaltope for Space 15 has dimension 29.
Below is a plot of the homogeneous linear system of causality and quasi-normalisation equations for the standard causaltope, put in reduced row echelon form:

\begin{center}
    \includegraphics[width=11cm]{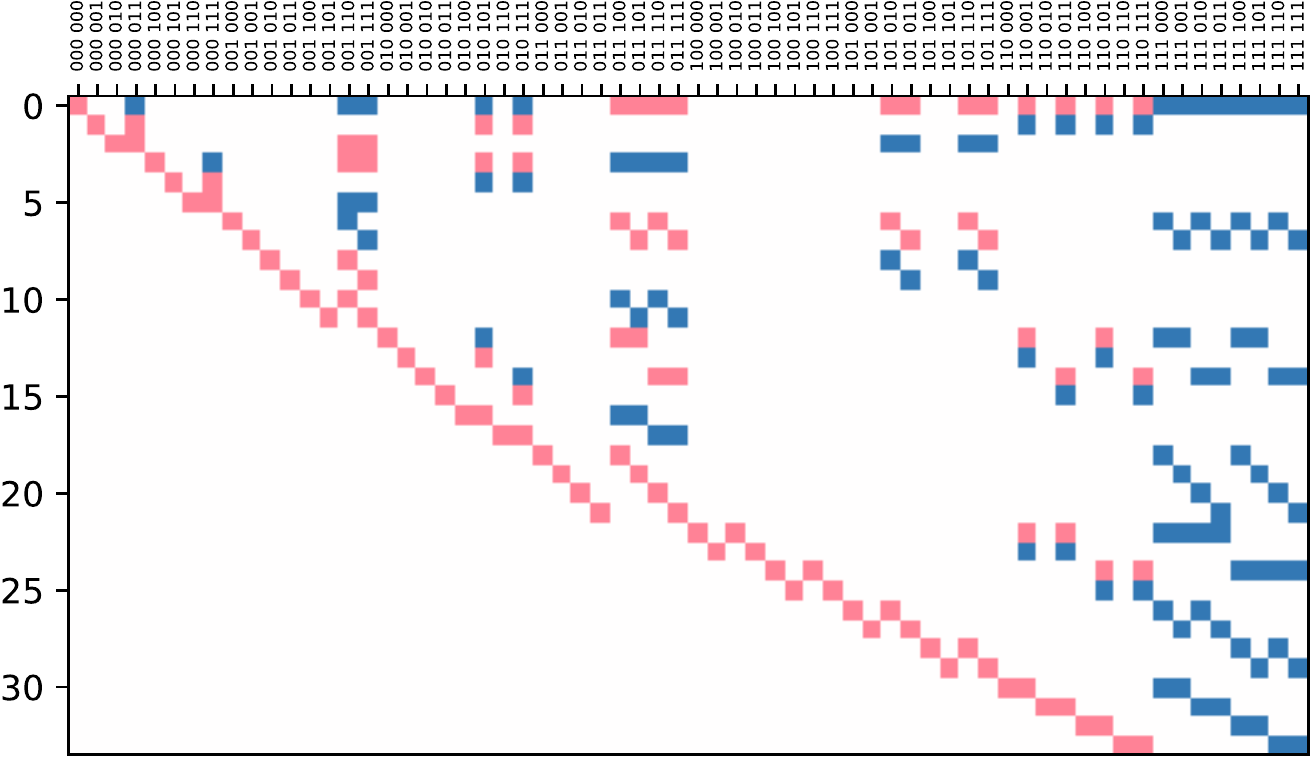}
\end{center}

\noindent Rows correspond to the 34 independent linear equations.
Columns in the plot correspond to entries of empirical models, indexed as $i_A i_B i_C$ $o_A o_B o_C$.
Coefficients in the equations are color-coded as white=0, red=+1 and blue=-1.

Space 15 has closest refinements in equivalence classes 4 and 6; 
it is the join of its (closest) refinements.
It has closest coarsenings in equivalence classes 23, 24 and 26; 
it is the meet of its (closest) coarsenings.
It has 128 causal functions, 64 of which are not causal for any of its refinements.
It is not a tight space: for event \ev{B}, a causal function must yield identical output values on input histories \hist{A/0,B/0} and \hist{B/0,C/1}, and it must also yield identical output values on input histories \hist{A/0,B/1}, \hist{A/1,B/1}, \hist{B/1,C/0} and \hist{B/1,C/1}.

The standard causaltope for Space 15 coincides with that of its subspace in equivalence class 4.
The standard causaltope for Space 15 is the meet of the standard causaltopes for its closest coarsenings.
For completeness, below is a plot of the full homogeneous linear system of causality and quasi-normalisation equations for the standard causaltope:

\begin{center}
    \includegraphics[width=12cm]{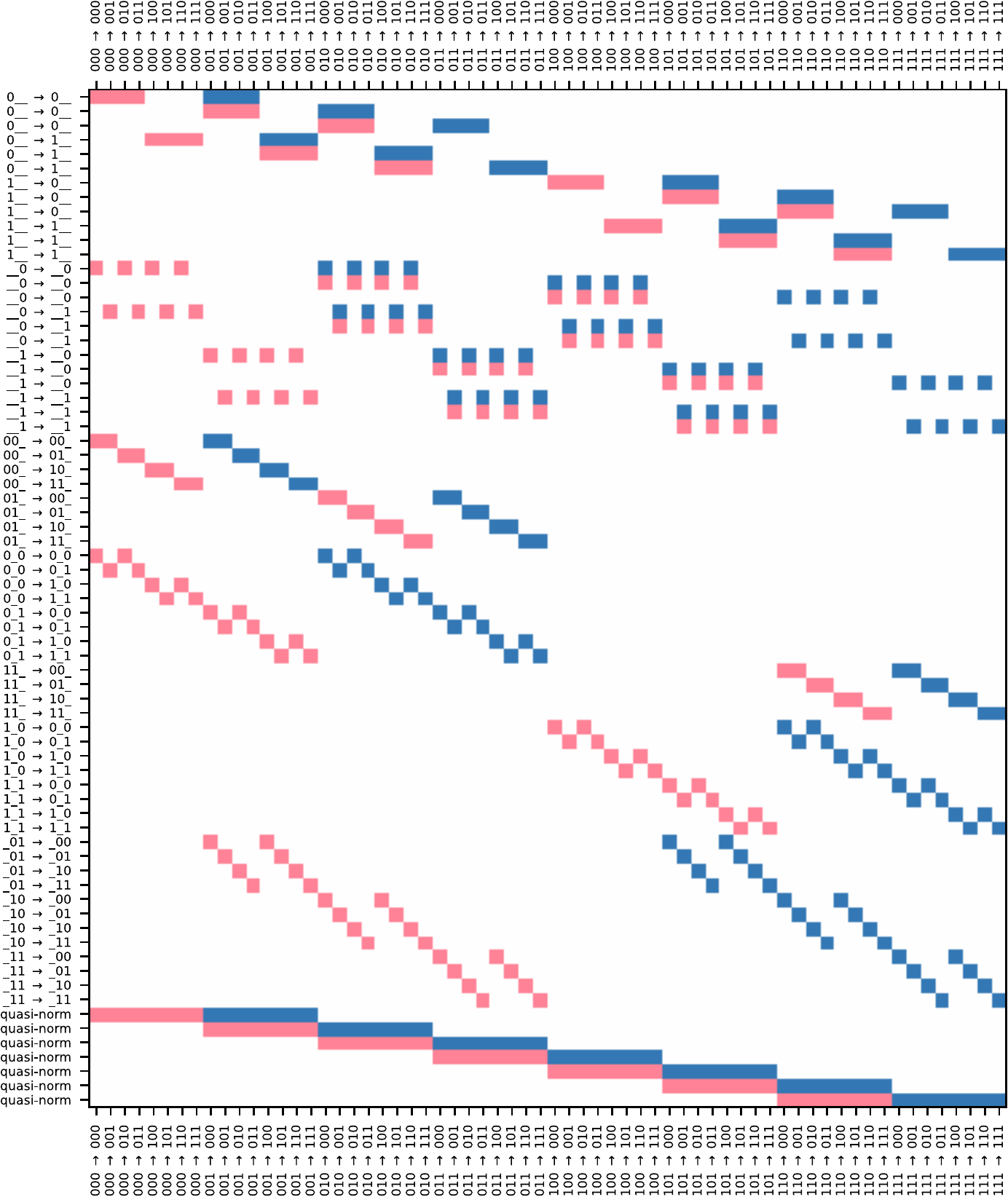}
\end{center}

\noindent Rows correspond to the 71 linear equations, of which 34 are independent.

\newpage
\subsection*{Space 16}

Space 16 is not induced by a causal order, but it is a refinement of the space in equivalence class 92 induced by the definite causal order $\total{\ev{A},\ev{B}}\vee\total{\ev{C},\ev{B}}$ (note that the space induced by the order is not the same as space 92).
Its equivalence class under event-input permutation symmetry contains 24 spaces.
Space 16 differs as follows from the space induced by causal order $\total{\ev{A},\ev{B}}\vee\total{\ev{C},\ev{B}}$:
\begin{itemize}
  \item The outputs at events \evset{\ev{A}, \ev{B}} are independent of the input at event \ev{C} when the inputs at events \evset{A, B} are given by \hist{A/0,B/0}, \hist{A/0,B/1} and \hist{A/1,B/0}.
  \item The outputs at events \evset{\ev{B}, \ev{C}} are independent of the input at event \ev{A} when the inputs at events \evset{B, C} are given by \hist{B/1,C/0} and \hist{B/1,C/1}.
\end{itemize}

\noindent Below are the histories and extended histories for space 16: 
\begin{center}
    \begin{tabular}{cc}
    \includegraphics[height=3.5cm]{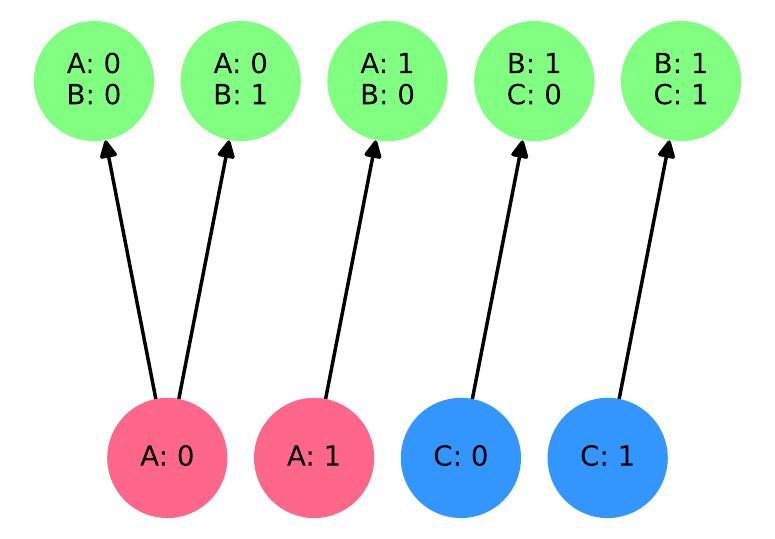}
    &
    \includegraphics[height=3.5cm]{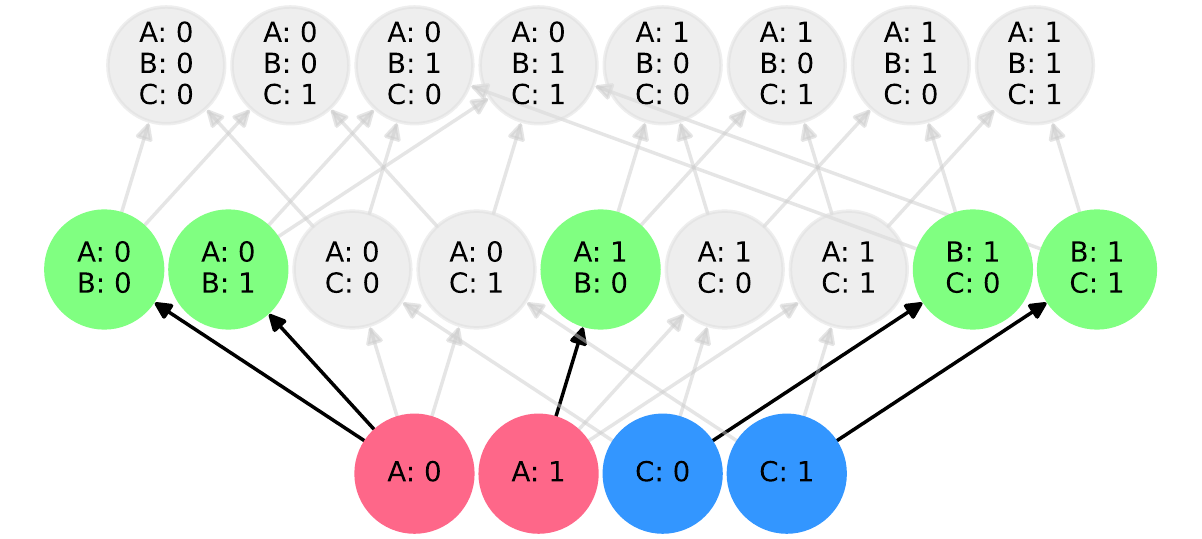}
    \\
    $\Theta_{16}$
    &
    $\Ext{\Theta_{16}}$
    \end{tabular}
\end{center}

\noindent The standard causaltope for Space 16 has dimension 30.
Below is a plot of the homogeneous linear system of causality and quasi-normalisation equations for the standard causaltope, put in reduced row echelon form:

\begin{center}
    \includegraphics[width=11cm]{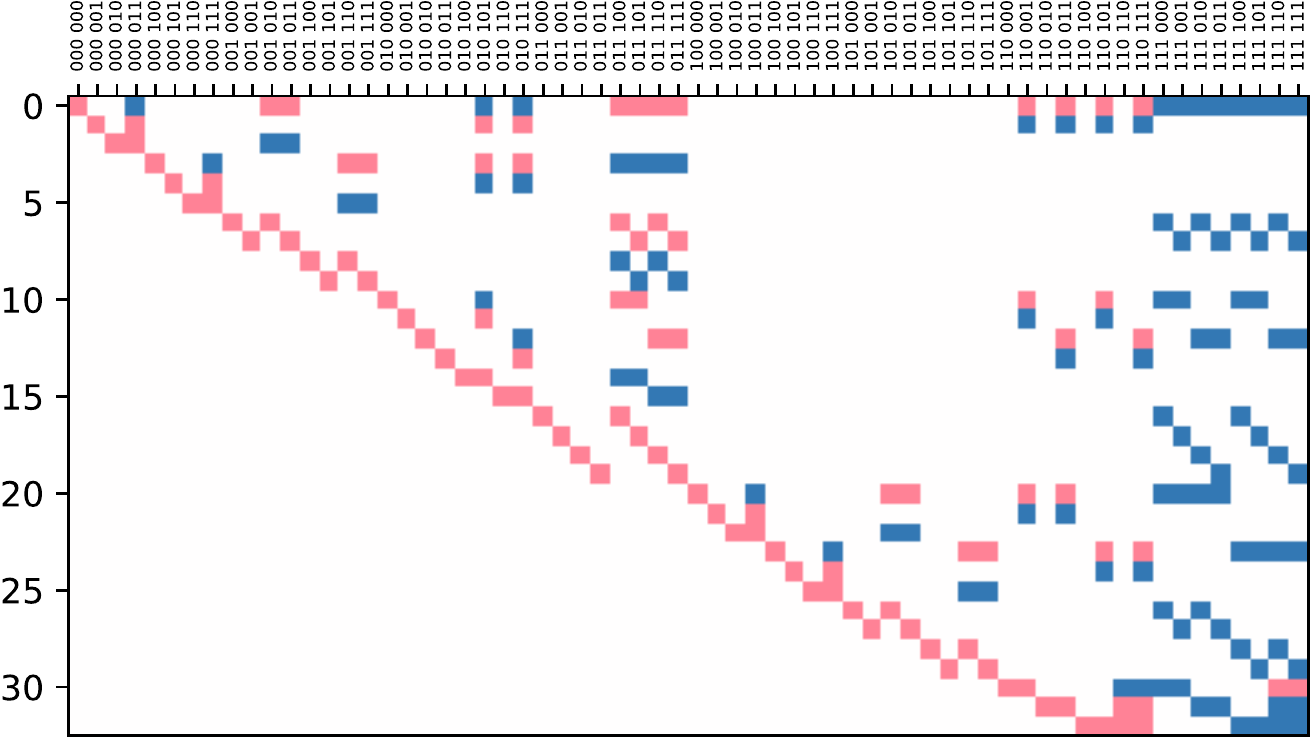}
\end{center}

\noindent Rows correspond to the 33 independent linear equations.
Columns in the plot correspond to entries of empirical models, indexed as $i_A i_B i_C$ $o_A o_B o_C$.
Coefficients in the equations are color-coded as white=0, red=+1 and blue=-1.

Space 16 has closest refinements in equivalence classes 10 and 14; 
it is the join of its (closest) refinements.
It has closest coarsenings in equivalence classes 29, 35, 39 and 41; 
it is the meet of its (closest) coarsenings.
It has 128 causal functions, all of which are causal for at least one of its refinements.
It is not a tight space: for event \ev{B}, a causal function must yield identical output values on input histories \hist{A/0,B/1}, \hist{B/1,C/0} and \hist{B/1,C/1}.

The standard causaltope for Space 16 has 1 more dimension than that of its subspace in equivalence class 14.
The standard causaltope for Space 16 is the meet of the standard causaltopes for its closest coarsenings.
For completeness, below is a plot of the full homogeneous linear system of causality and quasi-normalisation equations for the standard causaltope:

\begin{center}
    \includegraphics[width=12cm]{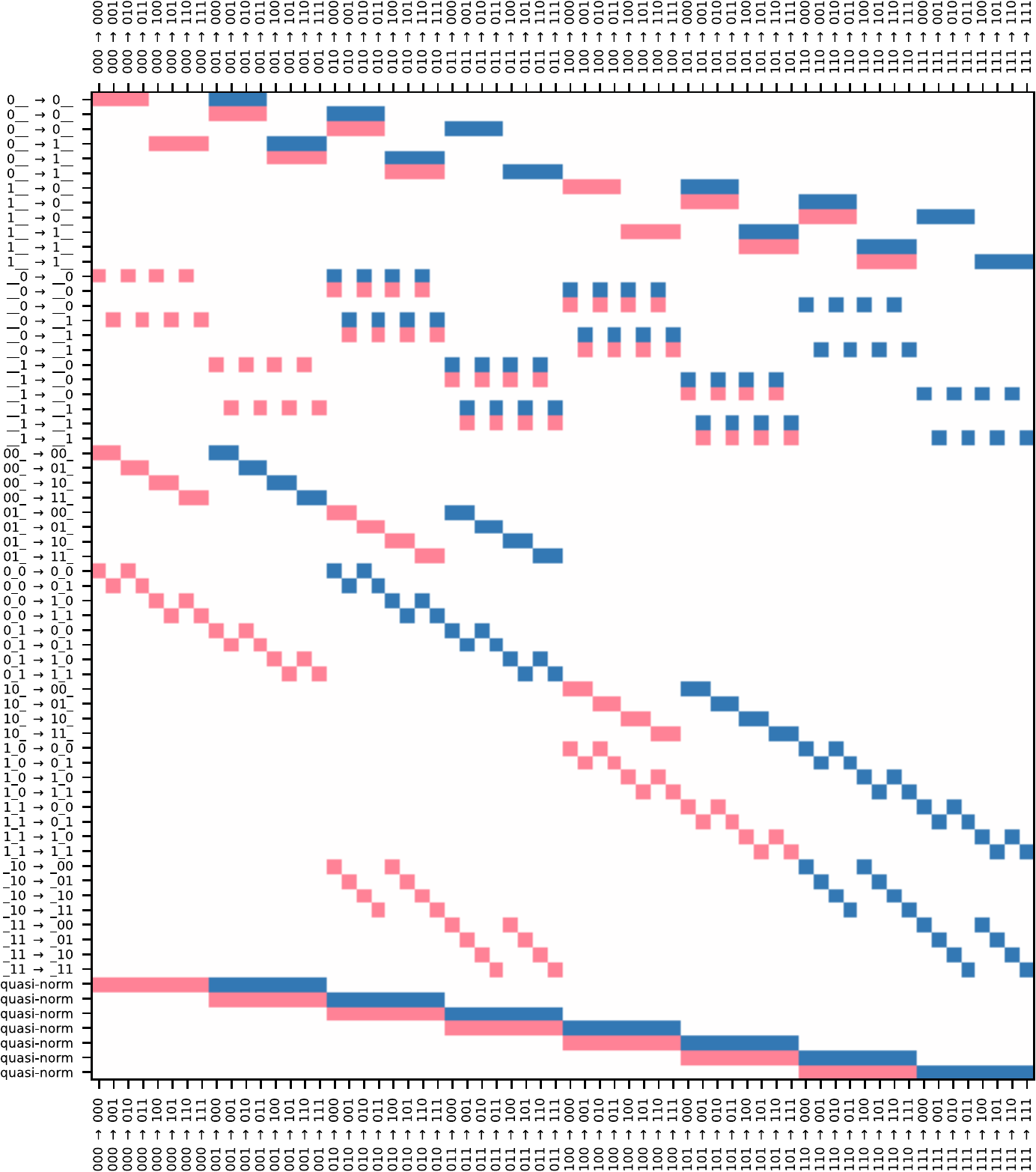}
\end{center}

\noindent Rows correspond to the 67 linear equations, of which 33 are independent.

\newpage
\subsection*{Space 17}

Space 17 is not induced by a causal order, but it is a refinement of the space 100 induced by the definite causal order $\total{\ev{A},\ev{B},\ev{C}}$.
Its equivalence class under event-input permutation symmetry contains 48 spaces.
Space 17 differs as follows from the space induced by causal order $\total{\ev{A},\ev{B},\ev{C}}$:
\begin{itemize}
  \item The outputs at events \evset{\ev{B}, \ev{C}} are independent of the input at event \ev{A} when the inputs at events \evset{B, C} are given by \hist{B/1,C/0} and \hist{B/1,C/1}.
  \item The outputs at events \evset{\ev{A}, \ev{C}} are independent of the input at event \ev{B} when the inputs at events \evset{A, C} are given by \hist{A/0,C/0}, \hist{A/1,C/0} and \hist{A/1,C/1}.
  \item The output at event \ev{C} is independent of the inputs at events \evset{\ev{A}, \ev{B}} when the input at event C is given by \hist{C/0}.
  \item The output at event \ev{B} is independent of the input at event \ev{A} when the input at event B is given by \hist{B/1}.
\end{itemize}

\noindent Below are the histories and extended histories for space 17: 
\begin{center}
    \begin{tabular}{cc}
    \includegraphics[height=3.5cm]{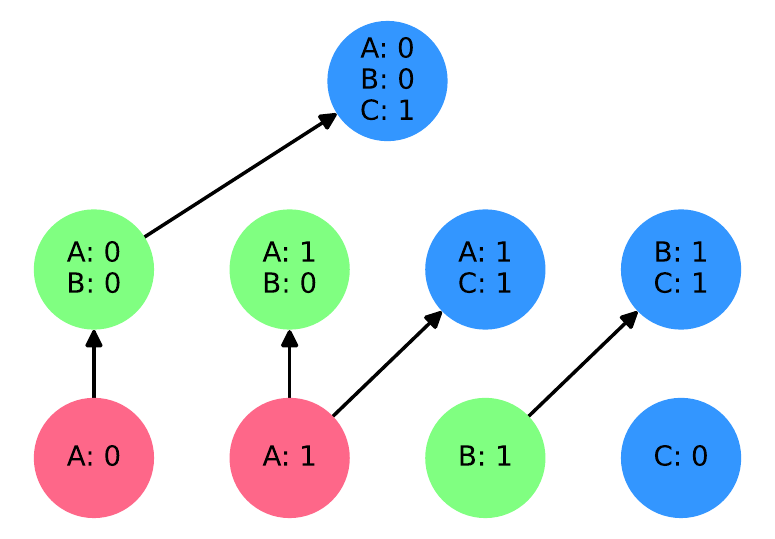}
    &
    \includegraphics[height=3.5cm]{svg-inkscape/space-ABC-unique-untight-17-ext-highlighted_svg-tex.pdf}
    \\
    $\Theta_{17}$
    &
    $\Ext{\Theta_{17}}$
    \end{tabular}
\end{center}

\noindent The standard causaltope for Space 17 has dimension 31.
Below is a plot of the homogeneous linear system of causality and quasi-normalisation equations for the standard causaltope, put in reduced row echelon form:

\begin{center}
    \includegraphics[width=11cm]{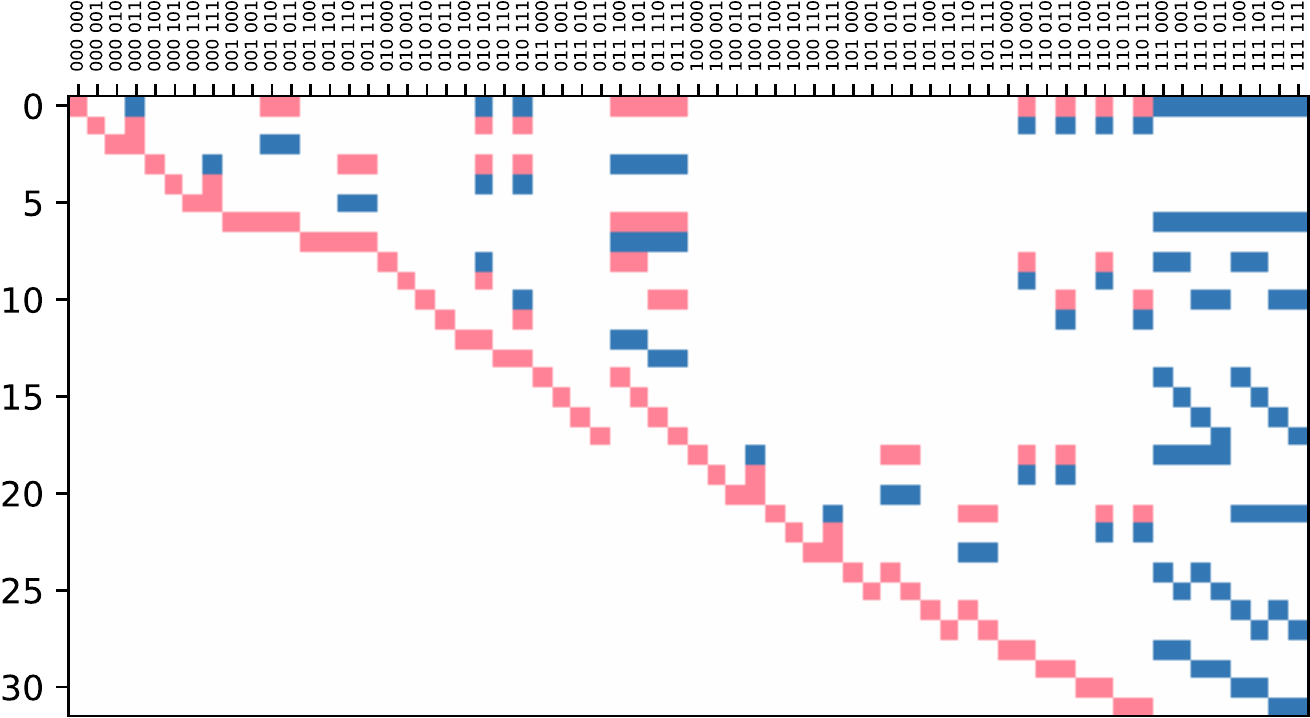}
\end{center}

\noindent Rows correspond to the 32 independent linear equations.
Columns in the plot correspond to entries of empirical models, indexed as $i_A i_B i_C$ $o_A o_B o_C$.
Coefficients in the equations are color-coded as white=0, red=+1 and blue=-1.

Space 17 has closest refinements in equivalence classes 11 and 12; 
it is the join of its (closest) refinements.
It has closest coarsenings in equivalence classes 28, 30, 32 and 43; 
it is the meet of its (closest) coarsenings.
It has 256 causal functions, 128 of which are not causal for any of its refinements.
It is not a tight space: for event \ev{C}, a causal function must yield identical output values on input histories \hist{A/1,C/1} and \hist{B/1,C/1}.

The standard causaltope for Space 17 has 2 more dimensions than those of its 2 subspaces in equivalence classes 11 and 12.
The standard causaltope for Space 17 is the meet of the standard causaltopes for its closest coarsenings.
For completeness, below is a plot of the full homogeneous linear system of causality and quasi-normalisation equations for the standard causaltope:

\begin{center}
    \includegraphics[width=12cm]{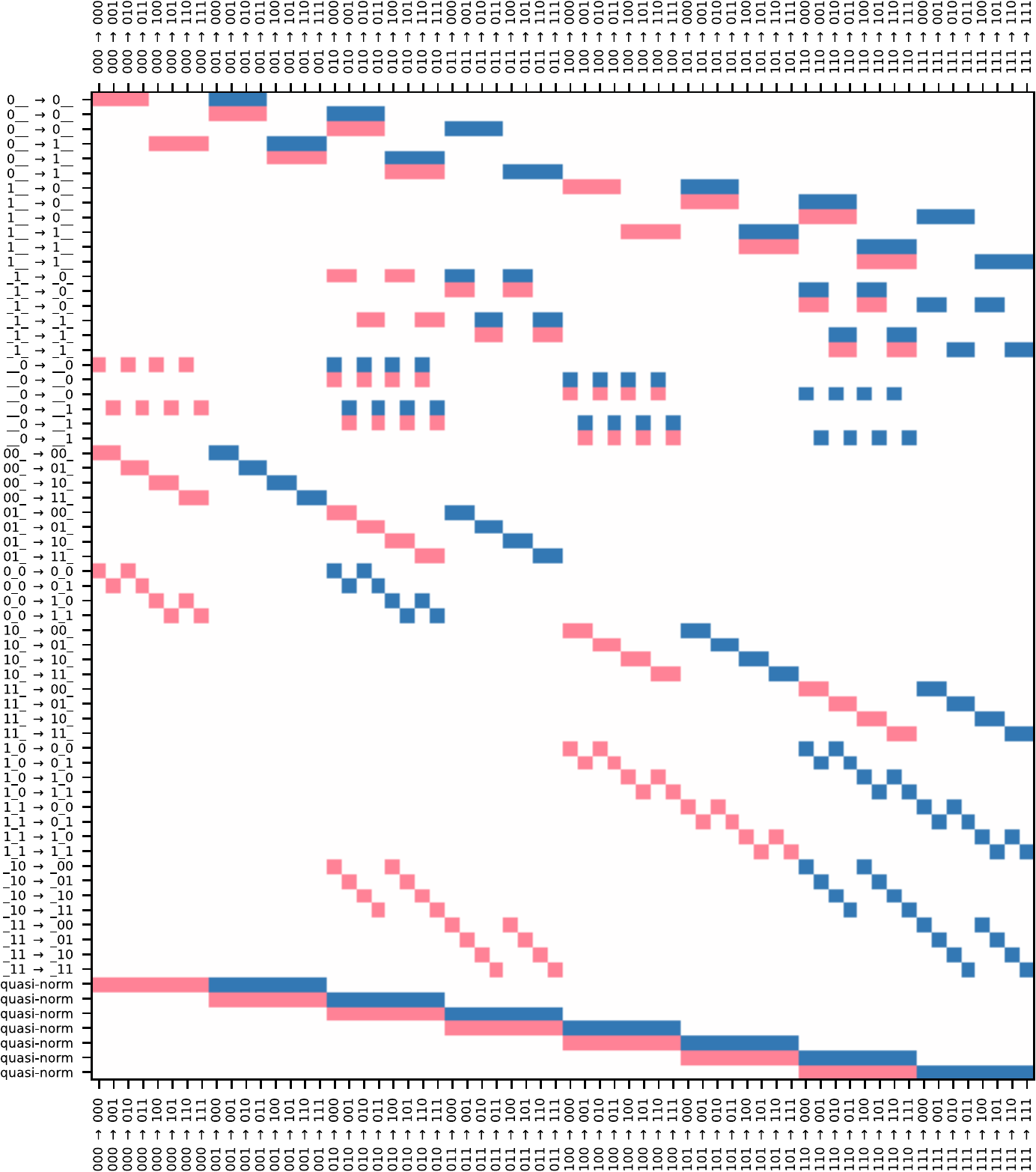}
\end{center}

\noindent Rows correspond to the 67 linear equations, of which 32 are independent.

\newpage
\subsection*{Space 18}

Space 18 is not induced by a causal order, but it is a refinement of the space 92 induced by the definite causal order $\total{\ev{A},\ev{C}}\vee\total{\ev{B},\ev{C}}$.
Its equivalence class under event-input permutation symmetry contains 6 spaces.
Space 18 differs as follows from the space induced by causal order $\total{\ev{A},\ev{C}}\vee\total{\ev{B},\ev{C}}$:
\begin{itemize}
  \item The outputs at events \evset{\ev{A}, \ev{C}} are independent of the input at event \ev{B} when the inputs at events \evset{A, C} are given by \hist{A/0,C/1} and \hist{A/1,C/1}.
  \item The outputs at events \evset{\ev{B}, \ev{C}} are independent of the input at event \ev{A} when the inputs at events \evset{B, C} are given by \hist{B/1,C/1} and \hist{B/0,C/1}.
  \item The output at event \ev{C} is independent of the inputs at events \evset{\ev{A}, \ev{B}} when the input at event C is given by \hist{C/1}.
\end{itemize}

\noindent Below are the histories and extended histories for space 18: 
\begin{center}
    \begin{tabular}{cc}
    \includegraphics[height=3.5cm]{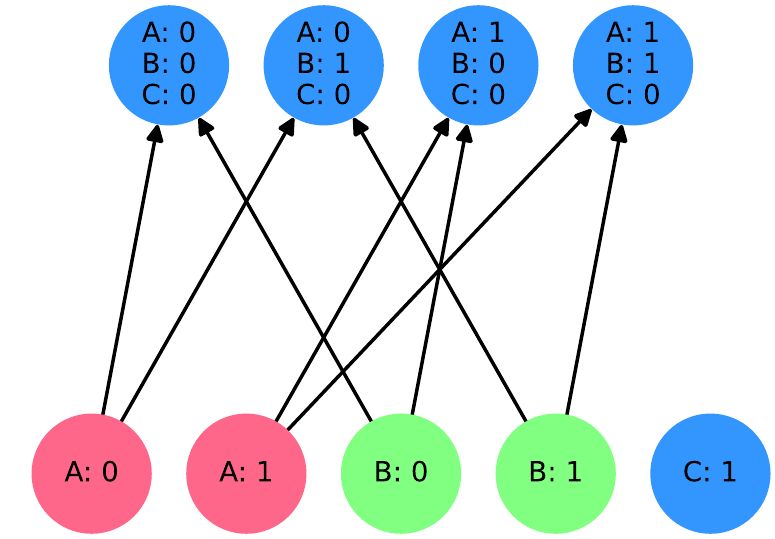}
    &
    \includegraphics[height=3.5cm]{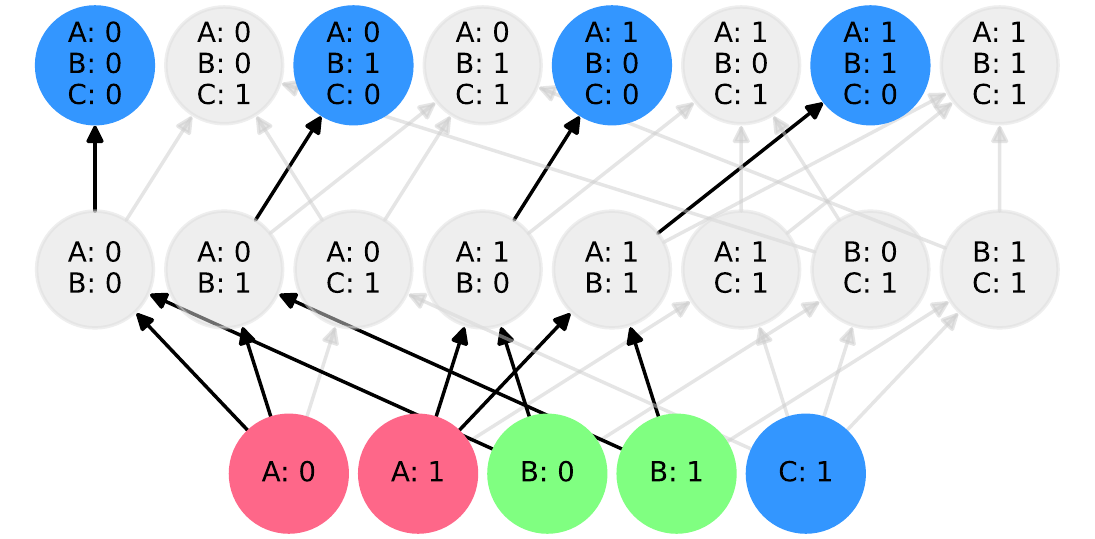}
    \\
    $\Theta_{18}$
    &
    $\Ext{\Theta_{18}}$
    \end{tabular}
\end{center}

\noindent The standard causaltope for Space 18 has dimension 33.
Below is a plot of the homogeneous linear system of causality and quasi-normalisation equations for the standard causaltope, put in reduced row echelon form:

\begin{center}
    \includegraphics[width=11cm]{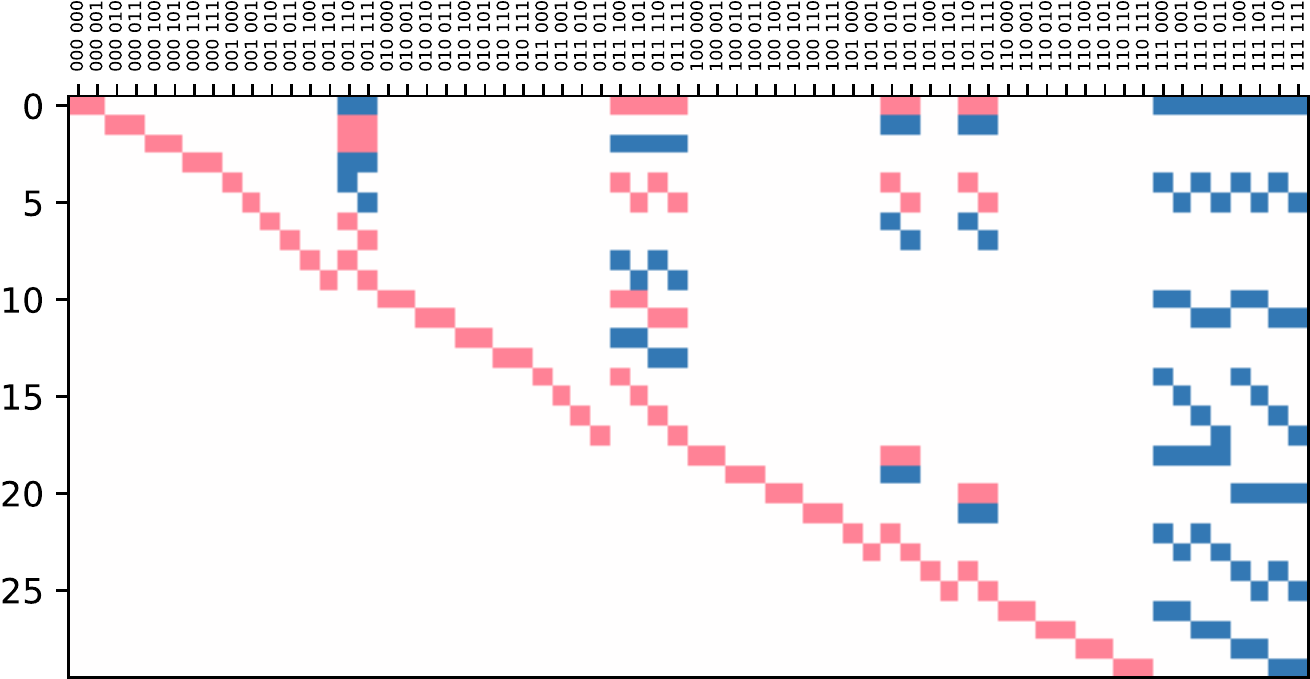}
\end{center}

\noindent Rows correspond to the 30 independent linear equations.
Columns in the plot correspond to entries of empirical models, indexed as $i_A i_B i_C$ $o_A o_B o_C$.
Coefficients in the equations are color-coded as white=0, red=+1 and blue=-1.

Space 18 has closest refinements in equivalence class 8; 
it is the join of its (closest) refinements.
It has closest coarsenings in equivalence classes 31 and 44; 
it is the meet of its (closest) coarsenings.
It has 512 causal functions, 448 of which are not causal for any of its refinements.
It is a tight space.

The standard causaltope for Space 18 has 2 more dimensions than those of its 4 subspaces in equivalence class 8.
The standard causaltope for Space 18 is the meet of the standard causaltopes for its closest coarsenings.
For completeness, below is a plot of the full homogeneous linear system of causality and quasi-normalisation equations for the standard causaltope:

\begin{center}
    \includegraphics[width=12cm]{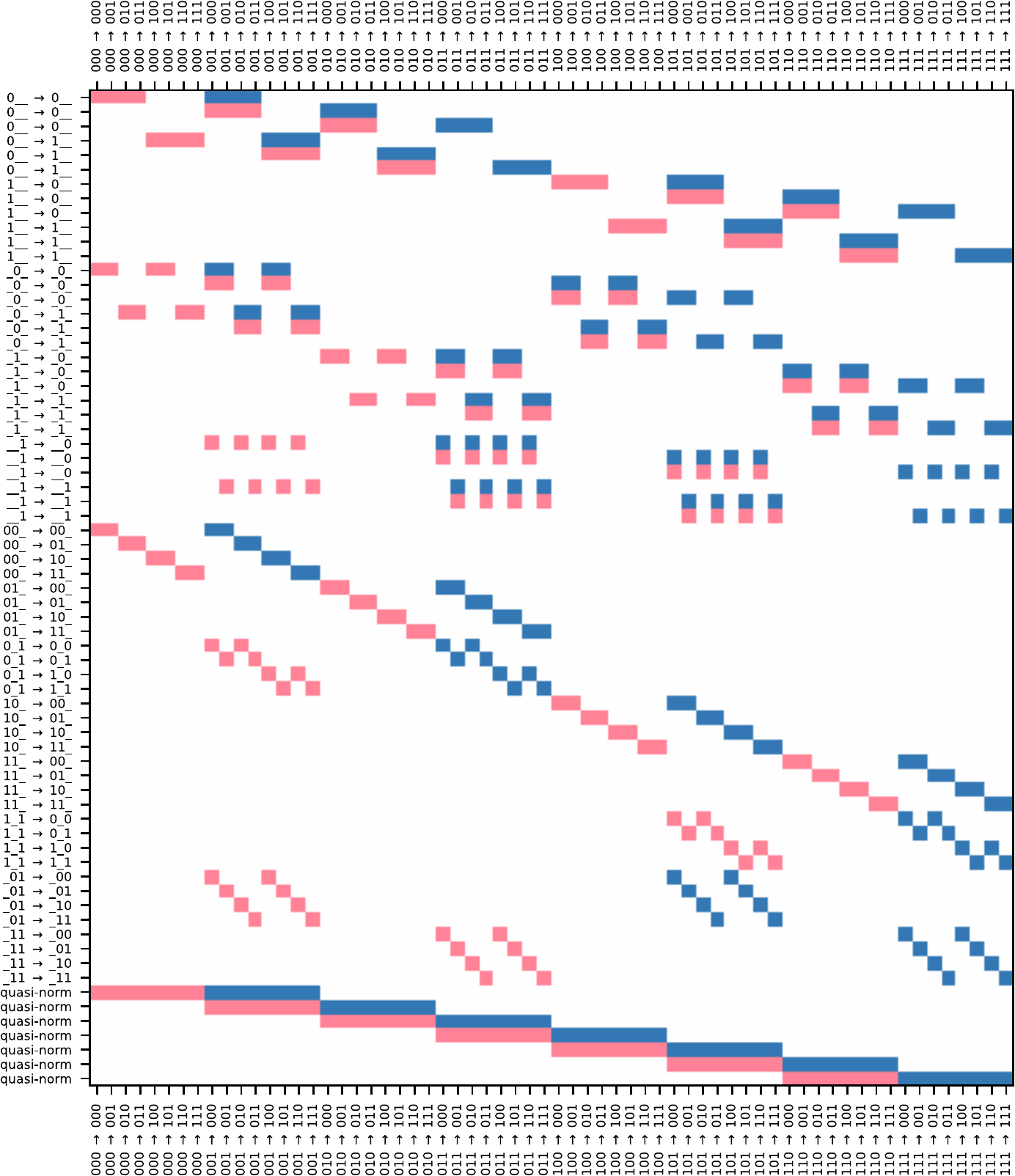}
\end{center}

\noindent Rows correspond to the 69 linear equations, of which 30 are independent.

\newpage
\subsection*{Space 19}

Space 19 is not induced by a causal order, but it is a refinement of the space 33 induced by the definite causal order $\total{\ev{A},\ev{B}}\vee\discrete{\ev{C}}$.
Its equivalence class under event-input permutation symmetry contains 24 spaces.
Space 19 differs as follows from the space induced by causal order $\total{\ev{A},\ev{B}}\vee\discrete{\ev{C}}$:
\begin{itemize}
  \item The outputs at events \evset{\ev{B}, \ev{C}} are independent of the input at event \ev{A} when the inputs at events \evset{B, C} are given by \hist{B/1,C/0}.
\end{itemize}

\noindent Below are the histories and extended histories for space 19: 
\begin{center}
    \begin{tabular}{cc}
    \includegraphics[height=3.5cm]{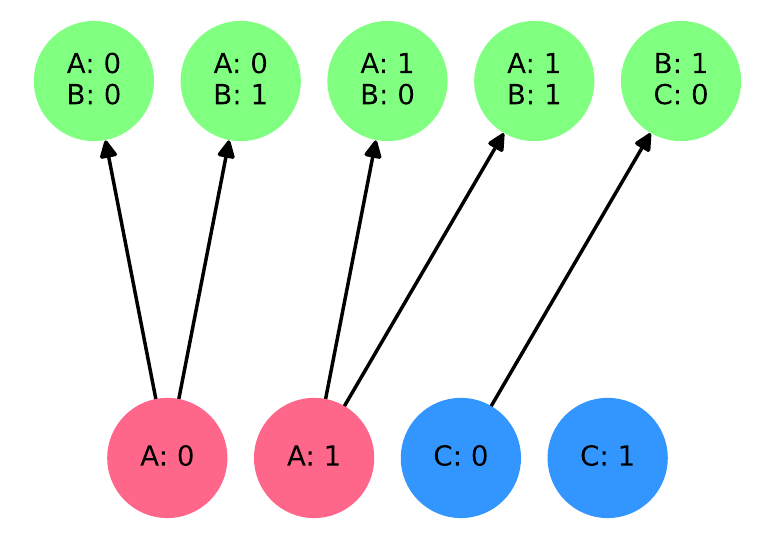}
    &
    \includegraphics[height=3.5cm]{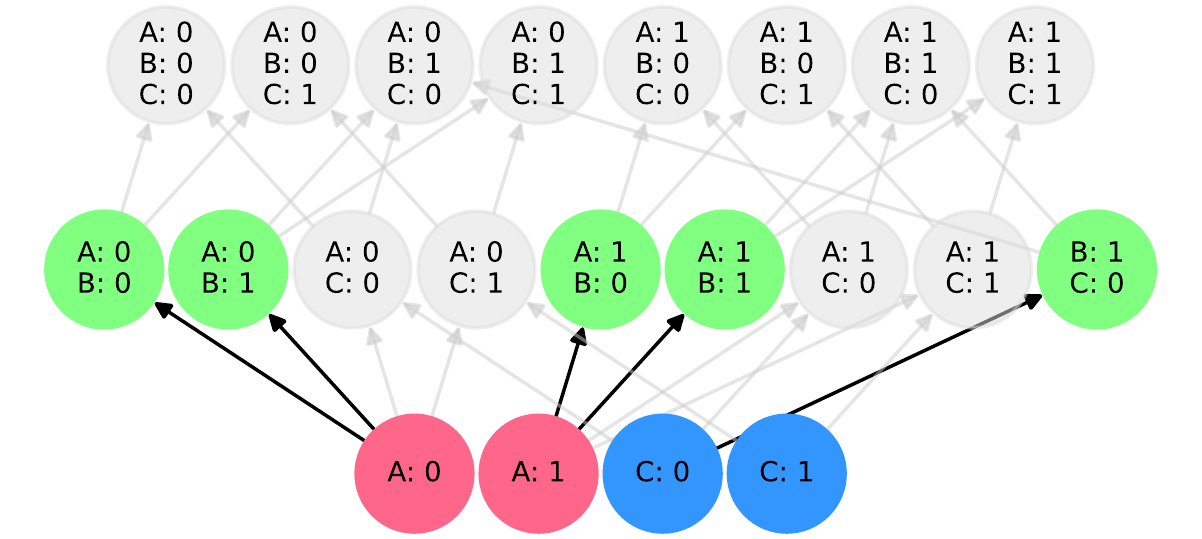}
    \\
    $\Theta_{19}$
    &
    $\Ext{\Theta_{19}}$
    \end{tabular}
\end{center}

\noindent The standard causaltope for Space 19 has dimension 30.
Below is a plot of the homogeneous linear system of causality and quasi-normalisation equations for the standard causaltope, put in reduced row echelon form:

\begin{center}
    \includegraphics[width=11cm]{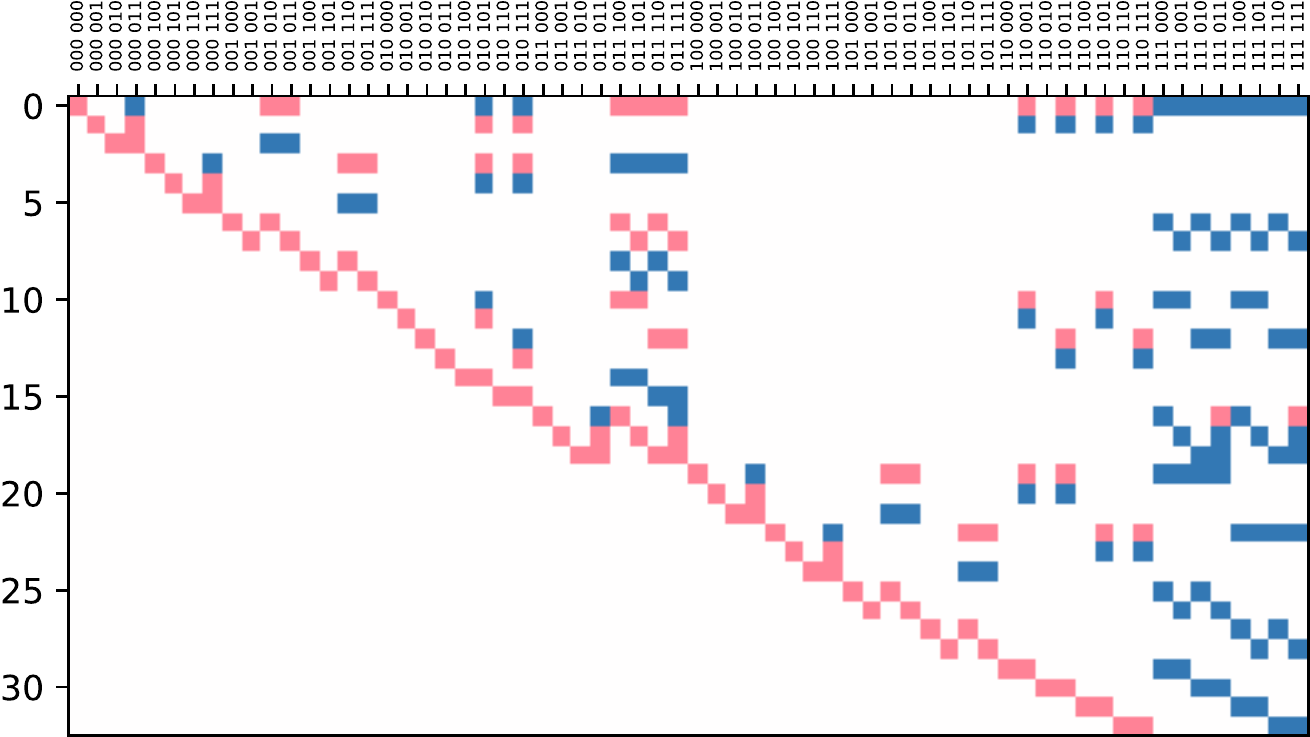}
\end{center}

\noindent Rows correspond to the 33 independent linear equations.
Columns in the plot correspond to entries of empirical models, indexed as $i_A i_B i_C$ $o_A o_B o_C$.
Coefficients in the equations are color-coded as white=0, red=+1 and blue=-1.

Space 19 has closest refinements in equivalence classes 9, 13 and 14; 
it is the join of its (closest) refinements.
It has closest coarsenings in equivalence classes 29, 33, 34 and 38; 
it is the meet of its (closest) coarsenings.
It has 128 causal functions, all of which are causal for at least one of its refinements.
It is not a tight space: for event \ev{B}, a causal function must yield identical output values on input histories \hist{A/0,B/1}, \hist{A/1,B/1} and \hist{B/1,C/0}.

The standard causaltope for Space 19 has 1 more dimension than that of its subspace in equivalence class 14.
The standard causaltope for Space 19 is the meet of the standard causaltopes for its closest coarsenings.
For completeness, below is a plot of the full homogeneous linear system of causality and quasi-normalisation equations for the standard causaltope:

\begin{center}
    \includegraphics[width=12cm]{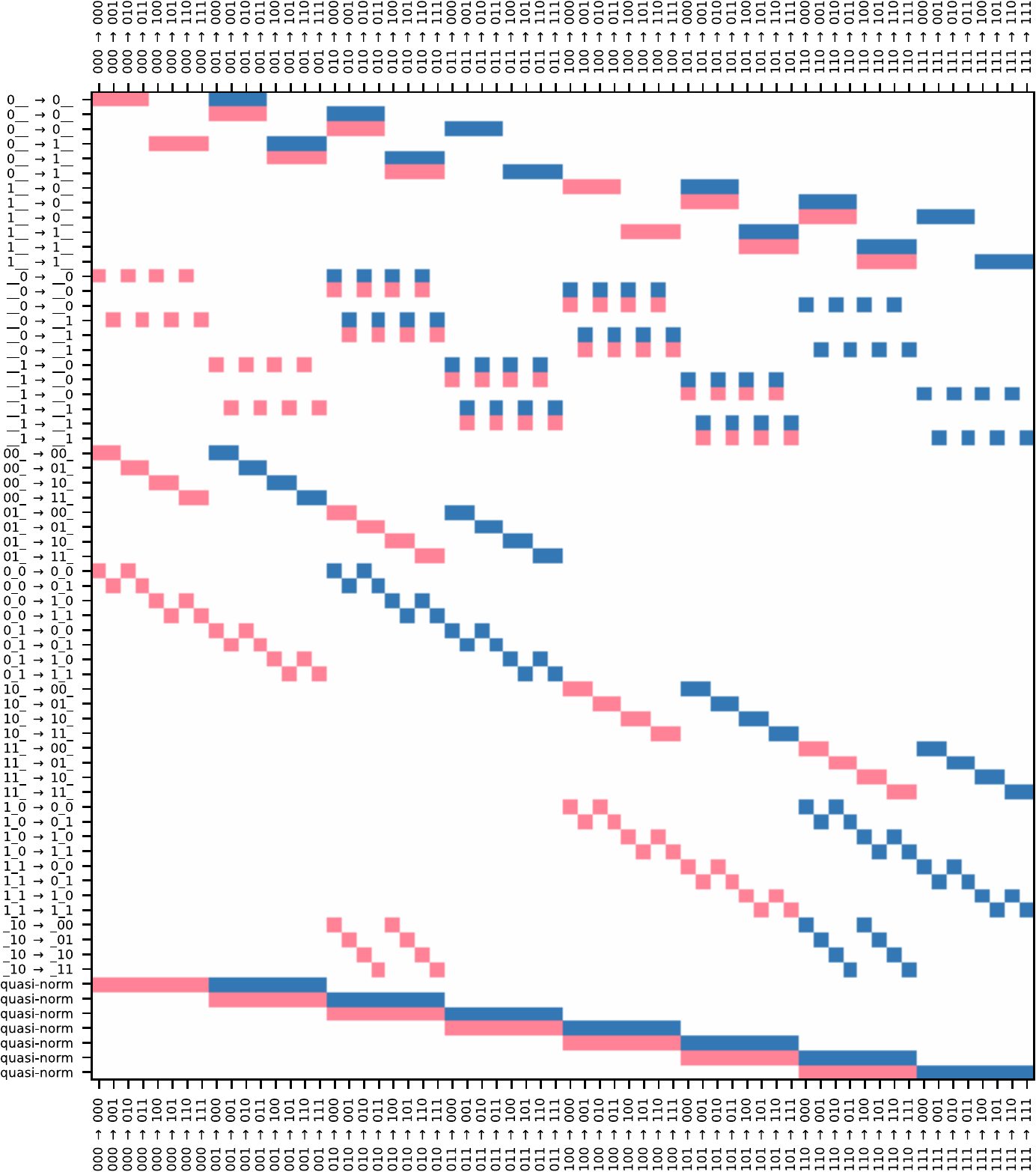}
\end{center}

\noindent Rows correspond to the 67 linear equations, of which 33 are independent.

\newpage
\subsection*{Space 20}

Space 20 is not induced by a causal order, but it is a refinement of the space 100 induced by the definite causal order $\total{\ev{A},\ev{B},\ev{C}}$.
Its equivalence class under event-input permutation symmetry contains 24 spaces.
Space 20 differs as follows from the space induced by causal order $\total{\ev{A},\ev{B},\ev{C}}$:
\begin{itemize}
  \item The outputs at events \evset{\ev{B}, \ev{C}} are independent of the input at event \ev{A} when the inputs at events \evset{B, C} are given by \hist{B/1,C/0}, \hist{B/0,C/0} and \hist{B/0,C/1}.
  \item The output at event \ev{B} is independent of the input at event \ev{A} when the input at event B is given by \hist{B/0}.
  \item The outputs at events \evset{\ev{A}, \ev{C}} are independent of the input at event \ev{B} when the inputs at events \evset{A, C} are given by \hist{A/0,C/0} and \hist{A/1,C/0}.
  \item The output at event \ev{C} is independent of the inputs at events \evset{\ev{A}, \ev{B}} when the input at event C is given by \hist{C/0}.
\end{itemize}

\noindent Below are the histories and extended histories for space 20: 
\begin{center}
    \begin{tabular}{cc}
    \includegraphics[height=3.5cm]{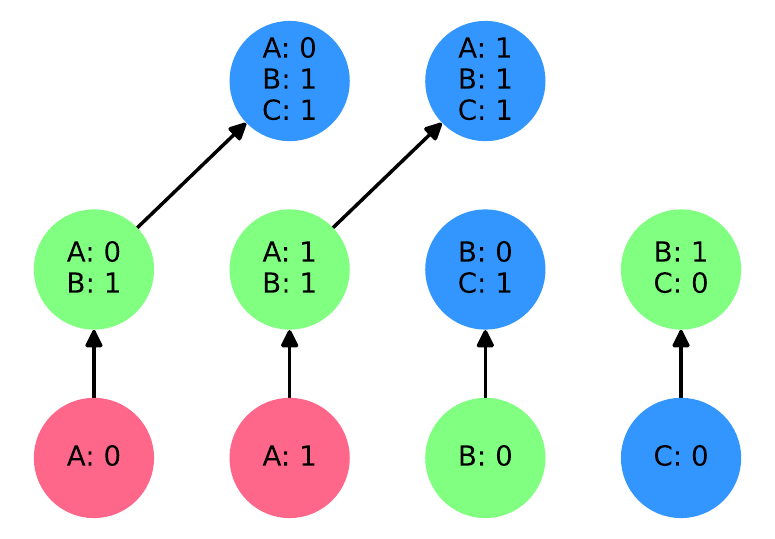}
    &
    \includegraphics[height=3.5cm]{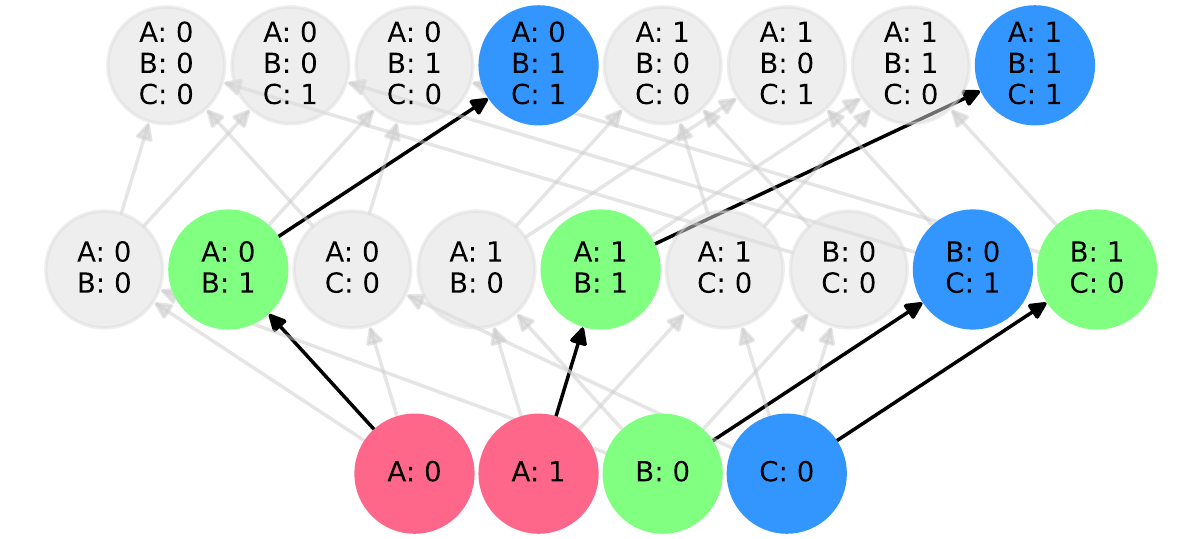}
    \\
    $\Theta_{20}$
    &
    $\Ext{\Theta_{20}}$
    \end{tabular}
\end{center}

\noindent The standard causaltope for Space 20 has dimension 31.
Below is a plot of the homogeneous linear system of causality and quasi-normalisation equations for the standard causaltope, put in reduced row echelon form:

\begin{center}
    \includegraphics[width=11cm]{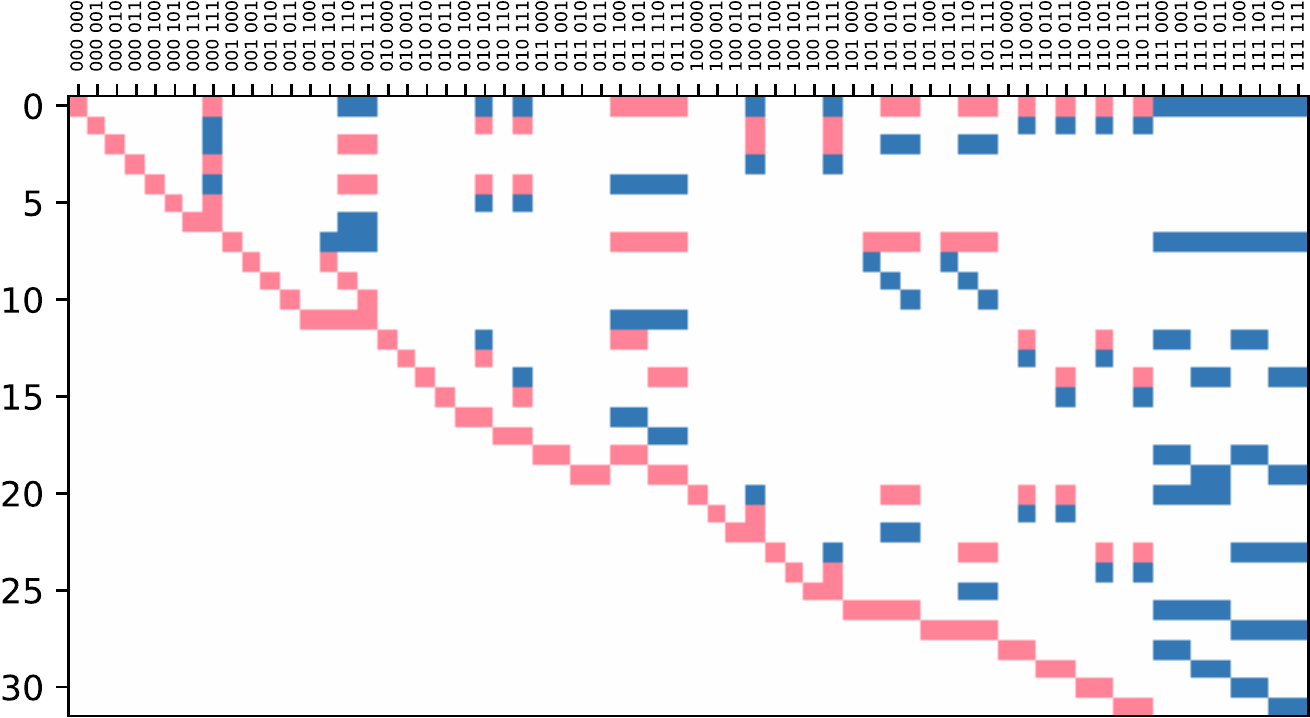}
\end{center}

\noindent Rows correspond to the 32 independent linear equations.
Columns in the plot correspond to entries of empirical models, indexed as $i_A i_B i_C$ $o_A o_B o_C$.
Coefficients in the equations are color-coded as white=0, red=+1 and blue=-1.

Space 20 has closest refinements in equivalence classes 8 and 11; 
it is the join of its (closest) refinements.
It has closest coarsenings in equivalence classes 28 and 31; 
it is the meet of its (closest) coarsenings.
It has 256 causal functions, 192 of which are not causal for any of its refinements.
It is not a tight space: for event \ev{B}, a causal function must yield identical output values on input histories \hist{A/0,B/1}, \hist{A/1,B/1} and \hist{B/1,C/0}.

The standard causaltope for Space 20 coincides with that of its subspace in equivalence class 8.
The standard causaltope for Space 20 is the meet of the standard causaltopes for its closest coarsenings.
For completeness, below is a plot of the full homogeneous linear system of causality and quasi-normalisation equations for the standard causaltope:

\begin{center}
    \includegraphics[width=12cm]{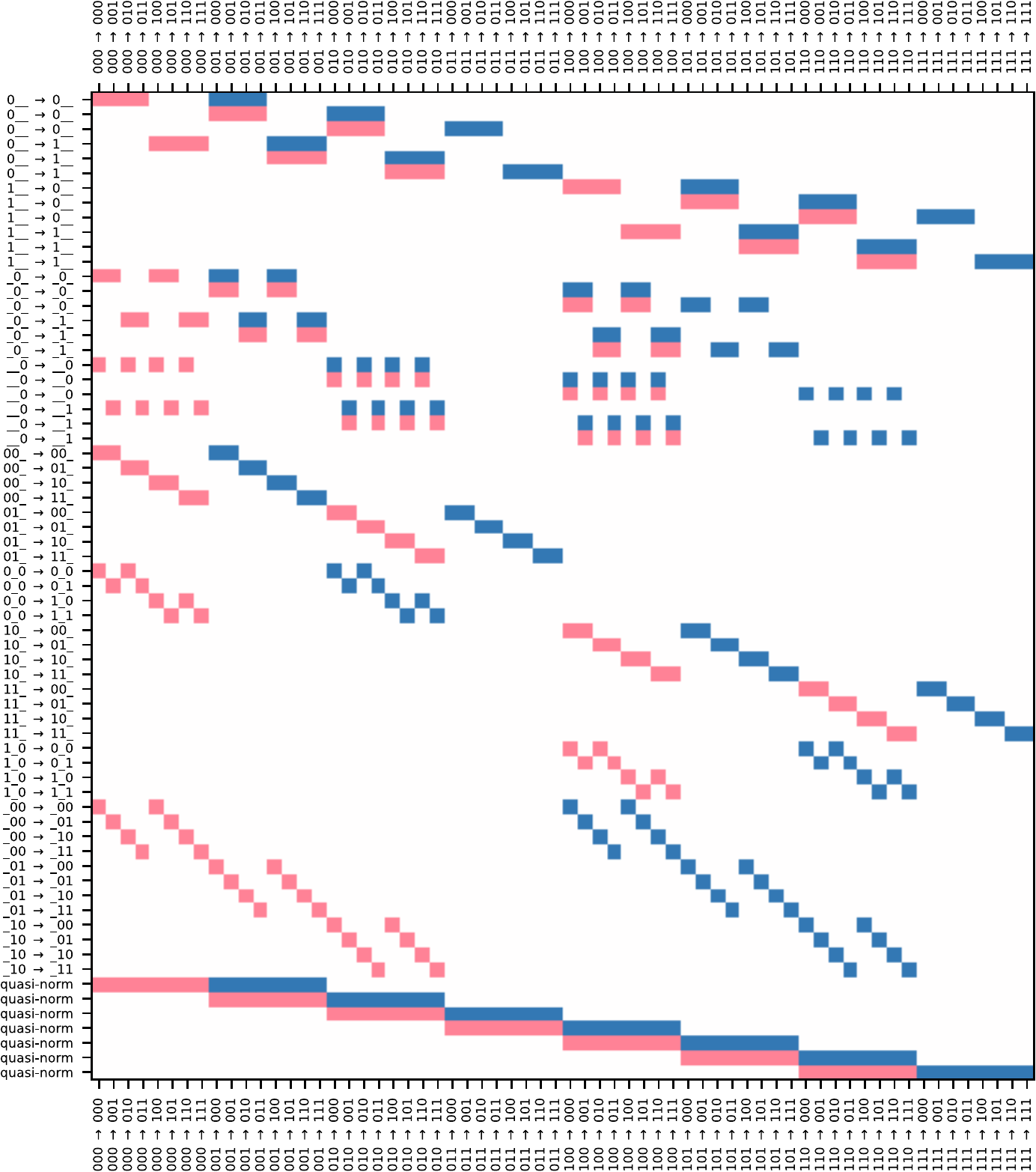}
\end{center}

\noindent Rows correspond to the 67 linear equations, of which 32 are independent.

\newpage
\subsection*{Space 21}

Space 21 is not induced by a causal order, but it is a refinement of the space induced by the indefinite causal order $\total{\ev{A},\{\ev{B},\ev{C}\}}$.
Its equivalence class under event-input permutation symmetry contains 24 spaces.
Space 21 differs as follows from the space induced by causal order $\total{\ev{A},\{\ev{B},\ev{C}\}}$:
\begin{itemize}
  \item The outputs at events \evset{\ev{A}, \ev{B}} are independent of the input at event \ev{C} when the inputs at events \evset{A, B} are given by \hist{A/0,B/0}, \hist{A/0,B/1} and \hist{A/1,B/0}.
  \item The outputs at events \evset{\ev{B}, \ev{C}} are independent of the input at event \ev{A} when the inputs at events \evset{B, C} are given by \hist{B/1,C/0}, \hist{B/0,C/0} and \hist{B/0,C/1}.
  \item The outputs at events \evset{\ev{A}, \ev{C}} are independent of the input at event \ev{B} when the inputs at events \evset{A, C} are given by \hist{A/0,C/0}, \hist{A/1,C/0} and \hist{A/1,C/1}.
  \item The output at event \ev{B} is independent of the inputs at events \evset{\ev{A}, \ev{C}} when the input at event B is given by \hist{B/0}.
  \item The output at event \ev{C} is independent of the inputs at events \evset{\ev{A}, \ev{B}} when the input at event C is given by \hist{C/0}.
\end{itemize}

\noindent Below are the histories and extended histories for space 21: 
\begin{center}
    \begin{tabular}{cc}
    \includegraphics[height=3.5cm]{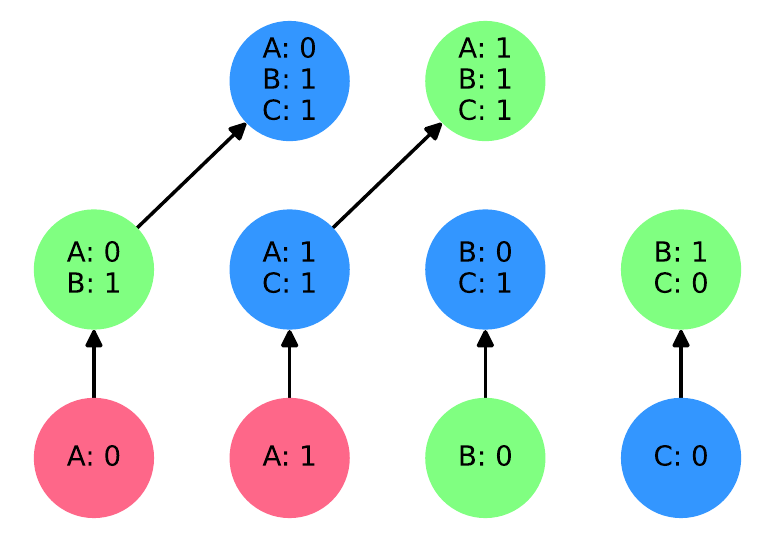}
    &
    \includegraphics[height=3.5cm]{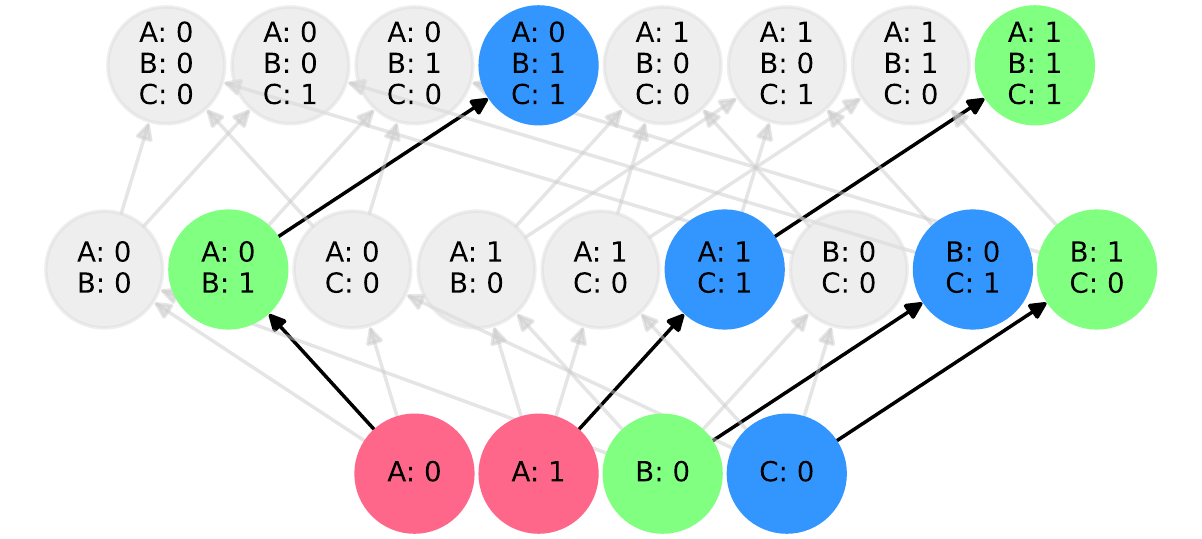}
    \\
    $\Theta_{21}$
    &
    $\Ext{\Theta_{21}}$
    \end{tabular}
\end{center}

\noindent The standard causaltope for Space 21 has dimension 31.
Below is a plot of the homogeneous linear system of causality and quasi-normalisation equations for the standard causaltope, put in reduced row echelon form:

\begin{center}
    \includegraphics[width=11cm]{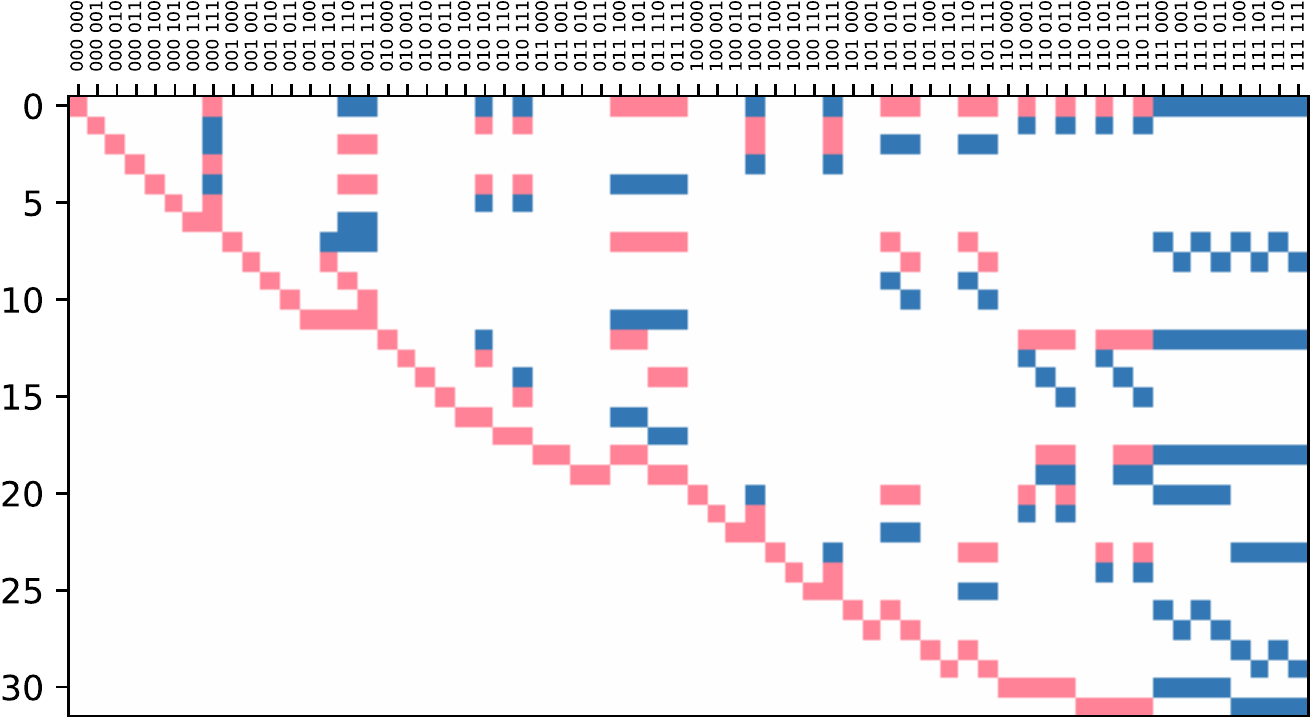}
\end{center}

\noindent Rows correspond to the 32 independent linear equations.
Columns in the plot correspond to entries of empirical models, indexed as $i_A i_B i_C$ $o_A o_B o_C$.
Coefficients in the equations are color-coded as white=0, red=+1 and blue=-1.

Space 21 has closest refinements in equivalence class 11; 
it is the join of its (closest) refinements.
It has closest coarsenings in equivalence class 32; 
it is the meet of its (closest) coarsenings.
It has 256 causal functions, 64 of which are not causal for any of its refinements.
It is not a tight space: for event \ev{B}, a causal function must yield identical output values on input histories \hist{A/0,B/1} and \hist{B/1,C/0}; for event \ev{C}, a causal function must yield identical output values on input histories \hist{A/1,C/1} and \hist{B/0,C/1}.

The standard causaltope for Space 21 has 2 more dimensions than those of its 2 subspaces in equivalence class 11.
The standard causaltope for Space 21 is the meet of the standard causaltopes for its closest coarsenings.
For completeness, below is a plot of the full homogeneous linear system of causality and quasi-normalisation equations for the standard causaltope:

\begin{center}
    \includegraphics[width=12cm]{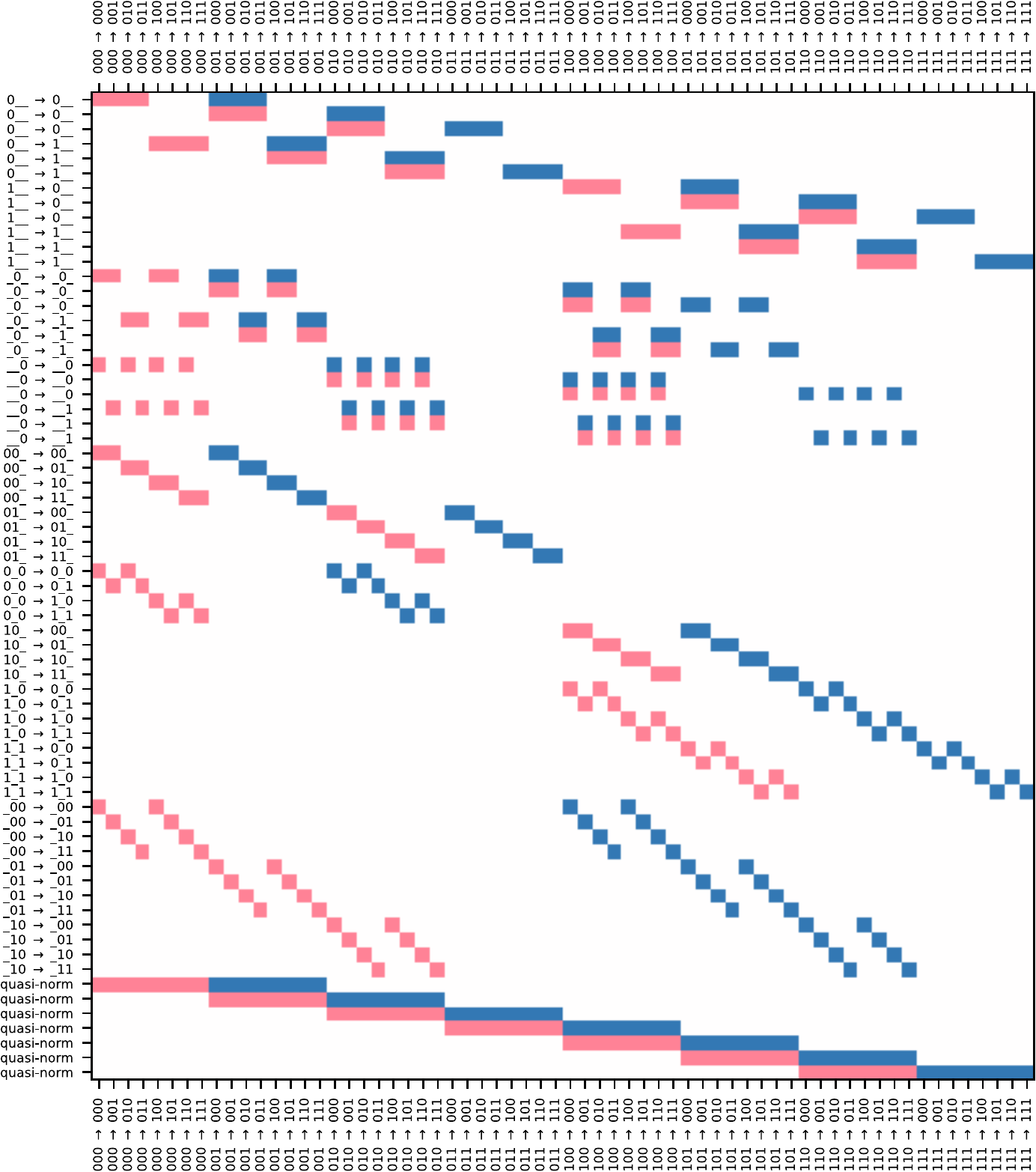}
\end{center}

\noindent Rows correspond to the 67 linear equations, of which 32 are independent.

\newpage
\subsection*{Space 22}

Space 22 is not induced by a causal order, but it is a refinement of the space 100 induced by the definite causal order $\total{\ev{A},\ev{B},\ev{C}}$.
Its equivalence class under event-input permutation symmetry contains 48 spaces.
Space 22 differs as follows from the space induced by causal order $\total{\ev{A},\ev{B},\ev{C}}$:
\begin{itemize}
  \item The outputs at events \evset{\ev{A}, \ev{C}} are independent of the input at event \ev{B} when the inputs at events \evset{A, C} are given by \hist{A/0,C/1}, \hist{A/1,C/0} and \hist{A/1,C/1}.
  \item The outputs at events \evset{\ev{B}, \ev{C}} are independent of the input at event \ev{A} when the inputs at events \evset{B, C} are given by \hist{B/1,C/1} and \hist{B/0,C/1}.
  \item The output at event \ev{C} is independent of the inputs at events \evset{\ev{A}, \ev{B}} when the input at event C is given by \hist{C/1}.
  \item The output at event \ev{B} is independent of the input at event \ev{A} when the input at event B is given by \hist{B/1}.
\end{itemize}

\noindent Below are the histories and extended histories for space 22: 
\begin{center}
    \begin{tabular}{cc}
    \includegraphics[height=3.5cm]{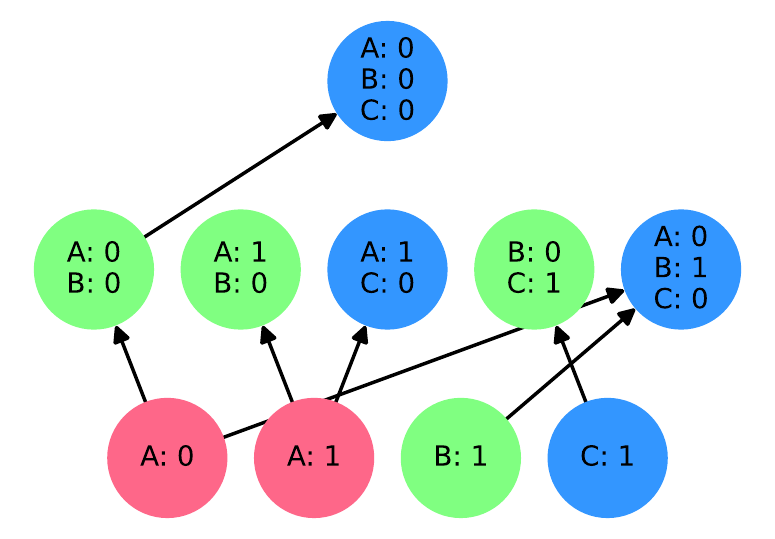}
    &
    \includegraphics[height=3.5cm]{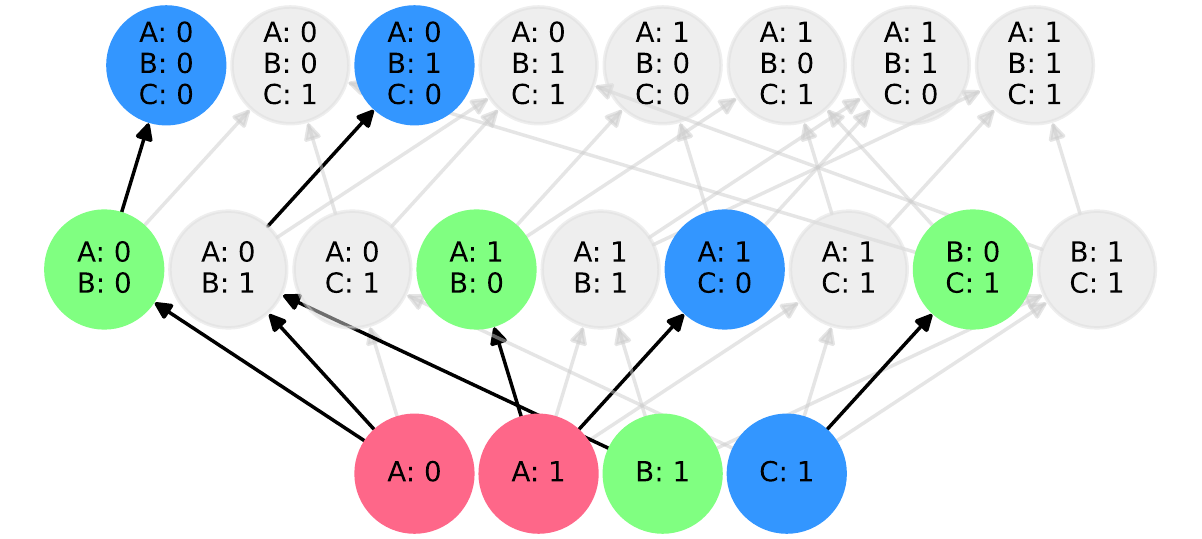}
    \\
    $\Theta_{22}$
    &
    $\Ext{\Theta_{22}}$
    \end{tabular}
\end{center}

\noindent The standard causaltope for Space 22 has dimension 31.
Below is a plot of the homogeneous linear system of causality and quasi-normalisation equations for the standard causaltope, put in reduced row echelon form:

\begin{center}
    \includegraphics[width=11cm]{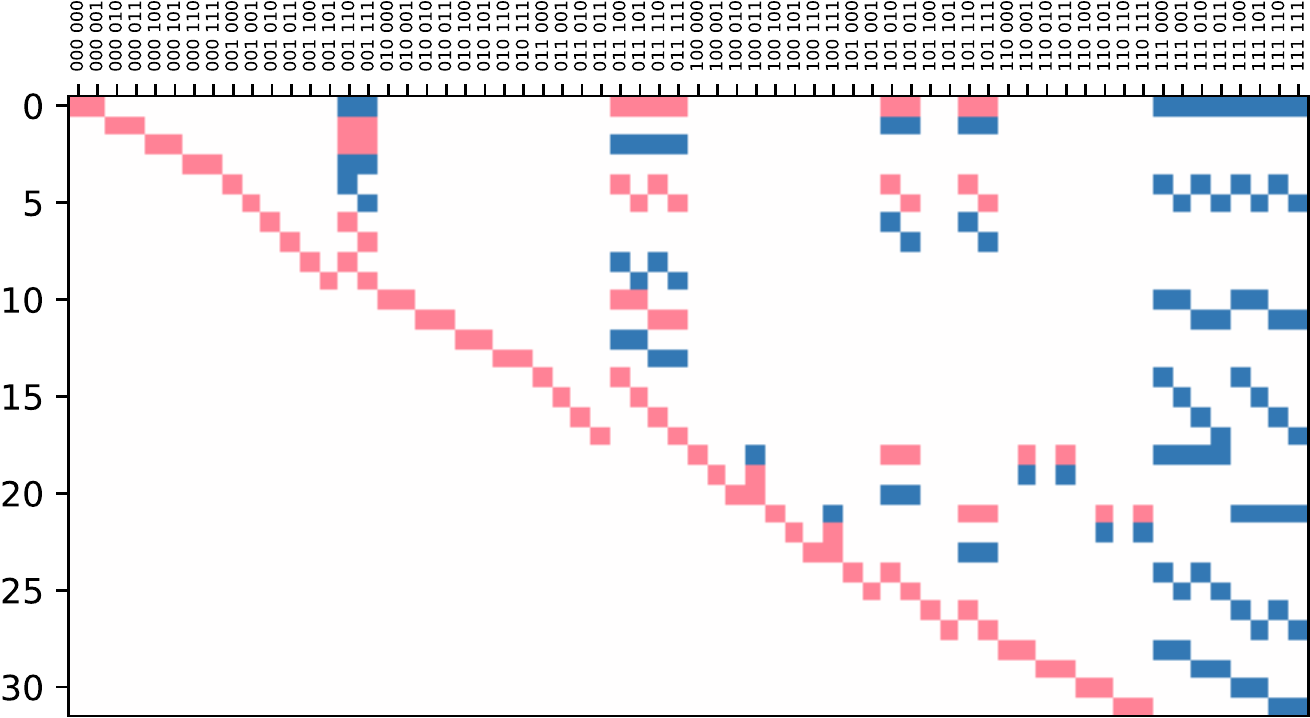}
\end{center}

\noindent Rows correspond to the 32 independent linear equations.
Columns in the plot correspond to entries of empirical models, indexed as $i_A i_B i_C$ $o_A o_B o_C$.
Coefficients in the equations are color-coded as white=0, red=+1 and blue=-1.

Space 22 has closest refinements in equivalence classes 8, 11 and 12; 
it is the join of its (closest) refinements.
It has closest coarsenings in equivalence classes 30, 31, 32 and 42; 
it is the meet of its (closest) coarsenings.
It has 256 causal functions, all of which are causal for at least one of its refinements.
It is not a tight space: for event \ev{B}, a causal function must yield identical output values on input histories \hist{A/0,B/0}, \hist{A/1,B/0} and \hist{B/0,C/1}.

The standard causaltope for Space 22 coincides with that of its subspace in equivalence class 8.
The standard causaltope for Space 22 is the meet of the standard causaltopes for its closest coarsenings.
For completeness, below is a plot of the full homogeneous linear system of causality and quasi-normalisation equations for the standard causaltope:

\begin{center}
    \includegraphics[width=12cm]{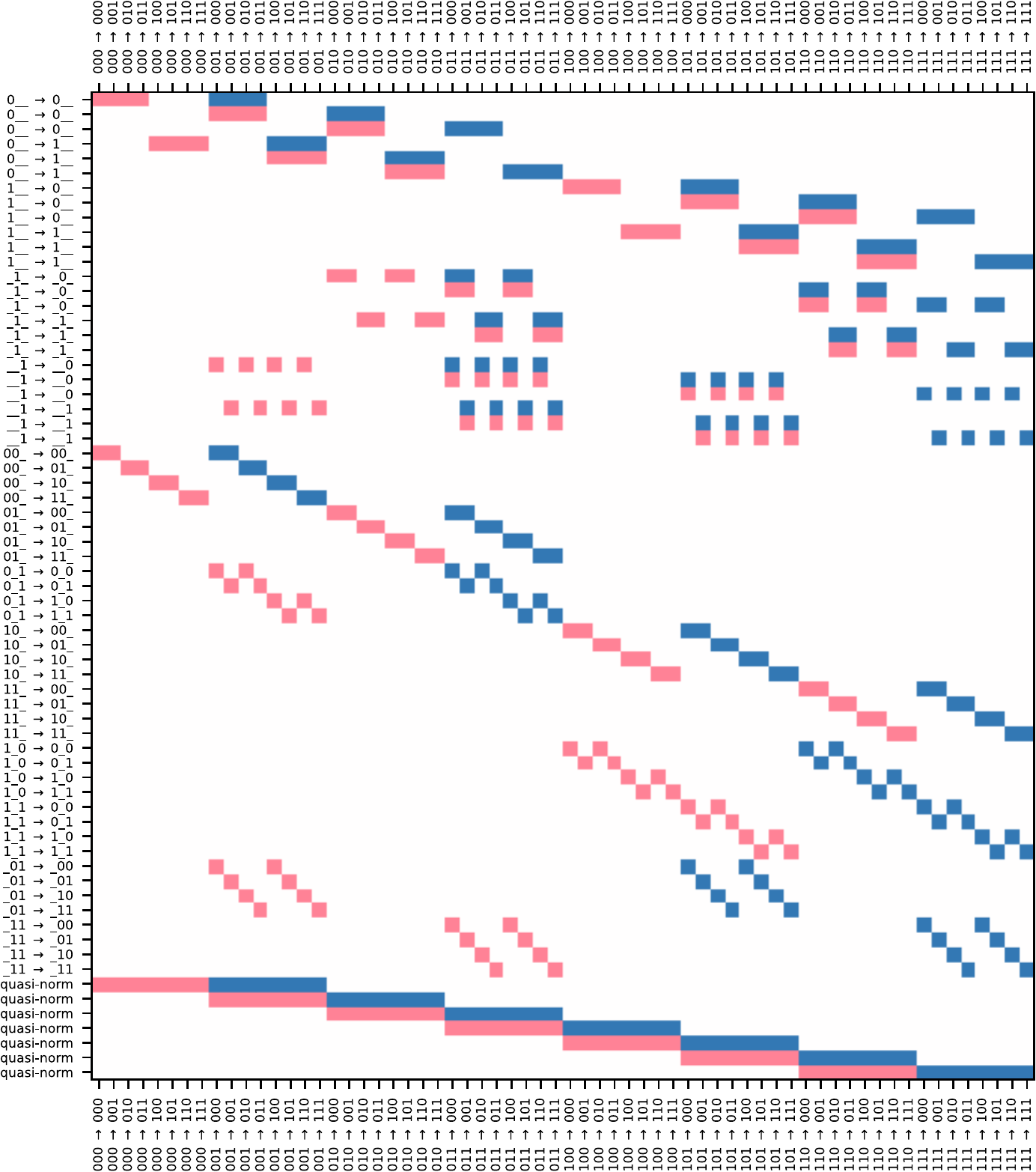}
\end{center}

\noindent Rows correspond to the 67 linear equations, of which 32 are independent.

\newpage
\subsection*{Space 23}

Space 23 is not induced by a causal order, but it is a refinement of the space in equivalence class 92 induced by the definite causal order $\total{\ev{A},\ev{B}}\vee\total{\ev{C},\ev{B}}$ (note that the space induced by the order is not the same as space 92).
Its equivalence class under event-input permutation symmetry contains 48 spaces.
Space 23 differs as follows from the space induced by causal order $\total{\ev{A},\ev{B}}\vee\total{\ev{C},\ev{B}}$:
\begin{itemize}
  \item The outputs at events \evset{\ev{A}, \ev{B}} are independent of the input at event \ev{C} when the inputs at events \evset{A, B} are given by \hist{A/0,B/0}, \hist{A/0,B/1} and \hist{A/1,B/0}.
  \item The outputs at events \evset{\ev{B}, \ev{C}} are independent of the input at event \ev{A} when the inputs at events \evset{B, C} are given by \hist{B/1,C/1} and \hist{B/0,C/1}.
\end{itemize}

\noindent Below are the histories and extended histories for space 23: 
\begin{center}
    \begin{tabular}{cc}
    \includegraphics[height=3.5cm]{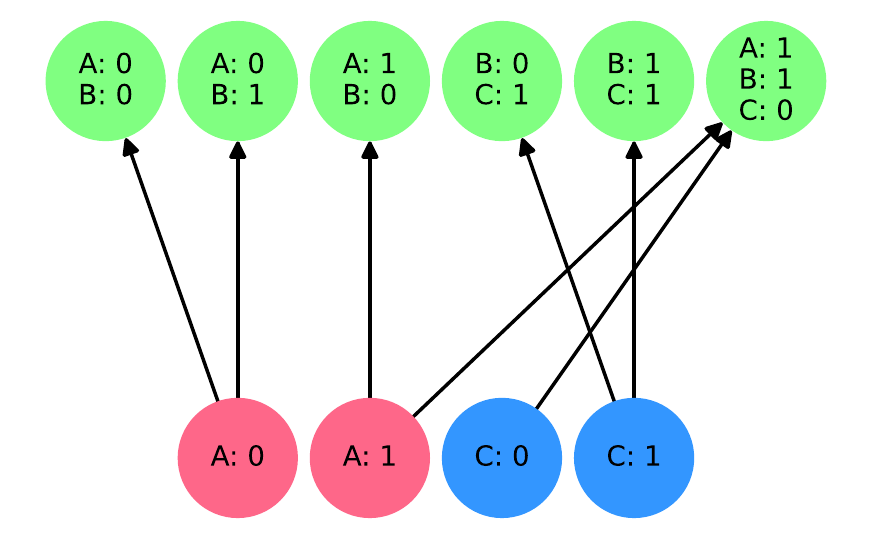}
    &
    \includegraphics[height=3.5cm]{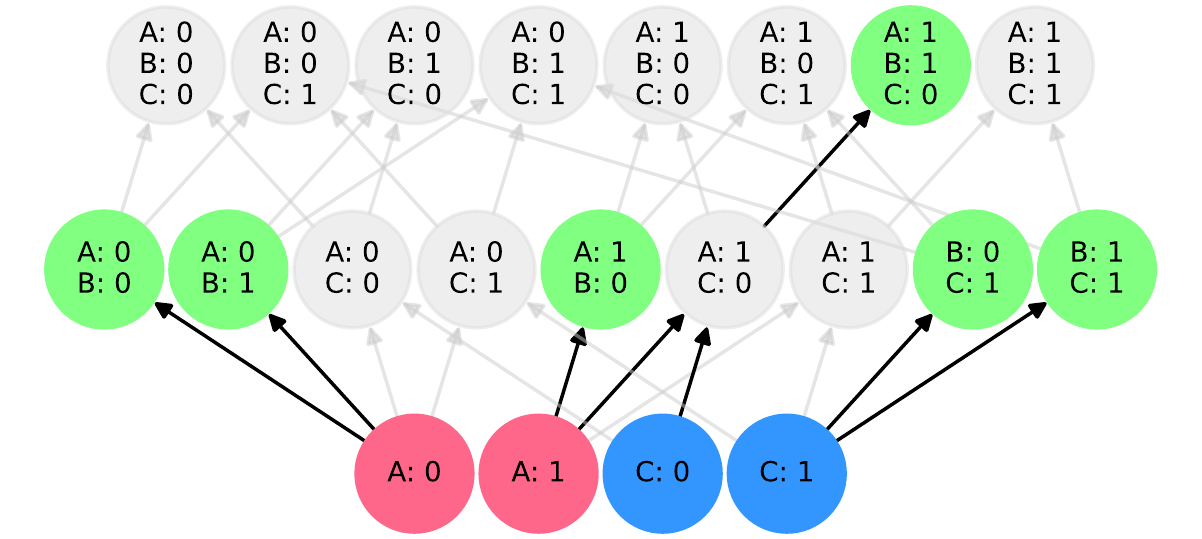}
    \\
    $\Theta_{23}$
    &
    $\Ext{\Theta_{23}}$
    \end{tabular}
\end{center}

\noindent The standard causaltope for Space 23 has dimension 30.
Below is a plot of the homogeneous linear system of causality and quasi-normalisation equations for the standard causaltope, put in reduced row echelon form:

\begin{center}
    \includegraphics[width=11cm]{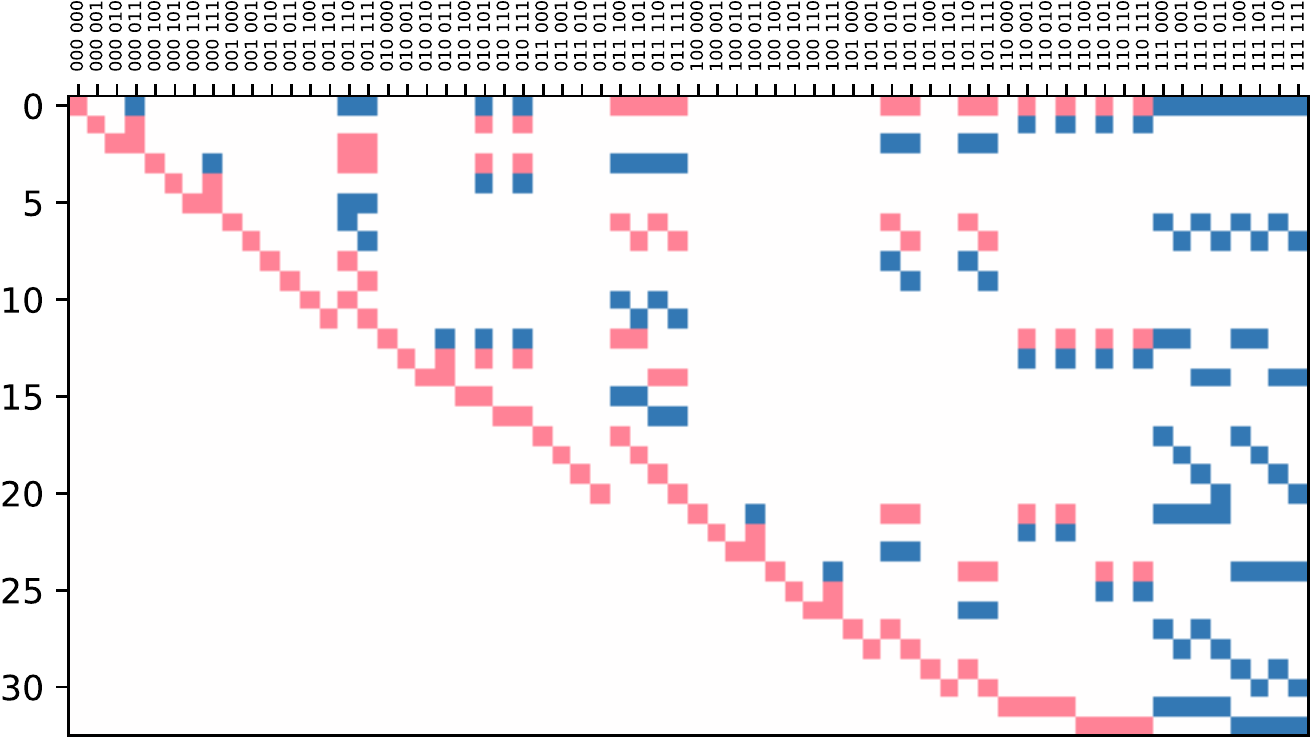}
\end{center}

\noindent Rows correspond to the 33 independent linear equations.
Columns in the plot correspond to entries of empirical models, indexed as $i_A i_B i_C$ $o_A o_B o_C$.
Coefficients in the equations are color-coded as white=0, red=+1 and blue=-1.

Space 23 has closest refinements in equivalence classes 10, 13 and 15; 
it is the join of its (closest) refinements.
It has closest coarsenings in equivalence classes 29, 34, 35, 36, 37 and 43; 
it is the meet of its (closest) coarsenings.
It has 128 causal functions, all of which are causal for at least one of its refinements.
It is not a tight space: for event \ev{B}, a causal function must yield identical output values on input histories \hist{A/0,B/1} and \hist{B/1,C/1}, and it must also yield identical output values on input histories \hist{A/0,B/0}, \hist{A/1,B/0} and \hist{B/0,C/1}.

The standard causaltope for Space 23 has 1 more dimension than that of its subspace in equivalence class 15.
The standard causaltope for Space 23 is the meet of the standard causaltopes for its closest coarsenings.
For completeness, below is a plot of the full homogeneous linear system of causality and quasi-normalisation equations for the standard causaltope:

\begin{center}
    \includegraphics[width=12cm]{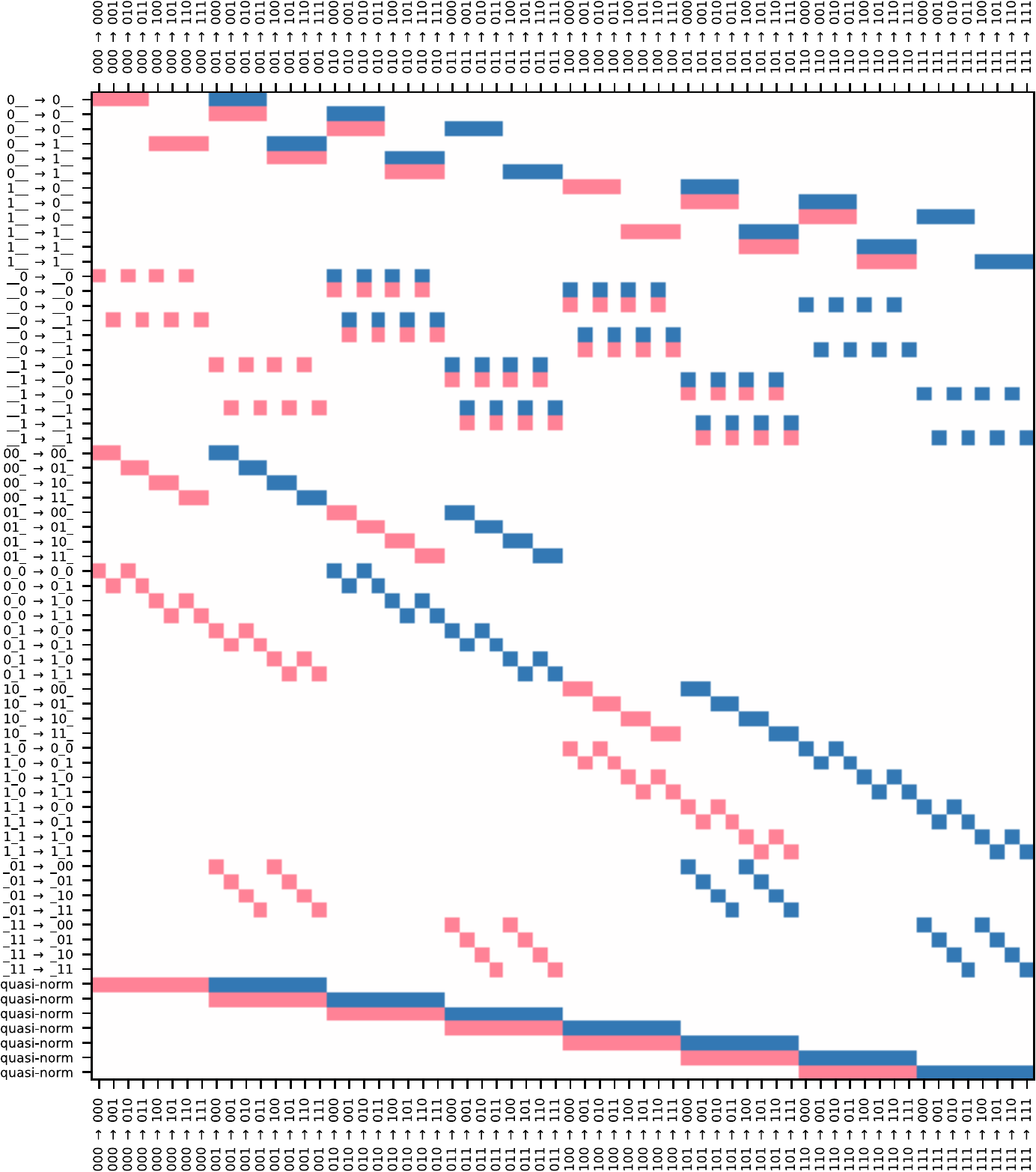}
\end{center}

\noindent Rows correspond to the 67 linear equations, of which 33 are independent.

\newpage
\subsection*{Space 24}

Space 24 is not induced by a causal order, but it is a refinement of the space in equivalence class 92 induced by the definite causal order $\total{\ev{A},\ev{B}}\vee\total{\ev{C},\ev{B}}$ (note that the space induced by the order is not the same as space 92).
Its equivalence class under event-input permutation symmetry contains 48 spaces.
Space 24 differs as follows from the space induced by causal order $\total{\ev{A},\ev{B}}\vee\total{\ev{C},\ev{B}}$:
\begin{itemize}
  \item The outputs at events \evset{\ev{A}, \ev{B}} are independent of the input at event \ev{C} when the inputs at events \evset{A, B} are given by \hist{A/0,B/0}, \hist{A/0,B/1} and \hist{A/1,B/0}.
  \item The outputs at events \evset{\ev{B}, \ev{C}} are independent of the input at event \ev{A} when the inputs at events \evset{B, C} are given by \hist{B/1,C/0} and \hist{B/0,C/1}.
\end{itemize}

\noindent Below are the histories and extended histories for space 24: 
\begin{center}
    \begin{tabular}{cc}
    \includegraphics[height=3.5cm]{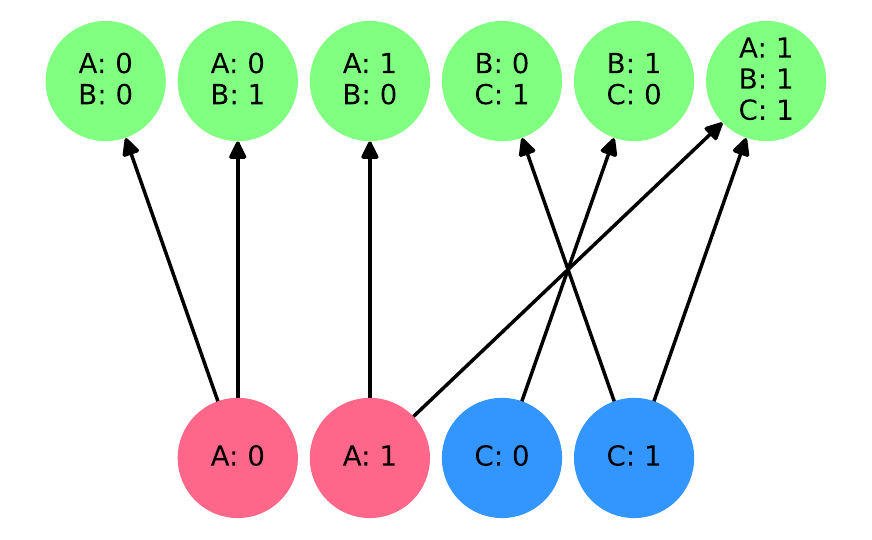}
    &
    \includegraphics[height=3.5cm]{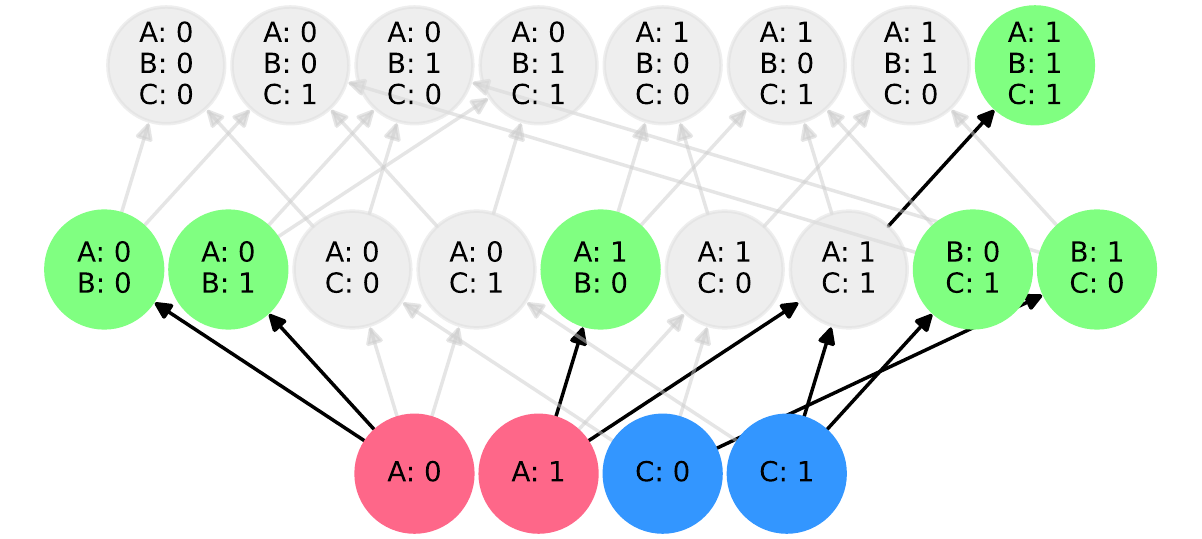}
    \\
    $\Theta_{24}$
    &
    $\Ext{\Theta_{24}}$
    \end{tabular}
\end{center}

\noindent The standard causaltope for Space 24 has dimension 30.
Below is a plot of the homogeneous linear system of causality and quasi-normalisation equations for the standard causaltope, put in reduced row echelon form:

\begin{center}
    \includegraphics[width=11cm]{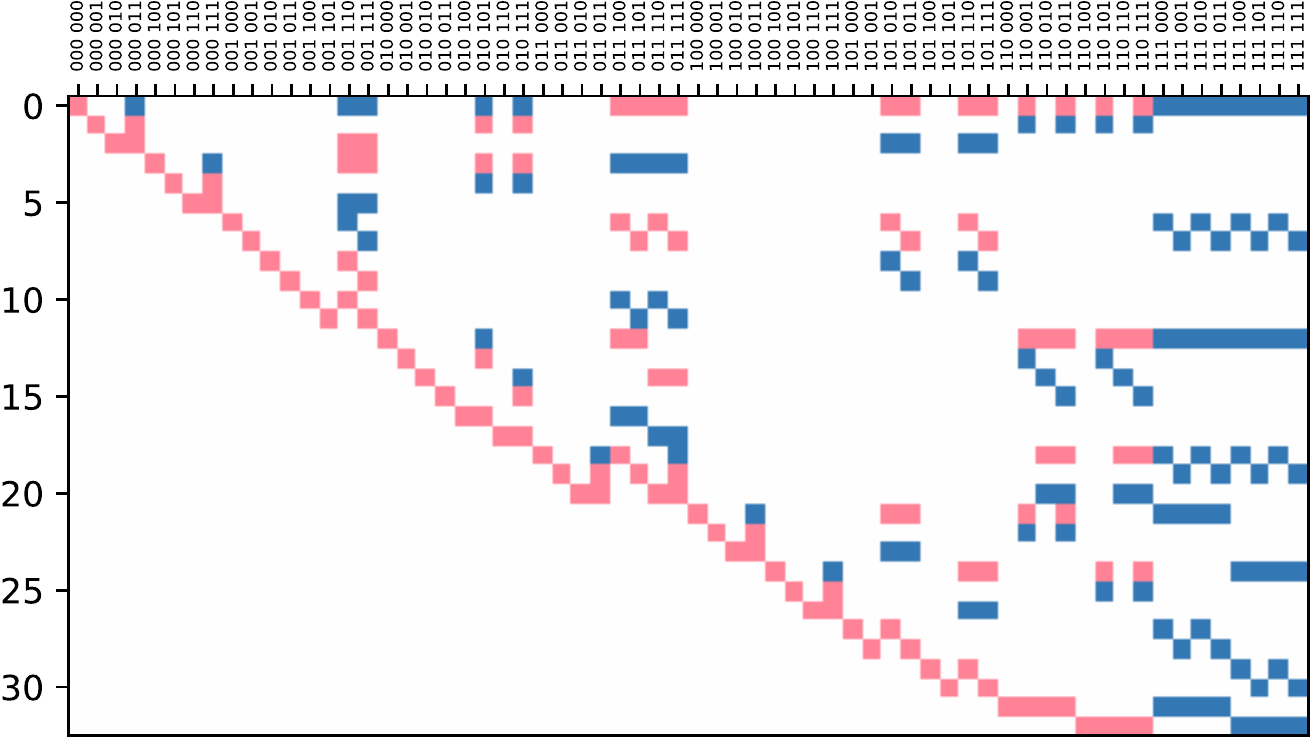}
\end{center}

\noindent Rows correspond to the 33 independent linear equations.
Columns in the plot correspond to entries of empirical models, indexed as $i_A i_B i_C$ $o_A o_B o_C$.
Coefficients in the equations are color-coded as white=0, red=+1 and blue=-1.

Space 24 has closest refinements in equivalence classes 9, 10 and 15; 
it is the join of its (closest) refinements.
It has closest coarsenings in equivalence classes 29, 34, 36, 39 and 40; 
it is the meet of its (closest) coarsenings.
It has 128 causal functions, all of which are causal for at least one of its refinements.
It is not a tight space: for event \ev{B}, a causal function must yield identical output values on input histories \hist{A/0,B/1} and \hist{B/1,C/0}, and it must also yield identical output values on input histories \hist{A/0,B/0}, \hist{A/1,B/0} and \hist{B/0,C/1}.

The standard causaltope for Space 24 has 1 more dimension than that of its subspace in equivalence class 15.
The standard causaltope for Space 24 is the meet of the standard causaltopes for its closest coarsenings.
For completeness, below is a plot of the full homogeneous linear system of causality and quasi-normalisation equations for the standard causaltope:

\begin{center}
    \includegraphics[width=12cm]{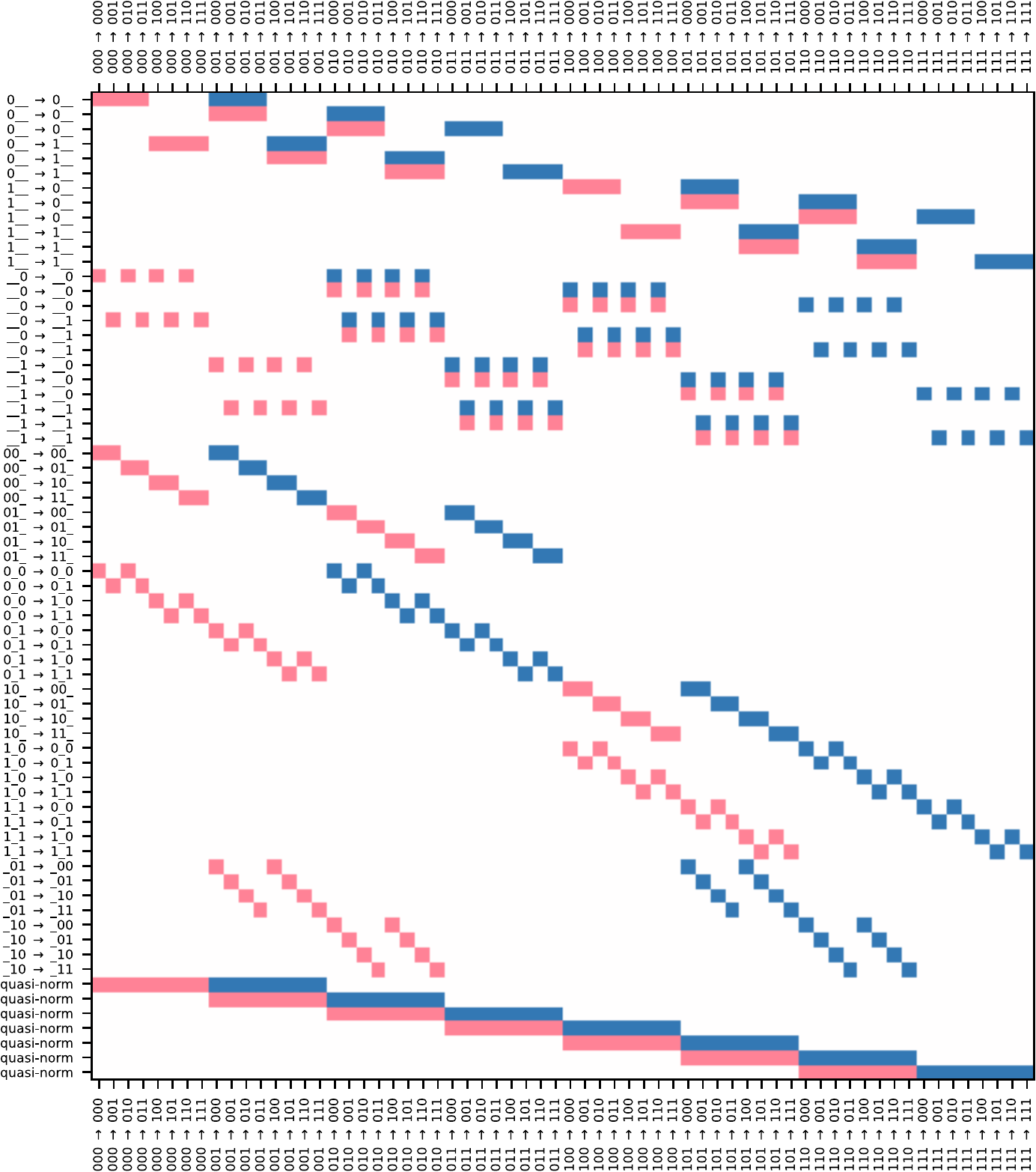}
\end{center}

\noindent Rows correspond to the 67 linear equations, of which 33 are independent.

\newpage
\subsection*{Space 25}

Space 25 is not induced by a causal order, but it is a refinement of the space 77 induced by the definite causal order $\total{\ev{A},\ev{B}}\vee\total{\ev{A},\ev{C}}$.
Its equivalence class under event-input permutation symmetry contains 12 spaces.
Space 25 differs as follows from the space induced by causal order $\total{\ev{A},\ev{B}}\vee\total{\ev{A},\ev{C}}$:
\begin{itemize}
  \item The outputs at events \evset{\ev{B}, \ev{C}} are independent of the input at event \ev{A} when the inputs at events \evset{B, C} are given by \hist{B/1,C/0}.
  \item The output at event \ev{C} is independent of the input at event \ev{A} when the input at event C is given by \hist{C/0}.
  \item The output at event \ev{B} is independent of the input at event \ev{A} when the input at event B is given by \hist{B/1}.
\end{itemize}

\noindent Below are the histories and extended histories for space 25: 
\begin{center}
    \begin{tabular}{cc}
    \includegraphics[height=3.5cm]{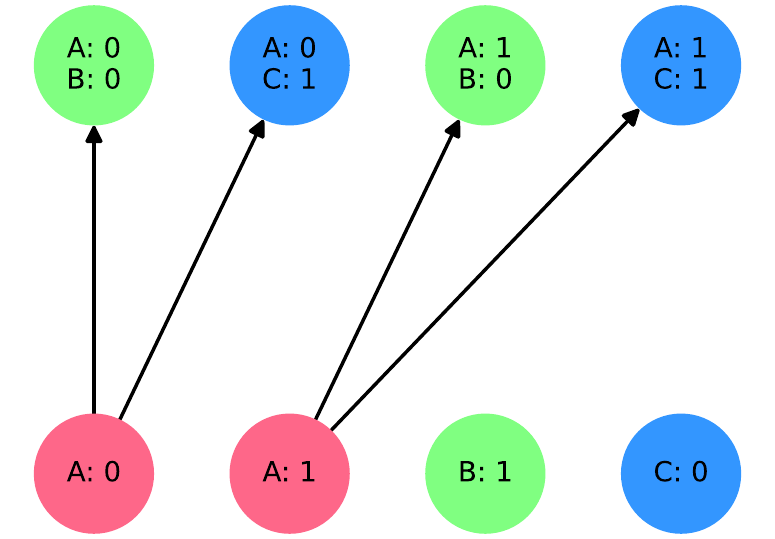}
    &
    \includegraphics[height=3.5cm]{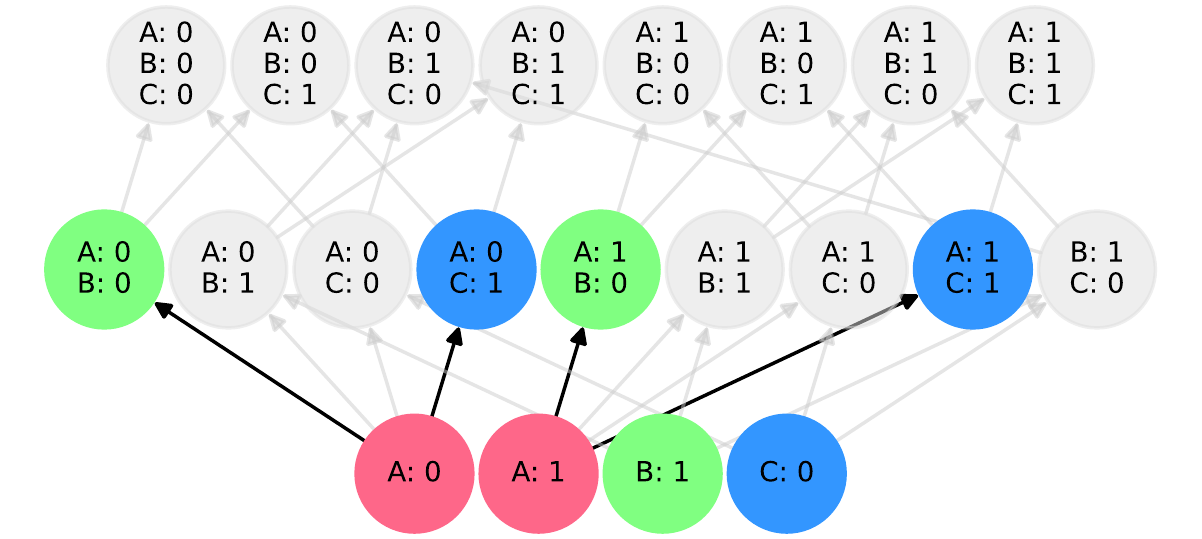}
    \\
    $\Theta_{25}$
    &
    $\Ext{\Theta_{25}}$
    \end{tabular}
\end{center}

\noindent The standard causaltope for Space 25 has dimension 31.
Below is a plot of the homogeneous linear system of causality and quasi-normalisation equations for the standard causaltope, put in reduced row echelon form:

\begin{center}
    \includegraphics[width=11cm]{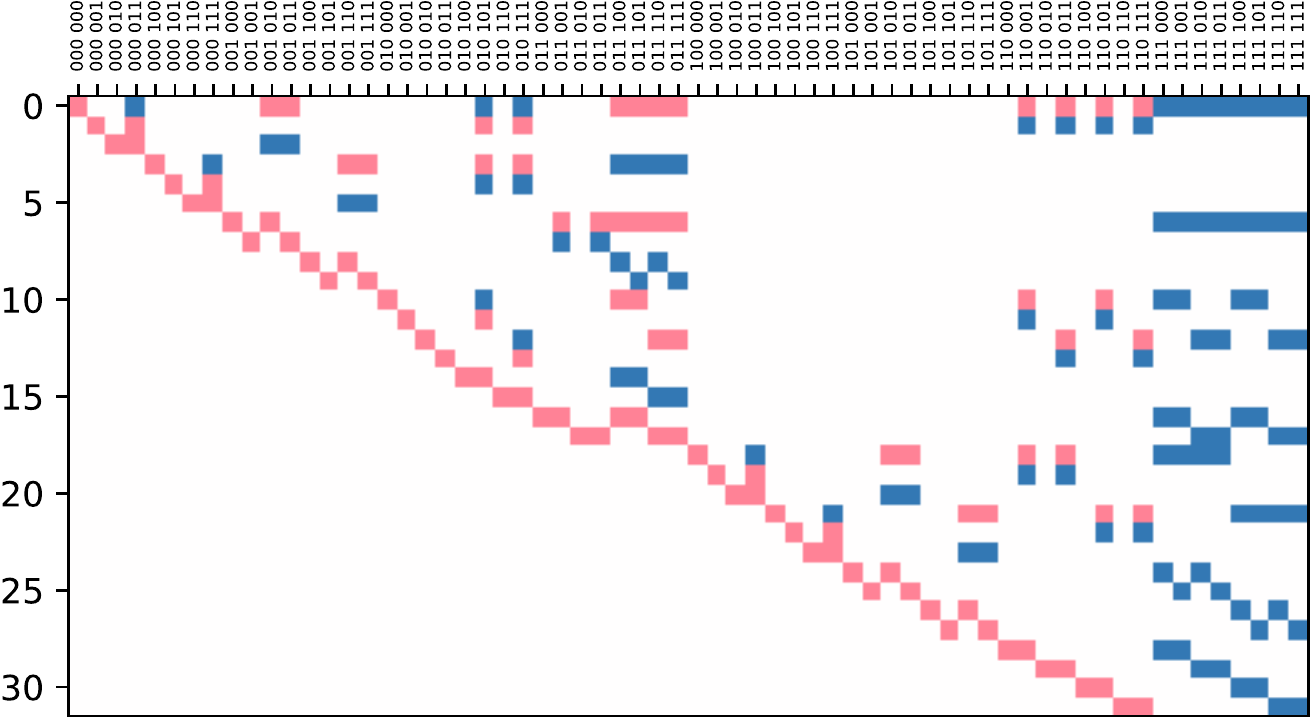}
\end{center}

\noindent Rows correspond to the 32 independent linear equations.
Columns in the plot correspond to entries of empirical models, indexed as $i_A i_B i_C$ $o_A o_B o_C$.
Coefficients in the equations are color-coded as white=0, red=+1 and blue=-1.

Space 25 has closest refinements in equivalence class 12; 
it is the join of its (closest) refinements.
It has closest coarsenings in equivalence classes 30 and 38; 
it is the meet of its (closest) coarsenings.
It has 256 causal functions, 128 of which are not causal for any of its refinements.
It is a tight space.

The standard causaltope for Space 25 has 2 more dimensions than those of its 2 subspaces in equivalence class 12.
The standard causaltope for Space 25 is the meet of the standard causaltopes for its closest coarsenings.
For completeness, below is a plot of the full homogeneous linear system of causality and quasi-normalisation equations for the standard causaltope:

\begin{center}
    \includegraphics[width=12cm]{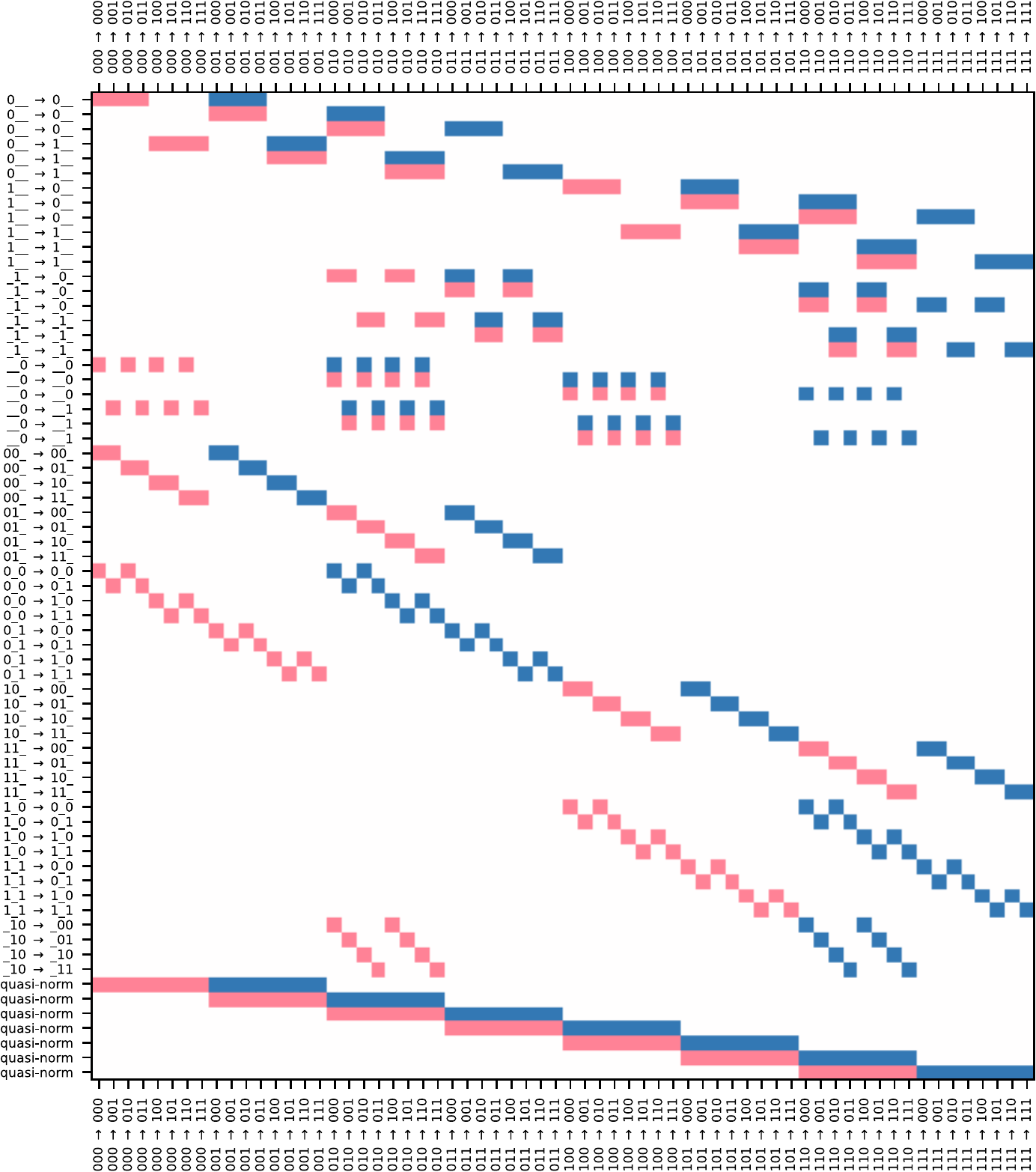}
\end{center}

\noindent Rows correspond to the 67 linear equations, of which 32 are independent.

\newpage
\subsection*{Space 26}

Space 26 is not induced by a causal order, but it is a refinement of the space in equivalence class 92 induced by the definite causal order $\total{\ev{A},\ev{B}}\vee\total{\ev{C},\ev{B}}$ (note that the space induced by the order is not the same as space 92).
Its equivalence class under event-input permutation symmetry contains 24 spaces.
Space 26 differs as follows from the space induced by causal order $\total{\ev{A},\ev{B}}\vee\total{\ev{C},\ev{B}}$:
\begin{itemize}
  \item The outputs at events \evset{\ev{A}, \ev{B}} are independent of the input at event \ev{C} when the inputs at events \evset{A, B} are given by \hist{A/0,B/0}, \hist{A/0,B/1} and \hist{A/1,B/1}.
  \item The outputs at events \evset{\ev{B}, \ev{C}} are independent of the input at event \ev{A} when the inputs at events \evset{B, C} are given by \hist{B/1,C/0} and \hist{B/1,C/1}.
\end{itemize}

\noindent Below are the histories and extended histories for space 26: 
\begin{center}
    \begin{tabular}{cc}
    \includegraphics[height=3.5cm]{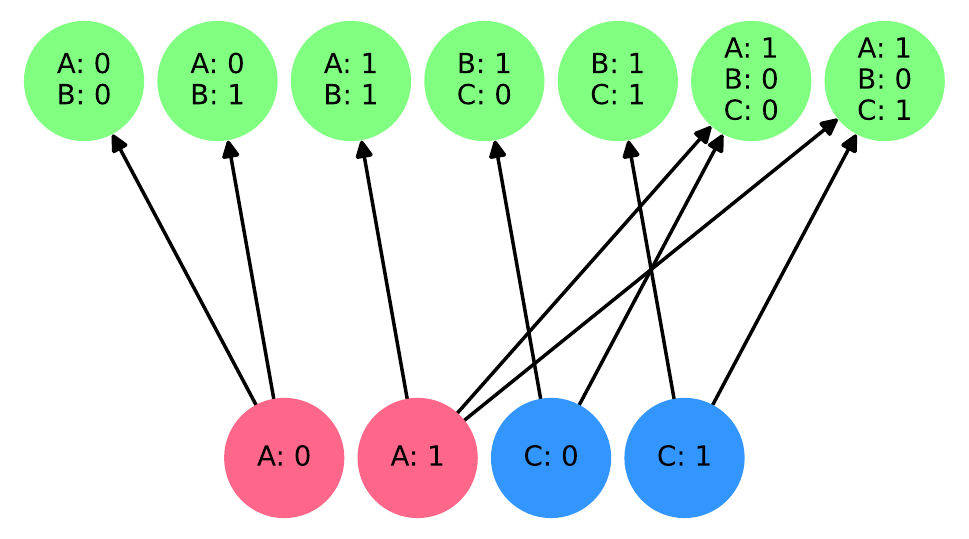}
    &
    \includegraphics[height=3.5cm]{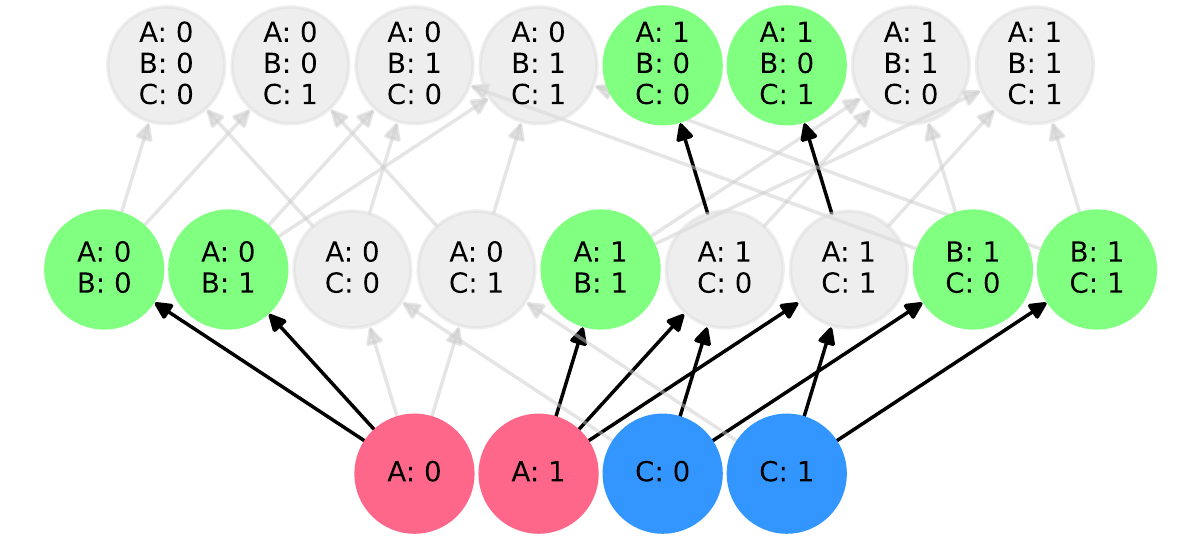}
    \\
    $\Theta_{26}$
    &
    $\Ext{\Theta_{26}}$
    \end{tabular}
\end{center}

\noindent The standard causaltope for Space 26 has dimension 31.
Below is a plot of the homogeneous linear system of causality and quasi-normalisation equations for the standard causaltope, put in reduced row echelon form:

\begin{center}
    \includegraphics[width=11cm]{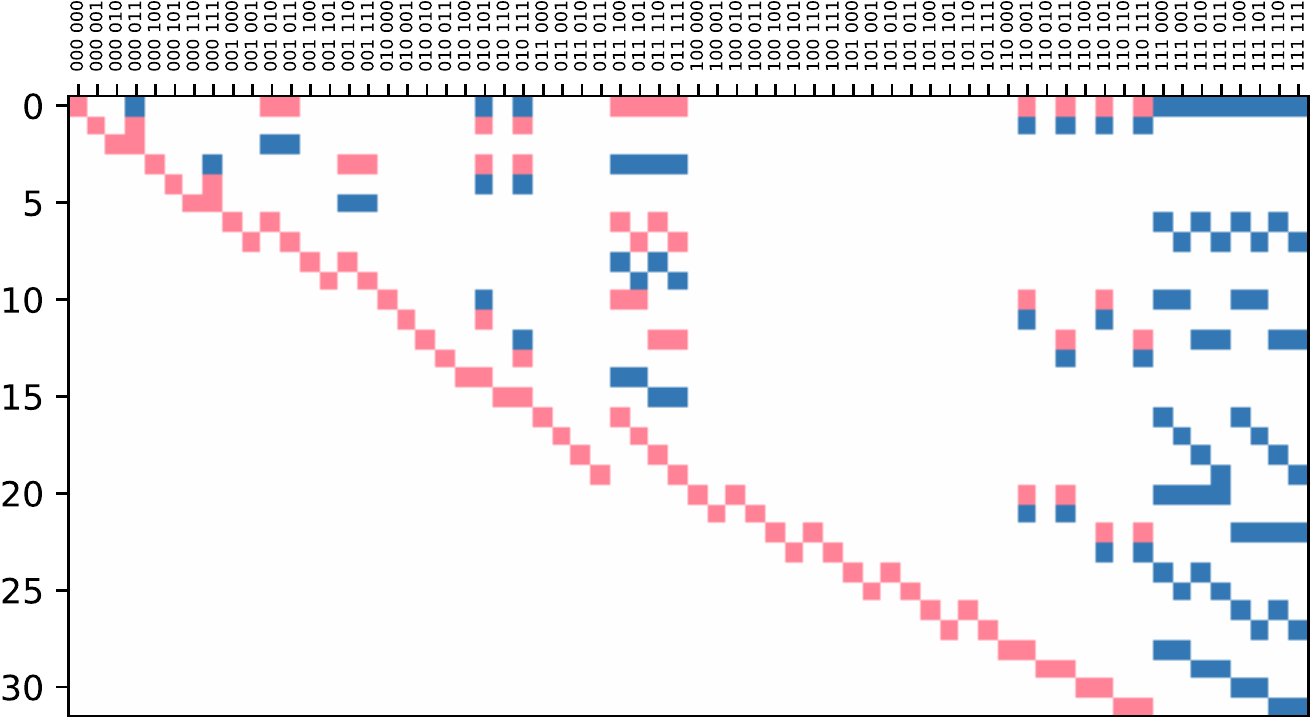}
\end{center}

\noindent Rows correspond to the 32 independent linear equations.
Columns in the plot correspond to entries of empirical models, indexed as $i_A i_B i_C$ $o_A o_B o_C$.
Coefficients in the equations are color-coded as white=0, red=+1 and blue=-1.

Space 26 has closest refinements in equivalence classes 8, 14 and 15; 
it is the join of its (closest) refinements.
It has closest coarsenings in equivalence classes 34, 35, 39 and 44; 
it is the meet of its (closest) coarsenings.
It has 256 causal functions, 64 of which are not causal for any of its refinements.
It is not a tight space: for event \ev{B}, a causal function must yield identical output values on input histories \hist{A/0,B/1}, \hist{A/1,B/1}, \hist{B/1,C/0} and \hist{B/1,C/1}.

The standard causaltope for Space 26 coincides with that of its subspace in equivalence class 8.
The standard causaltope for Space 26 is the meet of the standard causaltopes for its closest coarsenings.
For completeness, below is a plot of the full homogeneous linear system of causality and quasi-normalisation equations for the standard causaltope:

\begin{center}
    \includegraphics[width=12cm]{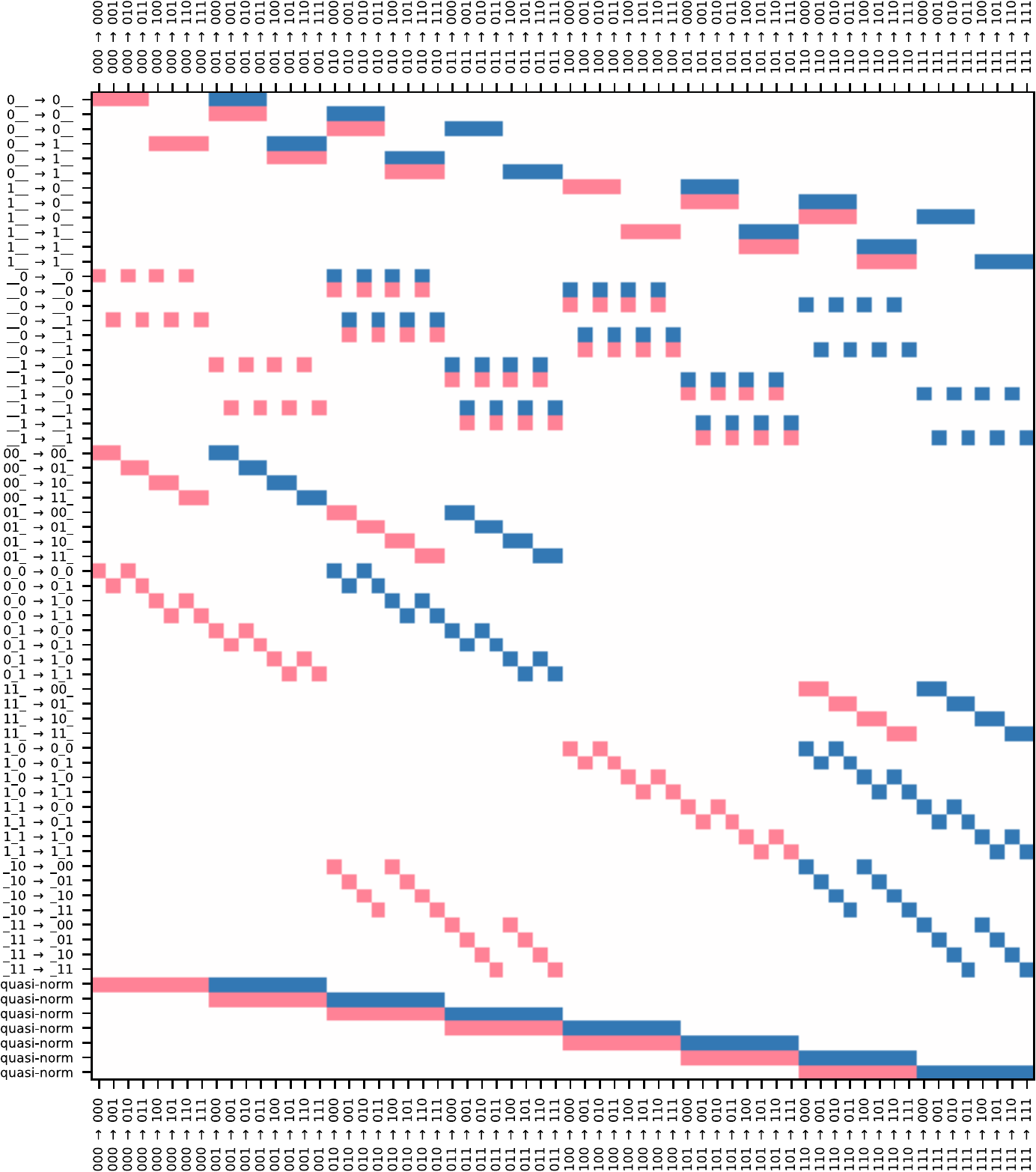}
\end{center}

\noindent Rows correspond to the 67 linear equations, of which 32 are independent.

\newpage
\subsection*{Space 27}

Space 27 is not induced by a causal order, but it is a refinement of the space 77 induced by the definite causal order $\total{\ev{A},\ev{B}}\vee\total{\ev{A},\ev{C}}$.
Its equivalence class under event-input permutation symmetry contains 12 spaces.
Space 27 differs as follows from the space induced by causal order $\total{\ev{A},\ev{B}}\vee\total{\ev{A},\ev{C}}$:
\begin{itemize}
  \item The outputs at events \evset{\ev{B}, \ev{C}} are independent of the input at event \ev{A} when the inputs at events \evset{B, C} are given by \hist{B/1,C/0} and \hist{B/1,C/1}.
  \item The output at event \ev{B} is independent of the input at event \ev{A} when the input at event B is given by \hist{B/1}.
\end{itemize}

\noindent Below are the histories and extended histories for space 27: 
\begin{center}
    \begin{tabular}{cc}
    \includegraphics[height=3.5cm]{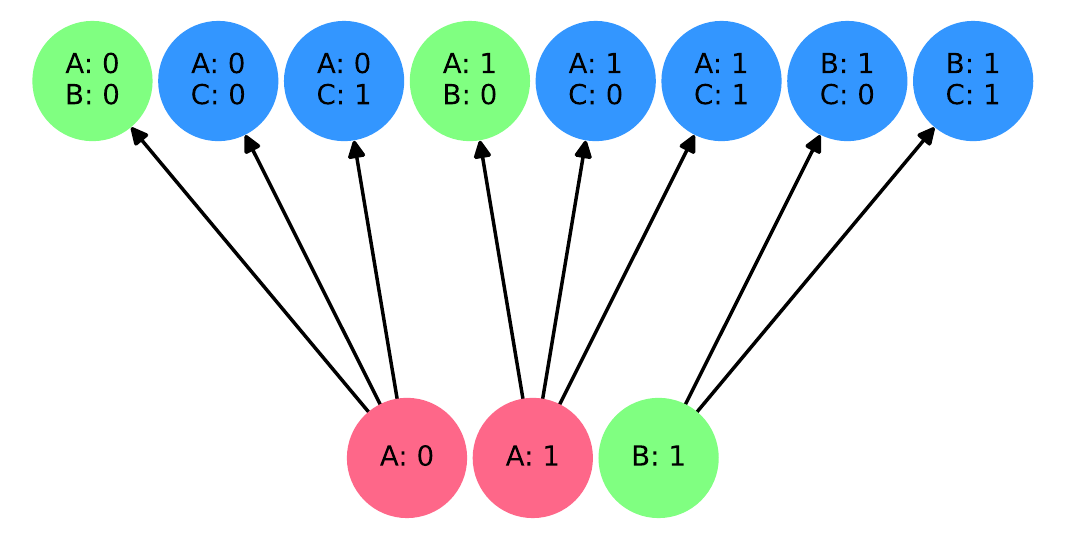}
    &
    \includegraphics[height=3.5cm]{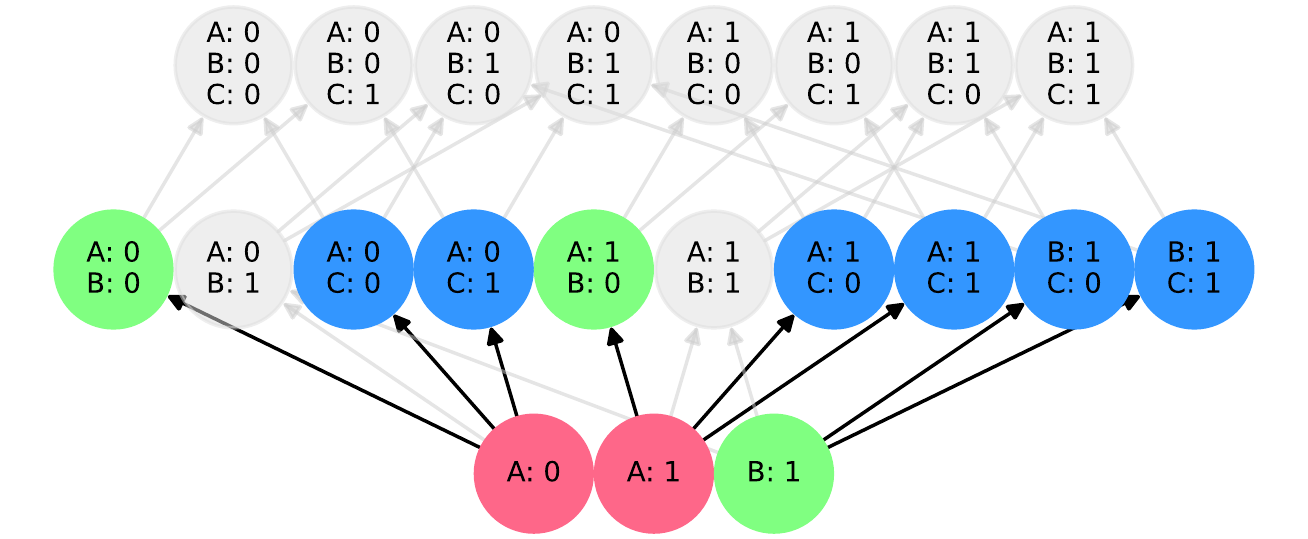}
    \\
    $\Theta_{27}$
    &
    $\Ext{\Theta_{27}}$
    \end{tabular}
\end{center}

\noindent The standard causaltope for Space 27 has dimension 29.
Below is a plot of the homogeneous linear system of causality and quasi-normalisation equations for the standard causaltope, put in reduced row echelon form:

\begin{center}
    \includegraphics[width=11cm]{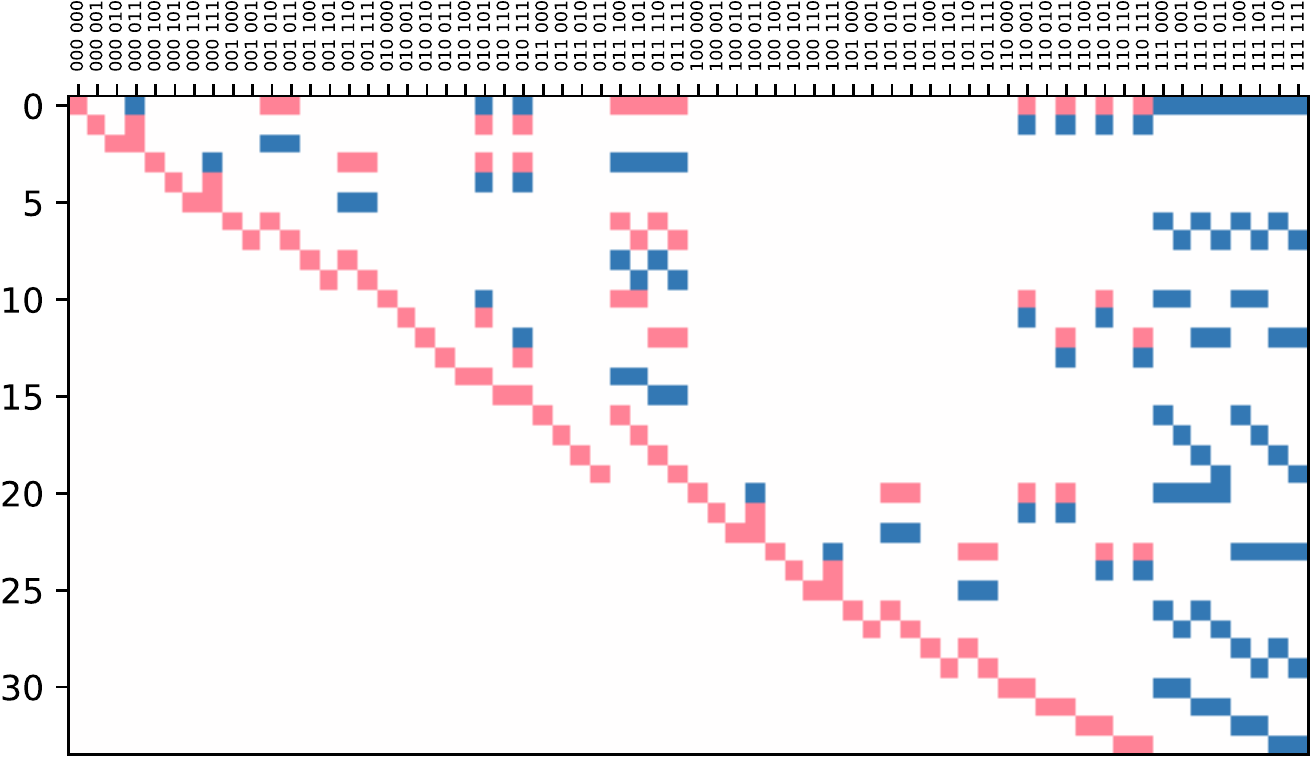}
\end{center}

\noindent Rows correspond to the 34 independent linear equations.
Columns in the plot correspond to entries of empirical models, indexed as $i_A i_B i_C$ $o_A o_B o_C$.
Coefficients in the equations are color-coded as white=0, red=+1 and blue=-1.

Space 27 has closest refinements in equivalence classes 12 and 13; 
it is the join of its (closest) refinements.
It has closest coarsenings in equivalence classes 38, 42 and 43; 
it is the meet of its (closest) coarsenings.
It has 128 causal functions, all of which are causal for at least one of its refinements.
It is not a tight space: for event \ev{C}, a causal function must yield identical output values on input histories \hist{A/0,C/0}, \hist{A/1,C/0} and \hist{B/1,C/0}, and it must also yield identical output values on input histories \hist{A/0,C/1}, \hist{A/1,C/1} and \hist{B/1,C/1}.

The standard causaltope for Space 27 coincides with that of its 2 subspaces in equivalence class 12.
The standard causaltope for Space 27 is the meet of the standard causaltopes for its closest coarsenings.
For completeness, below is a plot of the full homogeneous linear system of causality and quasi-normalisation equations for the standard causaltope:

\begin{center}
    \includegraphics[width=12cm]{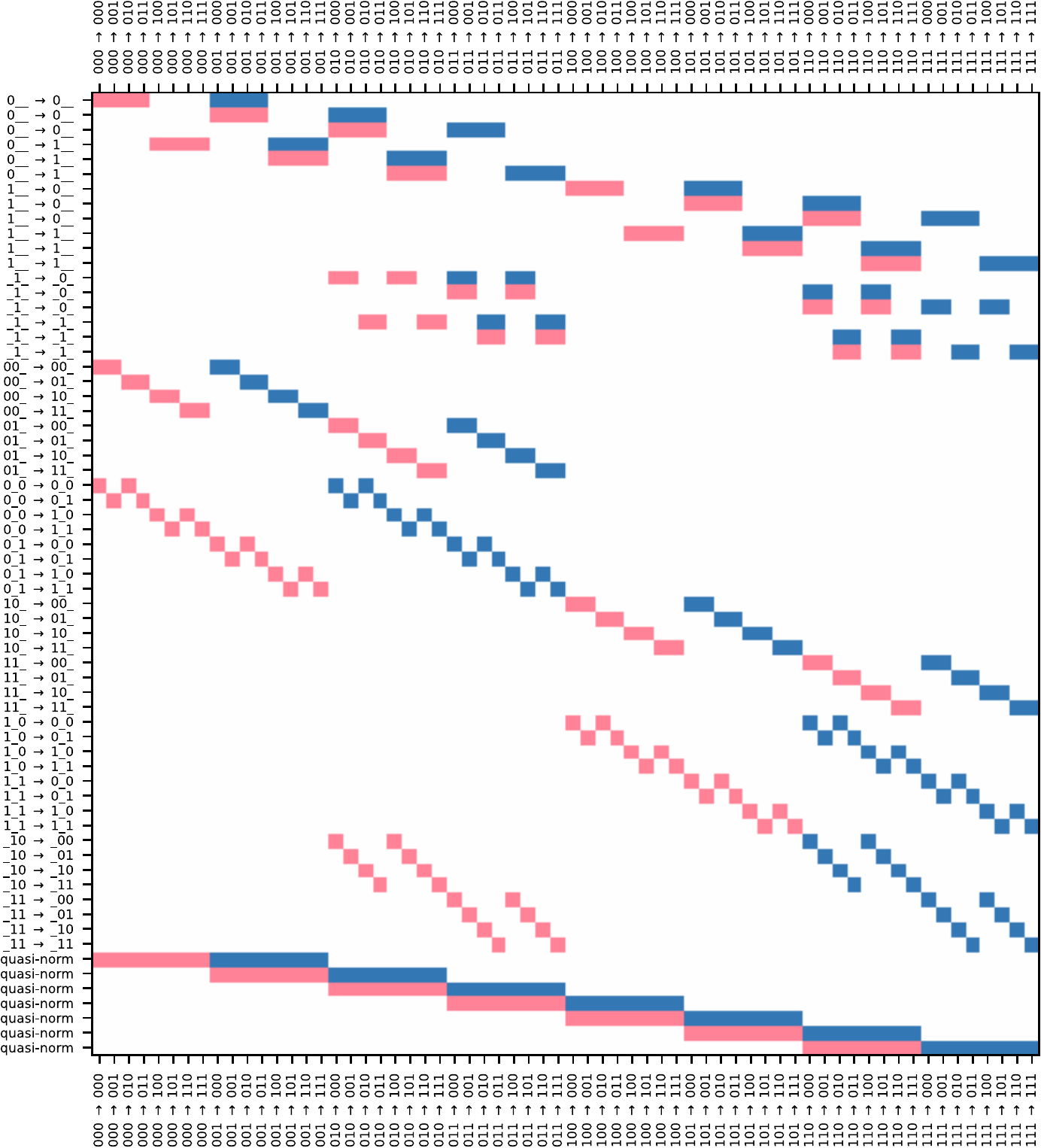}
\end{center}

\noindent Rows correspond to the 65 linear equations, of which 34 are independent.

\newpage
\subsection*{Space 28}

Space 28 is not induced by a causal order, but it is a refinement of the space 100 induced by the definite causal order $\total{\ev{A},\ev{B},\ev{C}}$.
Its equivalence class under event-input permutation symmetry contains 24 spaces.
Space 28 differs as follows from the space induced by causal order $\total{\ev{A},\ev{B},\ev{C}}$:
\begin{itemize}
  \item The outputs at events \evset{\ev{B}, \ev{C}} are independent of the input at event \ev{A} when the inputs at events \evset{B, C} are given by \hist{B/1,C/0} and \hist{B/1,C/1}.
  \item The outputs at events \evset{\ev{A}, \ev{C}} are independent of the input at event \ev{B} when the inputs at events \evset{A, C} are given by \hist{A/0,C/0} and \hist{A/1,C/0}.
  \item The output at event \ev{C} is independent of the inputs at events \evset{\ev{A}, \ev{B}} when the input at event C is given by \hist{C/0}.
  \item The output at event \ev{B} is independent of the input at event \ev{A} when the input at event B is given by \hist{B/1}.
\end{itemize}

\noindent Below are the histories and extended histories for space 28: 
\begin{center}
    \begin{tabular}{cc}
    \includegraphics[height=3.5cm]{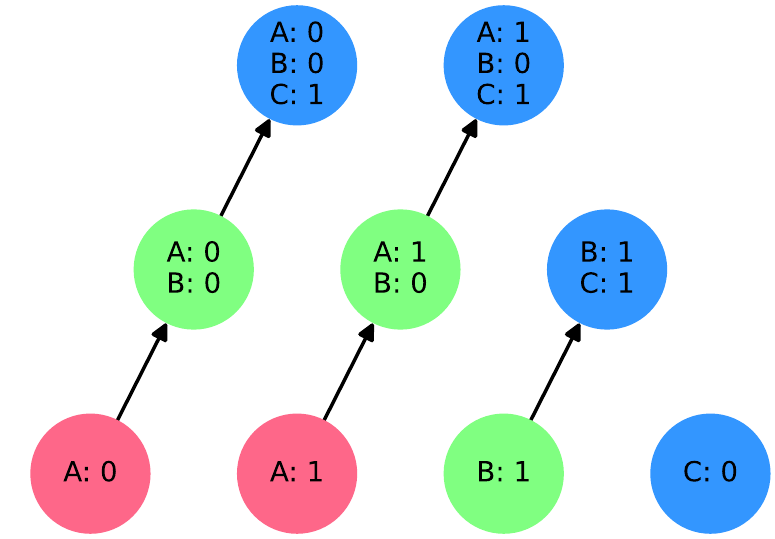}
    &
    \includegraphics[height=3.5cm]{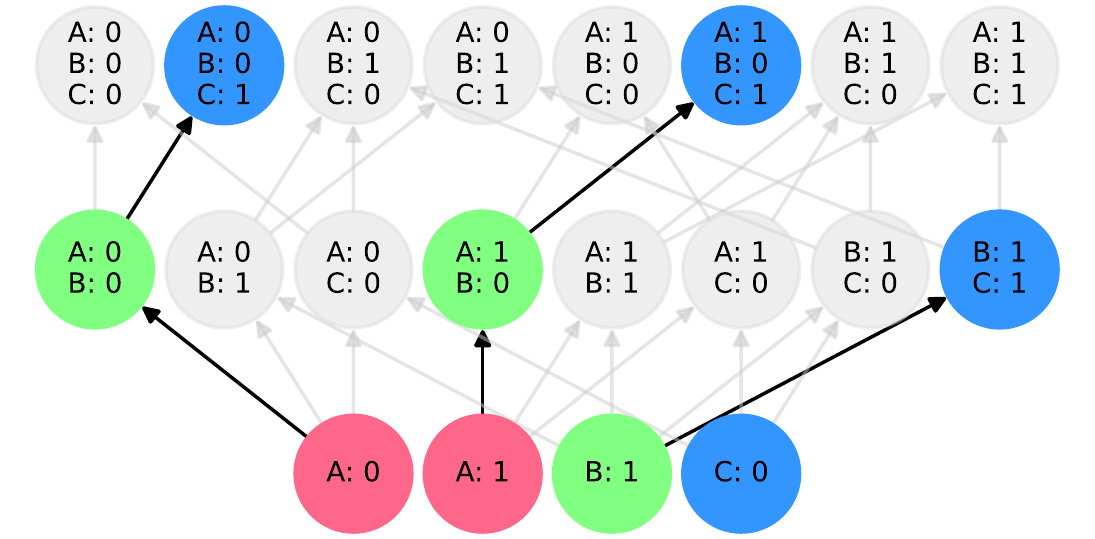}
    \\
    $\Theta_{28}$
    &
    $\Ext{\Theta_{28}}$
    \end{tabular}
\end{center}

\noindent The standard causaltope for Space 28 has dimension 33.
Below is a plot of the homogeneous linear system of causality and quasi-normalisation equations for the standard causaltope, put in reduced row echelon form:

\begin{center}
    \includegraphics[width=11cm]{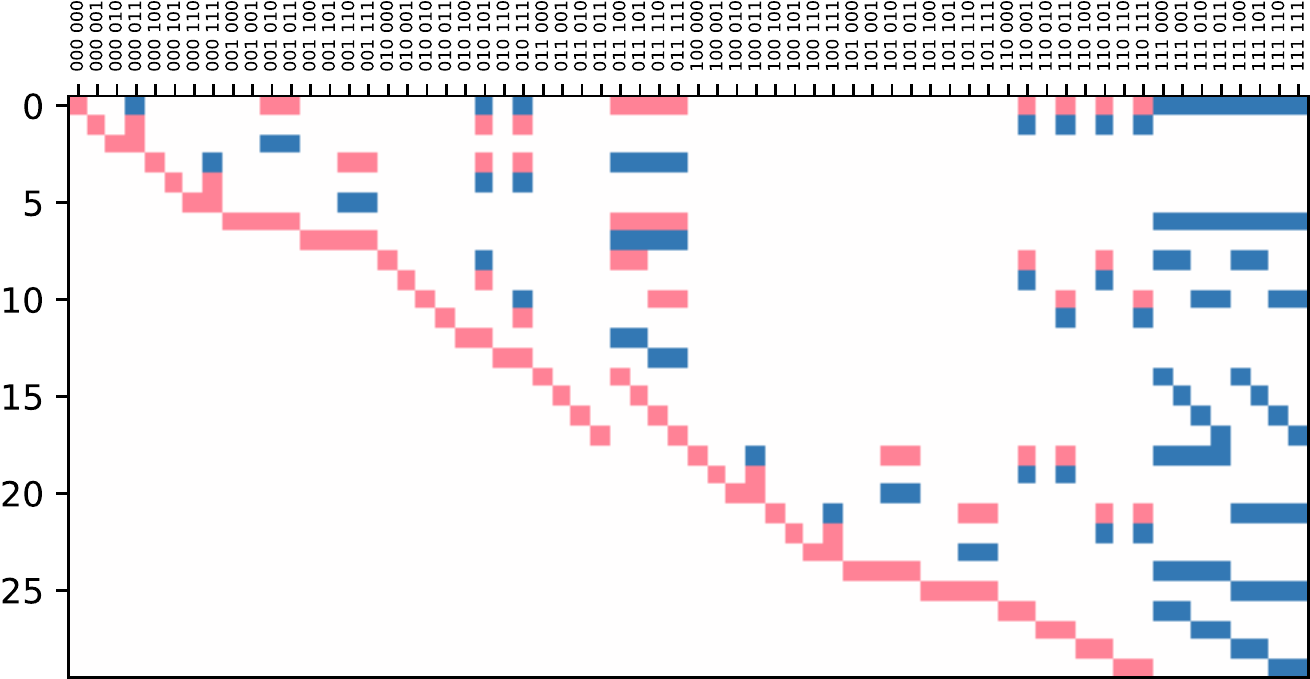}
\end{center}

\noindent Rows correspond to the 30 independent linear equations.
Columns in the plot correspond to entries of empirical models, indexed as $i_A i_B i_C$ $o_A o_B o_C$.
Coefficients in the equations are color-coded as white=0, red=+1 and blue=-1.

Space 28 has closest refinements in equivalence classes 17 and 20; 
it is the join of its (closest) refinements.
It has closest coarsenings in equivalence classes 45 and 54; 
it is the meet of its (closest) coarsenings.
It has 512 causal functions, 192 of which are not causal for any of its refinements.
It is a tight space.

The standard causaltope for Space 28 has 2 more dimensions than those of its 3 subspaces in equivalence classes 17 and 20.
The standard causaltope for Space 28 is the meet of the standard causaltopes for its closest coarsenings.
For completeness, below is a plot of the full homogeneous linear system of causality and quasi-normalisation equations for the standard causaltope:

\begin{center}
    \includegraphics[width=12cm]{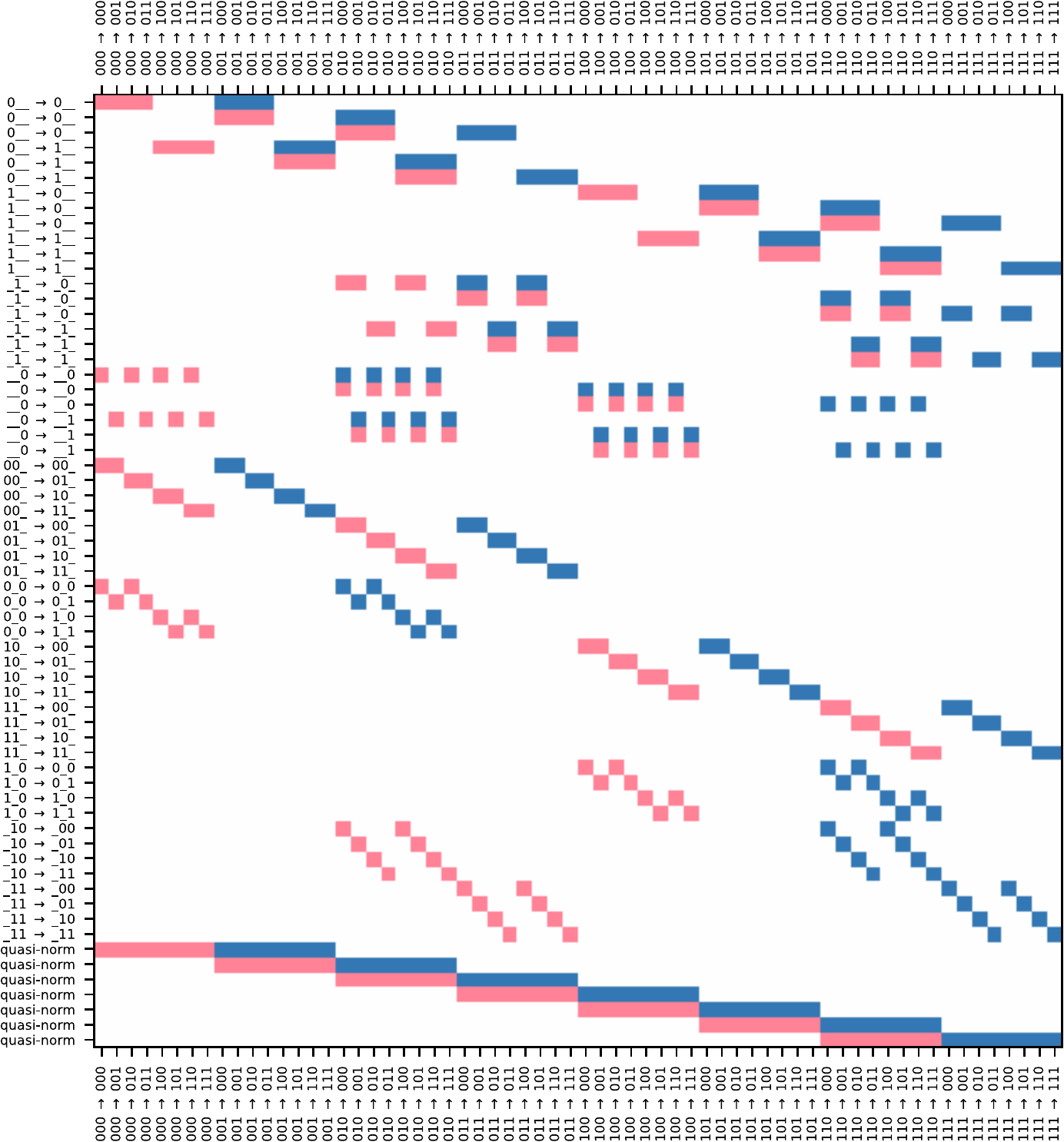}
\end{center}

\noindent Rows correspond to the 63 linear equations, of which 30 are independent.

\newpage
\subsection*{Space 29}

Space 29 is not induced by a causal order, but it is a refinement of the space in equivalence class 92 induced by the definite causal order $\total{\ev{A},\ev{B}}\vee\total{\ev{C},\ev{B}}$ (note that the space induced by the order is not the same as space 92).
Its equivalence class under event-input permutation symmetry contains 48 spaces.
Space 29 differs as follows from the space induced by causal order $\total{\ev{A},\ev{B}}\vee\total{\ev{C},\ev{B}}$:
\begin{itemize}
  \item The outputs at events \evset{\ev{A}, \ev{B}} are independent of the input at event \ev{C} when the inputs at events \evset{A, B} are given by \hist{A/0,B/0}, \hist{A/0,B/1} and \hist{A/1,B/0}.
  \item The outputs at events \evset{\ev{B}, \ev{C}} are independent of the input at event \ev{A} when the inputs at events \evset{B, C} are given by \hist{B/1,C/1}.
\end{itemize}

\noindent Below are the histories and extended histories for space 29: 
\begin{center}
    \begin{tabular}{cc}
    \includegraphics[height=3.5cm]{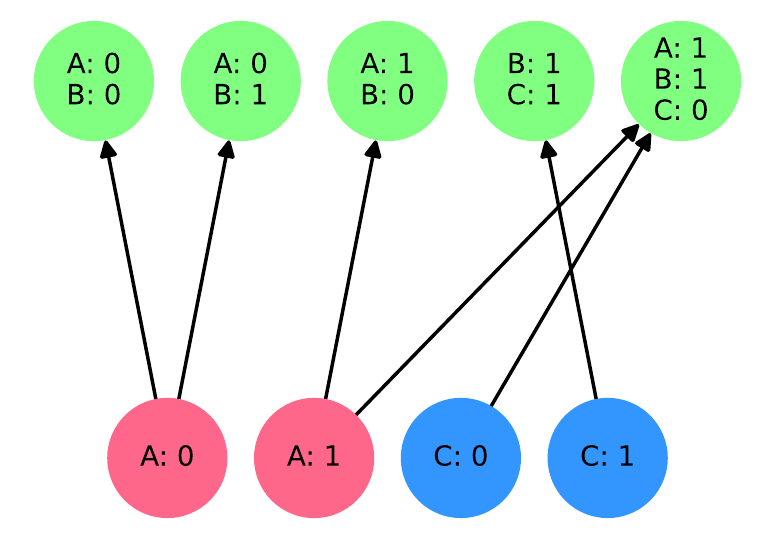}
    &
    \includegraphics[height=3.5cm]{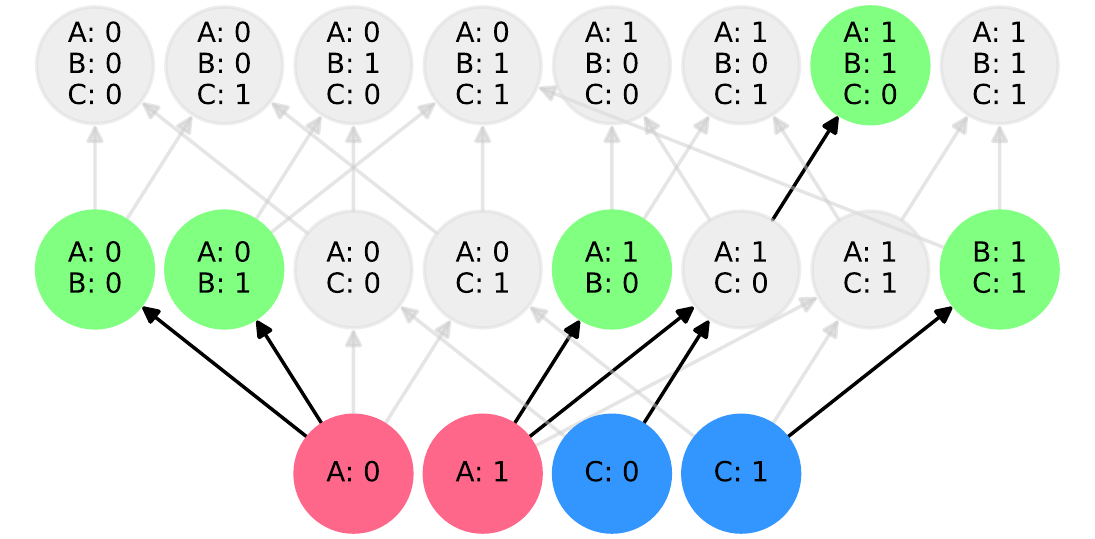}
    \\
    $\Theta_{29}$
    &
    $\Ext{\Theta_{29}}$
    \end{tabular}
\end{center}

\noindent The standard causaltope for Space 29 has dimension 32.
Below is a plot of the homogeneous linear system of causality and quasi-normalisation equations for the standard causaltope, put in reduced row echelon form:

\begin{center}
    \includegraphics[width=11cm]{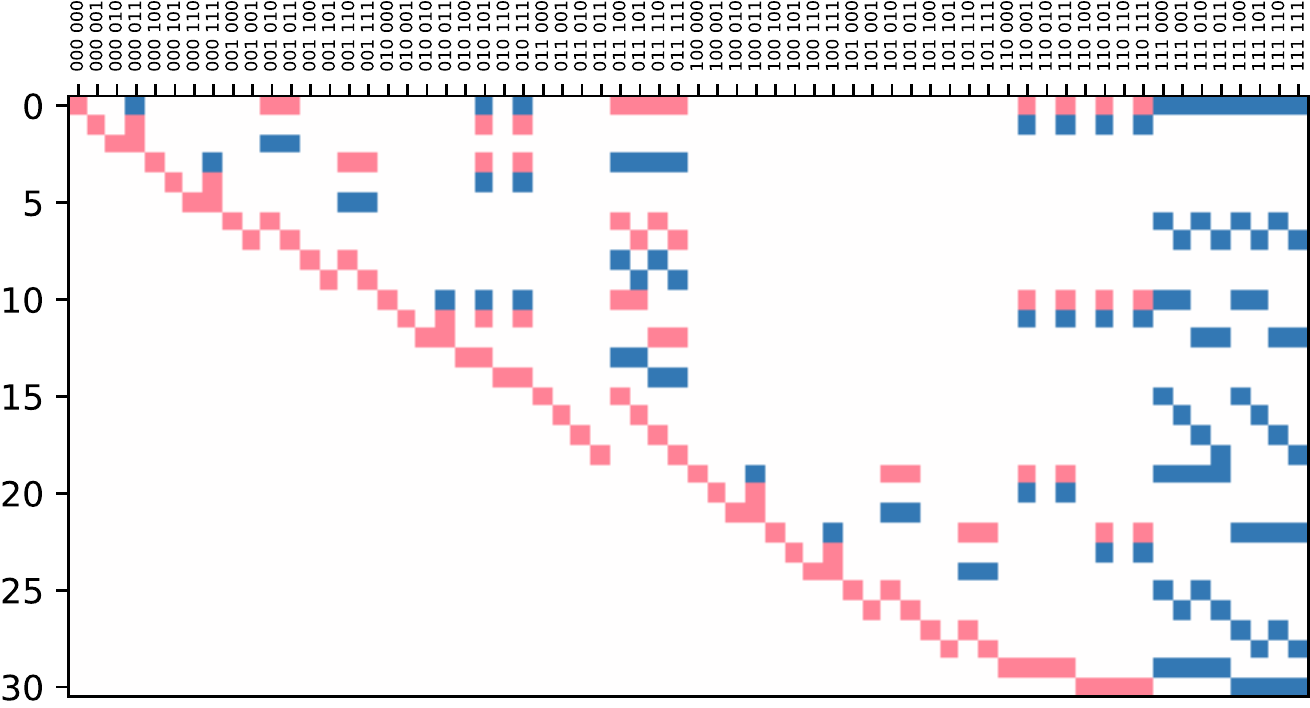}
\end{center}

\noindent Rows correspond to the 31 independent linear equations.
Columns in the plot correspond to entries of empirical models, indexed as $i_A i_B i_C$ $o_A o_B o_C$.
Coefficients in the equations are color-coded as white=0, red=+1 and blue=-1.

Space 29 has closest refinements in equivalence classes 16, 19, 23 and 24; 
it is the join of its (closest) refinements.
It has closest coarsenings in equivalence classes 46, 49, 50, 53 and 56; 
it is the meet of its (closest) coarsenings.
It has 256 causal functions, 64 of which are not causal for any of its refinements.
It is not a tight space: for event \ev{B}, a causal function must yield identical output values on input histories \hist{A/0,B/1} and \hist{B/1,C/1}.

The standard causaltope for Space 29 has 2 more dimensions than those of its 4 subspaces in equivalence classes 16, 19, 23 and 24.
The standard causaltope for Space 29 is the meet of the standard causaltopes for its closest coarsenings.
For completeness, below is a plot of the full homogeneous linear system of causality and quasi-normalisation equations for the standard causaltope:

\begin{center}
    \includegraphics[width=12cm]{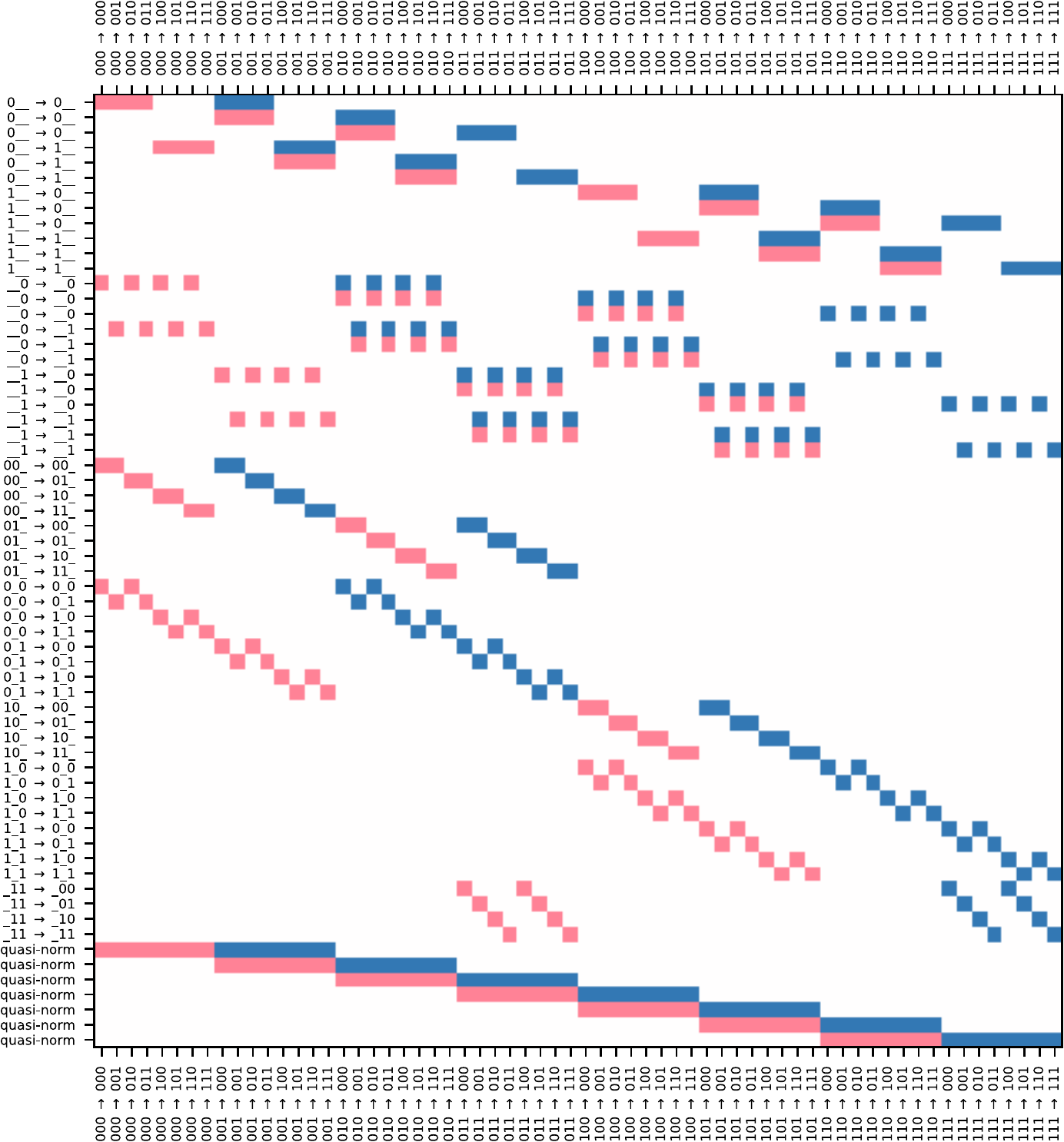}
\end{center}

\noindent Rows correspond to the 63 linear equations, of which 31 are independent.

\newpage
\subsection*{Space 30}

Space 30 is not induced by a causal order, but it is a refinement of the space 100 induced by the definite causal order $\total{\ev{A},\ev{B},\ev{C}}$.
Its equivalence class under event-input permutation symmetry contains 48 spaces.
Space 30 differs as follows from the space induced by causal order $\total{\ev{A},\ev{B},\ev{C}}$:
\begin{itemize}
  \item The outputs at events \evset{\ev{B}, \ev{C}} are independent of the input at event \ev{A} when the inputs at events \evset{B, C} are given by \hist{B/1,C/0}.
  \item The outputs at events \evset{\ev{A}, \ev{C}} are independent of the input at event \ev{B} when the inputs at events \evset{A, C} are given by \hist{A/0,C/0}, \hist{A/1,C/0} and \hist{A/1,C/1}.
  \item The output at event \ev{C} is independent of the inputs at events \evset{\ev{A}, \ev{B}} when the input at event C is given by \hist{C/0}.
  \item The output at event \ev{B} is independent of the input at event \ev{A} when the input at event B is given by \hist{B/1}.
\end{itemize}

\noindent Below are the histories and extended histories for space 30: 
\begin{center}
    \begin{tabular}{cc}
    \includegraphics[height=3.5cm]{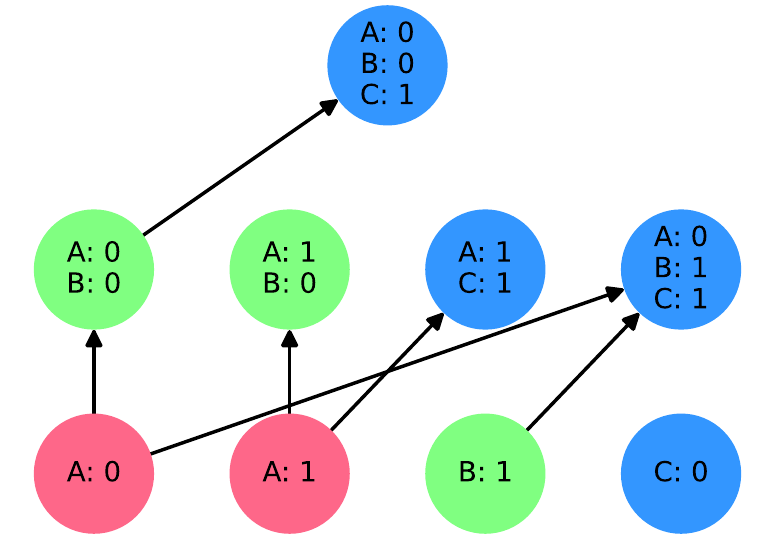}
    &
    \includegraphics[height=3.5cm]{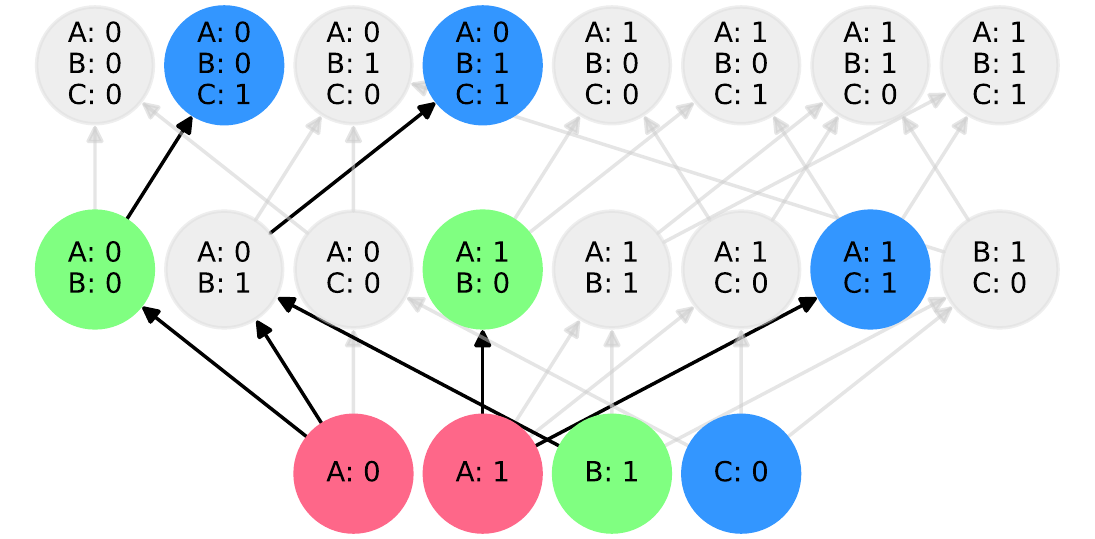}
    \\
    $\Theta_{30}$
    &
    $\Ext{\Theta_{30}}$
    \end{tabular}
\end{center}

\noindent The standard causaltope for Space 30 has dimension 33.
Below is a plot of the homogeneous linear system of causality and quasi-normalisation equations for the standard causaltope, put in reduced row echelon form:

\begin{center}
    \includegraphics[width=11cm]{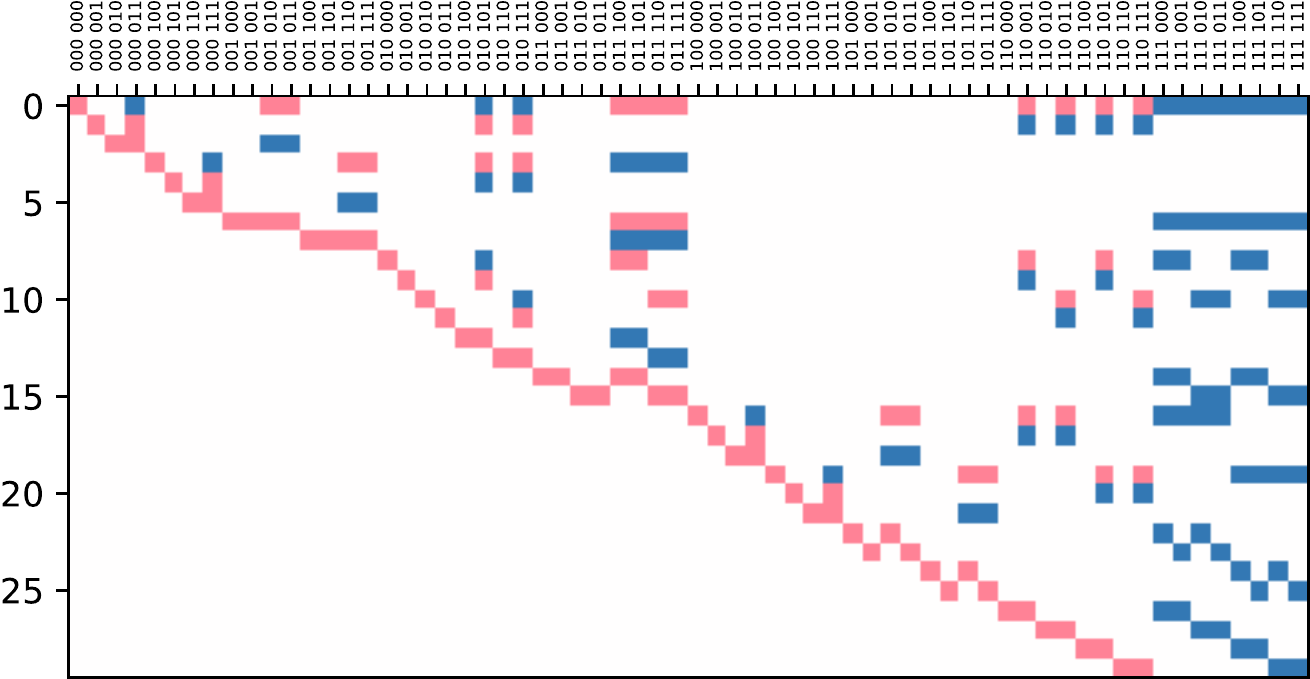}
\end{center}

\noindent Rows correspond to the 30 independent linear equations.
Columns in the plot correspond to entries of empirical models, indexed as $i_A i_B i_C$ $o_A o_B o_C$.
Coefficients in the equations are color-coded as white=0, red=+1 and blue=-1.

Space 30 has closest refinements in equivalence classes 17, 22 and 25; 
it is the join of its (closest) refinements.
It has closest coarsenings in equivalence classes 45, 47, 48 and 57; 
it is the meet of its (closest) coarsenings.
It has 512 causal functions, 64 of which are not causal for any of its refinements.
It is a tight space.

The standard causaltope for Space 30 has 2 more dimensions than those of its 3 subspaces in equivalence classes 17, 22 and 25.
The standard causaltope for Space 30 is the meet of the standard causaltopes for its closest coarsenings.
For completeness, below is a plot of the full homogeneous linear system of causality and quasi-normalisation equations for the standard causaltope:

\begin{center}
    \includegraphics[width=12cm]{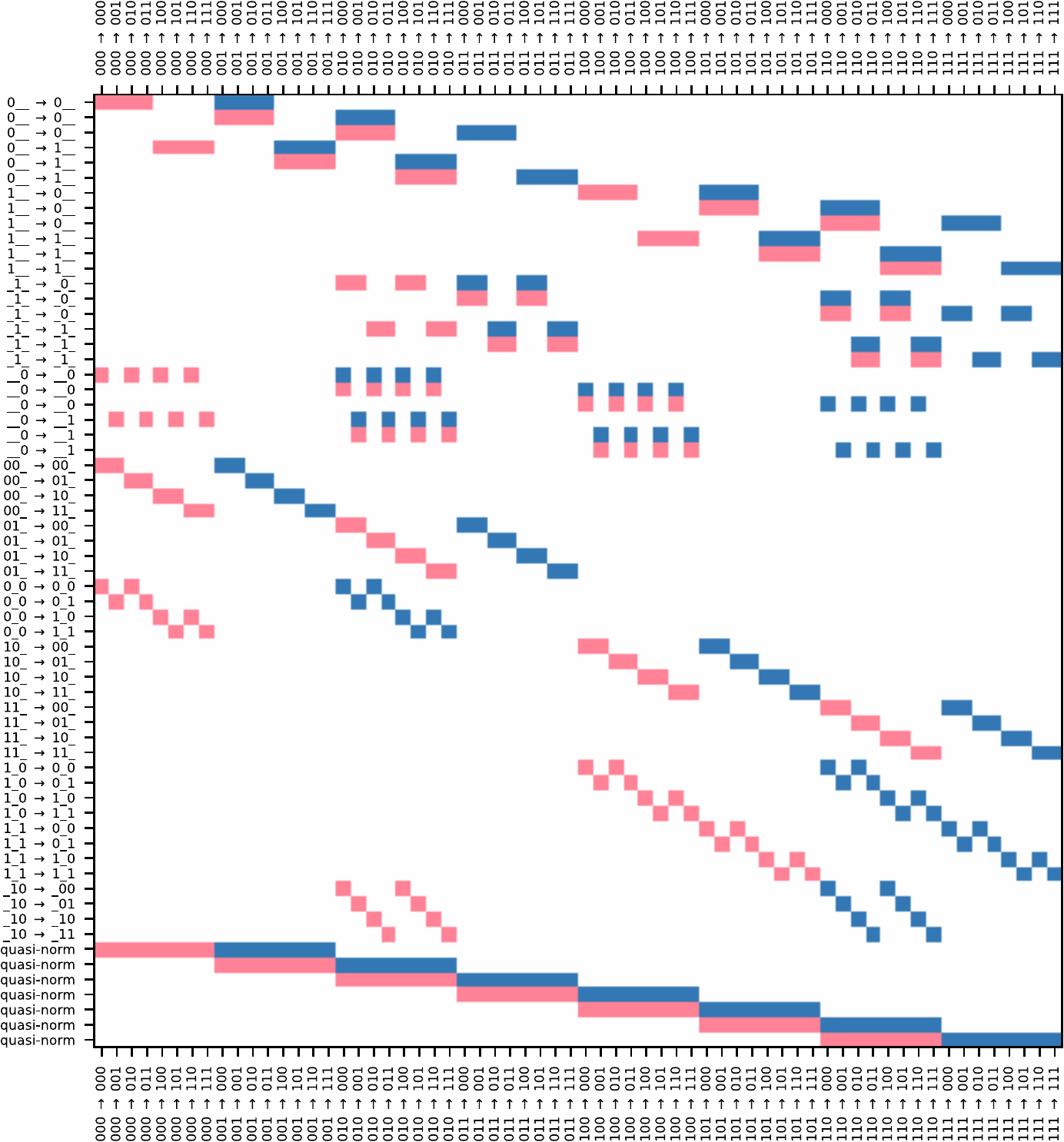}
\end{center}

\noindent Rows correspond to the 63 linear equations, of which 30 are independent.

\newpage
\subsection*{Space 31}

Space 31 is not induced by a causal order, but it is a refinement of the space 100 induced by the definite causal order $\total{\ev{A},\ev{B},\ev{C}}$.
Its equivalence class under event-input permutation symmetry contains 24 spaces.
Space 31 differs as follows from the space induced by causal order $\total{\ev{A},\ev{B},\ev{C}}$:
\begin{itemize}
  \item The outputs at events \evset{\ev{A}, \ev{C}} are independent of the input at event \ev{B} when the inputs at events \evset{A, C} are given by \hist{A/0,C/1} and \hist{A/1,C/1}.
  \item The outputs at events \evset{\ev{B}, \ev{C}} are independent of the input at event \ev{A} when the inputs at events \evset{B, C} are given by \hist{B/1,C/1} and \hist{B/0,C/1}.
  \item The output at event \ev{B} is independent of the input at event \ev{A} when the input at event B is given by \hist{B/0}.
  \item The output at event \ev{C} is independent of the inputs at events \evset{\ev{A}, \ev{B}} when the input at event C is given by \hist{C/1}.
\end{itemize}

\noindent Below are the histories and extended histories for space 31: 
\begin{center}
    \begin{tabular}{cc}
    \includegraphics[height=3.5cm]{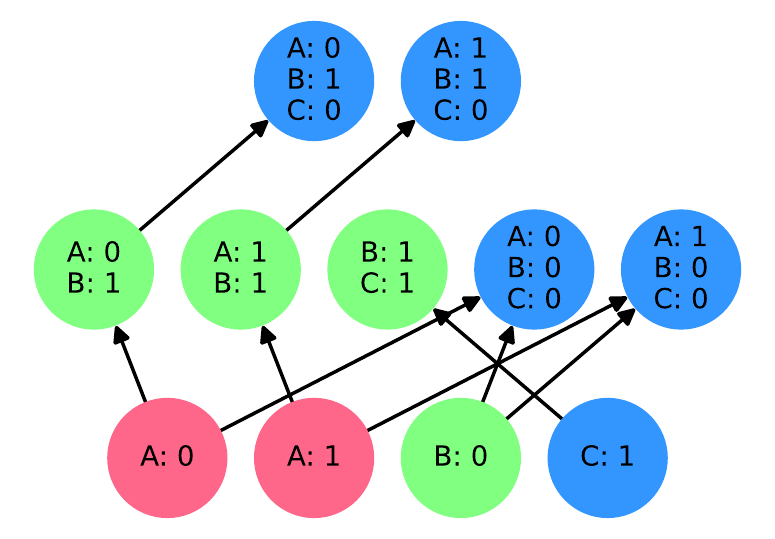}
    &
    \includegraphics[height=3.5cm]{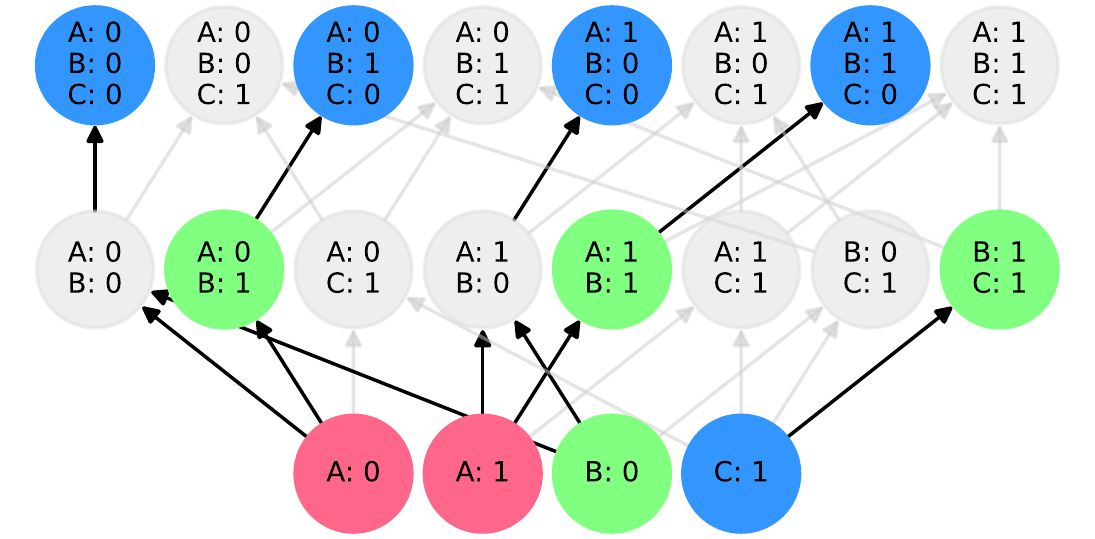}
    \\
    $\Theta_{31}$
    &
    $\Ext{\Theta_{31}}$
    \end{tabular}
\end{center}

\noindent The standard causaltope for Space 31 has dimension 33.
Below is a plot of the homogeneous linear system of causality and quasi-normalisation equations for the standard causaltope, put in reduced row echelon form:

\begin{center}
    \includegraphics[width=11cm]{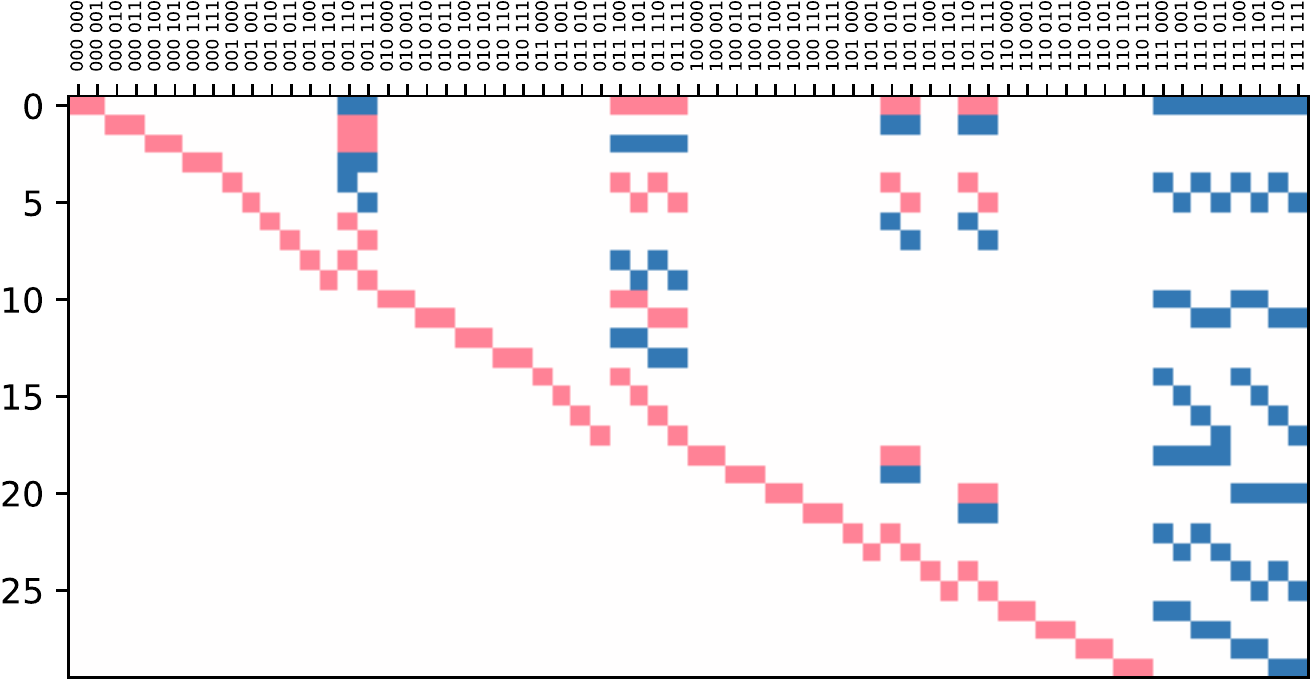}
\end{center}

\noindent Rows correspond to the 30 independent linear equations.
Columns in the plot correspond to entries of empirical models, indexed as $i_A i_B i_C$ $o_A o_B o_C$.
Coefficients in the equations are color-coded as white=0, red=+1 and blue=-1.

Space 31 has closest refinements in equivalence classes 18, 20 and 22; 
it is the join of its (closest) refinements.
It has closest coarsenings in equivalence classes 45 and 51; 
it is the meet of its (closest) coarsenings.
It has 512 causal functions, all of which are causal for at least one of its refinements.
It is not a tight space: for event \ev{B}, a causal function must yield identical output values on input histories \hist{A/0,B/1}, \hist{A/1,B/1} and \hist{B/1,C/1}.

The standard causaltope for Space 31 coincides with that of its subspace in equivalence class 18.
The standard causaltope for Space 31 is the meet of the standard causaltopes for its closest coarsenings.
For completeness, below is a plot of the full homogeneous linear system of causality and quasi-normalisation equations for the standard causaltope:

\begin{center}
    \includegraphics[width=12cm]{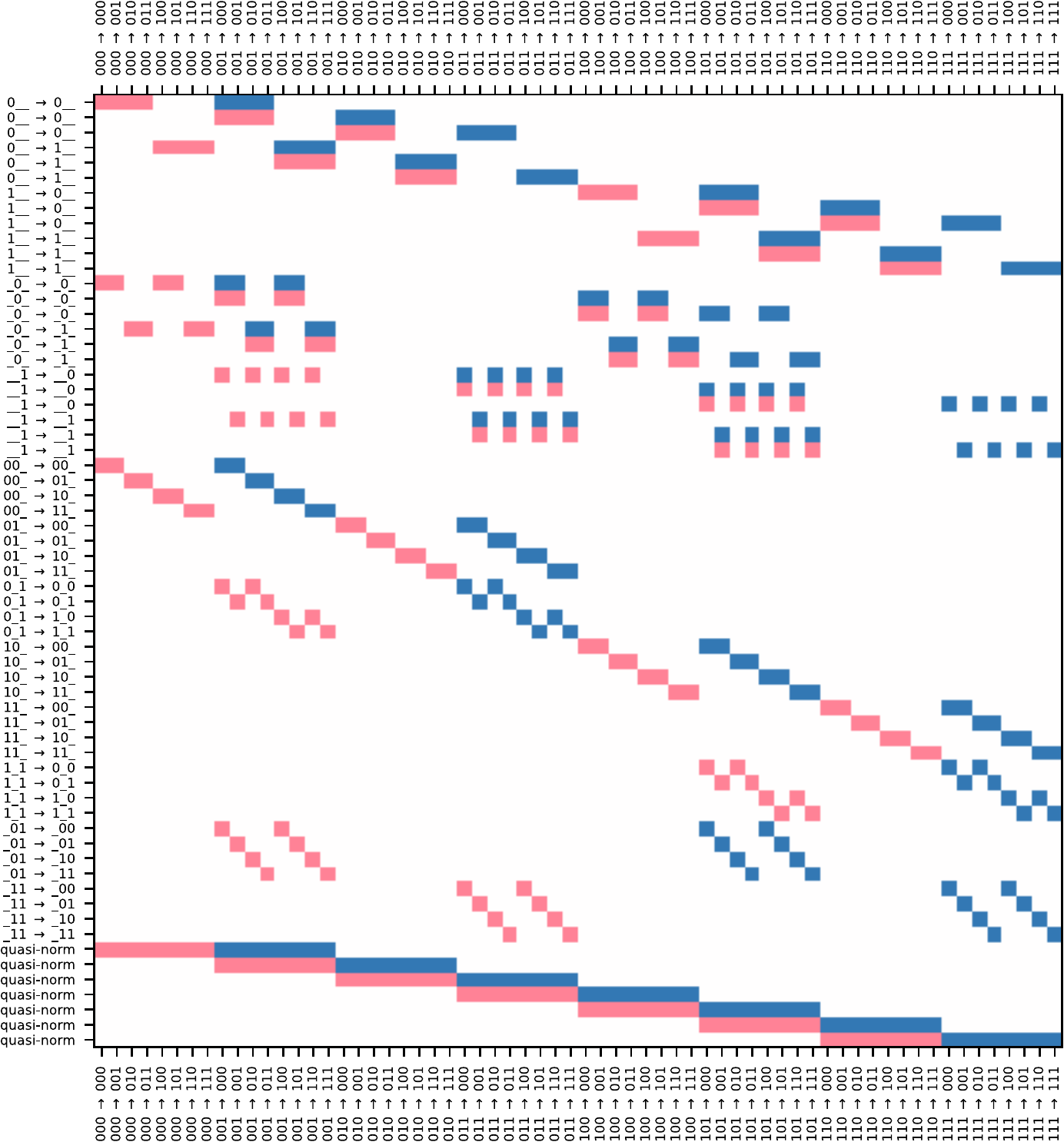}
\end{center}

\noindent Rows correspond to the 63 linear equations, of which 30 are independent.

\newpage
\subsection*{Space 32}

Space 32 is not induced by a causal order, but it is a refinement of the space induced by the indefinite causal order $\total{\ev{A},\{\ev{B},\ev{C}\}}$.
Its equivalence class under event-input permutation symmetry contains 48 spaces.
Space 32 differs as follows from the space induced by causal order $\total{\ev{A},\{\ev{B},\ev{C}\}}$:
\begin{itemize}
  \item The outputs at events \evset{\ev{A}, \ev{B}} are independent of the input at event \ev{C} when the inputs at events \evset{A, B} are given by \hist{A/0,B/0}, \hist{A/0,B/1} and \hist{A/1,B/0}.
  \item The outputs at events \evset{\ev{A}, \ev{C}} are independent of the input at event \ev{B} when the inputs at events \evset{A, C} are given by \hist{A/0,C/1}, \hist{A/1,C/0} and \hist{A/1,C/1}.
  \item The outputs at events \evset{\ev{B}, \ev{C}} are independent of the input at event \ev{A} when the inputs at events \evset{B, C} are given by \hist{B/1,C/1} and \hist{B/0,C/1}.
  \item The output at event \ev{B} is independent of the inputs at events \evset{\ev{A}, \ev{C}} when the input at event B is given by \hist{B/0}.
  \item The output at event \ev{C} is independent of the inputs at events \evset{\ev{A}, \ev{B}} when the input at event C is given by \hist{C/1}.
\end{itemize}

\noindent Below are the histories and extended histories for space 32: 
\begin{center}
    \begin{tabular}{cc}
    \includegraphics[height=3.5cm]{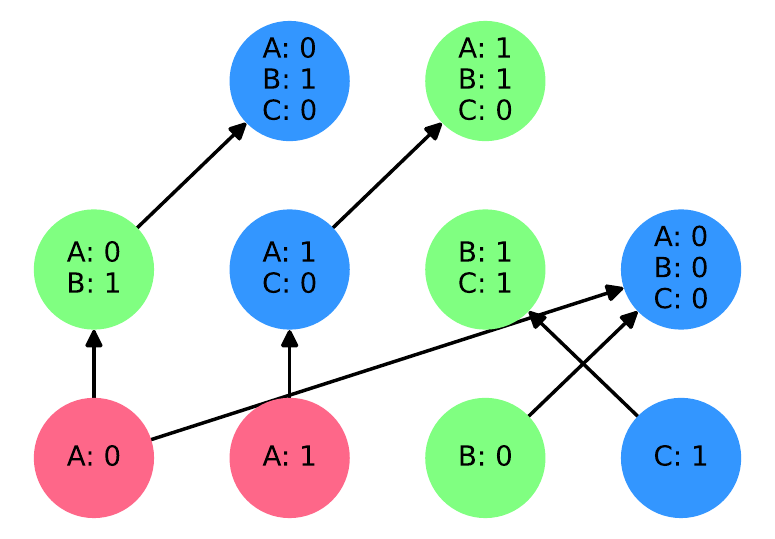}
    &
    \includegraphics[height=3.5cm]{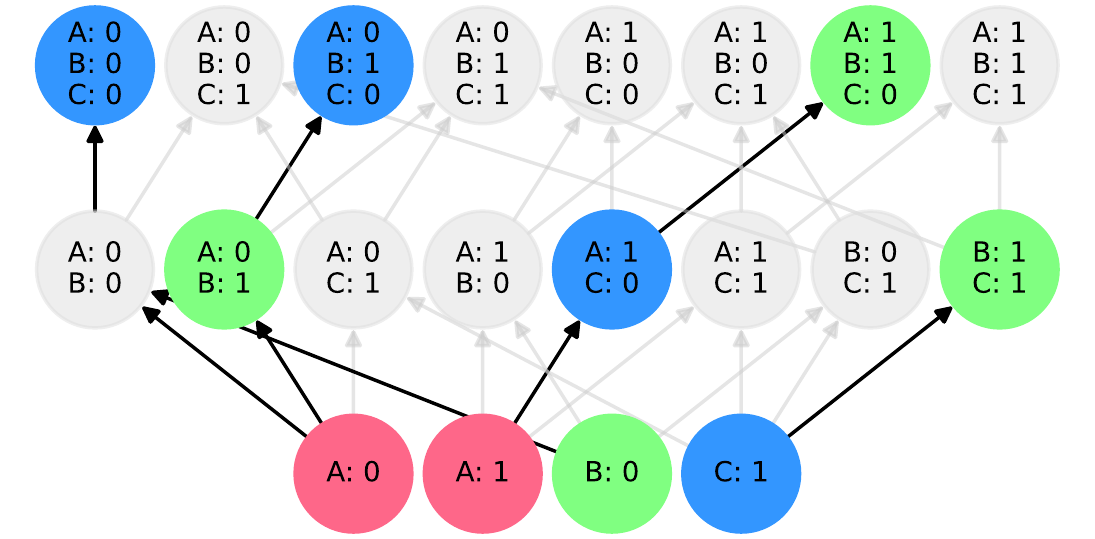}
    \\
    $\Theta_{32}$
    &
    $\Ext{\Theta_{32}}$
    \end{tabular}
\end{center}

\noindent The standard causaltope for Space 32 has dimension 33.
Below is a plot of the homogeneous linear system of causality and quasi-normalisation equations for the standard causaltope, put in reduced row echelon form:

\begin{center}
    \includegraphics[width=11cm]{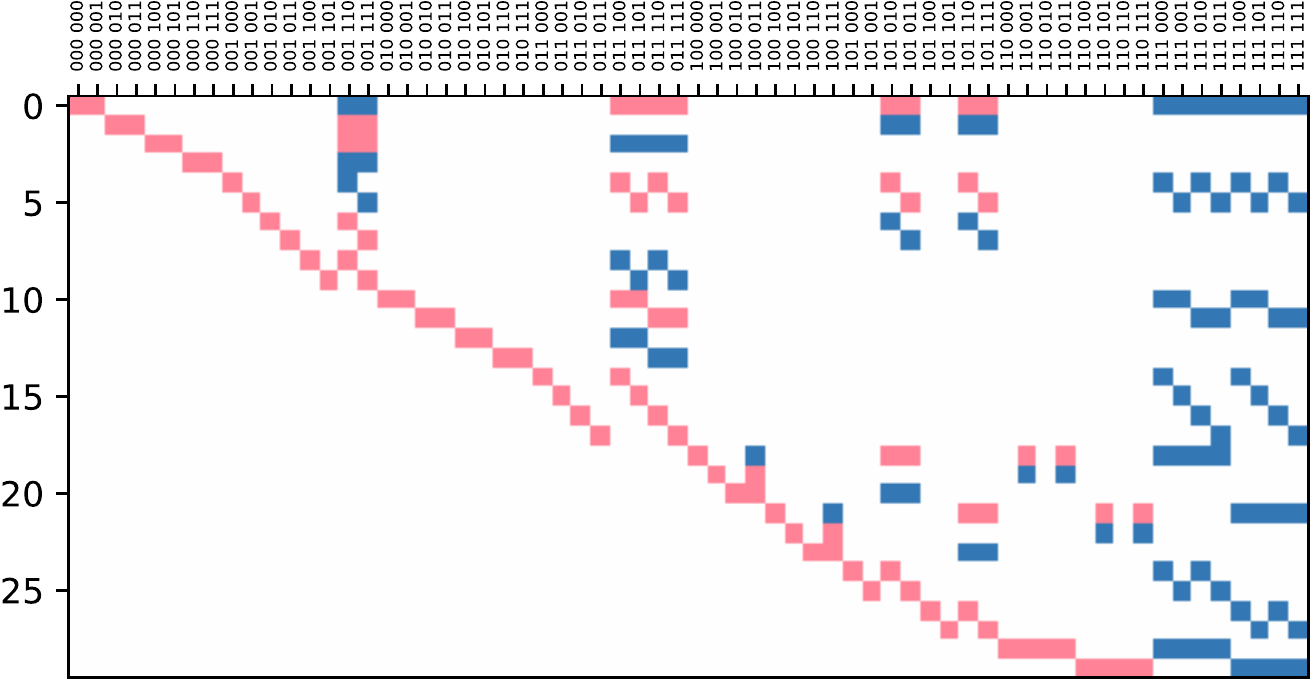}
\end{center}

\noindent Rows correspond to the 30 independent linear equations.
Columns in the plot correspond to entries of empirical models, indexed as $i_A i_B i_C$ $o_A o_B o_C$.
Coefficients in the equations are color-coded as white=0, red=+1 and blue=-1.

Space 32 has closest refinements in equivalence classes 17, 21 and 22; 
it is the join of its (closest) refinements.
It has closest coarsenings in equivalence classes 48 and 52; 
it is the meet of its (closest) coarsenings.
It has 512 causal functions, 192 of which are not causal for any of its refinements.
It is not a tight space: for event \ev{B}, a causal function must yield identical output values on input histories \hist{A/0,B/1} and \hist{B/1,C/1}.

The standard causaltope for Space 32 has 2 more dimensions than those of its 3 subspaces in equivalence classes 17, 21 and 22.
The standard causaltope for Space 32 is the meet of the standard causaltopes for its closest coarsenings.
For completeness, below is a plot of the full homogeneous linear system of causality and quasi-normalisation equations for the standard causaltope:

\begin{center}
    \includegraphics[width=12cm]{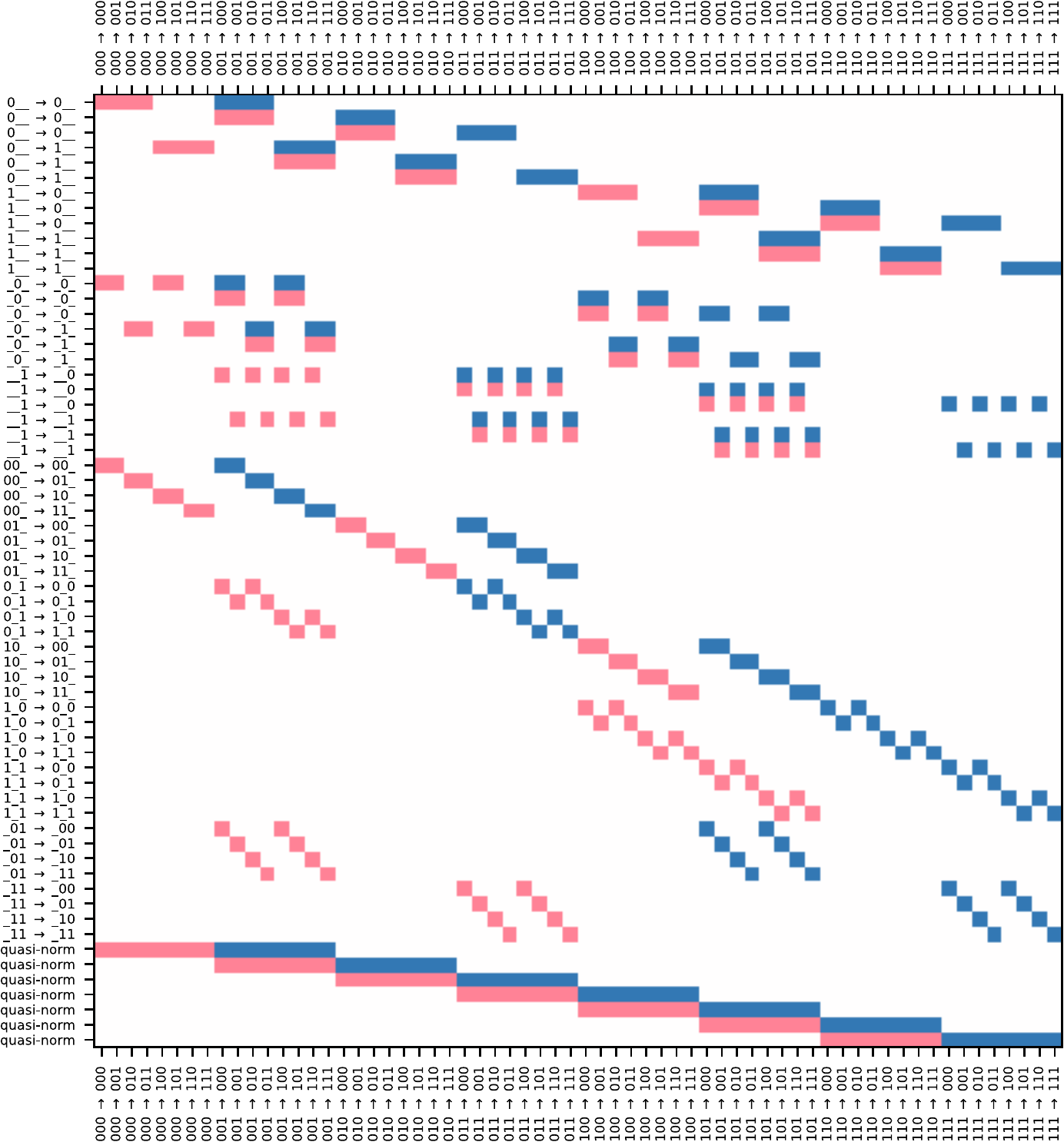}
\end{center}

\noindent Rows correspond to the 63 linear equations, of which 30 are independent.

\newpage
\subsection*{Space 33}

Space 33 is induced by the definite causal order $\total{\ev{A},\ev{B}}\vee\discrete{\ev{C}}$.
Its equivalence class under event-input permutation symmetry contains 6 spaces.

\noindent Below are the histories and extended histories for space 33: 
\begin{center}
    \begin{tabular}{cc}
    \includegraphics[height=3.5cm]{svg-inkscape/space-ABC-unique-tight-33-highlighted_svg-tex.pdf}
    &
    \includegraphics[height=3.5cm]{svg-inkscape/space-ABC-unique-tight-33-ext-highlighted_svg-tex.pdf}
    \\
    $\Theta_{33}$
    &
    $\Ext{\Theta_{33}}$
    \end{tabular}
\end{center}

\noindent The standard causaltope for Space 33 has dimension 32.
Below is a plot of the homogeneous linear system of causality and quasi-normalisation equations for the standard causaltope, put in reduced row echelon form:

\begin{center}
    \includegraphics[width=11cm]{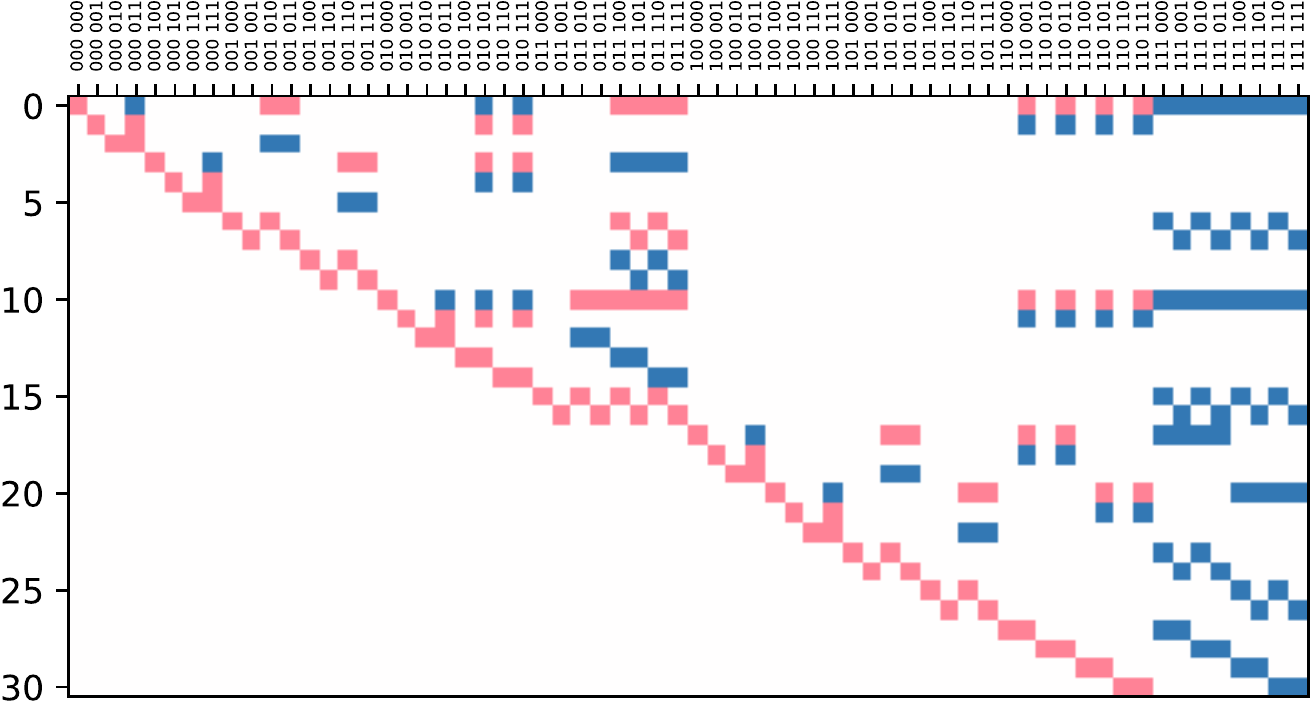}
\end{center}

\noindent Rows correspond to the 31 independent linear equations.
Columns in the plot correspond to entries of empirical models, indexed as $i_A i_B i_C$ $o_A o_B o_C$.
Coefficients in the equations are color-coded as white=0, red=+1 and blue=-1.

Space 33 has closest refinements in equivalence class 19; 
it is the join of its (closest) refinements.
It has closest coarsenings in equivalence classes 46 and 58; 
it is the meet of its (closest) coarsenings.
It has 256 causal functions, 64 of which are not causal for any of its refinements.
It is a tight space.

The standard causaltope for Space 33 has 2 more dimensions than those of its 4 subspaces in equivalence class 19.
The standard causaltope for Space 33 is the meet of the standard causaltopes for its closest coarsenings.
For completeness, below is a plot of the full homogeneous linear system of causality and quasi-normalisation equations for the standard causaltope:

\begin{center}
    \includegraphics[width=12cm]{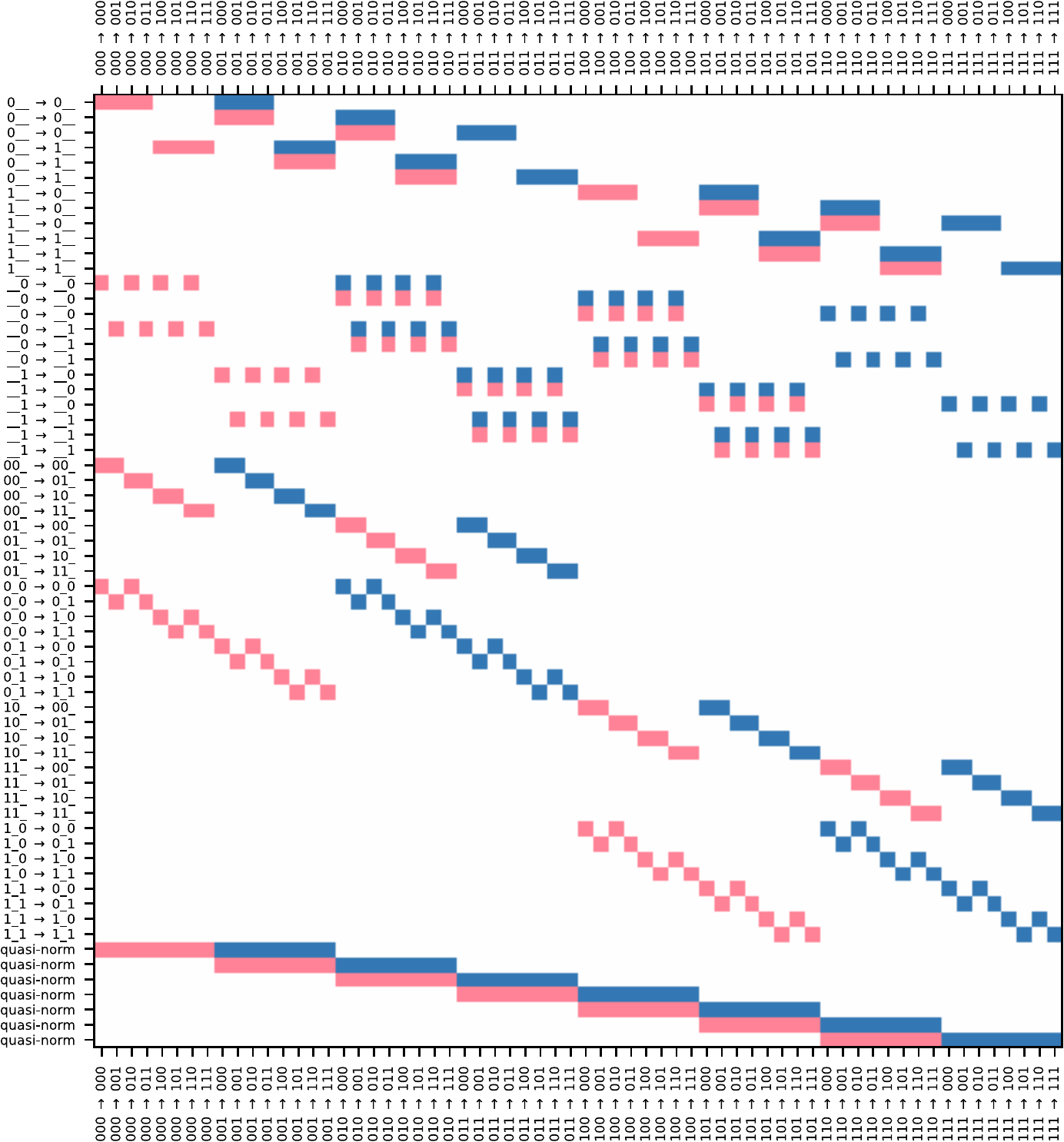}
\end{center}

\noindent Rows correspond to the 63 linear equations, of which 31 are independent.

\newpage
\subsection*{Space 34}

Space 34 is not induced by a causal order, but it is a refinement of the space in equivalence class 92 induced by the definite causal order $\total{\ev{A},\ev{B}}\vee\total{\ev{C},\ev{B}}$ (note that the space induced by the order is not the same as space 92).
Its equivalence class under event-input permutation symmetry contains 48 spaces.
Space 34 differs as follows from the space induced by causal order $\total{\ev{A},\ev{B}}\vee\total{\ev{C},\ev{B}}$:
\begin{itemize}
  \item The outputs at events \evset{\ev{A}, \ev{B}} are independent of the input at event \ev{C} when the inputs at events \evset{A, B} are given by \hist{A/0,B/0}, \hist{A/0,B/1} and \hist{A/1,B/0}.
  \item The outputs at events \evset{\ev{B}, \ev{C}} are independent of the input at event \ev{A} when the inputs at events \evset{B, C} are given by \hist{B/0,C/1}.
\end{itemize}

\noindent Below are the histories and extended histories for space 34: 
\begin{center}
    \begin{tabular}{cc}
    \includegraphics[height=3.5cm]{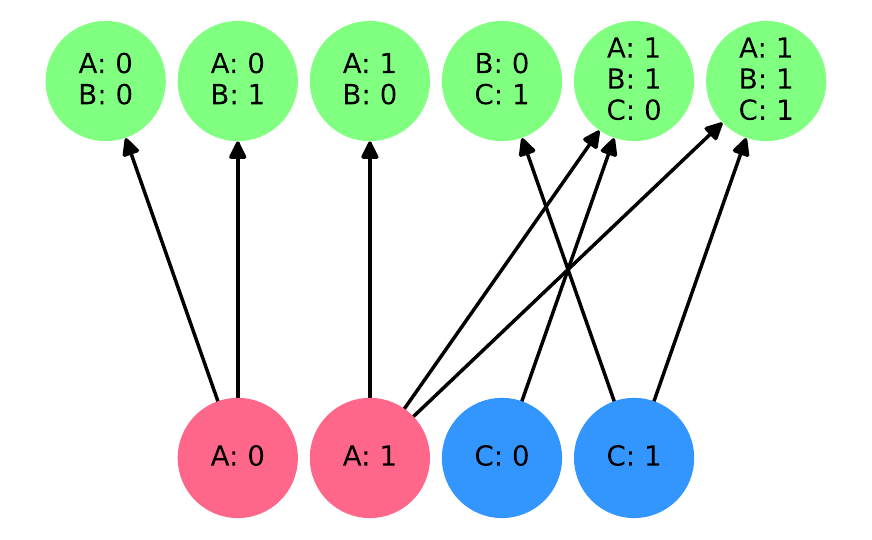}
    &
    \includegraphics[height=3.5cm]{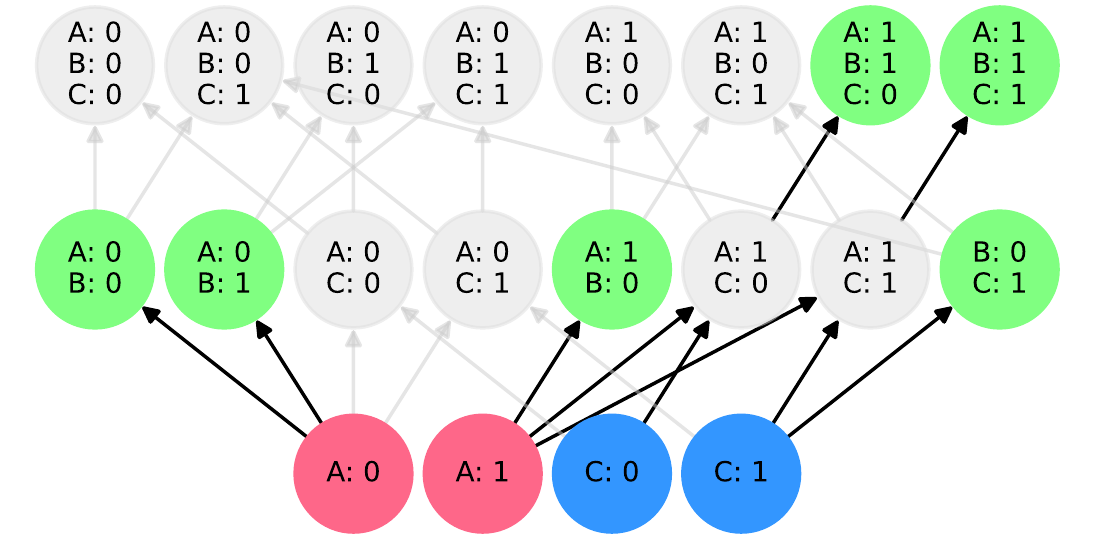}
    \\
    $\Theta_{34}$
    &
    $\Ext{\Theta_{34}}$
    \end{tabular}
\end{center}

\noindent The standard causaltope for Space 34 has dimension 32.
Below is a plot of the homogeneous linear system of causality and quasi-normalisation equations for the standard causaltope, put in reduced row echelon form:

\begin{center}
    \includegraphics[width=11cm]{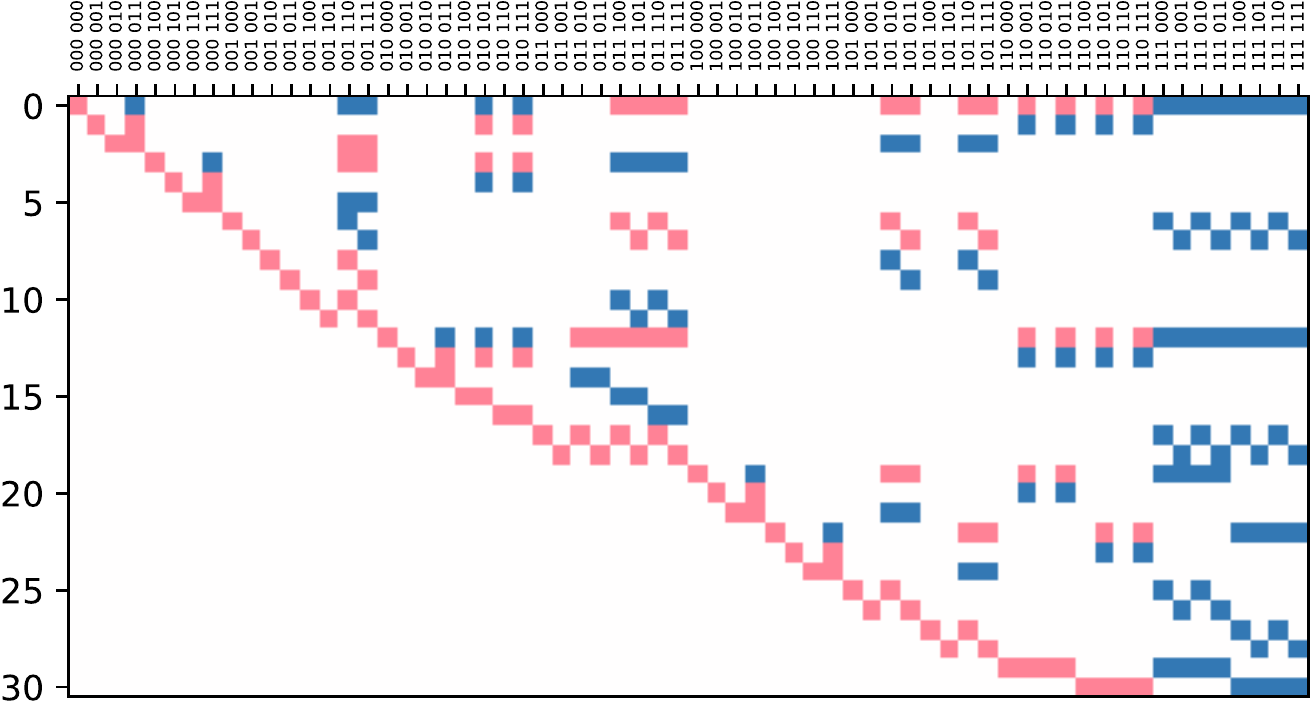}
\end{center}

\noindent Rows correspond to the 31 independent linear equations.
Columns in the plot correspond to entries of empirical models, indexed as $i_A i_B i_C$ $o_A o_B o_C$.
Coefficients in the equations are color-coded as white=0, red=+1 and blue=-1.

Space 34 has closest refinements in equivalence classes 19, 23, 24 and 26; 
it is the join of its (closest) refinements.
It has closest coarsenings in equivalence classes 46, 50, 53, 57 and 59; 
it is the meet of its (closest) coarsenings.
It has 256 causal functions, all of which are causal for at least one of its refinements.
It is not a tight space: for event \ev{B}, a causal function must yield identical output values on input histories \hist{A/0,B/0}, \hist{A/1,B/0} and \hist{B/0,C/1}.

The standard causaltope for Space 34 has 1 more dimension than that of its subspace in equivalence class 26.
The standard causaltope for Space 34 is the meet of the standard causaltopes for its closest coarsenings.
For completeness, below is a plot of the full homogeneous linear system of causality and quasi-normalisation equations for the standard causaltope:

\begin{center}
    \includegraphics[width=12cm]{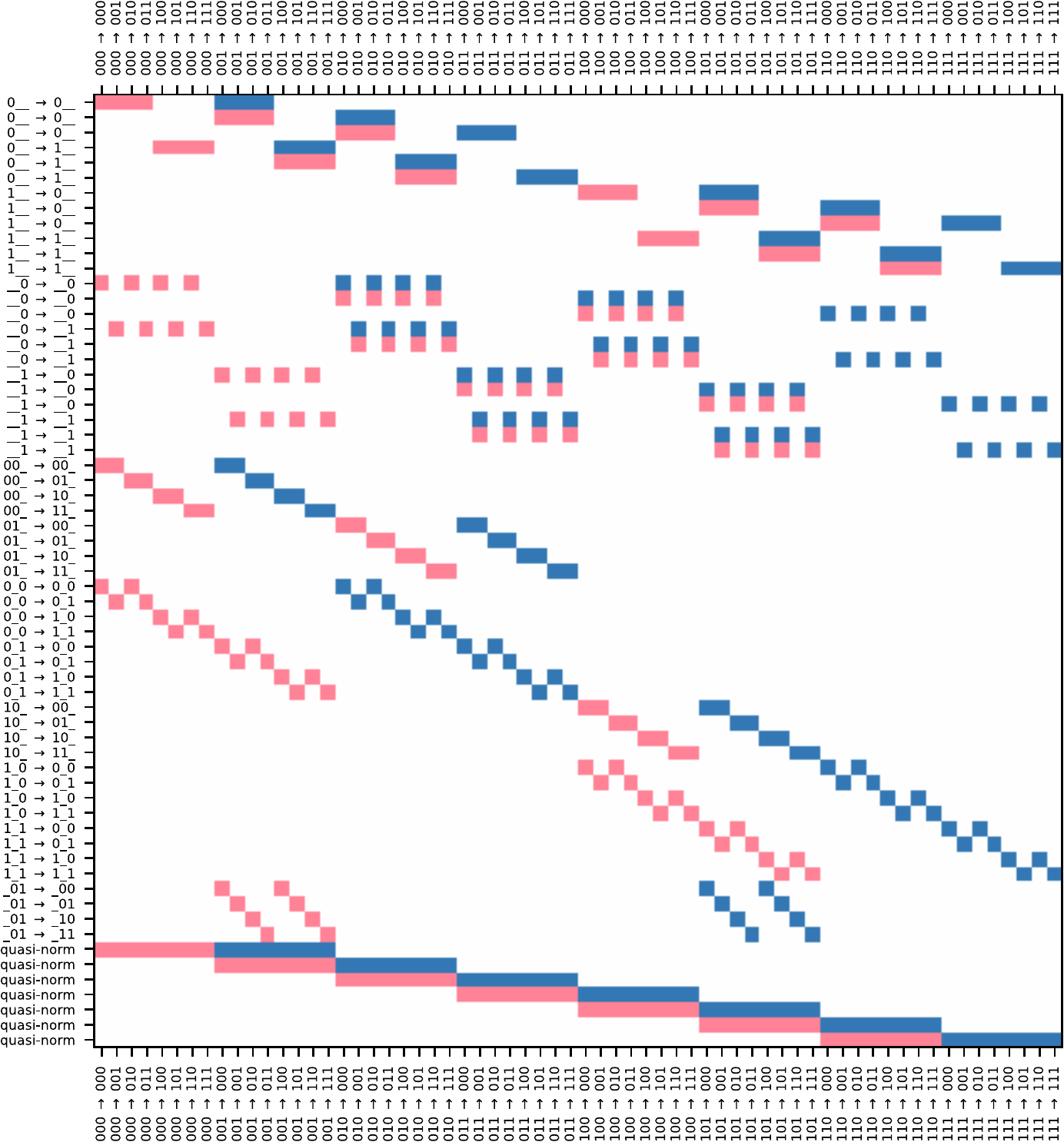}
\end{center}

\noindent Rows correspond to the 63 linear equations, of which 31 are independent.

\newpage
\subsection*{Space 35}

Space 35 is not induced by a causal order, but it is a refinement of the space in equivalence class 92 induced by the definite causal order $\total{\ev{A},\ev{B}}\vee\total{\ev{C},\ev{B}}$ (note that the space induced by the order is not the same as space 92).
Its equivalence class under event-input permutation symmetry contains 24 spaces.
Space 35 differs as follows from the space induced by causal order $\total{\ev{A},\ev{B}}\vee\total{\ev{C},\ev{B}}$:
\begin{itemize}
  \item The outputs at events \evset{\ev{A}, \ev{B}} are independent of the input at event \ev{C} when the inputs at events \evset{A, B} are given by \hist{A/0,B/0} and \hist{A/0,B/1}.
  \item The outputs at events \evset{\ev{B}, \ev{C}} are independent of the input at event \ev{A} when the inputs at events \evset{B, C} are given by \hist{B/1,C/0} and \hist{B/1,C/1}.
\end{itemize}

\noindent Below are the histories and extended histories for space 35: 
\begin{center}
    \begin{tabular}{cc}
    \includegraphics[height=3.5cm]{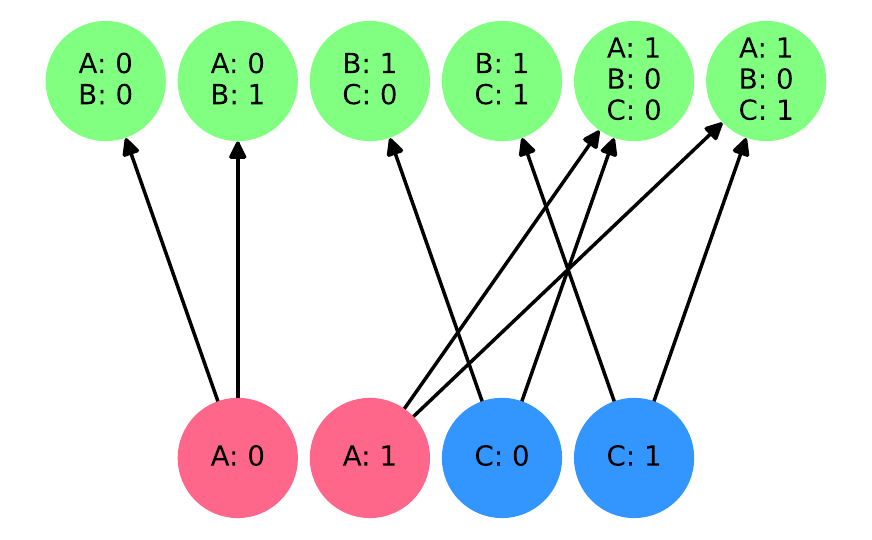}
    &
    \includegraphics[height=3.5cm]{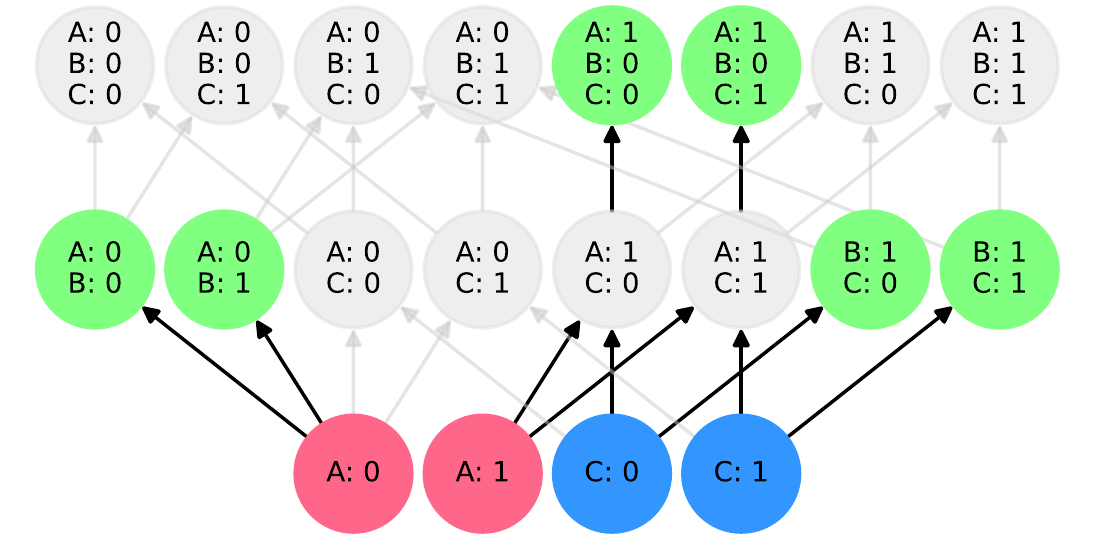}
    \\
    $\Theta_{35}$
    &
    $\Ext{\Theta_{35}}$
    \end{tabular}
\end{center}

\noindent The standard causaltope for Space 35 has dimension 32.
Below is a plot of the homogeneous linear system of causality and quasi-normalisation equations for the standard causaltope, put in reduced row echelon form:

\begin{center}
    \includegraphics[width=11cm]{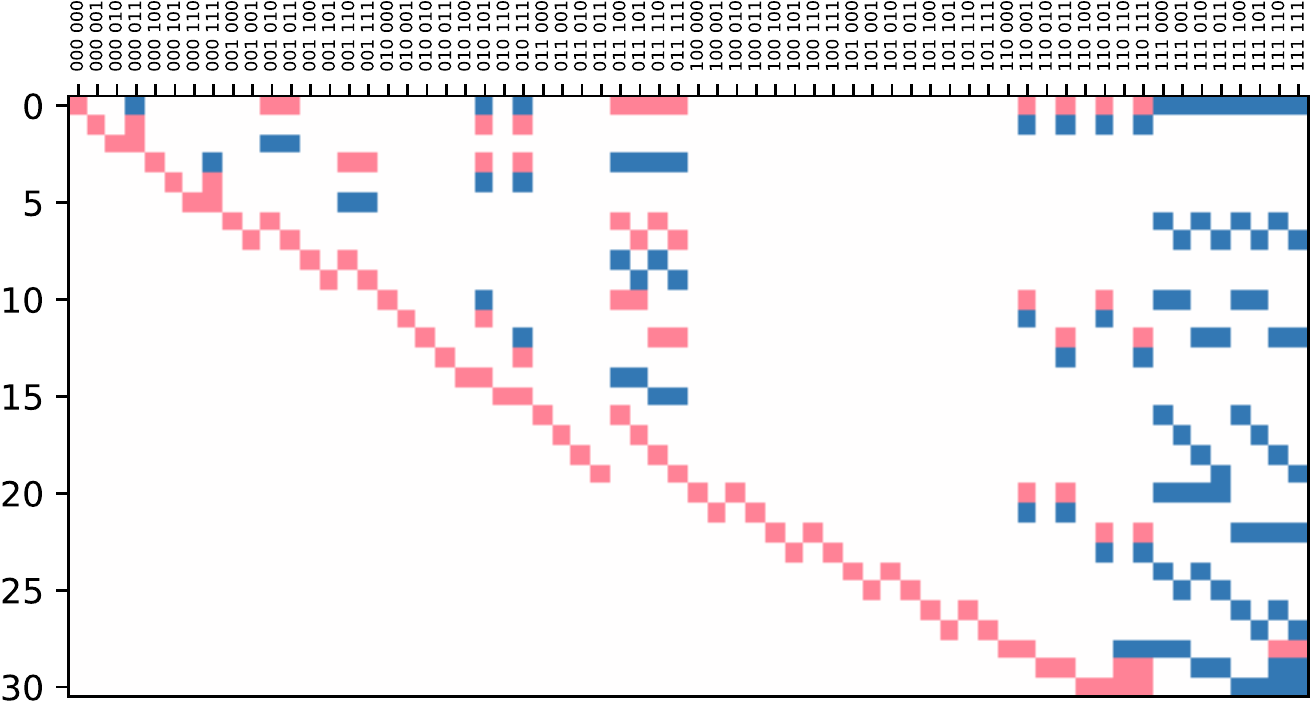}
\end{center}

\noindent Rows correspond to the 31 independent linear equations.
Columns in the plot correspond to entries of empirical models, indexed as $i_A i_B i_C$ $o_A o_B o_C$.
Coefficients in the equations are color-coded as white=0, red=+1 and blue=-1.

Space 35 has closest refinements in equivalence classes 16, 23 and 26; 
it is the join of its (closest) refinements.
It has closest coarsenings in equivalence classes 49, 53, 54 and 59; 
it is the meet of its (closest) coarsenings.
It has 256 causal functions, all of which are causal for at least one of its refinements.
It is not a tight space: for event \ev{B}, a causal function must yield identical output values on input histories \hist{A/0,B/1}, \hist{B/1,C/0} and \hist{B/1,C/1}.

The standard causaltope for Space 35 has 1 more dimension than that of its subspace in equivalence class 26.
The standard causaltope for Space 35 is the meet of the standard causaltopes for its closest coarsenings.
For completeness, below is a plot of the full homogeneous linear system of causality and quasi-normalisation equations for the standard causaltope:

\begin{center}
    \includegraphics[width=12cm]{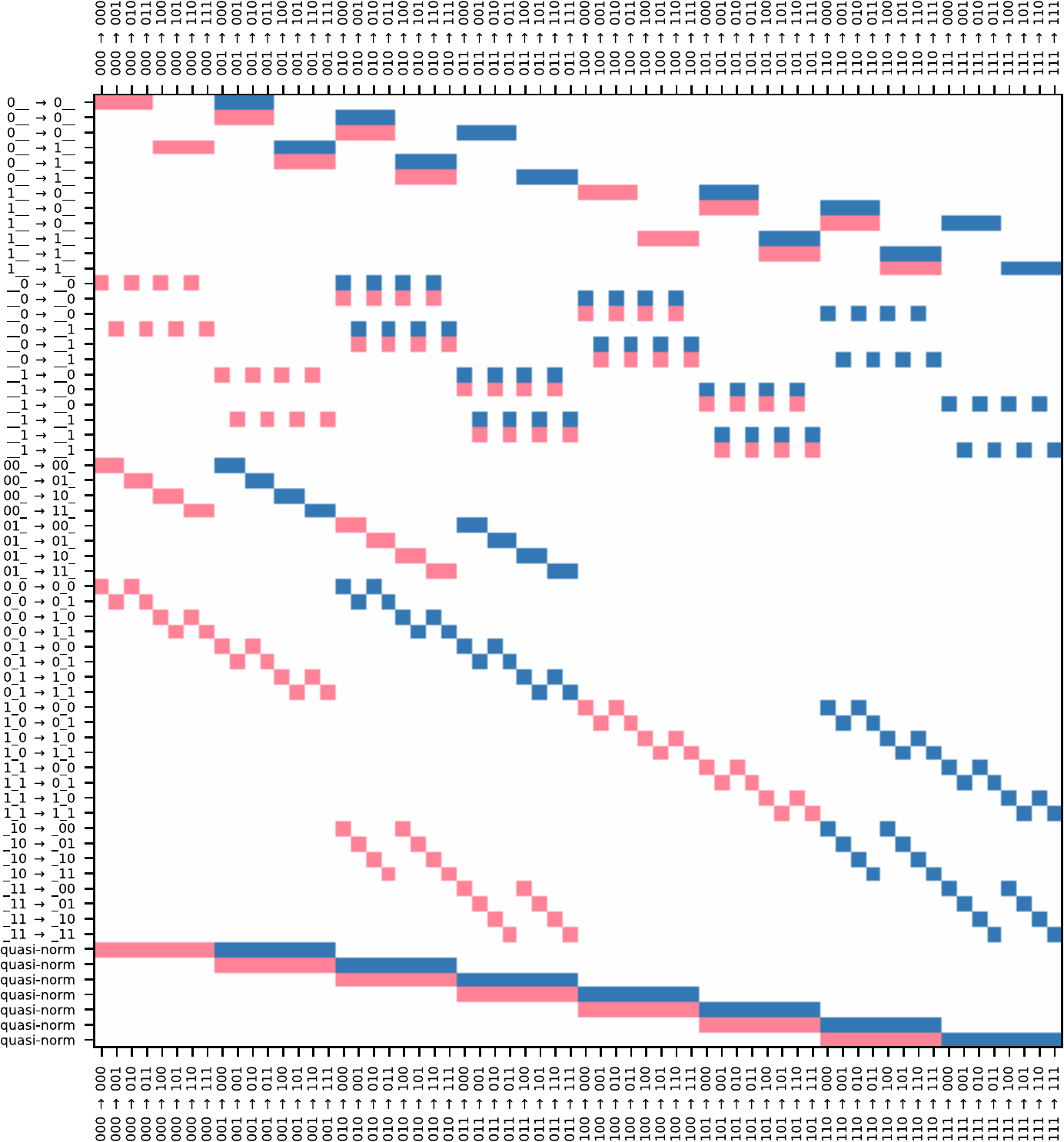}
\end{center}

\noindent Rows correspond to the 63 linear equations, of which 31 are independent.

\newpage
\subsection*{Space 36}

Space 36 is not induced by a causal order, but it is a refinement of the space 92 induced by the definite causal order $\total{\ev{A},\ev{C}}\vee\total{\ev{B},\ev{C}}$.
Its equivalence class under event-input permutation symmetry contains 24 spaces.
Space 36 differs as follows from the space induced by causal order $\total{\ev{A},\ev{C}}\vee\total{\ev{B},\ev{C}}$:
\begin{itemize}
  \item The outputs at events \evset{\ev{B}, \ev{C}} are independent of the input at event \ev{A} when the inputs at events \evset{B, C} are given by \hist{B/1,C/0} and \hist{B/0,C/1}.
  \item The outputs at events \evset{\ev{A}, \ev{C}} are independent of the input at event \ev{B} when the inputs at events \evset{A, C} are given by \hist{A/1,C/0} and \hist{A/1,C/1}.
\end{itemize}

\noindent Below are the histories and extended histories for space 36: 
\begin{center}
    \begin{tabular}{cc}
    \includegraphics[height=3.5cm]{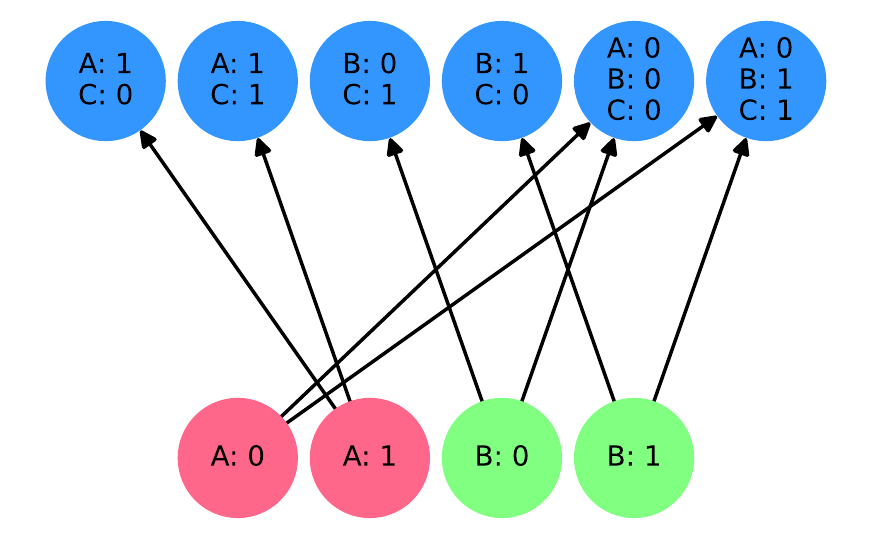}
    &
    \includegraphics[height=3.5cm]{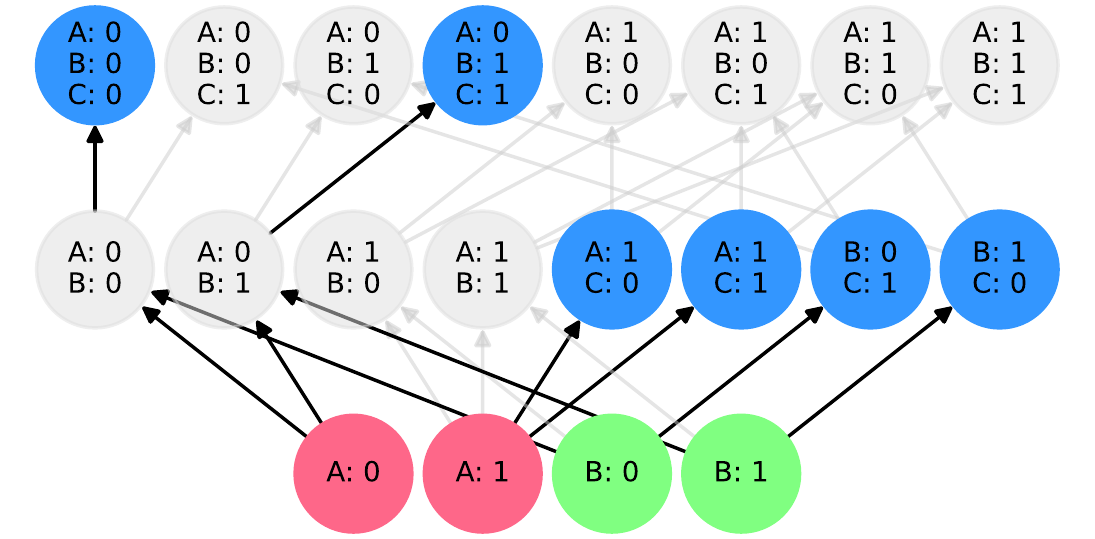}
    \\
    $\Theta_{36}$
    &
    $\Ext{\Theta_{36}}$
    \end{tabular}
\end{center}

\noindent The standard causaltope for Space 36 has dimension 32.
Below is a plot of the homogeneous linear system of causality and quasi-normalisation equations for the standard causaltope, put in reduced row echelon form:

\begin{center}
    \includegraphics[width=11cm]{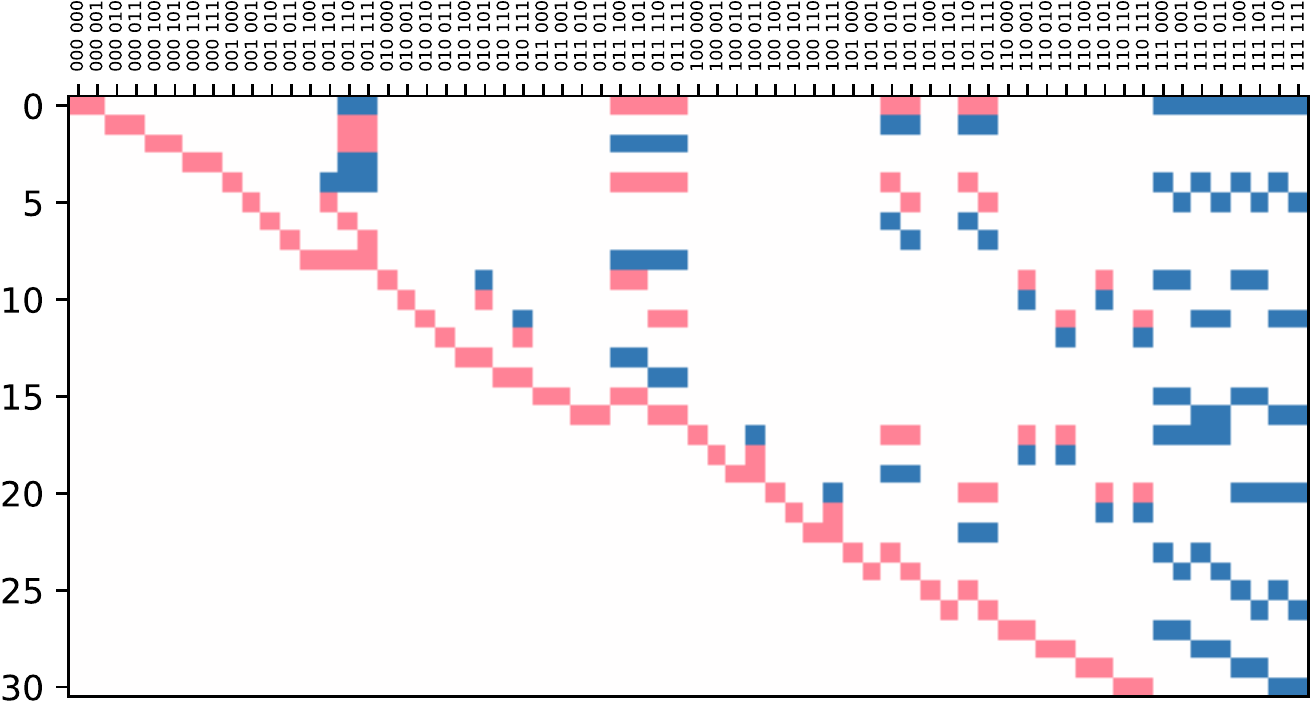}
\end{center}

\noindent Rows correspond to the 31 independent linear equations.
Columns in the plot correspond to entries of empirical models, indexed as $i_A i_B i_C$ $o_A o_B o_C$.
Coefficients in the equations are color-coded as white=0, red=+1 and blue=-1.

Space 36 has closest refinements in equivalence classes 23 and 24; 
it is the join of its (closest) refinements.
It has closest coarsenings in equivalence classes 50, 53 and 55; 
it is the meet of its (closest) coarsenings.
It has 256 causal functions, 64 of which are not causal for any of its refinements.
It is not a tight space: for event \ev{C}, a causal function must yield identical output values on input histories \hist{A/1,C/0} and \hist{B/1,C/0}, and it must also yield identical output values on input histories \hist{A/1,C/1} and \hist{B/0,C/1}.

The standard causaltope for Space 36 has 2 more dimensions than those of its 4 subspaces in equivalence classes 23 and 24.
The standard causaltope for Space 36 is the meet of the standard causaltopes for its closest coarsenings.
For completeness, below is a plot of the full homogeneous linear system of causality and quasi-normalisation equations for the standard causaltope:

\begin{center}
    \includegraphics[width=12cm]{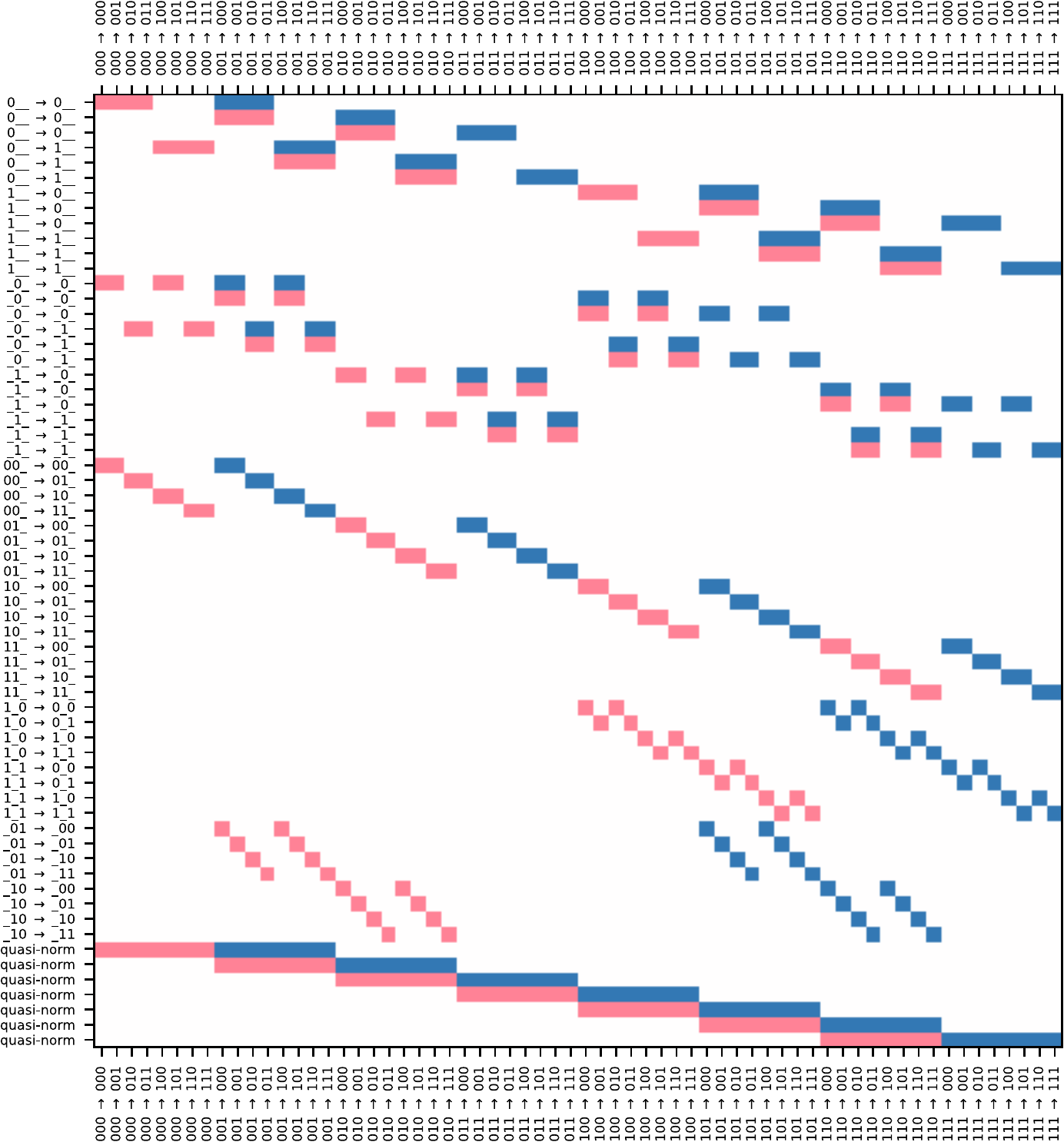}
\end{center}

\noindent Rows correspond to the 63 linear equations, of which 31 are independent.

\newpage
\subsection*{Space 37}

Space 37 is not induced by a causal order, but it is a refinement of the space in equivalence class 92 induced by the definite causal order $\total{\ev{A},\ev{B}}\vee\total{\ev{C},\ev{B}}$ (note that the space induced by the order is not the same as space 92).
Its equivalence class under event-input permutation symmetry contains 12 spaces.
Space 37 differs as follows from the space induced by causal order $\total{\ev{A},\ev{B}}\vee\total{\ev{C},\ev{B}}$:
\begin{itemize}
  \item The outputs at events \evset{\ev{A}, \ev{B}} are independent of the input at event \ev{C} when the inputs at events \evset{A, B} are given by \hist{A/0,B/0} and \hist{A/0,B/1}.
  \item The outputs at events \evset{\ev{B}, \ev{C}} are independent of the input at event \ev{A} when the inputs at events \evset{B, C} are given by \hist{B/1,C/1} and \hist{B/0,C/1}.
\end{itemize}

\noindent Below are the histories and extended histories for space 37: 
\begin{center}
    \begin{tabular}{cc}
    \includegraphics[height=3.5cm]{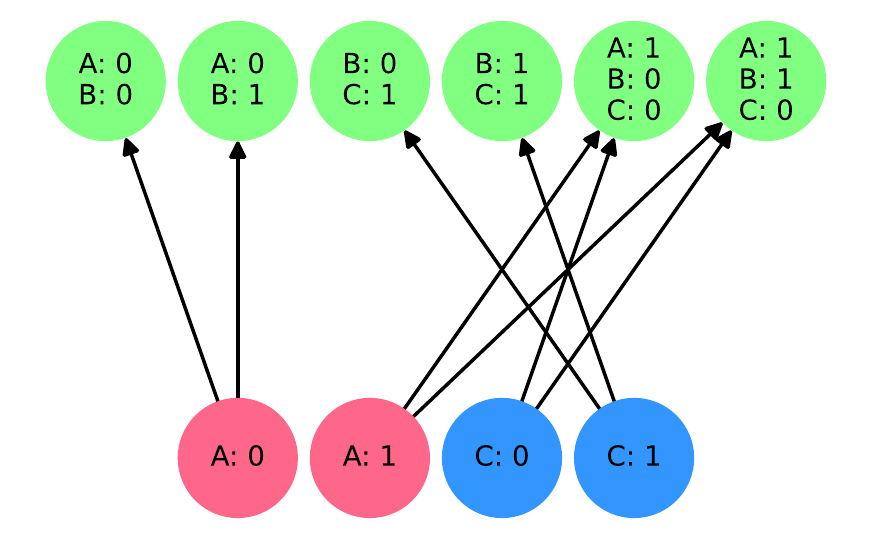}
    &
    \includegraphics[height=3.5cm]{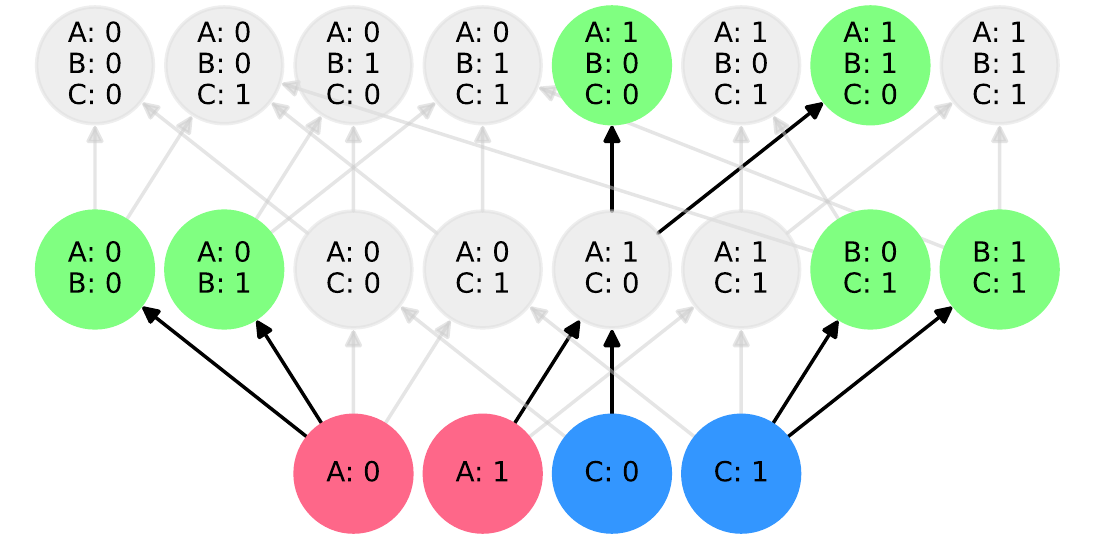}
    \\
    $\Theta_{37}$
    &
    $\Ext{\Theta_{37}}$
    \end{tabular}
\end{center}

\noindent The standard causaltope for Space 37 has dimension 32.
Below is a plot of the homogeneous linear system of causality and quasi-normalisation equations for the standard causaltope, put in reduced row echelon form:

\begin{center}
    \includegraphics[width=11cm]{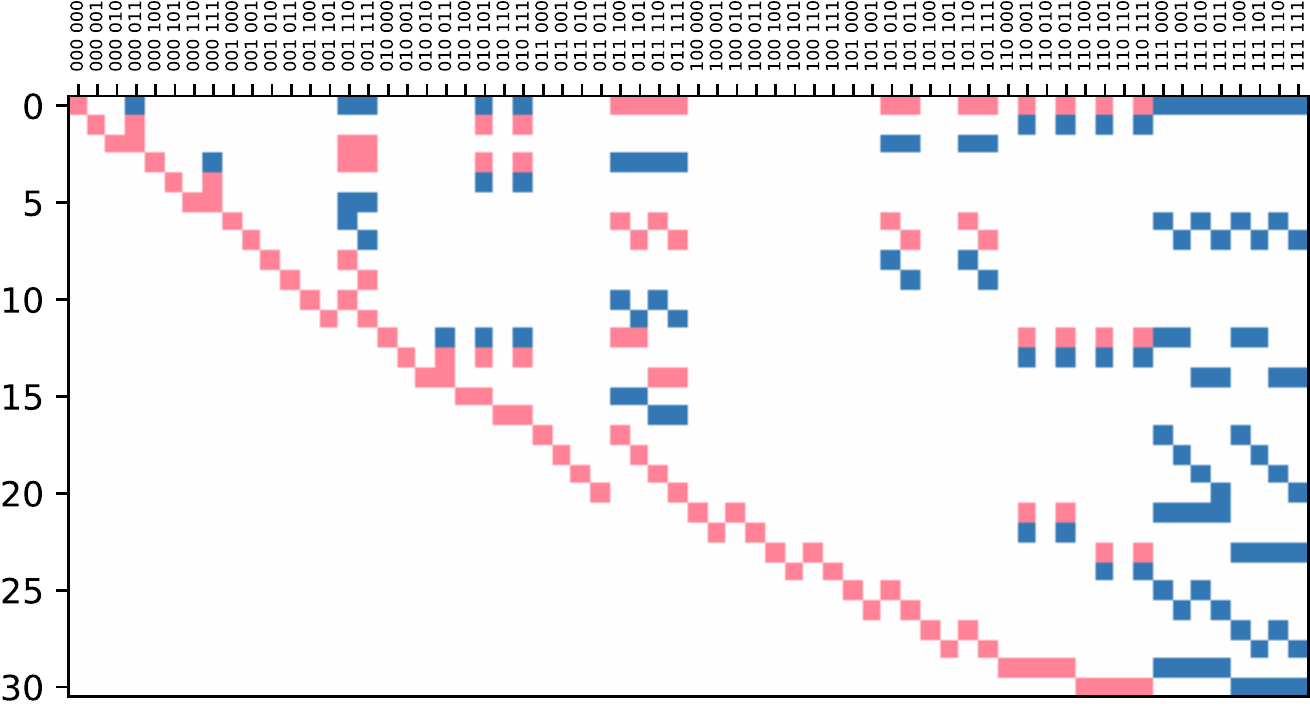}
\end{center}

\noindent Rows correspond to the 31 independent linear equations.
Columns in the plot correspond to entries of empirical models, indexed as $i_A i_B i_C$ $o_A o_B o_C$.
Coefficients in the equations are color-coded as white=0, red=+1 and blue=-1.

Space 37 has closest refinements in equivalence class 23; 
it is the join of its (closest) refinements.
It has closest coarsenings in equivalence classes 53 and 60; 
it is the meet of its (closest) coarsenings.
It has 256 causal functions, 64 of which are not causal for any of its refinements.
It is not a tight space: for event \ev{B}, a causal function must yield identical output values on input histories \hist{A/0,B/0} and \hist{B/0,C/1}, and it must also yield identical output values on input histories \hist{A/0,B/1} and \hist{B/1,C/1}.

The standard causaltope for Space 37 has 2 more dimensions than those of its 4 subspaces in equivalence class 23.
The standard causaltope for Space 37 is the meet of the standard causaltopes for its closest coarsenings.
For completeness, below is a plot of the full homogeneous linear system of causality and quasi-normalisation equations for the standard causaltope:

\begin{center}
    \includegraphics[width=12cm]{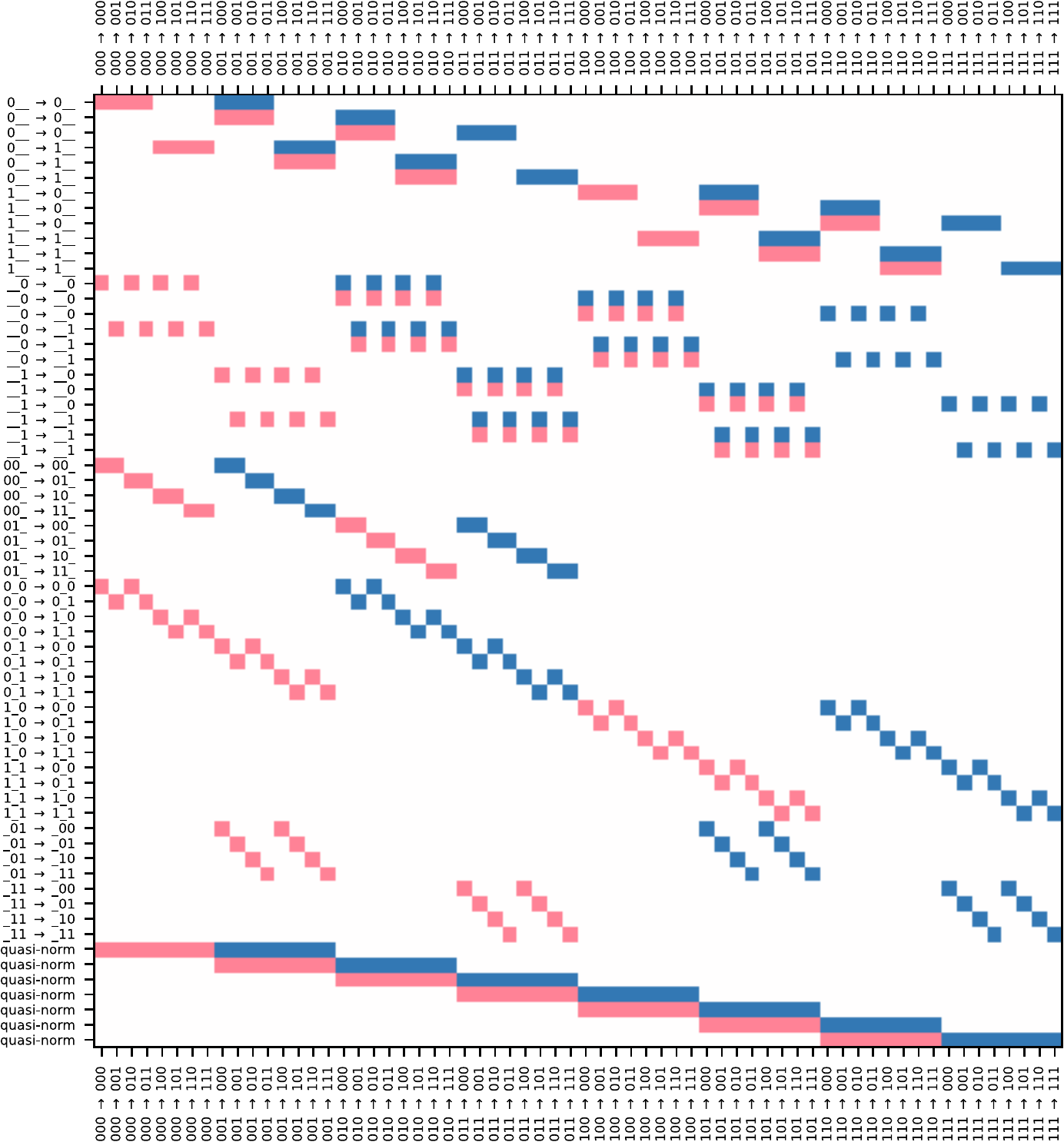}
\end{center}

\noindent Rows correspond to the 63 linear equations, of which 31 are independent.

\newpage
\subsection*{Space 38}

Space 38 is not induced by a causal order, but it is a refinement of the space 77 induced by the definite causal order $\total{\ev{A},\ev{B}}\vee\total{\ev{A},\ev{C}}$.
Its equivalence class under event-input permutation symmetry contains 24 spaces.
Space 38 differs as follows from the space induced by causal order $\total{\ev{A},\ev{B}}\vee\total{\ev{A},\ev{C}}$:
\begin{itemize}
  \item The outputs at events \evset{\ev{B}, \ev{C}} are independent of the input at event \ev{A} when the inputs at events \evset{B, C} are given by \hist{B/1,C/0}.
  \item The output at event \ev{C} is independent of the input at event \ev{A} when the input at event C is given by \hist{C/0}.
\end{itemize}

\noindent Below are the histories and extended histories for space 38: 
\begin{center}
    \begin{tabular}{cc}
    \includegraphics[height=3.5cm]{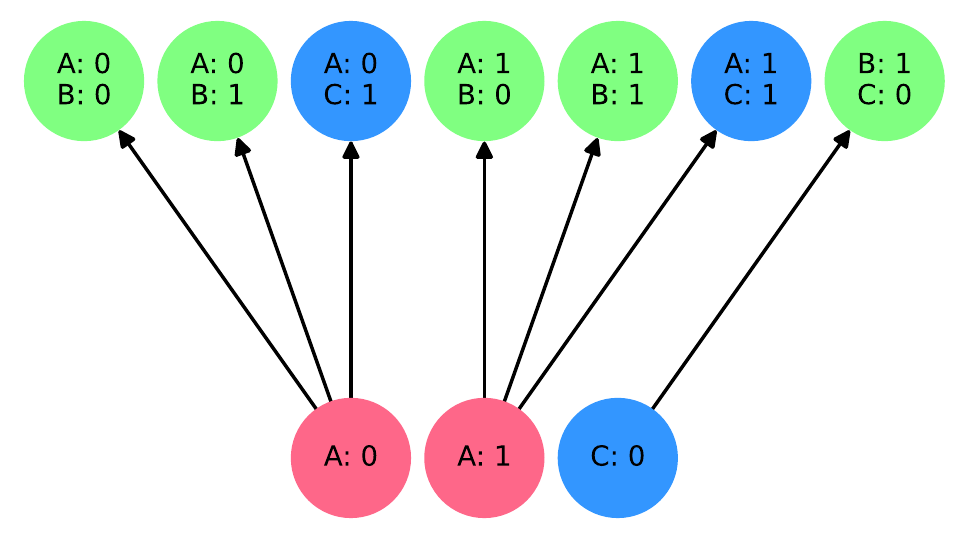}
    &
    \includegraphics[height=3.5cm]{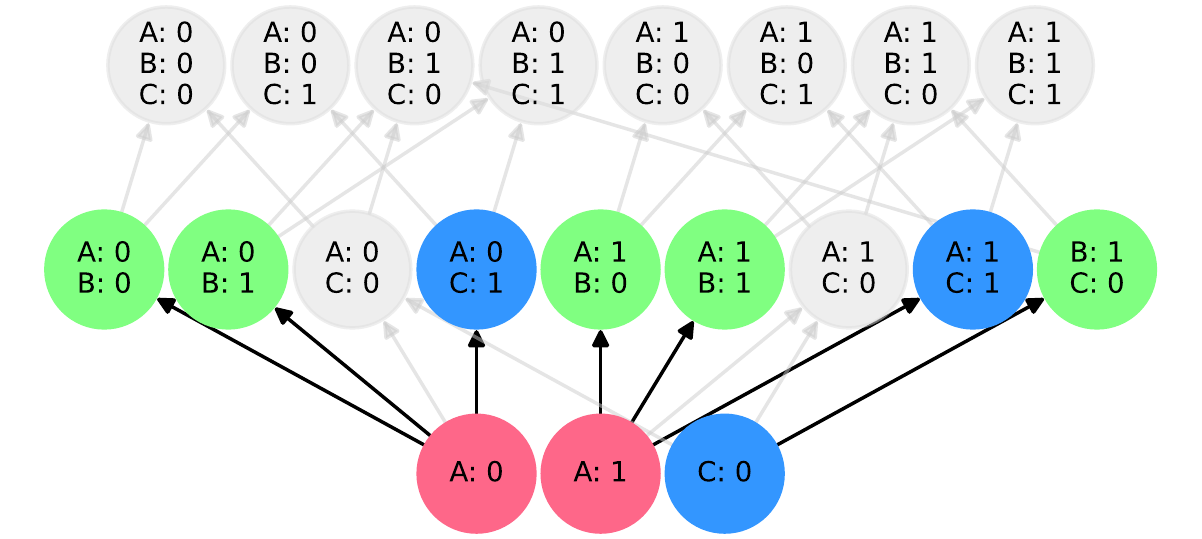}
    \\
    $\Theta_{38}$
    &
    $\Ext{\Theta_{38}}$
    \end{tabular}
\end{center}

\noindent The standard causaltope for Space 38 has dimension 31.
Below is a plot of the homogeneous linear system of causality and quasi-normalisation equations for the standard causaltope, put in reduced row echelon form:

\begin{center}
    \includegraphics[width=11cm]{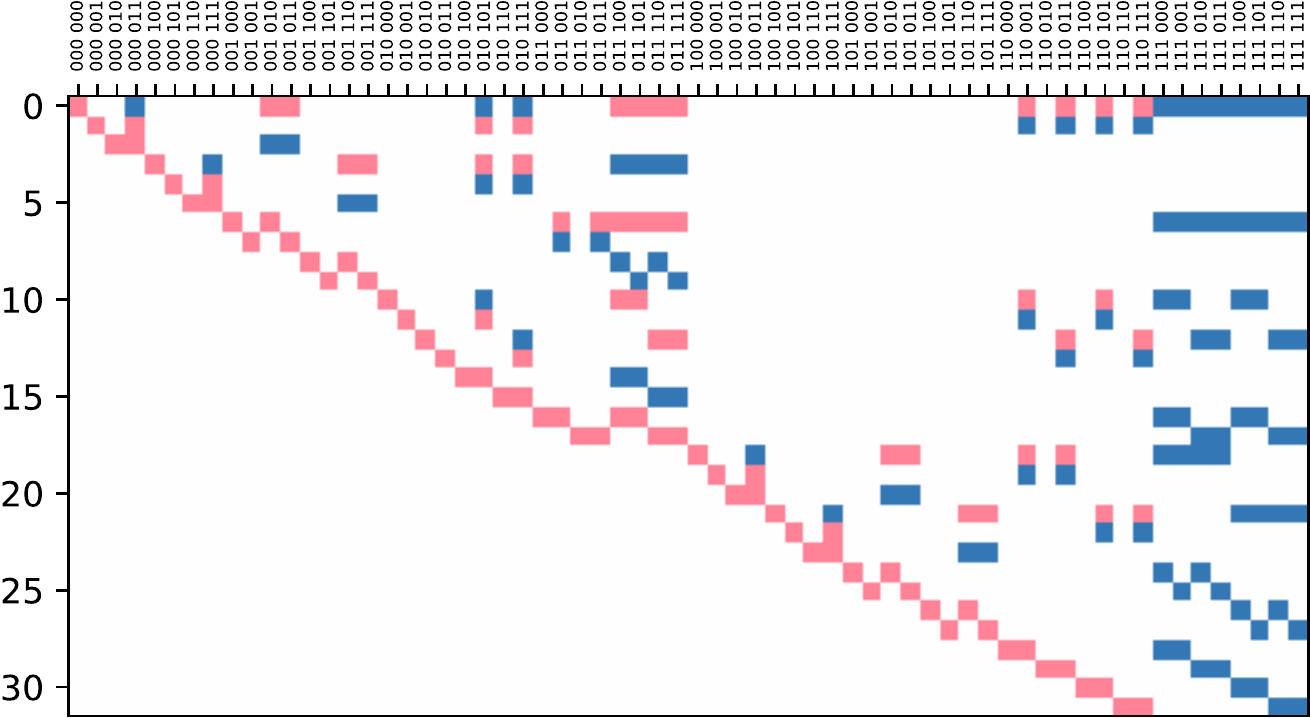}
\end{center}

\noindent Rows correspond to the 32 independent linear equations.
Columns in the plot correspond to entries of empirical models, indexed as $i_A i_B i_C$ $o_A o_B o_C$.
Coefficients in the equations are color-coded as white=0, red=+1 and blue=-1.

Space 38 has closest refinements in equivalence classes 19, 25 and 27; 
it is the join of its (closest) refinements.
It has closest coarsenings in equivalence classes 47, 56, 57 and 58; 
it is the meet of its (closest) coarsenings.
It has 256 causal functions, all of which are causal for at least one of its refinements.
It is not a tight space: for event \ev{B}, a causal function must yield identical output values on input histories \hist{A/0,B/1}, \hist{A/1,B/1} and \hist{B/1,C/0}.

The standard causaltope for Space 38 coincides with that of its subspace in equivalence class 25.
The standard causaltope for Space 38 is the meet of the standard causaltopes for its closest coarsenings.
For completeness, below is a plot of the full homogeneous linear system of causality and quasi-normalisation equations for the standard causaltope:

\begin{center}
    \includegraphics[width=12cm]{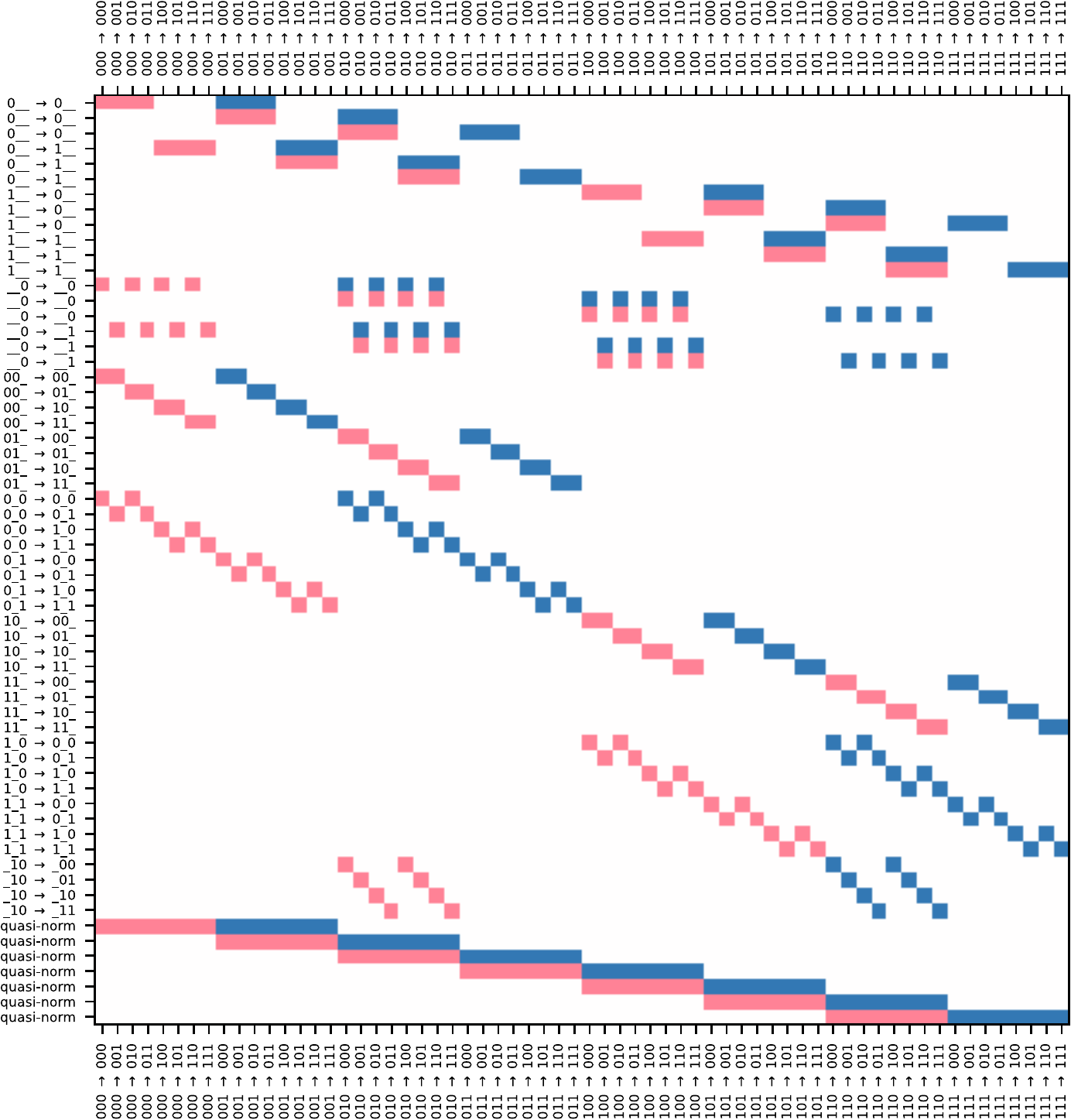}
\end{center}

\noindent Rows correspond to the 61 linear equations, of which 32 are independent.

\newpage
\subsection*{Space 39}

Space 39 is not induced by a causal order, but it is a refinement of the space 92 induced by the definite causal order $\total{\ev{A},\ev{C}}\vee\total{\ev{B},\ev{C}}$.
Its equivalence class under event-input permutation symmetry contains 24 spaces.
Space 39 differs as follows from the space induced by causal order $\total{\ev{A},\ev{C}}\vee\total{\ev{B},\ev{C}}$:
\begin{itemize}
  \item The outputs at events \evset{\ev{A}, \ev{C}} are independent of the input at event \ev{B} when the inputs at events \evset{A, C} are given by \hist{A/0,C/1} and \hist{A/1,C/0}.
  \item The outputs at events \evset{\ev{B}, \ev{C}} are independent of the input at event \ev{A} when the inputs at events \evset{B, C} are given by \hist{B/1,C/1} and \hist{B/0,C/1}.
\end{itemize}

\noindent Below are the histories and extended histories for space 39: 
\begin{center}
    \begin{tabular}{cc}
    \includegraphics[height=3.5cm]{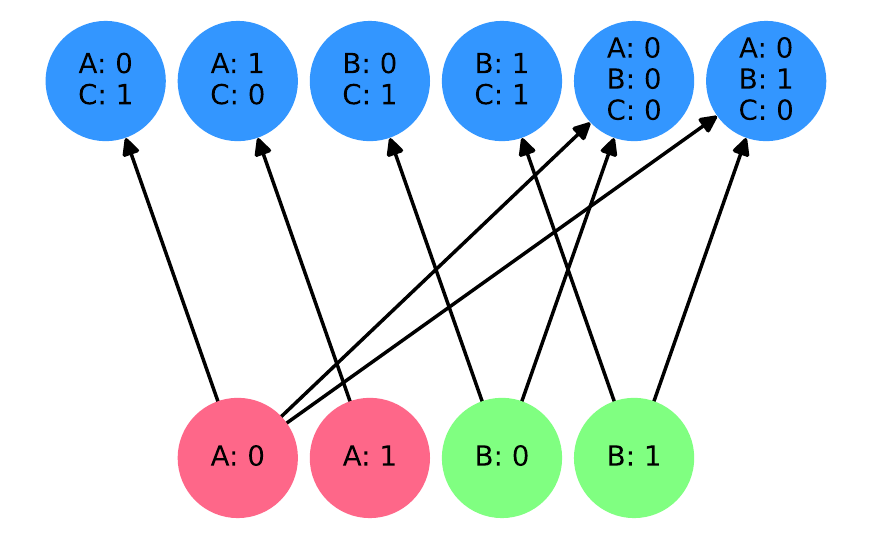}
    &
    \includegraphics[height=3.5cm]{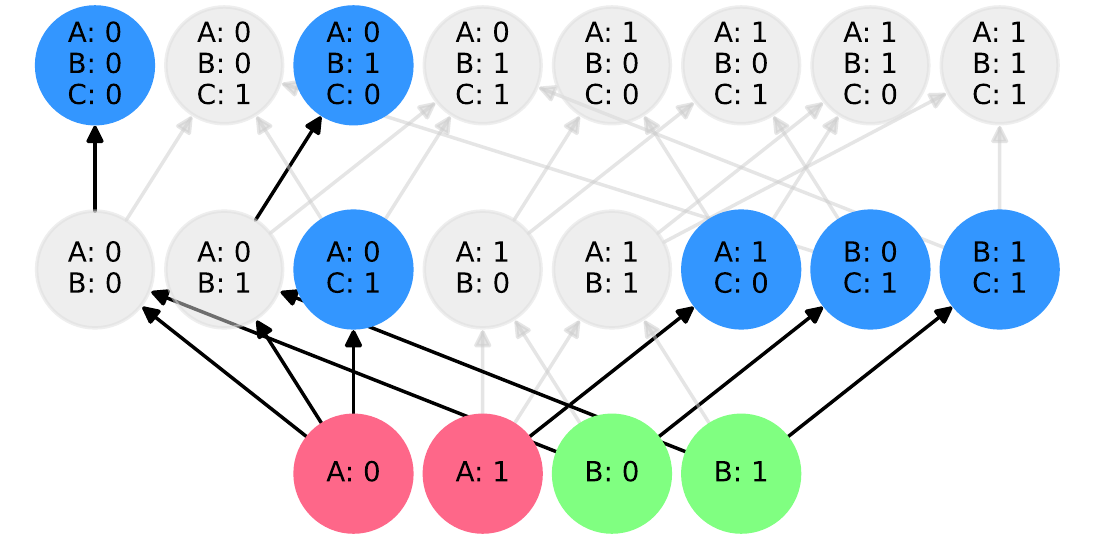}
    \\
    $\Theta_{39}$
    &
    $\Ext{\Theta_{39}}$
    \end{tabular}
\end{center}

\noindent The standard causaltope for Space 39 has dimension 32.
Below is a plot of the homogeneous linear system of causality and quasi-normalisation equations for the standard causaltope, put in reduced row echelon form:

\begin{center}
    \includegraphics[width=11cm]{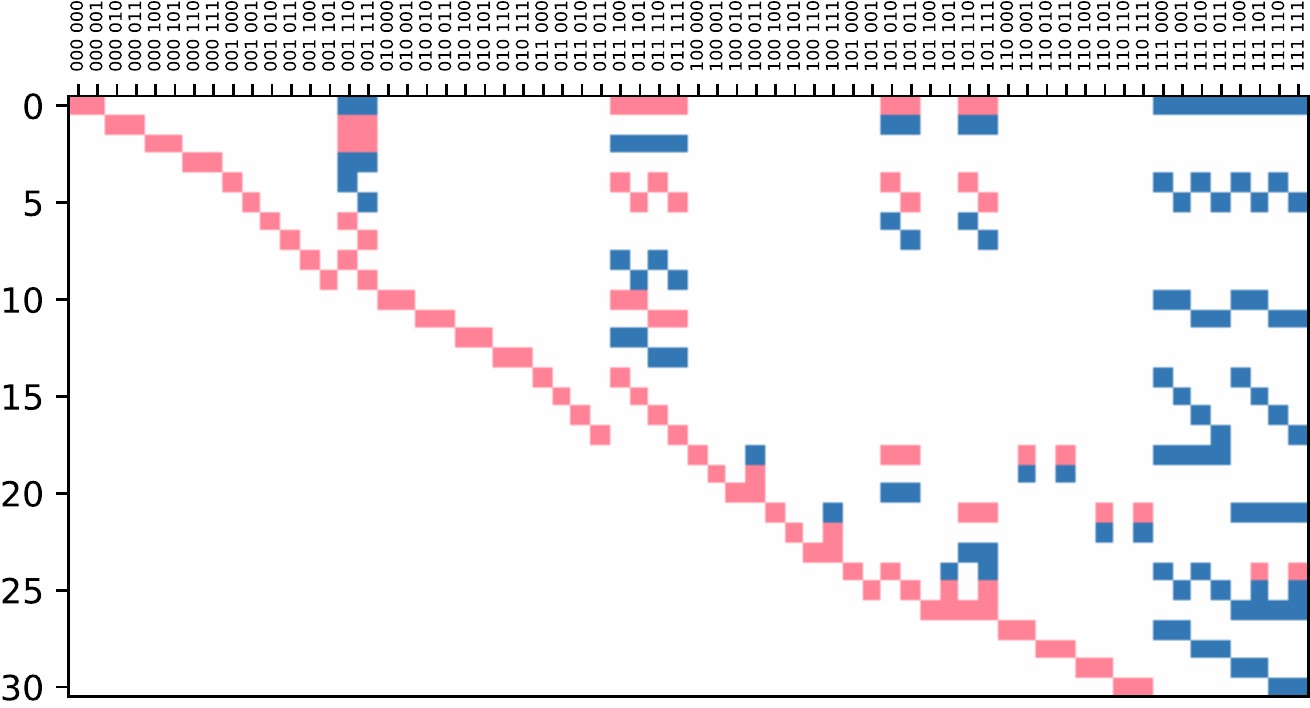}
\end{center}

\noindent Rows correspond to the 31 independent linear equations.
Columns in the plot correspond to entries of empirical models, indexed as $i_A i_B i_C$ $o_A o_B o_C$.
Coefficients in the equations are color-coded as white=0, red=+1 and blue=-1.

Space 39 has closest refinements in equivalence classes 16, 24 and 26; 
it is the join of its (closest) refinements.
It has closest coarsenings in equivalence classes 49, 50 and 59; 
it is the meet of its (closest) coarsenings.
It has 256 causal functions, all of which are causal for at least one of its refinements.
It is not a tight space: for event \ev{C}, a causal function must yield identical output values on input histories \hist{A/0,C/1}, \hist{B/0,C/1} and \hist{B/1,C/1}.

The standard causaltope for Space 39 has 1 more dimension than that of its subspace in equivalence class 26.
The standard causaltope for Space 39 is the meet of the standard causaltopes for its closest coarsenings.
For completeness, below is a plot of the full homogeneous linear system of causality and quasi-normalisation equations for the standard causaltope:

\begin{center}
    \includegraphics[width=12cm]{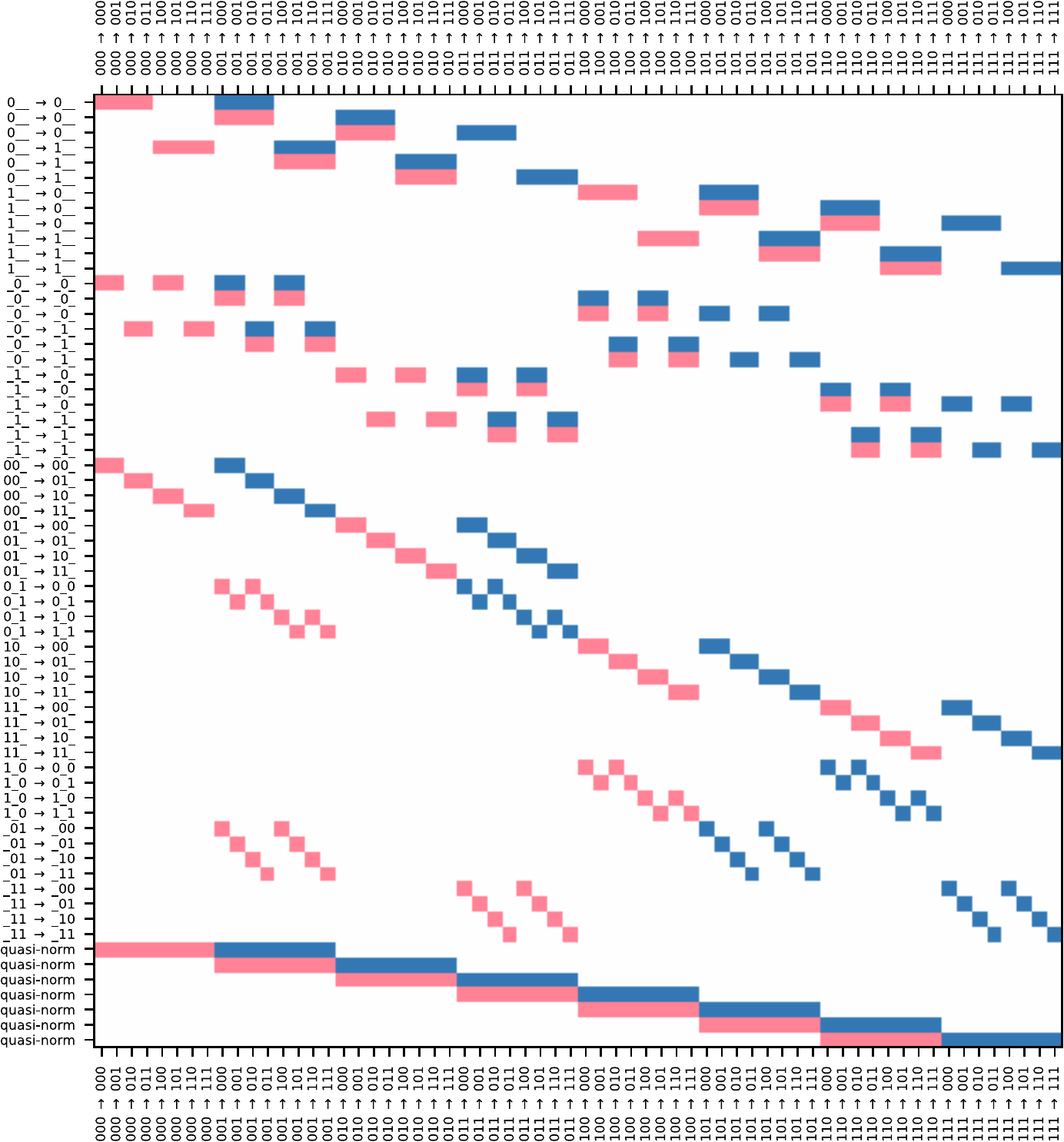}
\end{center}

\noindent Rows correspond to the 63 linear equations, of which 31 are independent.

\newpage
\subsection*{Space 40}

Space 40 is not induced by a causal order, but it is a refinement of the space 92 induced by the definite causal order $\total{\ev{A},\ev{C}}\vee\total{\ev{B},\ev{C}}$.
Its equivalence class under event-input permutation symmetry contains 12 spaces.
Space 40 differs as follows from the space induced by causal order $\total{\ev{A},\ev{C}}\vee\total{\ev{B},\ev{C}}$:
\begin{itemize}
  \item The outputs at events \evset{\ev{A}, \ev{C}} are independent of the input at event \ev{B} when the inputs at events \evset{A, C} are given by \hist{A/0,C/1} and \hist{A/1,C/0}.
  \item The outputs at events \evset{\ev{B}, \ev{C}} are independent of the input at event \ev{A} when the inputs at events \evset{B, C} are given by \hist{B/1,C/0} and \hist{B/0,C/1}.
\end{itemize}

\noindent Below are the histories and extended histories for space 40: 
\begin{center}
    \begin{tabular}{cc}
    \includegraphics[height=3.5cm]{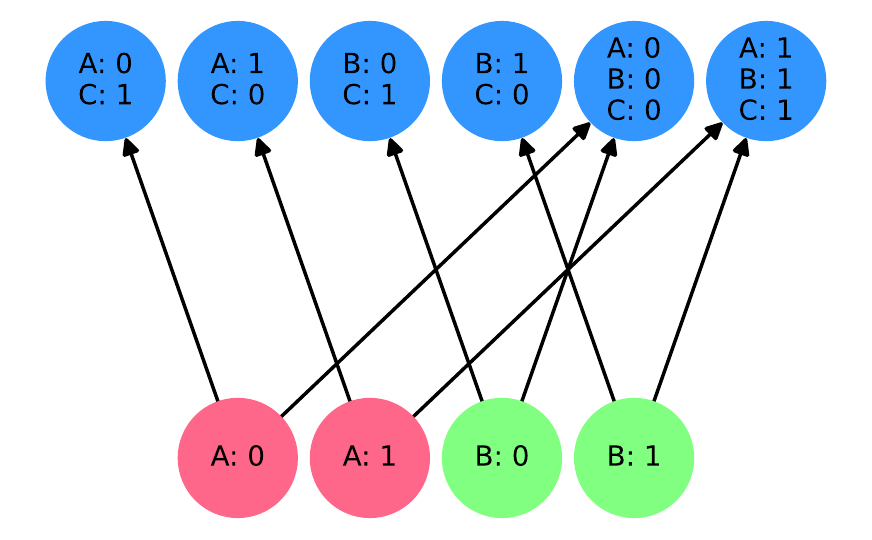}
    &
    \includegraphics[height=3.5cm]{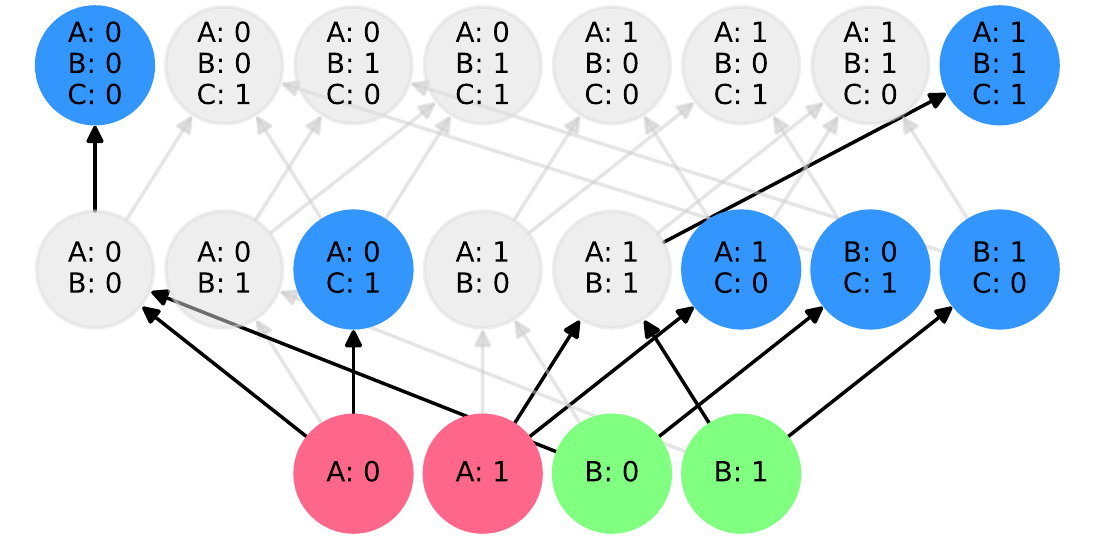}
    \\
    $\Theta_{40}$
    &
    $\Ext{\Theta_{40}}$
    \end{tabular}
\end{center}

\noindent The standard causaltope for Space 40 has dimension 32.
Below is a plot of the homogeneous linear system of causality and quasi-normalisation equations for the standard causaltope, put in reduced row echelon form:

\begin{center}
    \includegraphics[width=11cm]{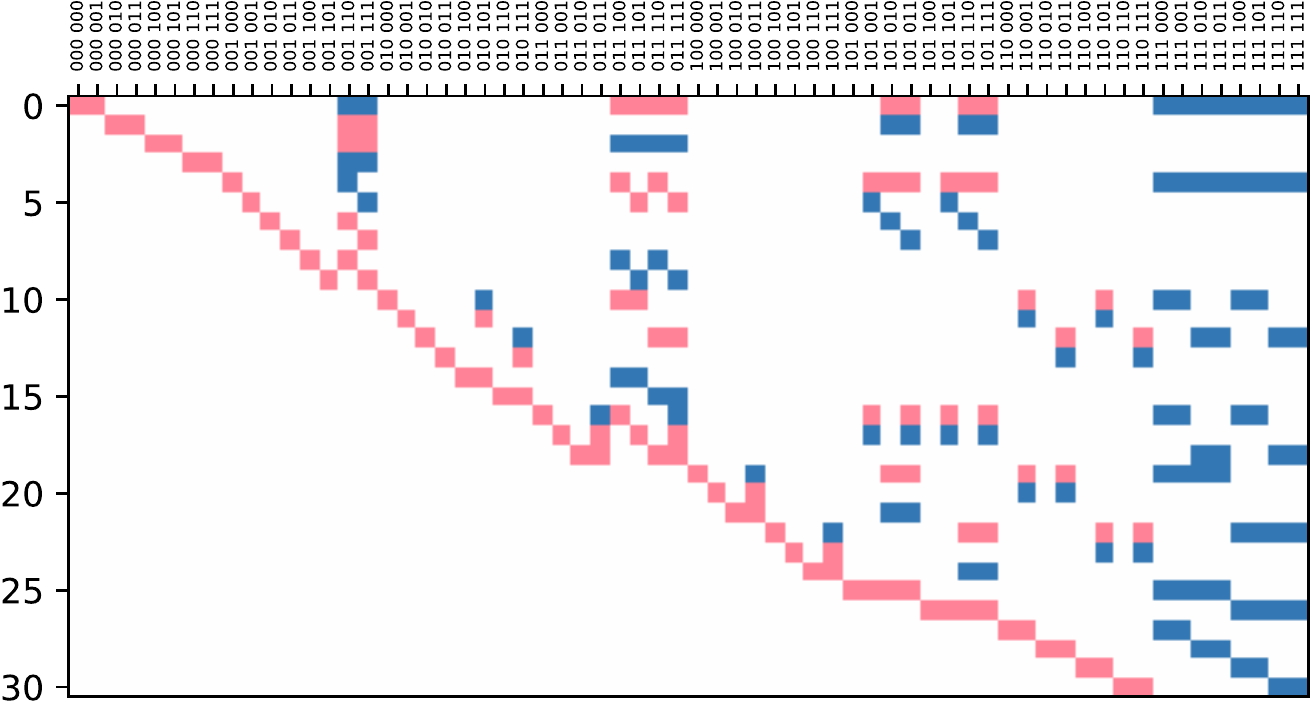}
\end{center}

\noindent Rows correspond to the 31 independent linear equations.
Columns in the plot correspond to entries of empirical models, indexed as $i_A i_B i_C$ $o_A o_B o_C$.
Coefficients in the equations are color-coded as white=0, red=+1 and blue=-1.

Space 40 has closest refinements in equivalence class 24; 
it is the join of its (closest) refinements.
It has closest coarsenings in equivalence class 50; 
it is the meet of its (closest) coarsenings.
It has 256 causal functions, 64 of which are not causal for any of its refinements.
It is not a tight space: for event \ev{C}, a causal function must yield identical output values on input histories \hist{A/0,C/1} and \hist{B/0,C/1}, and it must also yield identical output values on input histories \hist{A/1,C/0} and \hist{B/1,C/0}.

The standard causaltope for Space 40 has 2 more dimensions than those of its 4 subspaces in equivalence class 24.
The standard causaltope for Space 40 is the meet of the standard causaltopes for its closest coarsenings.
For completeness, below is a plot of the full homogeneous linear system of causality and quasi-normalisation equations for the standard causaltope:

\begin{center}
    \includegraphics[width=12cm]{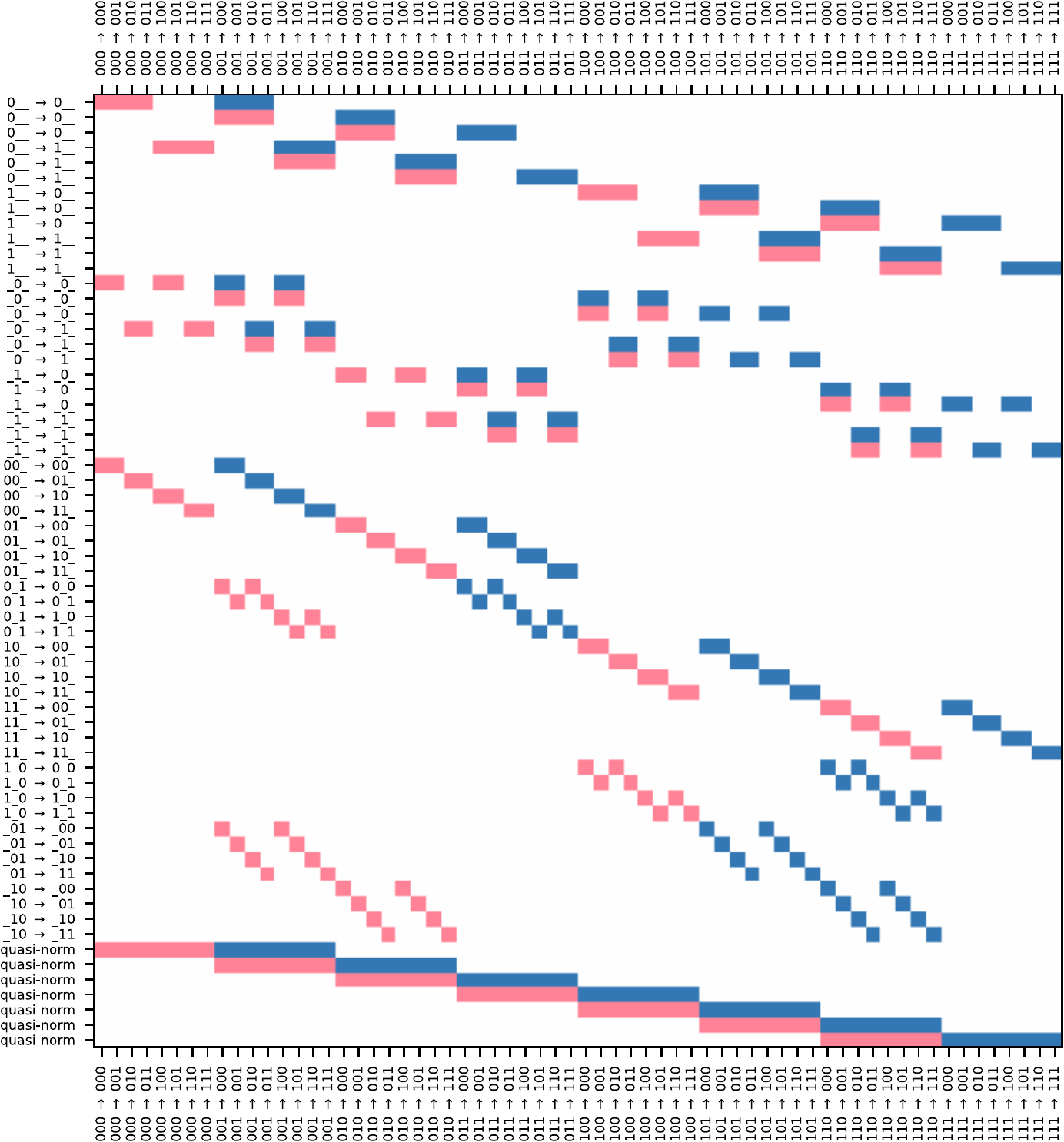}
\end{center}

\noindent Rows correspond to the 63 linear equations, of which 31 are independent.

\newpage
\subsection*{Space 41}

Space 41 is not induced by a causal order, but it is a refinement of the space 92 induced by the definite causal order $\total{\ev{A},\ev{C}}\vee\total{\ev{B},\ev{C}}$.
Its equivalence class under event-input permutation symmetry contains 6 spaces.
Space 41 differs as follows from the space induced by causal order $\total{\ev{A},\ev{C}}\vee\total{\ev{B},\ev{C}}$:
\begin{itemize}
  \item The outputs at events \evset{\ev{A}, \ev{C}} are independent of the input at event \ev{B} when the inputs at events \evset{A, C} are given by \hist{A/0,C/1} and \hist{A/1,C/1}.
  \item The outputs at events \evset{\ev{B}, \ev{C}} are independent of the input at event \ev{A} when the inputs at events \evset{B, C} are given by \hist{B/1,C/0} and \hist{B/0,C/0}.
\end{itemize}

\noindent Below are the histories and extended histories for space 41: 
\begin{center}
    \begin{tabular}{cc}
    \includegraphics[height=3.5cm]{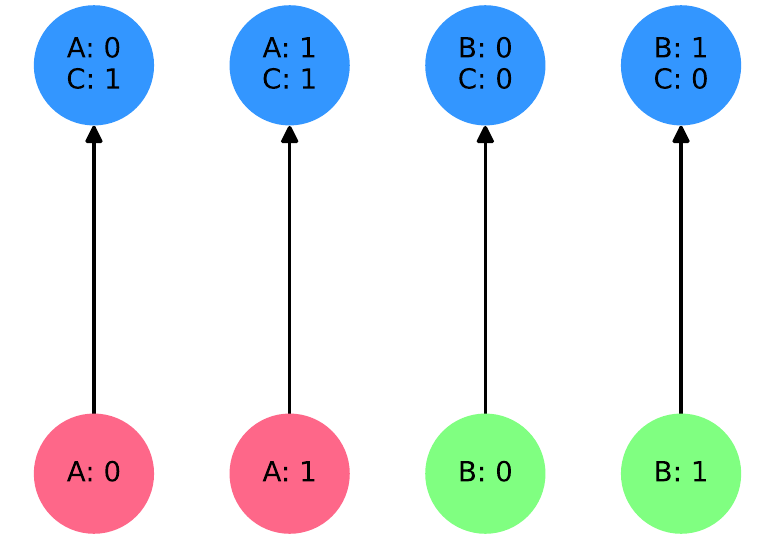}
    &
    \includegraphics[height=3.5cm]{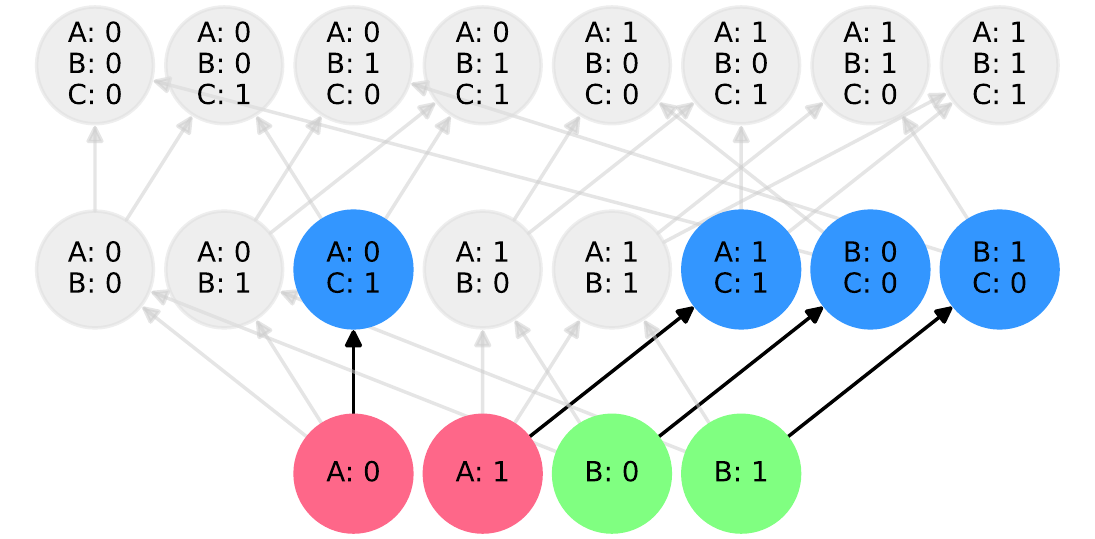}
    \\
    $\Theta_{41}$
    &
    $\Ext{\Theta_{41}}$
    \end{tabular}
\end{center}

\noindent The standard causaltope for Space 41 has dimension 32.
Below is a plot of the homogeneous linear system of causality and quasi-normalisation equations for the standard causaltope, put in reduced row echelon form:

\begin{center}
    \includegraphics[width=11cm]{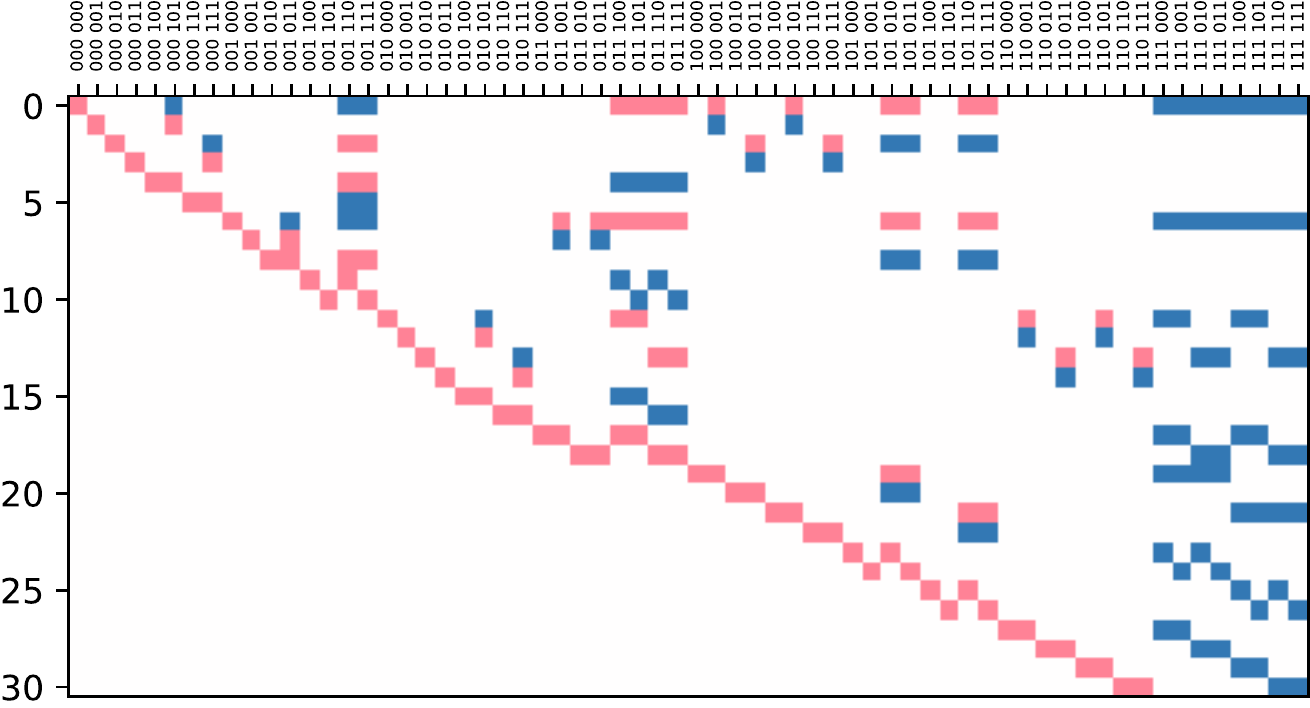}
\end{center}

\noindent Rows correspond to the 31 independent linear equations.
Columns in the plot correspond to entries of empirical models, indexed as $i_A i_B i_C$ $o_A o_B o_C$.
Coefficients in the equations are color-coded as white=0, red=+1 and blue=-1.

Space 41 has closest refinements in equivalence class 16; 
it is the join of its (closest) refinements.
It has closest coarsenings in equivalence class 49; 
it is the meet of its (closest) coarsenings.
It has 256 causal functions, 64 of which are not causal for any of its refinements.
It is a tight space.

The standard causaltope for Space 41 has 2 more dimensions than those of its 4 subspaces in equivalence class 16.
The standard causaltope for Space 41 is the meet of the standard causaltopes for its closest coarsenings.
For completeness, below is a plot of the full homogeneous linear system of causality and quasi-normalisation equations for the standard causaltope:

\begin{center}
    \includegraphics[width=12cm]{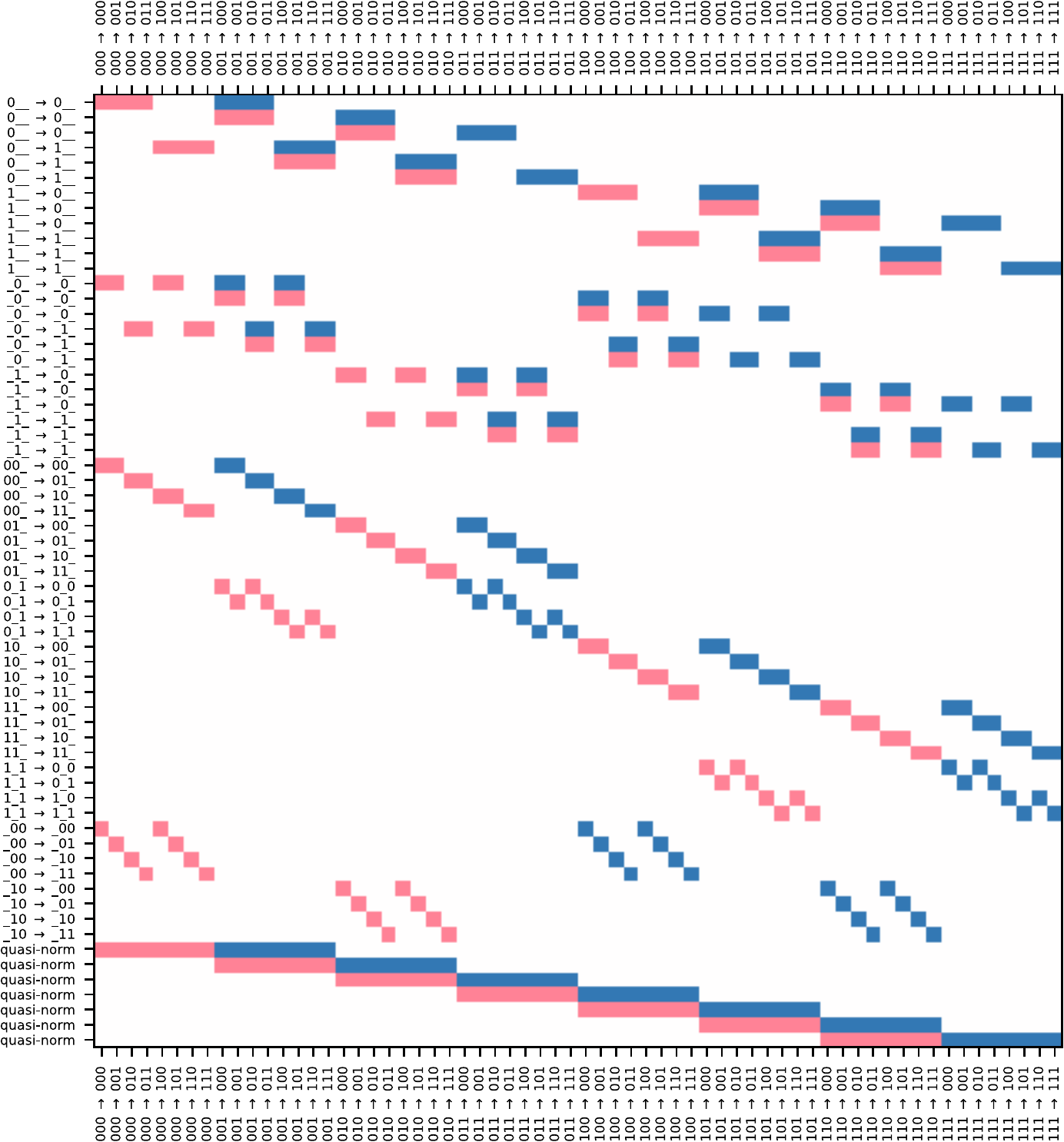}
\end{center}

\noindent Rows correspond to the 63 linear equations, of which 31 are independent.

\newpage
\subsection*{Space 42}

Space 42 is not induced by a causal order, but it is a refinement of the space 100 induced by the definite causal order $\total{\ev{A},\ev{B},\ev{C}}$.
Its equivalence class under event-input permutation symmetry contains 24 spaces.
Space 42 differs as follows from the space induced by causal order $\total{\ev{A},\ev{B},\ev{C}}$:
\begin{itemize}
  \item The outputs at events \evset{\ev{A}, \ev{C}} are independent of the input at event \ev{B} when the inputs at events \evset{A, C} are given by \hist{A/0,C/1}, \hist{A/0,C/0} and \hist{A/1,C/0}.
  \item The outputs at events \evset{\ev{B}, \ev{C}} are independent of the input at event \ev{A} when the inputs at events \evset{B, C} are given by \hist{B/1,C/0} and \hist{B/0,C/0}.
  \item The output at event \ev{C} is independent of the inputs at events \evset{\ev{A}, \ev{B}} when the input at event C is given by \hist{C/0}.
\end{itemize}

\noindent Below are the histories and extended histories for space 42: 
\begin{center}
    \begin{tabular}{cc}
    \includegraphics[height=3.5cm]{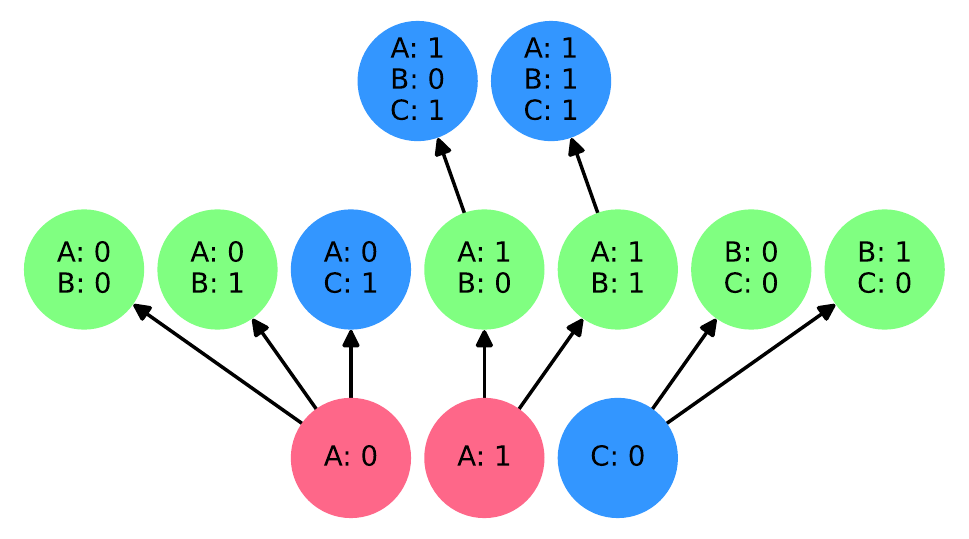}
    &
    \includegraphics[height=3.5cm]{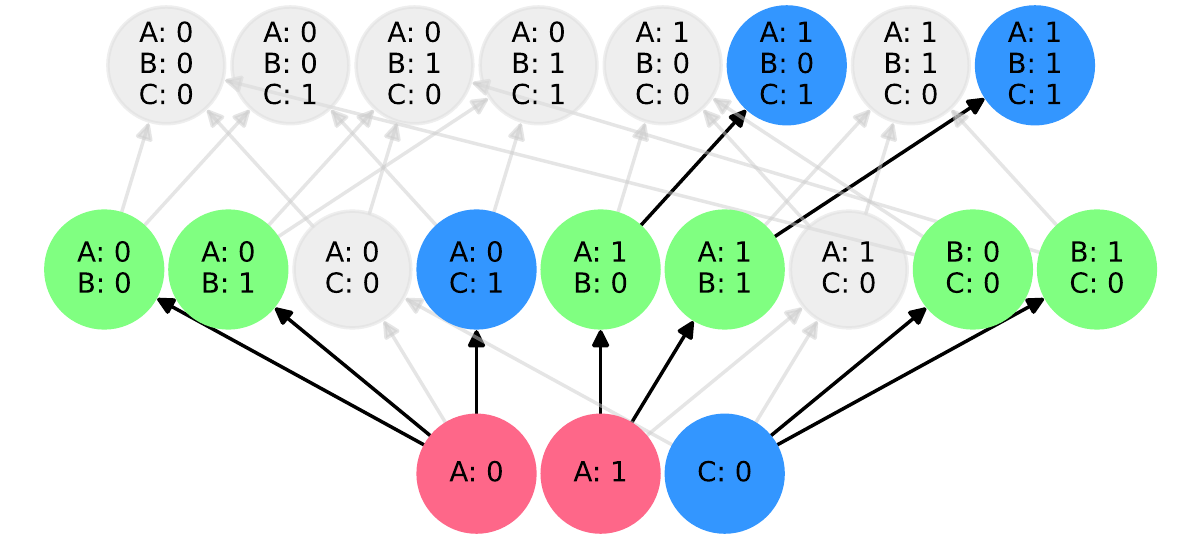}
    \\
    $\Theta_{42}$
    &
    $\Ext{\Theta_{42}}$
    \end{tabular}
\end{center}

\noindent The standard causaltope for Space 42 has dimension 31.
Below is a plot of the homogeneous linear system of causality and quasi-normalisation equations for the standard causaltope, put in reduced row echelon form:

\begin{center}
    \includegraphics[width=11cm]{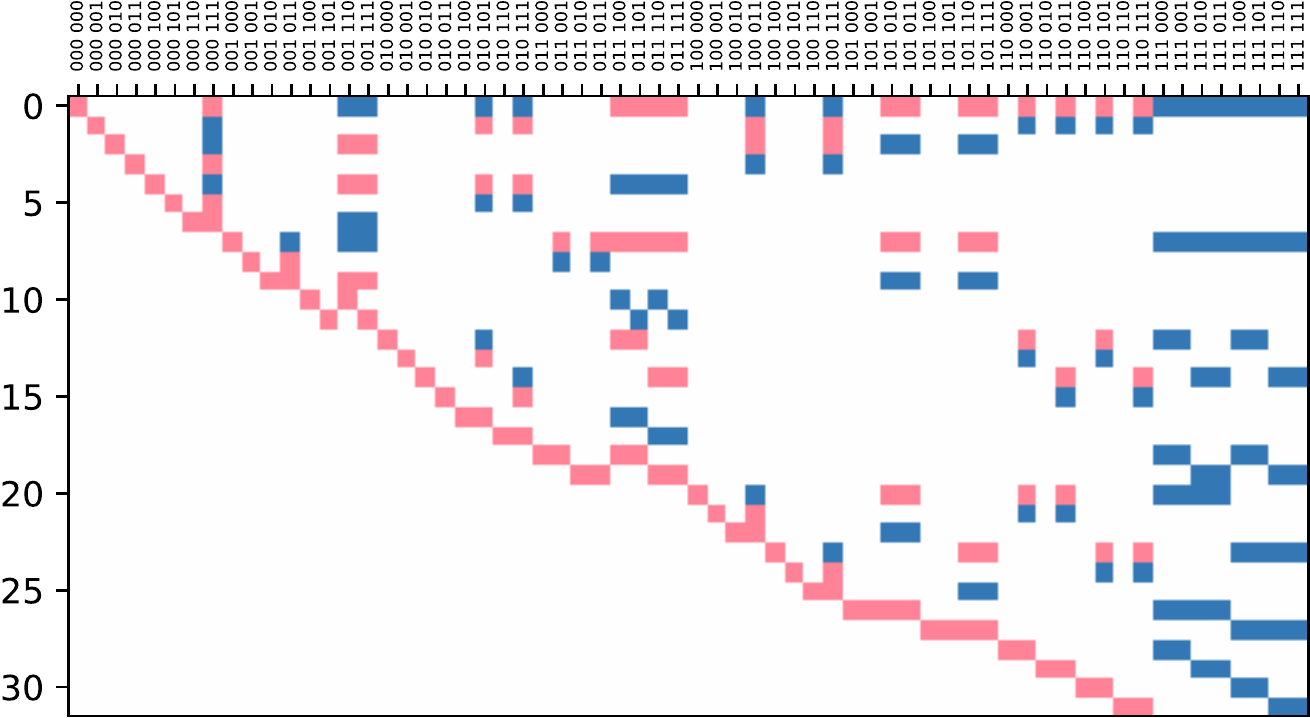}
\end{center}

\noindent Rows correspond to the 32 independent linear equations.
Columns in the plot correspond to entries of empirical models, indexed as $i_A i_B i_C$ $o_A o_B o_C$.
Coefficients in the equations are color-coded as white=0, red=+1 and blue=-1.

Space 42 has closest refinements in equivalence classes 22 and 27; 
it is the join of its (closest) refinements.
It has closest coarsenings in equivalence classes 47, 51 and 52; 
it is the meet of its (closest) coarsenings.
It has 256 causal functions, all of which are causal for at least one of its refinements.
It is not a tight space: for event \ev{B}, a causal function must yield identical output values on input histories \hist{A/0,B/0}, \hist{A/1,B/0} and \hist{B/0,C/0}, and it must also yield identical output values on input histories \hist{A/0,B/1}, \hist{A/1,B/1} and \hist{B/1,C/0}.

The standard causaltope for Space 42 coincides with that of its 2 subspaces in equivalence class 22.
The standard causaltope for Space 42 is the meet of the standard causaltopes for its closest coarsenings.
For completeness, below is a plot of the full homogeneous linear system of causality and quasi-normalisation equations for the standard causaltope:

\begin{center}
    \includegraphics[width=12cm]{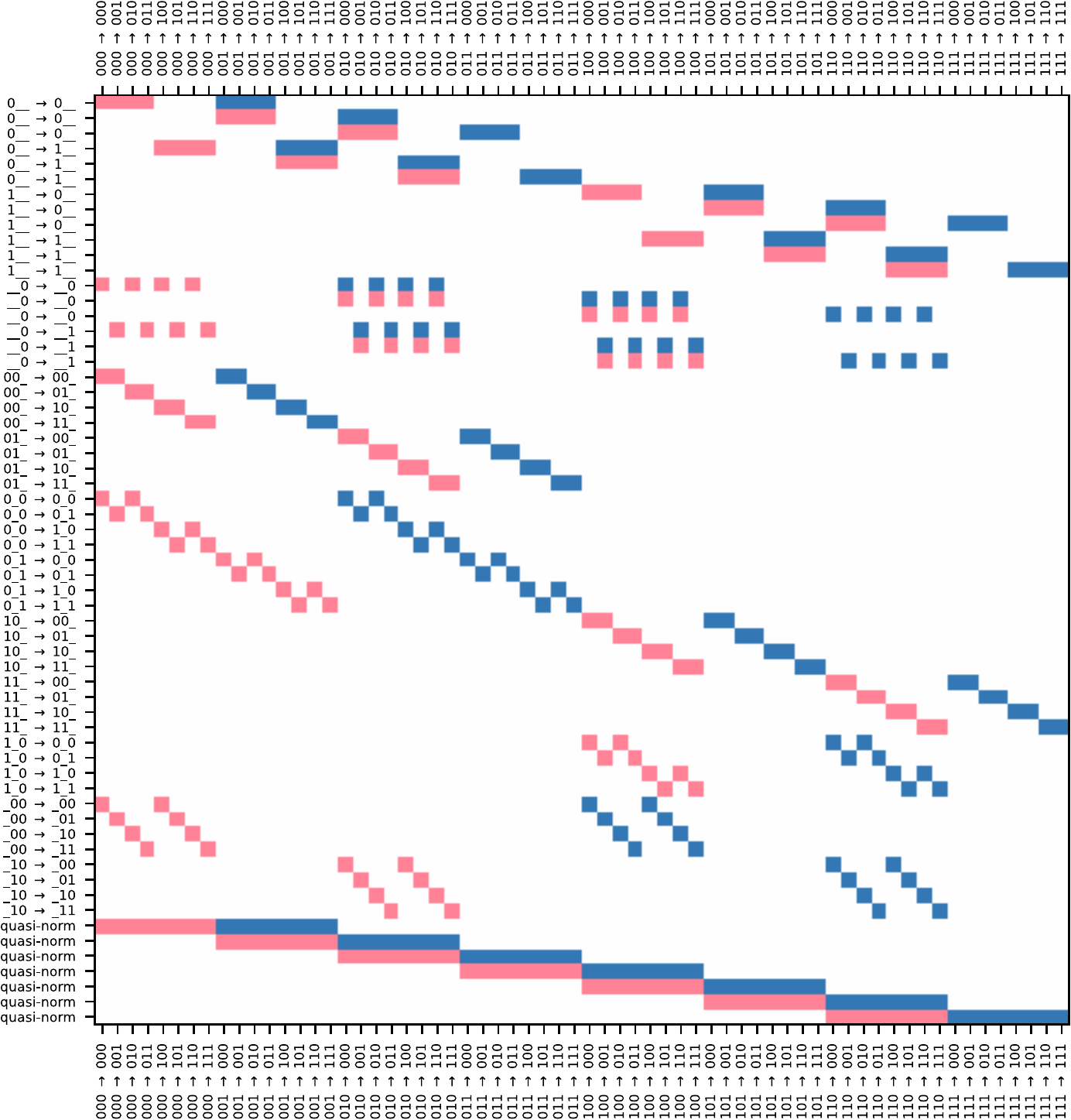}
\end{center}

\noindent Rows correspond to the 61 linear equations, of which 32 are independent.

\newpage
\subsection*{Space 43}

Space 43 is not induced by a causal order, but it is a refinement of the space in equivalence class 100 induced by the definite causal order $\total{\ev{A},\ev{C},\ev{B}}$ (note that the space induced by the order is not the same as space 100).
Its equivalence class under event-input permutation symmetry contains 48 spaces.
Space 43 differs as follows from the space induced by causal order $\total{\ev{A},\ev{C},\ev{B}}$:
\begin{itemize}
  \item The outputs at events \evset{\ev{A}, \ev{B}} are independent of the input at event \ev{C} when the inputs at events \evset{A, B} are given by \hist{A/0,B/0}, \hist{A/0,B/1} and \hist{A/1,B/0}.
  \item The outputs at events \evset{\ev{B}, \ev{C}} are independent of the input at event \ev{A} when the inputs at events \evset{B, C} are given by \hist{B/1,C/0} and \hist{B/0,C/0}.
  \item The output at event \ev{C} is independent of the input at event \ev{A} when the input at event C is given by \hist{C/0}.
\end{itemize}

\noindent Below are the histories and extended histories for space 43: 
\begin{center}
    \begin{tabular}{cc}
    \includegraphics[height=3.5cm]{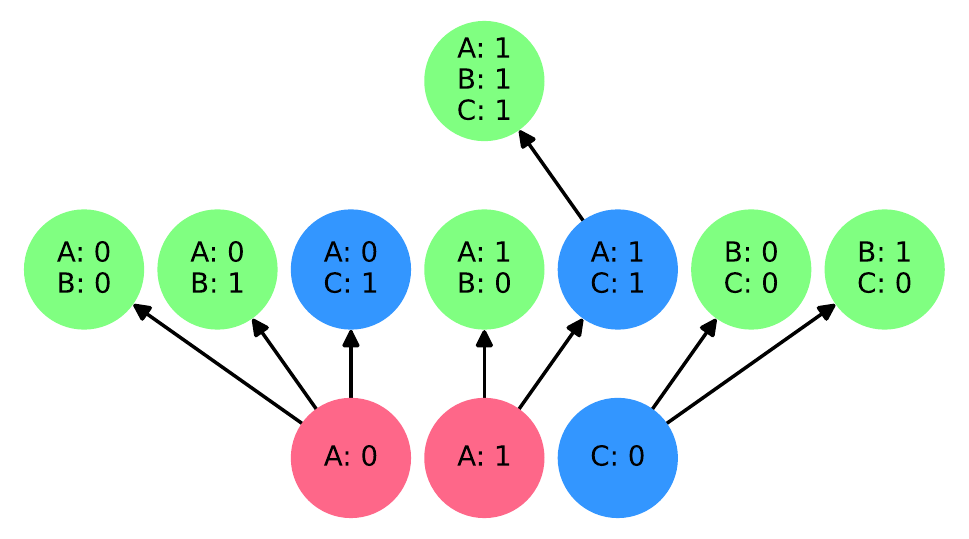}
    &
    \includegraphics[height=3.5cm]{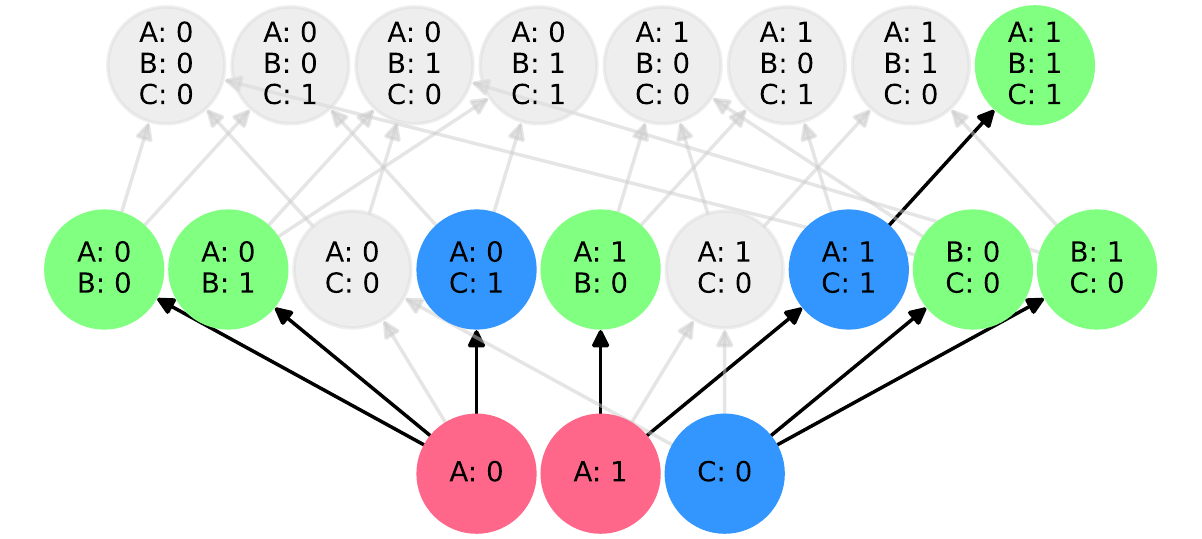}
    \\
    $\Theta_{43}$
    &
    $\Ext{\Theta_{43}}$
    \end{tabular}
\end{center}

\noindent The standard causaltope for Space 43 has dimension 31.
Below is a plot of the homogeneous linear system of causality and quasi-normalisation equations for the standard causaltope, put in reduced row echelon form:

\begin{center}
    \includegraphics[width=11cm]{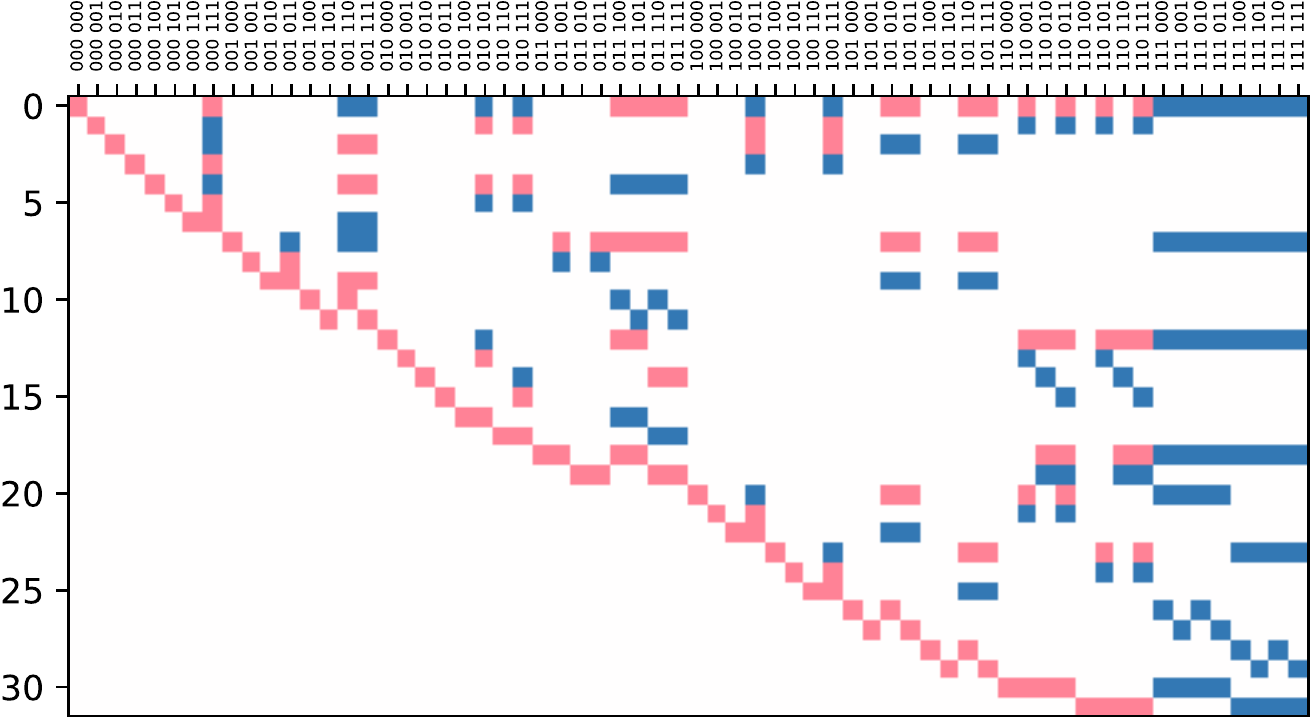}
\end{center}

\noindent Rows correspond to the 32 independent linear equations.
Columns in the plot correspond to entries of empirical models, indexed as $i_A i_B i_C$ $o_A o_B o_C$.
Coefficients in the equations are color-coded as white=0, red=+1 and blue=-1.

Space 43 has closest refinements in equivalence classes 17, 23 and 27; 
it is the join of its (closest) refinements.
It has closest coarsenings in equivalence classes 52, 54, 55, 56, 57 and 60; 
it is the meet of its (closest) coarsenings.
It has 256 causal functions, 64 of which are not causal for any of its refinements.
It is not a tight space: for event \ev{B}, a causal function must yield identical output values on input histories \hist{A/0,B/1} and \hist{B/1,C/0}, and it must also yield identical output values on input histories \hist{A/0,B/0}, \hist{A/1,B/0} and \hist{B/0,C/0}.

The standard causaltope for Space 43 coincides with that of its subspace in equivalence class 17.
The standard causaltope for Space 43 is the meet of the standard causaltopes for its closest coarsenings.
For completeness, below is a plot of the full homogeneous linear system of causality and quasi-normalisation equations for the standard causaltope:

\begin{center}
    \includegraphics[width=12cm]{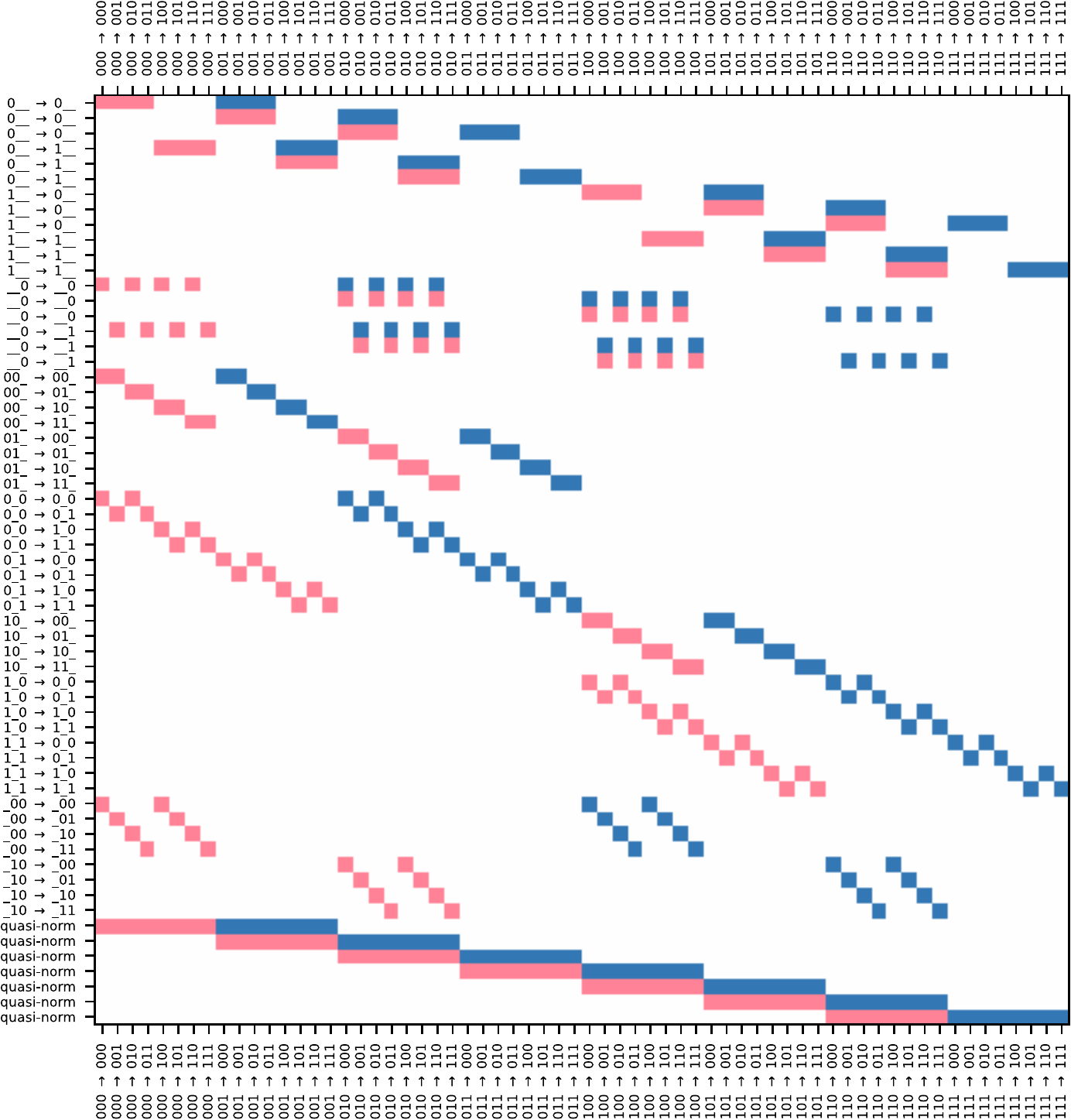}
\end{center}

\noindent Rows correspond to the 61 linear equations, of which 32 are independent.

\newpage
\subsection*{Space 44}

Space 44 is not induced by a causal order, but it is a refinement of the space 92 induced by the definite causal order $\total{\ev{A},\ev{C}}\vee\total{\ev{B},\ev{C}}$.
Its equivalence class under event-input permutation symmetry contains 6 spaces.
Space 44 differs as follows from the space induced by causal order $\total{\ev{A},\ev{C}}\vee\total{\ev{B},\ev{C}}$:
\begin{itemize}
  \item The outputs at events \evset{\ev{A}, \ev{C}} are independent of the input at event \ev{B} when the inputs at events \evset{A, C} are given by \hist{A/0,C/1} and \hist{A/1,C/1}.
  \item The outputs at events \evset{\ev{B}, \ev{C}} are independent of the input at event \ev{A} when the inputs at events \evset{B, C} are given by \hist{B/1,C/1} and \hist{B/0,C/1}.
\end{itemize}

\noindent Below are the histories and extended histories for space 44: 
\begin{center}
    \begin{tabular}{cc}
    \includegraphics[height=3.5cm]{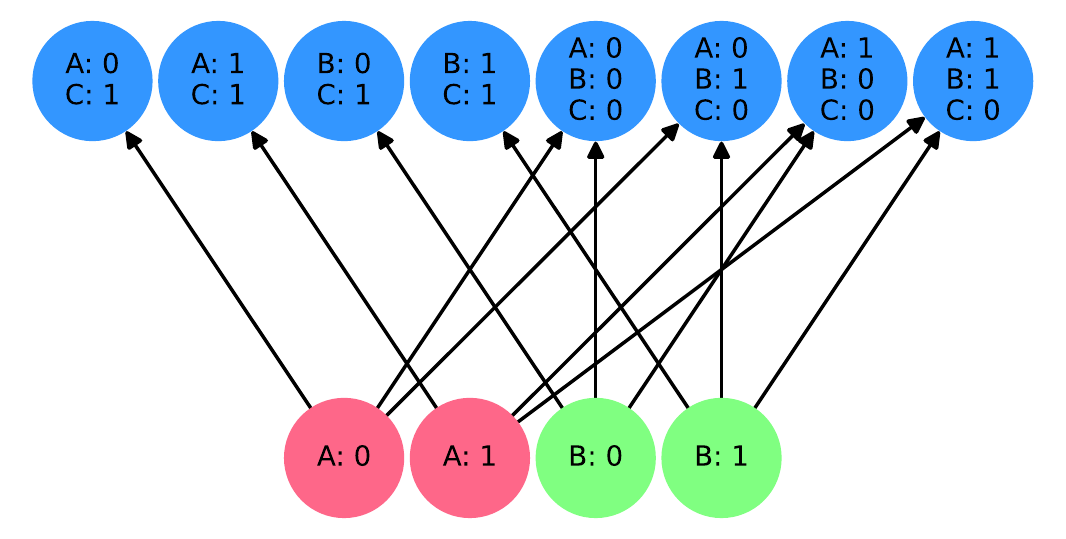}
    &
    \includegraphics[height=3.5cm]{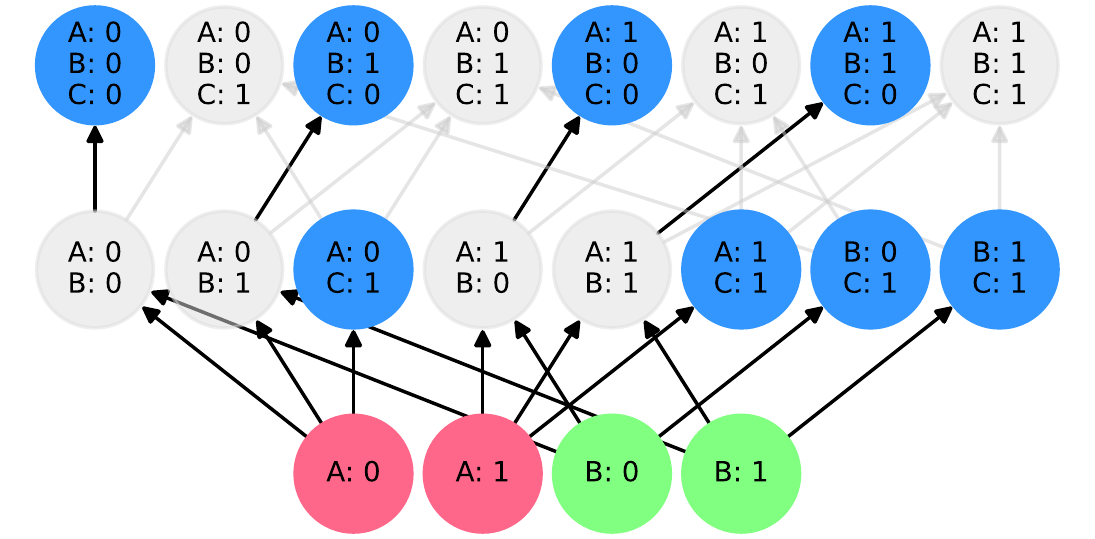}
    \\
    $\Theta_{44}$
    &
    $\Ext{\Theta_{44}}$
    \end{tabular}
\end{center}

\noindent The standard causaltope for Space 44 has dimension 33.
Below is a plot of the homogeneous linear system of causality and quasi-normalisation equations for the standard causaltope, put in reduced row echelon form:

\begin{center}
    \includegraphics[width=11cm]{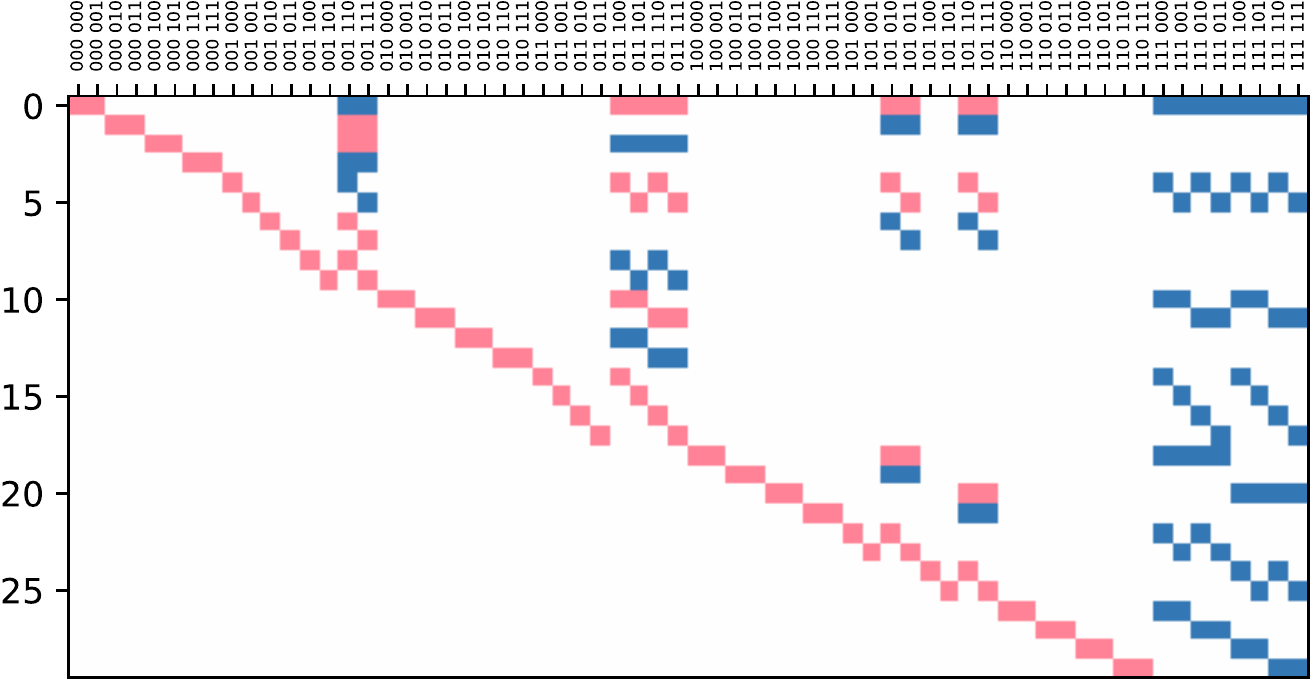}
\end{center}

\noindent Rows correspond to the 30 independent linear equations.
Columns in the plot correspond to entries of empirical models, indexed as $i_A i_B i_C$ $o_A o_B o_C$.
Coefficients in the equations are color-coded as white=0, red=+1 and blue=-1.

Space 44 has closest refinements in equivalence classes 18 and 26; 
it is the join of its (closest) refinements.
It has closest coarsenings in equivalence class 59; 
it is the meet of its (closest) coarsenings.
It has 512 causal functions, all of which are causal for at least one of its refinements.
It is not a tight space: for event \ev{C}, a causal function must yield identical output values on input histories \hist{A/0,C/1}, \hist{A/1,C/1}, \hist{B/0,C/1} and \hist{B/1,C/1}.

The standard causaltope for Space 44 coincides with that of its subspace in equivalence class 18.
The standard causaltope for Space 44 is the meet of the standard causaltopes for its closest coarsenings.
For completeness, below is a plot of the full homogeneous linear system of causality and quasi-normalisation equations for the standard causaltope:

\begin{center}
    \includegraphics[width=12cm]{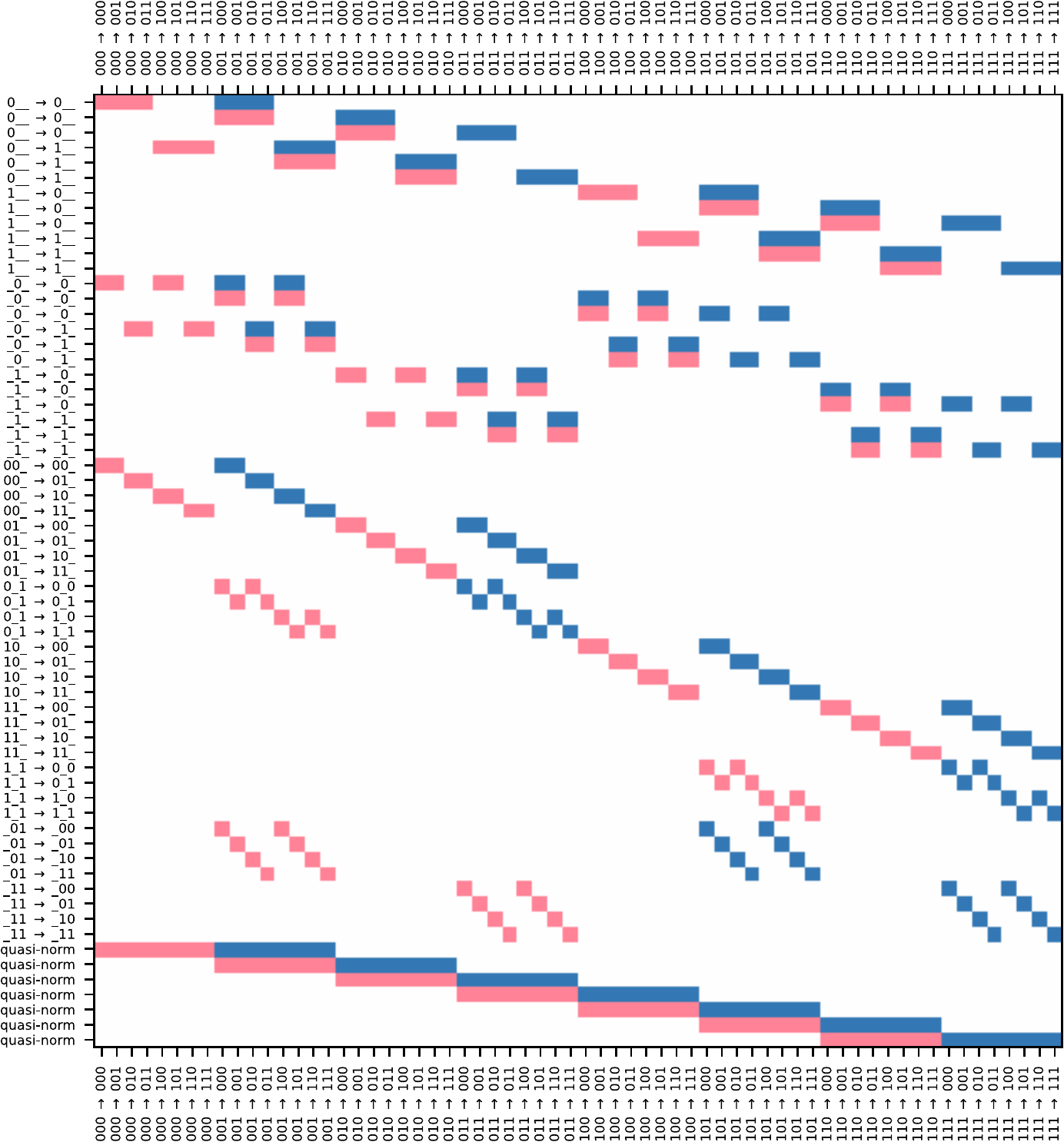}
\end{center}

\noindent Rows correspond to the 63 linear equations, of which 30 are independent.

\newpage
\subsection*{Space 45}

Space 45 is not induced by a causal order, but it is a refinement of the space 100 induced by the definite causal order $\total{\ev{A},\ev{B},\ev{C}}$.
Its equivalence class under event-input permutation symmetry contains 24 spaces.
Space 45 differs as follows from the space induced by causal order $\total{\ev{A},\ev{B},\ev{C}}$:
\begin{itemize}
  \item The outputs at events \evset{\ev{A}, \ev{C}} are independent of the input at event \ev{B} when the inputs at events \evset{A, C} are given by \hist{A/0,C/1} and \hist{A/1,C/1}.
  \item The outputs at events \evset{\ev{B}, \ev{C}} are independent of the input at event \ev{A} when the inputs at events \evset{B, C} are given by \hist{B/1,C/1}.
  \item The output at event \ev{C} is independent of the inputs at events \evset{\ev{A}, \ev{B}} when the input at event C is given by \hist{C/1}.
  \item The output at event \ev{B} is independent of the input at event \ev{A} when the input at event B is given by \hist{B/1}.
\end{itemize}

\noindent Below are the histories and extended histories for space 45: 
\begin{center}
    \begin{tabular}{cc}
    \includegraphics[height=3.5cm]{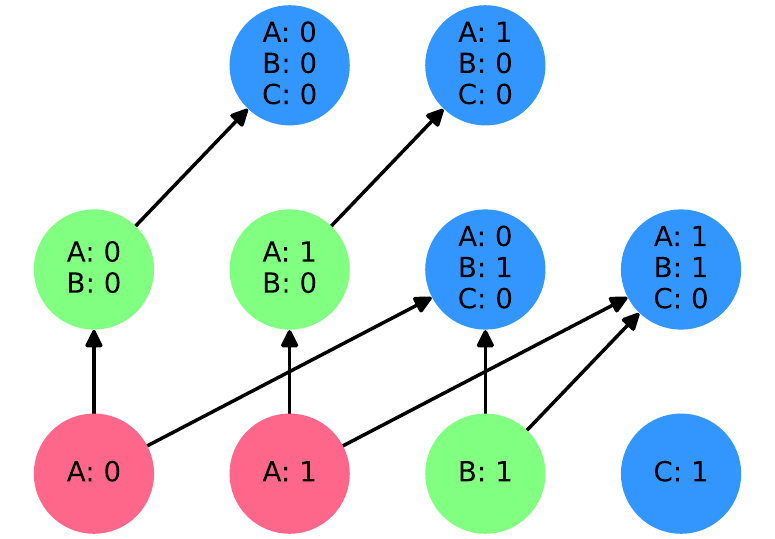}
    &
    \includegraphics[height=3.5cm]{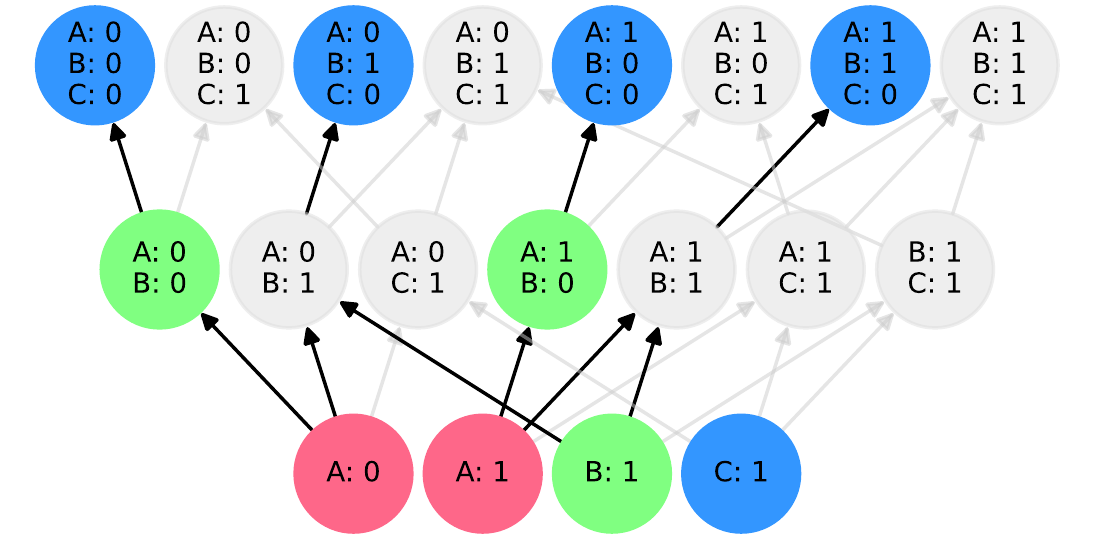}
    \\
    $\Theta_{45}$
    &
    $\Ext{\Theta_{45}}$
    \end{tabular}
\end{center}

\noindent The standard causaltope for Space 45 has dimension 35.
Below is a plot of the homogeneous linear system of causality and quasi-normalisation equations for the standard causaltope, put in reduced row echelon form:

\begin{center}
    \includegraphics[width=11cm]{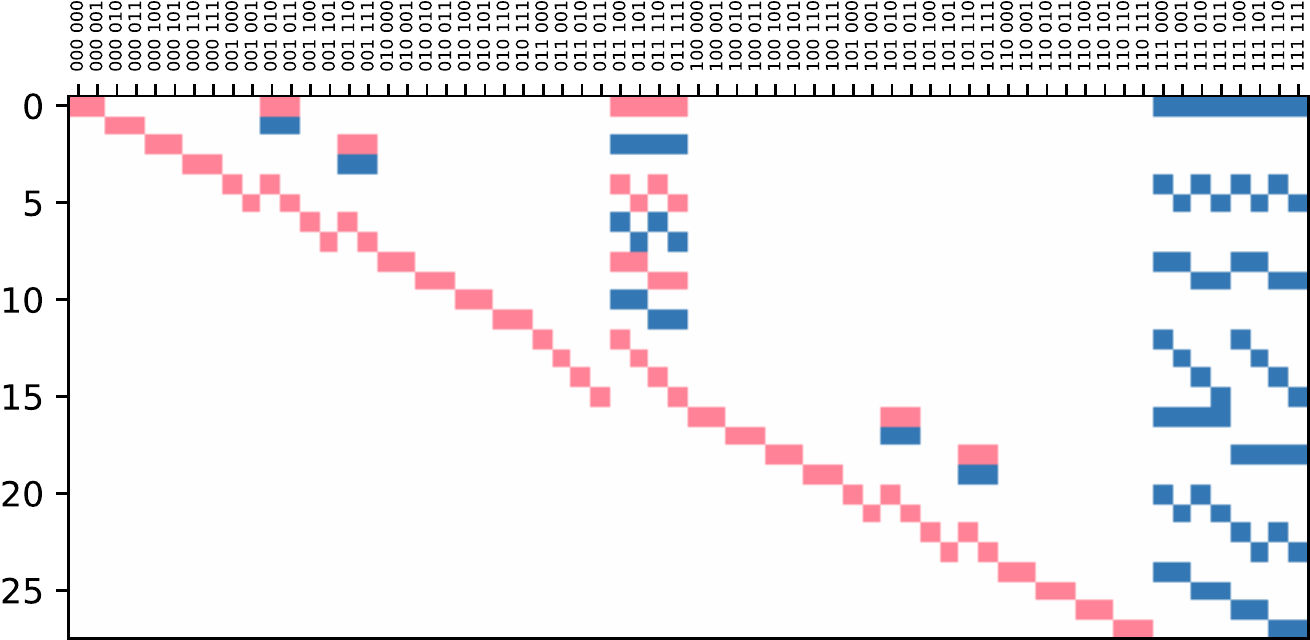}
\end{center}

\noindent Rows correspond to the 28 independent linear equations.
Columns in the plot correspond to entries of empirical models, indexed as $i_A i_B i_C$ $o_A o_B o_C$.
Coefficients in the equations are color-coded as white=0, red=+1 and blue=-1.

Space 45 has closest refinements in equivalence classes 28, 30 and 31; 
it is the join of its (closest) refinements.
It has closest coarsenings in equivalence classes 61 and 76; 
it is the meet of its (closest) coarsenings.
It has 1024 causal functions, 448 of which are not causal for any of its refinements.
It is a tight space.

The standard causaltope for Space 45 has 2 more dimensions than those of its 4 subspaces in equivalence classes 28, 30 and 31.
The standard causaltope for Space 45 is the meet of the standard causaltopes for its closest coarsenings.
For completeness, below is a plot of the full homogeneous linear system of causality and quasi-normalisation equations for the standard causaltope:

\begin{center}
    \includegraphics[width=12cm]{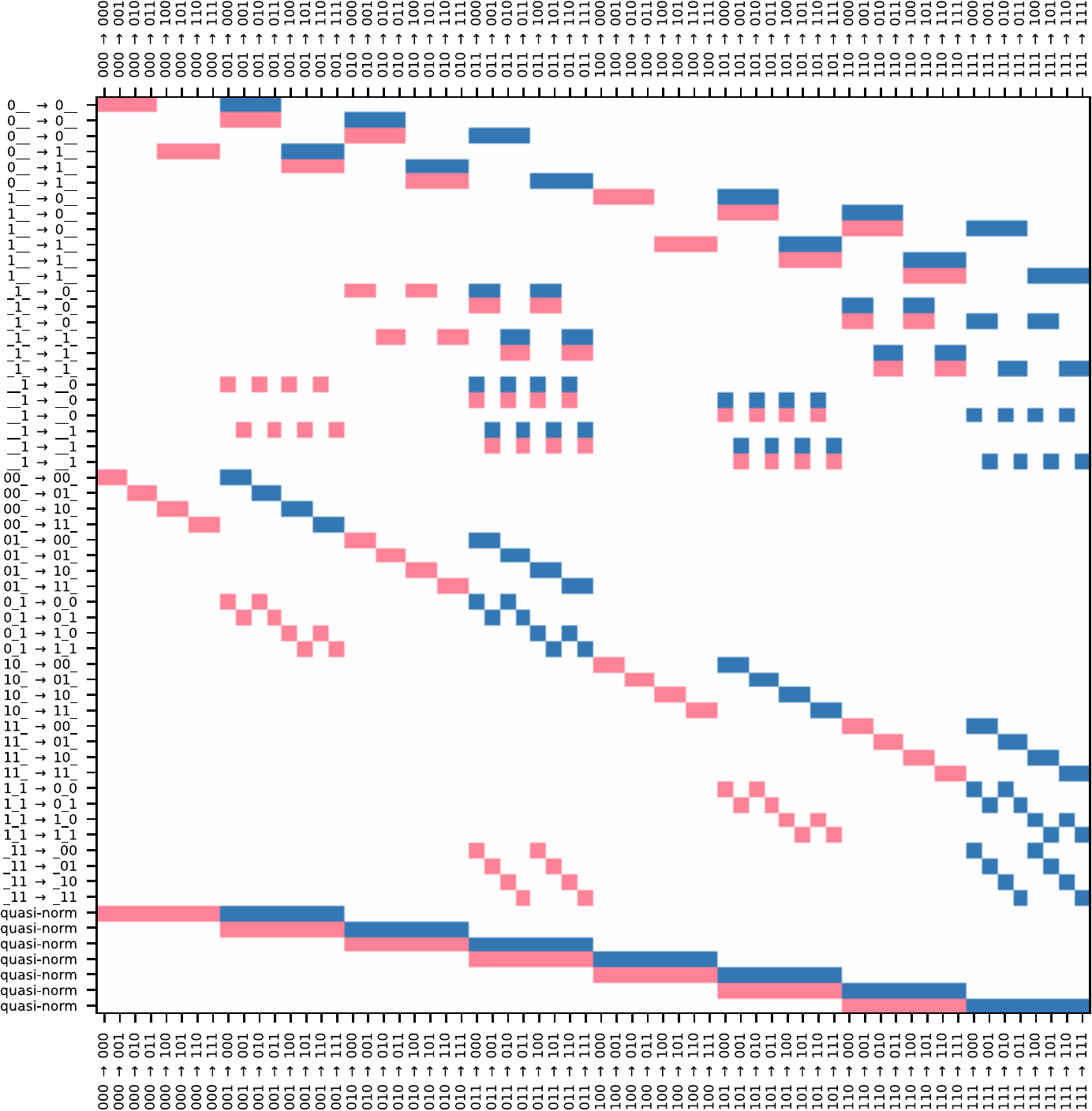}
\end{center}

\noindent Rows correspond to the 59 linear equations, of which 28 are independent.

\newpage
\subsection*{Space 46}

Space 46 is not induced by a causal order, but it is a refinement of the space in equivalence class 92 induced by the definite causal order $\total{\ev{A},\ev{B}}\vee\total{\ev{C},\ev{B}}$ (note that the space induced by the order is not the same as space 92).
Its equivalence class under event-input permutation symmetry contains 24 spaces.
Space 46 differs as follows from the space induced by causal order $\total{\ev{A},\ev{B}}\vee\total{\ev{C},\ev{B}}$:
\begin{itemize}
  \item The outputs at events \evset{\ev{A}, \ev{B}} are independent of the input at event \ev{C} when the inputs at events \evset{A, B} are given by \hist{A/0,B/0}, \hist{A/0,B/1} and \hist{A/1,B/0}.
\end{itemize}

\noindent Below are the histories and extended histories for space 46: 
\begin{center}
    \begin{tabular}{cc}
    \includegraphics[height=3.5cm]{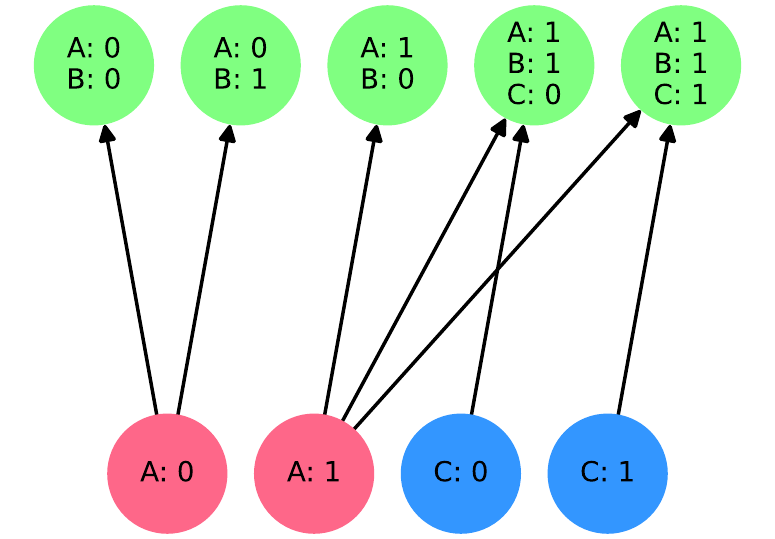}
    &
    \includegraphics[height=3.5cm]{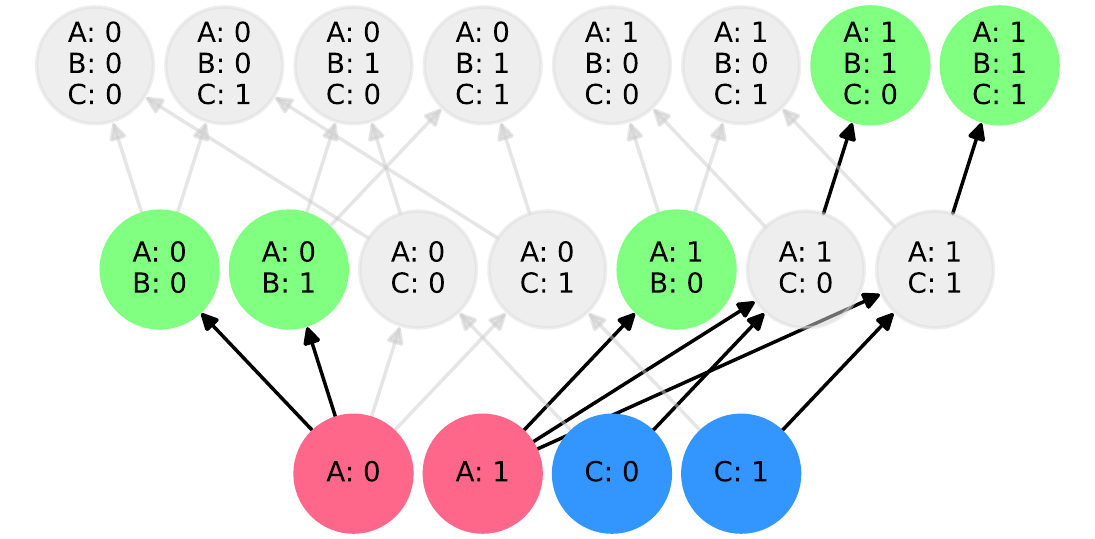}
    \\
    $\Theta_{46}$
    &
    $\Ext{\Theta_{46}}$
    \end{tabular}
\end{center}

\noindent The standard causaltope for Space 46 has dimension 34.
Below is a plot of the homogeneous linear system of causality and quasi-normalisation equations for the standard causaltope, put in reduced row echelon form:

\begin{center}
    \includegraphics[width=11cm]{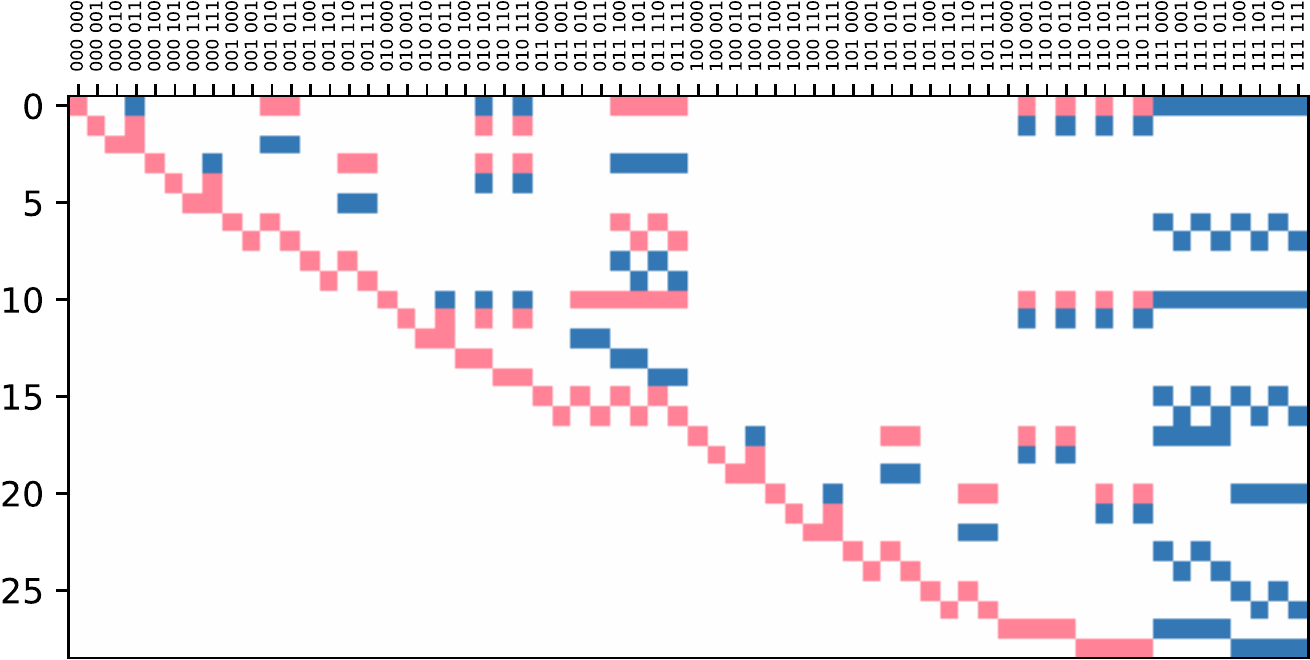}
\end{center}

\noindent Rows correspond to the 29 independent linear equations.
Columns in the plot correspond to entries of empirical models, indexed as $i_A i_B i_C$ $o_A o_B o_C$.
Coefficients in the equations are color-coded as white=0, red=+1 and blue=-1.

Space 46 has closest refinements in equivalence classes 29, 33 and 34; 
it is the join of its (closest) refinements.
It has closest coarsenings in equivalence classes 67, 68, 70 and 74; 
it is the meet of its (closest) coarsenings.
It has 512 causal functions, 64 of which are not causal for any of its refinements.
It is a tight space.

The standard causaltope for Space 46 has 2 more dimensions than those of its 5 subspaces in equivalence classes 29, 33 and 34.
The standard causaltope for Space 46 is the meet of the standard causaltopes for its closest coarsenings.
For completeness, below is a plot of the full homogeneous linear system of causality and quasi-normalisation equations for the standard causaltope:

\begin{center}
    \includegraphics[width=12cm]{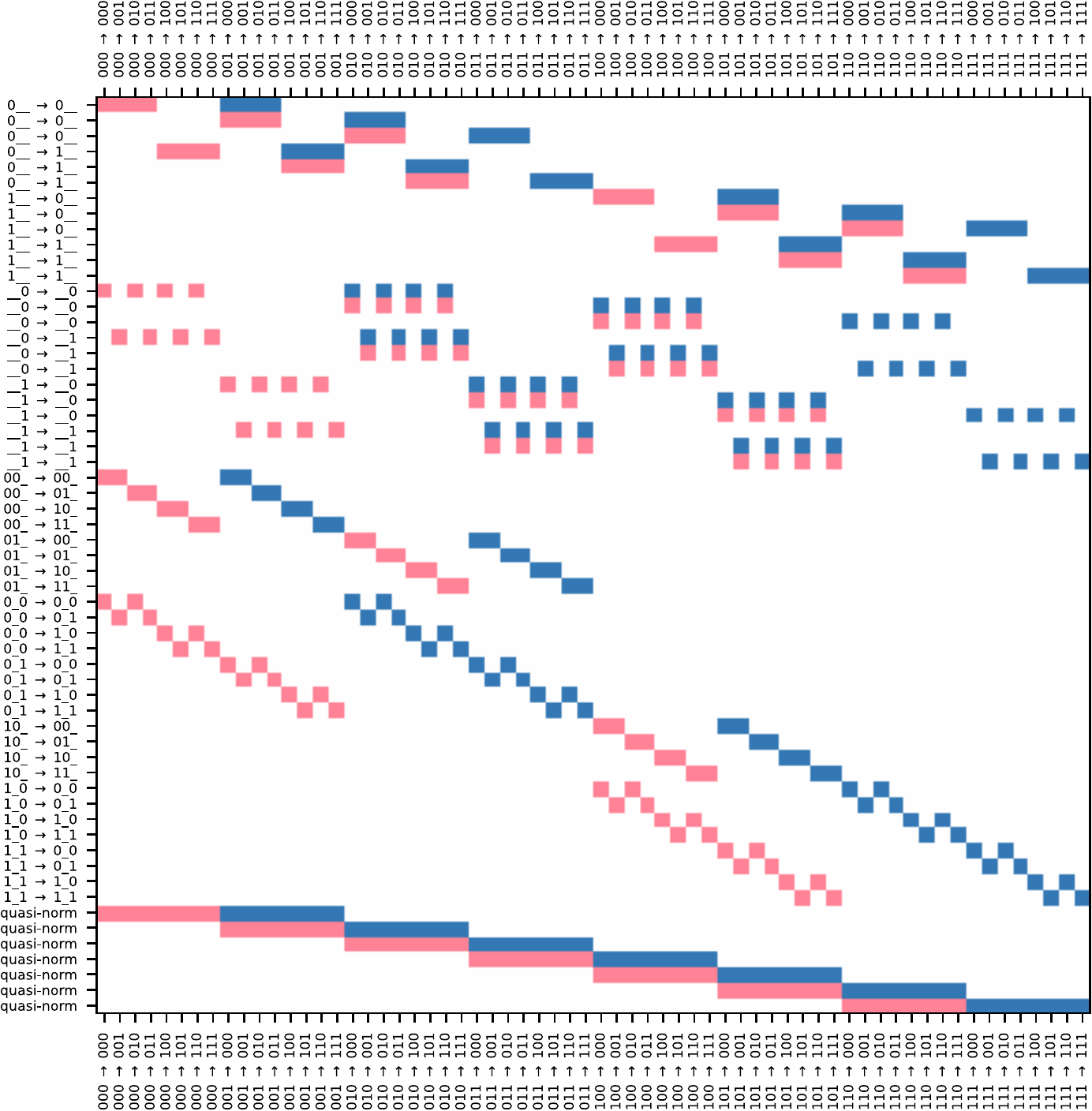}
\end{center}

\noindent Rows correspond to the 59 linear equations, of which 29 are independent.

\newpage
\subsection*{Space 47}

Space 47 is not induced by a causal order, but it is a refinement of the space 100 induced by the definite causal order $\total{\ev{A},\ev{B},\ev{C}}$.
Its equivalence class under event-input permutation symmetry contains 48 spaces.
Space 47 differs as follows from the space induced by causal order $\total{\ev{A},\ev{B},\ev{C}}$:
\begin{itemize}
  \item The outputs at events \evset{\ev{B}, \ev{C}} are independent of the input at event \ev{A} when the inputs at events \evset{B, C} are given by \hist{B/1,C/0}.
  \item The outputs at events \evset{\ev{A}, \ev{C}} are independent of the input at event \ev{B} when the inputs at events \evset{A, C} are given by \hist{A/0,C/0}, \hist{A/1,C/0} and \hist{A/1,C/1}.
  \item The output at event \ev{C} is independent of the inputs at events \evset{\ev{A}, \ev{B}} when the input at event C is given by \hist{C/0}.
\end{itemize}

\noindent Below are the histories and extended histories for space 47: 
\begin{center}
    \begin{tabular}{cc}
    \includegraphics[height=3.5cm]{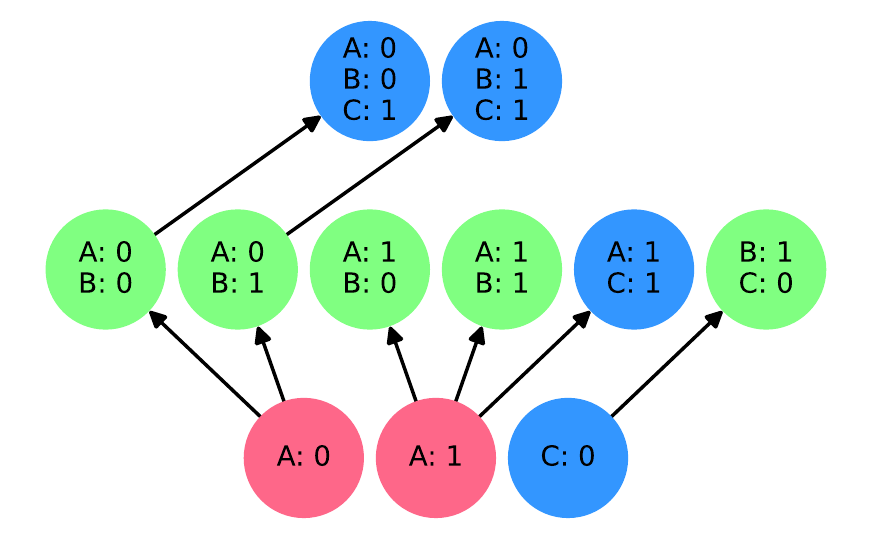}
    &
    \includegraphics[height=3.5cm]{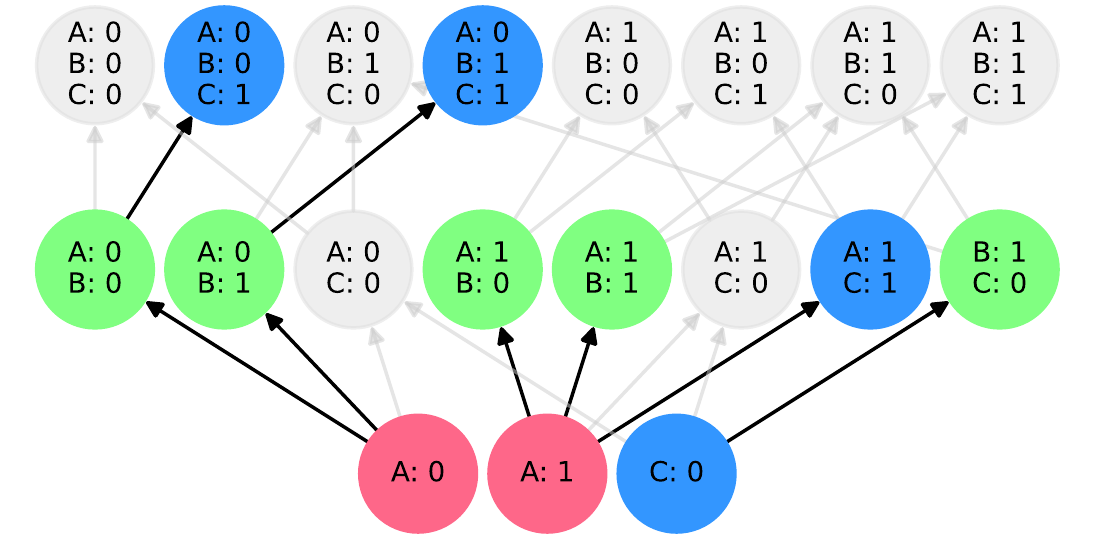}
    \\
    $\Theta_{47}$
    &
    $\Ext{\Theta_{47}}$
    \end{tabular}
\end{center}

\noindent The standard causaltope for Space 47 has dimension 33.
Below is a plot of the homogeneous linear system of causality and quasi-normalisation equations for the standard causaltope, put in reduced row echelon form:

\begin{center}
    \includegraphics[width=11cm]{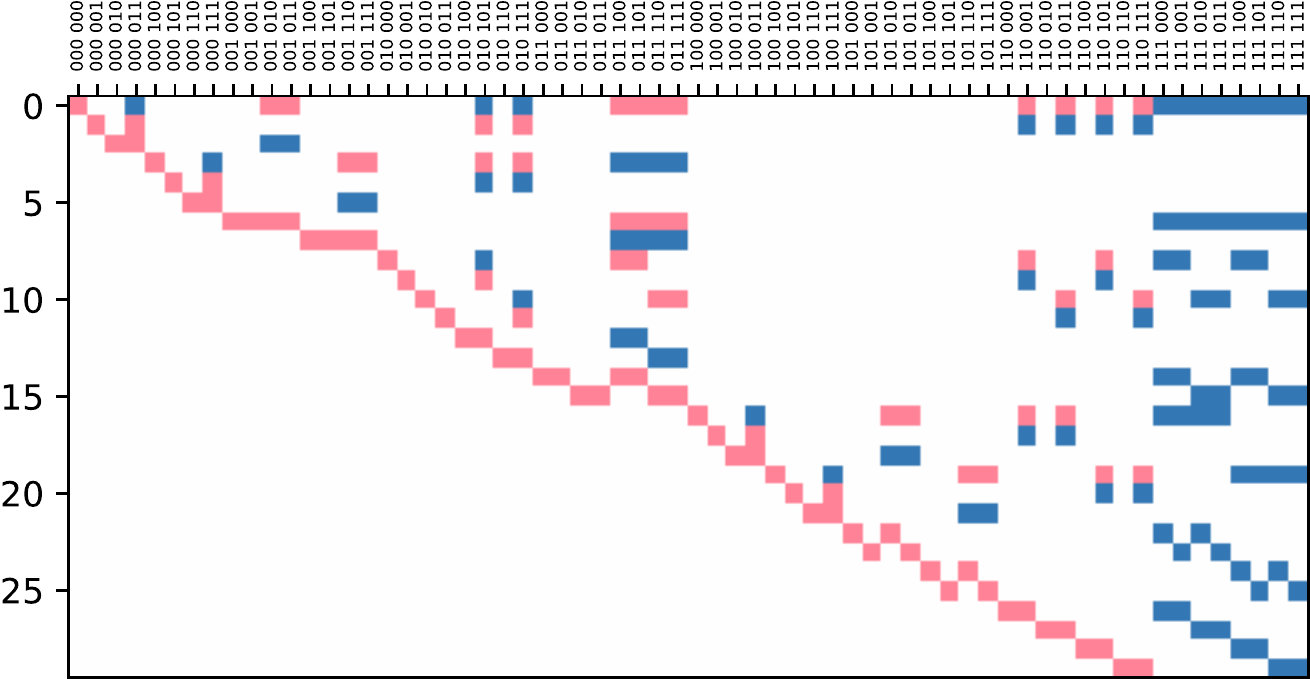}
\end{center}

\noindent Rows correspond to the 30 independent linear equations.
Columns in the plot correspond to entries of empirical models, indexed as $i_A i_B i_C$ $o_A o_B o_C$.
Coefficients in the equations are color-coded as white=0, red=+1 and blue=-1.

Space 47 has closest refinements in equivalence classes 30, 38 and 42; 
it is the join of its (closest) refinements.
It has closest coarsenings in equivalence classes 61, 62, 64 and 71; 
it is the meet of its (closest) coarsenings.
It has 512 causal functions, all of which are causal for at least one of its refinements.
It is not a tight space: for event \ev{B}, a causal function must yield identical output values on input histories \hist{A/0,B/1}, \hist{A/1,B/1} and \hist{B/1,C/0}.

The standard causaltope for Space 47 coincides with that of its subspace in equivalence class 30.
The standard causaltope for Space 47 is the meet of the standard causaltopes for its closest coarsenings.
For completeness, below is a plot of the full homogeneous linear system of causality and quasi-normalisation equations for the standard causaltope:

\begin{center}
    \includegraphics[width=12cm]{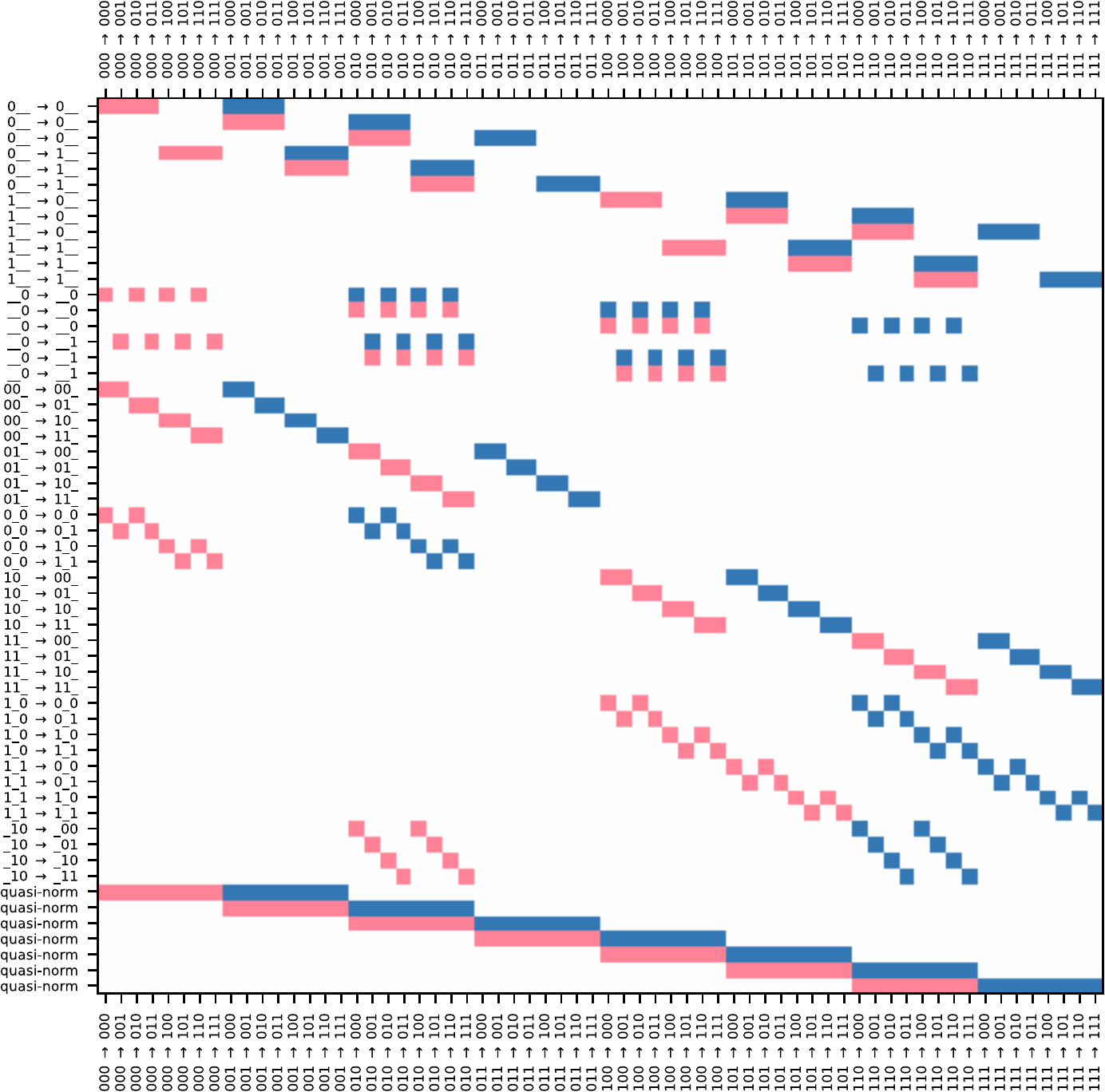}
\end{center}

\noindent Rows correspond to the 57 linear equations, of which 30 are independent.

\newpage
\subsection*{Space 48}

Space 48 is not induced by a causal order, but it is a refinement of the space induced by the indefinite causal order $\total{\ev{A},\{\ev{B},\ev{C}\}}$.
Its equivalence class under event-input permutation symmetry contains 24 spaces.
Space 48 differs as follows from the space induced by causal order $\total{\ev{A},\{\ev{B},\ev{C}\}}$:
\begin{itemize}
  \item The outputs at events \evset{\ev{A}, \ev{B}} are independent of the input at event \ev{C} when the inputs at events \evset{A, B} are given by \hist{A/0,B/0}, \hist{A/0,B/1} and \hist{A/1,B/0}.
  \item The outputs at events \evset{\ev{A}, \ev{C}} are independent of the input at event \ev{B} when the inputs at events \evset{A, C} are given by \hist{A/0,C/1}, \hist{A/1,C/0} and \hist{A/1,C/1}.
  \item The output at event \ev{B} is independent of the inputs at events \evset{\ev{A}, \ev{C}} when the input at event B is given by \hist{B/0}.
  \item The output at event \ev{C} is independent of the inputs at events \evset{\ev{A}, \ev{B}} when the input at event C is given by \hist{C/1}.
  \item The outputs at events \evset{\ev{B}, \ev{C}} are independent of the input at event \ev{A} when the inputs at events \evset{B, C} are given by \hist{B/0,C/1}.
\end{itemize}

\noindent Below are the histories and extended histories for space 48: 
\begin{center}
    \begin{tabular}{cc}
    \includegraphics[height=3.5cm]{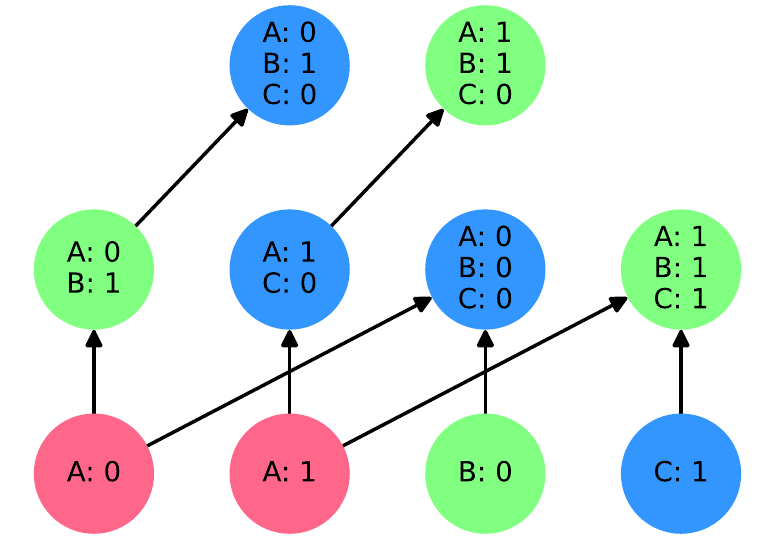}
    &
    \includegraphics[height=3.5cm]{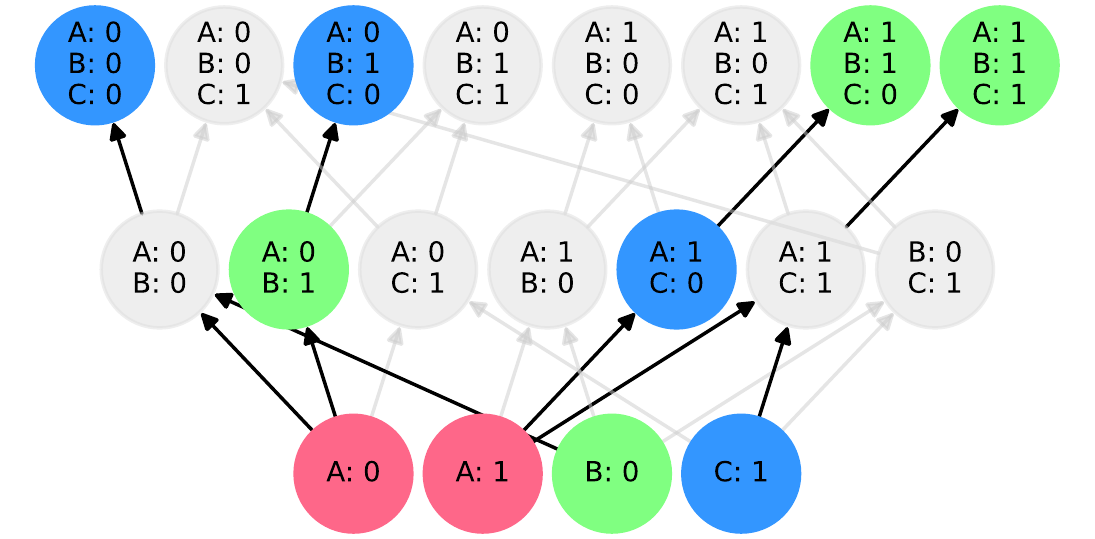}
    \\
    $\Theta_{48}$
    &
    $\Ext{\Theta_{48}}$
    \end{tabular}
\end{center}

\noindent The standard causaltope for Space 48 has dimension 35.
Below is a plot of the homogeneous linear system of causality and quasi-normalisation equations for the standard causaltope, put in reduced row echelon form:

\begin{center}
    \includegraphics[width=11cm]{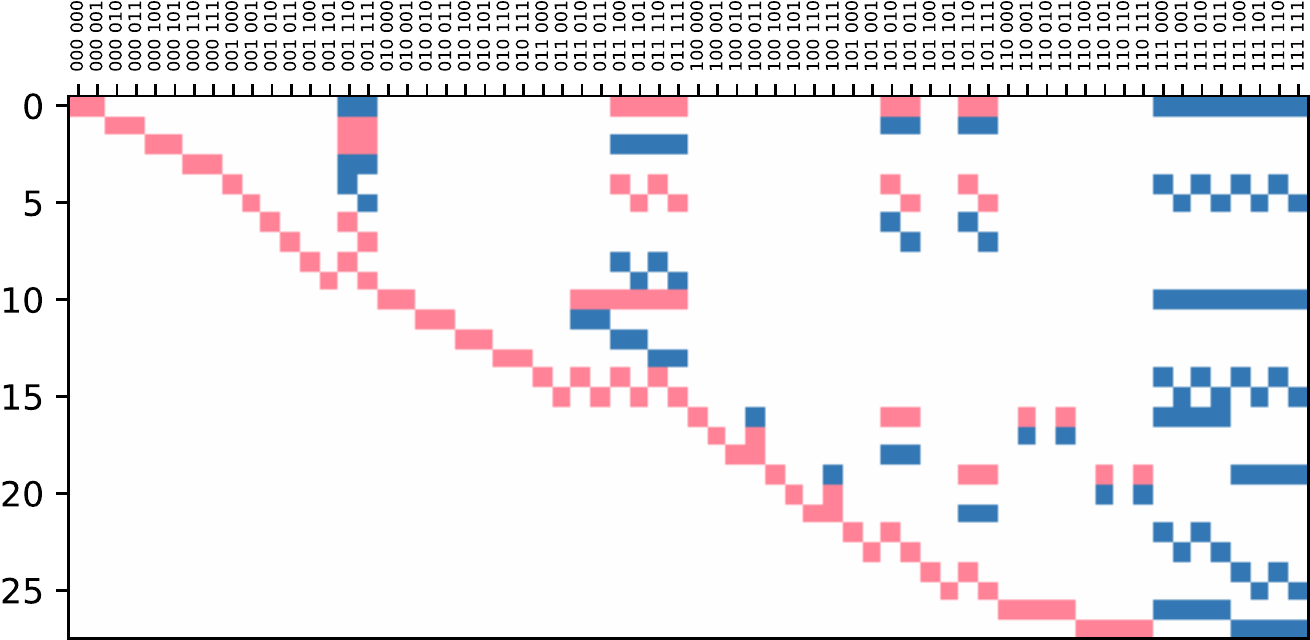}
\end{center}

\noindent Rows correspond to the 28 independent linear equations.
Columns in the plot correspond to entries of empirical models, indexed as $i_A i_B i_C$ $o_A o_B o_C$.
Coefficients in the equations are color-coded as white=0, red=+1 and blue=-1.

Space 48 has closest refinements in equivalence classes 30 and 32; 
it is the join of its (closest) refinements.
It has closest coarsenings in equivalence class 71; 
it is the meet of its (closest) coarsenings.
It has 1024 causal functions, 384 of which are not causal for any of its refinements.
It is a tight space.

The standard causaltope for Space 48 has 2 more dimensions than those of its 4 subspaces in equivalence classes 30 and 32.
The standard causaltope for Space 48 is the meet of the standard causaltopes for its closest coarsenings.
For completeness, below is a plot of the full homogeneous linear system of causality and quasi-normalisation equations for the standard causaltope:

\begin{center}
    \includegraphics[width=12cm]{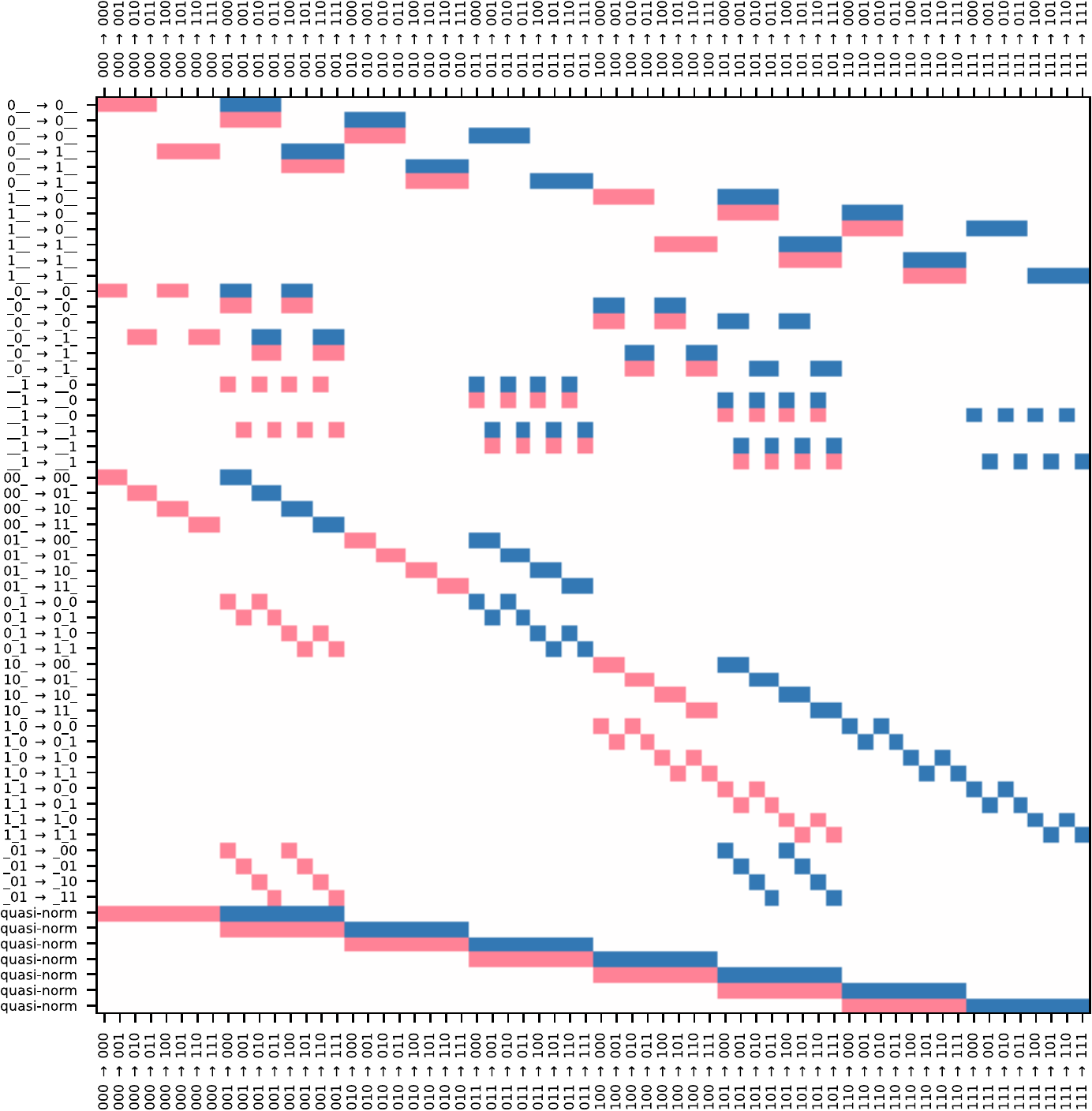}
\end{center}

\noindent Rows correspond to the 59 linear equations, of which 28 are independent.

\newpage
\subsection*{Space 49}

Space 49 is not induced by a causal order, but it is a refinement of the space 92 induced by the definite causal order $\total{\ev{A},\ev{C}}\vee\total{\ev{B},\ev{C}}$.
Its equivalence class under event-input permutation symmetry contains 24 spaces.
Space 49 differs as follows from the space induced by causal order $\total{\ev{A},\ev{C}}\vee\total{\ev{B},\ev{C}}$:
\begin{itemize}
  \item The outputs at events \evset{\ev{B}, \ev{C}} are independent of the input at event \ev{A} when the inputs at events \evset{B, C} are given by \hist{B/1,C/1} and \hist{B/0,C/1}.
  \item The outputs at events \evset{\ev{A}, \ev{C}} are independent of the input at event \ev{B} when the inputs at events \evset{A, C} are given by \hist{A/1,C/0}.
\end{itemize}

\noindent Below are the histories and extended histories for space 49: 
\begin{center}
    \begin{tabular}{cc}
    \includegraphics[height=3.5cm]{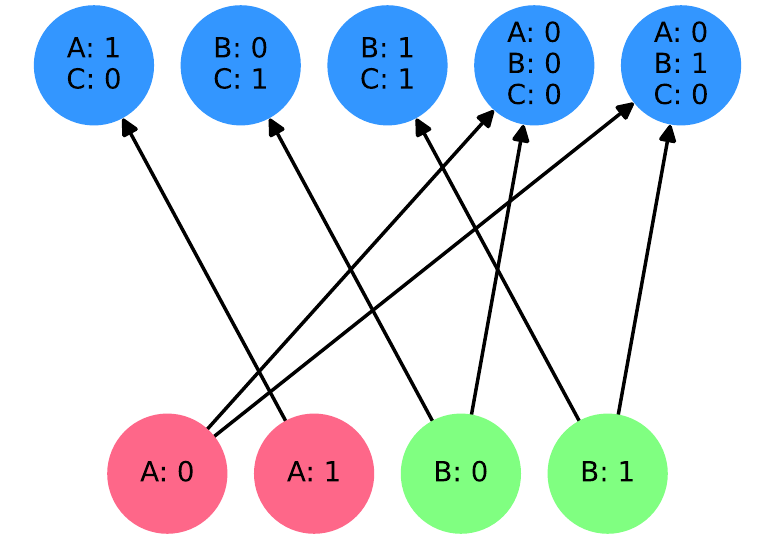}
    &
    \includegraphics[height=3.5cm]{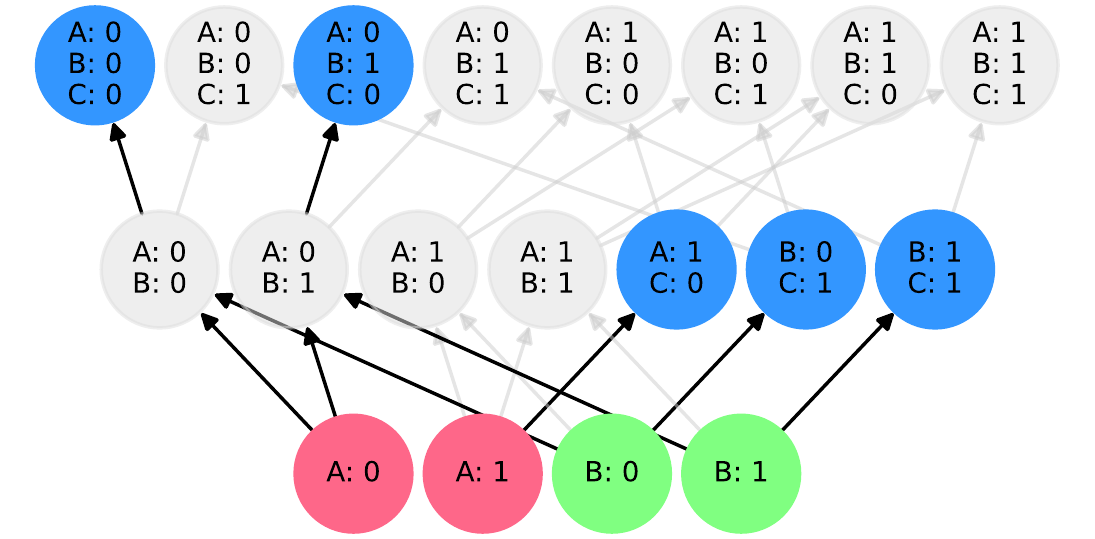}
    \\
    $\Theta_{49}$
    &
    $\Ext{\Theta_{49}}$
    \end{tabular}
\end{center}

\noindent The standard causaltope for Space 49 has dimension 34.
Below is a plot of the homogeneous linear system of causality and quasi-normalisation equations for the standard causaltope, put in reduced row echelon form:

\begin{center}
    \includegraphics[width=11cm]{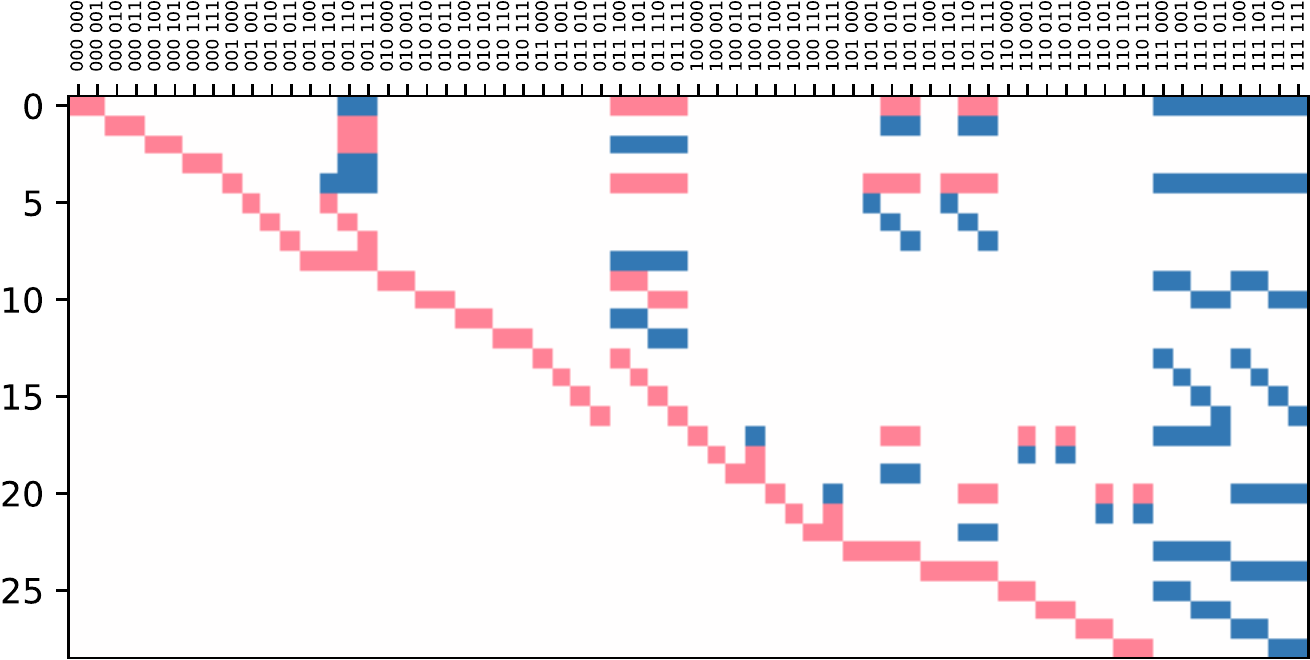}
\end{center}

\noindent Rows correspond to the 29 independent linear equations.
Columns in the plot correspond to entries of empirical models, indexed as $i_A i_B i_C$ $o_A o_B o_C$.
Coefficients in the equations are color-coded as white=0, red=+1 and blue=-1.

Space 49 has closest refinements in equivalence classes 29, 35, 39 and 41; 
it is the join of its (closest) refinements.
It has closest coarsenings in equivalence classes 66, 68 and 75; 
it is the meet of its (closest) coarsenings.
It has 512 causal functions, 192 of which are not causal for any of its refinements.
It is a tight space.

The standard causaltope for Space 49 has 2 more dimensions than those of its 5 subspaces in equivalence classes 29, 35, 39 and 41.
The standard causaltope for Space 49 is the meet of the standard causaltopes for its closest coarsenings.
For completeness, below is a plot of the full homogeneous linear system of causality and quasi-normalisation equations for the standard causaltope:

\begin{center}
    \includegraphics[width=12cm]{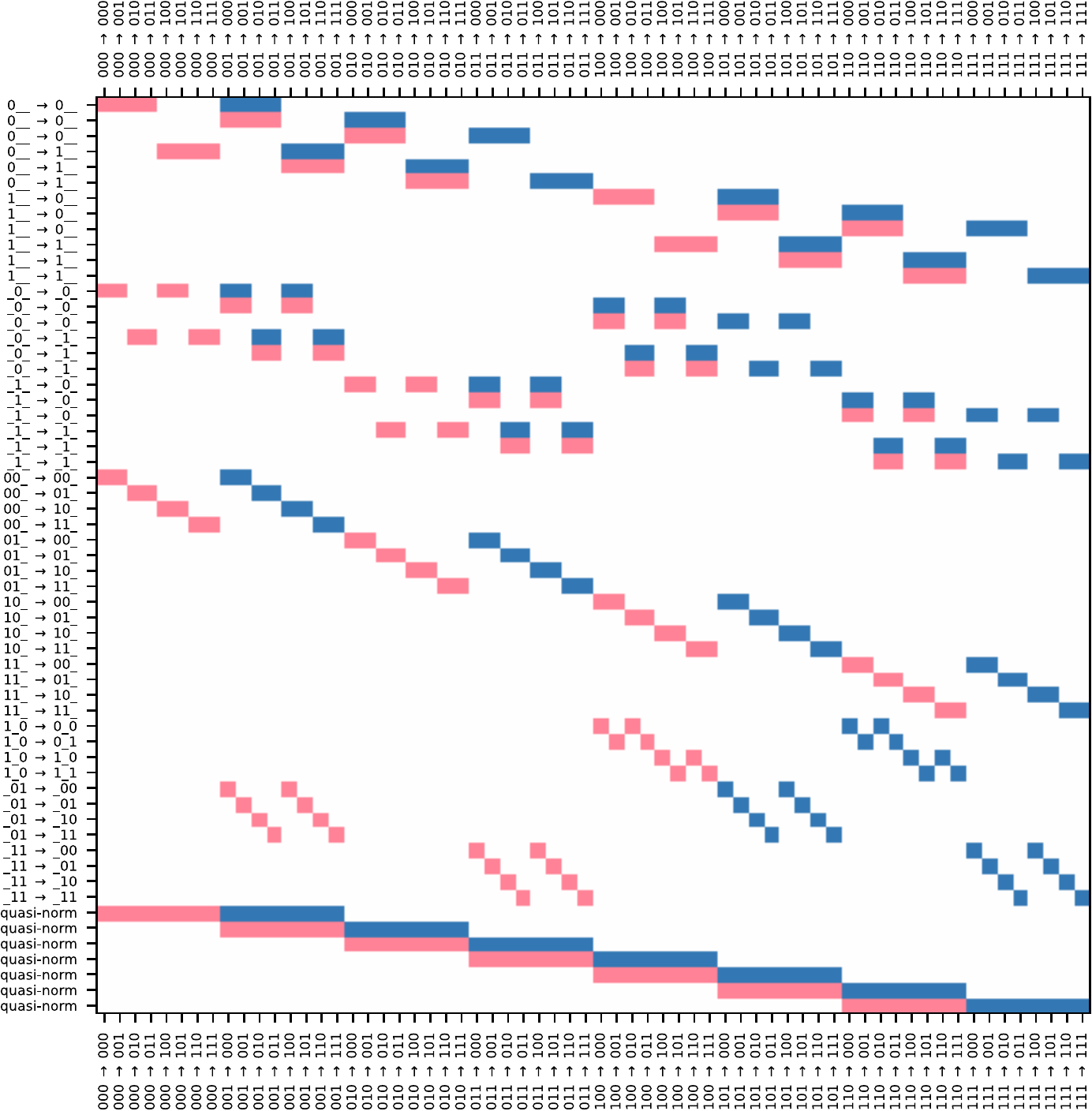}
\end{center}

\noindent Rows correspond to the 59 linear equations, of which 29 are independent.

\newpage
\subsection*{Space 50}

Space 50 is not induced by a causal order, but it is a refinement of the space 92 induced by the definite causal order $\total{\ev{A},\ev{C}}\vee\total{\ev{B},\ev{C}}$.
Its equivalence class under event-input permutation symmetry contains 48 spaces.
Space 50 differs as follows from the space induced by causal order $\total{\ev{A},\ev{C}}\vee\total{\ev{B},\ev{C}}$:
\begin{itemize}
  \item The outputs at events \evset{\ev{B}, \ev{C}} are independent of the input at event \ev{A} when the inputs at events \evset{B, C} are given by \hist{B/1,C/0} and \hist{B/0,C/1}.
  \item The outputs at events \evset{\ev{A}, \ev{C}} are independent of the input at event \ev{B} when the inputs at events \evset{A, C} are given by \hist{A/1,C/1}.
\end{itemize}

\noindent Below are the histories and extended histories for space 50: 
\begin{center}
    \begin{tabular}{cc}
    \includegraphics[height=3.5cm]{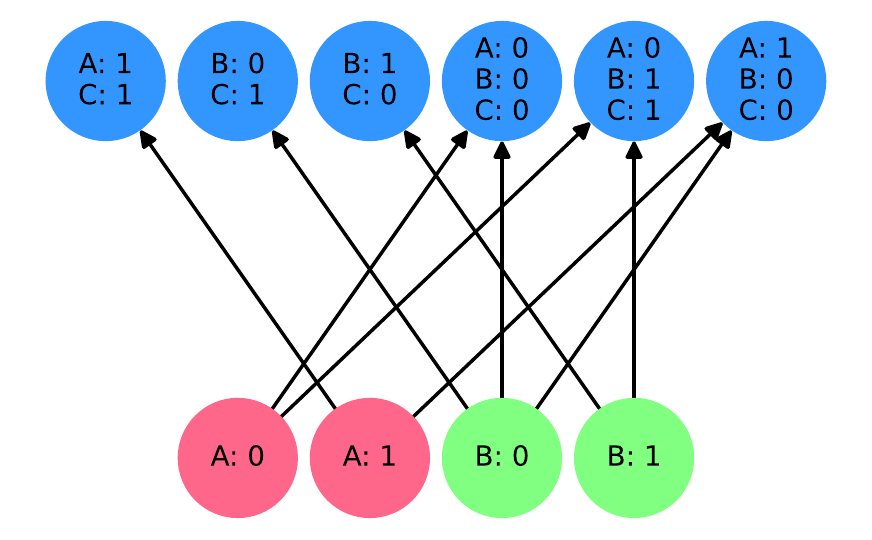}
    &
    \includegraphics[height=3.5cm]{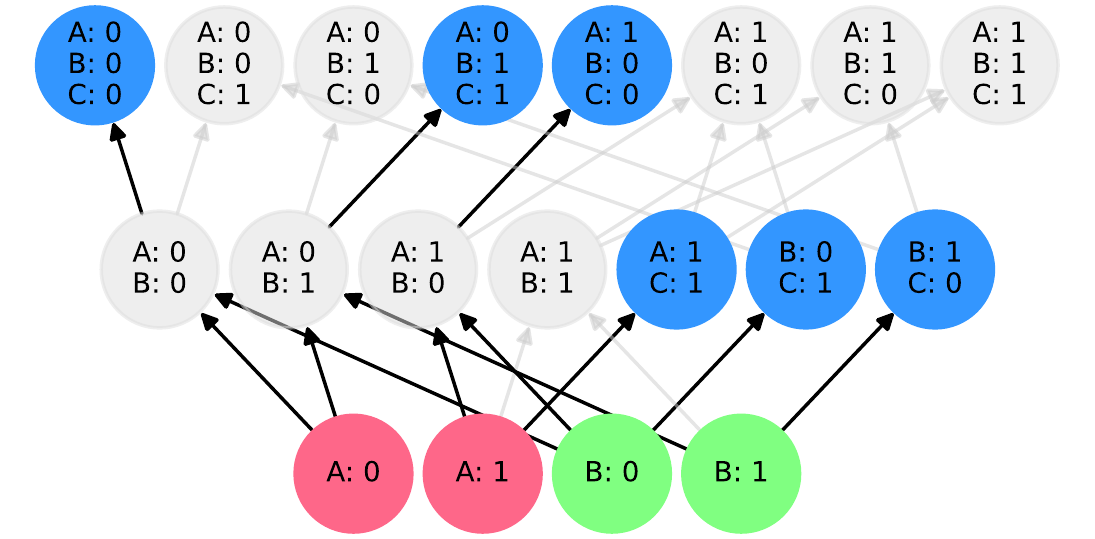}
    \\
    $\Theta_{50}$
    &
    $\Ext{\Theta_{50}}$
    \end{tabular}
\end{center}

\noindent The standard causaltope for Space 50 has dimension 34.
Below is a plot of the homogeneous linear system of causality and quasi-normalisation equations for the standard causaltope, put in reduced row echelon form:

\begin{center}
    \includegraphics[width=11cm]{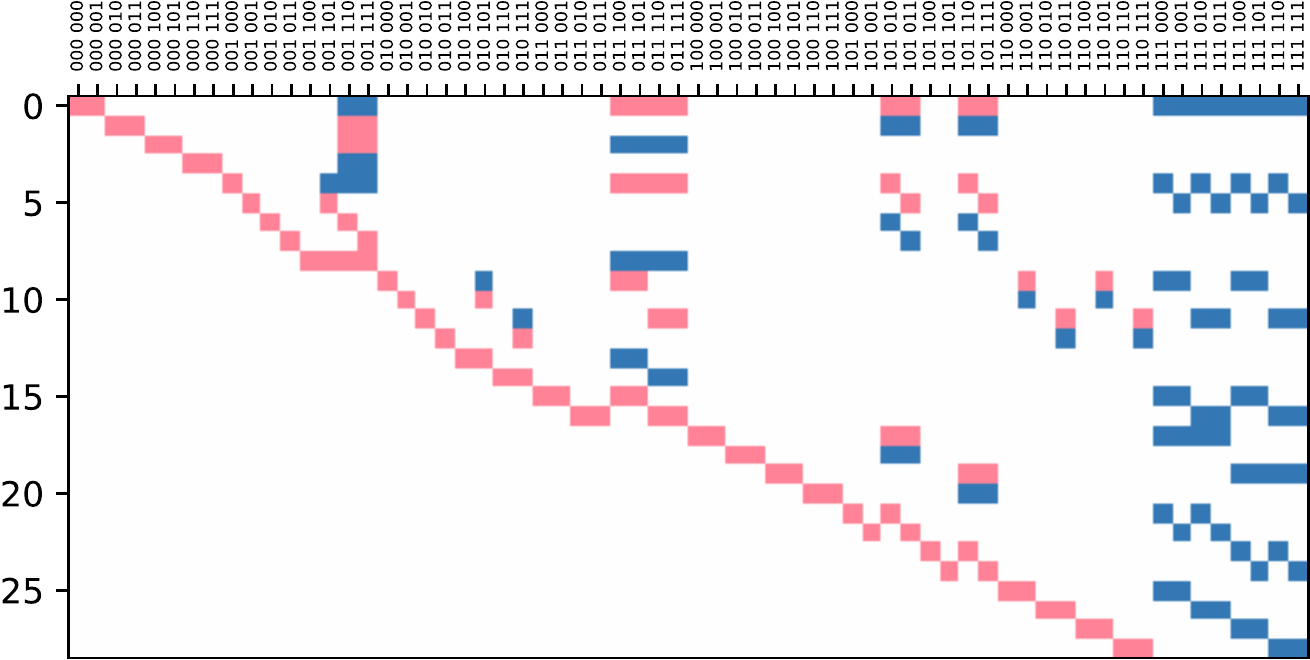}
\end{center}

\noindent Rows correspond to the 29 independent linear equations.
Columns in the plot correspond to entries of empirical models, indexed as $i_A i_B i_C$ $o_A o_B o_C$.
Coefficients in the equations are color-coded as white=0, red=+1 and blue=-1.

Space 50 has closest refinements in equivalence classes 29, 34, 36, 39 and 40; 
it is the join of its (closest) refinements.
It has closest coarsenings in equivalence classes 66, 69, 70 and 73; 
it is the meet of its (closest) coarsenings.
It has 512 causal functions, 128 of which are not causal for any of its refinements.
It is not a tight space: for event \ev{C}, a causal function must yield identical output values on input histories \hist{A/1,C/1} and \hist{B/0,C/1}.

The standard causaltope for Space 50 has 2 more dimensions than those of its 5 subspaces in equivalence classes 29, 34, 36, 39 and 40.
The standard causaltope for Space 50 is the meet of the standard causaltopes for its closest coarsenings.
For completeness, below is a plot of the full homogeneous linear system of causality and quasi-normalisation equations for the standard causaltope:

\begin{center}
    \includegraphics[width=12cm]{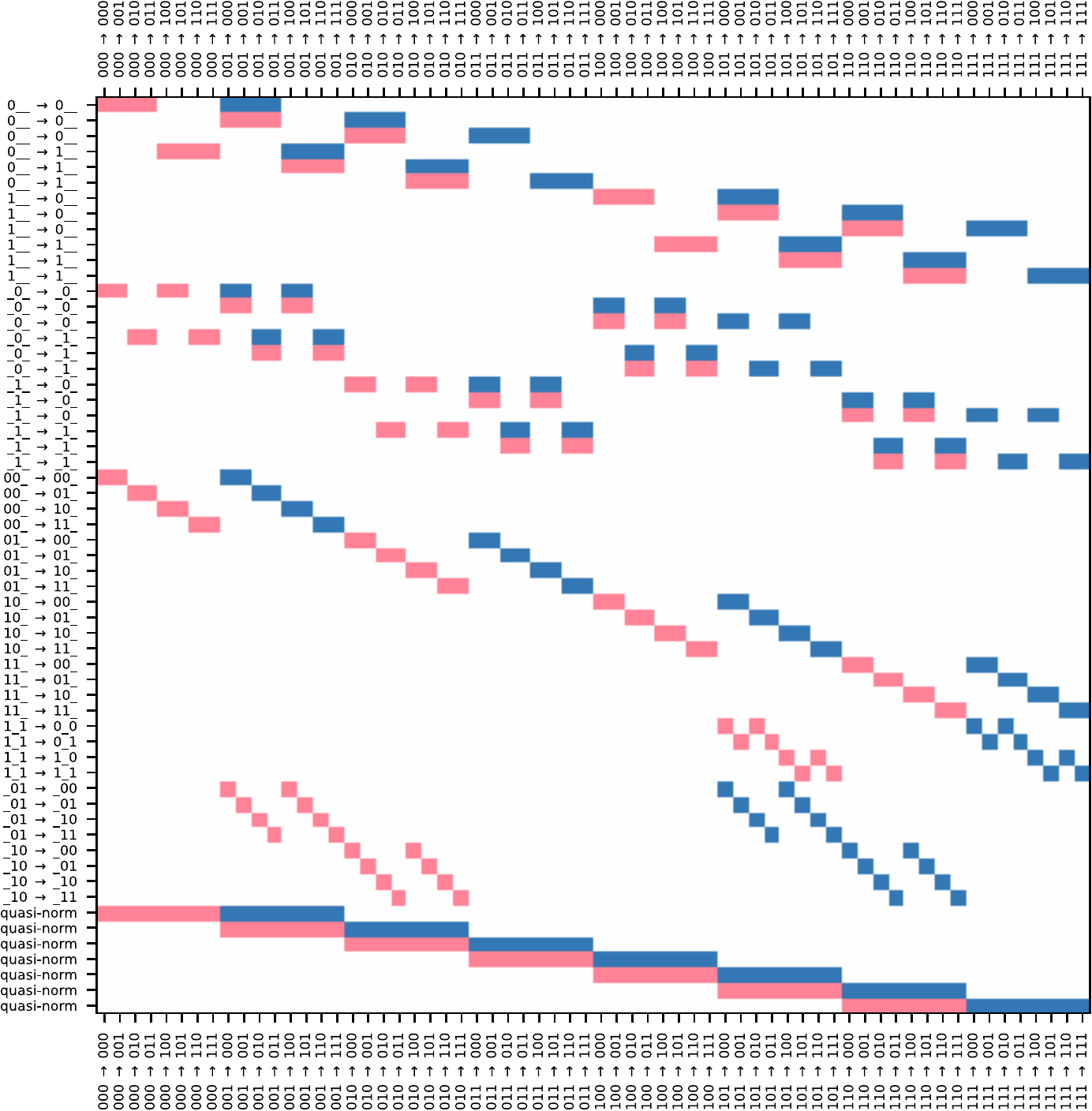}
\end{center}

\noindent Rows correspond to the 59 linear equations, of which 29 are independent.

\newpage
\subsection*{Space 51}

Space 51 is not induced by a causal order, but it is a refinement of the space 100 induced by the definite causal order $\total{\ev{A},\ev{B},\ev{C}}$.
Its equivalence class under event-input permutation symmetry contains 12 spaces.
Space 51 differs as follows from the space induced by causal order $\total{\ev{A},\ev{B},\ev{C}}$:
\begin{itemize}
  \item The outputs at events \evset{\ev{A}, \ev{C}} are independent of the input at event \ev{B} when the inputs at events \evset{A, C} are given by \hist{A/0,C/1} and \hist{A/1,C/1}.
  \item The outputs at events \evset{\ev{B}, \ev{C}} are independent of the input at event \ev{A} when the inputs at events \evset{B, C} are given by \hist{B/1,C/1} and \hist{B/0,C/1}.
  \item The output at event \ev{C} is independent of the inputs at events \evset{\ev{A}, \ev{B}} when the input at event C is given by \hist{C/1}.
\end{itemize}

\noindent Below are the histories and extended histories for space 51: 
\begin{center}
    \begin{tabular}{cc}
    \includegraphics[height=3.5cm]{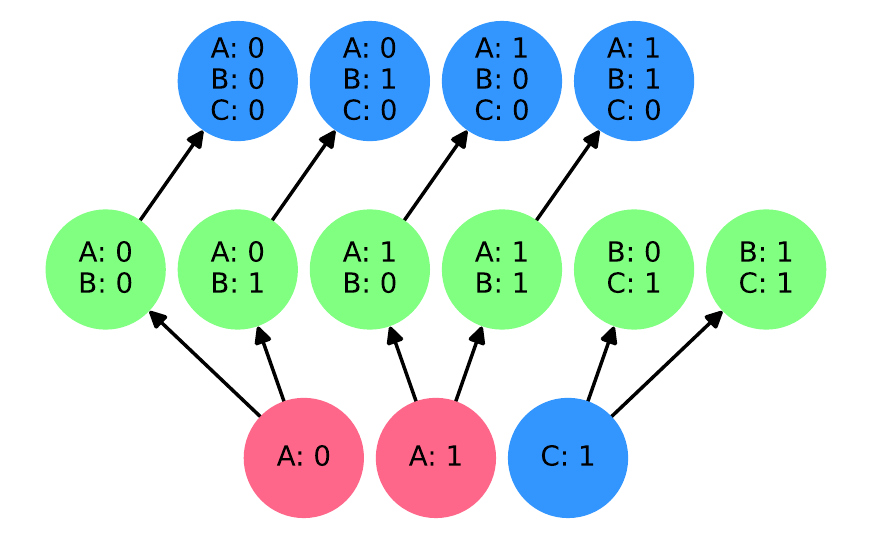}
    &
    \includegraphics[height=3.5cm]{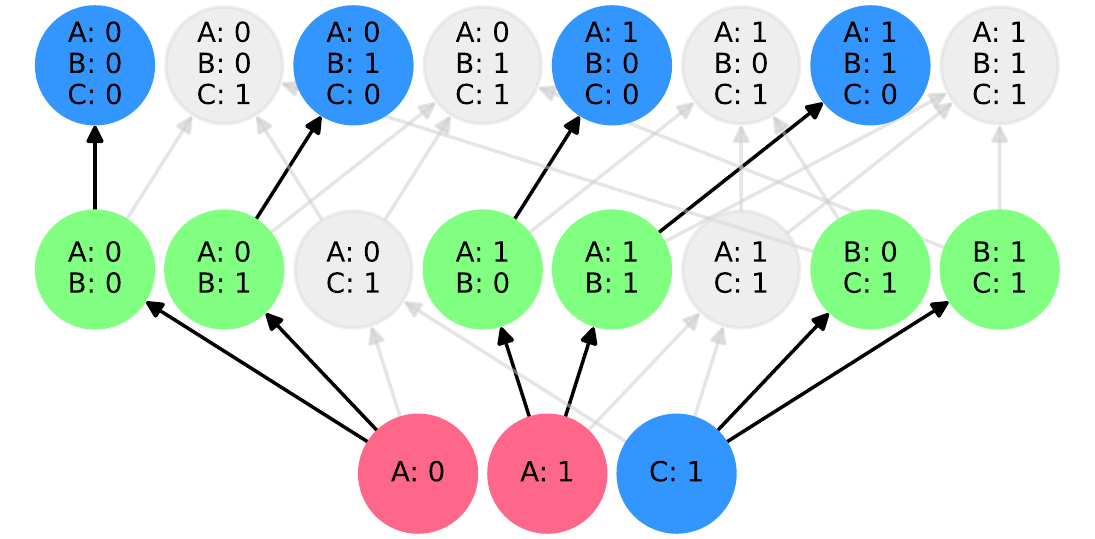}
    \\
    $\Theta_{51}$
    &
    $\Ext{\Theta_{51}}$
    \end{tabular}
\end{center}

\noindent The standard causaltope for Space 51 has dimension 33.
Below is a plot of the homogeneous linear system of causality and quasi-normalisation equations for the standard causaltope, put in reduced row echelon form:

\begin{center}
    \includegraphics[width=11cm]{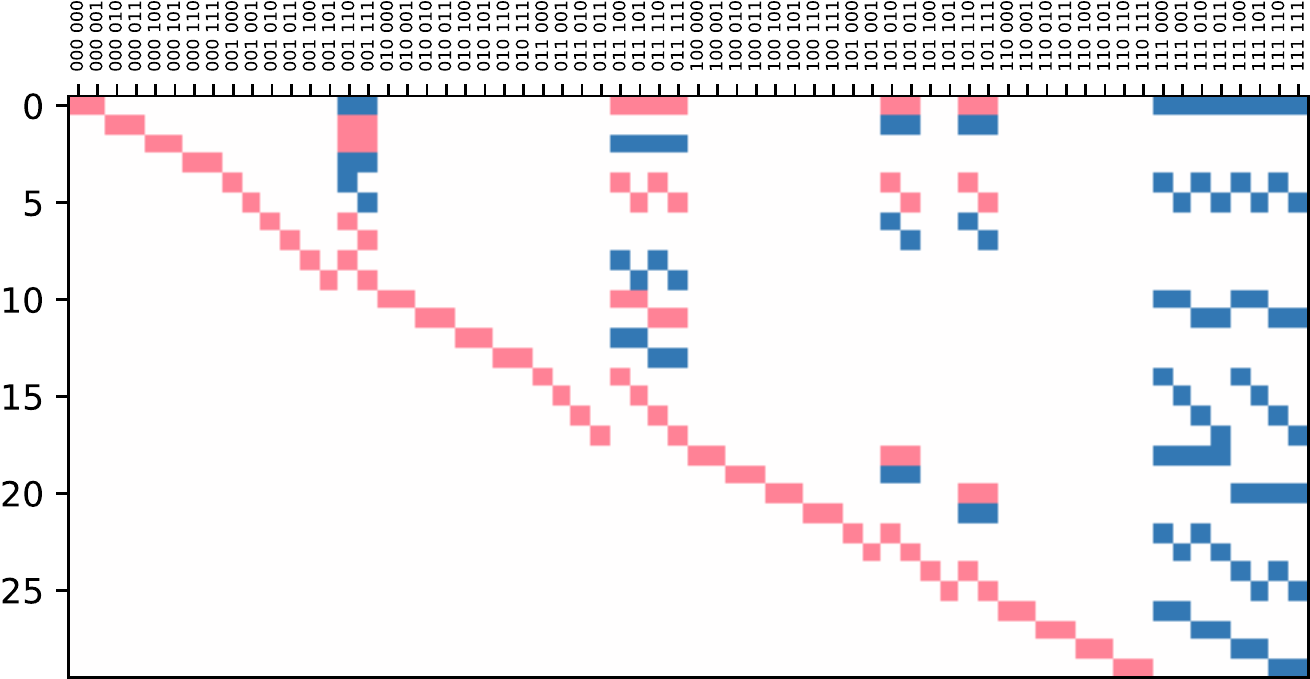}
\end{center}

\noindent Rows correspond to the 30 independent linear equations.
Columns in the plot correspond to entries of empirical models, indexed as $i_A i_B i_C$ $o_A o_B o_C$.
Coefficients in the equations are color-coded as white=0, red=+1 and blue=-1.

Space 51 has closest refinements in equivalence classes 31 and 42; 
it is the join of its (closest) refinements.
It has closest coarsenings in equivalence class 61; 
it is the meet of its (closest) coarsenings.
It has 512 causal functions, all of which are causal for at least one of its refinements.
It is not a tight space: for event \ev{B}, a causal function must yield identical output values on input histories \hist{A/0,B/0}, \hist{A/1,B/0} and \hist{B/0,C/1}, and it must also yield identical output values on input histories \hist{A/0,B/1}, \hist{A/1,B/1} and \hist{B/1,C/1}.

The standard causaltope for Space 51 coincides with that of its 2 subspaces in equivalence class 31.
The standard causaltope for Space 51 is the meet of the standard causaltopes for its closest coarsenings.
For completeness, below is a plot of the full homogeneous linear system of causality and quasi-normalisation equations for the standard causaltope:

\begin{center}
    \includegraphics[width=12cm]{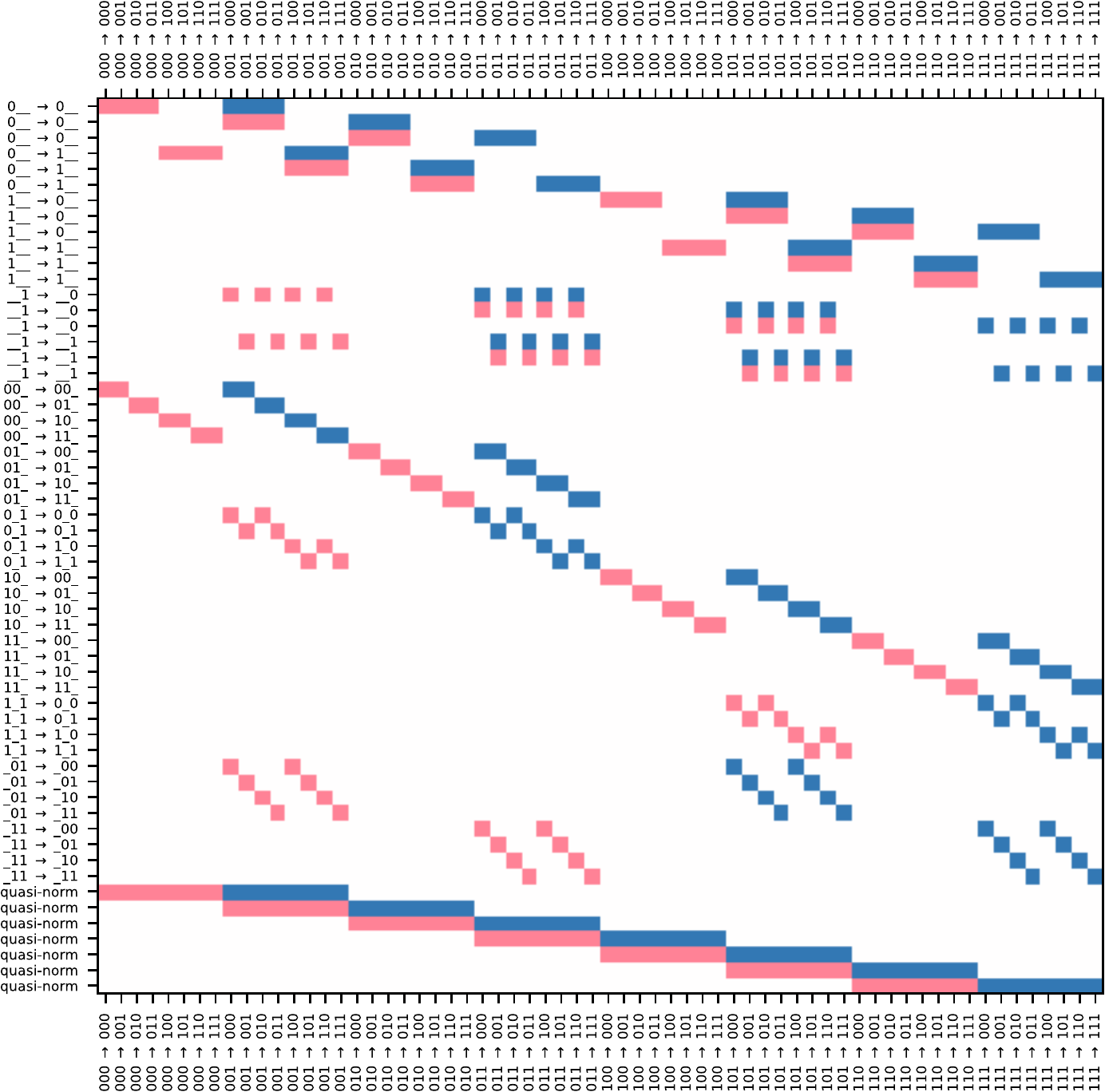}
\end{center}

\noindent Rows correspond to the 57 linear equations, of which 30 are independent.

\newpage
\subsection*{Space 52}

Space 52 is not induced by a causal order, but it is a refinement of the space induced by the indefinite causal order $\total{\ev{A},\{\ev{B},\ev{C}\}}$.
Its equivalence class under event-input permutation symmetry contains 48 spaces.
Space 52 differs as follows from the space induced by causal order $\total{\ev{A},\{\ev{B},\ev{C}\}}$:
\begin{itemize}
  \item The outputs at events \evset{\ev{A}, \ev{B}} are independent of the input at event \ev{C} when the inputs at events \evset{A, B} are given by \hist{A/0,B/0}, \hist{A/0,B/1} and \hist{A/1,B/0}.
  \item The outputs at events \evset{\ev{A}, \ev{C}} are independent of the input at event \ev{B} when the inputs at events \evset{A, C} are given by \hist{A/0,C/1}, \hist{A/1,C/0} and \hist{A/1,C/1}.
  \item The outputs at events \evset{\ev{B}, \ev{C}} are independent of the input at event \ev{A} when the inputs at events \evset{B, C} are given by \hist{B/1,C/1} and \hist{B/0,C/1}.
  \item The output at event \ev{C} is independent of the inputs at events \evset{\ev{A}, \ev{B}} when the input at event C is given by \hist{C/1}.
\end{itemize}

\noindent Below are the histories and extended histories for space 52: 
\begin{center}
    \begin{tabular}{cc}
    \includegraphics[height=3.5cm]{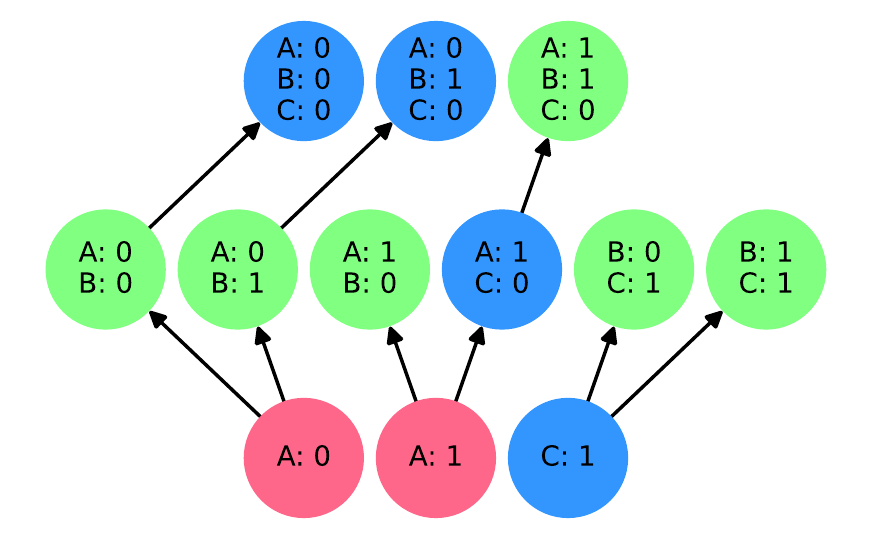}
    &
    \includegraphics[height=3.5cm]{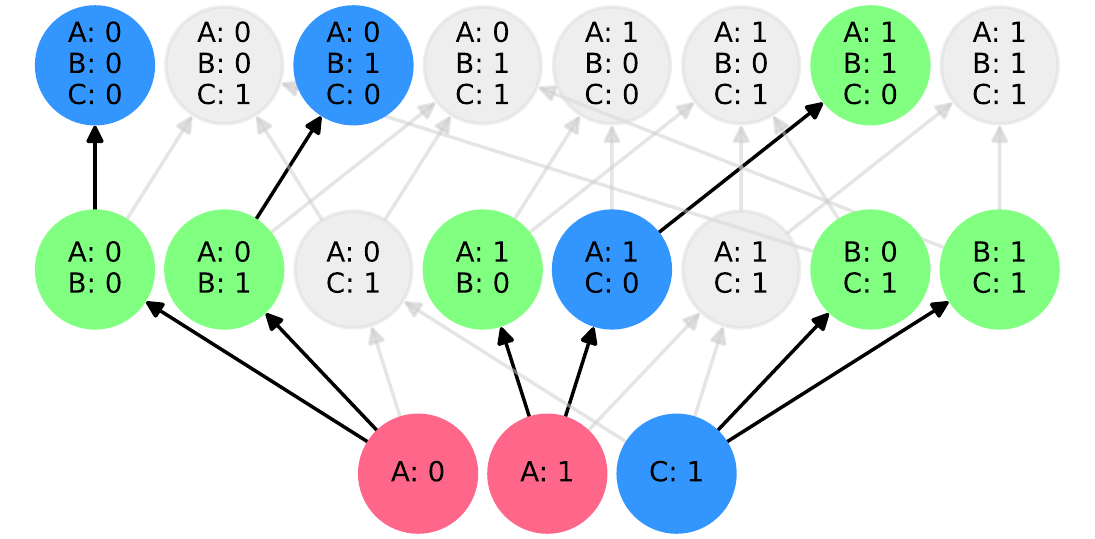}
    \\
    $\Theta_{52}$
    &
    $\Ext{\Theta_{52}}$
    \end{tabular}
\end{center}

\noindent The standard causaltope for Space 52 has dimension 33.
Below is a plot of the homogeneous linear system of causality and quasi-normalisation equations for the standard causaltope, put in reduced row echelon form:

\begin{center}
    \includegraphics[width=11cm]{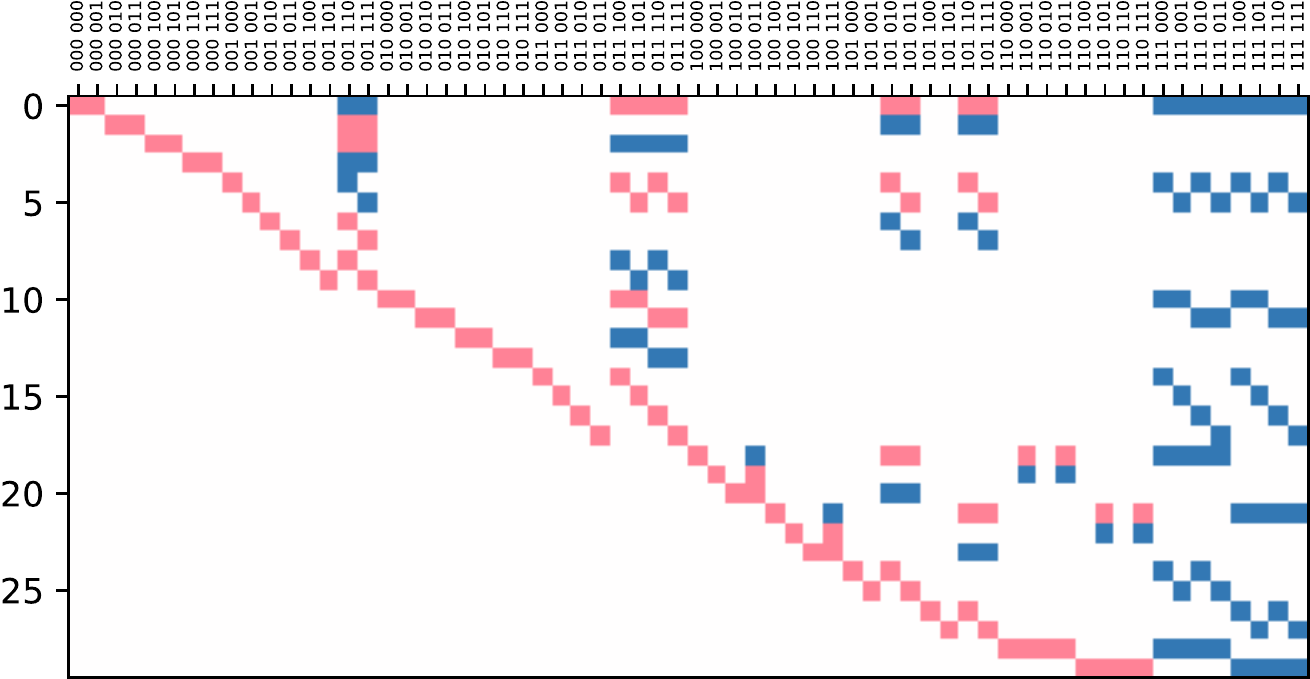}
\end{center}

\noindent Rows correspond to the 30 independent linear equations.
Columns in the plot correspond to entries of empirical models, indexed as $i_A i_B i_C$ $o_A o_B o_C$.
Coefficients in the equations are color-coded as white=0, red=+1 and blue=-1.

Space 52 has closest refinements in equivalence classes 32, 42 and 43; 
it is the join of its (closest) refinements.
It has closest coarsenings in equivalence classes 62, 71 and 72; 
it is the meet of its (closest) coarsenings.
It has 512 causal functions, all of which are causal for at least one of its refinements.
It is not a tight space: for event \ev{B}, a causal function must yield identical output values on input histories \hist{A/0,B/1} and \hist{B/1,C/1}, and it must also yield identical output values on input histories \hist{A/0,B/0}, \hist{A/1,B/0} and \hist{B/0,C/1}.

The standard causaltope for Space 52 coincides with that of its subspace in equivalence class 32.
The standard causaltope for Space 52 is the meet of the standard causaltopes for its closest coarsenings.
For completeness, below is a plot of the full homogeneous linear system of causality and quasi-normalisation equations for the standard causaltope:

\begin{center}
    \includegraphics[width=12cm]{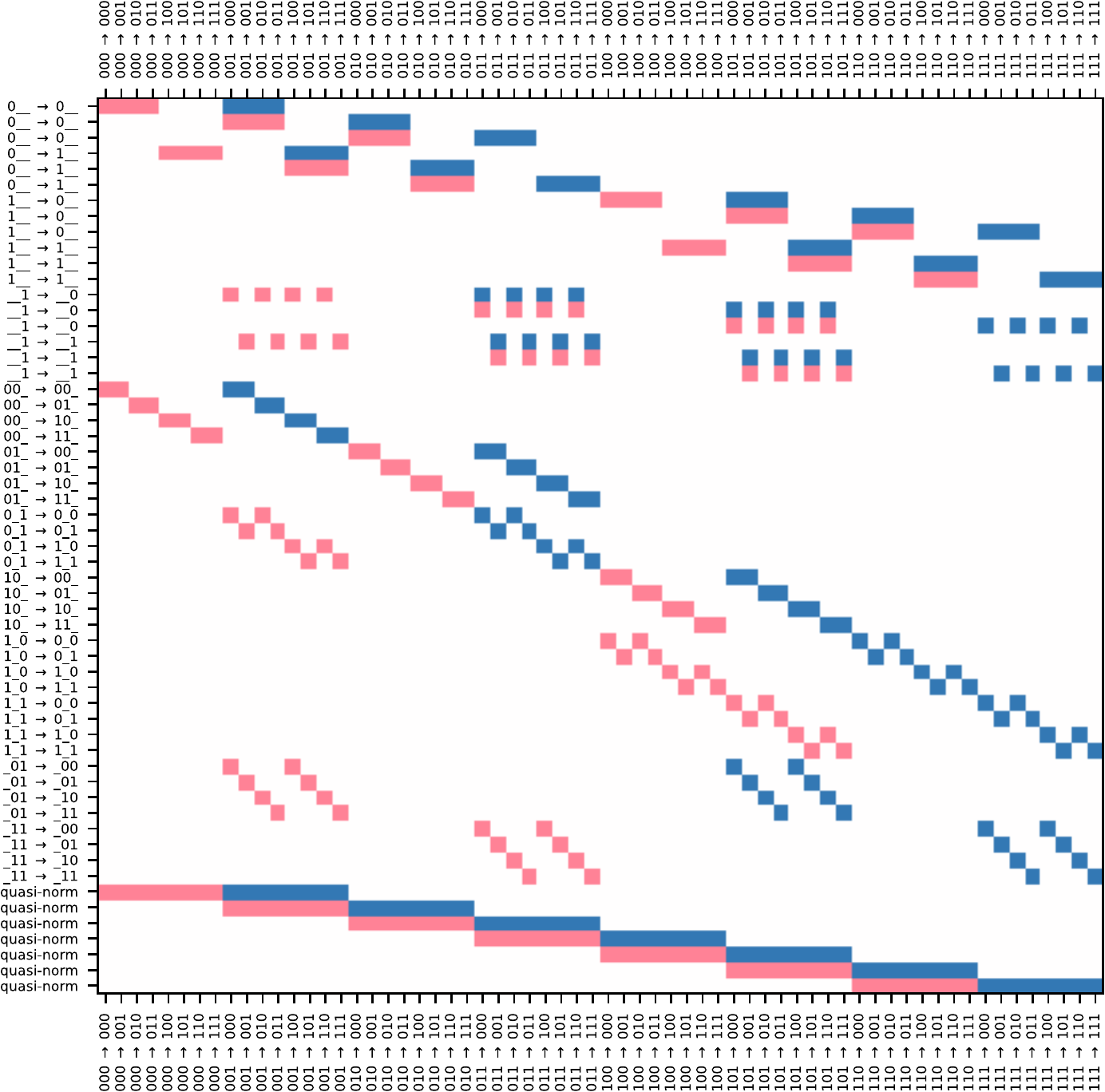}
\end{center}

\noindent Rows correspond to the 57 linear equations, of which 30 are independent.

\newpage
\subsection*{Space 53}

Space 53 is not induced by a causal order, but it is a refinement of the space in equivalence class 92 induced by the definite causal order $\total{\ev{A},\ev{B}}\vee\total{\ev{C},\ev{B}}$ (note that the space induced by the order is not the same as space 92).
Its equivalence class under event-input permutation symmetry contains 48 spaces.
Space 53 differs as follows from the space induced by causal order $\total{\ev{A},\ev{B}}\vee\total{\ev{C},\ev{B}}$:
\begin{itemize}
  \item The outputs at events \evset{\ev{A}, \ev{B}} are independent of the input at event \ev{C} when the inputs at events \evset{A, B} are given by \hist{A/0,B/0} and \hist{A/0,B/1}.
  \item The outputs at events \evset{\ev{B}, \ev{C}} are independent of the input at event \ev{A} when the inputs at events \evset{B, C} are given by \hist{B/1,C/1}.
\end{itemize}

\noindent Below are the histories and extended histories for space 53: 
\begin{center}
    \begin{tabular}{cc}
    \includegraphics[height=3.5cm]{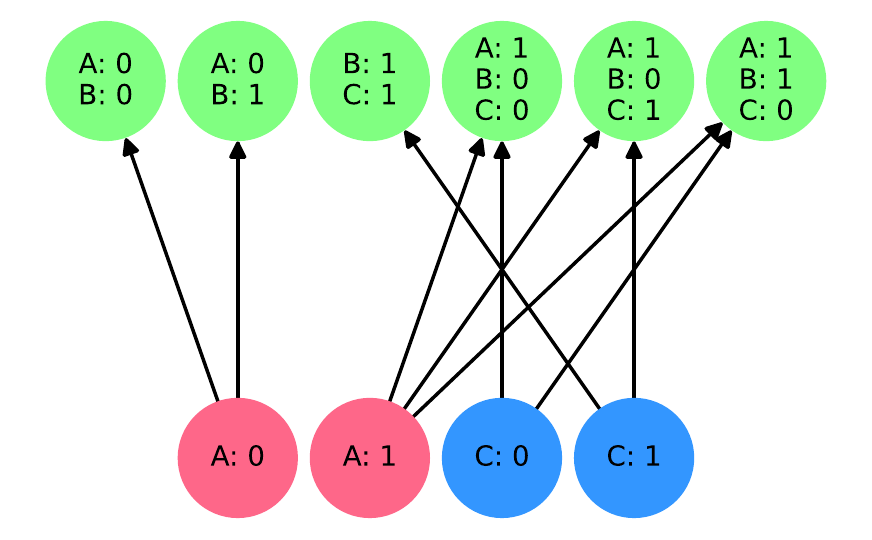}
    &
    \includegraphics[height=3.5cm]{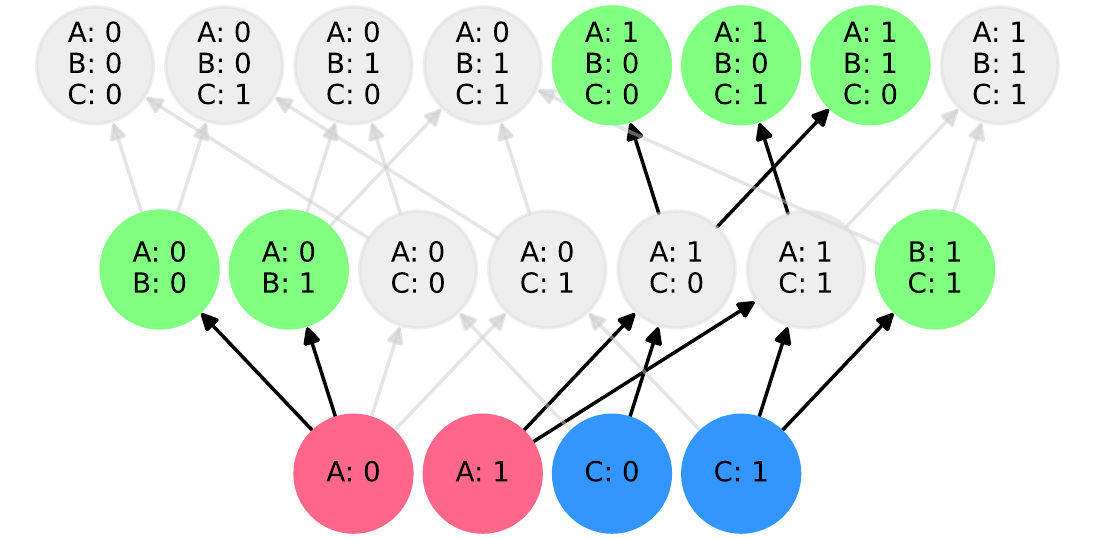}
    \\
    $\Theta_{53}$
    &
    $\Ext{\Theta_{53}}$
    \end{tabular}
\end{center}

\noindent The standard causaltope for Space 53 has dimension 34.
Below is a plot of the homogeneous linear system of causality and quasi-normalisation equations for the standard causaltope, put in reduced row echelon form:

\begin{center}
    \includegraphics[width=11cm]{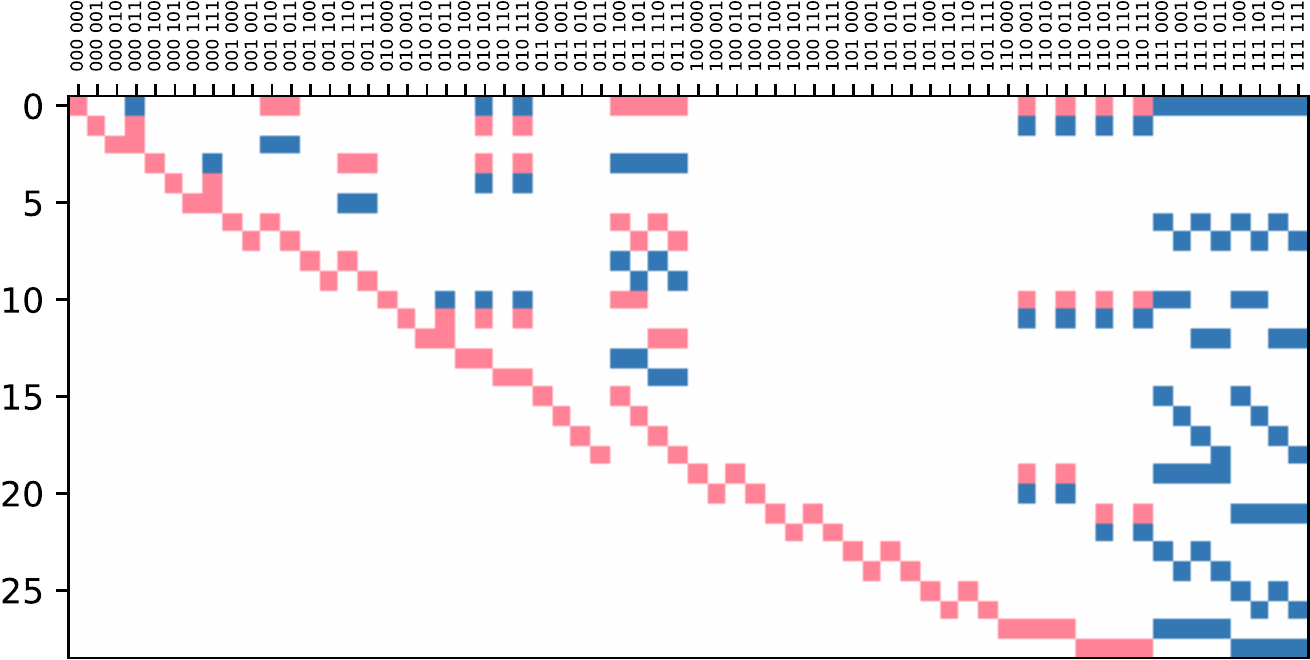}
\end{center}

\noindent Rows correspond to the 29 independent linear equations.
Columns in the plot correspond to entries of empirical models, indexed as $i_A i_B i_C$ $o_A o_B o_C$.
Coefficients in the equations are color-coded as white=0, red=+1 and blue=-1.

Space 53 has closest refinements in equivalence classes 29, 34, 35, 36 and 37; 
it is the join of its (closest) refinements.
It has closest coarsenings in equivalence classes 63, 65, 66, 67 and 69; 
it is the meet of its (closest) coarsenings.
It has 512 causal functions, 64 of which are not causal for any of its refinements.
It is not a tight space: for event \ev{B}, a causal function must yield identical output values on input histories \hist{A/0,B/1} and \hist{B/1,C/1}.

The standard causaltope for Space 53 has 2 more dimensions than those of its 5 subspaces in equivalence classes 29, 34, 35, 36 and 37.
The standard causaltope for Space 53 is the meet of the standard causaltopes for its closest coarsenings.
For completeness, below is a plot of the full homogeneous linear system of causality and quasi-normalisation equations for the standard causaltope:

\begin{center}
    \includegraphics[width=12cm]{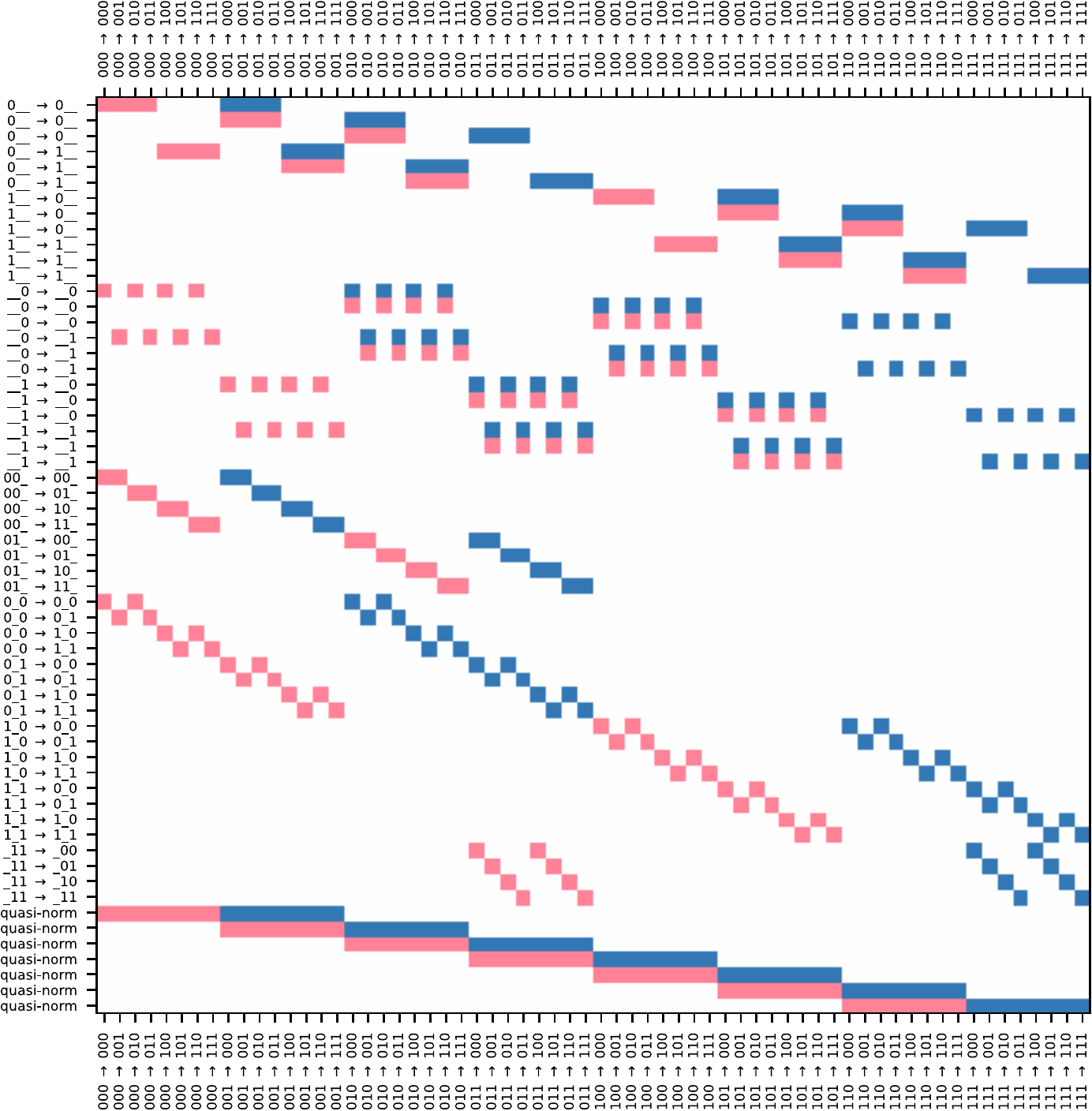}
\end{center}

\noindent Rows correspond to the 59 linear equations, of which 29 are independent.

\newpage
\subsection*{Space 54}

Space 54 is not induced by a causal order, but it is a refinement of the space in equivalence class 100 induced by the definite causal order $\total{\ev{C},\ev{A},\ev{B}}$ (note that the space induced by the order is not the same as space 100).
Its equivalence class under event-input permutation symmetry contains 24 spaces.
Space 54 differs as follows from the space induced by causal order $\total{\ev{C},\ev{A},\ev{B}}$:
\begin{itemize}
  \item The outputs at events \evset{\ev{A}, \ev{B}} are independent of the input at event \ev{C} when the inputs at events \evset{A, B} are given by \hist{A/0,B/0} and \hist{A/0,B/1}.
  \item The outputs at events \evset{\ev{B}, \ev{C}} are independent of the input at event \ev{A} when the inputs at events \evset{B, C} are given by \hist{B/1,C/0} and \hist{B/1,C/1}.
  \item The output at event \ev{A} is independent of the input at event \ev{C} when the input at event A is given by \hist{A/0}.
\end{itemize}

\noindent Below are the histories and extended histories for space 54: 
\begin{center}
    \begin{tabular}{cc}
    \includegraphics[height=3.5cm]{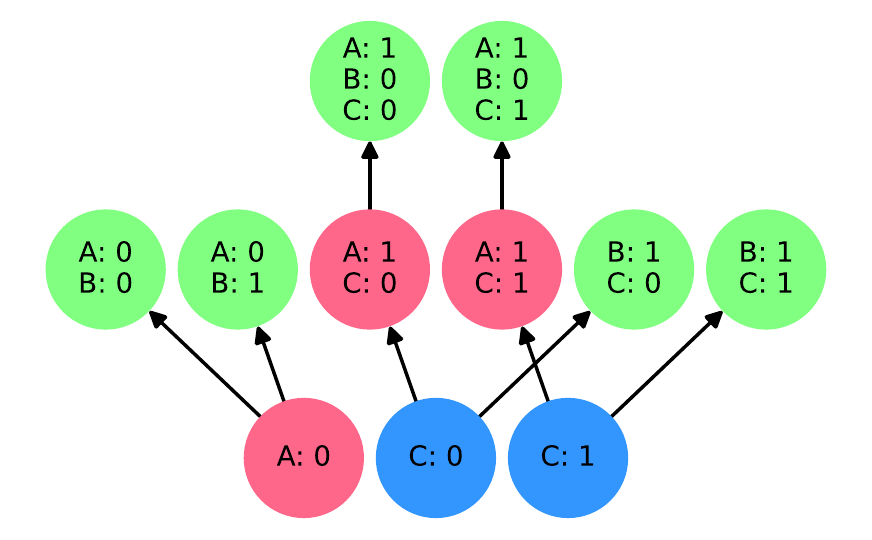}
    &
    \includegraphics[height=3.5cm]{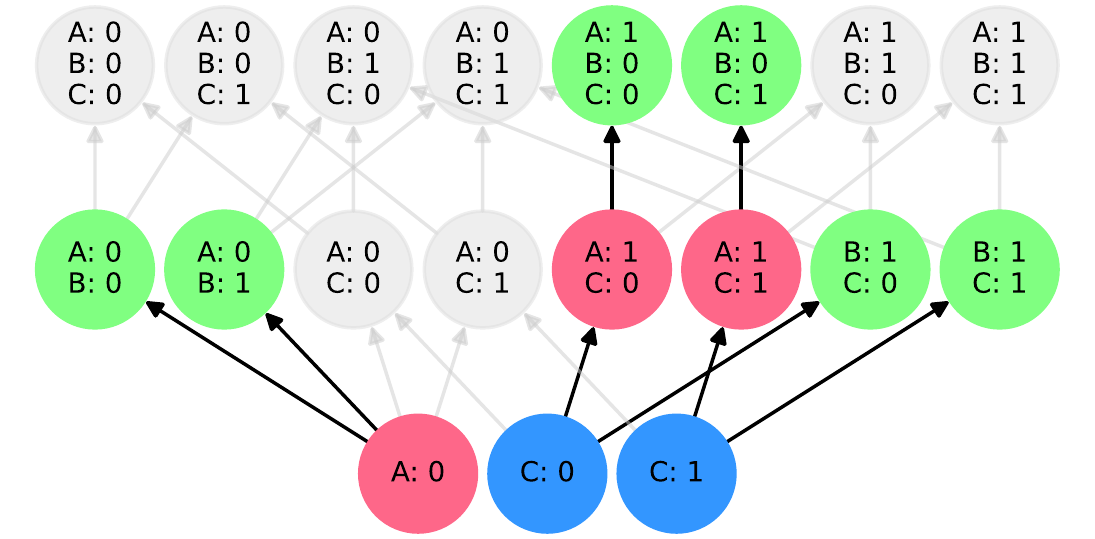}
    \\
    $\Theta_{54}$
    &
    $\Ext{\Theta_{54}}$
    \end{tabular}
\end{center}

\noindent The standard causaltope for Space 54 has dimension 33.
Below is a plot of the homogeneous linear system of causality and quasi-normalisation equations for the standard causaltope, put in reduced row echelon form:

\begin{center}
    \includegraphics[width=11cm]{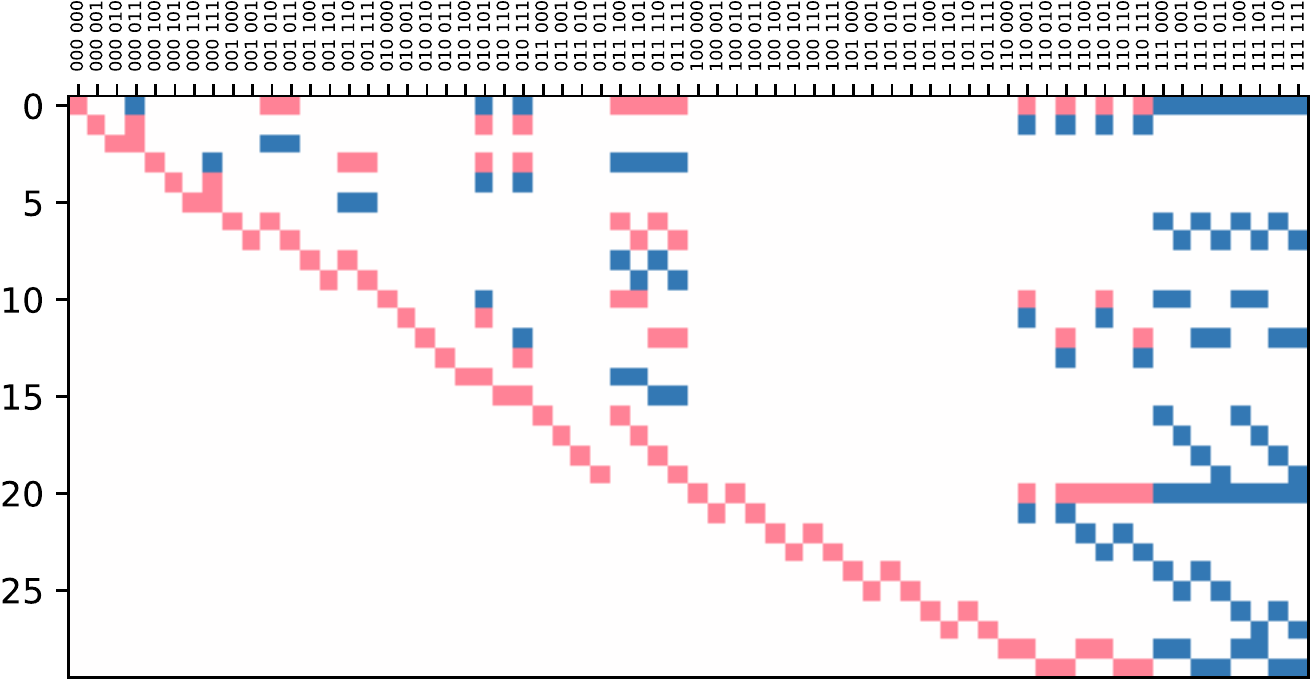}
\end{center}

\noindent Rows correspond to the 30 independent linear equations.
Columns in the plot correspond to entries of empirical models, indexed as $i_A i_B i_C$ $o_A o_B o_C$.
Coefficients in the equations are color-coded as white=0, red=+1 and blue=-1.

Space 54 has closest refinements in equivalence classes 28, 35 and 43; 
it is the join of its (closest) refinements.
It has closest coarsenings in equivalence classes 65, 75 and 76; 
it is the meet of its (closest) coarsenings.
It has 512 causal functions, 256 of which are not causal for any of its refinements.
It is not a tight space: for event \ev{B}, a causal function must yield identical output values on input histories \hist{A/0,B/1}, \hist{B/1,C/0} and \hist{B/1,C/1}.

The standard causaltope for Space 54 coincides with that of its subspace in equivalence class 28.
The standard causaltope for Space 54 is the meet of the standard causaltopes for its closest coarsenings.
For completeness, below is a plot of the full homogeneous linear system of causality and quasi-normalisation equations for the standard causaltope:

\begin{center}
    \includegraphics[width=12cm]{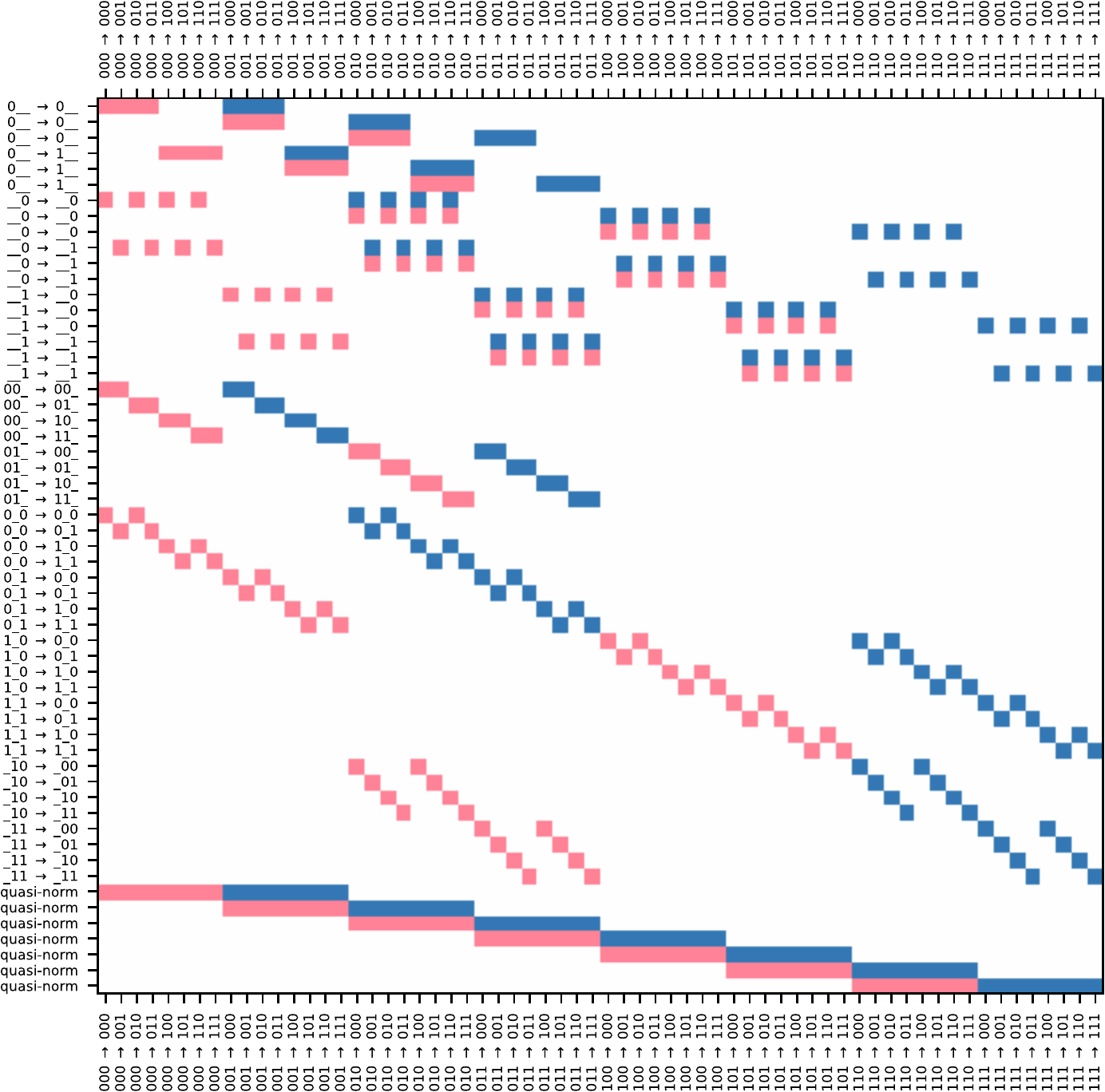}
\end{center}

\noindent Rows correspond to the 57 linear equations, of which 30 are independent.

\newpage
\subsection*{Space 55}

Space 55 is not induced by a causal order, but it is a refinement of the space in equivalence class 100 induced by the definite causal order $\total{\ev{C},\ev{A},\ev{B}}$ (note that the space induced by the order is not the same as space 100).
Its equivalence class under event-input permutation symmetry contains 24 spaces.
Space 55 differs as follows from the space induced by causal order $\total{\ev{C},\ev{A},\ev{B}}$:
\begin{itemize}
  \item The outputs at events \evset{\ev{A}, \ev{B}} are independent of the input at event \ev{C} when the inputs at events \evset{A, B} are given by \hist{A/0,B/0} and \hist{A/0,B/1}.
  \item The outputs at events \evset{\ev{B}, \ev{C}} are independent of the input at event \ev{A} when the inputs at events \evset{B, C} are given by \hist{B/1,C/0} and \hist{B/0,C/1}.
  \item The output at event \ev{A} is independent of the input at event \ev{C} when the input at event A is given by \hist{A/0}.
\end{itemize}

\noindent Below are the histories and extended histories for space 55: 
\begin{center}
    \begin{tabular}{cc}
    \includegraphics[height=3.5cm]{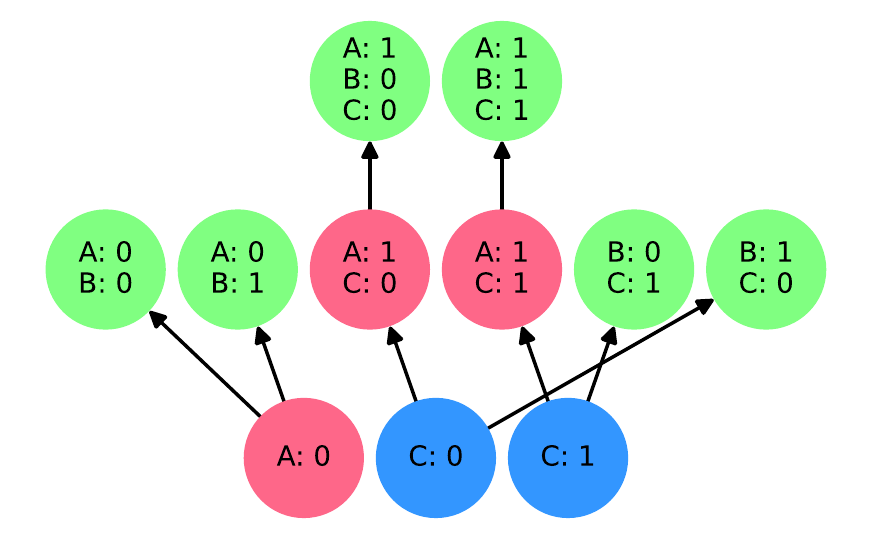}
    &
    \includegraphics[height=3.5cm]{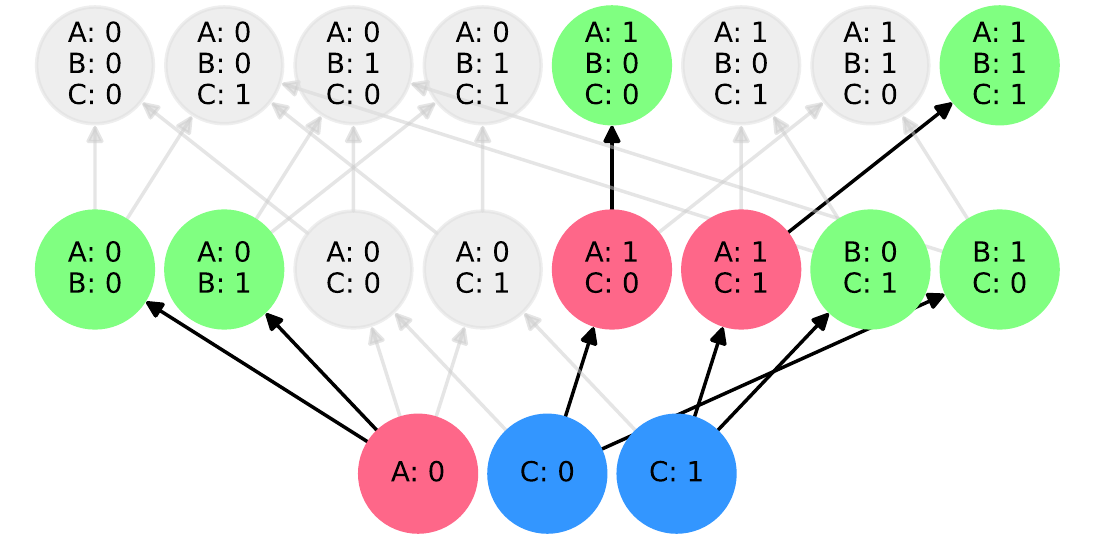}
    \\
    $\Theta_{55}$
    &
    $\Ext{\Theta_{55}}$
    \end{tabular}
\end{center}

\noindent The standard causaltope for Space 55 has dimension 33.
Below is a plot of the homogeneous linear system of causality and quasi-normalisation equations for the standard causaltope, put in reduced row echelon form:

\begin{center}
    \includegraphics[width=11cm]{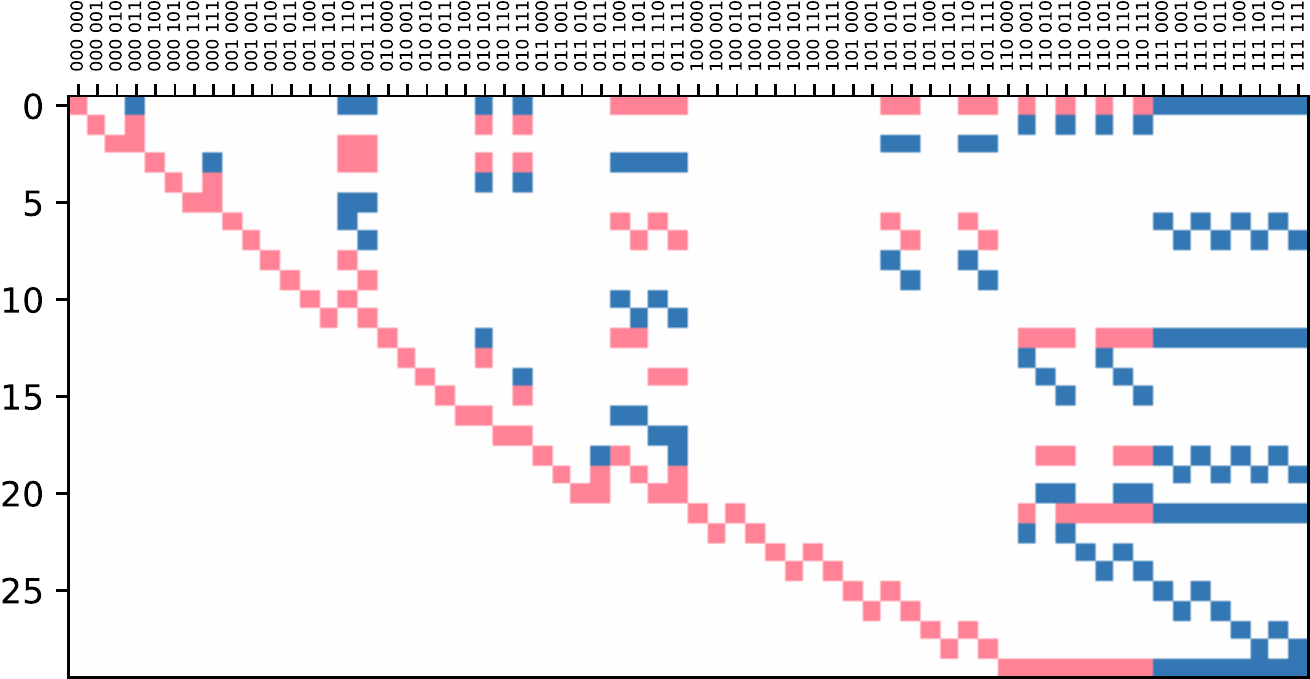}
\end{center}

\noindent Rows correspond to the 30 independent linear equations.
Columns in the plot correspond to entries of empirical models, indexed as $i_A i_B i_C$ $o_A o_B o_C$.
Coefficients in the equations are color-coded as white=0, red=+1 and blue=-1.

Space 55 has closest refinements in equivalence classes 36 and 43; 
it is the join of its (closest) refinements.
It has closest coarsenings in equivalence classes 65 and 73; 
it is the meet of its (closest) coarsenings.
It has 512 causal functions, 192 of which are not causal for any of its refinements.
It is not a tight space: for event \ev{B}, a causal function must yield identical output values on input histories \hist{A/0,B/0} and \hist{B/0,C/1}, and it must also yield identical output values on input histories \hist{A/0,B/1} and \hist{B/1,C/0}.

The standard causaltope for Space 55 has 1 more dimension than that of its subspace in equivalence class 36.
The standard causaltope for Space 55 is the meet of the standard causaltopes for its closest coarsenings.
For completeness, below is a plot of the full homogeneous linear system of causality and quasi-normalisation equations for the standard causaltope:

\begin{center}
    \includegraphics[width=12cm]{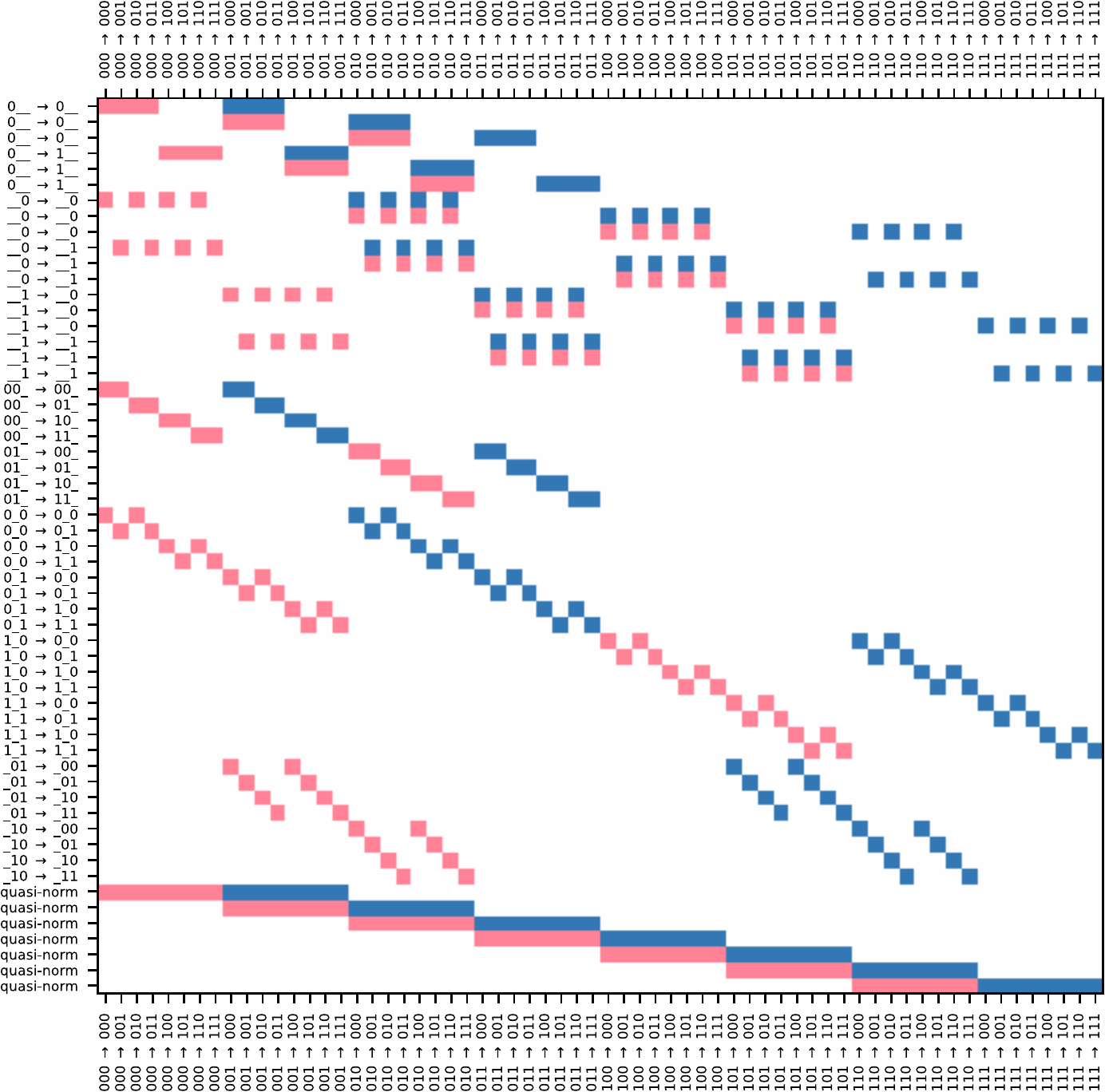}
\end{center}

\noindent Rows correspond to the 57 linear equations, of which 30 are independent.

\newpage
\subsection*{Space 56}

Space 56 is not induced by a causal order, but it is a refinement of the space in equivalence class 100 induced by the definite causal order $\total{\ev{A},\ev{C},\ev{B}}$ (note that the space induced by the order is not the same as space 100).
Its equivalence class under event-input permutation symmetry contains 48 spaces.
Space 56 differs as follows from the space induced by causal order $\total{\ev{A},\ev{C},\ev{B}}$:
\begin{itemize}
  \item The outputs at events \evset{\ev{A}, \ev{B}} are independent of the input at event \ev{C} when the inputs at events \evset{A, B} are given by \hist{A/0,B/0}, \hist{A/0,B/1} and \hist{A/1,B/0}.
  \item The outputs at events \evset{\ev{B}, \ev{C}} are independent of the input at event \ev{A} when the inputs at events \evset{B, C} are given by \hist{B/1,C/0}.
  \item The output at event \ev{C} is independent of the input at event \ev{A} when the input at event C is given by \hist{C/0}.
\end{itemize}

\noindent Below are the histories and extended histories for space 56: 
\begin{center}
    \begin{tabular}{cc}
    \includegraphics[height=3.5cm]{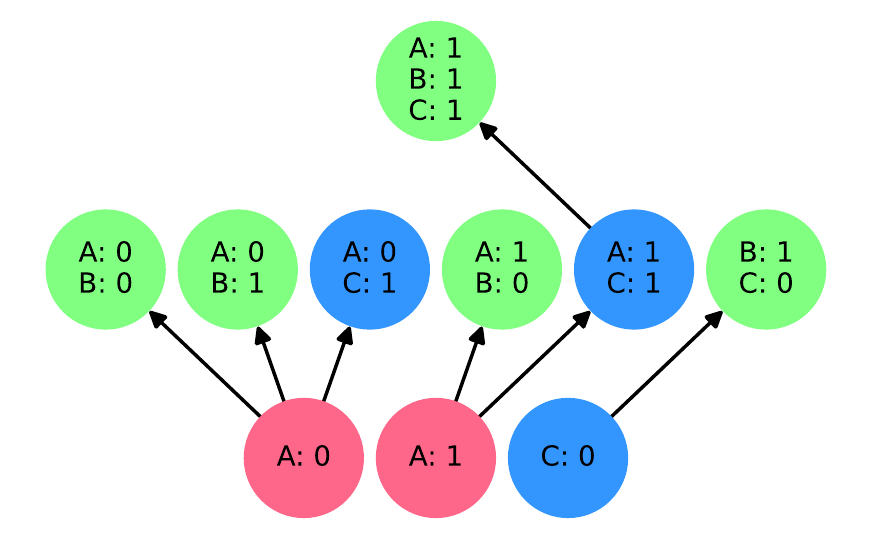}
    &
    \includegraphics[height=3.5cm]{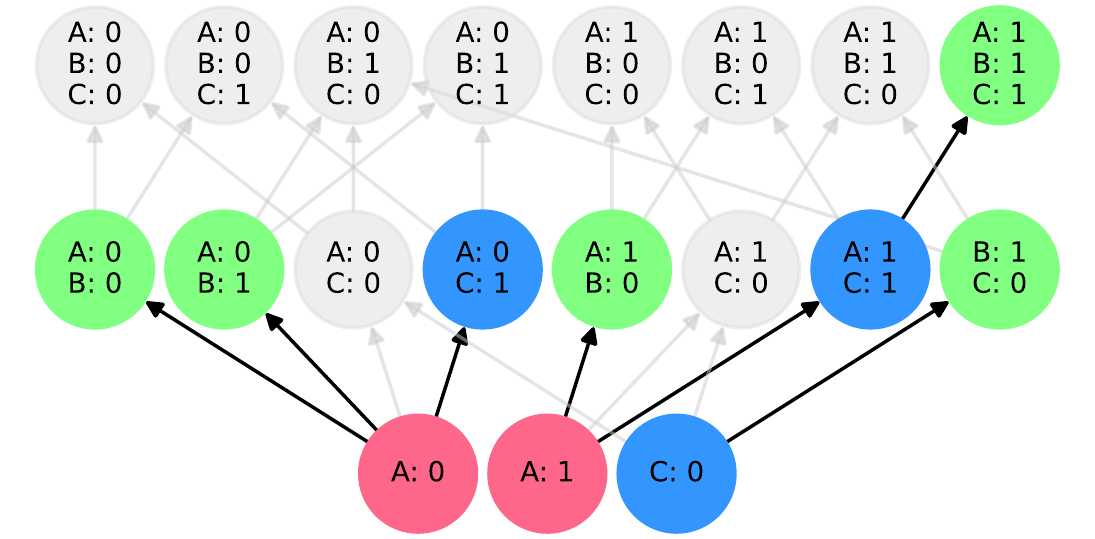}
    \\
    $\Theta_{56}$
    &
    $\Ext{\Theta_{56}}$
    \end{tabular}
\end{center}

\noindent The standard causaltope for Space 56 has dimension 33.
Below is a plot of the homogeneous linear system of causality and quasi-normalisation equations for the standard causaltope, put in reduced row echelon form:

\begin{center}
    \includegraphics[width=11cm]{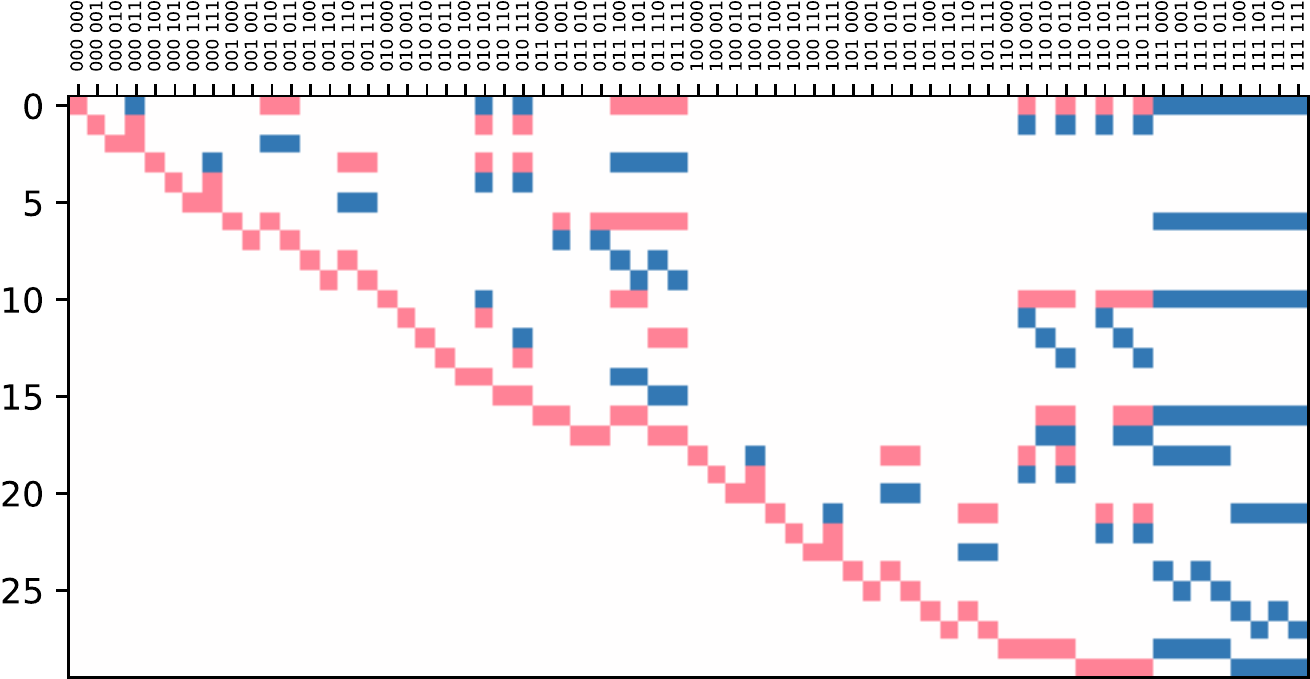}
\end{center}

\noindent Rows correspond to the 30 independent linear equations.
Columns in the plot correspond to entries of empirical models, indexed as $i_A i_B i_C$ $o_A o_B o_C$.
Coefficients in the equations are color-coded as white=0, red=+1 and blue=-1.

Space 56 has closest refinements in equivalence classes 29, 38 and 43; 
it is the join of its (closest) refinements.
It has closest coarsenings in equivalence classes 62, 63, 73, 74 and 75; 
it is the meet of its (closest) coarsenings.
It has 512 causal functions, 64 of which are not causal for any of its refinements.
It is not a tight space: for event \ev{B}, a causal function must yield identical output values on input histories \hist{A/0,B/1} and \hist{B/1,C/0}.

The standard causaltope for Space 56 has 1 more dimension than that of its subspace in equivalence class 29.
The standard causaltope for Space 56 is the meet of the standard causaltopes for its closest coarsenings.
For completeness, below is a plot of the full homogeneous linear system of causality and quasi-normalisation equations for the standard causaltope:

\begin{center}
    \includegraphics[width=12cm]{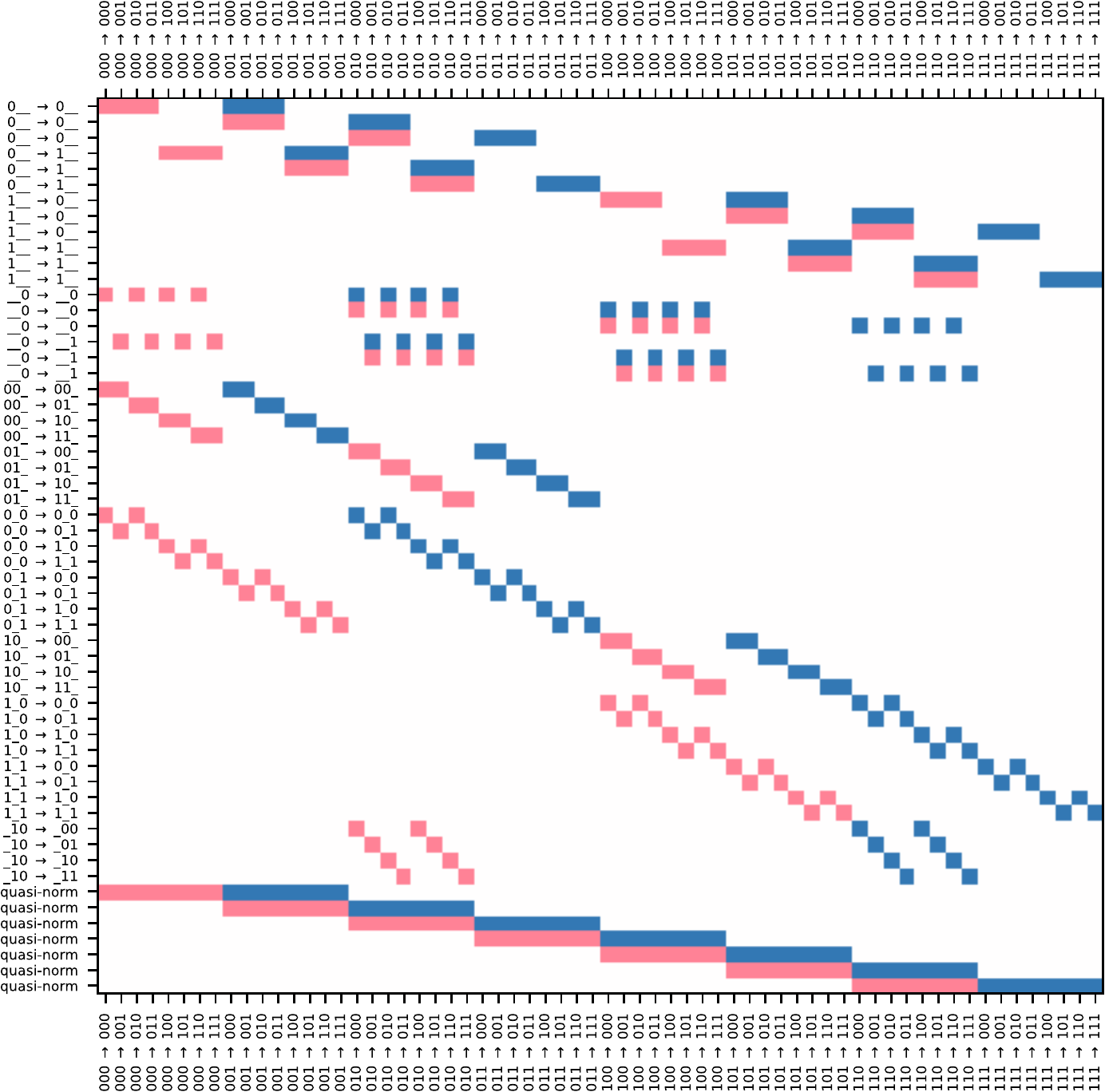}
\end{center}

\noindent Rows correspond to the 57 linear equations, of which 30 are independent.

\newpage
\subsection*{Space 57}

Space 57 is not induced by a causal order, but it is a refinement of the space 100 induced by the definite causal order $\total{\ev{A},\ev{B},\ev{C}}$.
Its equivalence class under event-input permutation symmetry contains 48 spaces.
Space 57 differs as follows from the space induced by causal order $\total{\ev{A},\ev{B},\ev{C}}$:
\begin{itemize}
  \item The outputs at events \evset{\ev{A}, \ev{C}} are independent of the input at event \ev{B} when the inputs at events \evset{A, C} are given by \hist{A/0,C/1}, \hist{A/1,C/0} and \hist{A/1,C/1}.
  \item The outputs at events \evset{\ev{B}, \ev{C}} are independent of the input at event \ev{A} when the inputs at events \evset{B, C} are given by \hist{B/1,C/1}.
  \item The output at event \ev{B} is independent of the input at event \ev{A} when the input at event B is given by \hist{B/1}.
\end{itemize}

\noindent Below are the histories and extended histories for space 57: 
\begin{center}
    \begin{tabular}{cc}
    \includegraphics[height=3.5cm]{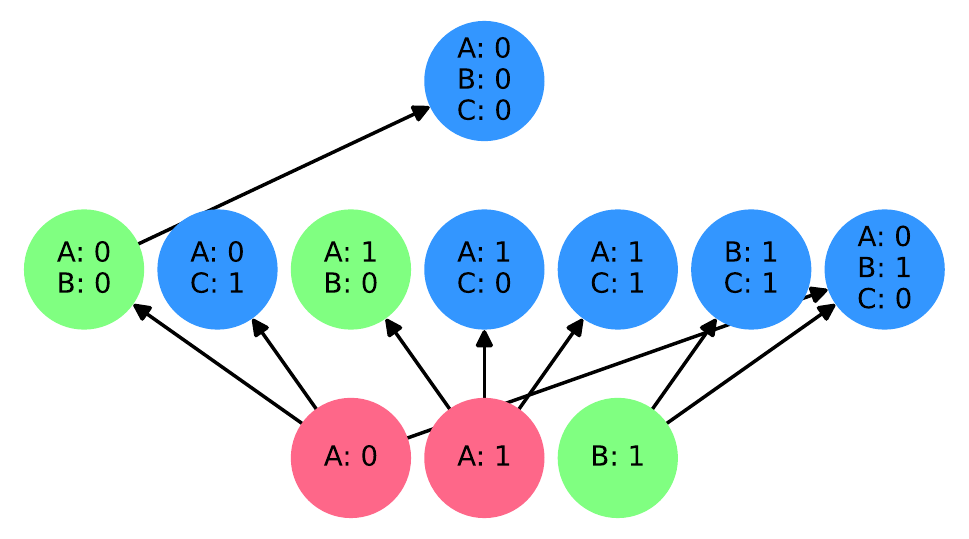}
    &
    \includegraphics[height=3.5cm]{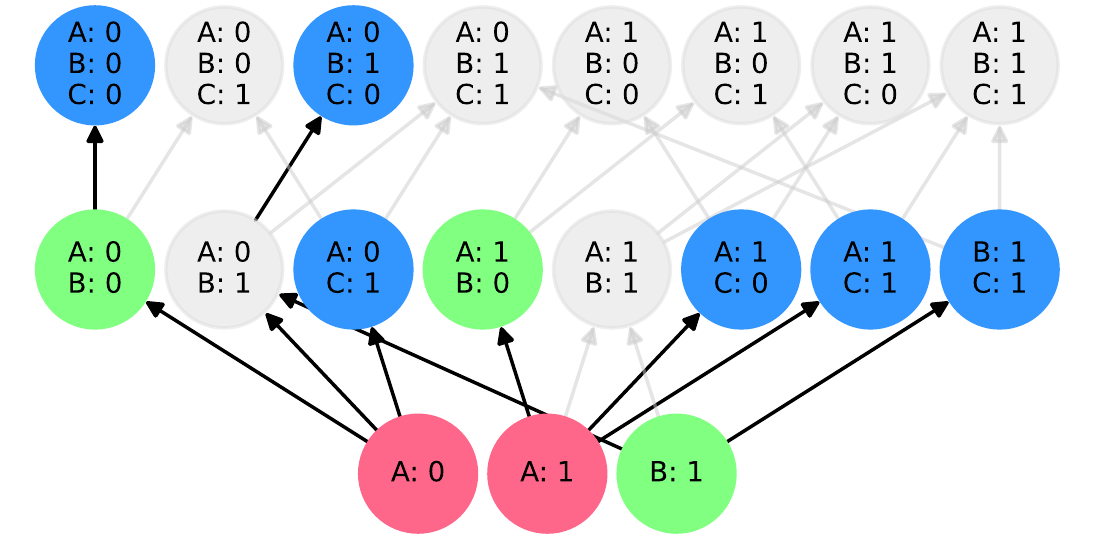}
    \\
    $\Theta_{57}$
    &
    $\Ext{\Theta_{57}}$
    \end{tabular}
\end{center}

\noindent The standard causaltope for Space 57 has dimension 33.
Below is a plot of the homogeneous linear system of causality and quasi-normalisation equations for the standard causaltope, put in reduced row echelon form:

\begin{center}
    \includegraphics[width=11cm]{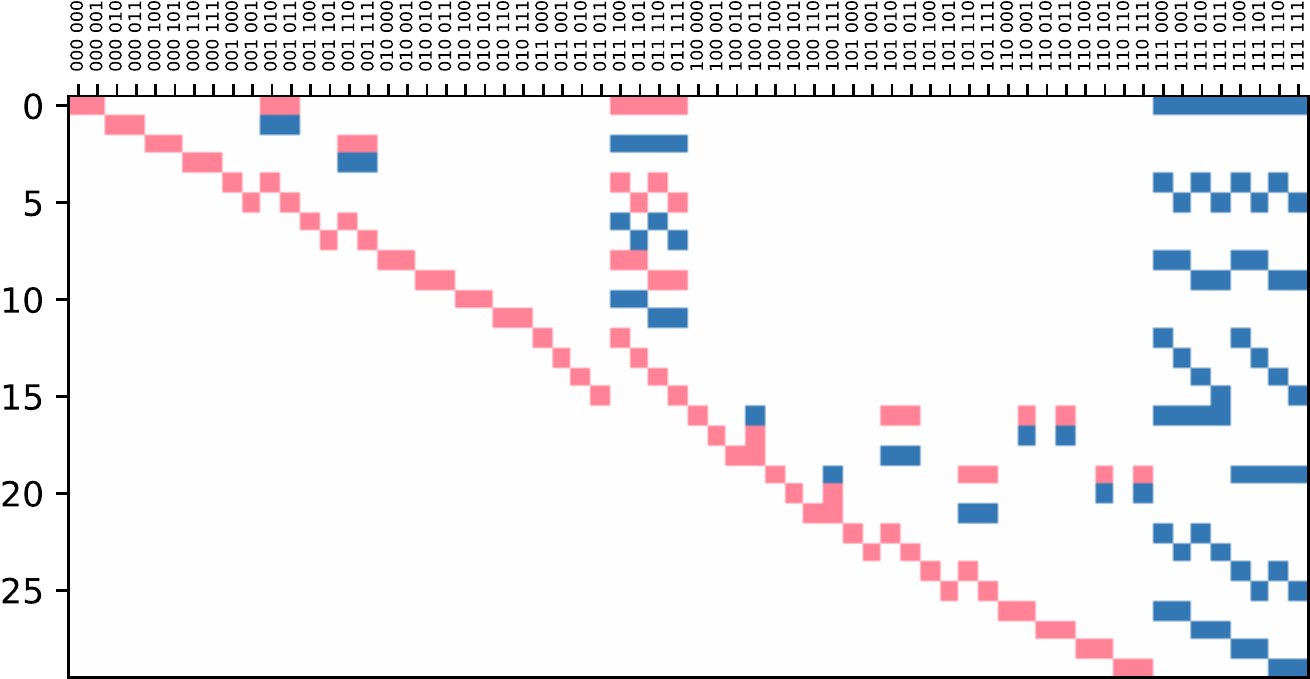}
\end{center}

\noindent Rows correspond to the 30 independent linear equations.
Columns in the plot correspond to entries of empirical models, indexed as $i_A i_B i_C$ $o_A o_B o_C$.
Coefficients in the equations are color-coded as white=0, red=+1 and blue=-1.

Space 57 has closest refinements in equivalence classes 30, 34, 38 and 43; 
it is the join of its (closest) refinements.
It has closest coarsenings in equivalence classes 63, 71, 73, 74 and 76; 
it is the meet of its (closest) coarsenings.
It has 512 causal functions, all of which are causal for at least one of its refinements.
It is not a tight space: for event \ev{C}, a causal function must yield identical output values on input histories \hist{A/0,C/1}, \hist{A/1,C/1} and \hist{B/1,C/1}.

The standard causaltope for Space 57 coincides with that of its subspace in equivalence class 30.
The standard causaltope for Space 57 is the meet of the standard causaltopes for its closest coarsenings.
For completeness, below is a plot of the full homogeneous linear system of causality and quasi-normalisation equations for the standard causaltope:

\begin{center}
    \includegraphics[width=12cm]{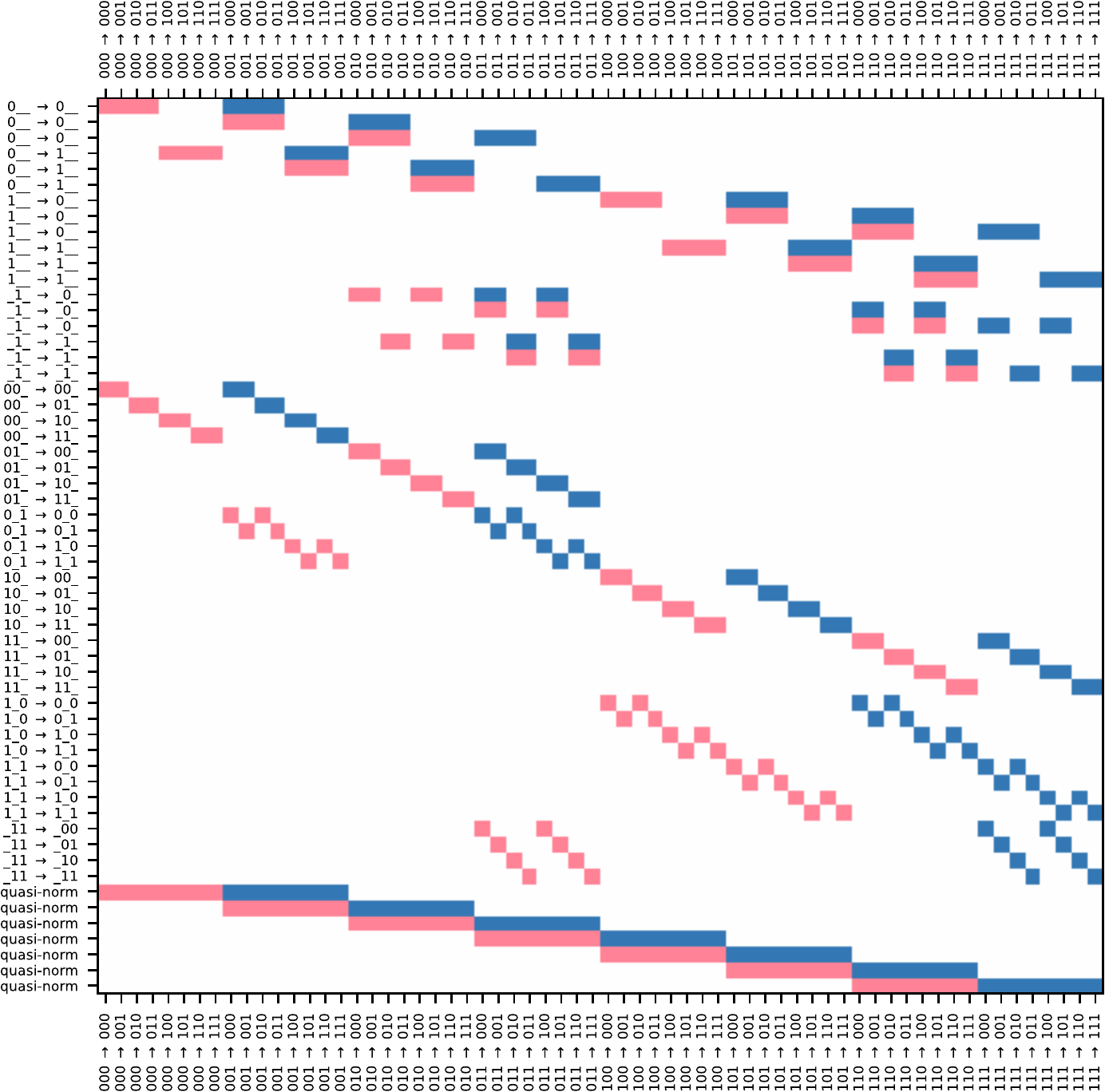}
\end{center}

\noindent Rows correspond to the 57 linear equations, of which 30 are independent.

\newpage
\subsection*{Space 58}

Space 58 is not induced by a causal order, but it is a refinement of the space 77 induced by the definite causal order $\total{\ev{A},\ev{B}}\vee\total{\ev{A},\ev{C}}$.
Its equivalence class under event-input permutation symmetry contains 12 spaces.
Space 58 differs as follows from the space induced by causal order $\total{\ev{A},\ev{B}}\vee\total{\ev{A},\ev{C}}$:
\begin{itemize}
  \item The output at event \ev{C} is independent of the input at event \ev{A} when the input at event C is given by \hist{C/0}.
\end{itemize}

\noindent Below are the histories and extended histories for space 58: 
\begin{center}
    \begin{tabular}{cc}
    \includegraphics[height=3.5cm]{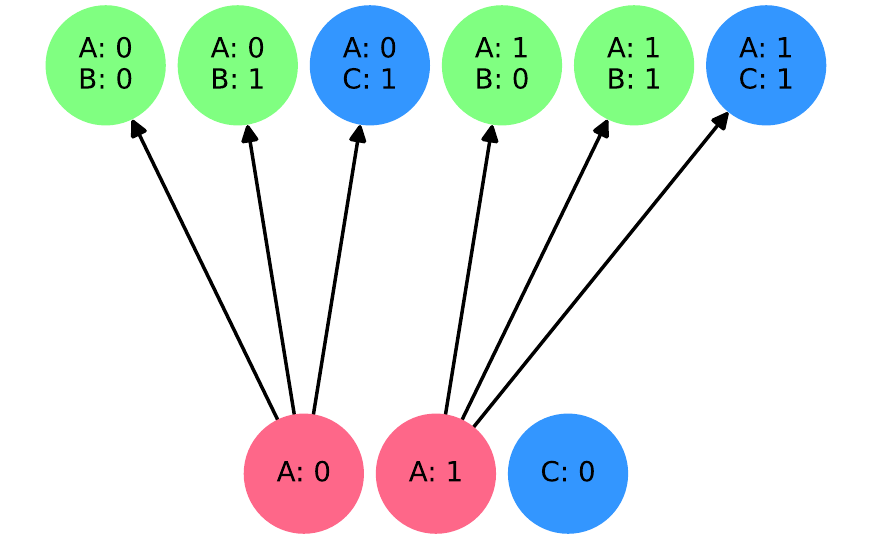}
    &
    \includegraphics[height=3.5cm]{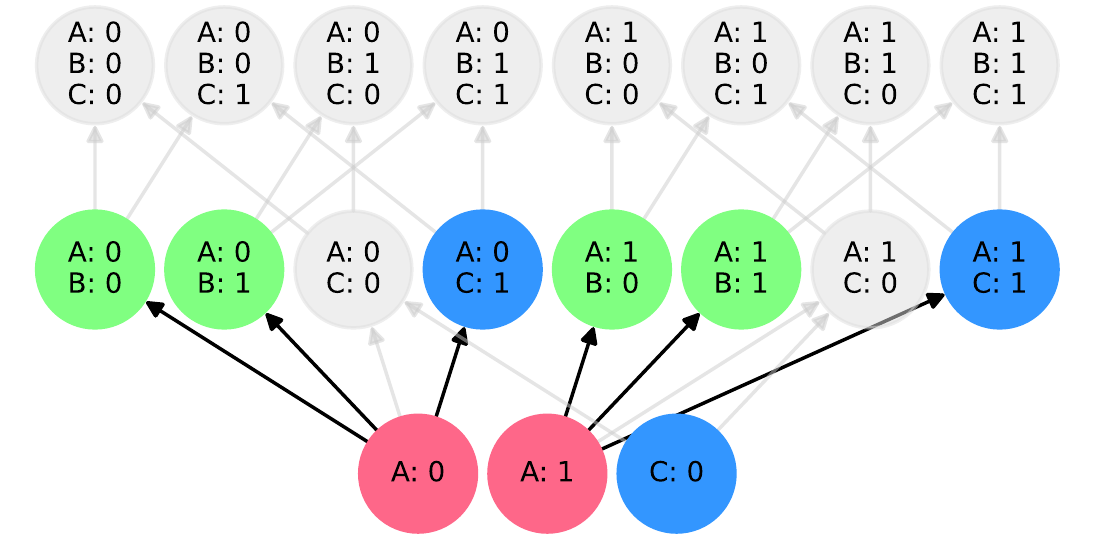}
    \\
    $\Theta_{58}$
    &
    $\Ext{\Theta_{58}}$
    \end{tabular}
\end{center}

\noindent The standard causaltope for Space 58 has dimension 33.
Below is a plot of the homogeneous linear system of causality and quasi-normalisation equations for the standard causaltope, put in reduced row echelon form:

\begin{center}
    \includegraphics[width=11cm]{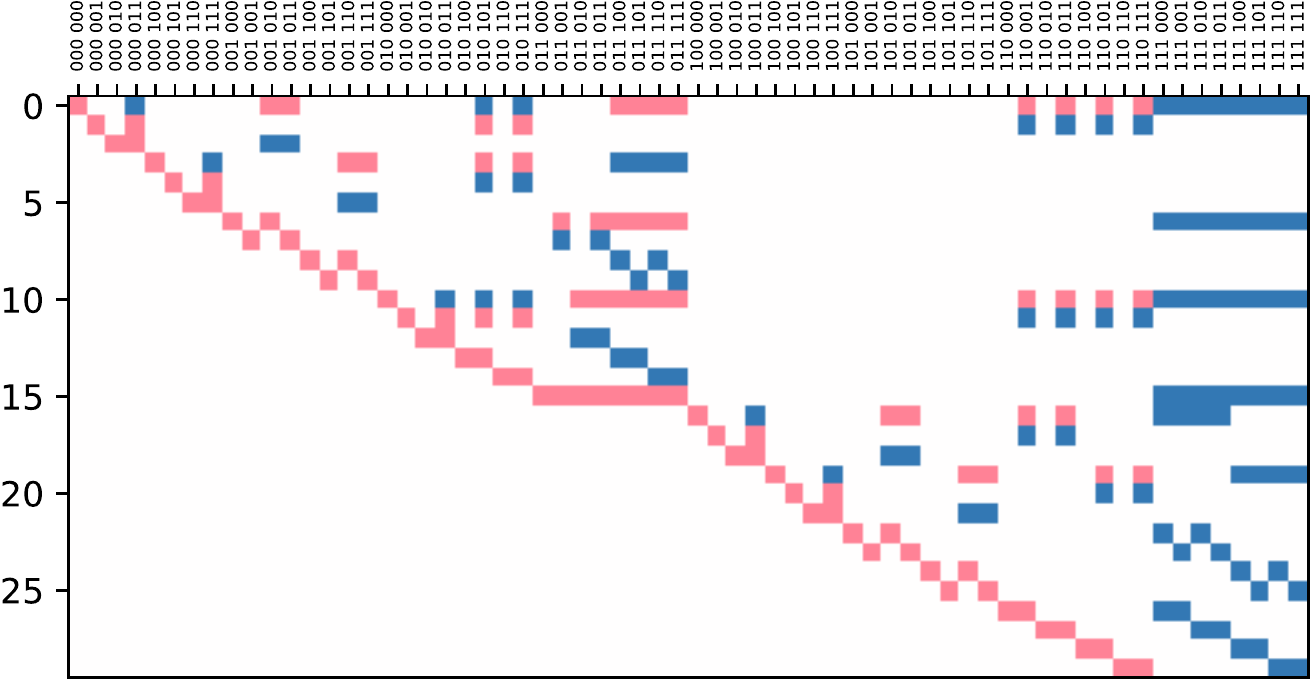}
\end{center}

\noindent Rows correspond to the 30 independent linear equations.
Columns in the plot correspond to entries of empirical models, indexed as $i_A i_B i_C$ $o_A o_B o_C$.
Coefficients in the equations are color-coded as white=0, red=+1 and blue=-1.

Space 58 has closest refinements in equivalence classes 33 and 38; 
it is the join of its (closest) refinements.
It has closest coarsenings in equivalence classes 64, 74 and 77; 
it is the meet of its (closest) coarsenings.
It has 512 causal functions, 128 of which are not causal for any of its refinements.
It is a tight space.

The standard causaltope for Space 58 has 1 more dimension than that of its subspace in equivalence class 33.
The standard causaltope for Space 58 is the meet of the standard causaltopes for its closest coarsenings.
For completeness, below is a plot of the full homogeneous linear system of causality and quasi-normalisation equations for the standard causaltope:

\begin{center}
    \includegraphics[width=12cm]{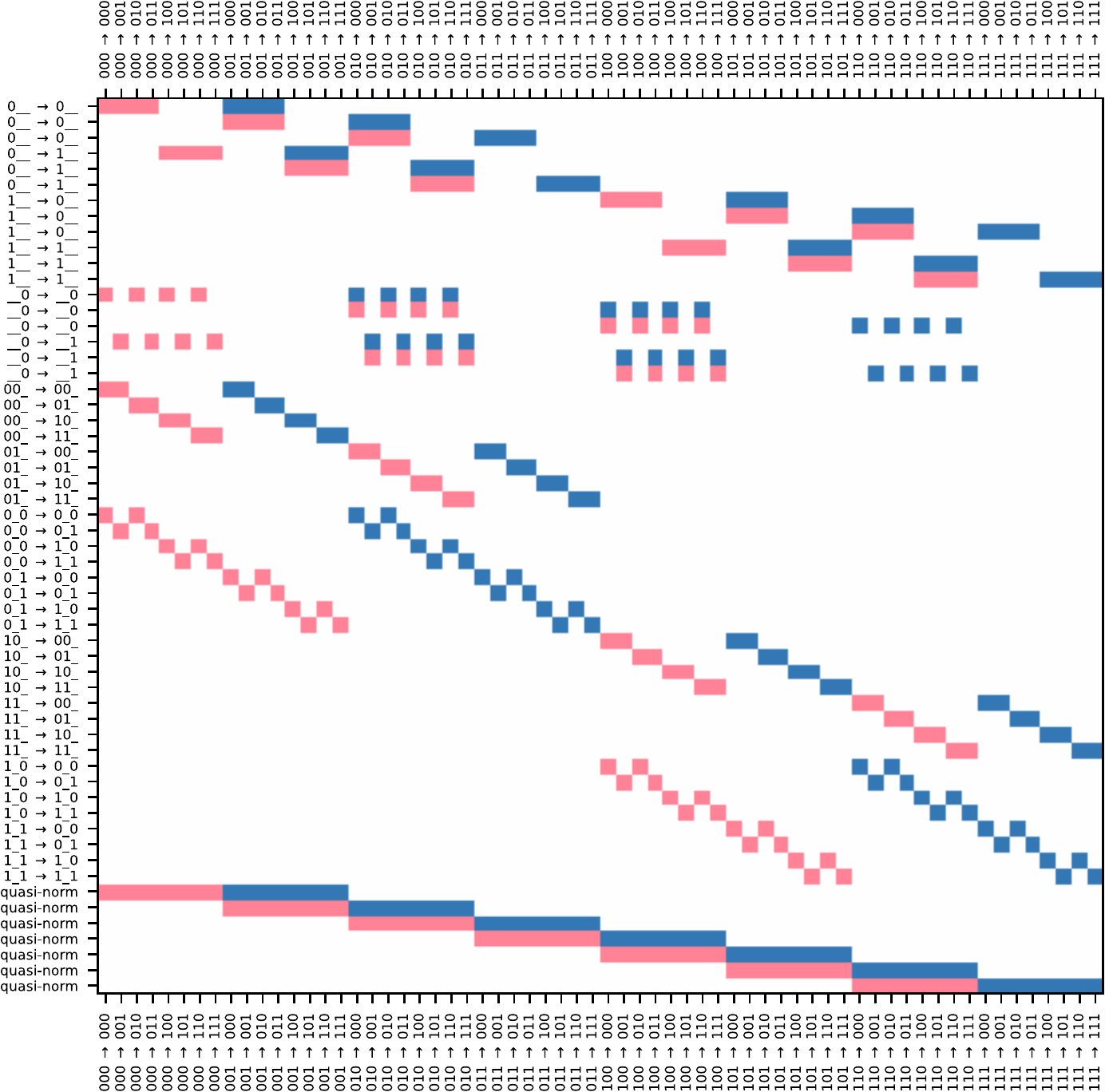}
\end{center}

\noindent Rows correspond to the 57 linear equations, of which 30 are independent.

\newpage
\subsection*{Space 59}

Space 59 is not induced by a causal order, but it is a refinement of the space 92 induced by the definite causal order $\total{\ev{A},\ev{C}}\vee\total{\ev{B},\ev{C}}$.
Its equivalence class under event-input permutation symmetry contains 24 spaces.
Space 59 differs as follows from the space induced by causal order $\total{\ev{A},\ev{C}}\vee\total{\ev{B},\ev{C}}$:
\begin{itemize}
  \item The outputs at events \evset{\ev{A}, \ev{C}} are independent of the input at event \ev{B} when the inputs at events \evset{A, C} are given by \hist{A/0,C/1}.
  \item The outputs at events \evset{\ev{B}, \ev{C}} are independent of the input at event \ev{A} when the inputs at events \evset{B, C} are given by \hist{B/1,C/1} and \hist{B/0,C/1}.
\end{itemize}

\noindent Below are the histories and extended histories for space 59: 
\begin{center}
    \begin{tabular}{cc}
    \includegraphics[height=3.5cm]{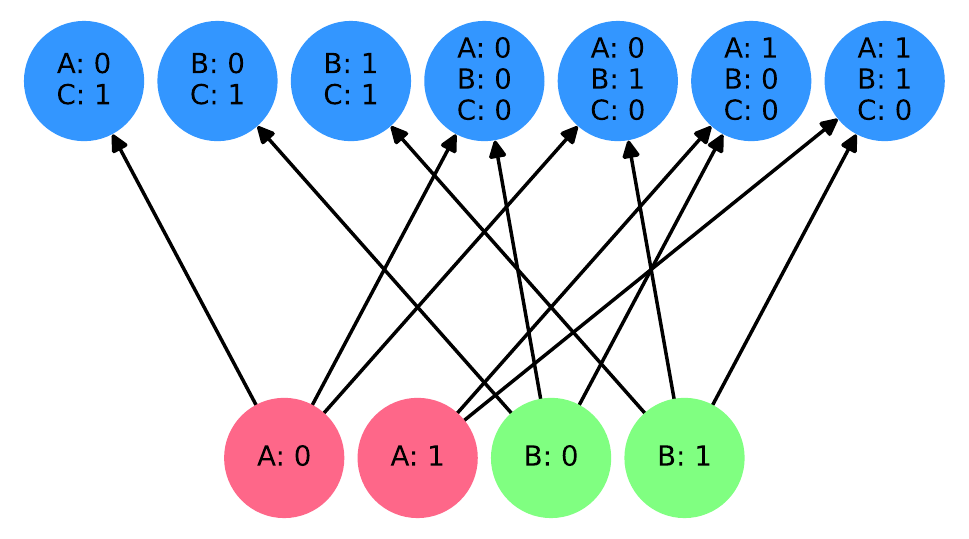}
    &
    \includegraphics[height=3.5cm]{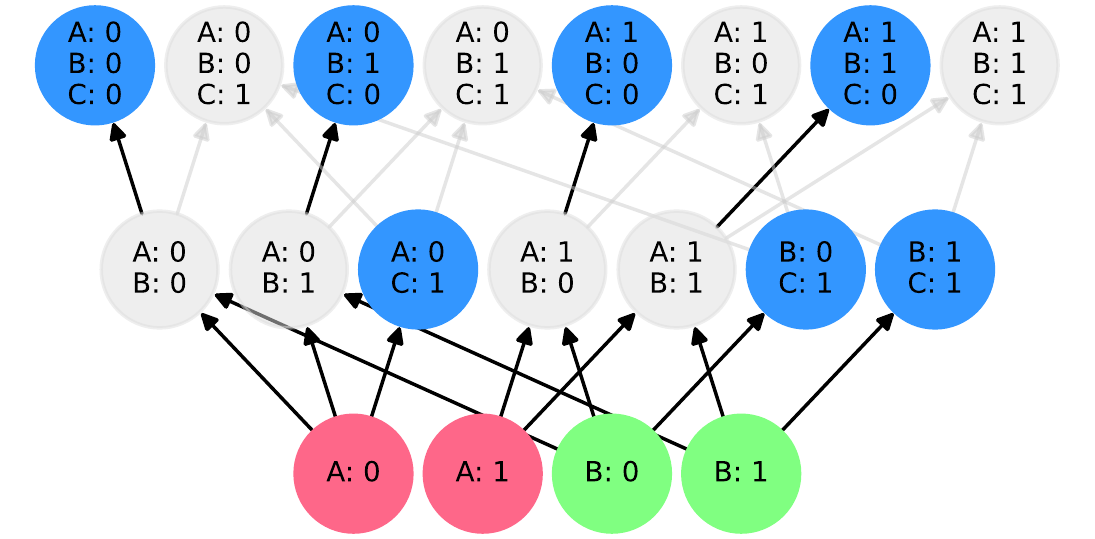}
    \\
    $\Theta_{59}$
    &
    $\Ext{\Theta_{59}}$
    \end{tabular}
\end{center}

\noindent The standard causaltope for Space 59 has dimension 34.
Below is a plot of the homogeneous linear system of causality and quasi-normalisation equations for the standard causaltope, put in reduced row echelon form:

\begin{center}
    \includegraphics[width=11cm]{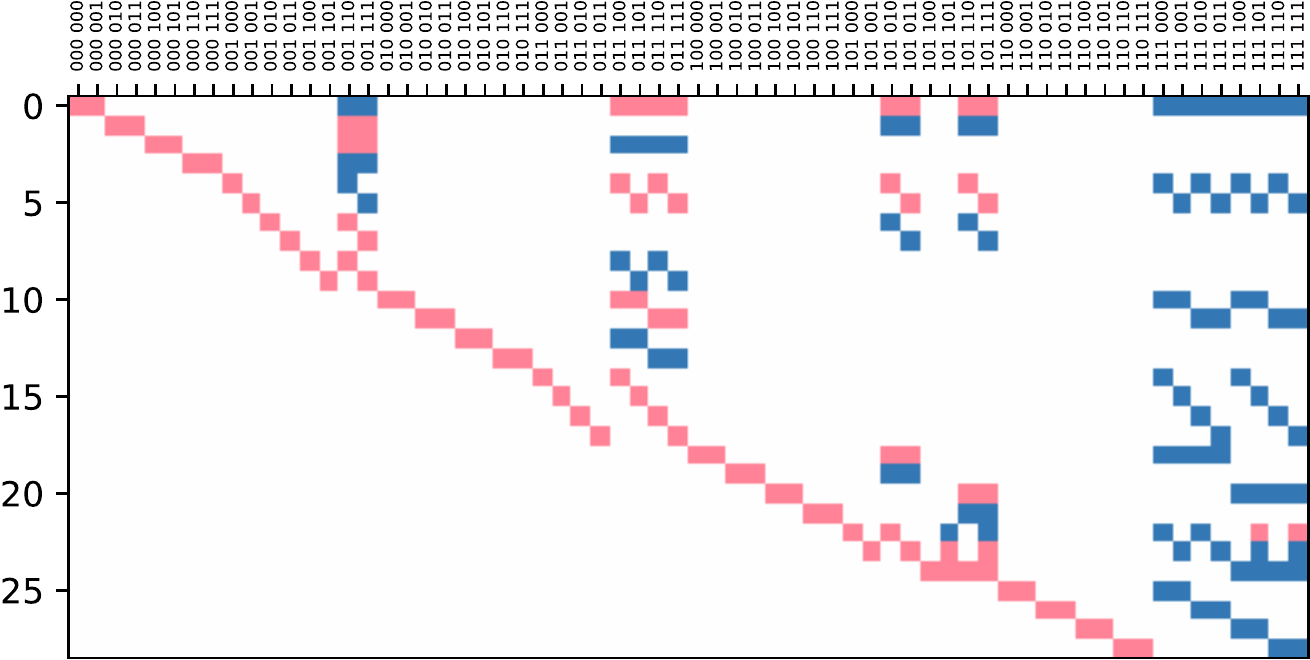}
\end{center}

\noindent Rows correspond to the 29 independent linear equations.
Columns in the plot correspond to entries of empirical models, indexed as $i_A i_B i_C$ $o_A o_B o_C$.
Coefficients in the equations are color-coded as white=0, red=+1 and blue=-1.

Space 59 has closest refinements in equivalence classes 34, 35, 39 and 44; 
it is the join of its (closest) refinements.
It has closest coarsenings in equivalence classes 68, 69 and 76; 
it is the meet of its (closest) coarsenings.
It has 512 causal functions, all of which are causal for at least one of its refinements.
It is not a tight space: for event \ev{C}, a causal function must yield identical output values on input histories \hist{A/0,C/1}, \hist{B/0,C/1} and \hist{B/1,C/1}.

The standard causaltope for Space 59 has 1 more dimension than that of its subspace in equivalence class 44.
The standard causaltope for Space 59 is the meet of the standard causaltopes for its closest coarsenings.
For completeness, below is a plot of the full homogeneous linear system of causality and quasi-normalisation equations for the standard causaltope:

\begin{center}
    \includegraphics[width=12cm]{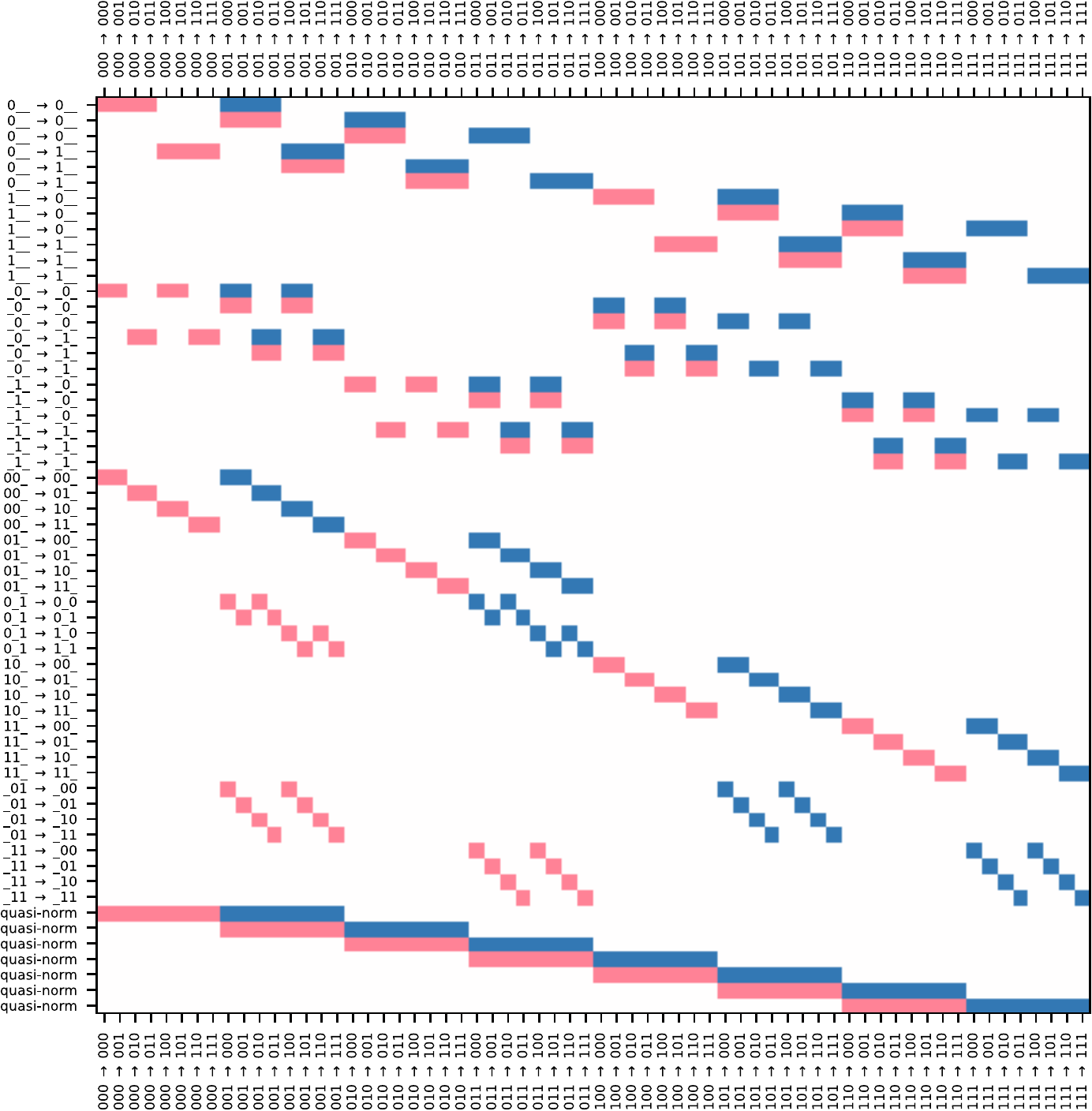}
\end{center}

\noindent Rows correspond to the 59 linear equations, of which 29 are independent.

\newpage
\subsection*{Space 60}

Space 60 is not induced by a causal order, but it is a refinement of the space in equivalence class 100 induced by the definite causal order $\total{\ev{A},\ev{C},\ev{B}}$ (note that the space induced by the order is not the same as space 100).
Its equivalence class under event-input permutation symmetry contains 24 spaces.
Space 60 differs as follows from the space induced by causal order $\total{\ev{A},\ev{C},\ev{B}}$:
\begin{itemize}
  \item The outputs at events \evset{\ev{A}, \ev{B}} are independent of the input at event \ev{C} when the inputs at events \evset{A, B} are given by \hist{A/0,B/0} and \hist{A/0,B/1}.
  \item The outputs at events \evset{\ev{B}, \ev{C}} are independent of the input at event \ev{A} when the inputs at events \evset{B, C} are given by \hist{B/1,C/0} and \hist{B/0,C/0}.
  \item The output at event \ev{C} is independent of the input at event \ev{A} when the input at event C is given by \hist{C/0}.
\end{itemize}

\noindent Below are the histories and extended histories for space 60: 
\begin{center}
    \begin{tabular}{cc}
    \includegraphics[height=3.5cm]{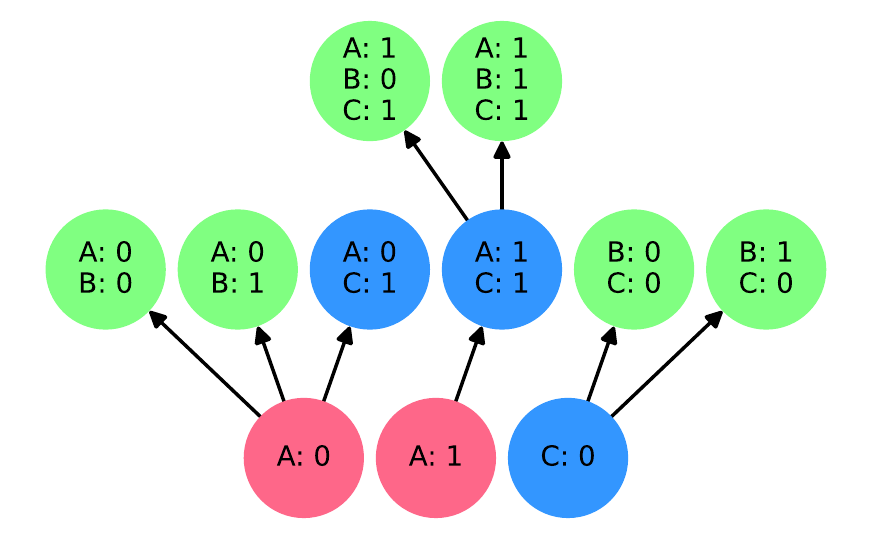}
    &
    \includegraphics[height=3.5cm]{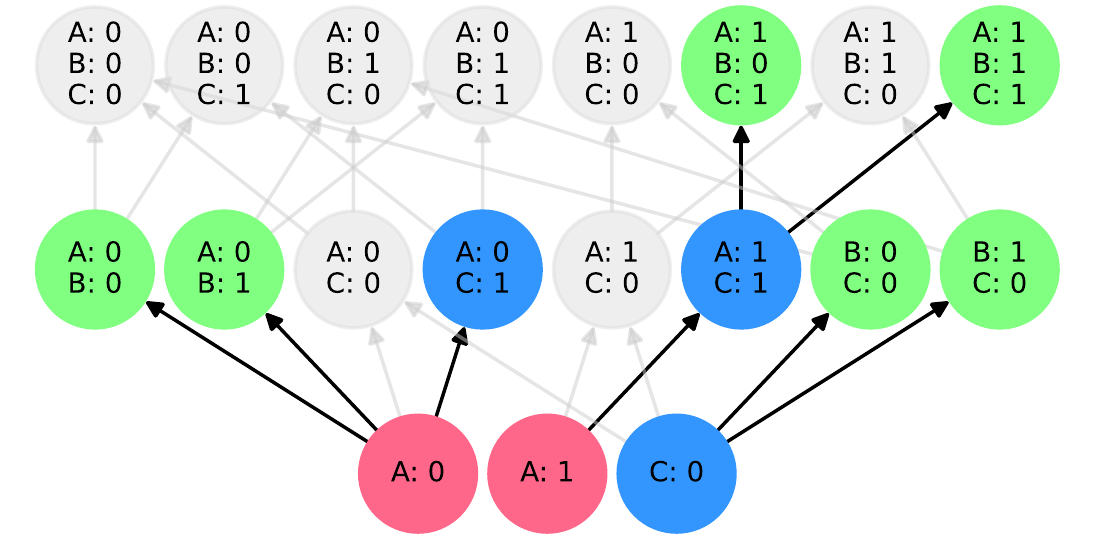}
    \\
    $\Theta_{60}$
    &
    $\Ext{\Theta_{60}}$
    \end{tabular}
\end{center}

\noindent The standard causaltope for Space 60 has dimension 33.
Below is a plot of the homogeneous linear system of causality and quasi-normalisation equations for the standard causaltope, put in reduced row echelon form:

\begin{center}
    \includegraphics[width=11cm]{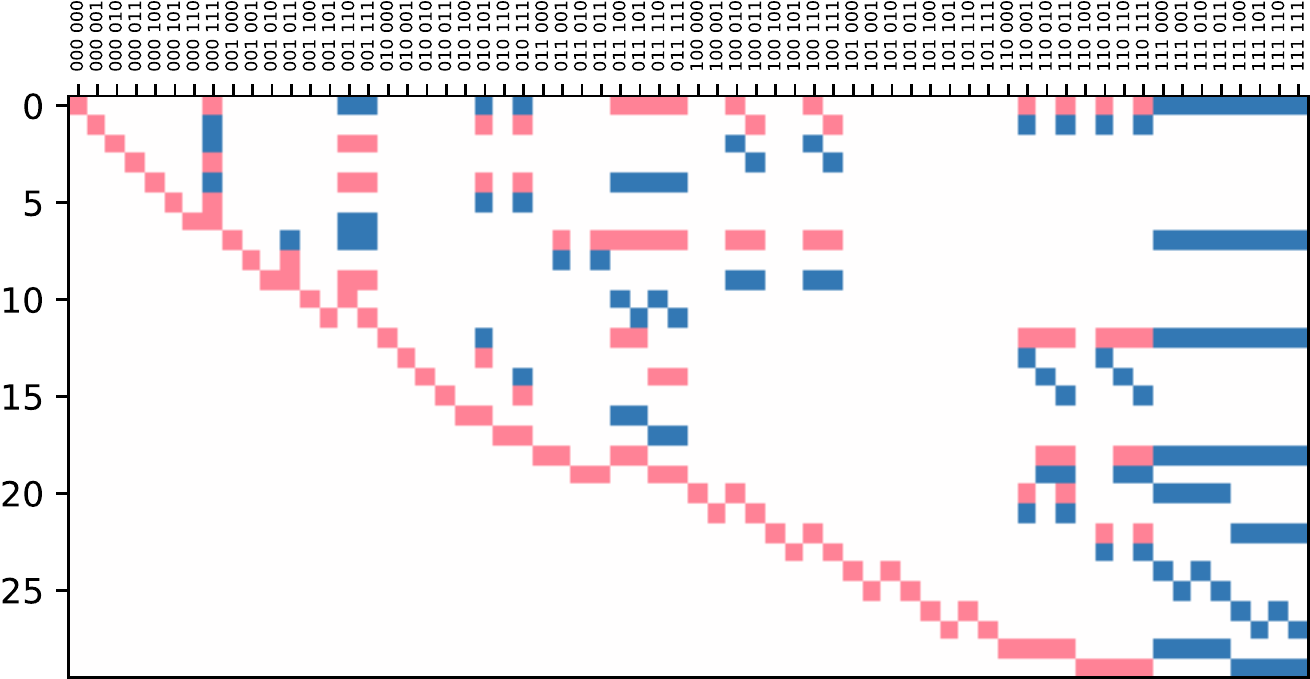}
\end{center}

\noindent Rows correspond to the 30 independent linear equations.
Columns in the plot correspond to entries of empirical models, indexed as $i_A i_B i_C$ $o_A o_B o_C$.
Coefficients in the equations are color-coded as white=0, red=+1 and blue=-1.

Space 60 has closest refinements in equivalence classes 37 and 43; 
it is the join of its (closest) refinements.
It has closest coarsenings in equivalence classes 63, 65 and 72; 
it is the meet of its (closest) coarsenings.
It has 512 causal functions, 192 of which are not causal for any of its refinements.
It is not a tight space: for event \ev{B}, a causal function must yield identical output values on input histories \hist{A/0,B/0} and \hist{B/0,C/0}, and it must also yield identical output values on input histories \hist{A/0,B/1} and \hist{B/1,C/0}.

The standard causaltope for Space 60 has 1 more dimension than that of its subspace in equivalence class 37.
The standard causaltope for Space 60 is the meet of the standard causaltopes for its closest coarsenings.
For completeness, below is a plot of the full homogeneous linear system of causality and quasi-normalisation equations for the standard causaltope:

\begin{center}
    \includegraphics[width=12cm]{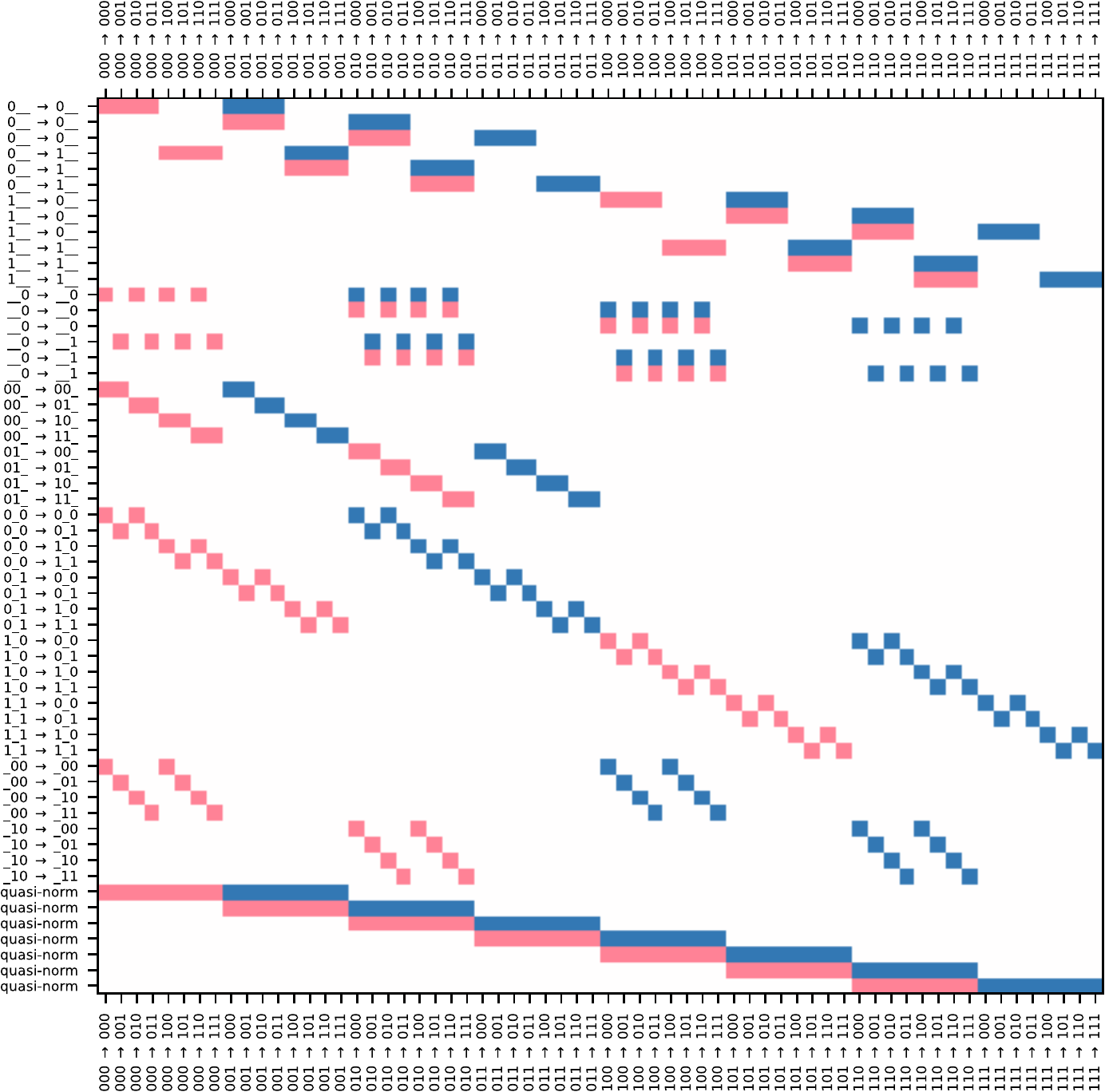}
\end{center}

\noindent Rows correspond to the 57 linear equations, of which 30 are independent.

\newpage
\subsection*{Space 61}

Space 61 is not induced by a causal order, but it is a refinement of the space 100 induced by the definite causal order $\total{\ev{A},\ev{B},\ev{C}}$.
Its equivalence class under event-input permutation symmetry contains 24 spaces.
Space 61 differs as follows from the space induced by causal order $\total{\ev{A},\ev{B},\ev{C}}$:
\begin{itemize}
  \item The outputs at events \evset{\ev{A}, \ev{C}} are independent of the input at event \ev{B} when the inputs at events \evset{A, C} are given by \hist{A/0,C/1} and \hist{A/1,C/1}.
  \item The outputs at events \evset{\ev{B}, \ev{C}} are independent of the input at event \ev{A} when the inputs at events \evset{B, C} are given by \hist{B/1,C/1}.
  \item The output at event \ev{C} is independent of the inputs at events \evset{\ev{A}, \ev{B}} when the input at event C is given by \hist{C/1}.
\end{itemize}

\noindent Below are the histories and extended histories for space 61: 
\begin{center}
    \begin{tabular}{cc}
    \includegraphics[height=3.5cm]{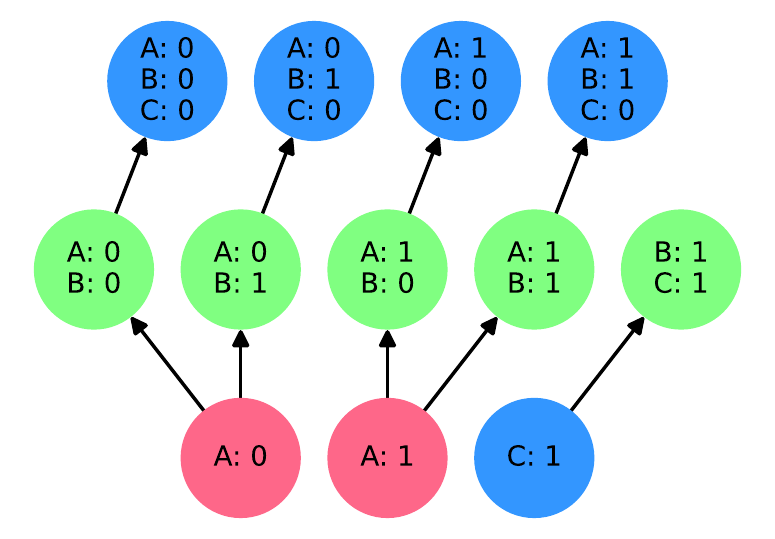}
    &
    \includegraphics[height=3.5cm]{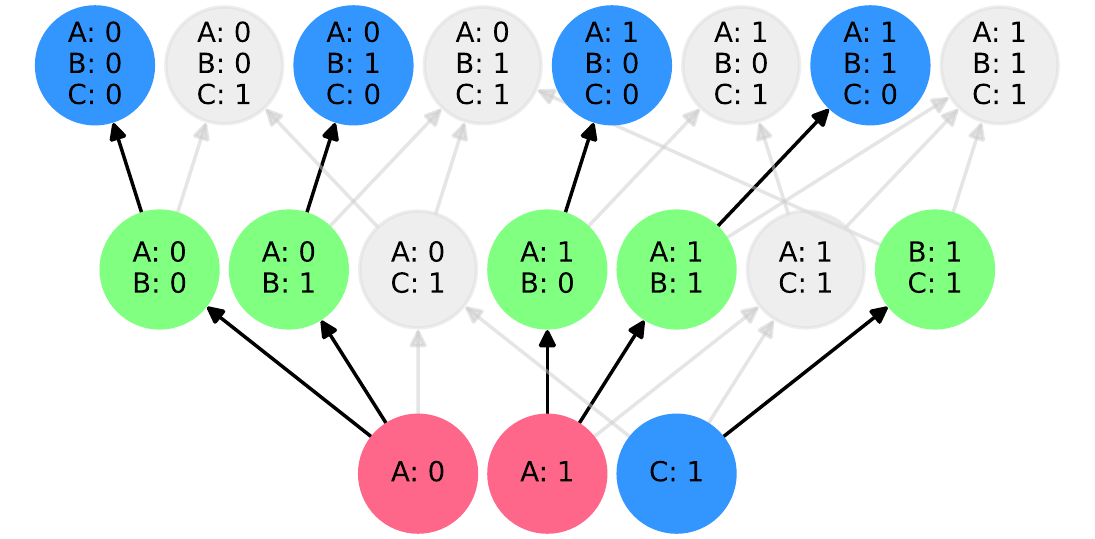}
    \\
    $\Theta_{61}$
    &
    $\Ext{\Theta_{61}}$
    \end{tabular}
\end{center}

\noindent The standard causaltope for Space 61 has dimension 35.
Below is a plot of the homogeneous linear system of causality and quasi-normalisation equations for the standard causaltope, put in reduced row echelon form:

\begin{center}
    \includegraphics[width=11cm]{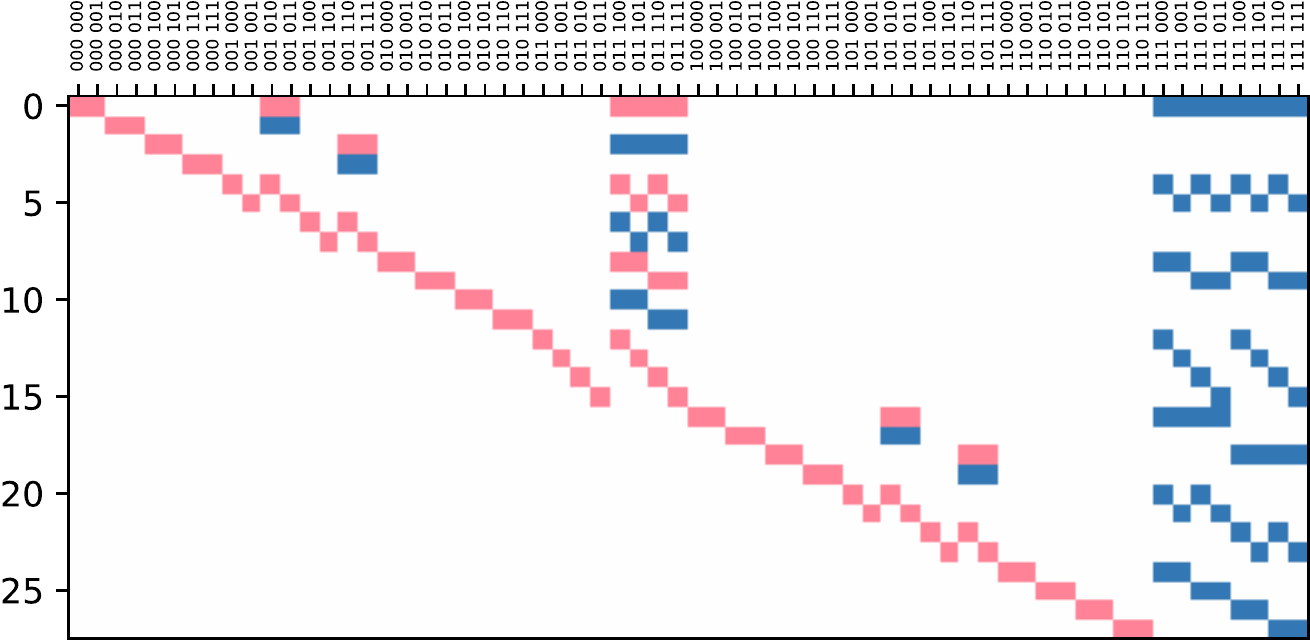}
\end{center}

\noindent Rows correspond to the 28 independent linear equations.
Columns in the plot correspond to entries of empirical models, indexed as $i_A i_B i_C$ $o_A o_B o_C$.
Coefficients in the equations are color-coded as white=0, red=+1 and blue=-1.

Space 61 has closest refinements in equivalence classes 45, 47 and 51; 
it is the join of its (closest) refinements.
It has closest coarsenings in equivalence class 78; 
it does not arise as a nontrivial meet in the hierarchy.
It has 1024 causal functions, all of which are causal for at least one of its refinements.
It is not a tight space: for event \ev{B}, a causal function must yield identical output values on input histories \hist{A/0,B/1}, \hist{A/1,B/1} and \hist{B/1,C/1}.

The standard causaltope for Space 61 coincides with that of its subspace in equivalence class 45.
The standard causaltope for Space 61 has 2 dimensions fewer than the meet of the standard causaltopes for its closest coarsenings.
For completeness, below is a plot of the full homogeneous linear system of causality and quasi-normalisation equations for the standard causaltope:

\begin{center}
    \includegraphics[width=12cm]{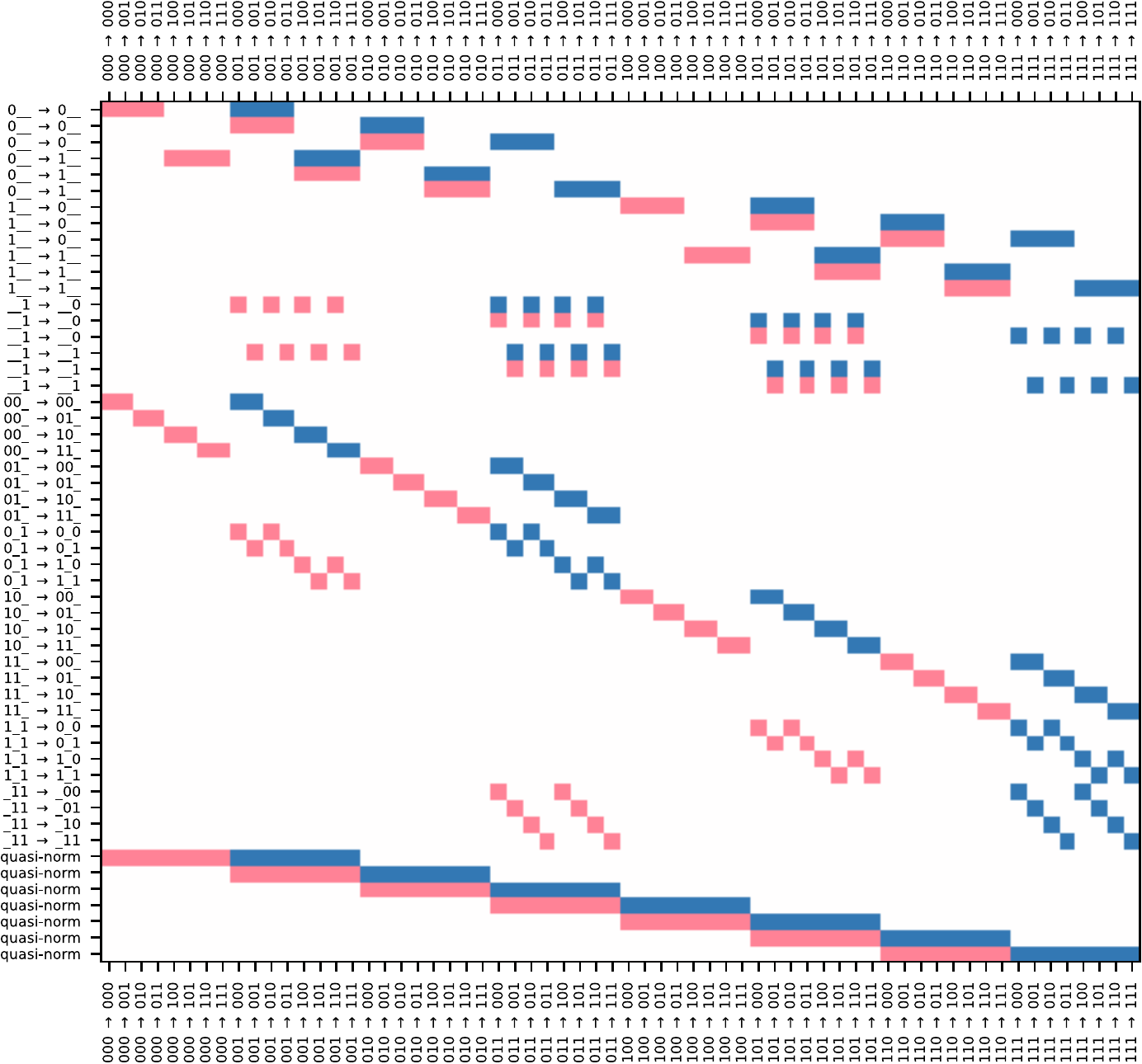}
\end{center}

\noindent Rows correspond to the 53 linear equations, of which 28 are independent.

\newpage
\subsection*{Space 62}

Space 62 is not induced by a causal order, but it is a refinement of the space induced by the indefinite causal order $\total{\ev{A},\{\ev{B},\ev{C}\}}$.
Its equivalence class under event-input permutation symmetry contains 48 spaces.
Space 62 differs as follows from the space induced by causal order $\total{\ev{A},\{\ev{B},\ev{C}\}}$:
\begin{itemize}
  \item The outputs at events \evset{\ev{A}, \ev{B}} are independent of the input at event \ev{C} when the inputs at events \evset{A, B} are given by \hist{A/0,B/0}, \hist{A/0,B/1} and \hist{A/1,B/0}.
  \item The outputs at events \evset{\ev{A}, \ev{C}} are independent of the input at event \ev{B} when the inputs at events \evset{A, C} are given by \hist{A/0,C/1}, \hist{A/1,C/0} and \hist{A/1,C/1}.
  \item The outputs at events \evset{\ev{B}, \ev{C}} are independent of the input at event \ev{A} when the inputs at events \evset{B, C} are given by \hist{B/1,C/1}.
  \item The output at event \ev{C} is independent of the inputs at events \evset{\ev{A}, \ev{B}} when the input at event C is given by \hist{C/1}.
\end{itemize}

\noindent Below are the histories and extended histories for space 62: 
\begin{center}
    \begin{tabular}{cc}
    \includegraphics[height=3.5cm]{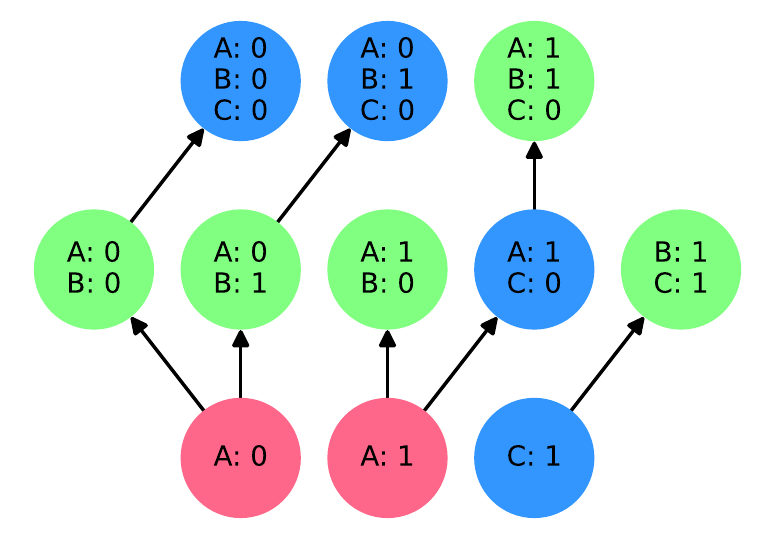}
    &
    \includegraphics[height=3.5cm]{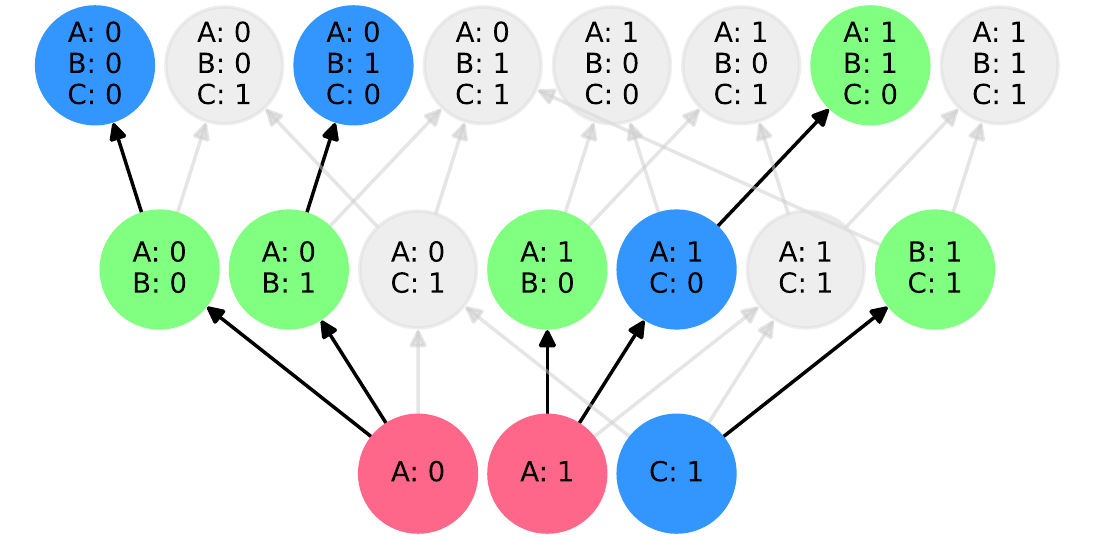}
    \\
    $\Theta_{62}$
    &
    $\Ext{\Theta_{62}}$
    \end{tabular}
\end{center}

\noindent The standard causaltope for Space 62 has dimension 35.
Below is a plot of the homogeneous linear system of causality and quasi-normalisation equations for the standard causaltope, put in reduced row echelon form:

\begin{center}
    \includegraphics[width=11cm]{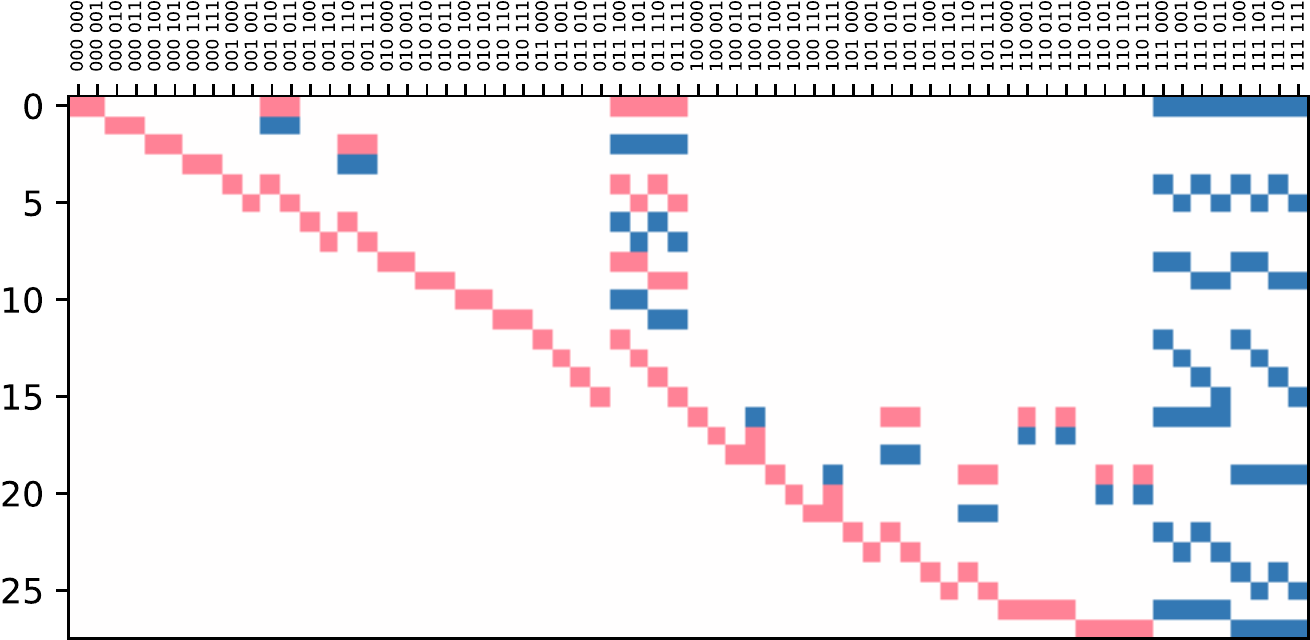}
\end{center}

\noindent Rows correspond to the 28 independent linear equations.
Columns in the plot correspond to entries of empirical models, indexed as $i_A i_B i_C$ $o_A o_B o_C$.
Coefficients in the equations are color-coded as white=0, red=+1 and blue=-1.

Space 62 has closest refinements in equivalence classes 47, 52 and 56; 
it is the join of its (closest) refinements.
It has closest coarsenings in equivalence classes 79 and 85; 
it is the meet of its (closest) coarsenings.
It has 1024 causal functions, 192 of which are not causal for any of its refinements.
It is not a tight space: for event \ev{B}, a causal function must yield identical output values on input histories \hist{A/0,B/1} and \hist{B/1,C/1}.

The standard causaltope for Space 62 has 2 more dimensions than those of its 3 subspaces in equivalence classes 47, 52 and 56.
The standard causaltope for Space 62 is the meet of the standard causaltopes for its closest coarsenings.
For completeness, below is a plot of the full homogeneous linear system of causality and quasi-normalisation equations for the standard causaltope:

\begin{center}
    \includegraphics[width=12cm]{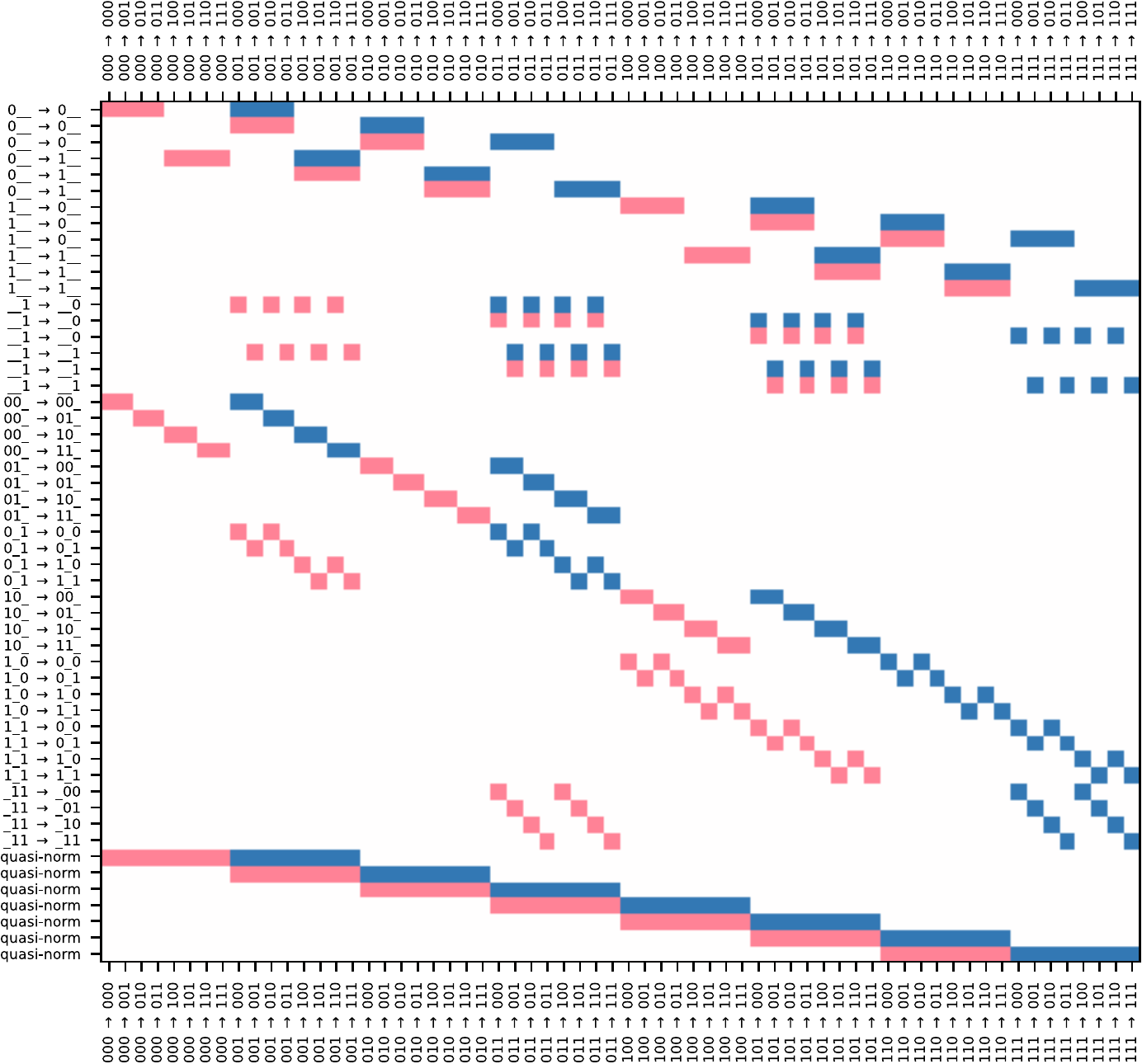}
\end{center}

\noindent Rows correspond to the 53 linear equations, of which 28 are independent.

\newpage
\subsection*{Space 63}

Space 63 is not induced by a causal order, but it is a refinement of the space 100 induced by the definite causal order $\total{\ev{A},\ev{B},\ev{C}}$.
Its equivalence class under event-input permutation symmetry contains 48 spaces.
Space 63 differs as follows from the space induced by causal order $\total{\ev{A},\ev{B},\ev{C}}$:
\begin{itemize}
  \item The outputs at events \evset{\ev{B}, \ev{C}} are independent of the input at event \ev{A} when the inputs at events \evset{B, C} are given by \hist{B/1,C/1}.
  \item The outputs at events \evset{\ev{A}, \ev{C}} are independent of the input at event \ev{B} when the inputs at events \evset{A, C} are given by \hist{A/1,C/0} and \hist{A/1,C/1}.
  \item The output at event \ev{B} is independent of the input at event \ev{A} when the input at event B is given by \hist{B/1}.
\end{itemize}

\noindent Below are the histories and extended histories for space 63: 
\begin{center}
    \begin{tabular}{cc}
    \includegraphics[height=3.5cm]{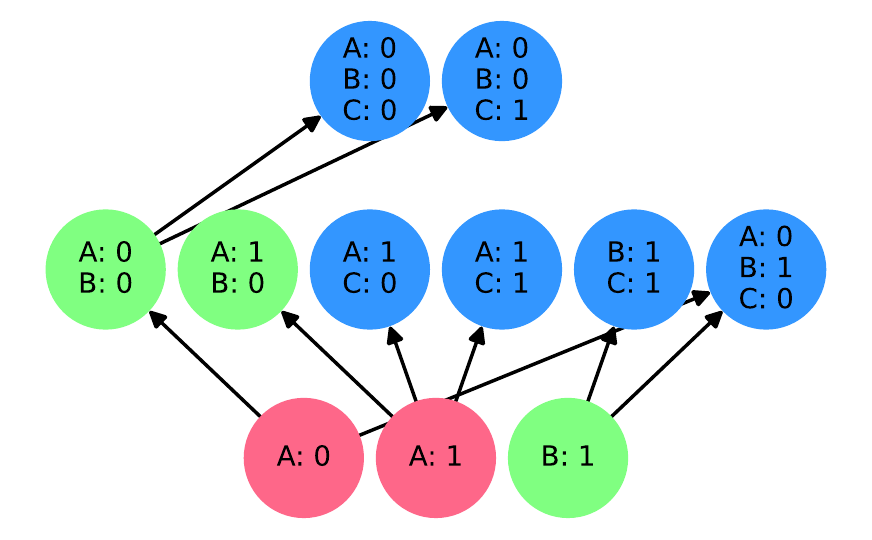}
    &
    \includegraphics[height=3.5cm]{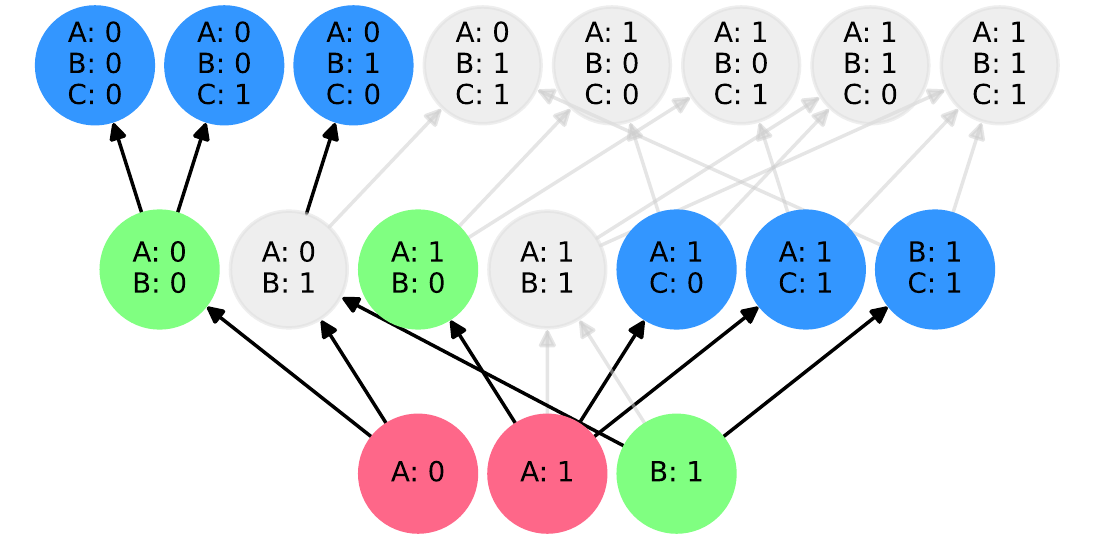}
    \\
    $\Theta_{63}$
    &
    $\Ext{\Theta_{63}}$
    \end{tabular}
\end{center}

\noindent The standard causaltope for Space 63 has dimension 35.
Below is a plot of the homogeneous linear system of causality and quasi-normalisation equations for the standard causaltope, put in reduced row echelon form:

\begin{center}
    \includegraphics[width=11cm]{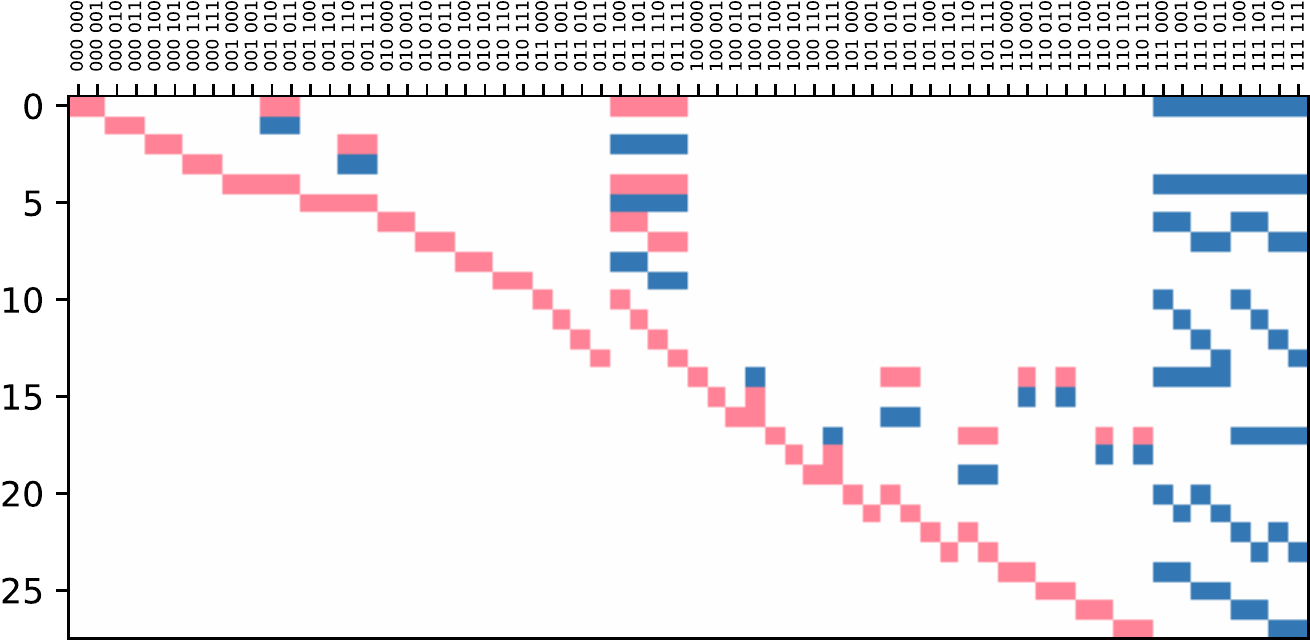}
\end{center}

\noindent Rows correspond to the 28 independent linear equations.
Columns in the plot correspond to entries of empirical models, indexed as $i_A i_B i_C$ $o_A o_B o_C$.
Coefficients in the equations are color-coded as white=0, red=+1 and blue=-1.

Space 63 has closest refinements in equivalence classes 53, 56, 57 and 60; 
it is the join of its (closest) refinements.
It has closest coarsenings in equivalence classes 80, 82, 83 and 85; 
it is the meet of its (closest) coarsenings.
It has 1024 causal functions, 384 of which are not causal for any of its refinements.
It is not a tight space: for event \ev{C}, a causal function must yield identical output values on input histories \hist{A/1,C/1} and \hist{B/1,C/1}.

The standard causaltope for Space 63 has 1 more dimension than that of its subspace in equivalence class 53.
The standard causaltope for Space 63 is the meet of the standard causaltopes for its closest coarsenings.
For completeness, below is a plot of the full homogeneous linear system of causality and quasi-normalisation equations for the standard causaltope:

\begin{center}
    \includegraphics[width=12cm]{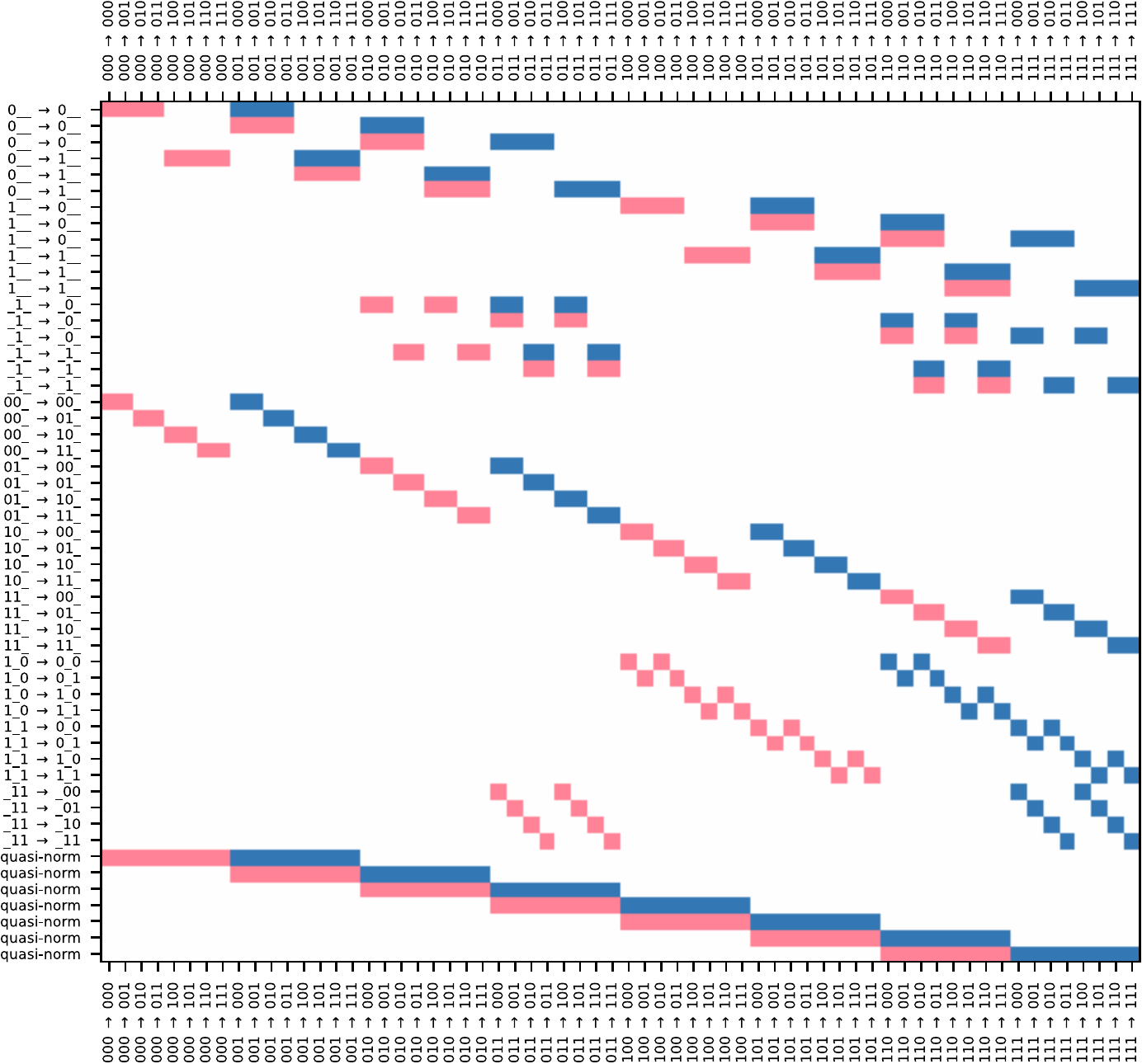}
\end{center}

\noindent Rows correspond to the 53 linear equations, of which 28 are independent.

\newpage
\subsection*{Space 64}

Space 64 is not induced by a causal order, but it is a refinement of the space 100 induced by the definite causal order $\total{\ev{A},\ev{B},\ev{C}}$.
Its equivalence class under event-input permutation symmetry contains 24 spaces.
Space 64 differs as follows from the space induced by causal order $\total{\ev{A},\ev{B},\ev{C}}$:
\begin{itemize}
  \item The outputs at events \evset{\ev{A}, \ev{C}} are independent of the input at event \ev{B} when the inputs at events \evset{A, C} are given by \hist{A/0,C/0}, \hist{A/1,C/0} and \hist{A/1,C/1}.
  \item The output at event \ev{C} is independent of the inputs at events \evset{\ev{A}, \ev{B}} when the input at event C is given by \hist{C/0}.
\end{itemize}

\noindent Below are the histories and extended histories for space 64: 
\begin{center}
    \begin{tabular}{cc}
    \includegraphics[height=3.5cm]{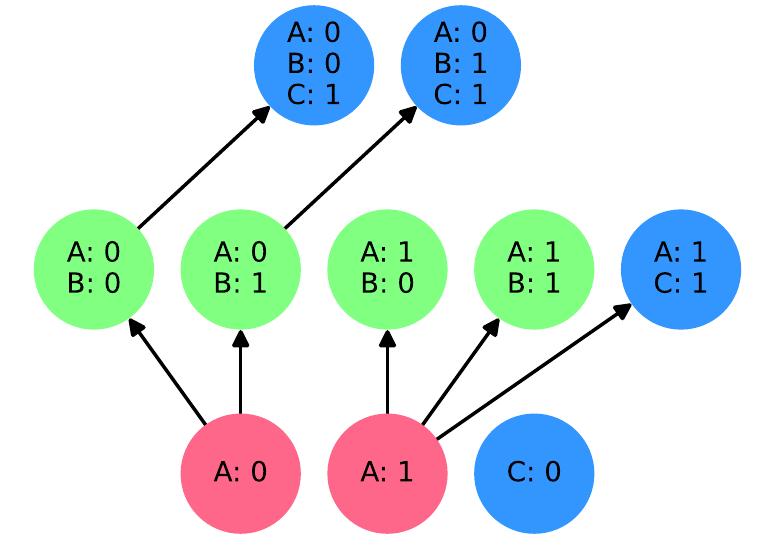}
    &
    \includegraphics[height=3.5cm]{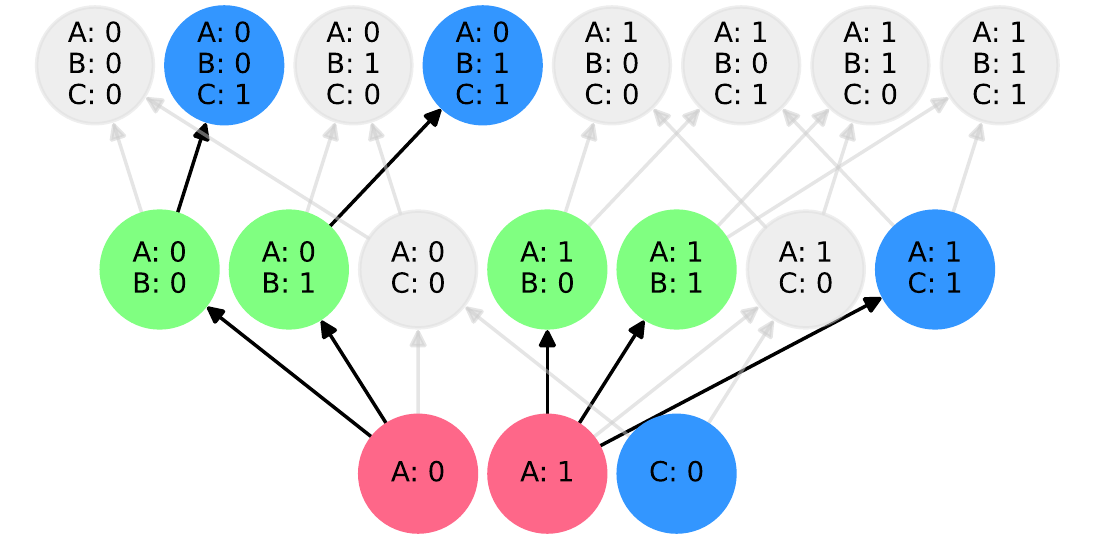}
    \\
    $\Theta_{64}$
    &
    $\Ext{\Theta_{64}}$
    \end{tabular}
\end{center}

\noindent The standard causaltope for Space 64 has dimension 35.
Below is a plot of the homogeneous linear system of causality and quasi-normalisation equations for the standard causaltope, put in reduced row echelon form:

\begin{center}
    \includegraphics[width=11cm]{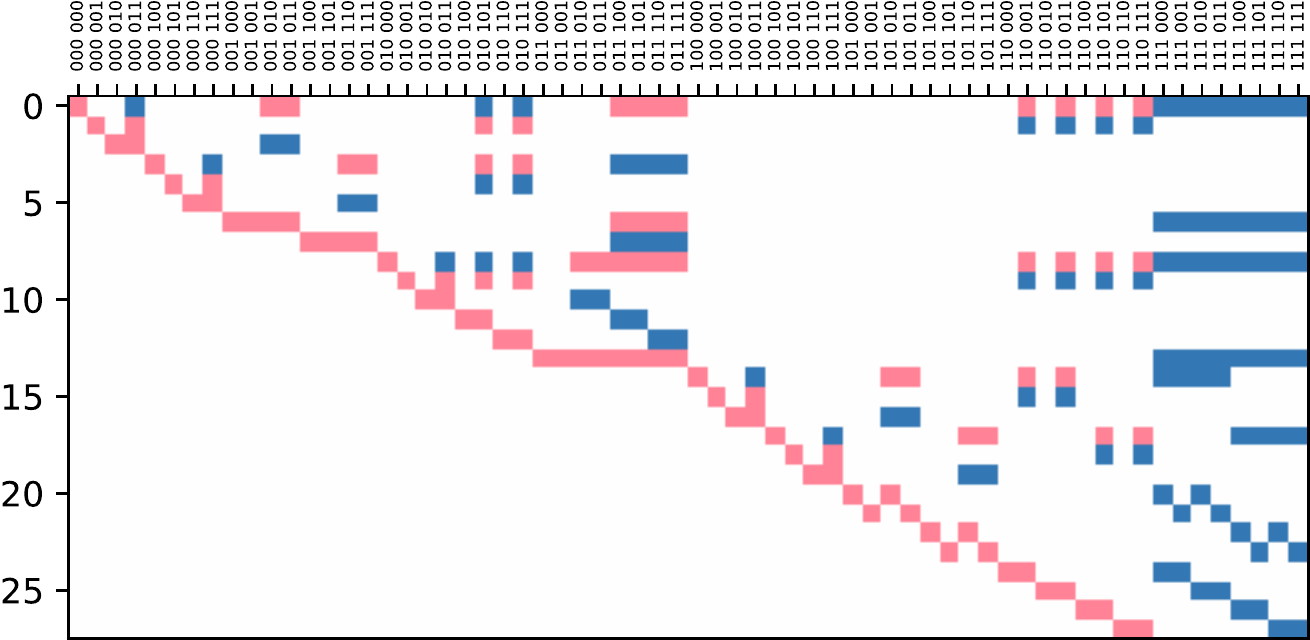}
\end{center}

\noindent Rows correspond to the 28 independent linear equations.
Columns in the plot correspond to entries of empirical models, indexed as $i_A i_B i_C$ $o_A o_B o_C$.
Coefficients in the equations are color-coded as white=0, red=+1 and blue=-1.

Space 64 has closest refinements in equivalence classes 47 and 58; 
it is the join of its (closest) refinements.
It has closest coarsenings in equivalence classes 78, 79 and 88; 
it is the meet of its (closest) coarsenings.
It has 1024 causal functions, 256 of which are not causal for any of its refinements.
It is a tight space.

The standard causaltope for Space 64 has 2 more dimensions than those of its 3 subspaces in equivalence classes 47 and 58.
The standard causaltope for Space 64 is the meet of the standard causaltopes for its closest coarsenings.
For completeness, below is a plot of the full homogeneous linear system of causality and quasi-normalisation equations for the standard causaltope:

\begin{center}
    \includegraphics[width=12cm]{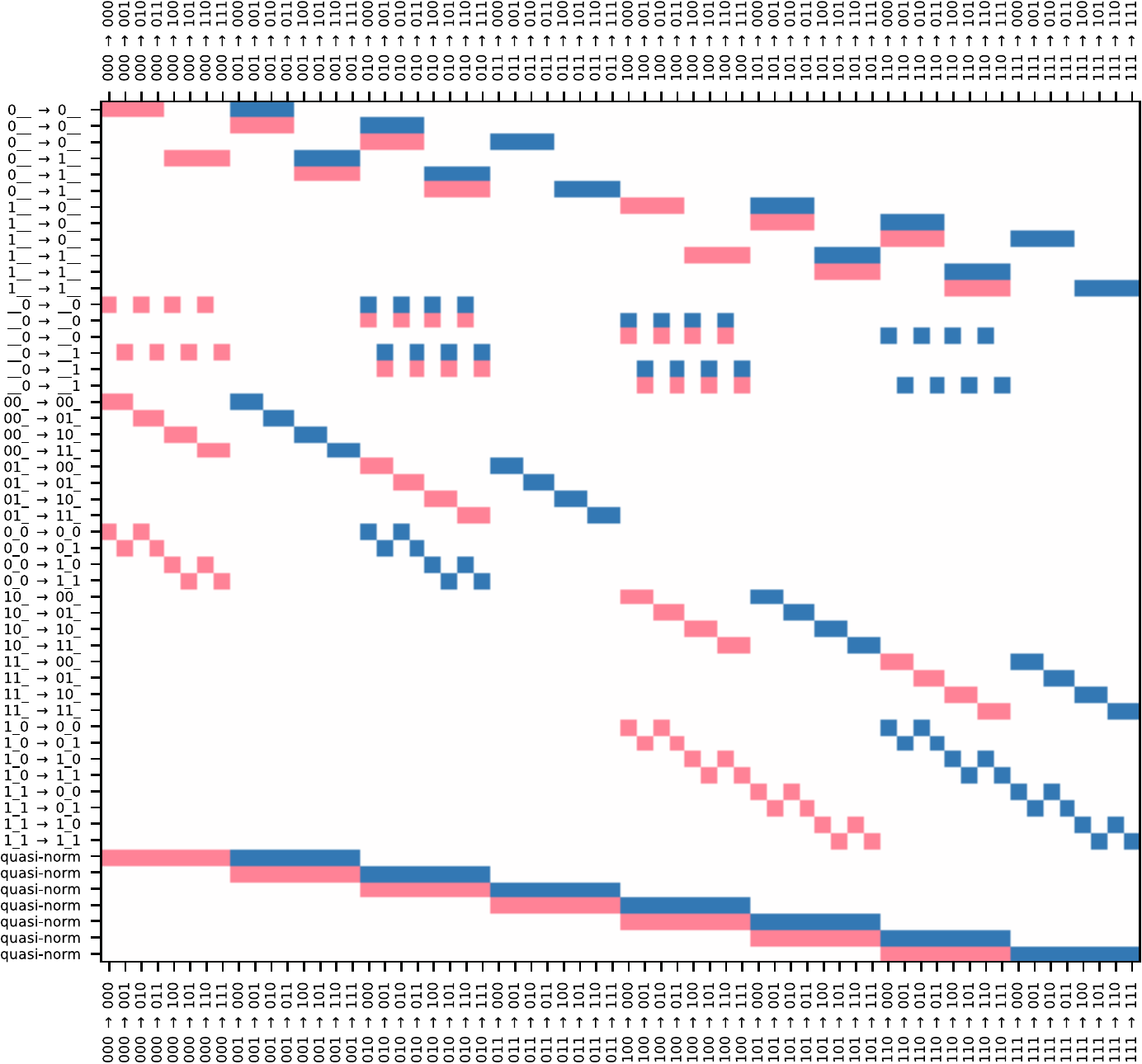}
\end{center}

\noindent Rows correspond to the 53 linear equations, of which 28 are independent.

\newpage
\subsection*{Space 65}

Space 65 is not induced by a causal order, but it is a refinement of the space 100 induced by the definite causal order $\total{\ev{A},\ev{B},\ev{C}}$.
Its equivalence class under event-input permutation symmetry contains 48 spaces.
Space 65 differs as follows from the space induced by causal order $\total{\ev{A},\ev{B},\ev{C}}$:
\begin{itemize}
  \item The outputs at events \evset{\ev{B}, \ev{C}} are independent of the input at event \ev{A} when the inputs at events \evset{B, C} are given by \hist{B/1,C/0} and \hist{B/1,C/1}.
  \item The output at event \ev{B} is independent of the input at event \ev{A} when the input at event B is given by \hist{B/1}.
  \item The outputs at events \evset{\ev{A}, \ev{C}} are independent of the input at event \ev{B} when the inputs at events \evset{A, C} are given by \hist{A/1,C/1}.
\end{itemize}

\noindent Below are the histories and extended histories for space 65: 
\begin{center}
    \begin{tabular}{cc}
    \includegraphics[height=3.5cm]{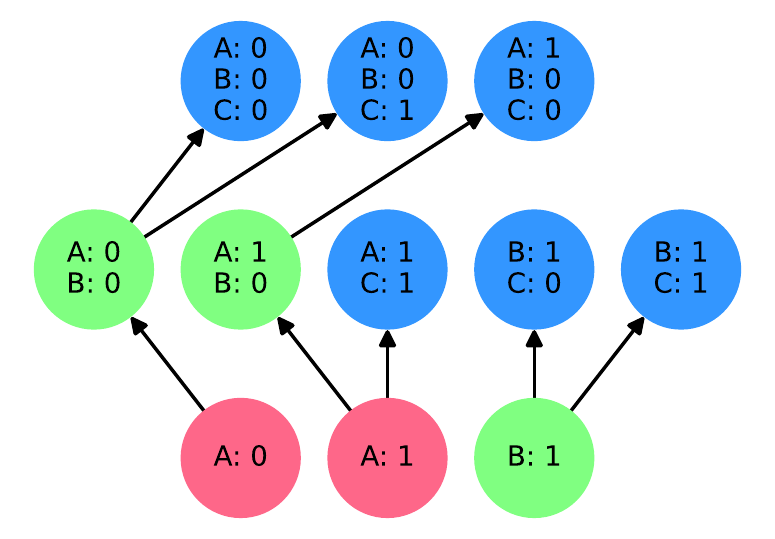}
    &
    \includegraphics[height=3.5cm]{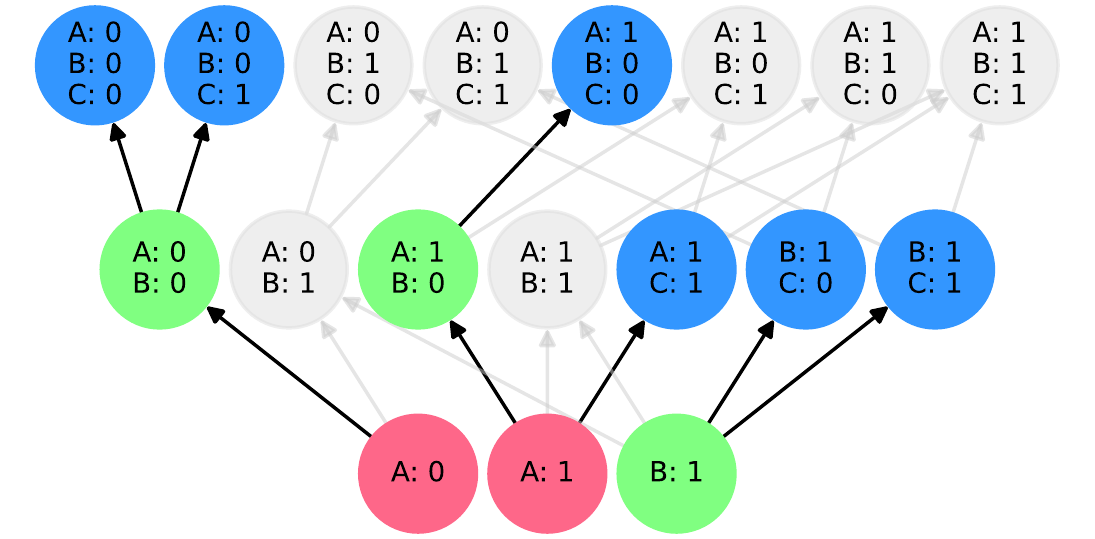}
    \\
    $\Theta_{65}$
    &
    $\Ext{\Theta_{65}}$
    \end{tabular}
\end{center}

\noindent The standard causaltope for Space 65 has dimension 35.
Below is a plot of the homogeneous linear system of causality and quasi-normalisation equations for the standard causaltope, put in reduced row echelon form:

\begin{center}
    \includegraphics[width=11cm]{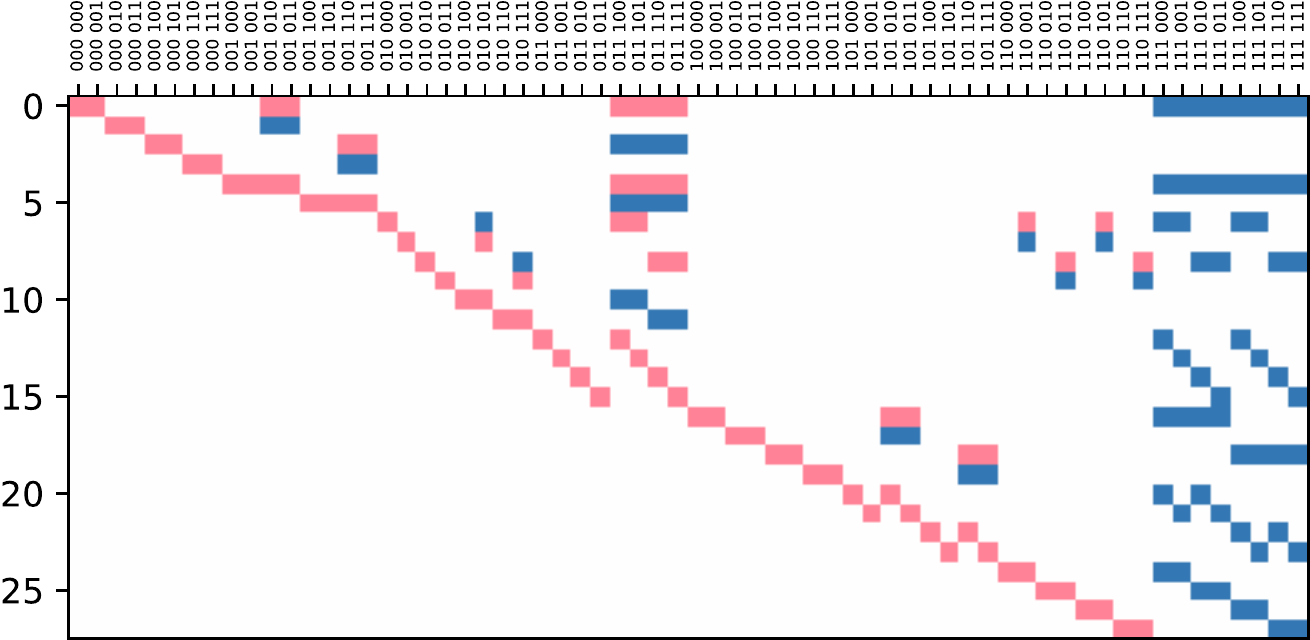}
\end{center}

\noindent Rows correspond to the 28 independent linear equations.
Columns in the plot correspond to entries of empirical models, indexed as $i_A i_B i_C$ $o_A o_B o_C$.
Coefficients in the equations are color-coded as white=0, red=+1 and blue=-1.

Space 65 has closest refinements in equivalence classes 53, 54, 55 and 60; 
it is the join of its (closest) refinements.
It has closest coarsenings in equivalence classes 80, 81 and 82; 
it is the meet of its (closest) coarsenings.
It has 1024 causal functions, 256 of which are not causal for any of its refinements.
It is not a tight space: for event \ev{C}, a causal function must yield identical output values on input histories \hist{A/1,C/1} and \hist{B/1,C/1}.

The standard causaltope for Space 65 has 1 more dimension than that of its subspace in equivalence class 53.
The standard causaltope for Space 65 is the meet of the standard causaltopes for its closest coarsenings.
For completeness, below is a plot of the full homogeneous linear system of causality and quasi-normalisation equations for the standard causaltope:

\begin{center}
    \includegraphics[width=12cm]{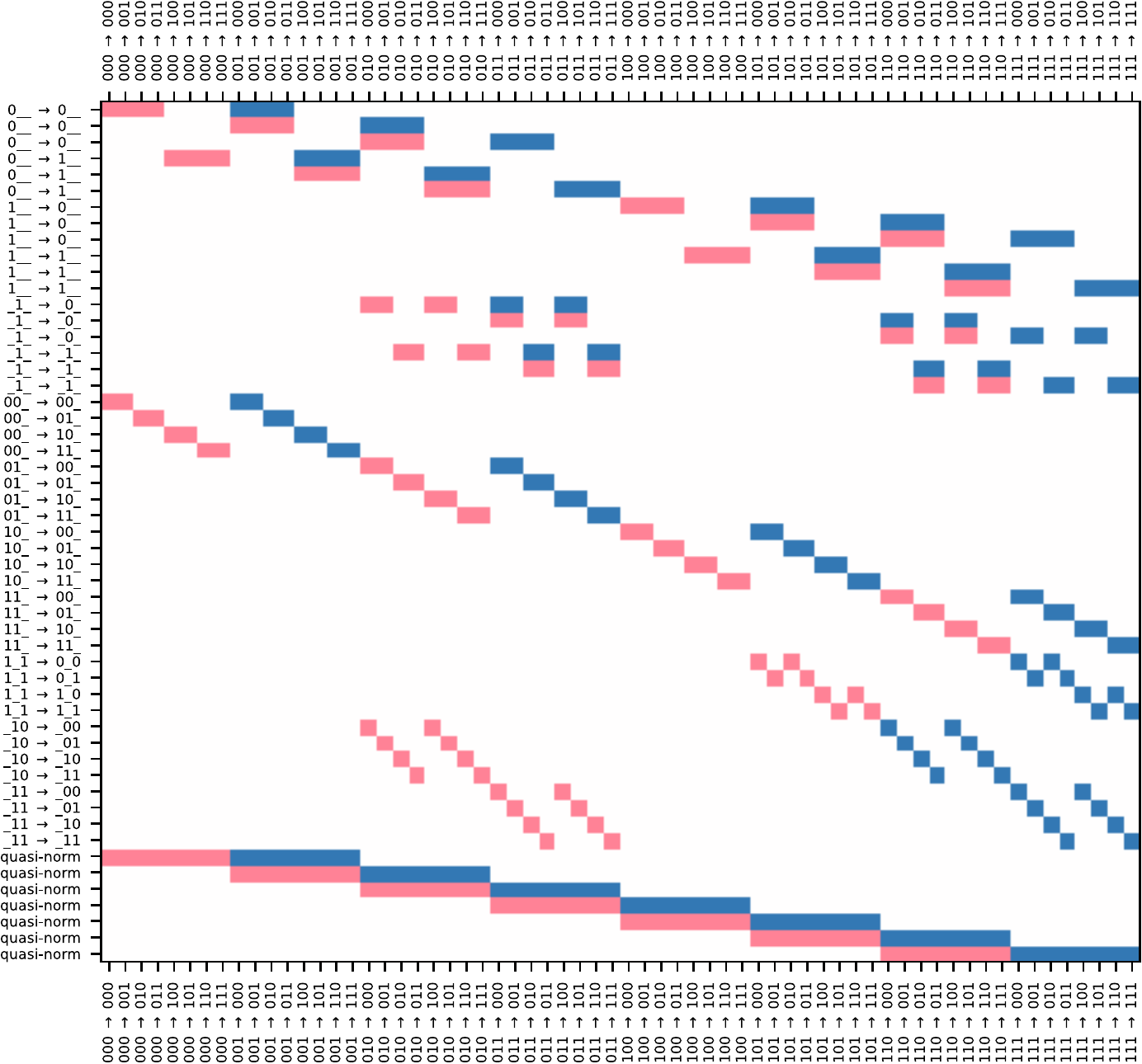}
\end{center}

\noindent Rows correspond to the 53 linear equations, of which 28 are independent.

\newpage
\subsection*{Space 66}

Space 66 is not induced by a causal order, but it is a refinement of the space 92 induced by the definite causal order $\total{\ev{A},\ev{C}}\vee\total{\ev{B},\ev{C}}$.
Its equivalence class under event-input permutation symmetry contains 24 spaces.
Space 66 differs as follows from the space induced by causal order $\total{\ev{A},\ev{C}}\vee\total{\ev{B},\ev{C}}$:
\begin{itemize}
  \item The outputs at events \evset{\ev{B}, \ev{C}} are independent of the input at event \ev{A} when the inputs at events \evset{B, C} are given by \hist{B/1,C/1}.
  \item The outputs at events \evset{\ev{A}, \ev{C}} are independent of the input at event \ev{B} when the inputs at events \evset{A, C} are given by \hist{A/1,C/0}.
\end{itemize}

\noindent Below are the histories and extended histories for space 66: 
\begin{center}
    \begin{tabular}{cc}
    \includegraphics[height=3.5cm]{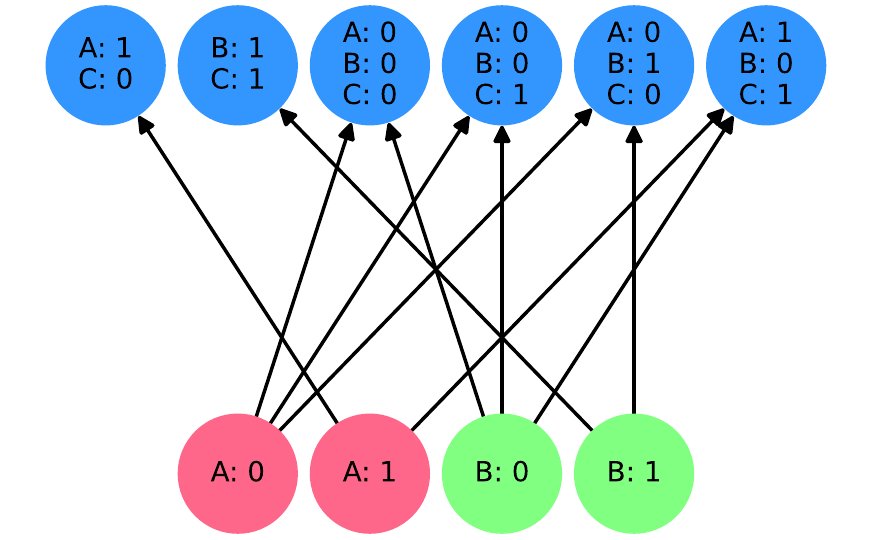}
    &
    \includegraphics[height=3.5cm]{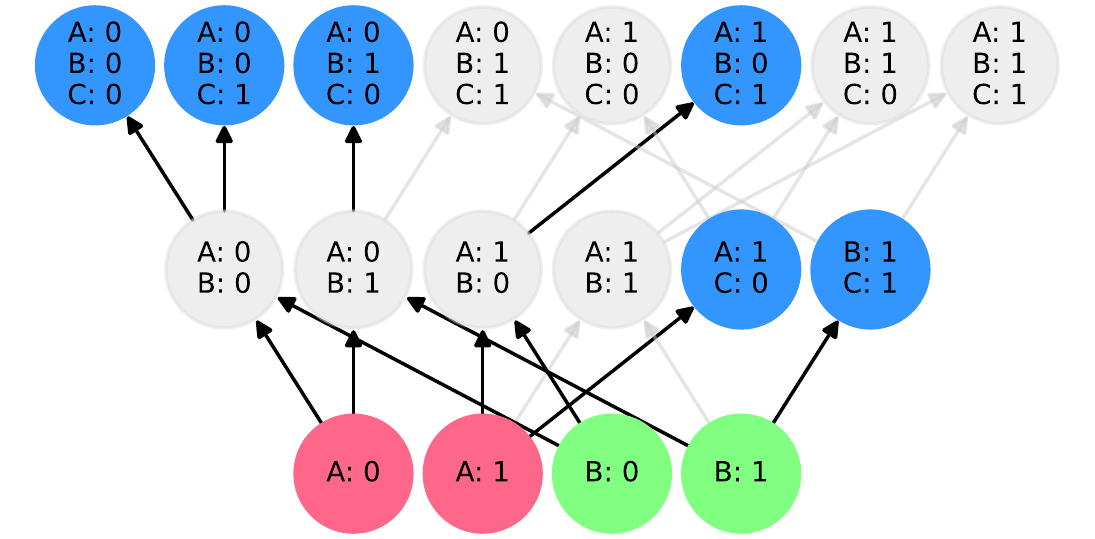}
    \\
    $\Theta_{66}$
    &
    $\Ext{\Theta_{66}}$
    \end{tabular}
\end{center}

\noindent The standard causaltope for Space 66 has dimension 36.
Below is a plot of the homogeneous linear system of causality and quasi-normalisation equations for the standard causaltope, put in reduced row echelon form:

\begin{center}
    \includegraphics[width=11cm]{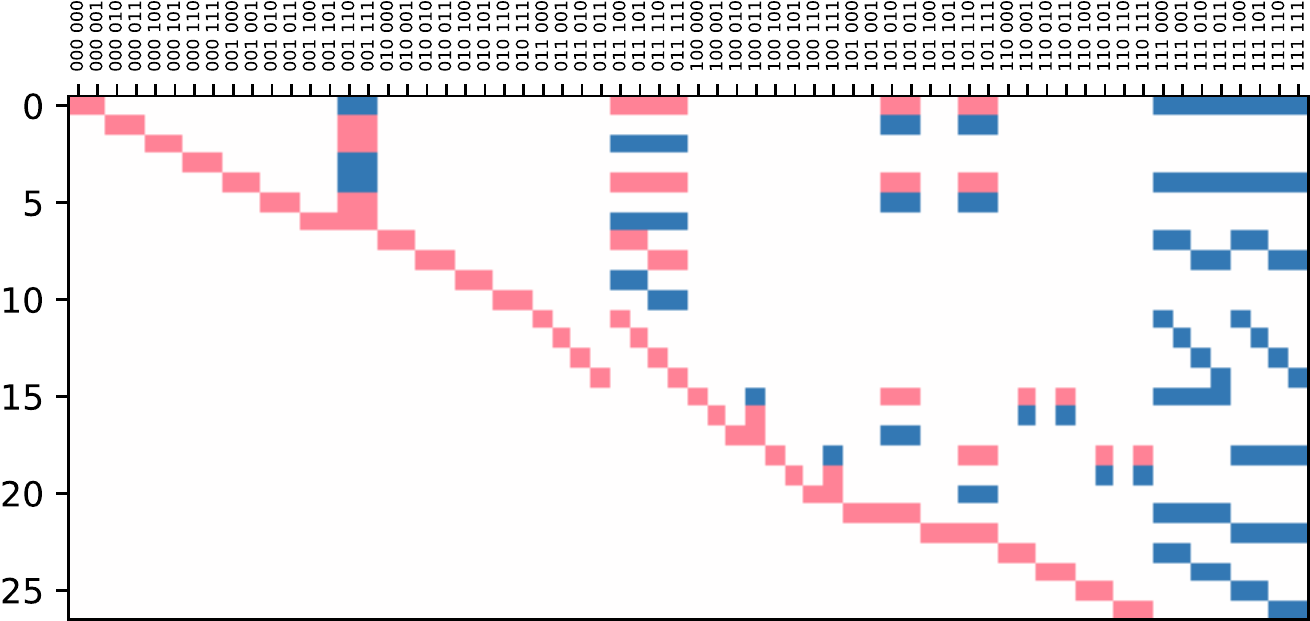}
\end{center}

\noindent Rows correspond to the 27 independent linear equations.
Columns in the plot correspond to entries of empirical models, indexed as $i_A i_B i_C$ $o_A o_B o_C$.
Coefficients in the equations are color-coded as white=0, red=+1 and blue=-1.

Space 66 has closest refinements in equivalence classes 49, 50 and 53; 
it is the join of its (closest) refinements.
It has closest coarsenings in equivalence classes 80 and 84; 
it is the meet of its (closest) coarsenings.
It has 1024 causal functions, 192 of which are not causal for any of its refinements.
It is a tight space.

The standard causaltope for Space 66 has 2 more dimensions than those of its 6 subspaces in equivalence classes 49, 50 and 53.
The standard causaltope for Space 66 is the meet of the standard causaltopes for its closest coarsenings.
For completeness, below is a plot of the full homogeneous linear system of causality and quasi-normalisation equations for the standard causaltope:

\begin{center}
    \includegraphics[width=12cm]{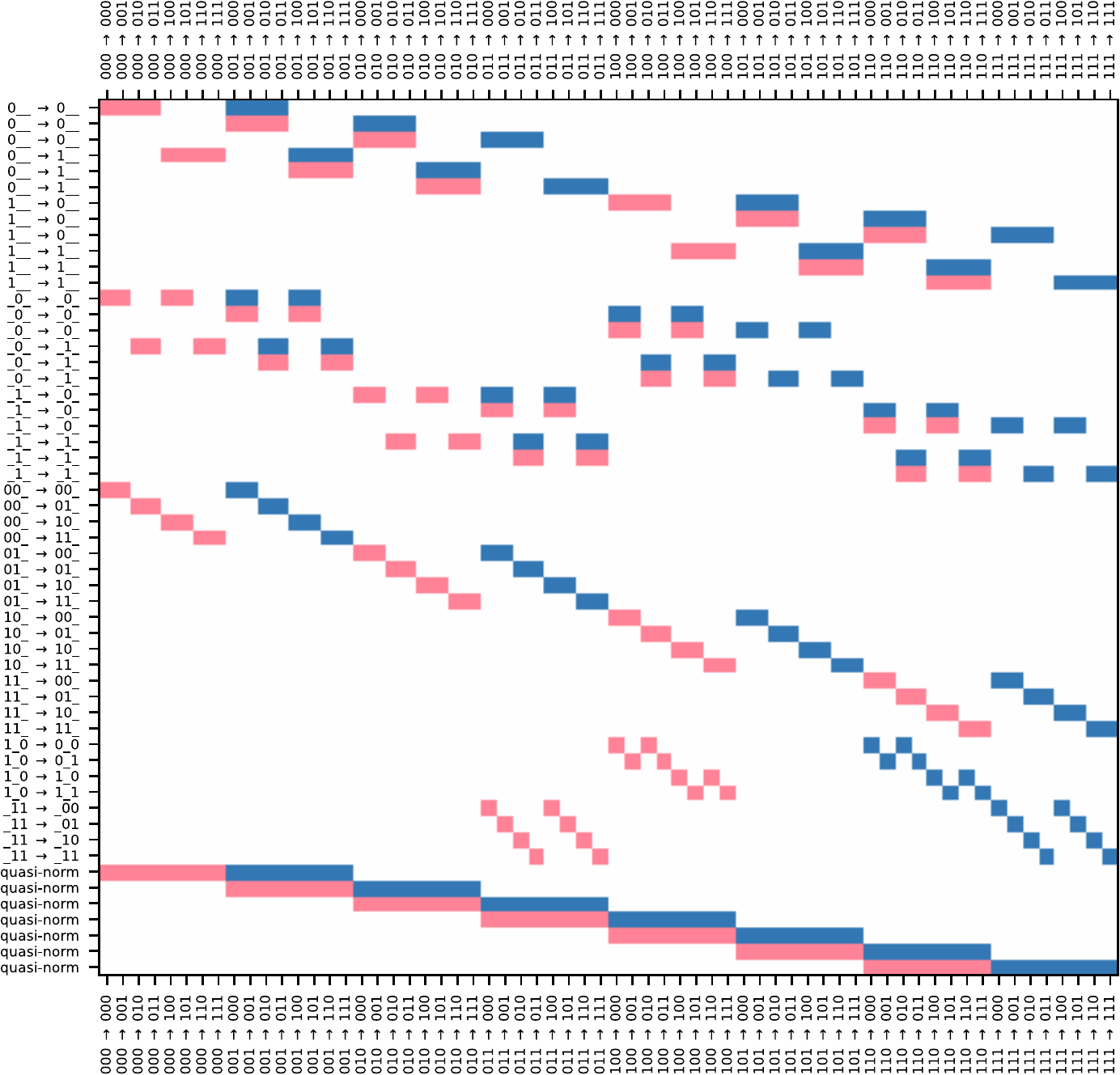}
\end{center}

\noindent Rows correspond to the 55 linear equations, of which 27 are independent.

\newpage
\subsection*{Space 67}

Space 67 is not induced by a causal order, but it is a refinement of the space 92 induced by the definite causal order $\total{\ev{A},\ev{C}}\vee\total{\ev{B},\ev{C}}$.
Its equivalence class under event-input permutation symmetry contains 12 spaces.
Space 67 differs as follows from the space induced by causal order $\total{\ev{A},\ev{C}}\vee\total{\ev{B},\ev{C}}$:
\begin{itemize}
  \item The outputs at events \evset{\ev{B}, \ev{C}} are independent of the input at event \ev{A} when the inputs at events \evset{B, C} are given by \hist{B/1,C/0} and \hist{B/1,C/1}.
\end{itemize}

\noindent Below are the histories and extended histories for space 67: 
\begin{center}
    \begin{tabular}{cc}
    \includegraphics[height=3.5cm]{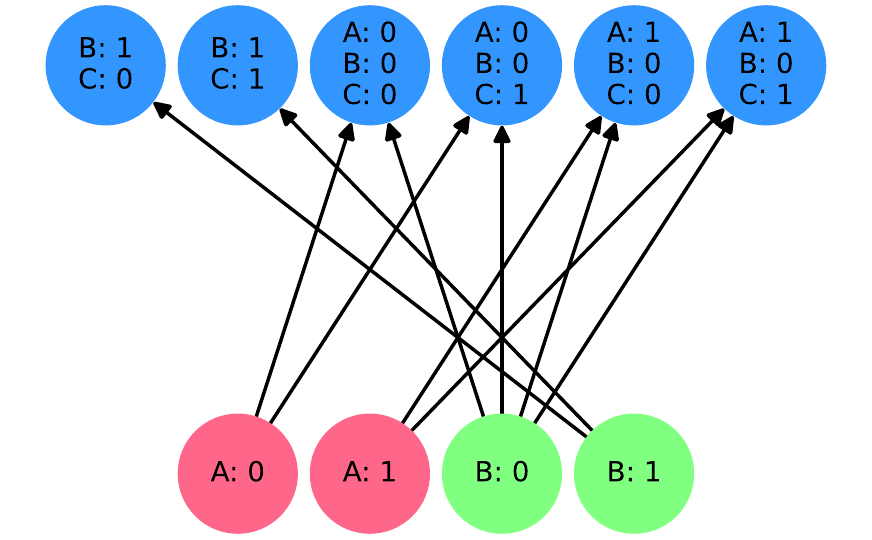}
    &
    \includegraphics[height=3.5cm]{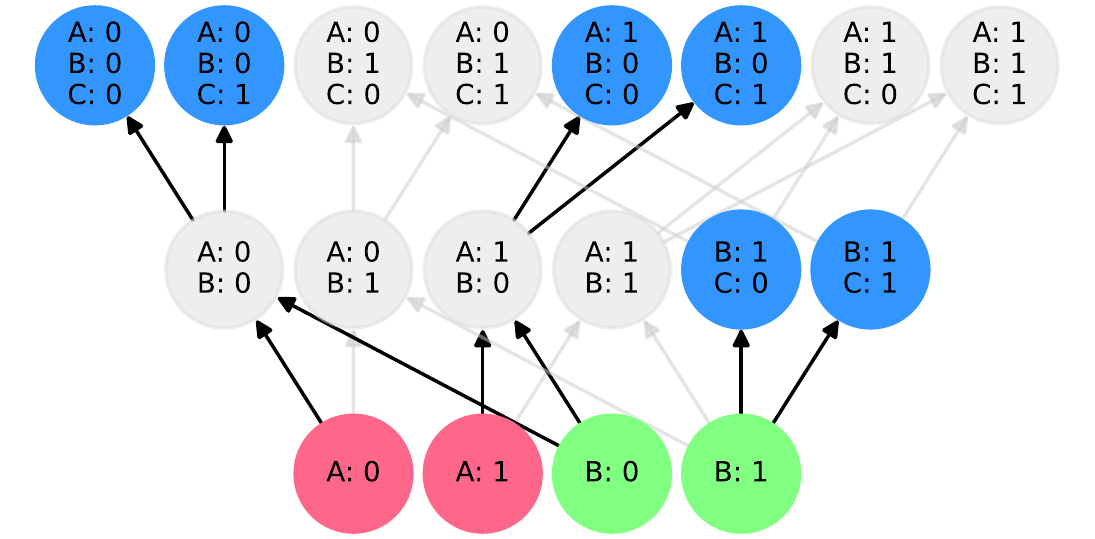}
    \\
    $\Theta_{67}$
    &
    $\Ext{\Theta_{67}}$
    \end{tabular}
\end{center}

\noindent The standard causaltope for Space 67 has dimension 36.
Below is a plot of the homogeneous linear system of causality and quasi-normalisation equations for the standard causaltope, put in reduced row echelon form:

\begin{center}
    \includegraphics[width=11cm]{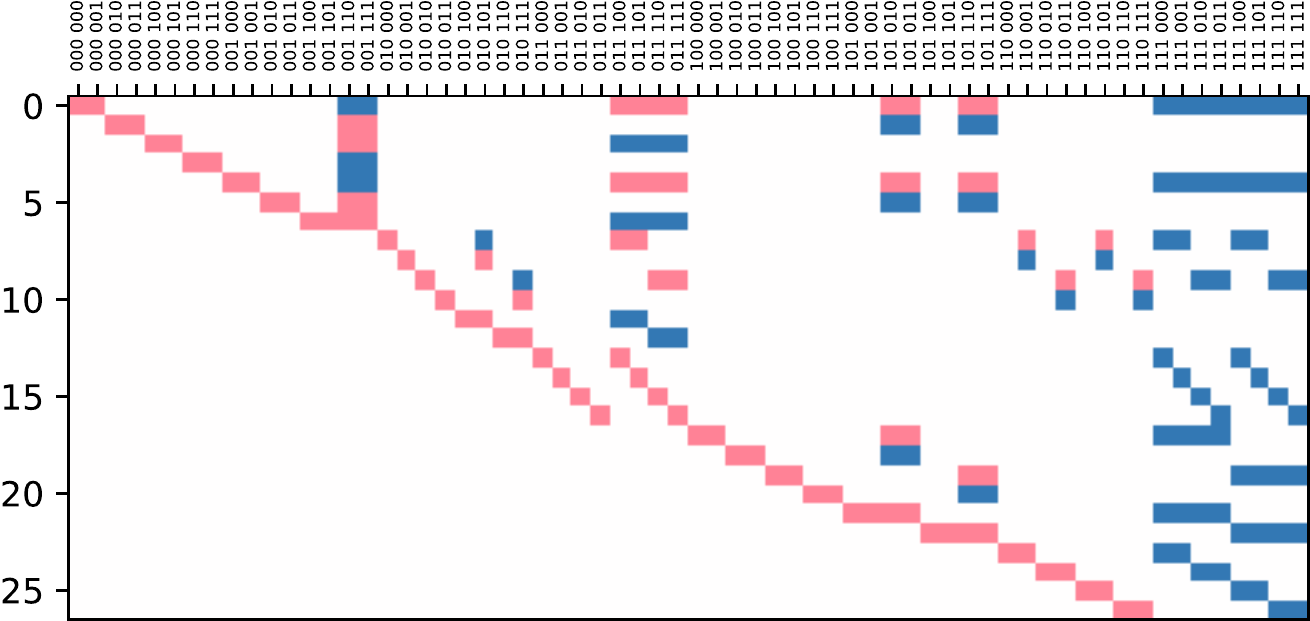}
\end{center}

\noindent Rows correspond to the 27 independent linear equations.
Columns in the plot correspond to entries of empirical models, indexed as $i_A i_B i_C$ $o_A o_B o_C$.
Coefficients in the equations are color-coded as white=0, red=+1 and blue=-1.

Space 67 has closest refinements in equivalence classes 46 and 53; 
it is the join of its (closest) refinements.
It has closest coarsenings in equivalence classes 81, 83 and 84; 
it is the meet of its (closest) coarsenings.
It has 1024 causal functions, 256 of which are not causal for any of its refinements.
It is a tight space.

The standard causaltope for Space 67 has 2 more dimensions than those of its 6 subspaces in equivalence classes 46 and 53.
The standard causaltope for Space 67 is the meet of the standard causaltopes for its closest coarsenings.
For completeness, below is a plot of the full homogeneous linear system of causality and quasi-normalisation equations for the standard causaltope:

\begin{center}
    \includegraphics[width=12cm]{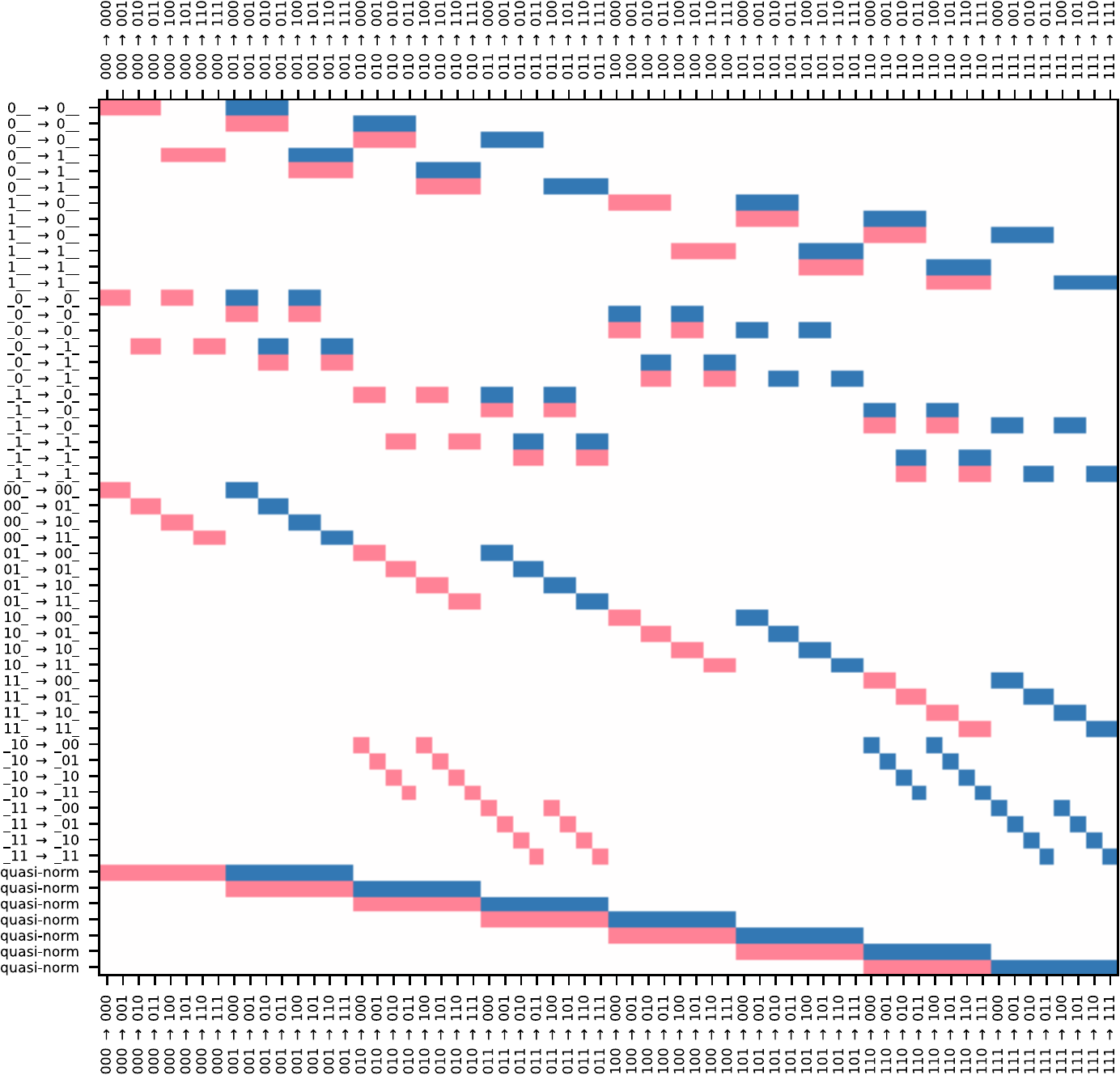}
\end{center}

\noindent Rows correspond to the 55 linear equations, of which 27 are independent.

\newpage
\subsection*{Space 68}

Space 68 is not induced by a causal order, but it is a refinement of the space 92 induced by the definite causal order $\total{\ev{A},\ev{C}}\vee\total{\ev{B},\ev{C}}$.
Its equivalence class under event-input permutation symmetry contains 12 spaces.
Space 68 differs as follows from the space induced by causal order $\total{\ev{A},\ev{C}}\vee\total{\ev{B},\ev{C}}$:
\begin{itemize}
  \item The outputs at events \evset{\ev{B}, \ev{C}} are independent of the input at event \ev{A} when the inputs at events \evset{B, C} are given by \hist{B/1,C/1} and \hist{B/0,C/1}.
\end{itemize}

\noindent Below are the histories and extended histories for space 68: 
\begin{center}
    \begin{tabular}{cc}
    \includegraphics[height=3.5cm]{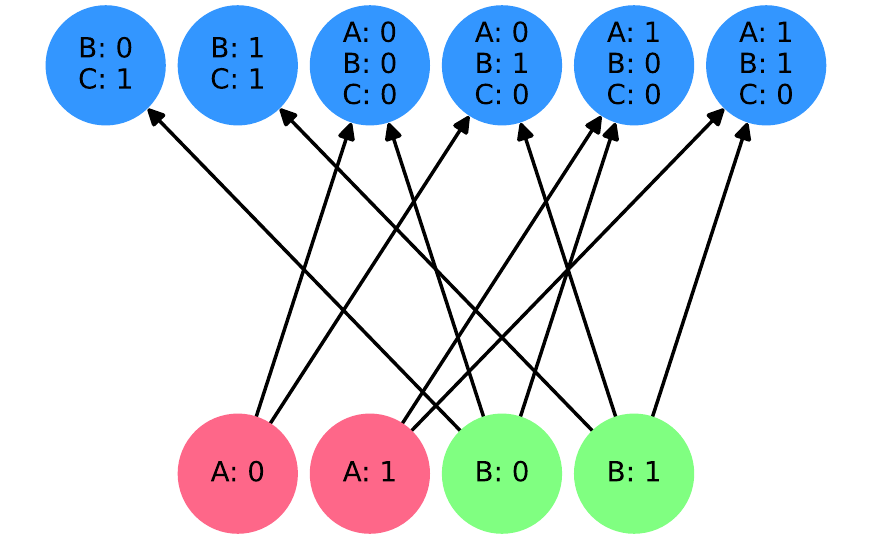}
    &
    \includegraphics[height=3.5cm]{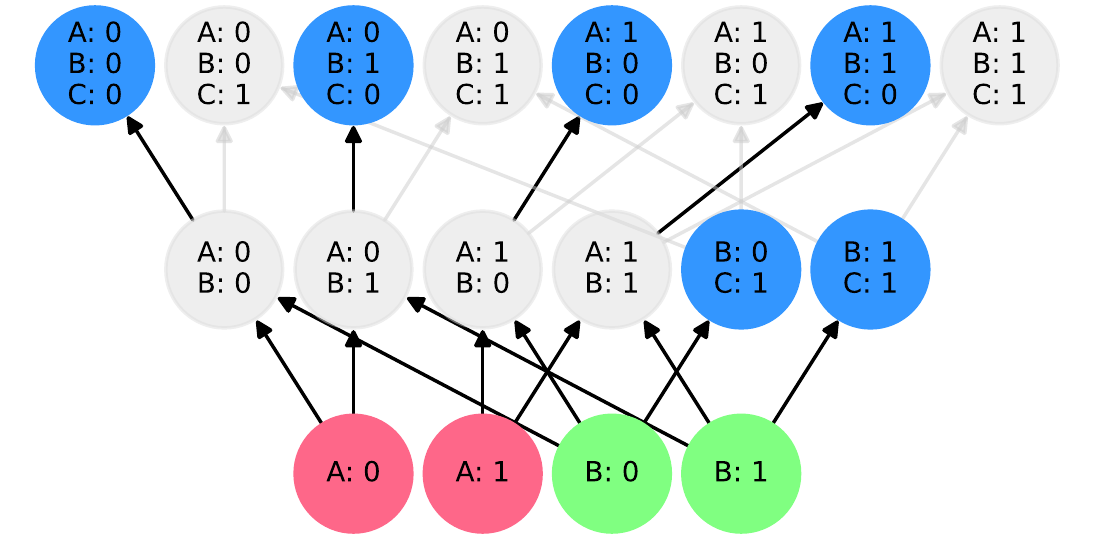}
    \\
    $\Theta_{68}$
    &
    $\Ext{\Theta_{68}}$
    \end{tabular}
\end{center}

\noindent The standard causaltope for Space 68 has dimension 36.
Below is a plot of the homogeneous linear system of causality and quasi-normalisation equations for the standard causaltope, put in reduced row echelon form:

\begin{center}
    \includegraphics[width=11cm]{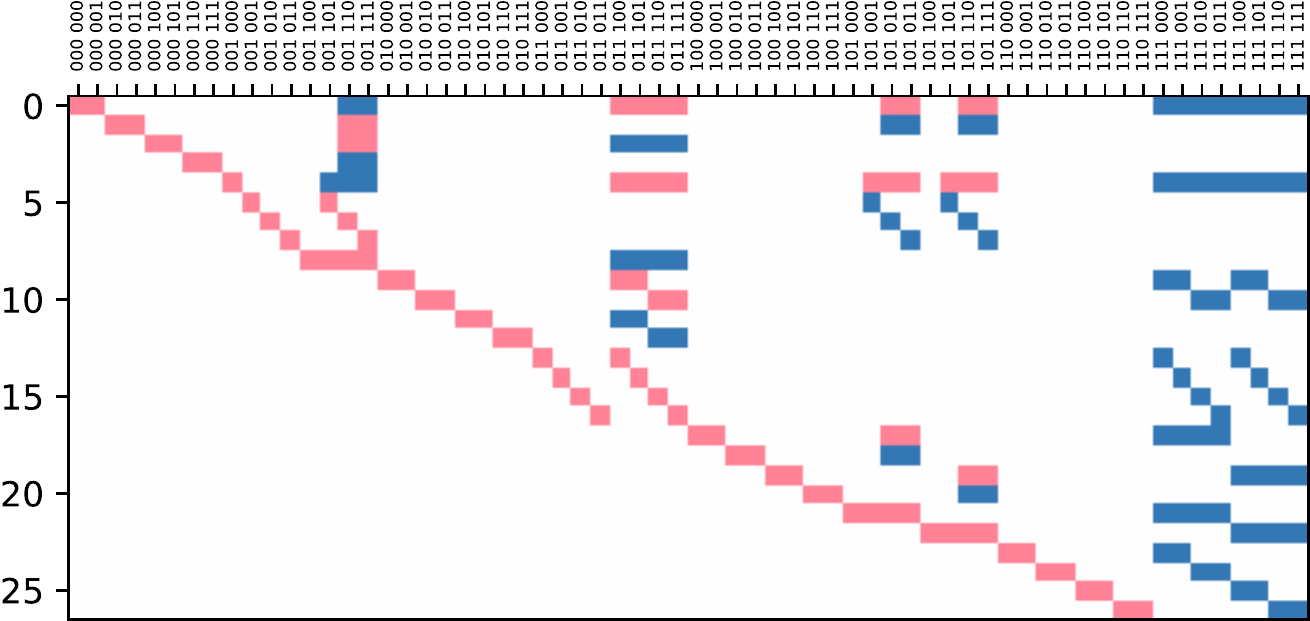}
\end{center}

\noindent Rows correspond to the 27 independent linear equations.
Columns in the plot correspond to entries of empirical models, indexed as $i_A i_B i_C$ $o_A o_B o_C$.
Coefficients in the equations are color-coded as white=0, red=+1 and blue=-1.

Space 68 has closest refinements in equivalence classes 46, 49 and 59; 
it is the join of its (closest) refinements.
It has closest coarsenings in equivalence classes 84 and 86; 
it is the meet of its (closest) coarsenings.
It has 1024 causal functions, 128 of which are not causal for any of its refinements.
It is a tight space.

The standard causaltope for Space 68 has 2 more dimensions than those of its 6 subspaces in equivalence classes 46, 49 and 59.
The standard causaltope for Space 68 is the meet of the standard causaltopes for its closest coarsenings.
For completeness, below is a plot of the full homogeneous linear system of causality and quasi-normalisation equations for the standard causaltope:

\begin{center}
    \includegraphics[width=12cm]{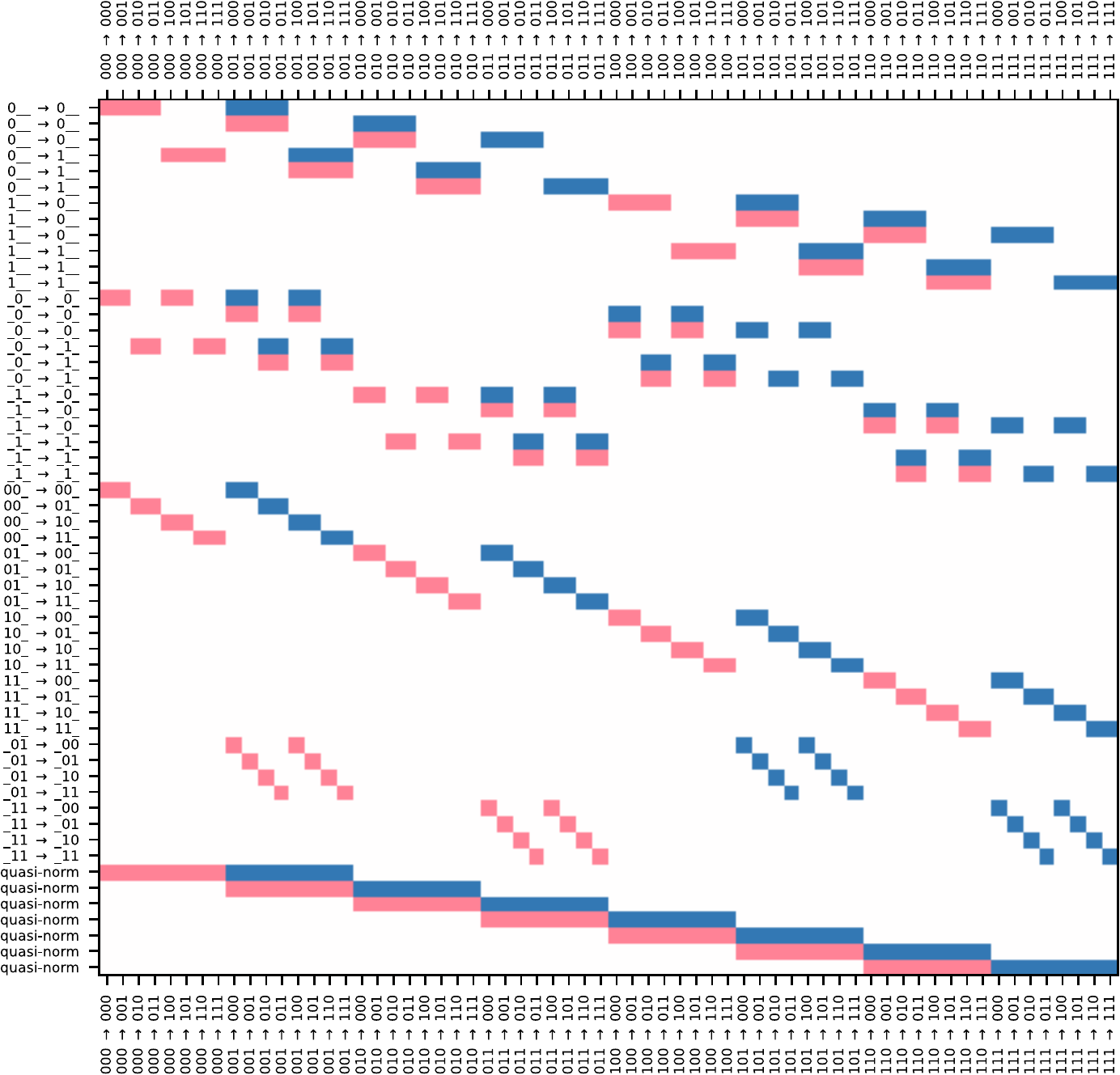}
\end{center}

\noindent Rows correspond to the 55 linear equations, of which 27 are independent.

\newpage
\subsection*{Space 69}

Space 69 is not induced by a causal order, but it is a refinement of the space 92 induced by the definite causal order $\total{\ev{A},\ev{C}}\vee\total{\ev{B},\ev{C}}$.
Its equivalence class under event-input permutation symmetry contains 24 spaces.
Space 69 differs as follows from the space induced by causal order $\total{\ev{A},\ev{C}}\vee\total{\ev{B},\ev{C}}$:
\begin{itemize}
  \item The outputs at events \evset{\ev{B}, \ev{C}} are independent of the input at event \ev{A} when the inputs at events \evset{B, C} are given by \hist{B/0,C/1}.
  \item The outputs at events \evset{\ev{A}, \ev{C}} are independent of the input at event \ev{B} when the inputs at events \evset{A, C} are given by \hist{A/1,C/1}.
\end{itemize}

\noindent Below are the histories and extended histories for space 69: 
\begin{center}
    \begin{tabular}{cc}
    \includegraphics[height=3.5cm]{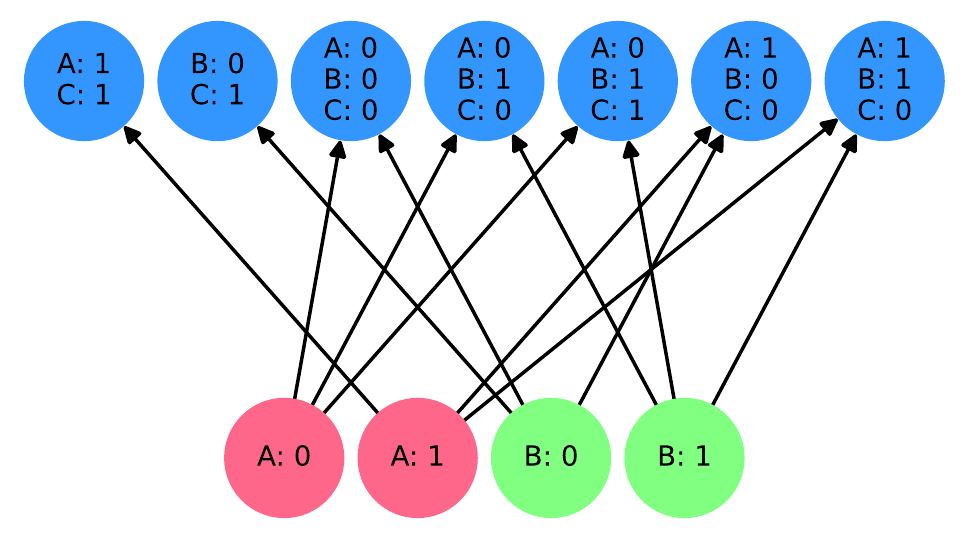}
    &
    \includegraphics[height=3.5cm]{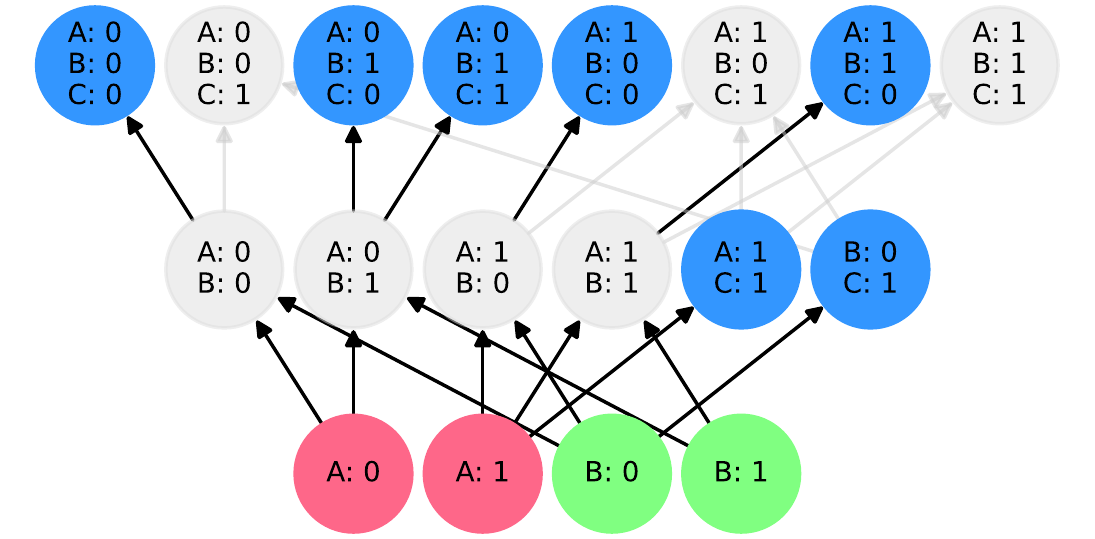}
    \\
    $\Theta_{69}$
    &
    $\Ext{\Theta_{69}}$
    \end{tabular}
\end{center}

\noindent The standard causaltope for Space 69 has dimension 36.
Below is a plot of the homogeneous linear system of causality and quasi-normalisation equations for the standard causaltope, put in reduced row echelon form:

\begin{center}
    \includegraphics[width=11cm]{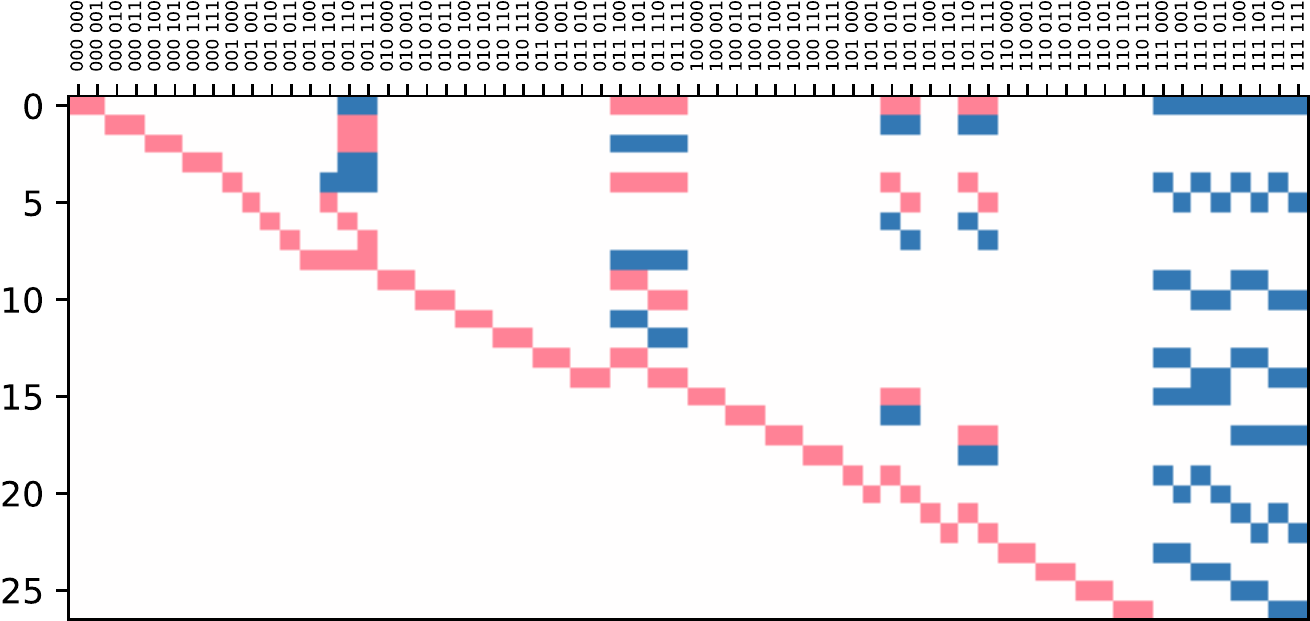}
\end{center}

\noindent Rows correspond to the 27 independent linear equations.
Columns in the plot correspond to entries of empirical models, indexed as $i_A i_B i_C$ $o_A o_B o_C$.
Coefficients in the equations are color-coded as white=0, red=+1 and blue=-1.

Space 69 has closest refinements in equivalence classes 50, 53 and 59; 
it is the join of its (closest) refinements.
It has closest coarsenings in equivalence classes 82 and 84; 
it is the meet of its (closest) coarsenings.
It has 1024 causal functions, 256 of which are not causal for any of its refinements.
It is not a tight space: for event \ev{C}, a causal function must yield identical output values on input histories \hist{A/1,C/1} and \hist{B/0,C/1}.

The standard causaltope for Space 69 has 2 more dimensions than those of its 6 subspaces in equivalence classes 50, 53 and 59.
The standard causaltope for Space 69 is the meet of the standard causaltopes for its closest coarsenings.
For completeness, below is a plot of the full homogeneous linear system of causality and quasi-normalisation equations for the standard causaltope:

\begin{center}
    \includegraphics[width=12cm]{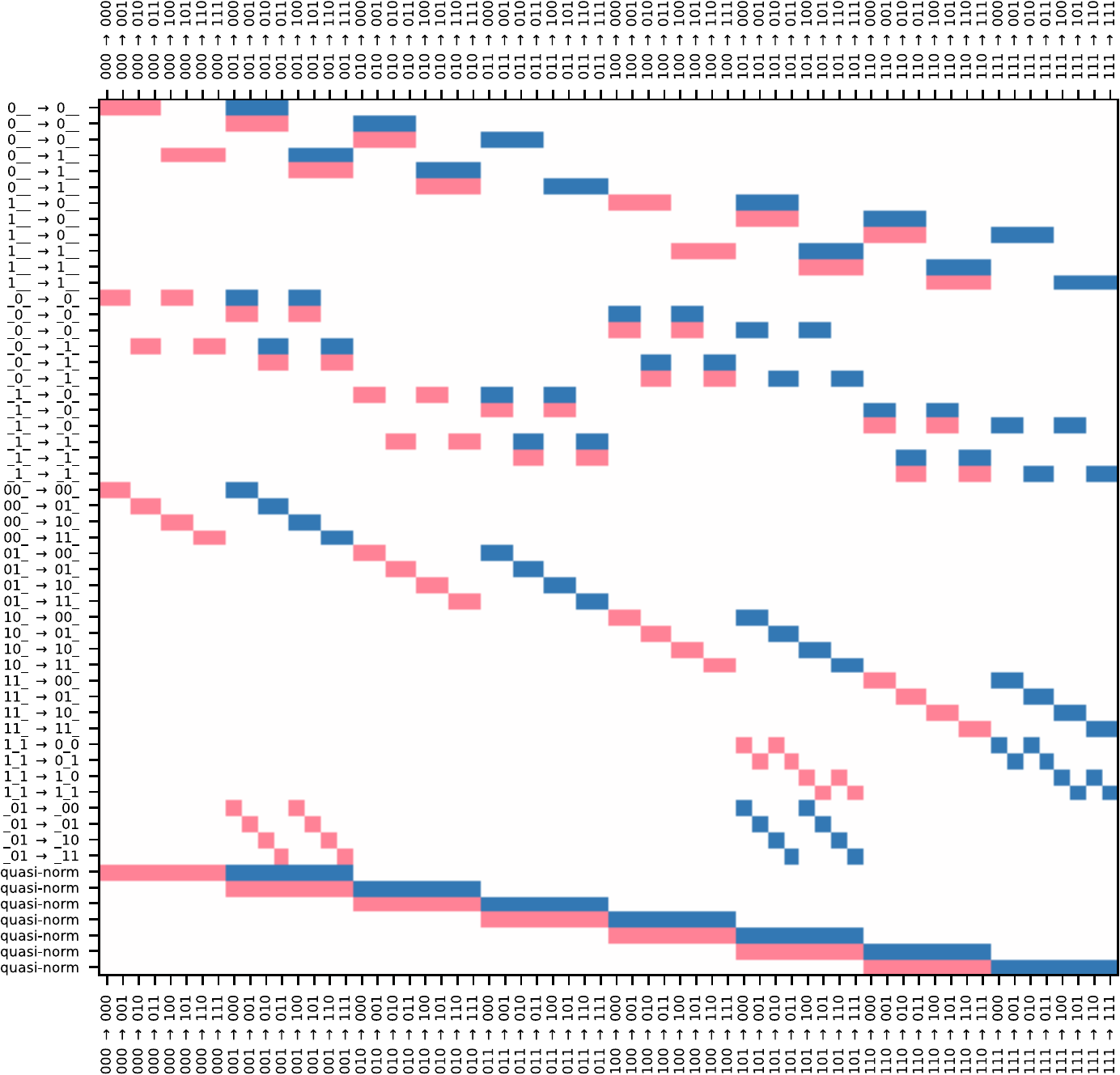}
\end{center}

\noindent Rows correspond to the 55 linear equations, of which 27 are independent.

\newpage
\subsection*{Space 70}

Space 70 is not induced by a causal order, but it is a refinement of the space 92 induced by the definite causal order $\total{\ev{A},\ev{C}}\vee\total{\ev{B},\ev{C}}$.
Its equivalence class under event-input permutation symmetry contains 12 spaces.
Space 70 differs as follows from the space induced by causal order $\total{\ev{A},\ev{C}}\vee\total{\ev{B},\ev{C}}$:
\begin{itemize}
  \item The outputs at events \evset{\ev{B}, \ev{C}} are independent of the input at event \ev{A} when the inputs at events \evset{B, C} are given by \hist{B/1,C/0} and \hist{B/0,C/1}.
\end{itemize}

\noindent Below are the histories and extended histories for space 70: 
\begin{center}
    \begin{tabular}{cc}
    \includegraphics[height=3.5cm]{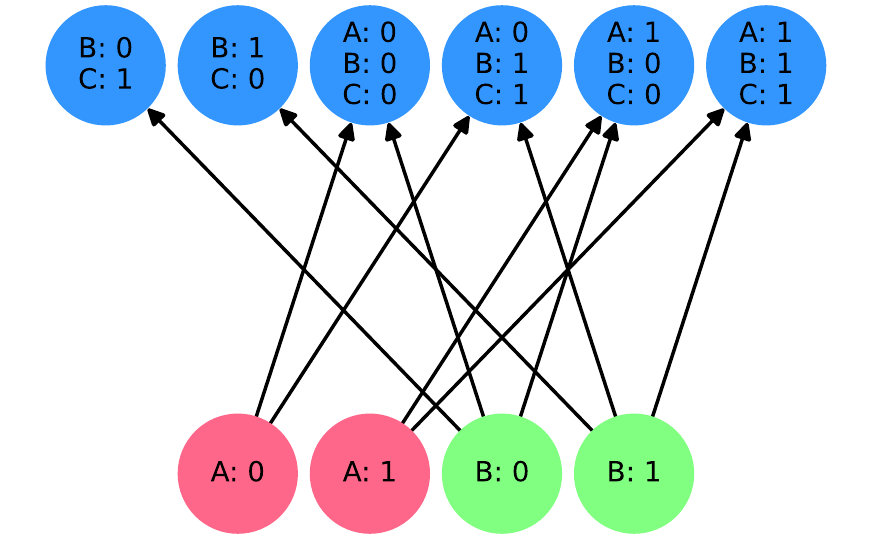}
    &
    \includegraphics[height=3.5cm]{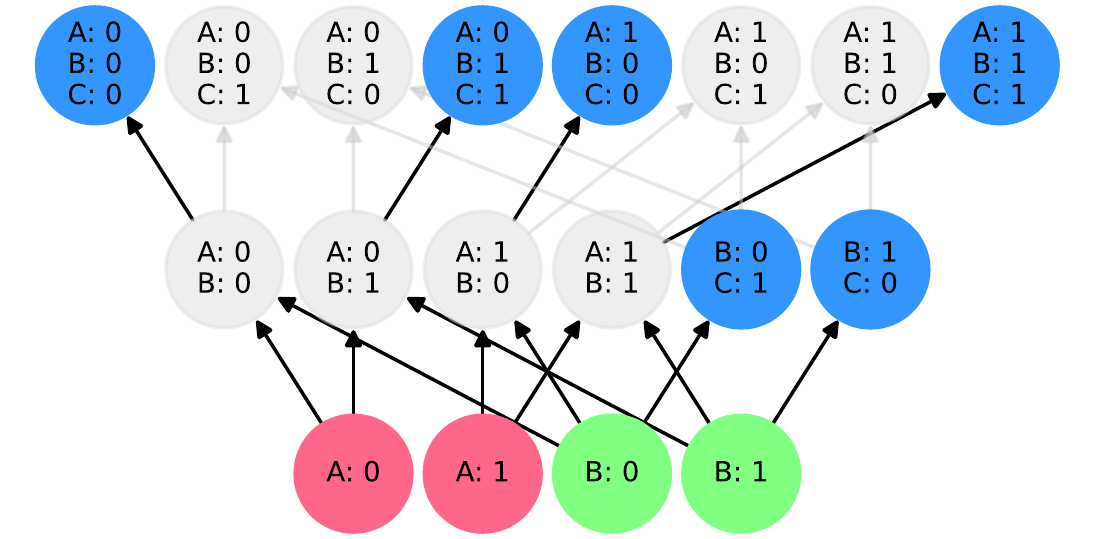}
    \\
    $\Theta_{70}$
    &
    $\Ext{\Theta_{70}}$
    \end{tabular}
\end{center}

\noindent The standard causaltope for Space 70 has dimension 36.
Below is a plot of the homogeneous linear system of causality and quasi-normalisation equations for the standard causaltope, put in reduced row echelon form:

\begin{center}
    \includegraphics[width=11cm]{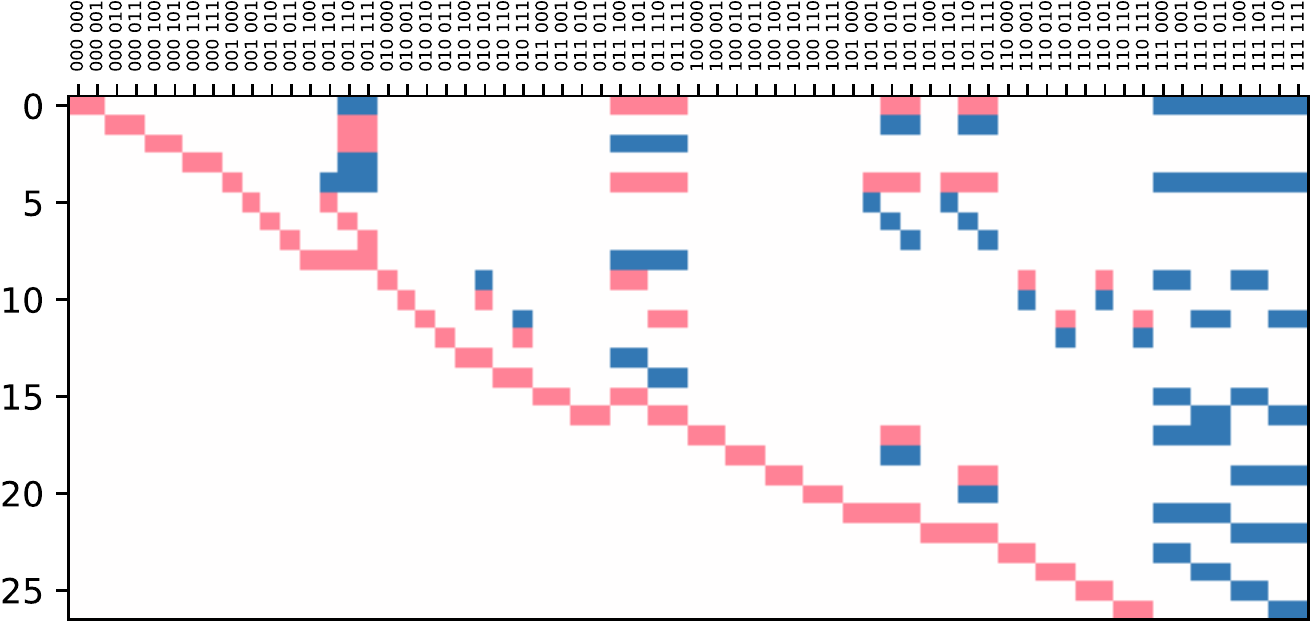}
\end{center}

\noindent Rows correspond to the 27 independent linear equations.
Columns in the plot correspond to entries of empirical models, indexed as $i_A i_B i_C$ $o_A o_B o_C$.
Coefficients in the equations are color-coded as white=0, red=+1 and blue=-1.

Space 70 has closest refinements in equivalence classes 46 and 50; 
it is the join of its (closest) refinements.
It has closest coarsenings in equivalence classes 84 and 87; 
it is the meet of its (closest) coarsenings.
It has 1024 causal functions, 128 of which are not causal for any of its refinements.
It is a tight space.

The standard causaltope for Space 70 has 2 more dimensions than those of its 6 subspaces in equivalence classes 46 and 50.
The standard causaltope for Space 70 is the meet of the standard causaltopes for its closest coarsenings.
For completeness, below is a plot of the full homogeneous linear system of causality and quasi-normalisation equations for the standard causaltope:

\begin{center}
    \includegraphics[width=12cm]{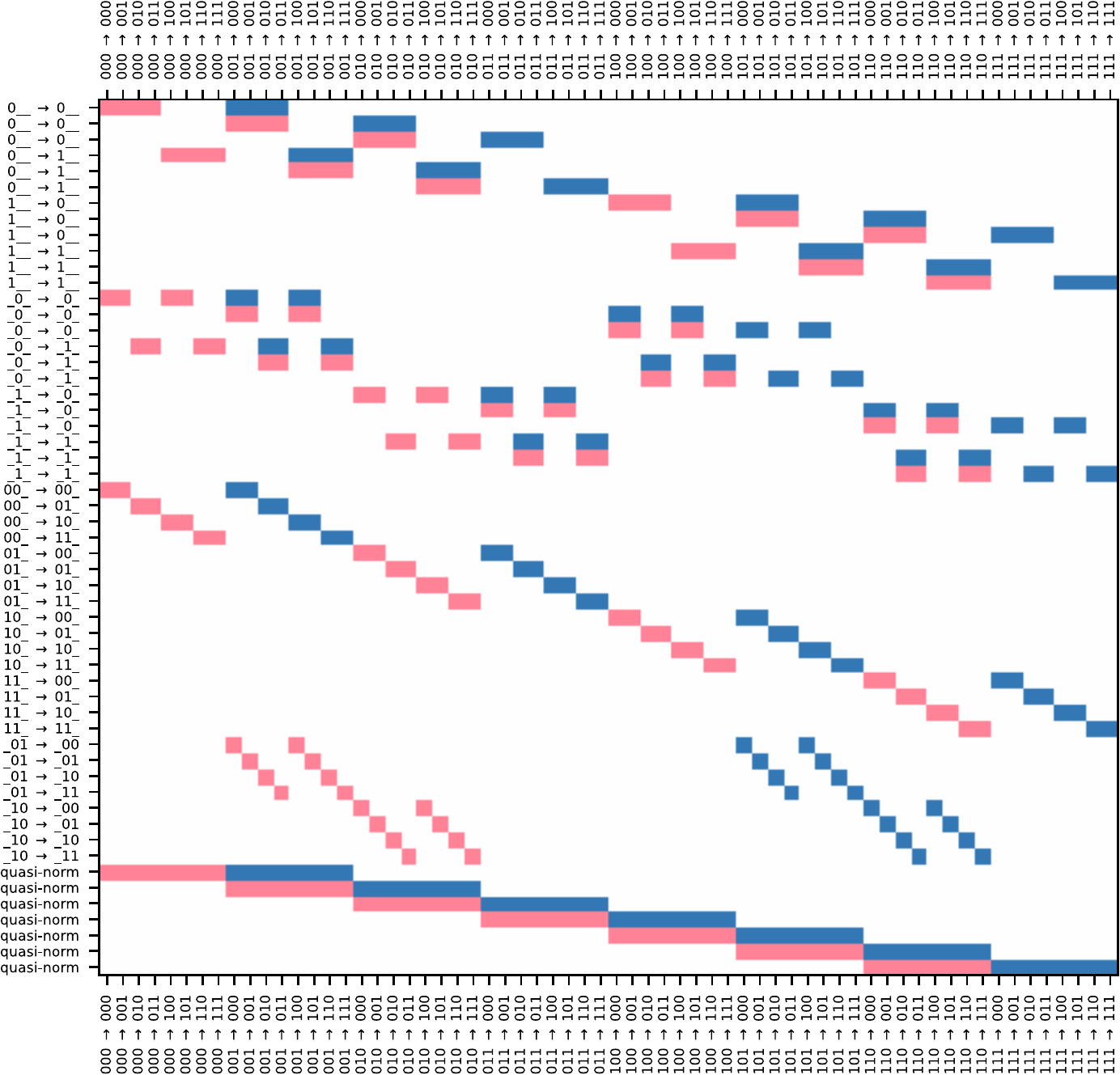}
\end{center}

\noindent Rows correspond to the 55 linear equations, of which 27 are independent.

\newpage
\subsection*{Space 71}

Space 71 is not induced by a causal order, but it is a refinement of the space induced by the indefinite causal order $\total{\ev{A},\{\ev{B},\ev{C}\}}$.
Its equivalence class under event-input permutation symmetry contains 48 spaces.
Space 71 differs as follows from the space induced by causal order $\total{\ev{A},\{\ev{B},\ev{C}\}}$:
\begin{itemize}
  \item The outputs at events \evset{\ev{A}, \ev{B}} are independent of the input at event \ev{C} when the inputs at events \evset{A, B} are given by \hist{A/0,B/0}, \hist{A/0,B/1} and \hist{A/1,B/0}.
  \item The outputs at events \evset{\ev{A}, \ev{C}} are independent of the input at event \ev{B} when the inputs at events \evset{A, C} are given by \hist{A/0,C/1}, \hist{A/1,C/0} and \hist{A/1,C/1}.
  \item The output at event \ev{C} is independent of the inputs at events \evset{\ev{A}, \ev{B}} when the input at event C is given by \hist{C/1}.
  \item The outputs at events \evset{\ev{B}, \ev{C}} are independent of the input at event \ev{A} when the inputs at events \evset{B, C} are given by \hist{B/0,C/1}.
\end{itemize}

\noindent Below are the histories and extended histories for space 71: 
\begin{center}
    \begin{tabular}{cc}
    \includegraphics[height=3.5cm]{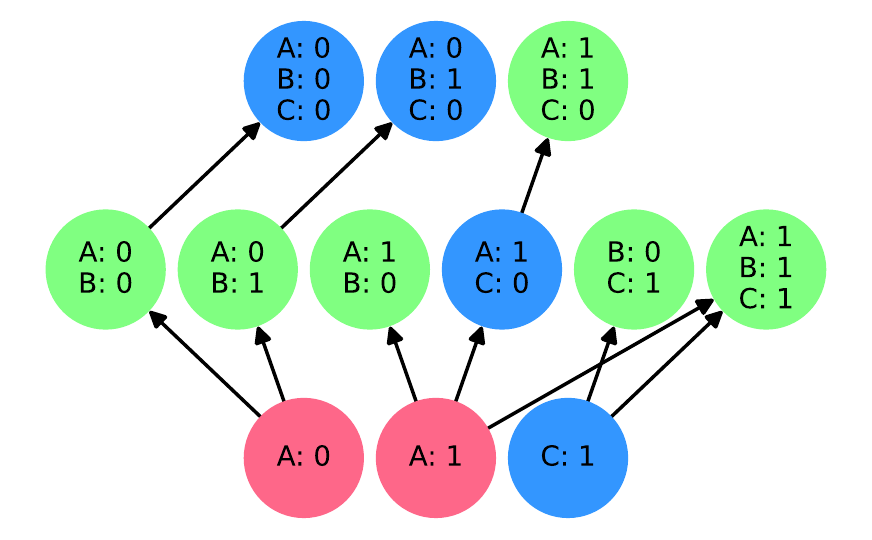}
    &
    \includegraphics[height=3.5cm]{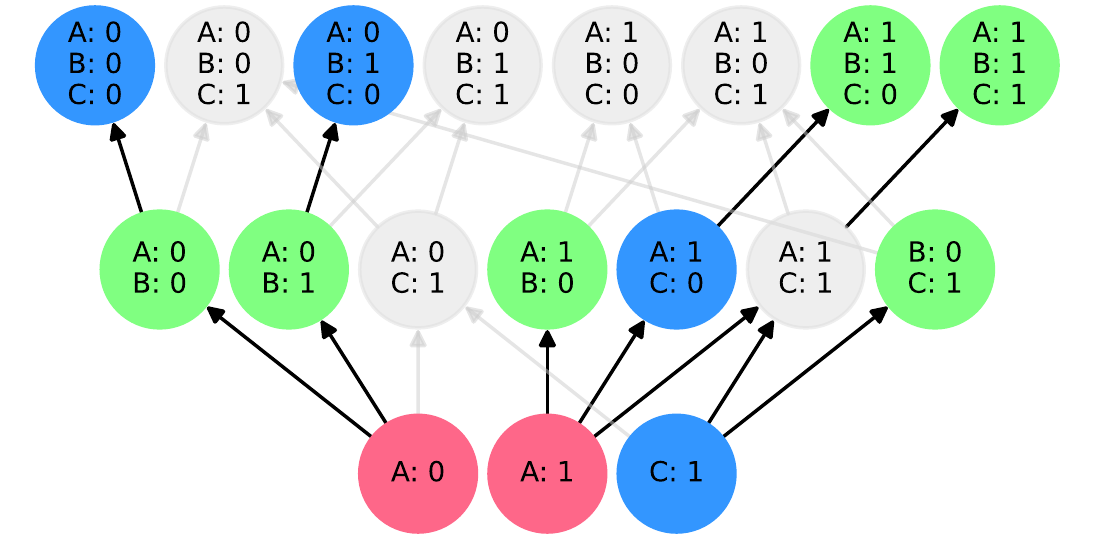}
    \\
    $\Theta_{71}$
    &
    $\Ext{\Theta_{71}}$
    \end{tabular}
\end{center}

\noindent The standard causaltope for Space 71 has dimension 35.
Below is a plot of the homogeneous linear system of causality and quasi-normalisation equations for the standard causaltope, put in reduced row echelon form:

\begin{center}
    \includegraphics[width=11cm]{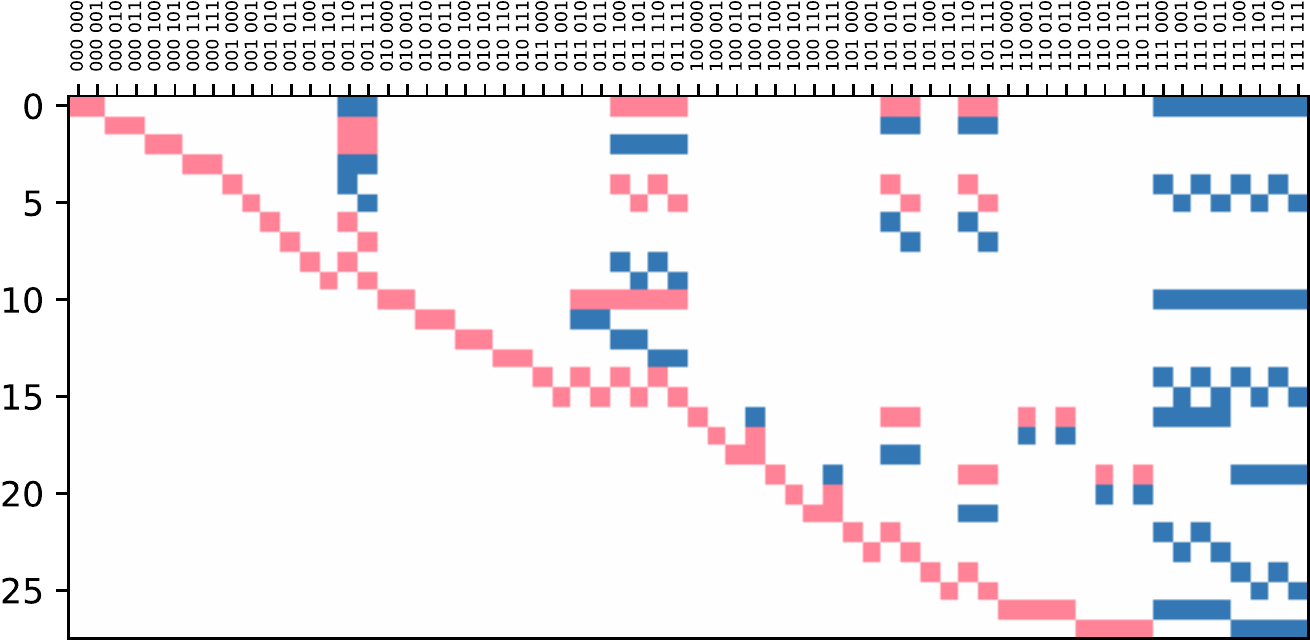}
\end{center}

\noindent Rows correspond to the 28 independent linear equations.
Columns in the plot correspond to entries of empirical models, indexed as $i_A i_B i_C$ $o_A o_B o_C$.
Coefficients in the equations are color-coded as white=0, red=+1 and blue=-1.

Space 71 has closest refinements in equivalence classes 47, 48, 52 and 57; 
it is the join of its (closest) refinements.
It has closest coarsenings in equivalence classes 79 and 85; 
it is the meet of its (closest) coarsenings.
It has 1024 causal functions, all of which are causal for at least one of its refinements.
It is not a tight space: for event \ev{B}, a causal function must yield identical output values on input histories \hist{A/0,B/0}, \hist{A/1,B/0} and \hist{B/0,C/1}.

The standard causaltope for Space 71 coincides with that of its subspace in equivalence class 48.
The standard causaltope for Space 71 is the meet of the standard causaltopes for its closest coarsenings.
For completeness, below is a plot of the full homogeneous linear system of causality and quasi-normalisation equations for the standard causaltope:

\begin{center}
    \includegraphics[width=12cm]{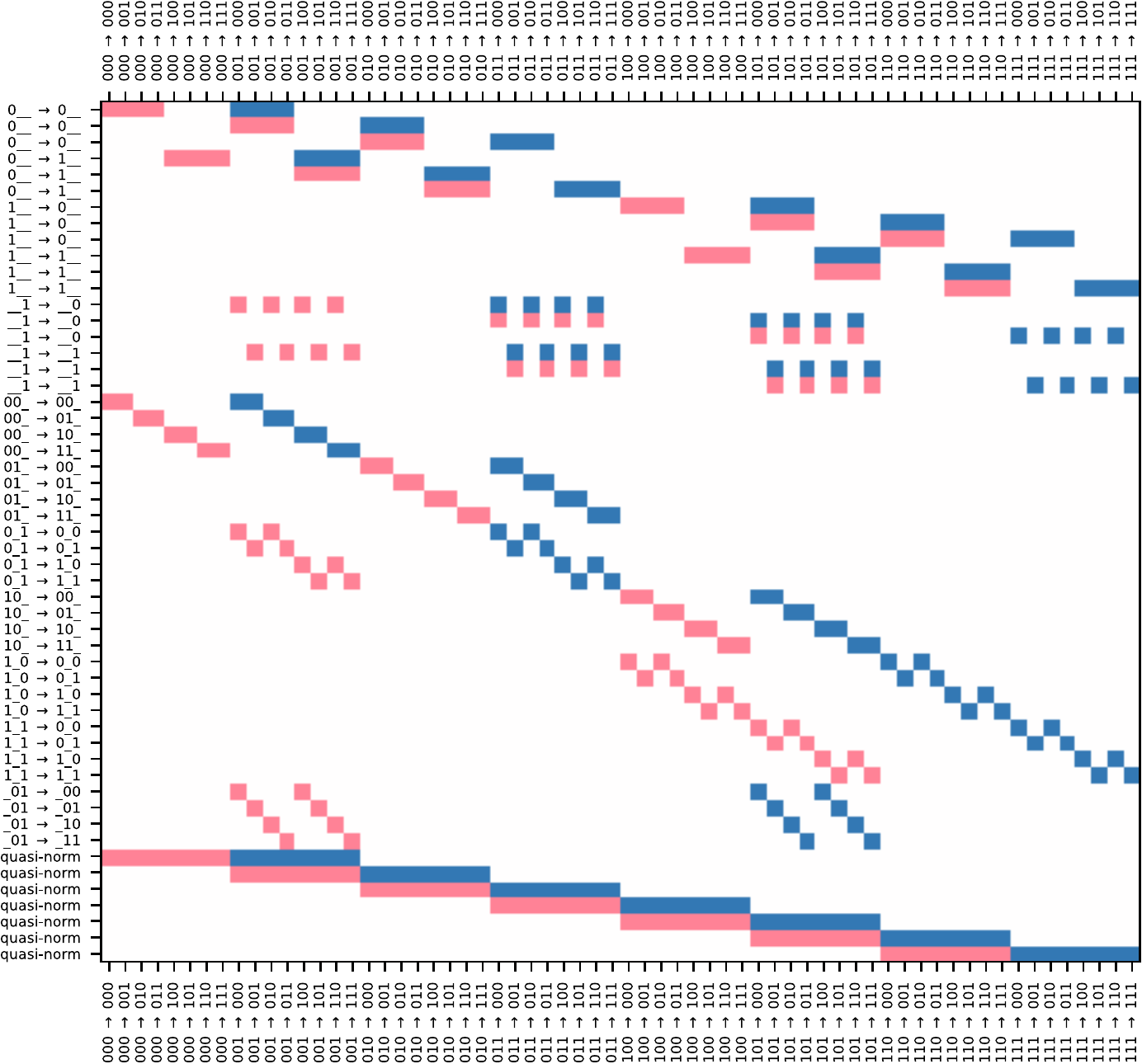}
\end{center}

\noindent Rows correspond to the 53 linear equations, of which 28 are independent.

\newpage
\subsection*{Space 72}

Space 72 is not induced by a causal order, but it is a refinement of the space induced by the indefinite causal order $\total{\ev{A},\{\ev{B},\ev{C}\}}$.
Its equivalence class under event-input permutation symmetry contains 24 spaces.
Space 72 differs as follows from the space induced by causal order $\total{\ev{A},\{\ev{B},\ev{C}\}}$:
\begin{itemize}
  \item The outputs at events \evset{\ev{A}, \ev{B}} are independent of the input at event \ev{C} when the inputs at events \evset{A, B} are given by \hist{A/0,B/0}, \hist{A/0,B/1} and \hist{A/1,B/1}.
  \item The outputs at events \evset{\ev{B}, \ev{C}} are independent of the input at event \ev{A} when the inputs at events \evset{B, C} are given by \hist{B/1,C/0} and \hist{B/1,C/1}.
  \item The outputs at events \evset{\ev{A}, \ev{C}} are independent of the input at event \ev{B} when the inputs at events \evset{A, C} are given by \hist{A/1,C/0} and \hist{A/1,C/1}.
  \item The output at event \ev{B} is independent of the inputs at events \evset{\ev{A}, \ev{C}} when the input at event B is given by \hist{B/1}.
\end{itemize}

\noindent Below are the histories and extended histories for space 72: 
\begin{center}
    \begin{tabular}{cc}
    \includegraphics[height=3.5cm]{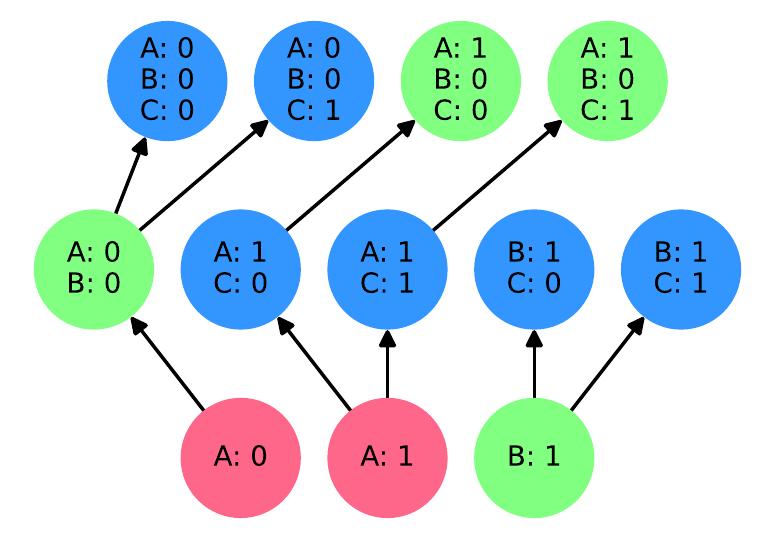}
    &
    \includegraphics[height=3.5cm]{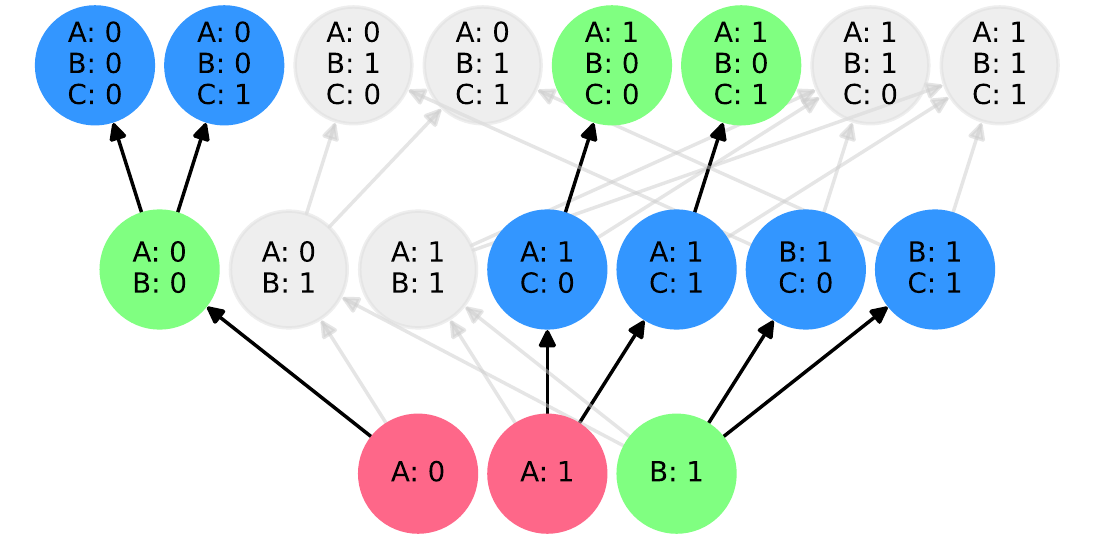}
    \\
    $\Theta_{72}$
    &
    $\Ext{\Theta_{72}}$
    \end{tabular}
\end{center}

\noindent The standard causaltope for Space 72 has dimension 35.
Below is a plot of the homogeneous linear system of causality and quasi-normalisation equations for the standard causaltope, put in reduced row echelon form:

\begin{center}
    \includegraphics[width=11cm]{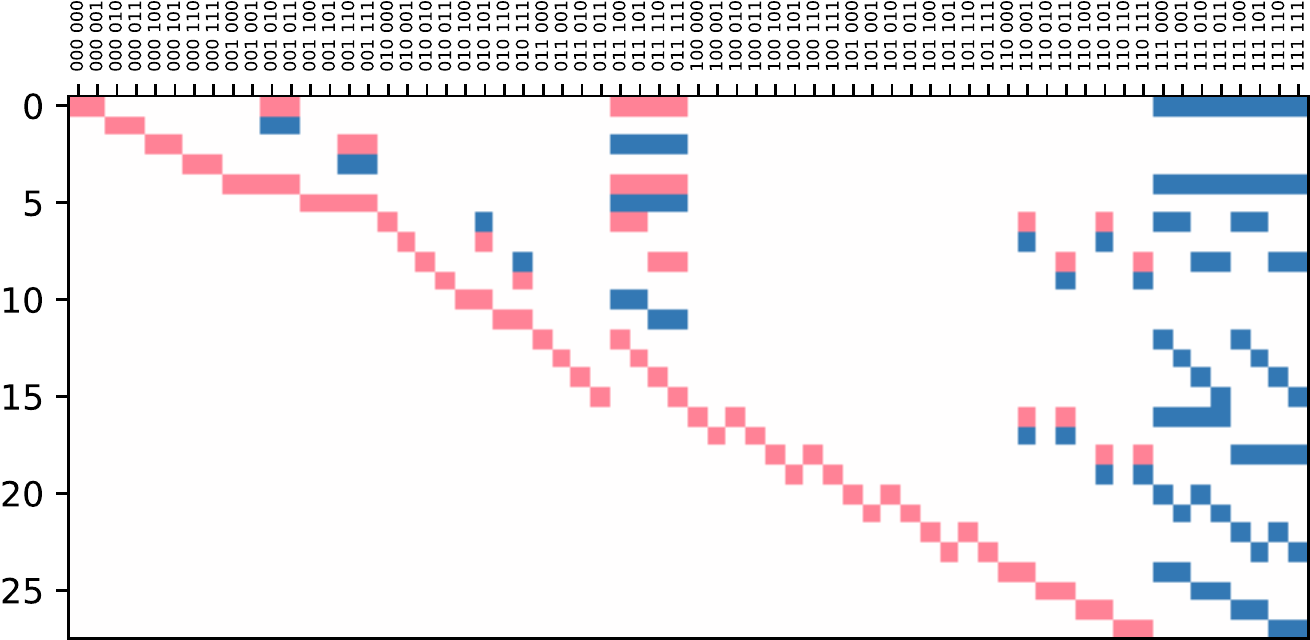}
\end{center}

\noindent Rows correspond to the 28 independent linear equations.
Columns in the plot correspond to entries of empirical models, indexed as $i_A i_B i_C$ $o_A o_B o_C$.
Coefficients in the equations are color-coded as white=0, red=+1 and blue=-1.

Space 72 has closest refinements in equivalence classes 52 and 60; 
it is the join of its (closest) refinements.
It has closest coarsenings in equivalence class 85; 
it is the meet of its (closest) coarsenings.
It has 1024 causal functions, 384 of which are not causal for any of its refinements.
It is not a tight space: for event \ev{C}, a causal function must yield identical output values on input histories \hist{A/1,C/0} and \hist{B/1,C/0}, and it must also yield identical output values on input histories \hist{A/1,C/1} and \hist{B/1,C/1}.

The standard causaltope for Space 72 has 2 more dimensions than those of its 3 subspaces in equivalence classes 52 and 60.
The standard causaltope for Space 72 is the meet of the standard causaltopes for its closest coarsenings.
For completeness, below is a plot of the full homogeneous linear system of causality and quasi-normalisation equations for the standard causaltope:

\begin{center}
    \includegraphics[width=12cm]{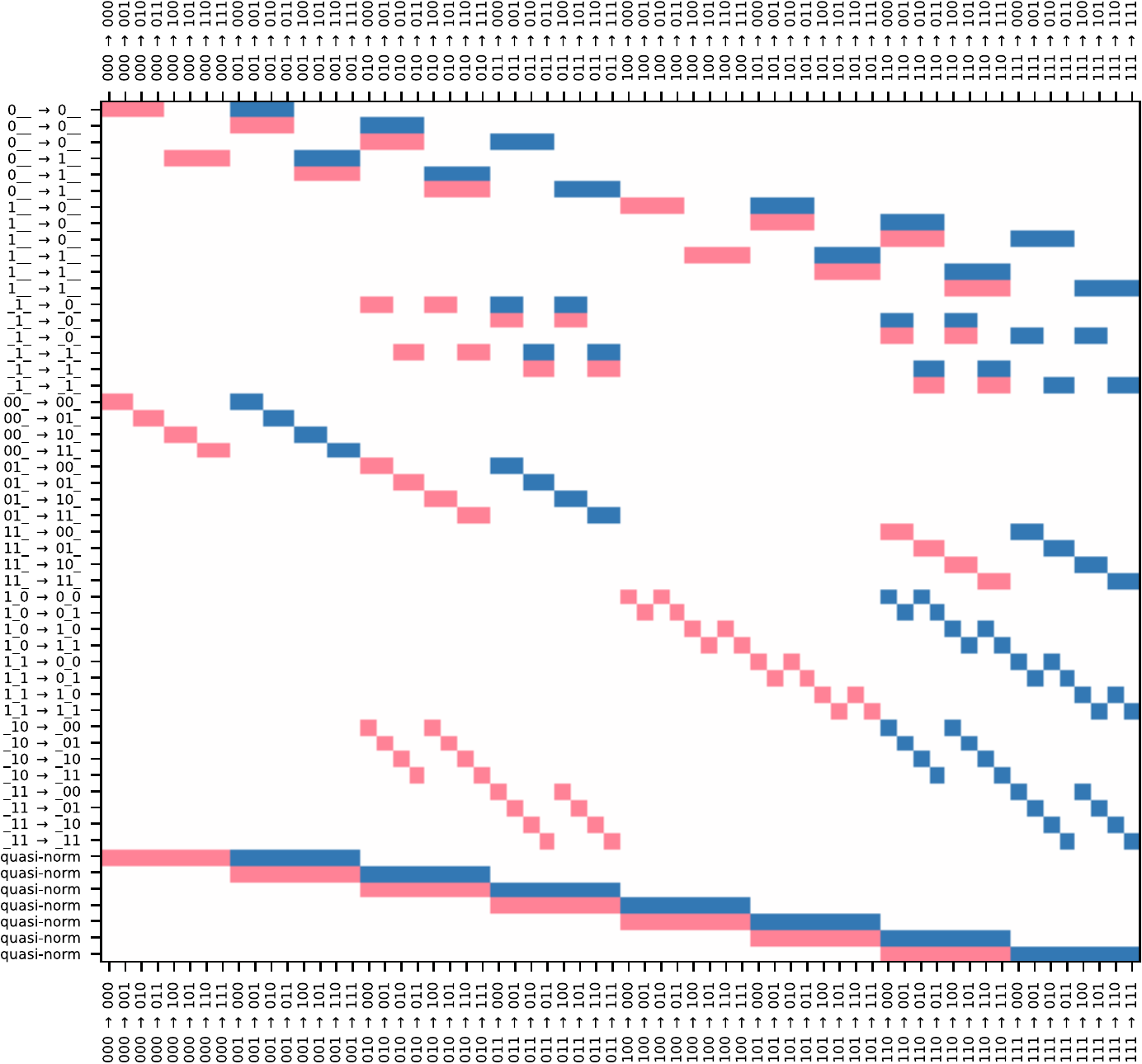}
\end{center}

\noindent Rows correspond to the 53 linear equations, of which 28 are independent.

\newpage
\subsection*{Space 73}

Space 73 is not induced by a causal order, but it is a refinement of the space 100 induced by the definite causal order $\total{\ev{A},\ev{B},\ev{C}}$.
Its equivalence class under event-input permutation symmetry contains 48 spaces.
Space 73 differs as follows from the space induced by causal order $\total{\ev{A},\ev{B},\ev{C}}$:
\begin{itemize}
  \item The outputs at events \evset{\ev{A}, \ev{C}} are independent of the input at event \ev{B} when the inputs at events \evset{A, C} are given by \hist{A/0,C/1} and \hist{A/1,C/0}.
  \item The outputs at events \evset{\ev{B}, \ev{C}} are independent of the input at event \ev{A} when the inputs at events \evset{B, C} are given by \hist{B/1,C/1}.
  \item The output at event \ev{B} is independent of the input at event \ev{A} when the input at event B is given by \hist{B/1}.
\end{itemize}

\noindent Below are the histories and extended histories for space 73: 
\begin{center}
    \begin{tabular}{cc}
    \includegraphics[height=3.5cm]{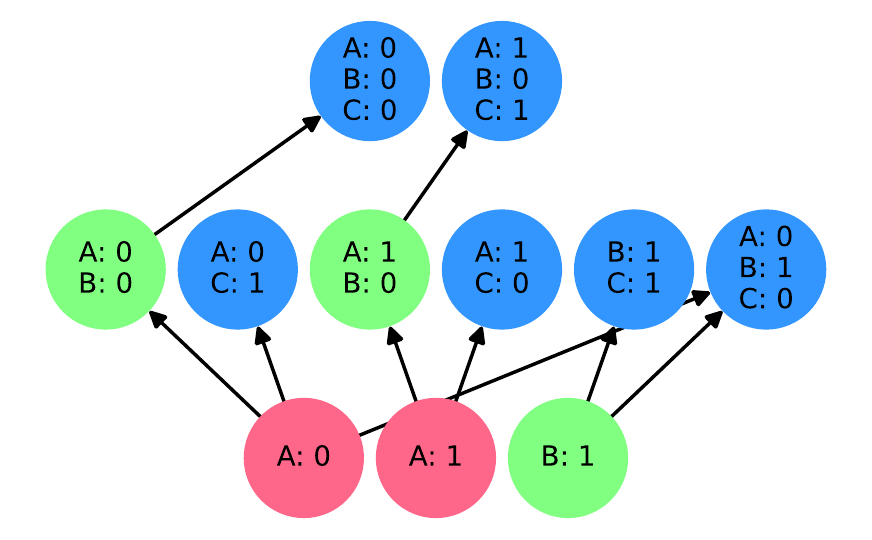}
    &
    \includegraphics[height=3.5cm]{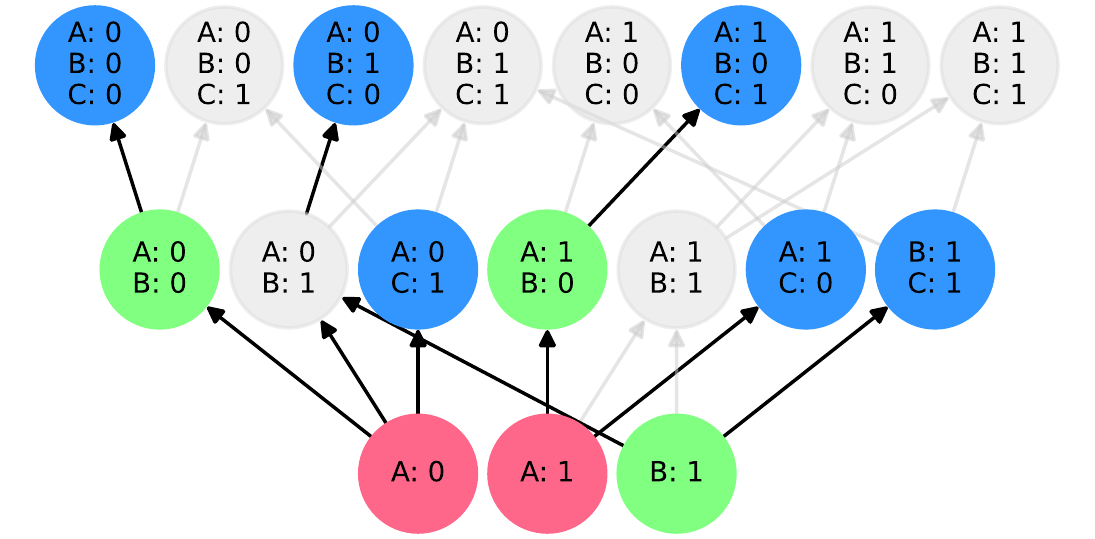}
    \\
    $\Theta_{73}$
    &
    $\Ext{\Theta_{73}}$
    \end{tabular}
\end{center}

\noindent The standard causaltope for Space 73 has dimension 35.
Below is a plot of the homogeneous linear system of causality and quasi-normalisation equations for the standard causaltope, put in reduced row echelon form:

\begin{center}
    \includegraphics[width=11cm]{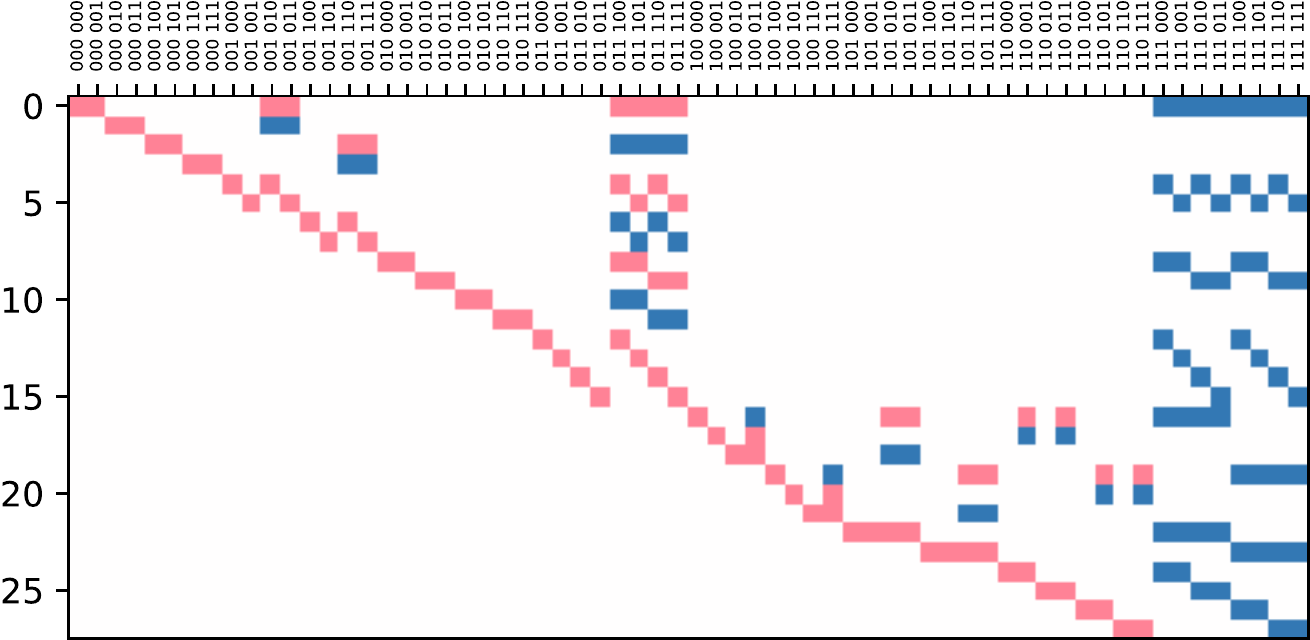}
\end{center}

\noindent Rows correspond to the 28 independent linear equations.
Columns in the plot correspond to entries of empirical models, indexed as $i_A i_B i_C$ $o_A o_B o_C$.
Coefficients in the equations are color-coded as white=0, red=+1 and blue=-1.

Space 73 has closest refinements in equivalence classes 50, 55, 56 and 57; 
it is the join of its (closest) refinements.
It has closest coarsenings in equivalence classes 80, 82 and 87; 
it is the meet of its (closest) coarsenings.
It has 1024 causal functions, 192 of which are not causal for any of its refinements.
It is not a tight space: for event \ev{C}, a causal function must yield identical output values on input histories \hist{A/0,C/1} and \hist{B/1,C/1}.

The standard causaltope for Space 73 has 1 more dimension than that of its subspace in equivalence class 50.
The standard causaltope for Space 73 is the meet of the standard causaltopes for its closest coarsenings.
For completeness, below is a plot of the full homogeneous linear system of causality and quasi-normalisation equations for the standard causaltope:

\begin{center}
    \includegraphics[width=12cm]{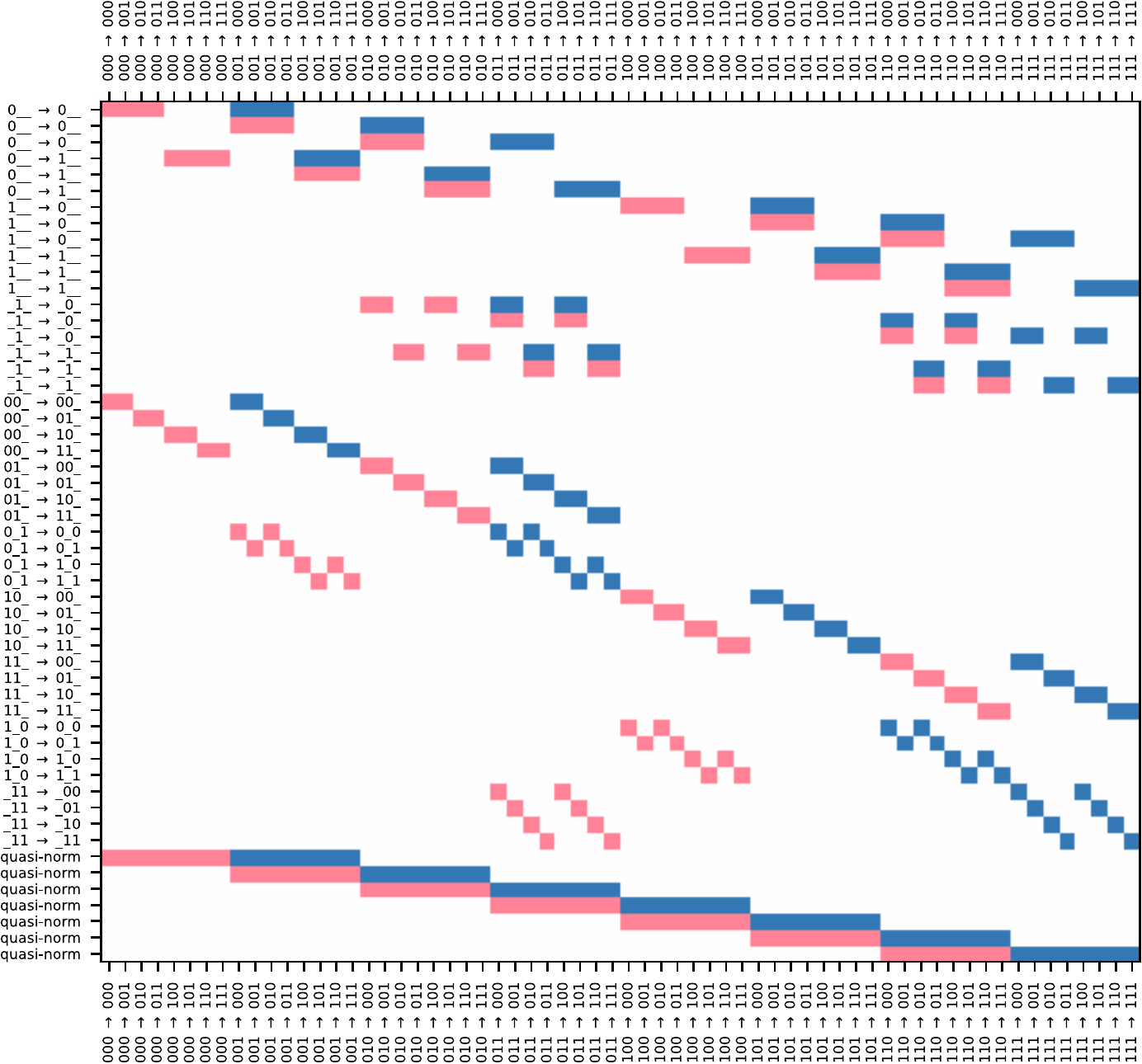}
\end{center}

\noindent Rows correspond to the 53 linear equations, of which 28 are independent.

\newpage
\subsection*{Space 74}

Space 74 is not induced by a causal order, but it is a refinement of the space in equivalence class 100 induced by the definite causal order $\total{\ev{A},\ev{C},\ev{B}}$ (note that the space induced by the order is not the same as space 100).
Its equivalence class under event-input permutation symmetry contains 48 spaces.
Space 74 differs as follows from the space induced by causal order $\total{\ev{A},\ev{C},\ev{B}}$:
\begin{itemize}
  \item The outputs at events \evset{\ev{A}, \ev{B}} are independent of the input at event \ev{C} when the inputs at events \evset{A, B} are given by \hist{A/0,B/0}, \hist{A/0,B/1} and \hist{A/1,B/0}.
  \item The output at event \ev{C} is independent of the input at event \ev{A} when the input at event C is given by \hist{C/0}.
\end{itemize}

\noindent Below are the histories and extended histories for space 74: 
\begin{center}
    \begin{tabular}{cc}
    \includegraphics[height=3.5cm]{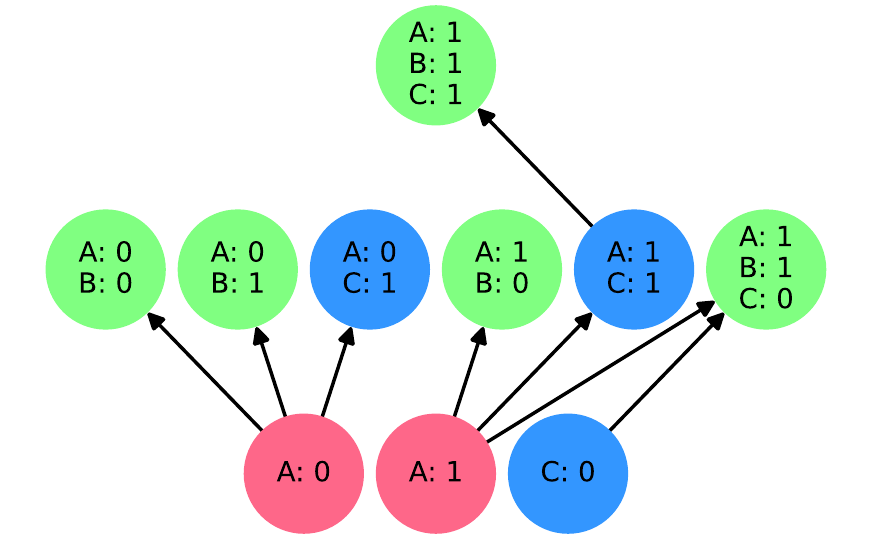}
    &
    \includegraphics[height=3.5cm]{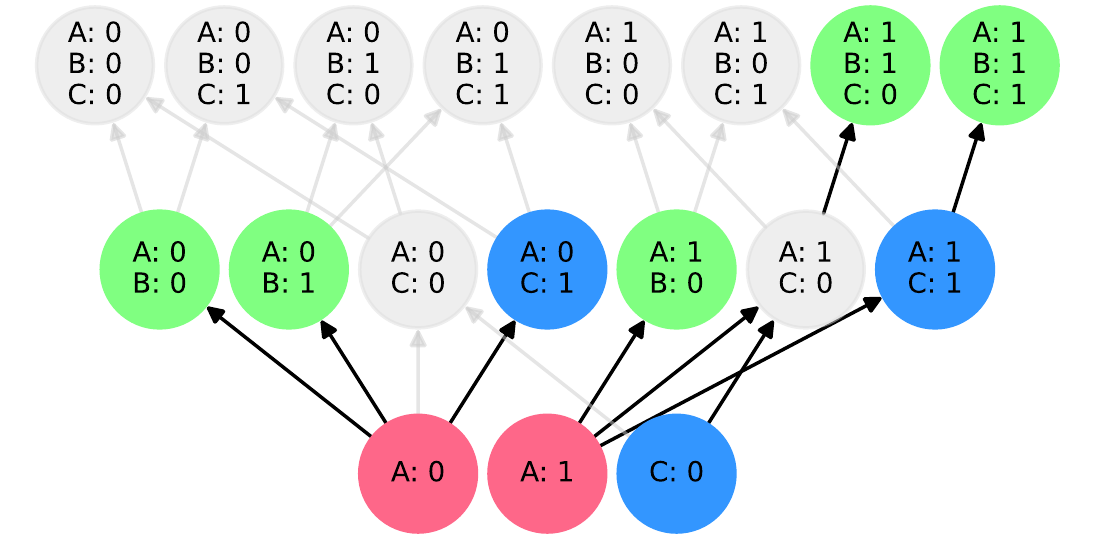}
    \\
    $\Theta_{74}$
    &
    $\Ext{\Theta_{74}}$
    \end{tabular}
\end{center}

\noindent The standard causaltope for Space 74 has dimension 35.
Below is a plot of the homogeneous linear system of causality and quasi-normalisation equations for the standard causaltope, put in reduced row echelon form:

\begin{center}
    \includegraphics[width=11cm]{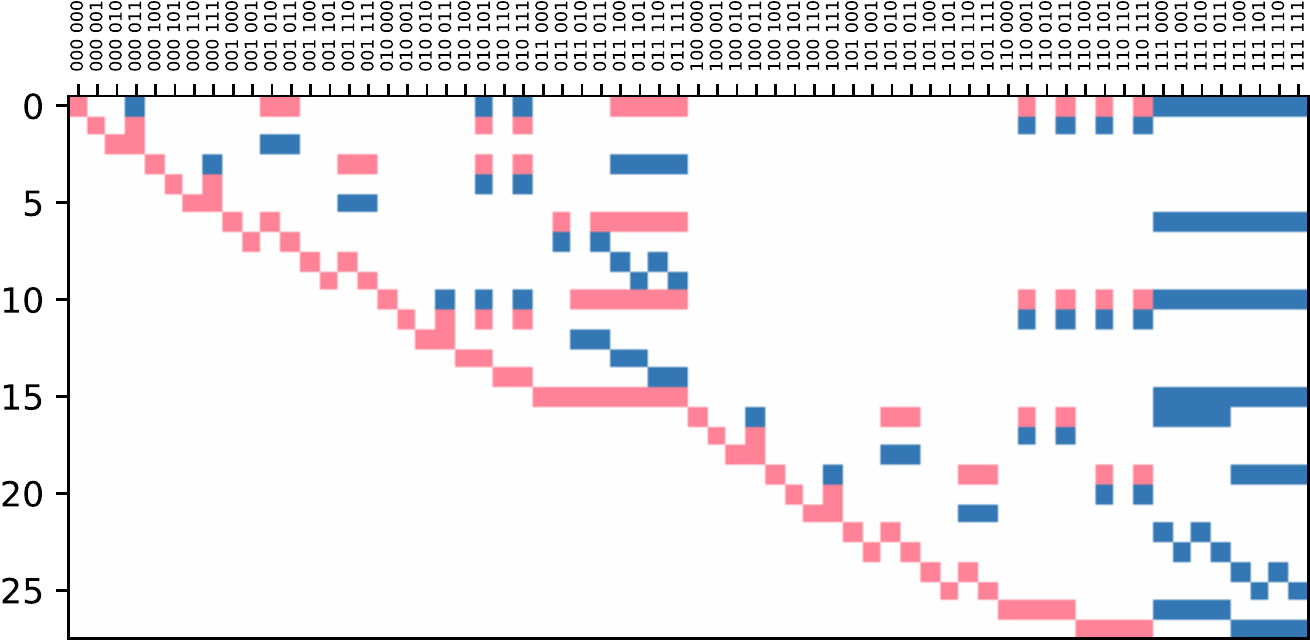}
\end{center}

\noindent Rows correspond to the 28 independent linear equations.
Columns in the plot correspond to entries of empirical models, indexed as $i_A i_B i_C$ $o_A o_B o_C$.
Coefficients in the equations are color-coded as white=0, red=+1 and blue=-1.

Space 74 has closest refinements in equivalence classes 46, 56, 57 and 58; 
it is the join of its (closest) refinements.
It has closest coarsenings in equivalence classes 79, 83, 86, 87 and 88; 
it is the meet of its (closest) coarsenings.
It has 1024 causal functions, 128 of which are not causal for any of its refinements.
It is a tight space.

The standard causaltope for Space 74 has 1 more dimension than that of its subspace in equivalence class 46.
The standard causaltope for Space 74 is the meet of the standard causaltopes for its closest coarsenings.
For completeness, below is a plot of the full homogeneous linear system of causality and quasi-normalisation equations for the standard causaltope:

\begin{center}
    \includegraphics[width=12cm]{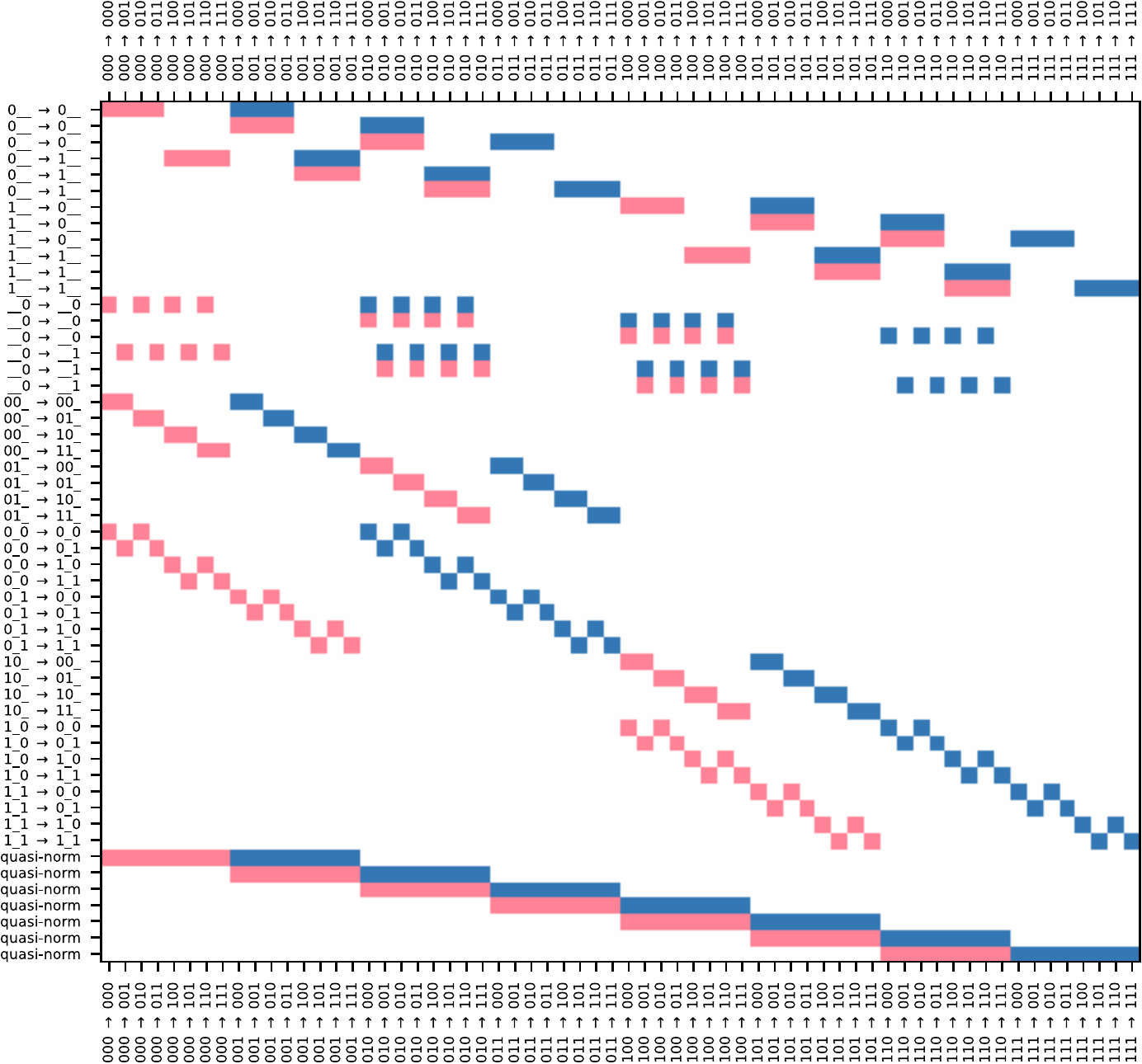}
\end{center}

\noindent Rows correspond to the 53 linear equations, of which 28 are independent.

\newpage
\subsection*{Space 75}

Space 75 is not induced by a causal order, but it is a refinement of the space 100 induced by the definite causal order $\total{\ev{A},\ev{B},\ev{C}}$.
Its equivalence class under event-input permutation symmetry contains 24 spaces.
Space 75 differs as follows from the space induced by causal order $\total{\ev{A},\ev{B},\ev{C}}$:
\begin{itemize}
  \item The outputs at events \evset{\ev{A}, \ev{C}} are independent of the input at event \ev{B} when the inputs at events \evset{A, C} are given by \hist{A/0,C/1} and \hist{A/1,C/1}.
  \item The outputs at events \evset{\ev{B}, \ev{C}} are independent of the input at event \ev{A} when the inputs at events \evset{B, C} are given by \hist{B/1,C/0}.
  \item The output at event \ev{B} is independent of the input at event \ev{A} when the input at event B is given by \hist{B/1}.
\end{itemize}

\noindent Below are the histories and extended histories for space 75: 
\begin{center}
    \begin{tabular}{cc}
    \includegraphics[height=3.5cm]{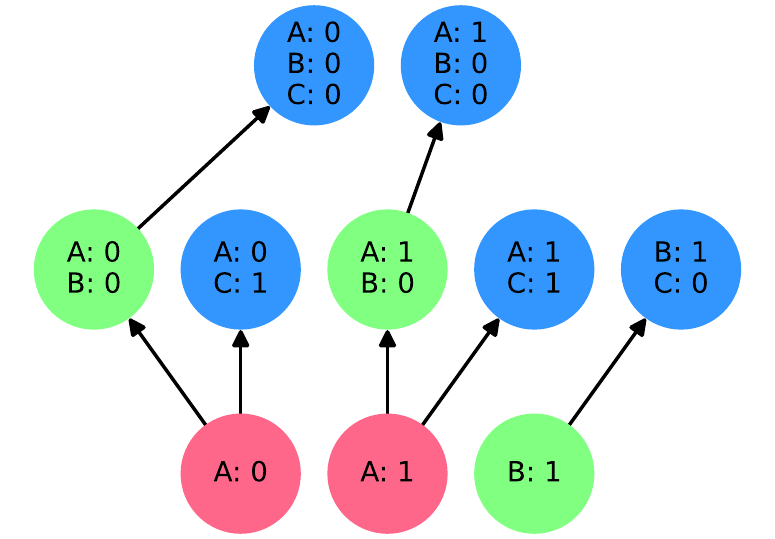}
    &
    \includegraphics[height=3.5cm]{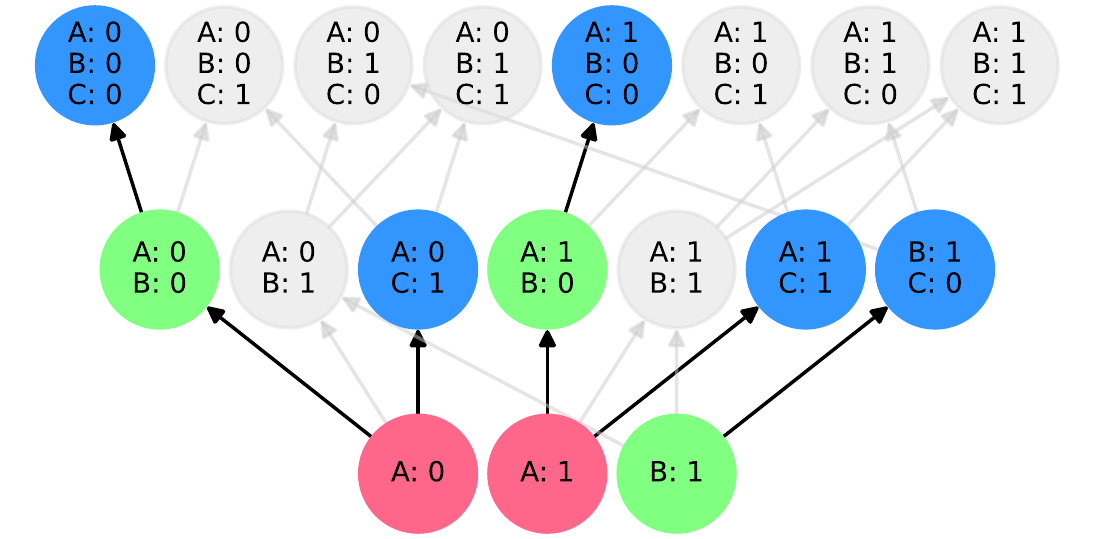}
    \\
    $\Theta_{75}$
    &
    $\Ext{\Theta_{75}}$
    \end{tabular}
\end{center}

\noindent The standard causaltope for Space 75 has dimension 35.
Below is a plot of the homogeneous linear system of causality and quasi-normalisation equations for the standard causaltope, put in reduced row echelon form:

\begin{center}
    \includegraphics[width=11cm]{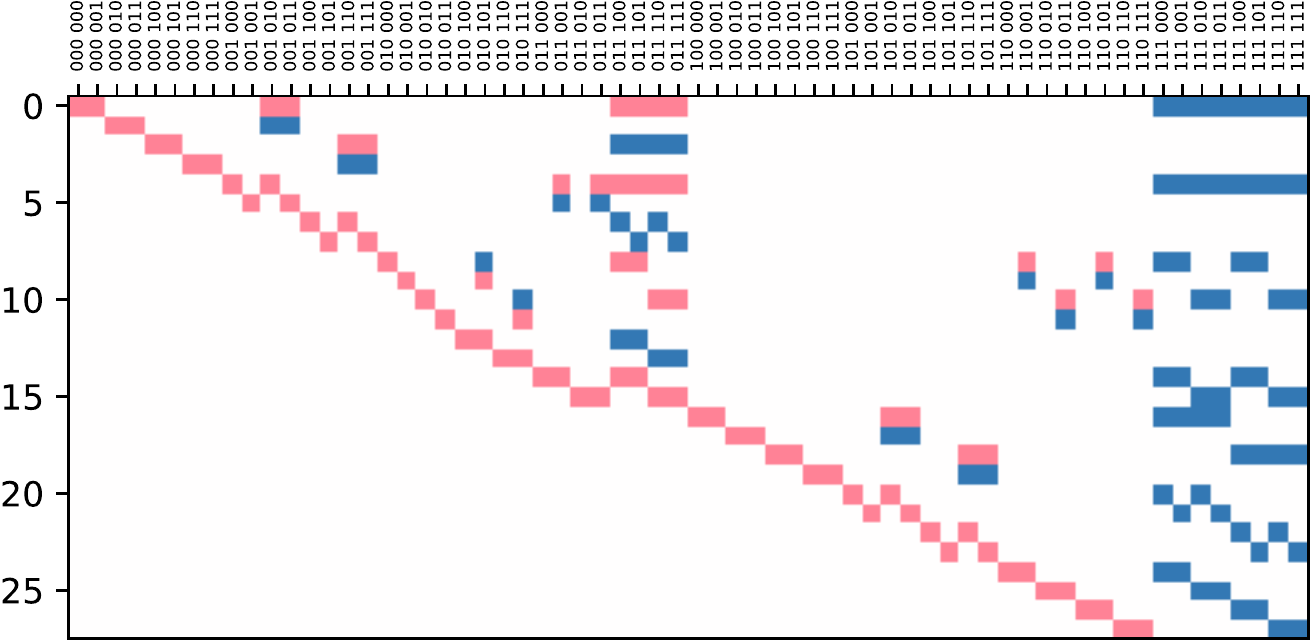}
\end{center}

\noindent Rows correspond to the 28 independent linear equations.
Columns in the plot correspond to entries of empirical models, indexed as $i_A i_B i_C$ $o_A o_B o_C$.
Coefficients in the equations are color-coded as white=0, red=+1 and blue=-1.

Space 75 has closest refinements in equivalence classes 49, 54 and 56; 
it is the join of its (closest) refinements.
It has closest coarsenings in equivalence classes 80 and 86; 
it is the meet of its (closest) coarsenings.
It has 1024 causal functions, 320 of which are not causal for any of its refinements.
It is a tight space.

The standard causaltope for Space 75 has 1 more dimension than that of its subspace in equivalence class 49.
The standard causaltope for Space 75 is the meet of the standard causaltopes for its closest coarsenings.
For completeness, below is a plot of the full homogeneous linear system of causality and quasi-normalisation equations for the standard causaltope:

\begin{center}
    \includegraphics[width=12cm]{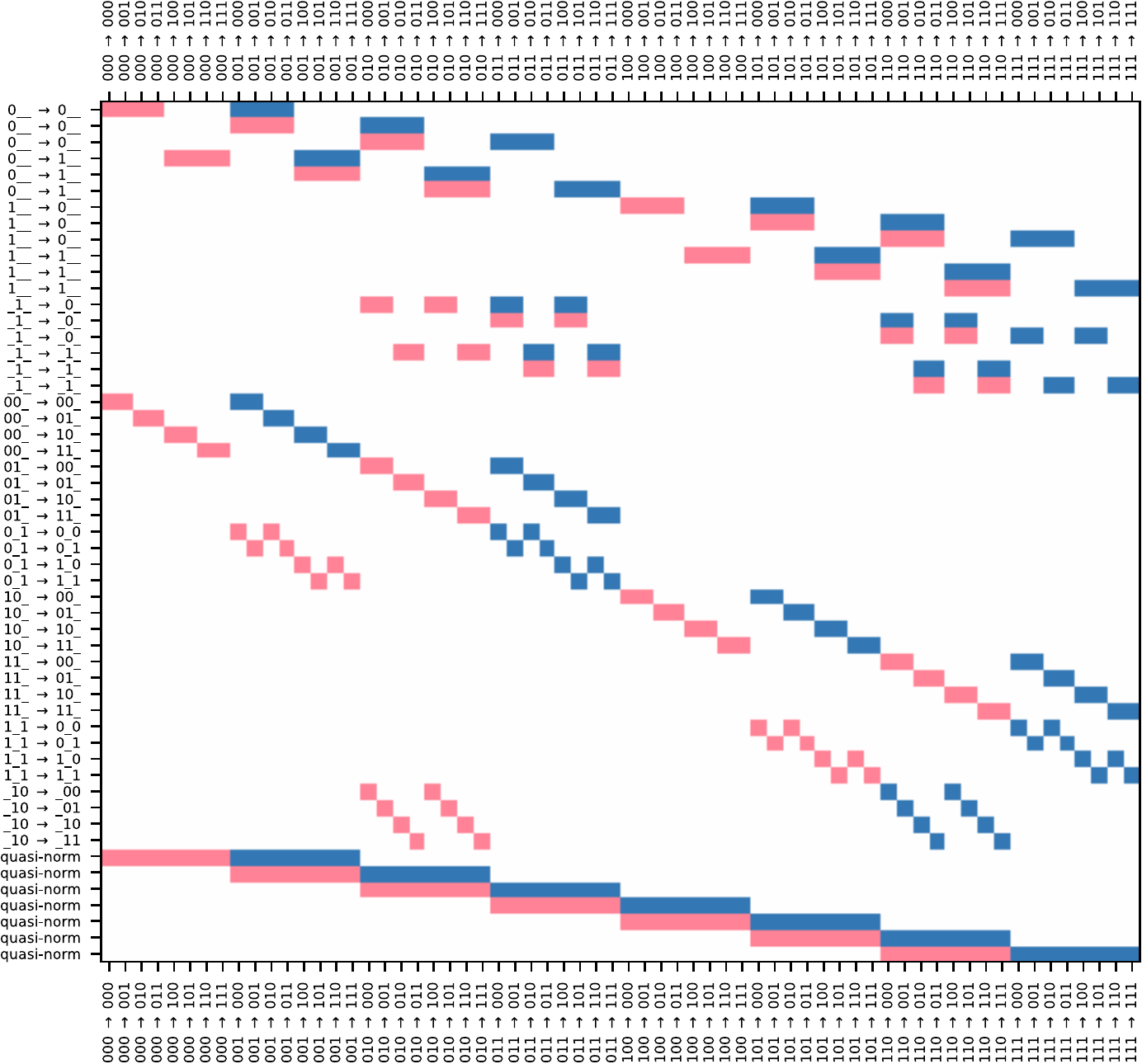}
\end{center}

\noindent Rows correspond to the 53 linear equations, of which 28 are independent.

\newpage
\subsection*{Space 76}

Space 76 is not induced by a causal order, but it is a refinement of the space 100 induced by the definite causal order $\total{\ev{A},\ev{B},\ev{C}}$.
Its equivalence class under event-input permutation symmetry contains 24 spaces.
Space 76 differs as follows from the space induced by causal order $\total{\ev{A},\ev{B},\ev{C}}$:
\begin{itemize}
  \item The outputs at events \evset{\ev{A}, \ev{C}} are independent of the input at event \ev{B} when the inputs at events \evset{A, C} are given by \hist{A/0,C/1} and \hist{A/1,C/1}.
  \item The outputs at events \evset{\ev{B}, \ev{C}} are independent of the input at event \ev{A} when the inputs at events \evset{B, C} are given by \hist{B/1,C/1}.
  \item The output at event \ev{B} is independent of the input at event \ev{A} when the input at event B is given by \hist{B/1}.
\end{itemize}

\noindent Below are the histories and extended histories for space 76: 
\begin{center}
    \begin{tabular}{cc}
    \includegraphics[height=3.5cm]{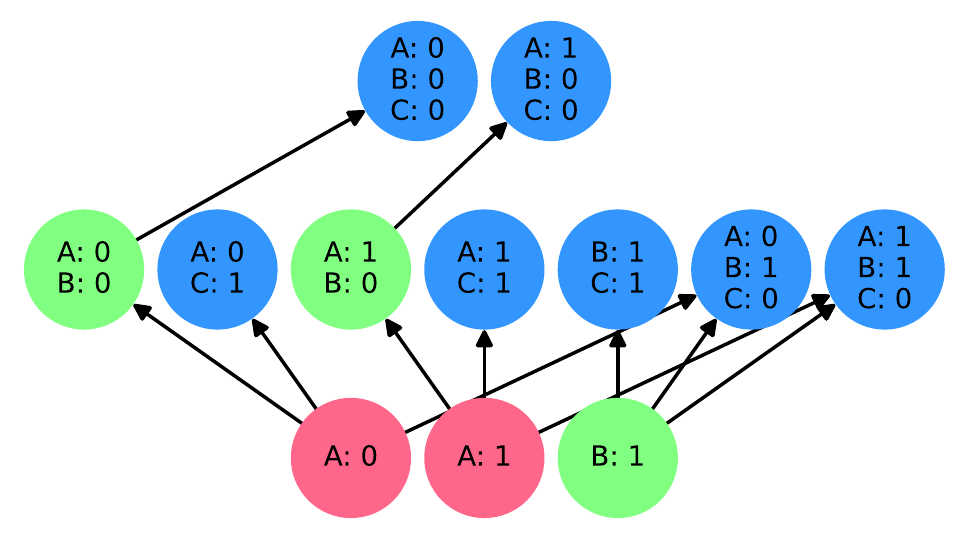}
    &
    \includegraphics[height=3.5cm]{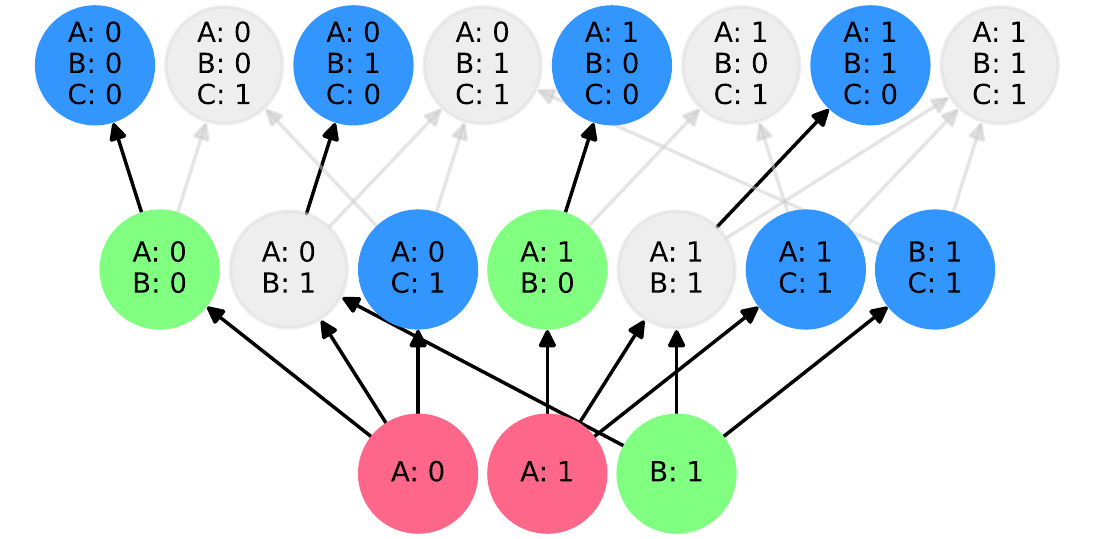}
    \\
    $\Theta_{76}$
    &
    $\Ext{\Theta_{76}}$
    \end{tabular}
\end{center}

\noindent The standard causaltope for Space 76 has dimension 35.
Below is a plot of the homogeneous linear system of causality and quasi-normalisation equations for the standard causaltope, put in reduced row echelon form:

\begin{center}
    \includegraphics[width=11cm]{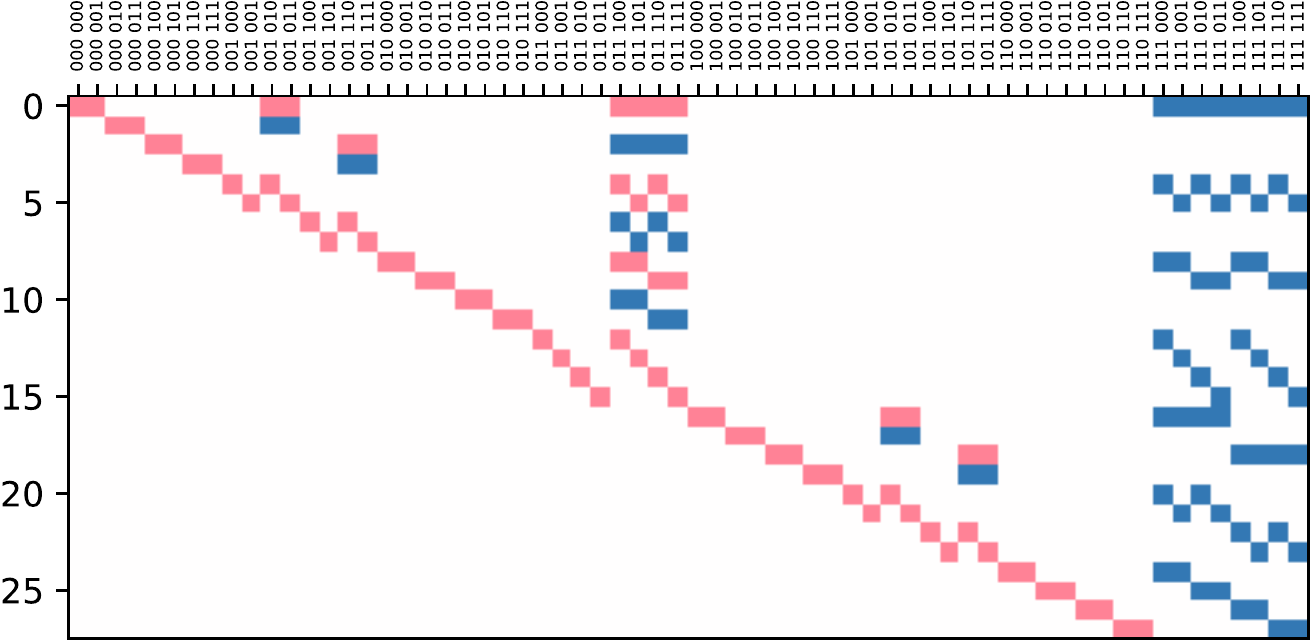}
\end{center}

\noindent Rows correspond to the 28 independent linear equations.
Columns in the plot correspond to entries of empirical models, indexed as $i_A i_B i_C$ $o_A o_B o_C$.
Coefficients in the equations are color-coded as white=0, red=+1 and blue=-1.

Space 76 has closest refinements in equivalence classes 45, 54, 57 and 59; 
it is the join of its (closest) refinements.
It has closest coarsenings in equivalence classes 82 and 86; 
it is the meet of its (closest) coarsenings.
It has 1024 causal functions, all of which are causal for at least one of its refinements.
It is not a tight space: for event \ev{C}, a causal function must yield identical output values on input histories \hist{A/0,C/1}, \hist{A/1,C/1} and \hist{B/1,C/1}.

The standard causaltope for Space 76 coincides with that of its subspace in equivalence class 45.
The standard causaltope for Space 76 is the meet of the standard causaltopes for its closest coarsenings.
For completeness, below is a plot of the full homogeneous linear system of causality and quasi-normalisation equations for the standard causaltope:

\begin{center}
    \includegraphics[width=12cm]{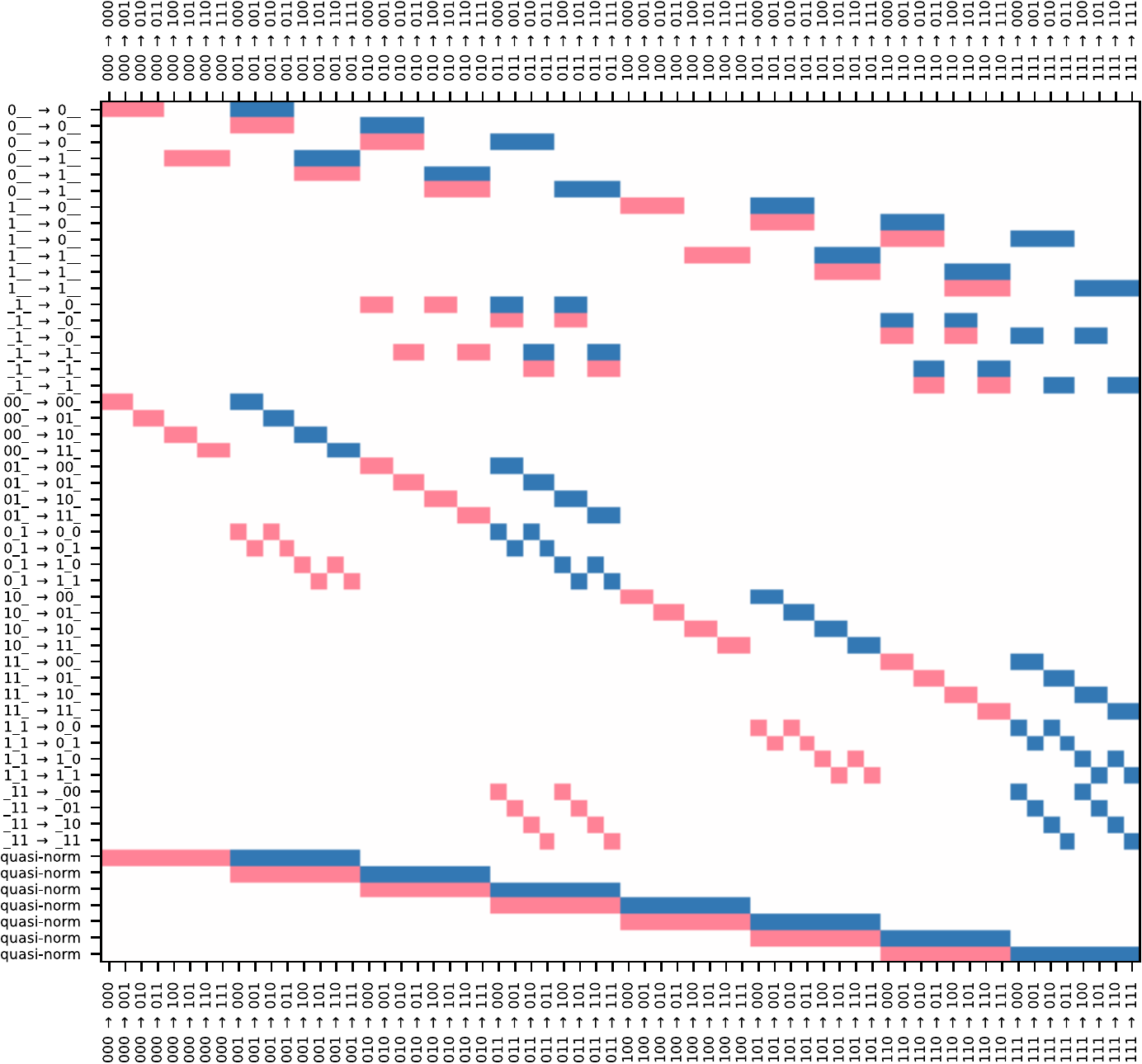}
\end{center}

\noindent Rows correspond to the 53 linear equations, of which 28 are independent.

\newpage
\subsection*{Space 77}

Space 77 is induced by the definite causal order $\total{\ev{A},\ev{B}}\vee\total{\ev{A},\ev{C}}$.
Its equivalence class under event-input permutation symmetry contains 3 spaces.

\noindent Below are the histories and extended histories for space 77: 
\begin{center}
    \begin{tabular}{cc}
    \includegraphics[height=3.5cm]{svg-inkscape/space-ABC-unique-tight-77-highlighted_svg-tex.pdf}
    &
    \includegraphics[height=3.5cm]{svg-inkscape/space-ABC-unique-tight-77-ext-highlighted_svg-tex.pdf}
    \\
    $\Theta_{77}$
    &
    $\Ext{\Theta_{77}}$
    \end{tabular}
\end{center}

\noindent The standard causaltope for Space 77 has dimension 34.
Below is a plot of the homogeneous linear system of causality and quasi-normalisation equations for the standard causaltope, put in reduced row echelon form:

\begin{center}
    \includegraphics[width=11cm]{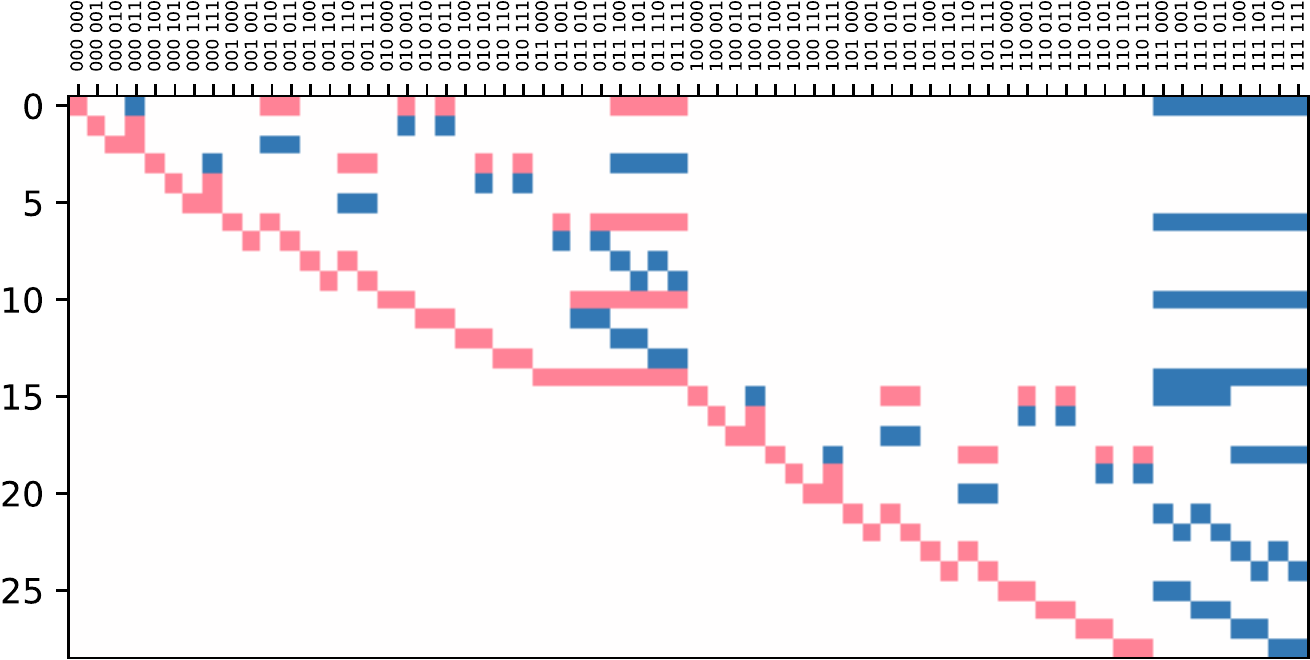}
\end{center}

\noindent Rows correspond to the 29 independent linear equations.
Columns in the plot correspond to entries of empirical models, indexed as $i_A i_B i_C$ $o_A o_B o_C$.
Coefficients in the equations are color-coded as white=0, red=+1 and blue=-1.

Space 77 has closest refinements in equivalence class 58; 
it is the join of its (closest) refinements.
It has closest coarsenings in equivalence class 88; 
it is the meet of its (closest) coarsenings.
It has 1024 causal functions, 320 of which are not causal for any of its refinements.
It is a tight space.

The standard causaltope for Space 77 has 1 more dimension than those of its 4 subspaces in equivalence class 58.
The standard causaltope for Space 77 is the meet of the standard causaltopes for its closest coarsenings.
For completeness, below is a plot of the full homogeneous linear system of causality and quasi-normalisation equations for the standard causaltope:

\begin{center}
    \includegraphics[width=12cm]{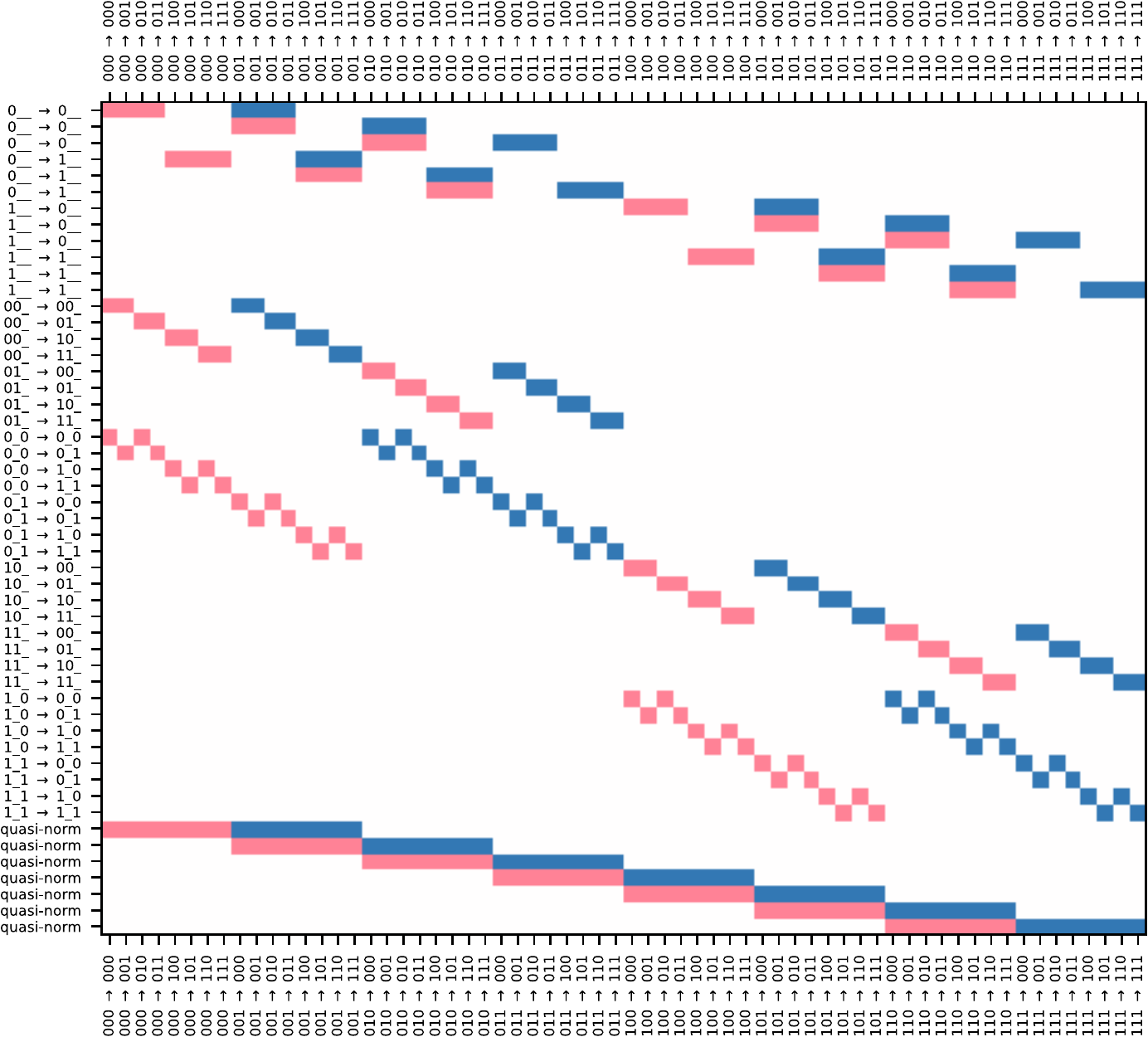}
\end{center}

\noindent Rows correspond to the 51 linear equations, of which 29 are independent.

\newpage
\subsection*{Space 78}

Space 78 is not induced by a causal order, but it is a refinement of the space 100 induced by the definite causal order $\total{\ev{A},\ev{B},\ev{C}}$.
Its equivalence class under event-input permutation symmetry contains 12 spaces.
Space 78 differs as follows from the space induced by causal order $\total{\ev{A},\ev{B},\ev{C}}$:
\begin{itemize}
  \item The outputs at events \evset{\ev{A}, \ev{C}} are independent of the input at event \ev{B} when the inputs at events \evset{A, C} are given by \hist{A/0,C/1} and \hist{A/1,C/1}.
  \item The output at event \ev{C} is independent of the inputs at events \evset{\ev{A}, \ev{B}} when the input at event C is given by \hist{C/1}.
\end{itemize}

\noindent Below are the histories and extended histories for space 78: 
\begin{center}
    \begin{tabular}{cc}
    \includegraphics[height=3.5cm]{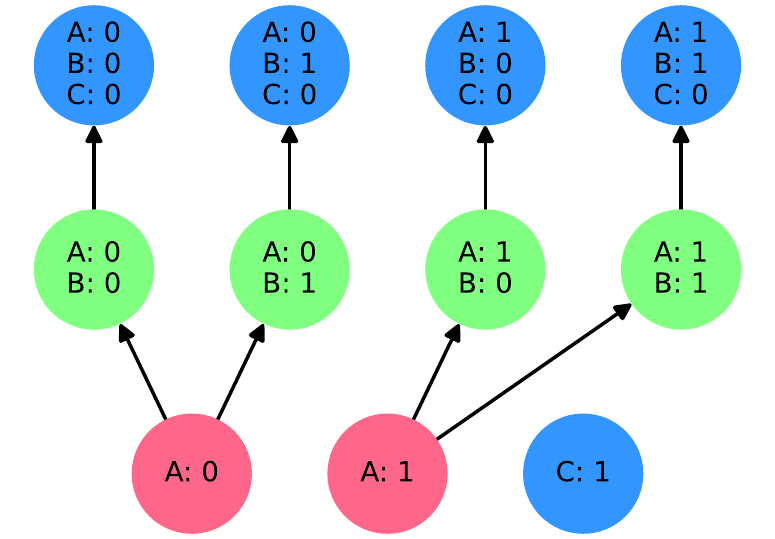}
    &
    \includegraphics[height=3.5cm]{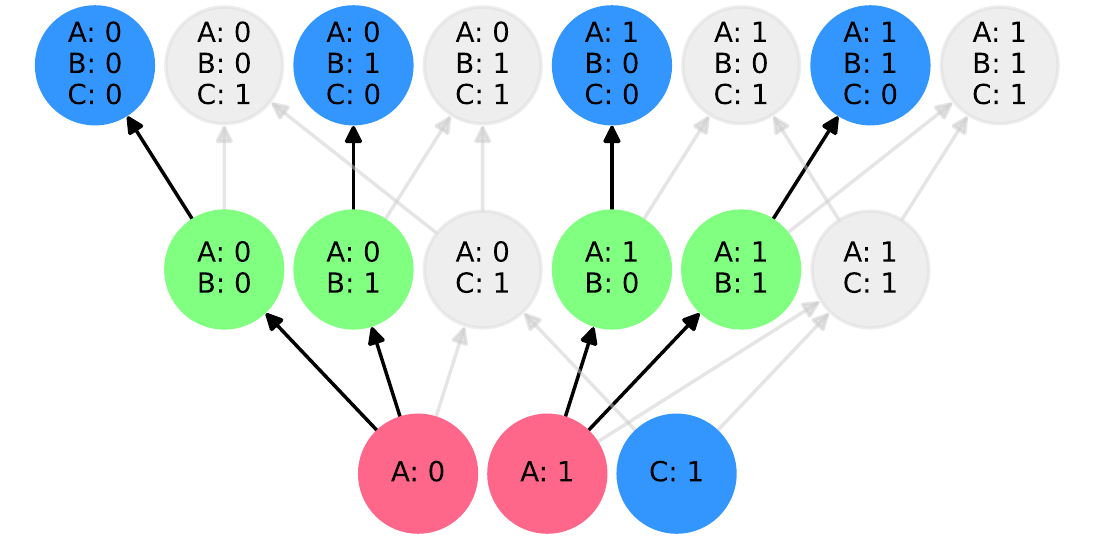}
    \\
    $\Theta_{78}$
    &
    $\Ext{\Theta_{78}}$
    \end{tabular}
\end{center}

\noindent The standard causaltope for Space 78 has dimension 37.
Below is a plot of the homogeneous linear system of causality and quasi-normalisation equations for the standard causaltope, put in reduced row echelon form:

\begin{center}
    \includegraphics[width=11cm]{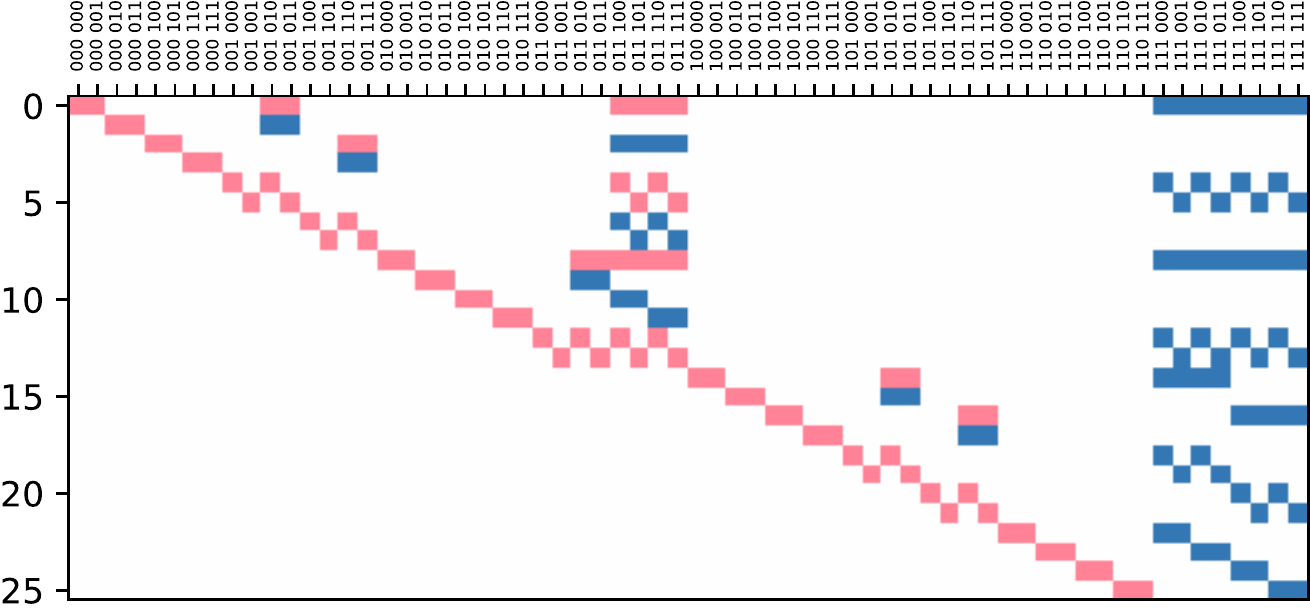}
\end{center}

\noindent Rows correspond to the 26 independent linear equations.
Columns in the plot correspond to entries of empirical models, indexed as $i_A i_B i_C$ $o_A o_B o_C$.
Coefficients in the equations are color-coded as white=0, red=+1 and blue=-1.

Space 78 has closest refinements in equivalence classes 61 and 64; 
it is the join of its (closest) refinements.
It has closest coarsenings in equivalence class 94; 
it does not arise as a nontrivial meet in the hierarchy.
It has 2048 causal functions, 640 of which are not causal for any of its refinements.
It is a tight space.

The standard causaltope for Space 78 has 2 more dimensions than those of its 4 subspaces in equivalence classes 61 and 64.
The standard causaltope for Space 78 has 1 dimension fewer than the meet of the standard causaltopes for its closest coarsenings.
For completeness, below is a plot of the full homogeneous linear system of causality and quasi-normalisation equations for the standard causaltope:

\begin{center}
    \includegraphics[width=12cm]{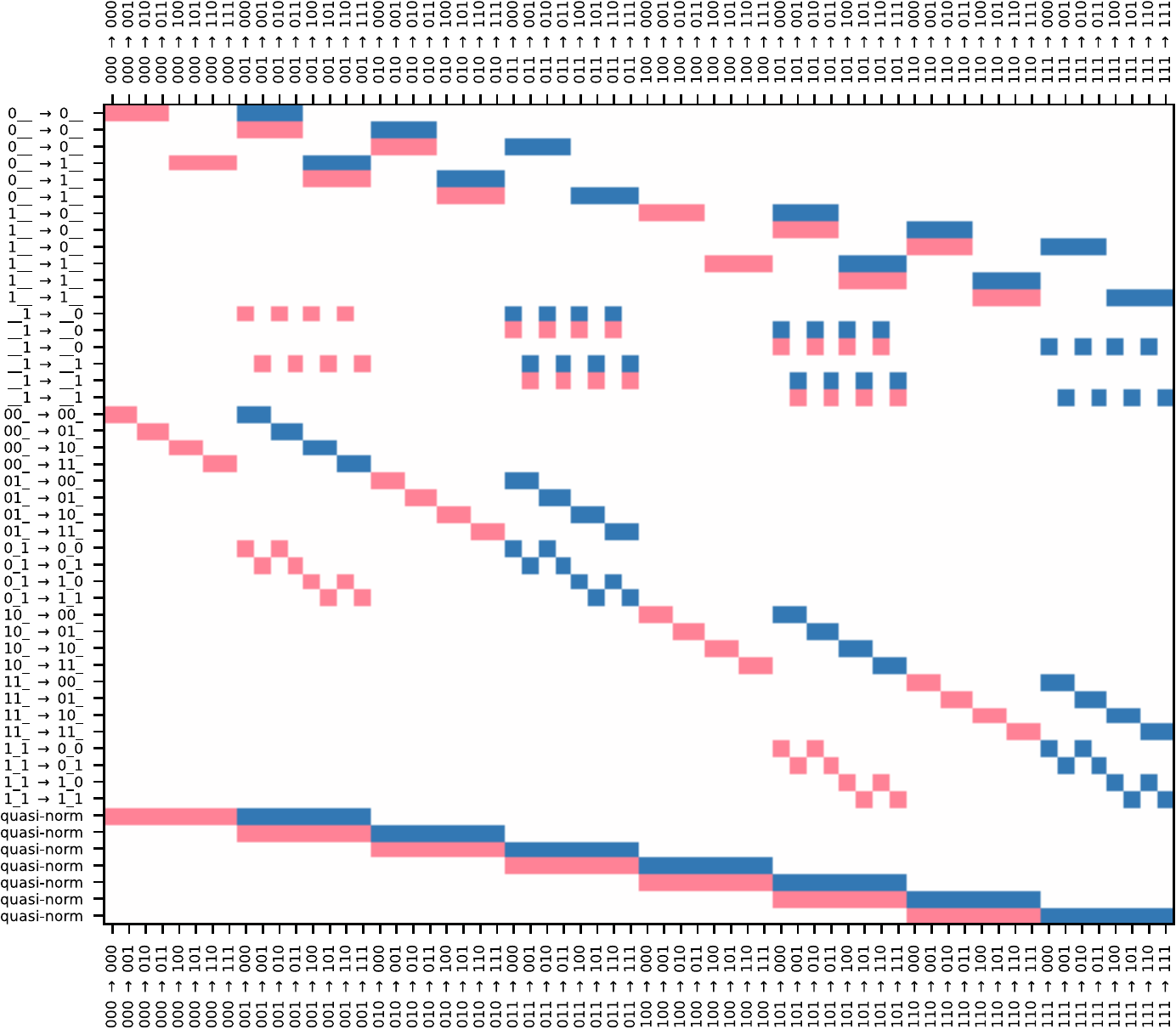}
\end{center}

\noindent Rows correspond to the 49 linear equations, of which 26 are independent.

\newpage
\subsection*{Space 79}

Space 79 is not induced by a causal order, but it is a refinement of the space induced by the indefinite causal order $\total{\ev{A},\{\ev{B},\ev{C}\}}$.
Its equivalence class under event-input permutation symmetry contains 48 spaces.
Space 79 differs as follows from the space induced by causal order $\total{\ev{A},\{\ev{B},\ev{C}\}}$:
\begin{itemize}
  \item The outputs at events \evset{\ev{A}, \ev{B}} are independent of the input at event \ev{C} when the inputs at events \evset{A, B} are given by \hist{A/0,B/0}, \hist{A/0,B/1} and \hist{A/1,B/0}.
  \item The outputs at events \evset{\ev{A}, \ev{C}} are independent of the input at event \ev{B} when the inputs at events \evset{A, C} are given by \hist{A/0,C/1}, \hist{A/1,C/0} and \hist{A/1,C/1}.
  \item The output at event \ev{C} is independent of the inputs at events \evset{\ev{A}, \ev{B}} when the input at event C is given by \hist{C/1}.
\end{itemize}

\noindent Below are the histories and extended histories for space 79: 
\begin{center}
    \begin{tabular}{cc}
    \includegraphics[height=3.5cm]{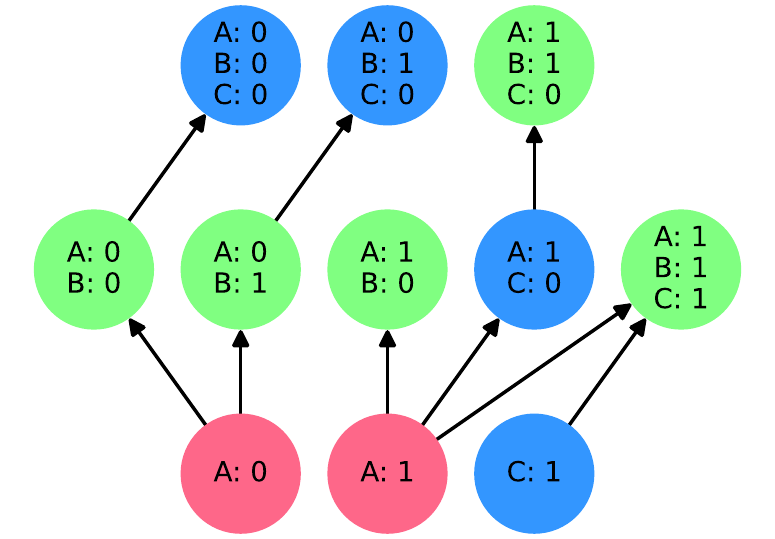}
    &
    \includegraphics[height=3.5cm]{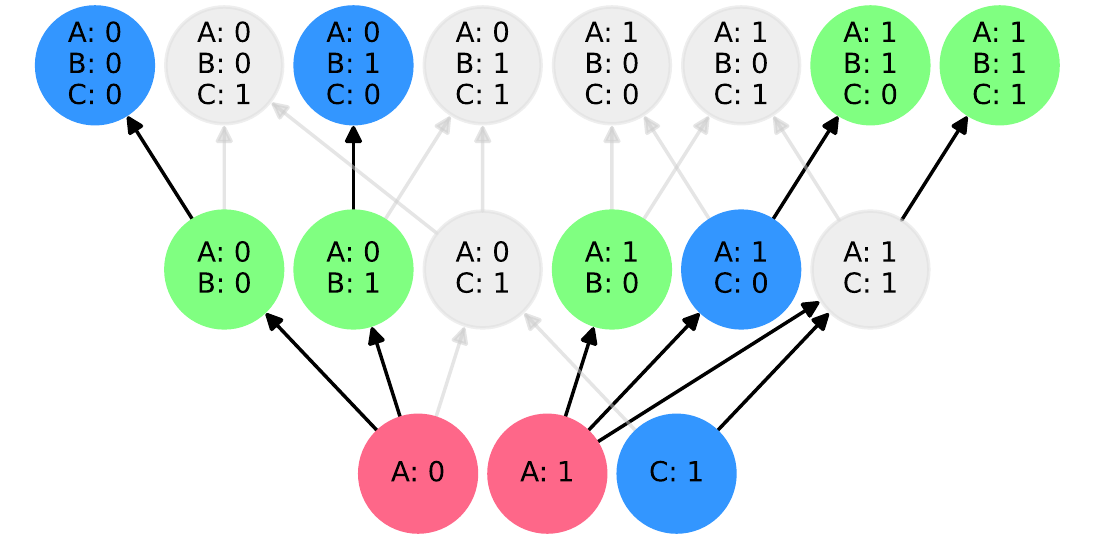}
    \\
    $\Theta_{79}$
    &
    $\Ext{\Theta_{79}}$
    \end{tabular}
\end{center}

\noindent The standard causaltope for Space 79 has dimension 37.
Below is a plot of the homogeneous linear system of causality and quasi-normalisation equations for the standard causaltope, put in reduced row echelon form:

\begin{center}
    \includegraphics[width=11cm]{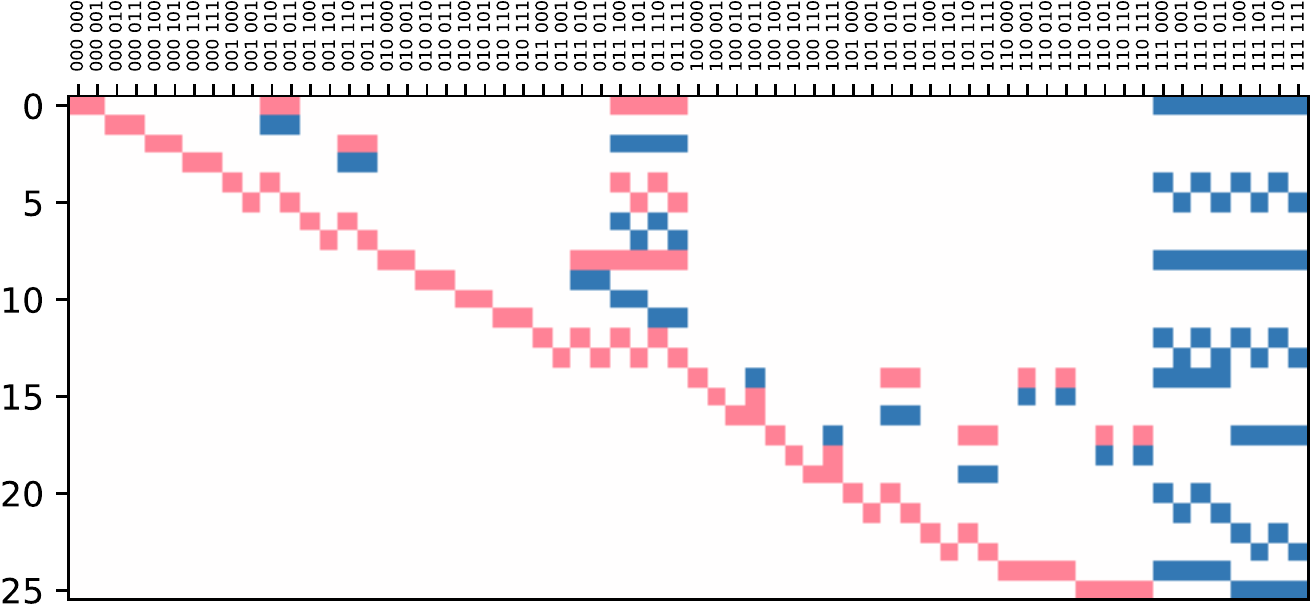}
\end{center}

\noindent Rows correspond to the 26 independent linear equations.
Columns in the plot correspond to entries of empirical models, indexed as $i_A i_B i_C$ $o_A o_B o_C$.
Coefficients in the equations are color-coded as white=0, red=+1 and blue=-1.

Space 79 has closest refinements in equivalence classes 62, 64, 71 and 74; 
it is the join of its (closest) refinements.
It has closest coarsenings in equivalence classes 93 and 96; 
it is the meet of its (closest) coarsenings.
It has 2048 causal functions, 192 of which are not causal for any of its refinements.
It is a tight space.

The standard causaltope for Space 79 has 2 more dimensions than those of its 4 subspaces in equivalence classes 62, 64, 71 and 74.
The standard causaltope for Space 79 is the meet of the standard causaltopes for its closest coarsenings.
For completeness, below is a plot of the full homogeneous linear system of causality and quasi-normalisation equations for the standard causaltope:

\begin{center}
    \includegraphics[width=12cm]{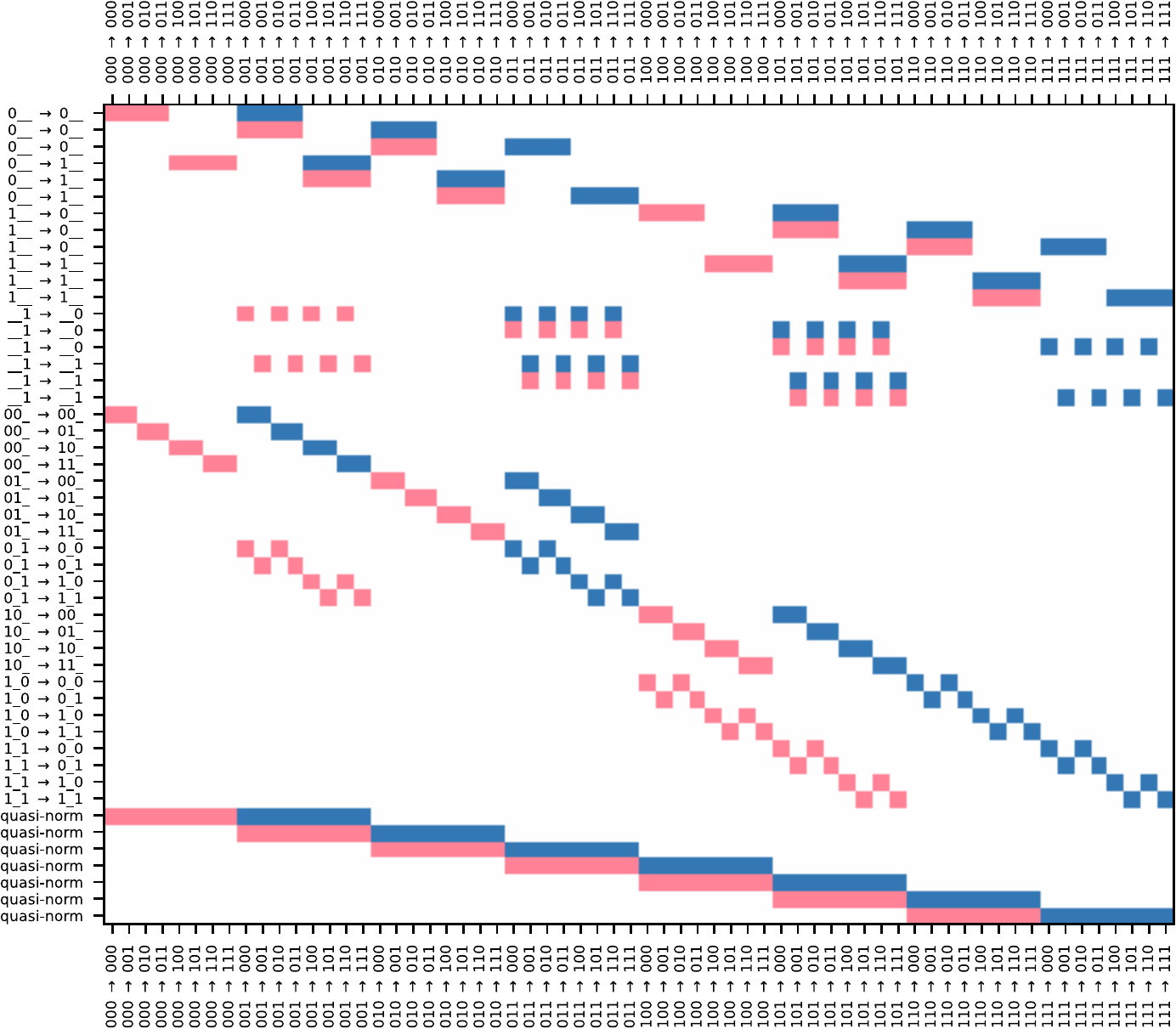}
\end{center}

\noindent Rows correspond to the 49 linear equations, of which 26 are independent.

\newpage
\subsection*{Space 80}

Space 80 is not induced by a causal order, but it is a refinement of the space 100 induced by the definite causal order $\total{\ev{A},\ev{B},\ev{C}}$.
Its equivalence class under event-input permutation symmetry contains 48 spaces.
Space 80 differs as follows from the space induced by causal order $\total{\ev{A},\ev{B},\ev{C}}$:
\begin{itemize}
  \item The outputs at events \evset{\ev{B}, \ev{C}} are independent of the input at event \ev{A} when the inputs at events \evset{B, C} are given by \hist{B/1,C/1}.
  \item The outputs at events \evset{\ev{A}, \ev{C}} are independent of the input at event \ev{B} when the inputs at events \evset{A, C} are given by \hist{A/1,C/0}.
  \item The output at event \ev{B} is independent of the input at event \ev{A} when the input at event B is given by \hist{B/1}.
\end{itemize}

\noindent Below are the histories and extended histories for space 80: 
\begin{center}
    \begin{tabular}{cc}
    \includegraphics[height=3.5cm]{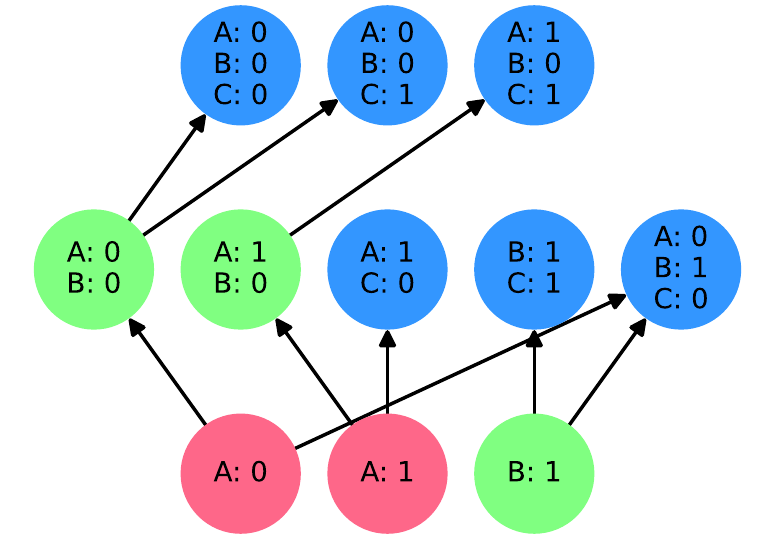}
    &
    \includegraphics[height=3.5cm]{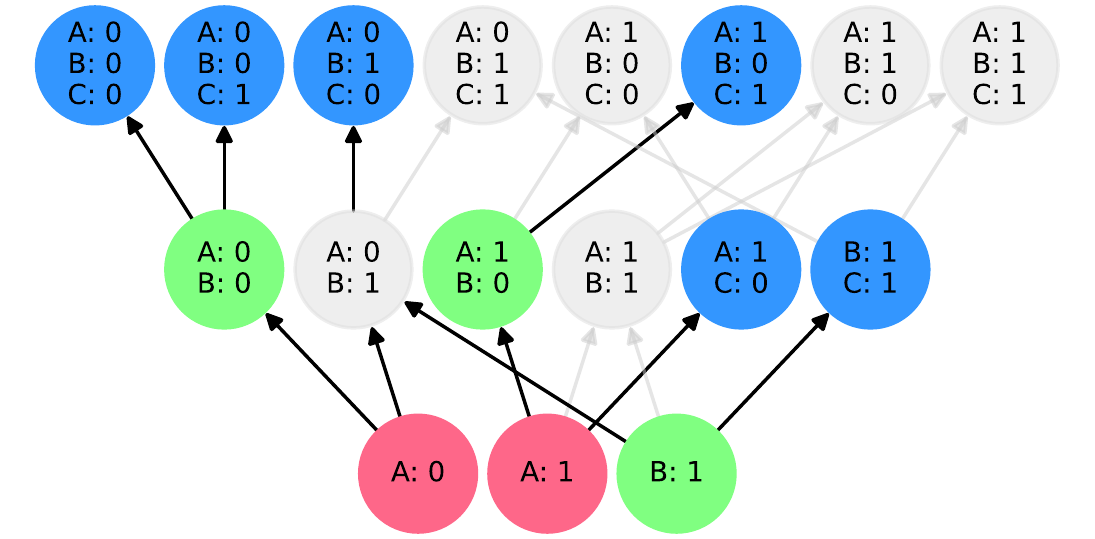}
    \\
    $\Theta_{80}$
    &
    $\Ext{\Theta_{80}}$
    \end{tabular}
\end{center}

\noindent The standard causaltope for Space 80 has dimension 37.
Below is a plot of the homogeneous linear system of causality and quasi-normalisation equations for the standard causaltope, put in reduced row echelon form:

\begin{center}
    \includegraphics[width=11cm]{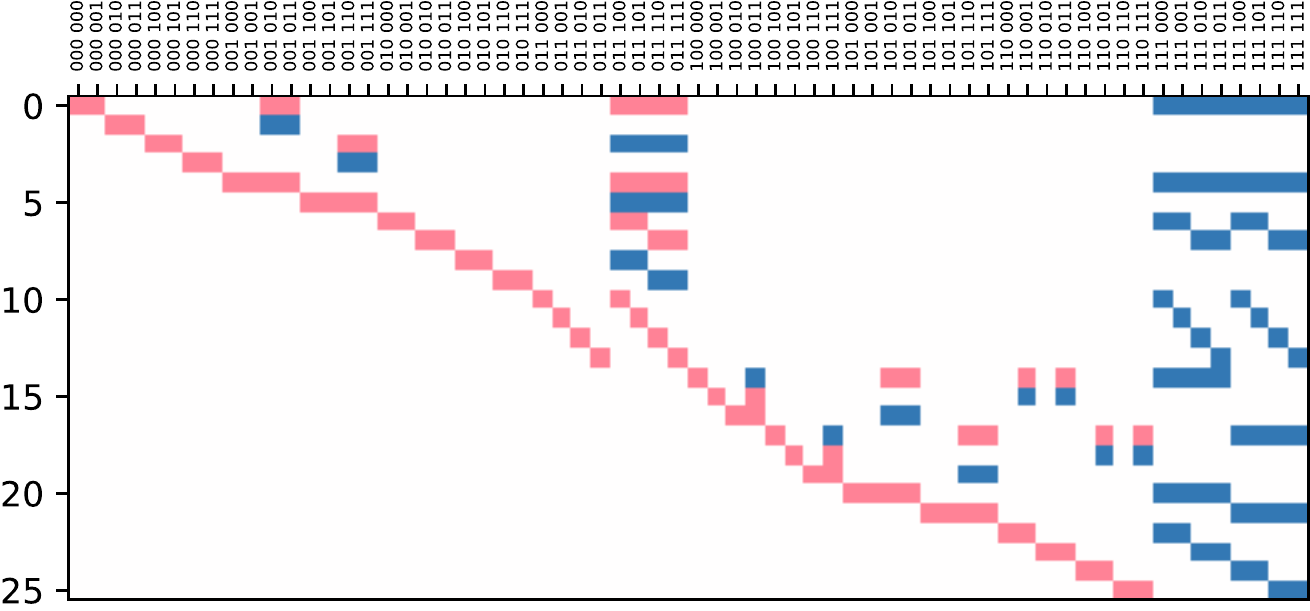}
\end{center}

\noindent Rows correspond to the 26 independent linear equations.
Columns in the plot correspond to entries of empirical models, indexed as $i_A i_B i_C$ $o_A o_B o_C$.
Coefficients in the equations are color-coded as white=0, red=+1 and blue=-1.

Space 80 has closest refinements in equivalence classes 63, 65, 66, 73 and 75; 
it is the join of its (closest) refinements.
It has closest coarsenings in equivalence classes 89 and 90; 
it is the meet of its (closest) coarsenings.
It has 2048 causal functions, 192 of which are not causal for any of its refinements.
It is a tight space.

The standard causaltope for Space 80 has 1 more dimension than that of its subspace in equivalence class 66.
The standard causaltope for Space 80 is the meet of the standard causaltopes for its closest coarsenings.
For completeness, below is a plot of the full homogeneous linear system of causality and quasi-normalisation equations for the standard causaltope:

\begin{center}
    \includegraphics[width=12cm]{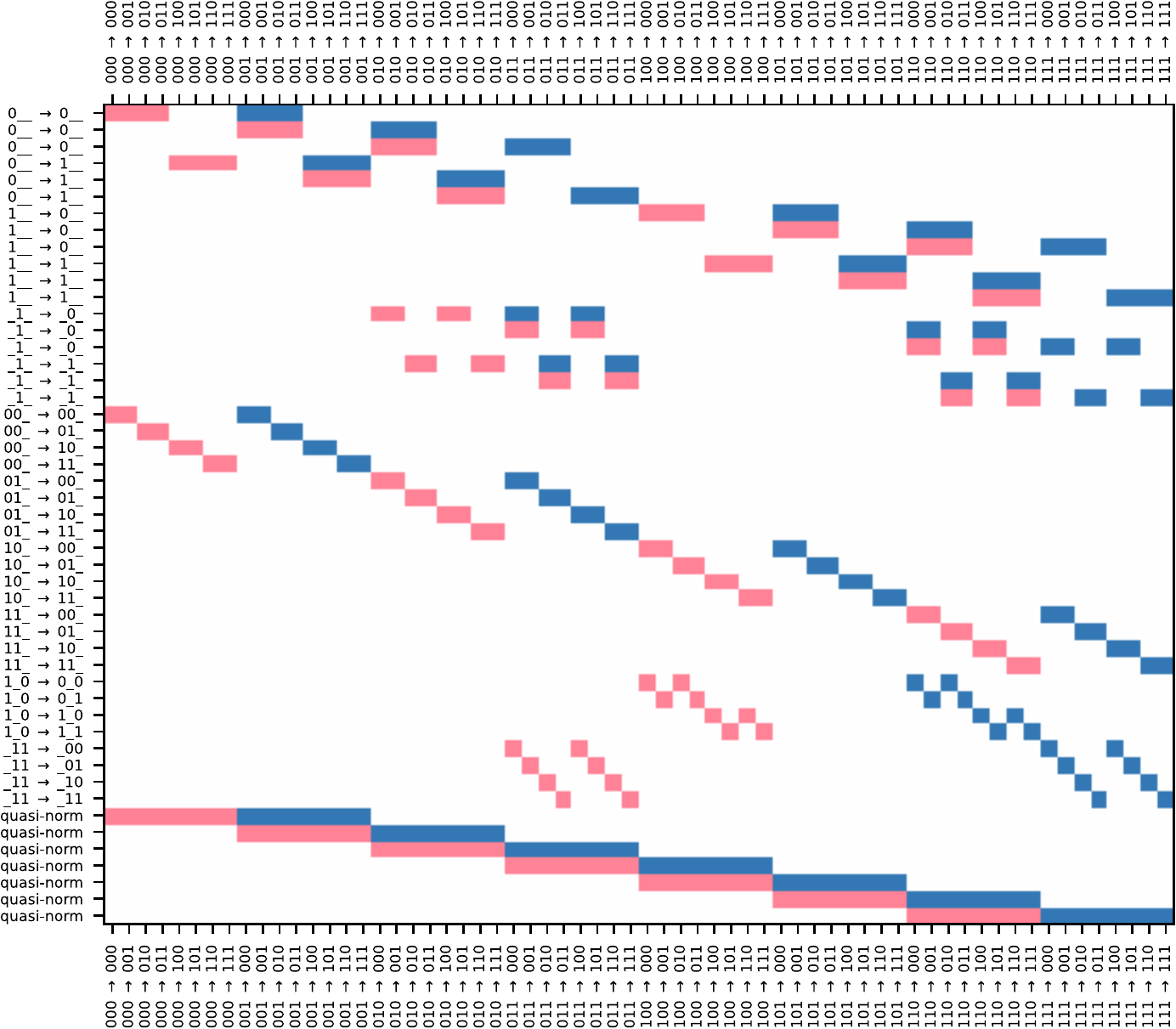}
\end{center}

\noindent Rows correspond to the 49 linear equations, of which 26 are independent.

\newpage
\subsection*{Space 81}

Space 81 is not induced by a causal order, but it is a refinement of the space 100 induced by the definite causal order $\total{\ev{A},\ev{B},\ev{C}}$.
Its equivalence class under event-input permutation symmetry contains 12 spaces.
Space 81 differs as follows from the space induced by causal order $\total{\ev{A},\ev{B},\ev{C}}$:
\begin{itemize}
  \item The outputs at events \evset{\ev{B}, \ev{C}} are independent of the input at event \ev{A} when the inputs at events \evset{B, C} are given by \hist{B/1,C/0} and \hist{B/1,C/1}.
  \item The output at event \ev{B} is independent of the input at event \ev{A} when the input at event B is given by \hist{B/1}.
\end{itemize}

\noindent Below are the histories and extended histories for space 81: 
\begin{center}
    \begin{tabular}{cc}
    \includegraphics[height=3.5cm]{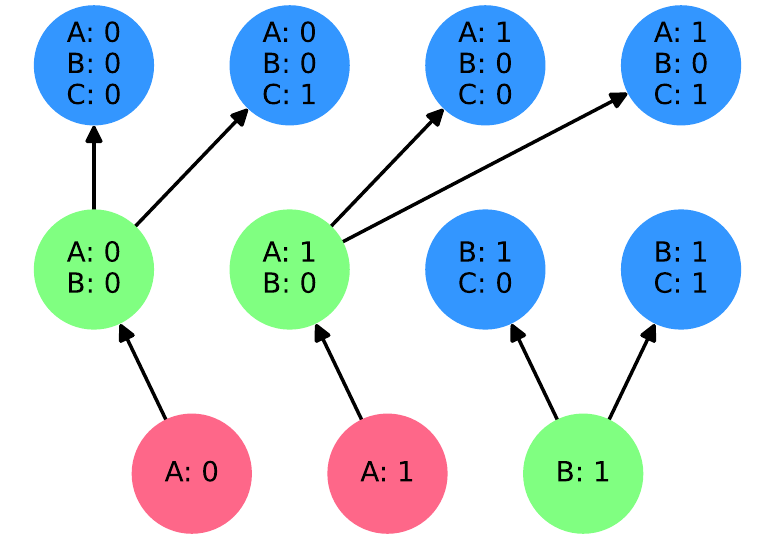}
    &
    \includegraphics[height=3.5cm]{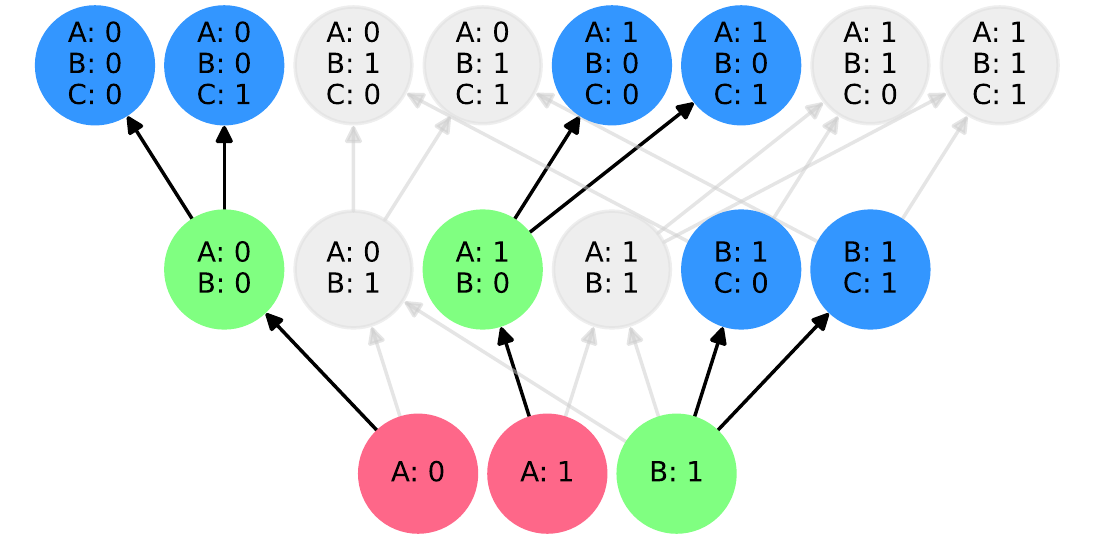}
    \\
    $\Theta_{81}$
    &
    $\Ext{\Theta_{81}}$
    \end{tabular}
\end{center}

\noindent The standard causaltope for Space 81 has dimension 37.
Below is a plot of the homogeneous linear system of causality and quasi-normalisation equations for the standard causaltope, put in reduced row echelon form:

\begin{center}
    \includegraphics[width=11cm]{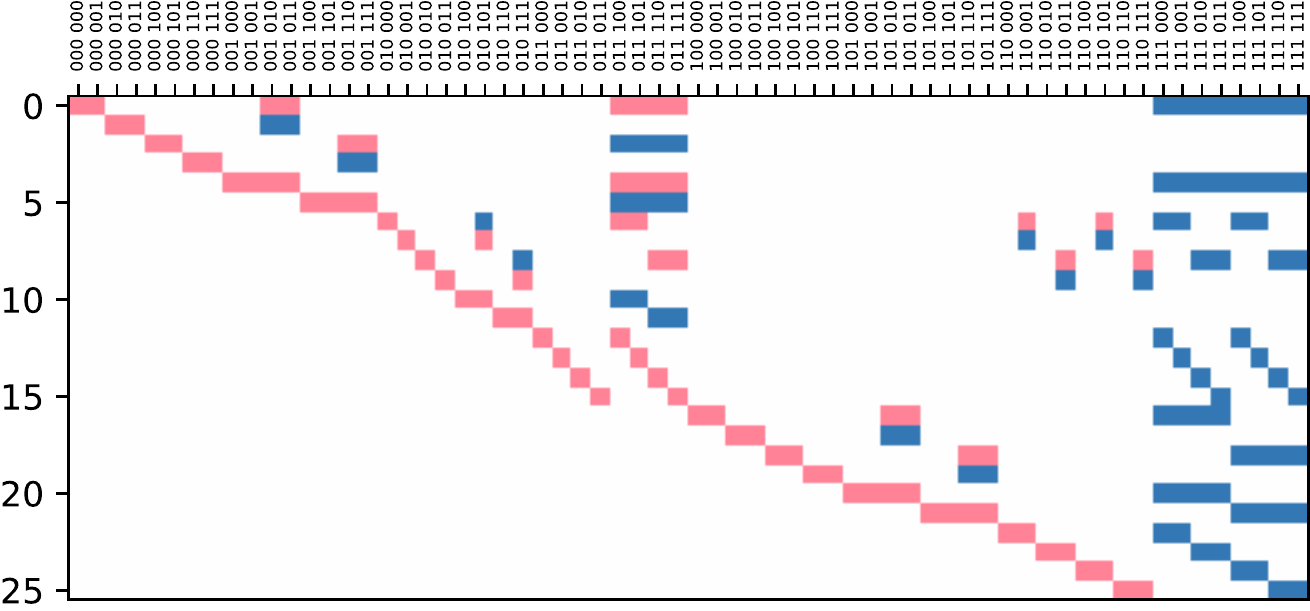}
\end{center}

\noindent Rows correspond to the 26 independent linear equations.
Columns in the plot correspond to entries of empirical models, indexed as $i_A i_B i_C$ $o_A o_B o_C$.
Coefficients in the equations are color-coded as white=0, red=+1 and blue=-1.

Space 81 has closest refinements in equivalence classes 65 and 67; 
it is the join of its (closest) refinements.
It has closest coarsenings in equivalence class 89; 
it is the meet of its (closest) coarsenings.
It has 2048 causal functions, 256 of which are not causal for any of its refinements.
It is a tight space.

The standard causaltope for Space 81 has 1 more dimension than that of its subspace in equivalence class 67.
The standard causaltope for Space 81 is the meet of the standard causaltopes for its closest coarsenings.
For completeness, below is a plot of the full homogeneous linear system of causality and quasi-normalisation equations for the standard causaltope:

\begin{center}
    \includegraphics[width=12cm]{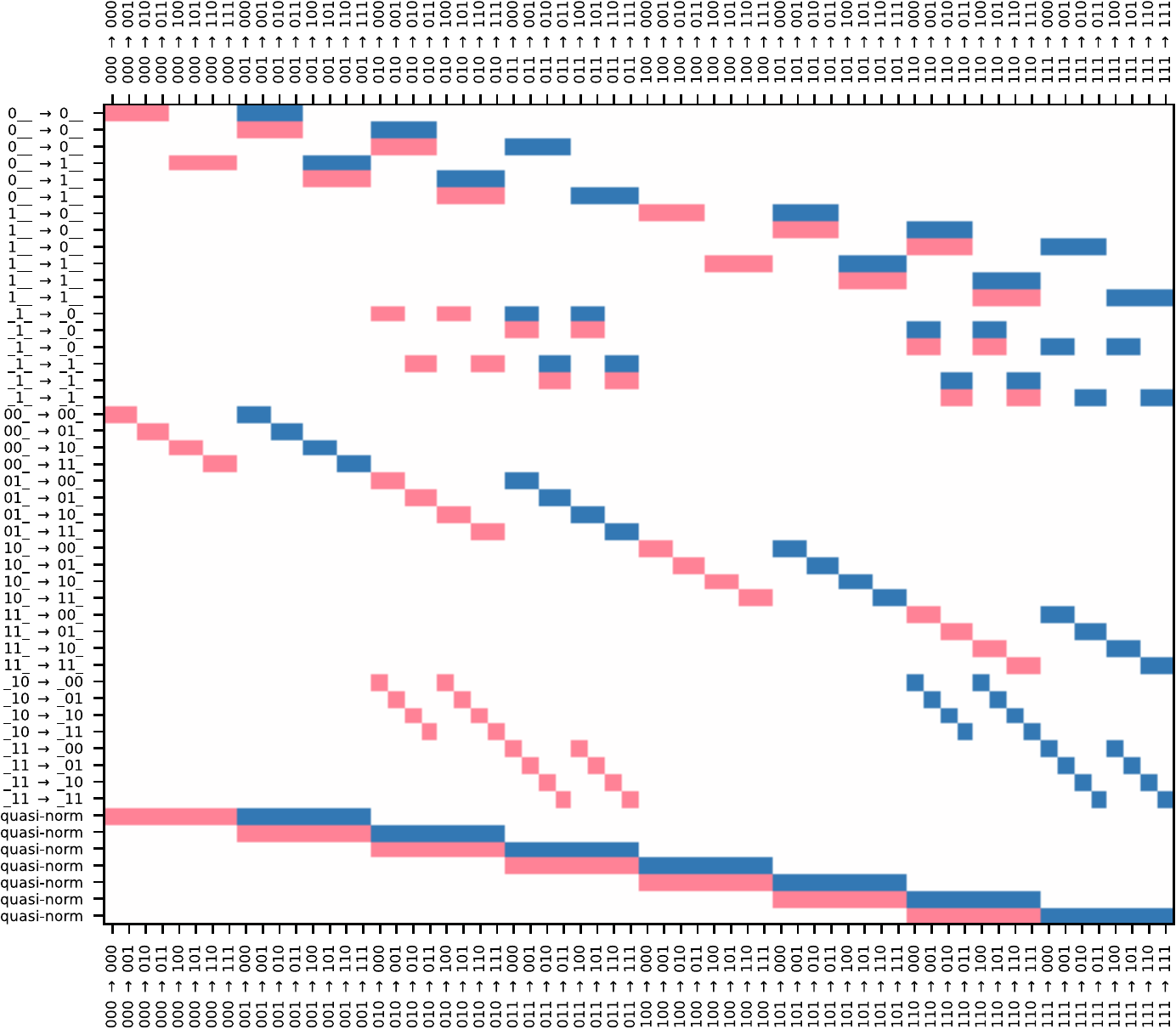}
\end{center}

\noindent Rows correspond to the 49 linear equations, of which 26 are independent.

\newpage
\subsection*{Space 82}

Space 82 is not induced by a causal order, but it is a refinement of the space 100 induced by the definite causal order $\total{\ev{A},\ev{B},\ev{C}}$.
Its equivalence class under event-input permutation symmetry contains 48 spaces.
Space 82 differs as follows from the space induced by causal order $\total{\ev{A},\ev{B},\ev{C}}$:
\begin{itemize}
  \item The outputs at events \evset{\ev{B}, \ev{C}} are independent of the input at event \ev{A} when the inputs at events \evset{B, C} are given by \hist{B/1,C/1}.
  \item The output at event \ev{B} is independent of the input at event \ev{A} when the input at event B is given by \hist{B/1}.
  \item The outputs at events \evset{\ev{A}, \ev{C}} are independent of the input at event \ev{B} when the inputs at events \evset{A, C} are given by \hist{A/1,C/1}.
\end{itemize}

\noindent Below are the histories and extended histories for space 82: 
\begin{center}
    \begin{tabular}{cc}
    \includegraphics[height=3.5cm]{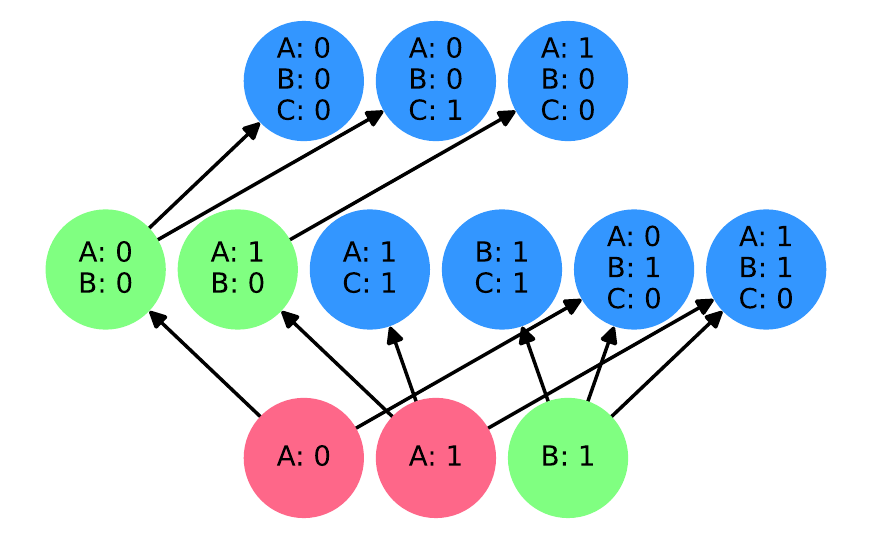}
    &
    \includegraphics[height=3.5cm]{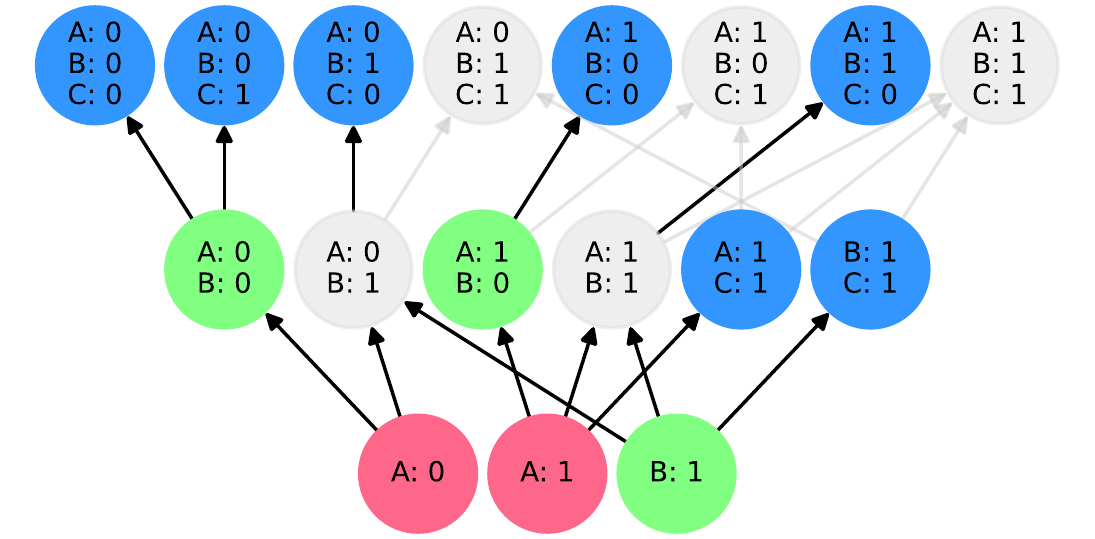}
    \\
    $\Theta_{82}$
    &
    $\Ext{\Theta_{82}}$
    \end{tabular}
\end{center}

\noindent The standard causaltope for Space 82 has dimension 37.
Below is a plot of the homogeneous linear system of causality and quasi-normalisation equations for the standard causaltope, put in reduced row echelon form:

\begin{center}
    \includegraphics[width=11cm]{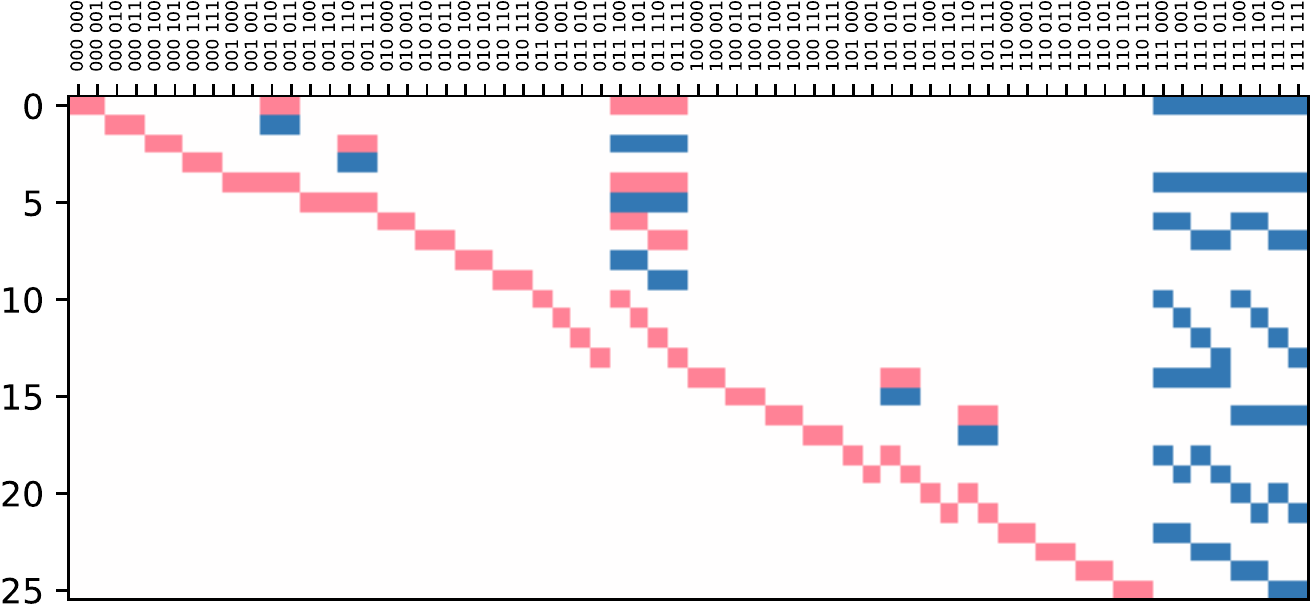}
\end{center}

\noindent Rows correspond to the 26 independent linear equations.
Columns in the plot correspond to entries of empirical models, indexed as $i_A i_B i_C$ $o_A o_B o_C$.
Coefficients in the equations are color-coded as white=0, red=+1 and blue=-1.

Space 82 has closest refinements in equivalence classes 63, 65, 69, 73 and 76; 
it is the join of its (closest) refinements.
It has closest coarsenings in equivalence classes 89 and 90; 
it is the meet of its (closest) coarsenings.
It has 2048 causal functions, 256 of which are not causal for any of its refinements.
It is not a tight space: for event \ev{C}, a causal function must yield identical output values on input histories \hist{A/1,C/1} and \hist{B/1,C/1}.

The standard causaltope for Space 82 has 1 more dimension than that of its subspace in equivalence class 69.
The standard causaltope for Space 82 is the meet of the standard causaltopes for its closest coarsenings.
For completeness, below is a plot of the full homogeneous linear system of causality and quasi-normalisation equations for the standard causaltope:

\begin{center}
    \includegraphics[width=12cm]{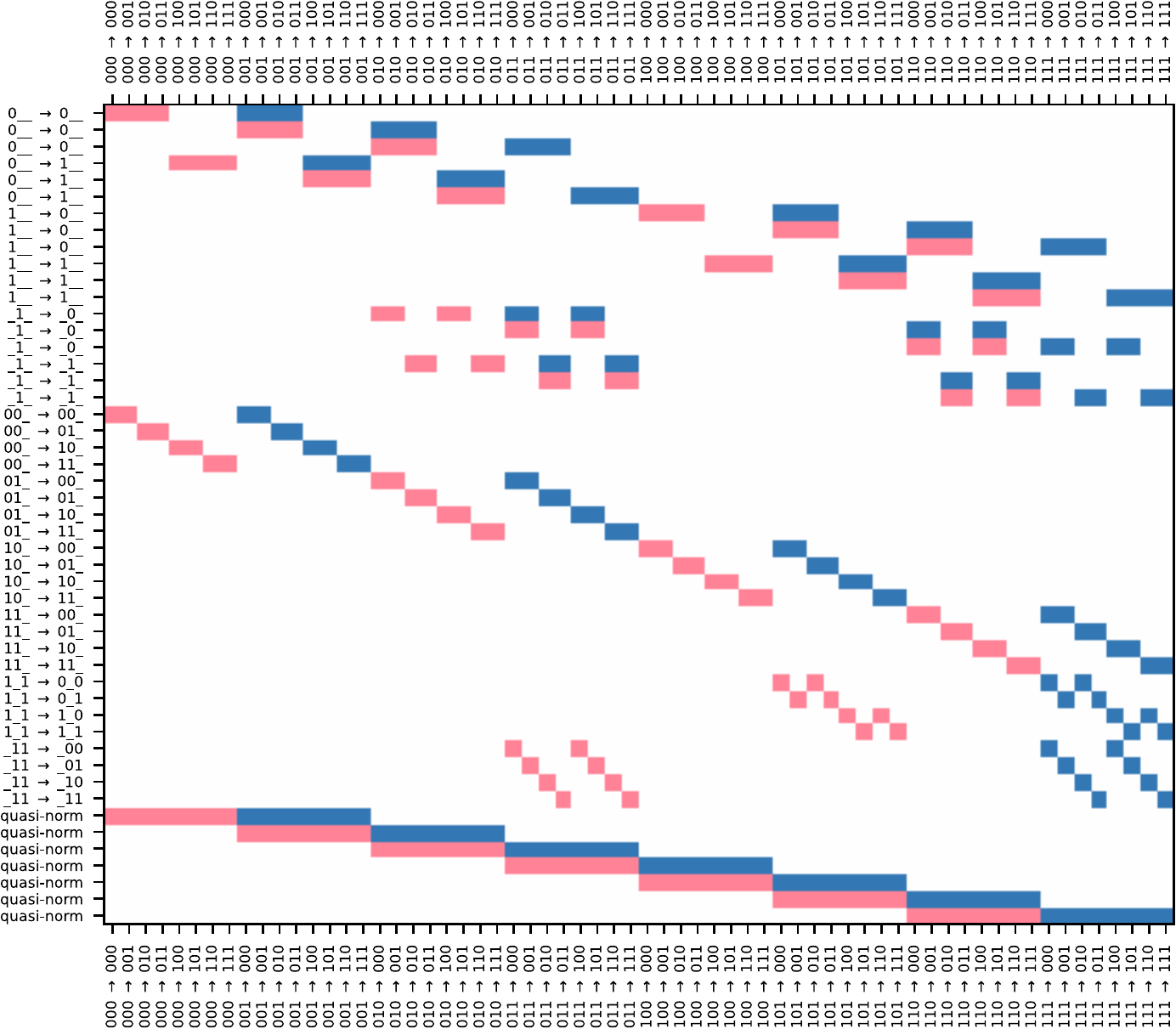}
\end{center}

\noindent Rows correspond to the 49 linear equations, of which 26 are independent.

\newpage
\subsection*{Space 83}

Space 83 is not induced by a causal order, but it is a refinement of the space 100 induced by the definite causal order $\total{\ev{A},\ev{B},\ev{C}}$.
Its equivalence class under event-input permutation symmetry contains 24 spaces.
Space 83 differs as follows from the space induced by causal order $\total{\ev{A},\ev{B},\ev{C}}$:
\begin{itemize}
  \item The outputs at events \evset{\ev{A}, \ev{C}} are independent of the input at event \ev{B} when the inputs at events \evset{A, C} are given by \hist{A/1,C/0} and \hist{A/1,C/1}.
  \item The output at event \ev{B} is independent of the input at event \ev{A} when the input at event B is given by \hist{B/1}.
\end{itemize}

\noindent Below are the histories and extended histories for space 83: 
\begin{center}
    \begin{tabular}{cc}
    \includegraphics[height=3.5cm]{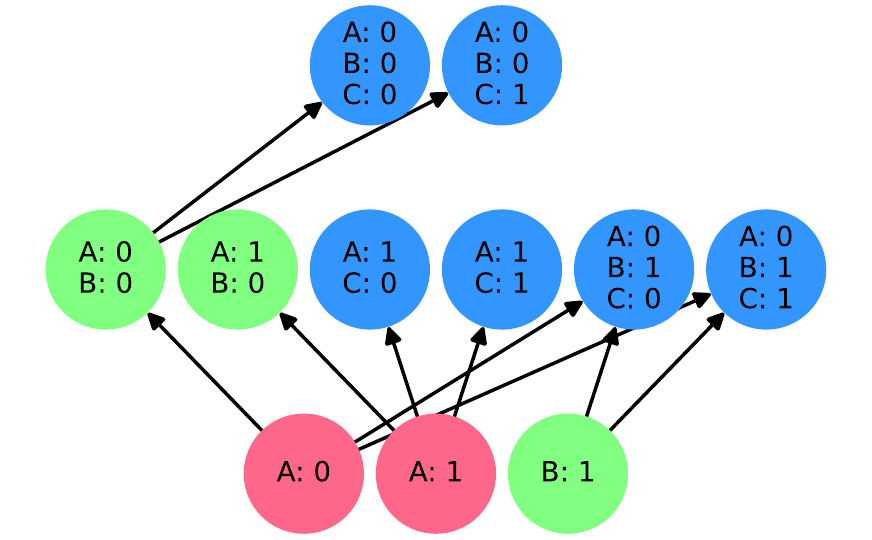}
    &
    \includegraphics[height=3.5cm]{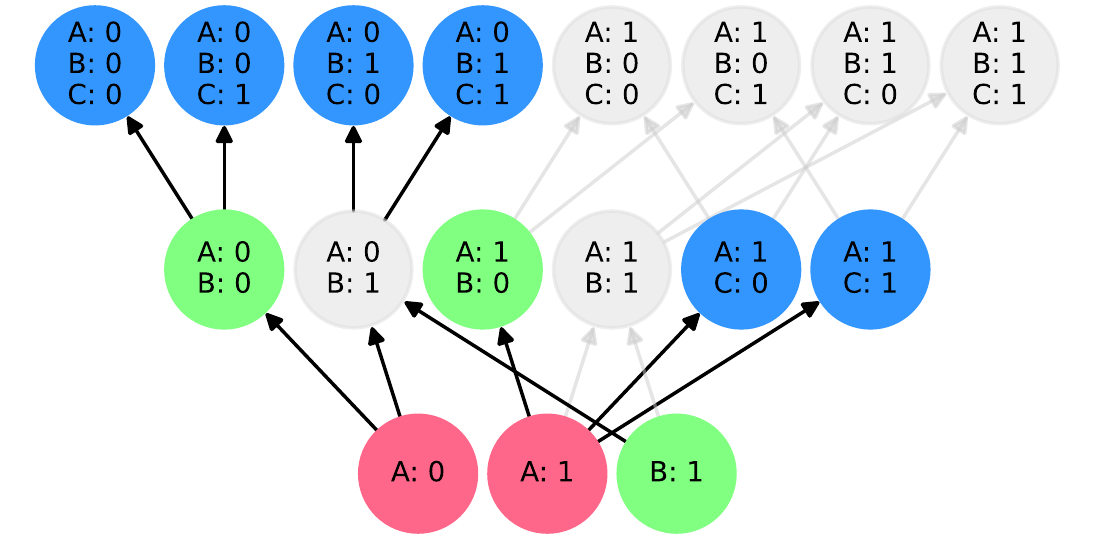}
    \\
    $\Theta_{83}$
    &
    $\Ext{\Theta_{83}}$
    \end{tabular}
\end{center}

\noindent The standard causaltope for Space 83 has dimension 37.
Below is a plot of the homogeneous linear system of causality and quasi-normalisation equations for the standard causaltope, put in reduced row echelon form:

\begin{center}
    \includegraphics[width=11cm]{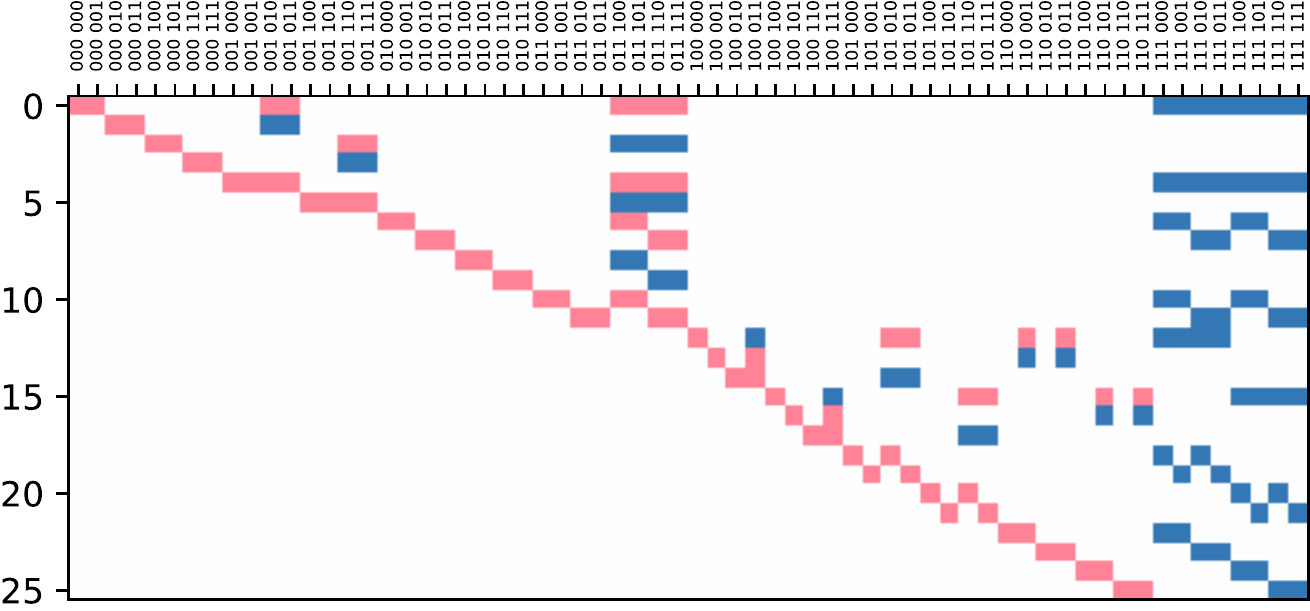}
\end{center}

\noindent Rows correspond to the 26 independent linear equations.
Columns in the plot correspond to entries of empirical models, indexed as $i_A i_B i_C$ $o_A o_B o_C$.
Coefficients in the equations are color-coded as white=0, red=+1 and blue=-1.

Space 83 has closest refinements in equivalence classes 63, 67 and 74; 
it is the join of its (closest) refinements.
It has closest coarsenings in equivalence classes 90, 91 and 93; 
it is the meet of its (closest) coarsenings.
It has 2048 causal functions, 320 of which are not causal for any of its refinements.
It is a tight space.

The standard causaltope for Space 83 has 1 more dimension than that of its subspace in equivalence class 67.
The standard causaltope for Space 83 is the meet of the standard causaltopes for its closest coarsenings.
For completeness, below is a plot of the full homogeneous linear system of causality and quasi-normalisation equations for the standard causaltope:

\begin{center}
    \includegraphics[width=12cm]{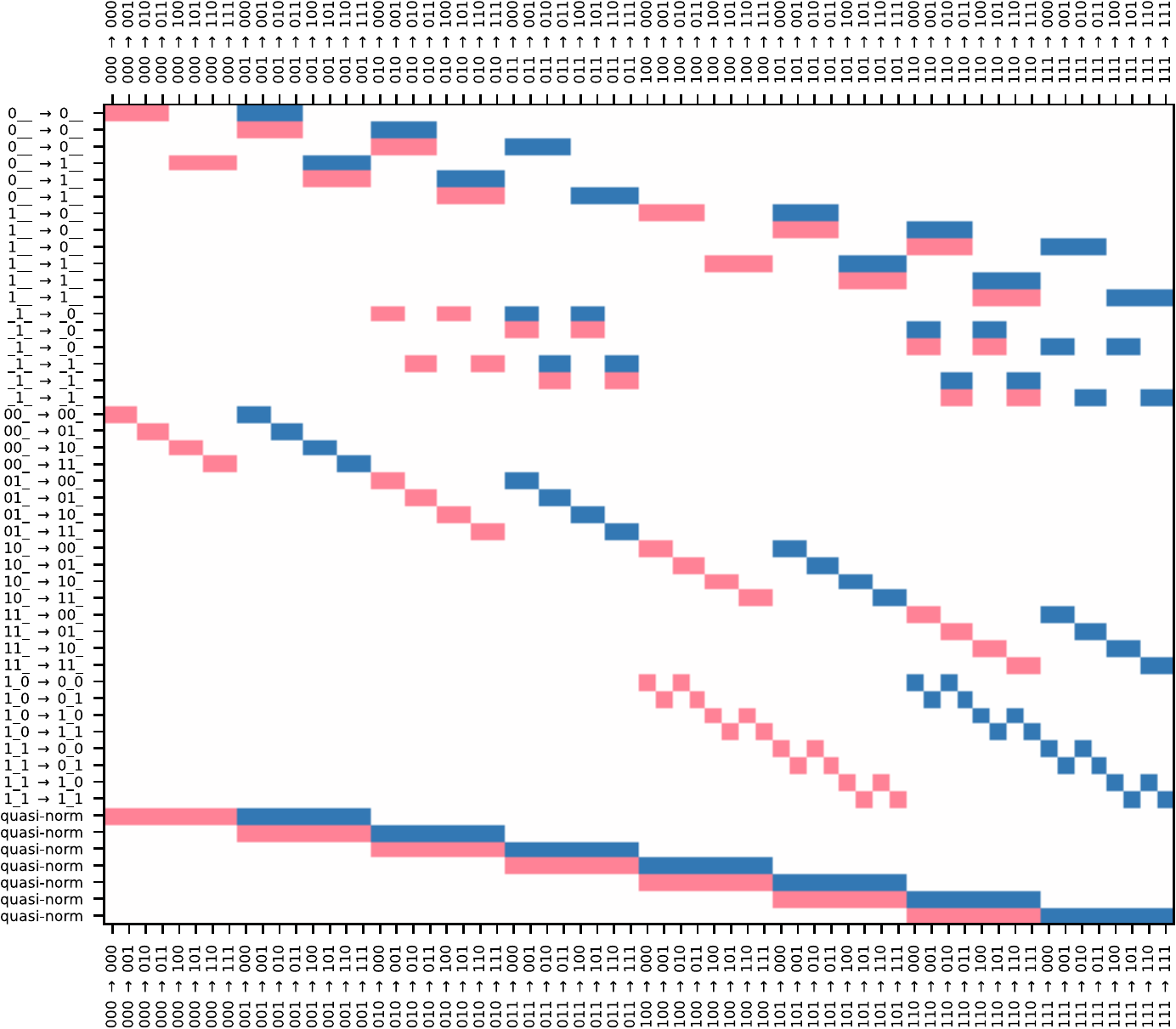}
\end{center}

\noindent Rows correspond to the 49 linear equations, of which 26 are independent.

\newpage
\subsection*{Space 84}

Space 84 is not induced by a causal order, but it is a refinement of the space 92 induced by the definite causal order $\total{\ev{A},\ev{C}}\vee\total{\ev{B},\ev{C}}$.
Its equivalence class under event-input permutation symmetry contains 24 spaces.
Space 84 differs as follows from the space induced by causal order $\total{\ev{A},\ev{C}}\vee\total{\ev{B},\ev{C}}$:
\begin{itemize}
  \item The outputs at events \evset{\ev{B}, \ev{C}} are independent of the input at event \ev{A} when the inputs at events \evset{B, C} are given by \hist{B/1,C/1}.
\end{itemize}

\noindent Below are the histories and extended histories for space 84: 
\begin{center}
    \begin{tabular}{cc}
    \includegraphics[height=3.5cm]{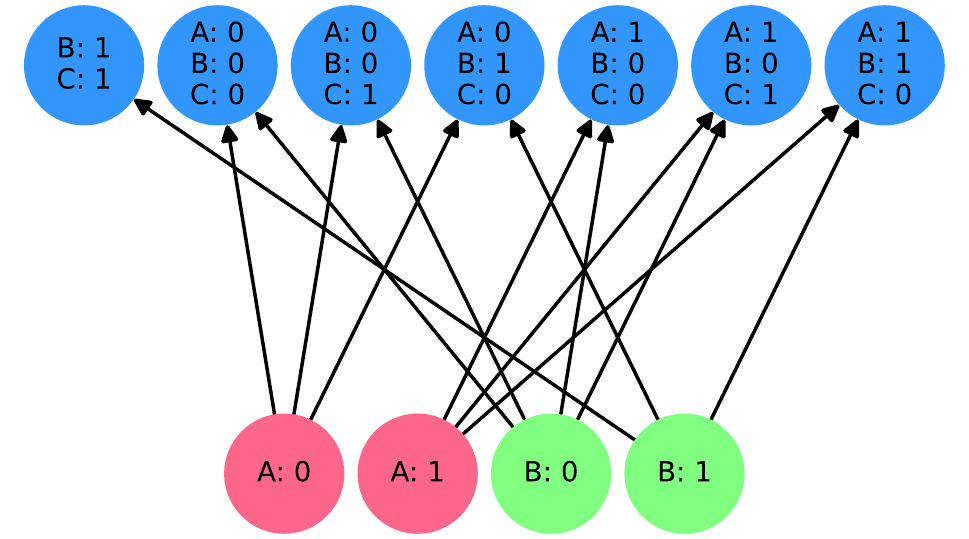}
    &
    \includegraphics[height=3.5cm]{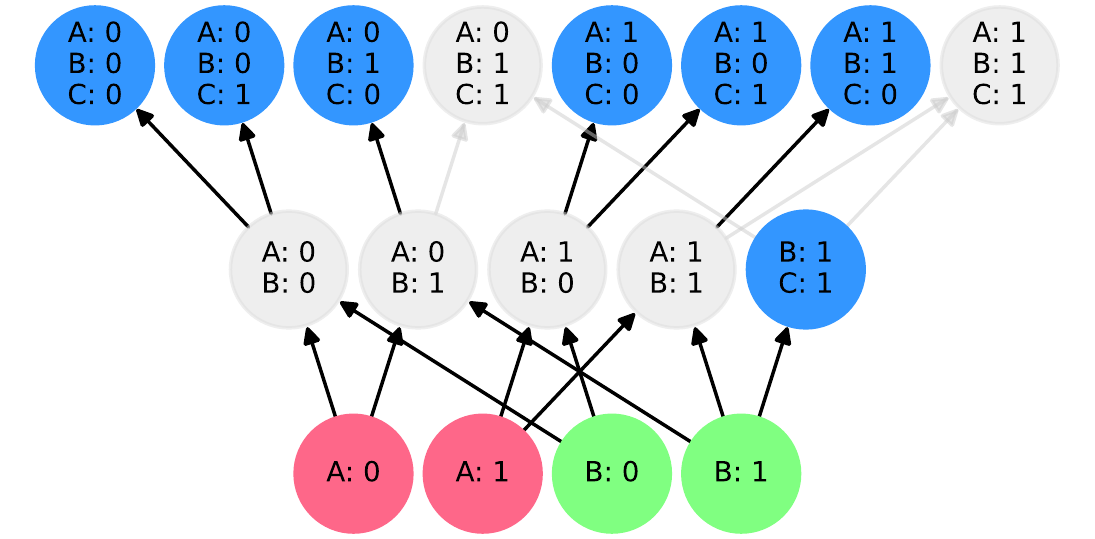}
    \\
    $\Theta_{84}$
    &
    $\Ext{\Theta_{84}}$
    \end{tabular}
\end{center}

\noindent The standard causaltope for Space 84 has dimension 38.
Below is a plot of the homogeneous linear system of causality and quasi-normalisation equations for the standard causaltope, put in reduced row echelon form:

\begin{center}
    \includegraphics[width=11cm]{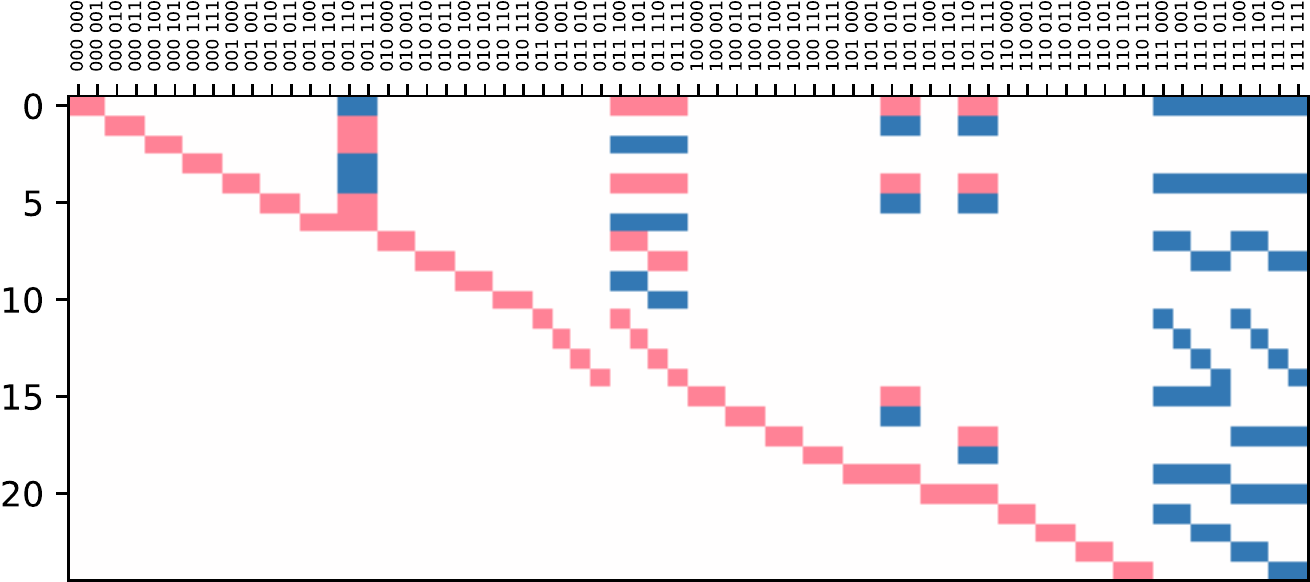}
\end{center}

\noindent Rows correspond to the 25 independent linear equations.
Columns in the plot correspond to entries of empirical models, indexed as $i_A i_B i_C$ $o_A o_B o_C$.
Coefficients in the equations are color-coded as white=0, red=+1 and blue=-1.

Space 84 has closest refinements in equivalence classes 66, 67, 68, 69 and 70; 
it is the join of its (closest) refinements.
It has closest coarsenings in equivalence classes 89, 90 and 92; 
it is the meet of its (closest) coarsenings.
It has 2048 causal functions, 128 of which are not causal for any of its refinements.
It is a tight space.

The standard causaltope for Space 84 has 2 more dimensions than those of its 7 subspaces in equivalence classes 66, 67, 68, 69 and 70.
The standard causaltope for Space 84 is the meet of the standard causaltopes for its closest coarsenings.
For completeness, below is a plot of the full homogeneous linear system of causality and quasi-normalisation equations for the standard causaltope:

\begin{center}
    \includegraphics[width=12cm]{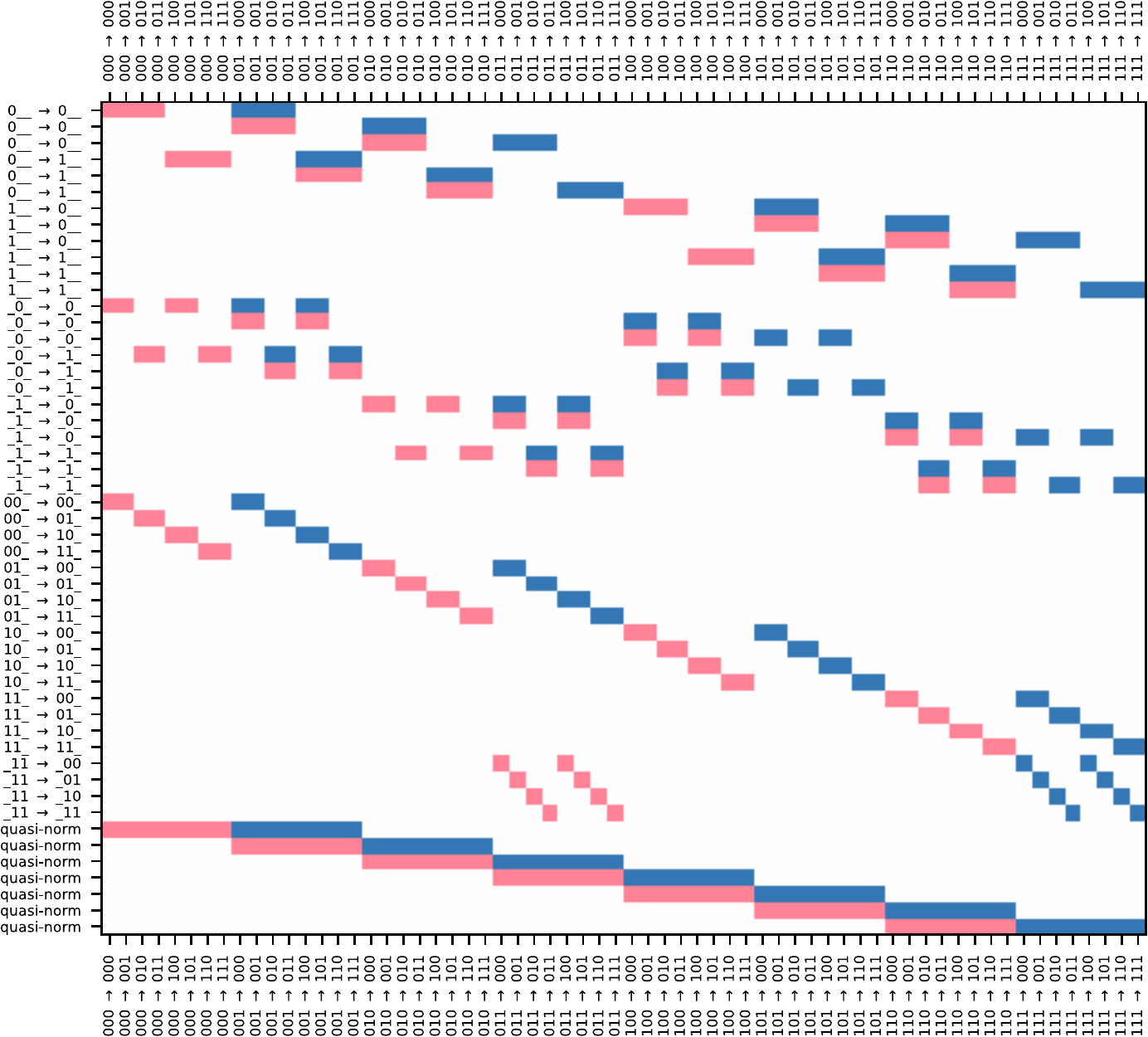}
\end{center}

\noindent Rows correspond to the 51 linear equations, of which 25 are independent.

\newpage
\subsection*{Space 85}

Space 85 is not induced by a causal order, but it is a refinement of the space induced by the indefinite causal order $\total{\ev{A},\{\ev{B},\ev{C}\}}$.
Its equivalence class under event-input permutation symmetry contains 48 spaces.
Space 85 differs as follows from the space induced by causal order $\total{\ev{A},\{\ev{B},\ev{C}\}}$:
\begin{itemize}
  \item The outputs at events \evset{\ev{A}, \ev{B}} are independent of the input at event \ev{C} when the inputs at events \evset{A, B} are given by \hist{A/0,B/0}, \hist{A/0,B/1} and \hist{A/1,B/0}.
  \item The output at event \ev{B} is independent of the inputs at events \evset{\ev{A}, \ev{C}} when the input at event B is given by \hist{B/0}.
  \item The outputs at events \evset{\ev{A}, \ev{C}} are independent of the input at event \ev{B} when the inputs at events \evset{A, C} are given by \hist{A/1,C/0} and \hist{A/1,C/1}.
  \item The outputs at events \evset{\ev{B}, \ev{C}} are independent of the input at event \ev{A} when the inputs at events \evset{B, C} are given by \hist{B/0,C/1}.
\end{itemize}

\noindent Below are the histories and extended histories for space 85: 
\begin{center}
    \begin{tabular}{cc}
    \includegraphics[height=3.5cm]{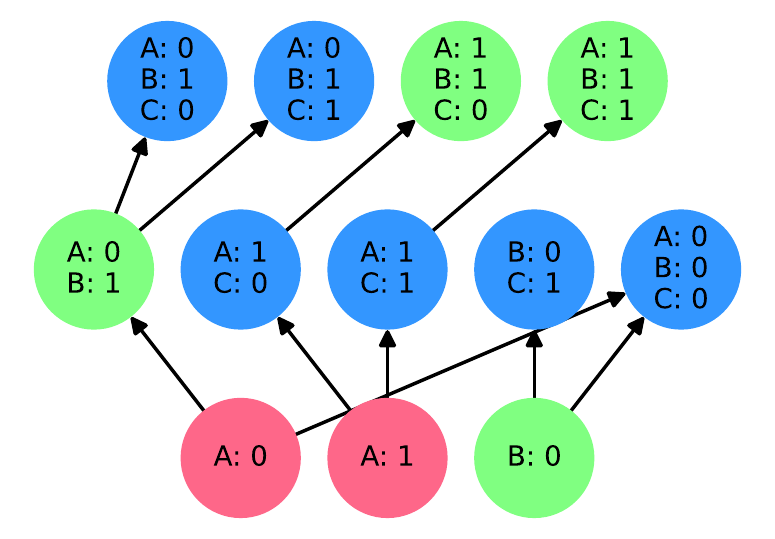}
    &
    \includegraphics[height=3.5cm]{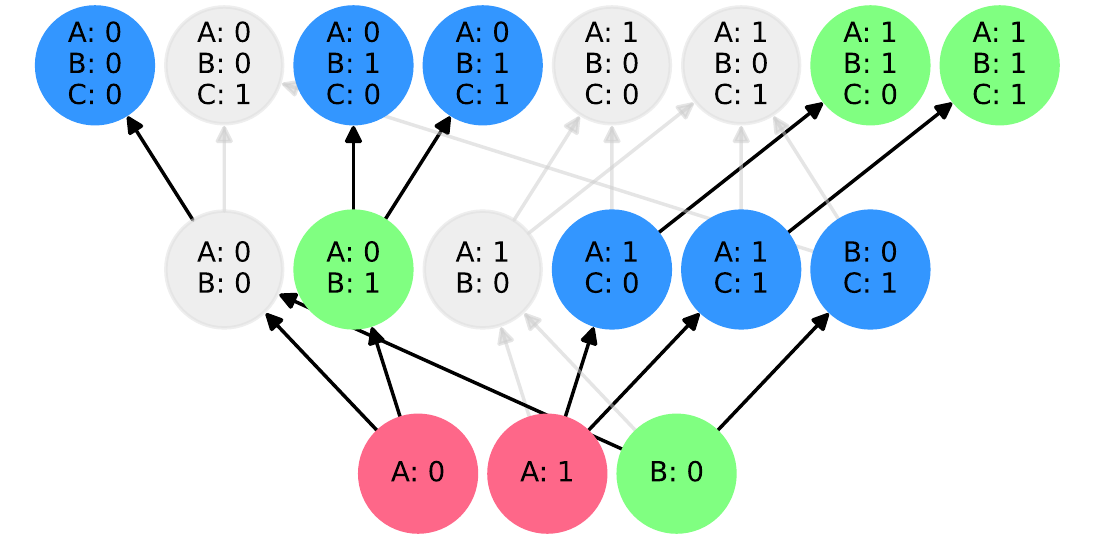}
    \\
    $\Theta_{85}$
    &
    $\Ext{\Theta_{85}}$
    \end{tabular}
\end{center}

\noindent The standard causaltope for Space 85 has dimension 37.
Below is a plot of the homogeneous linear system of causality and quasi-normalisation equations for the standard causaltope, put in reduced row echelon form:

\begin{center}
    \includegraphics[width=11cm]{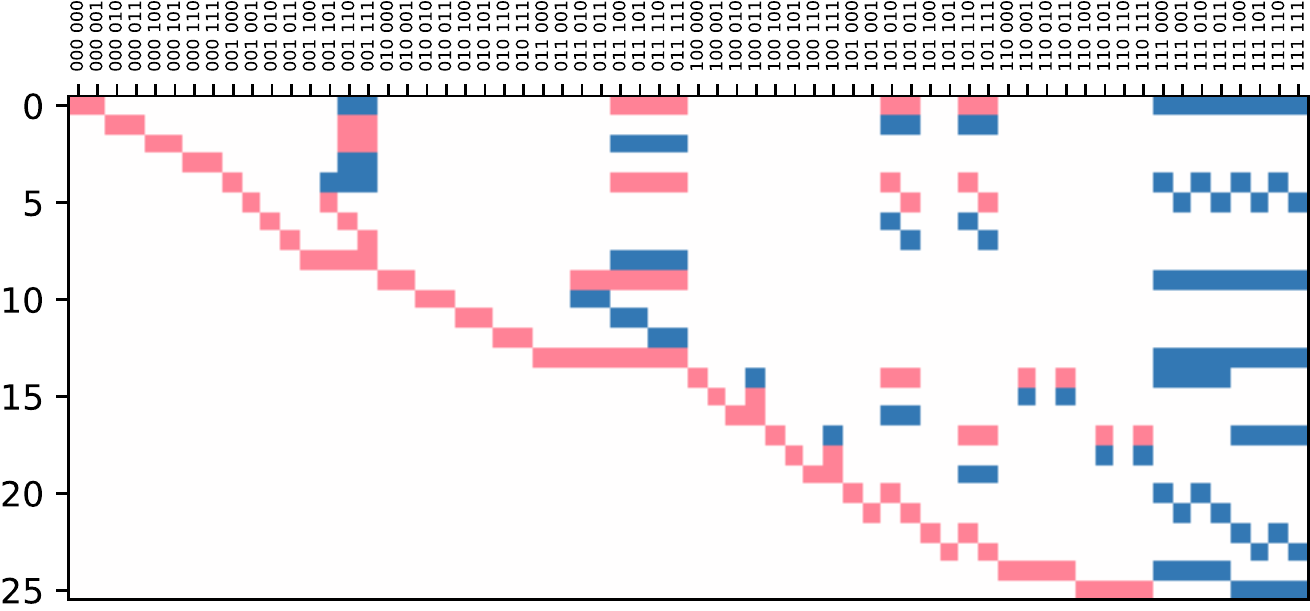}
\end{center}

\noindent Rows correspond to the 26 independent linear equations.
Columns in the plot correspond to entries of empirical models, indexed as $i_A i_B i_C$ $o_A o_B o_C$.
Coefficients in the equations are color-coded as white=0, red=+1 and blue=-1.

Space 85 has closest refinements in equivalence classes 62, 63, 71 and 72; 
it is the join of its (closest) refinements.
It has closest coarsenings in equivalence class 93; 
it does not arise as a nontrivial meet in the hierarchy.
It has 2048 causal functions, 640 of which are not causal for any of its refinements.
It is not a tight space: for event \ev{C}, a causal function must yield identical output values on input histories \hist{A/1,C/1} and \hist{B/0,C/1}.

The standard causaltope for Space 85 has 2 more dimensions than those of its 4 subspaces in equivalence classes 62, 63, 71 and 72.
The standard causaltope for Space 85 has 2 dimensions fewer than the meet of the standard causaltopes for its closest coarsenings.
For completeness, below is a plot of the full homogeneous linear system of causality and quasi-normalisation equations for the standard causaltope:

\begin{center}
    \includegraphics[width=12cm]{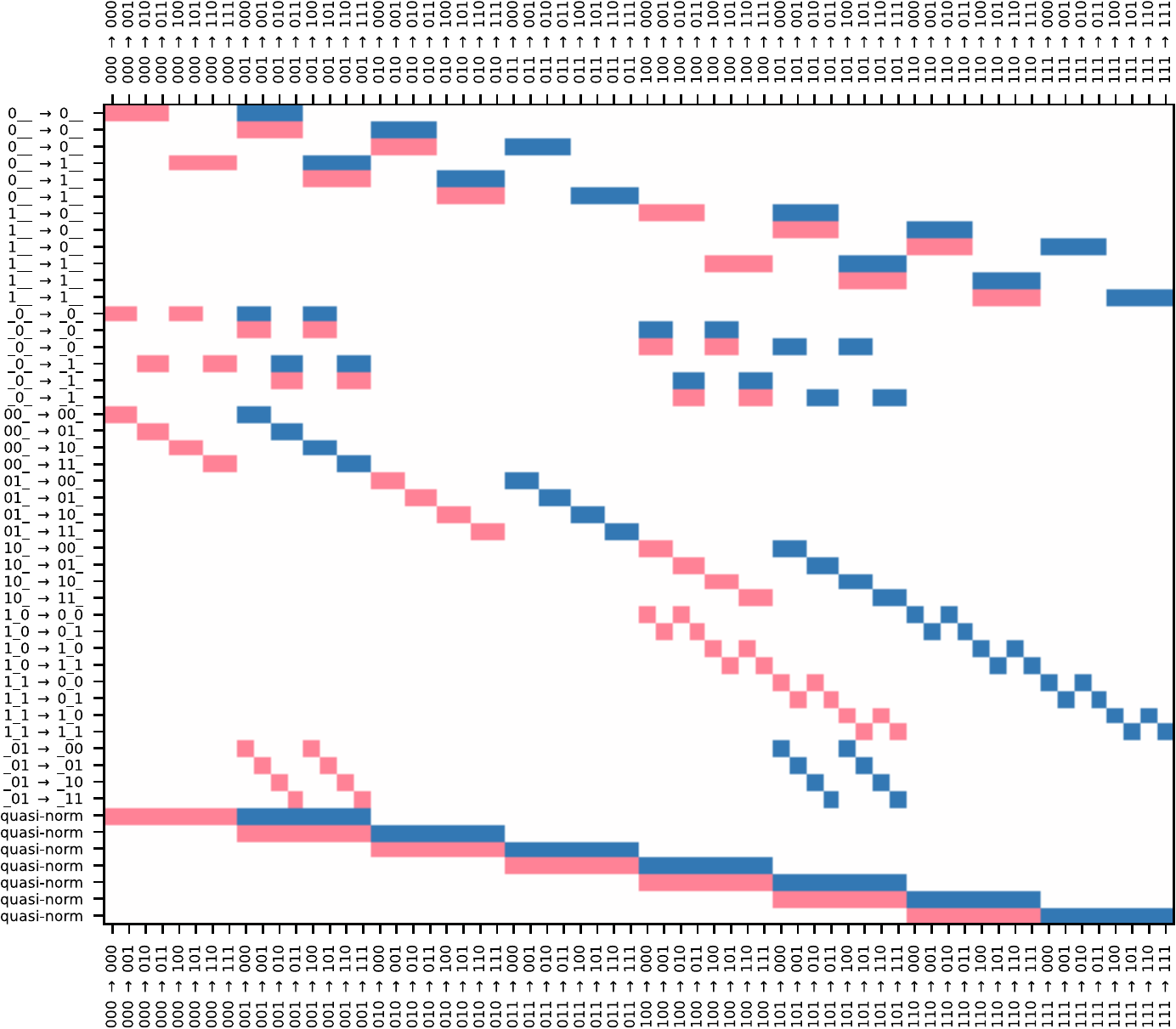}
\end{center}

\noindent Rows correspond to the 49 linear equations, of which 26 are independent.

\newpage
\subsection*{Space 86}

Space 86 is not induced by a causal order, but it is a refinement of the space in equivalence class 100 induced by the definite causal order $\total{\ev{B},\ev{A},\ev{B}}$ (note that the space induced by the order is not the same as space 100).
Its equivalence class under event-input permutation symmetry contains 24 spaces.
Space 86 differs as follows from the space induced by causal order $\total{\ev{B},\ev{A},\ev{B}}$:
\begin{itemize}
  \item The outputs at events \evset{\ev{B}, \ev{C}} are independent of the input at event \ev{A} when the inputs at events \evset{B, C} are given by \hist{B/1,C/1} and \hist{B/0,C/1}.
  \item The output at event \ev{A} is independent of the input at event \ev{B} when the input at event A is given by \hist{A/0}.
\end{itemize}

\noindent Below are the histories and extended histories for space 86: 
\begin{center}
    \begin{tabular}{cc}
    \includegraphics[height=3.5cm]{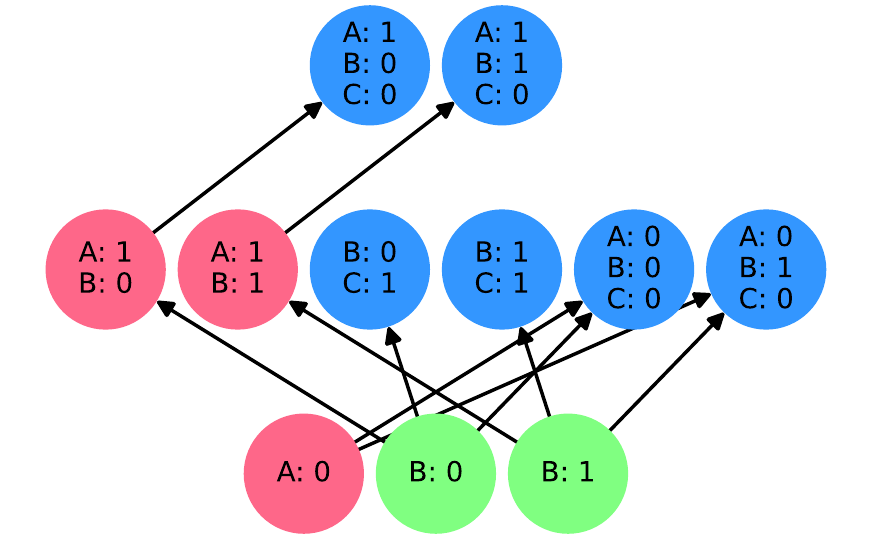}
    &
    \includegraphics[height=3.5cm]{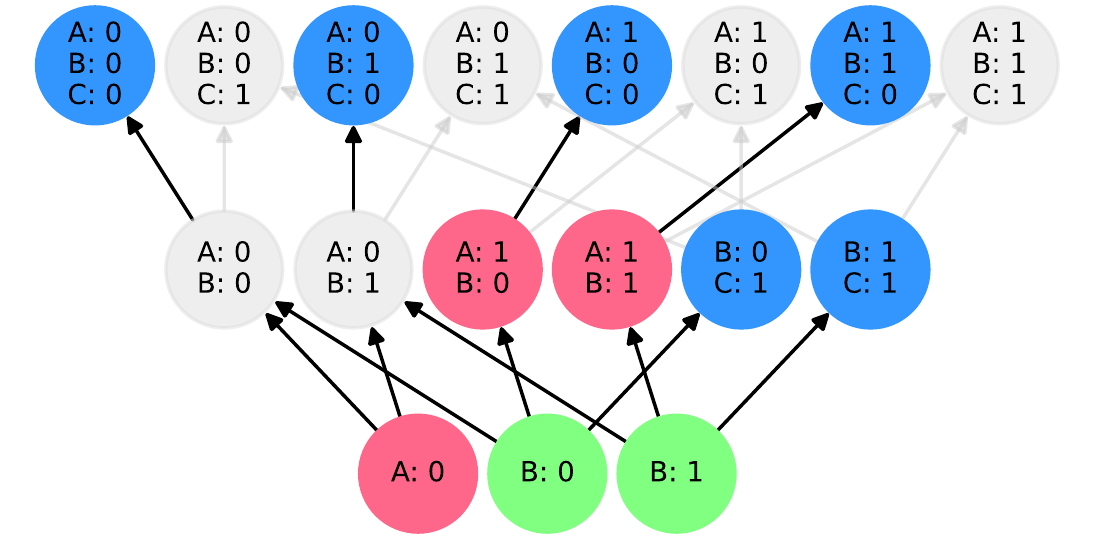}
    \\
    $\Theta_{86}$
    &
    $\Ext{\Theta_{86}}$
    \end{tabular}
\end{center}

\noindent The standard causaltope for Space 86 has dimension 37.
Below is a plot of the homogeneous linear system of causality and quasi-normalisation equations for the standard causaltope, put in reduced row echelon form:

\begin{center}
    \includegraphics[width=11cm]{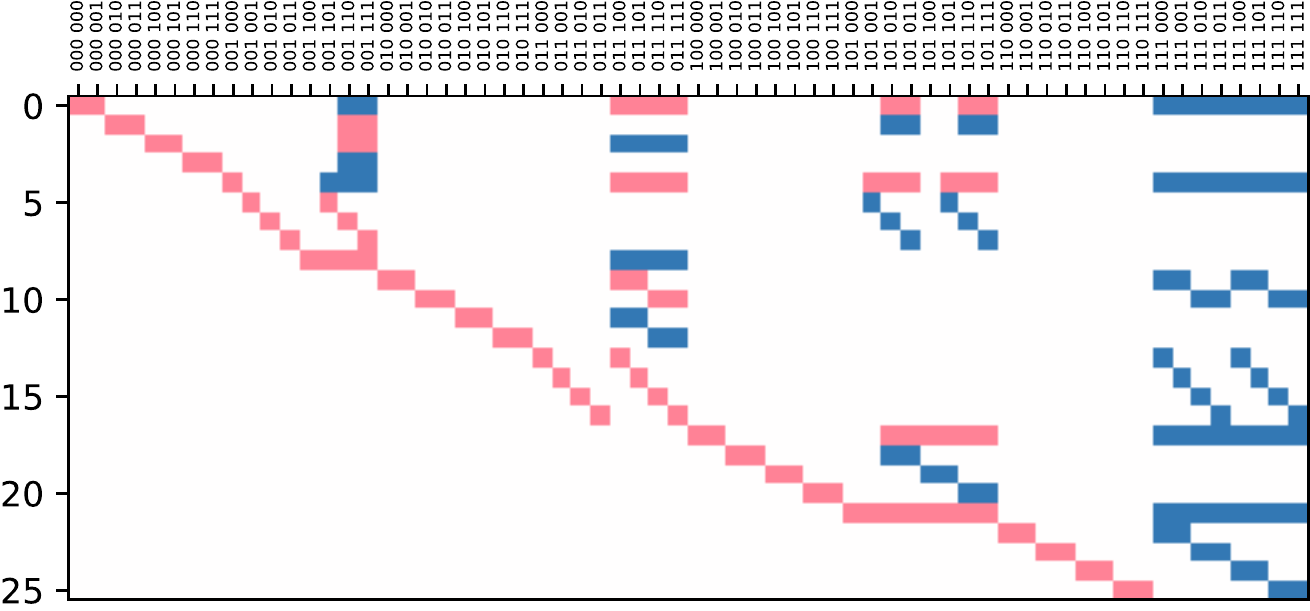}
\end{center}

\noindent Rows correspond to the 26 independent linear equations.
Columns in the plot correspond to entries of empirical models, indexed as $i_A i_B i_C$ $o_A o_B o_C$.
Coefficients in the equations are color-coded as white=0, red=+1 and blue=-1.

Space 86 has closest refinements in equivalence classes 68, 74, 75 and 76; 
it is the join of its (closest) refinements.
It has closest coarsenings in equivalence classes 90 and 94; 
it is the meet of its (closest) coarsenings.
It has 2048 causal functions, 1024 of which are not causal for any of its refinements.
It is a tight space.

The standard causaltope for Space 86 has 1 more dimension than that of its subspace in equivalence class 68.
The standard causaltope for Space 86 is the meet of the standard causaltopes for its closest coarsenings.
For completeness, below is a plot of the full homogeneous linear system of causality and quasi-normalisation equations for the standard causaltope:

\begin{center}
    \includegraphics[width=12cm]{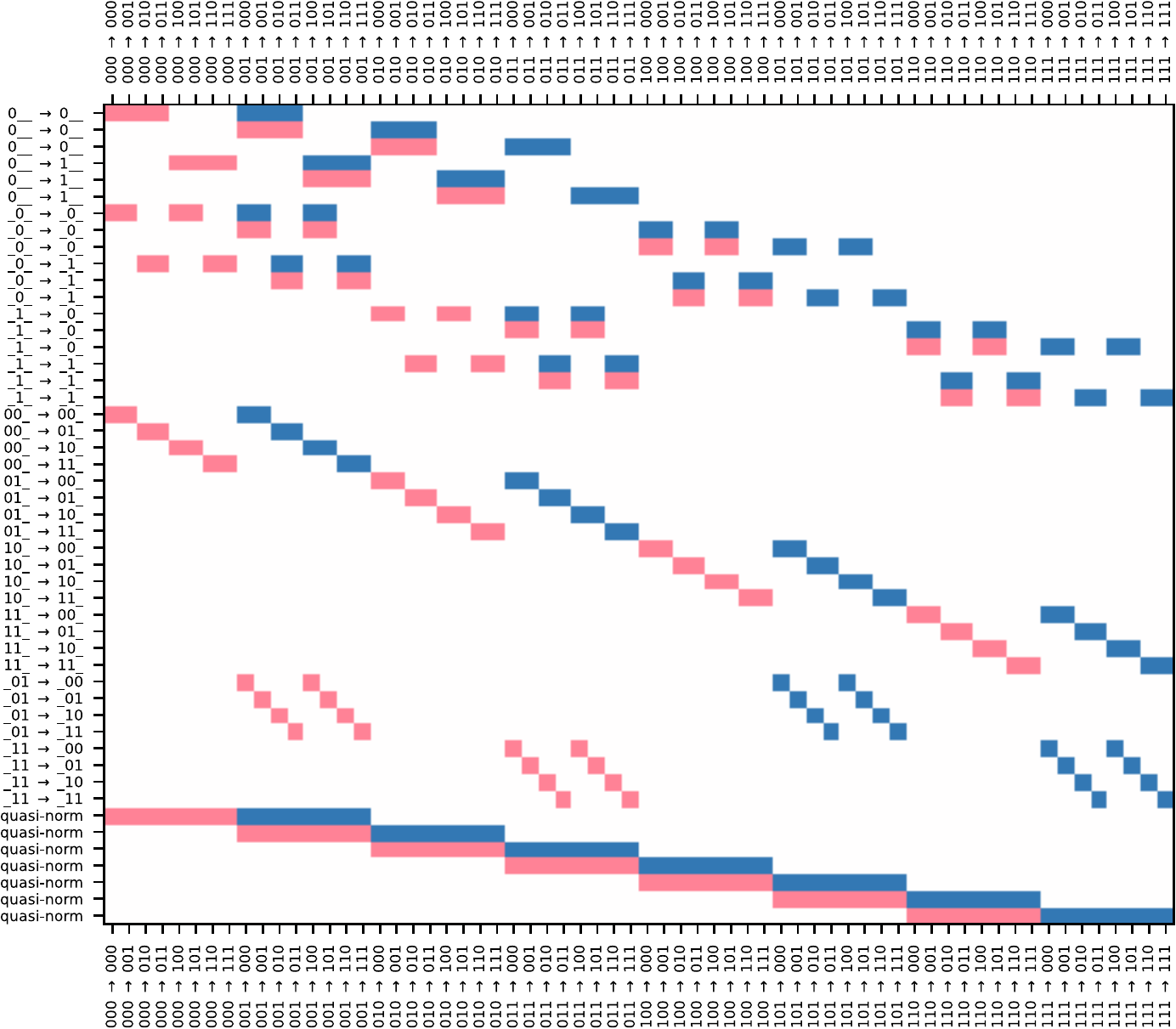}
\end{center}

\noindent Rows correspond to the 49 linear equations, of which 26 are independent.

\newpage
\subsection*{Space 87}

Space 87 is not induced by a causal order, but it is a refinement of the space in equivalence class 100 induced by the definite causal order $\total{\ev{B},\ev{A},\ev{B}}$ (note that the space induced by the order is not the same as space 100).
Its equivalence class under event-input permutation symmetry contains 24 spaces.
Space 87 differs as follows from the space induced by causal order $\total{\ev{B},\ev{A},\ev{B}}$:
\begin{itemize}
  \item The outputs at events \evset{\ev{B}, \ev{C}} are independent of the input at event \ev{A} when the inputs at events \evset{B, C} are given by \hist{B/1,C/0} and \hist{B/0,C/1}.
  \item The output at event \ev{A} is independent of the input at event \ev{B} when the input at event A is given by \hist{A/0}.
\end{itemize}

\noindent Below are the histories and extended histories for space 87: 
\begin{center}
    \begin{tabular}{cc}
    \includegraphics[height=3.5cm]{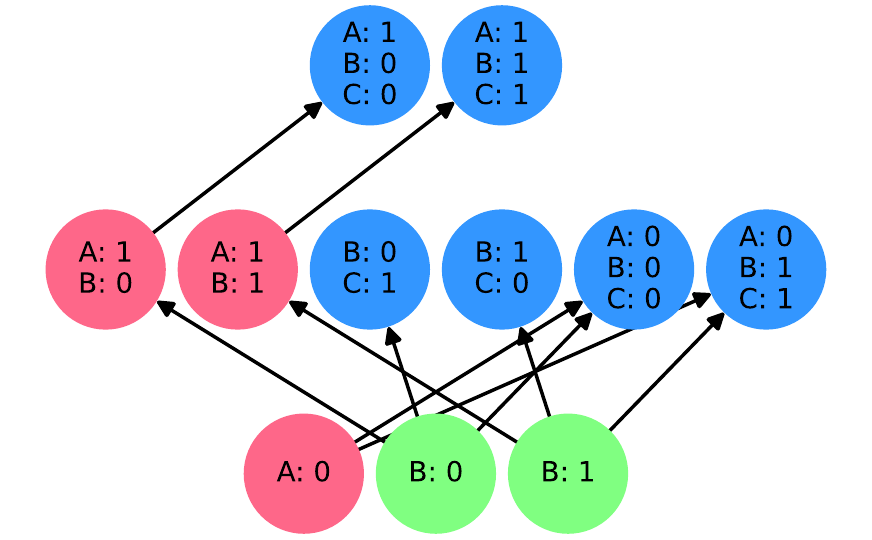}
    &
    \includegraphics[height=3.5cm]{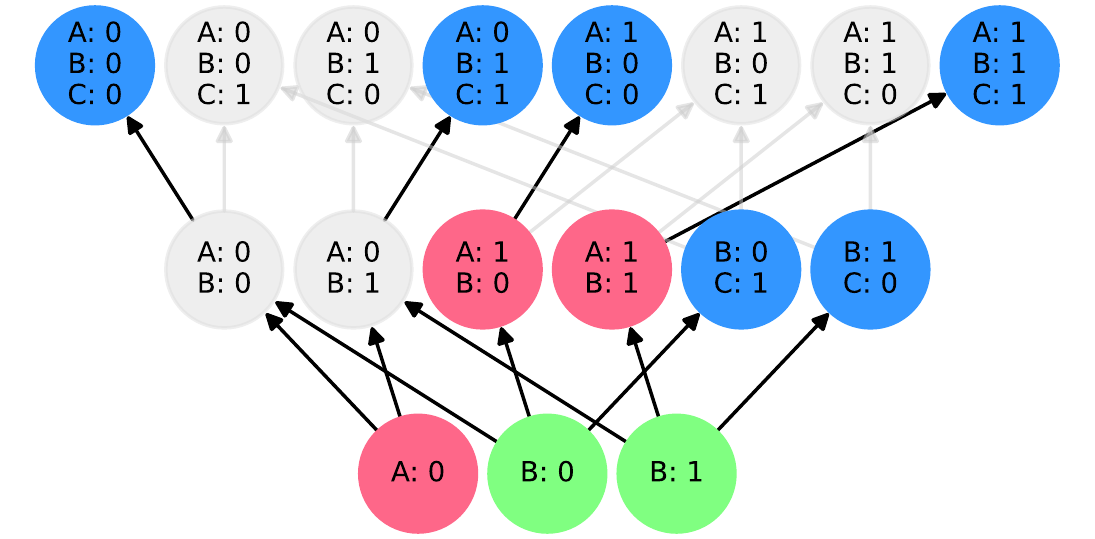}
    \\
    $\Theta_{87}$
    &
    $\Ext{\Theta_{87}}$
    \end{tabular}
\end{center}

\noindent The standard causaltope for Space 87 has dimension 37.
Below is a plot of the homogeneous linear system of causality and quasi-normalisation equations for the standard causaltope, put in reduced row echelon form:

\begin{center}
    \includegraphics[width=11cm]{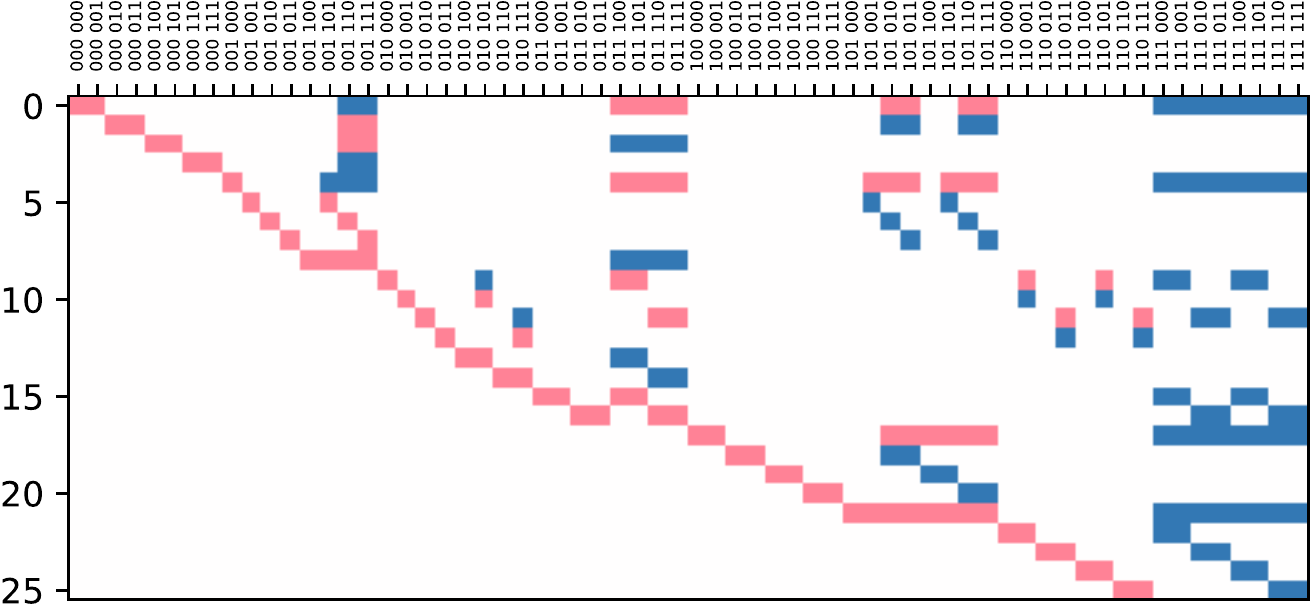}
\end{center}

\noindent Rows correspond to the 26 independent linear equations.
Columns in the plot correspond to entries of empirical models, indexed as $i_A i_B i_C$ $o_A o_B o_C$.
Coefficients in the equations are color-coded as white=0, red=+1 and blue=-1.

Space 87 has closest refinements in equivalence classes 70, 73 and 74; 
it is the join of its (closest) refinements.
It has closest coarsenings in equivalence classes 90 and 95; 
it is the meet of its (closest) coarsenings.
It has 2048 causal functions, 512 of which are not causal for any of its refinements.
It is a tight space.

The standard causaltope for Space 87 has 1 more dimension than that of its subspace in equivalence class 70.
The standard causaltope for Space 87 is the meet of the standard causaltopes for its closest coarsenings.
For completeness, below is a plot of the full homogeneous linear system of causality and quasi-normalisation equations for the standard causaltope:

\begin{center}
    \includegraphics[width=12cm]{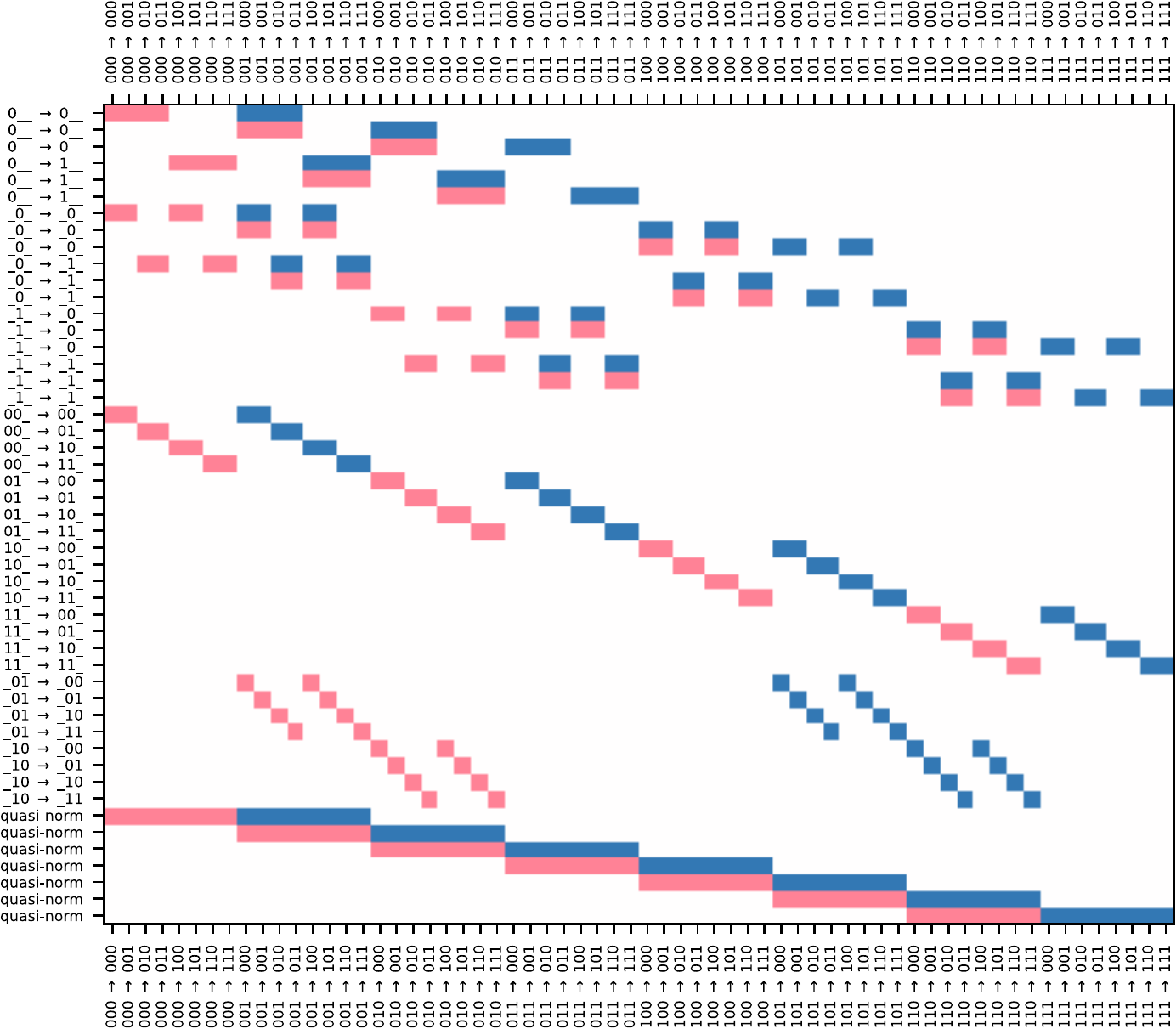}
\end{center}

\noindent Rows correspond to the 49 linear equations, of which 26 are independent.

\newpage
\subsection*{Space 88}

Space 88 is not induced by a causal order, but it is a refinement of the space 100 induced by the definite causal order $\total{\ev{A},\ev{B},\ev{C}}$.
Its equivalence class under event-input permutation symmetry contains 24 spaces.
Space 88 differs as follows from the space induced by causal order $\total{\ev{A},\ev{B},\ev{C}}$:
\begin{itemize}
  \item The outputs at events \evset{\ev{A}, \ev{C}} are independent of the input at event \ev{B} when the inputs at events \evset{A, C} are given by \hist{A/0,C/1}, \hist{A/1,C/0} and \hist{A/1,C/1}.
\end{itemize}

\noindent Below are the histories and extended histories for space 88: 
\begin{center}
    \begin{tabular}{cc}
    \includegraphics[height=3.5cm]{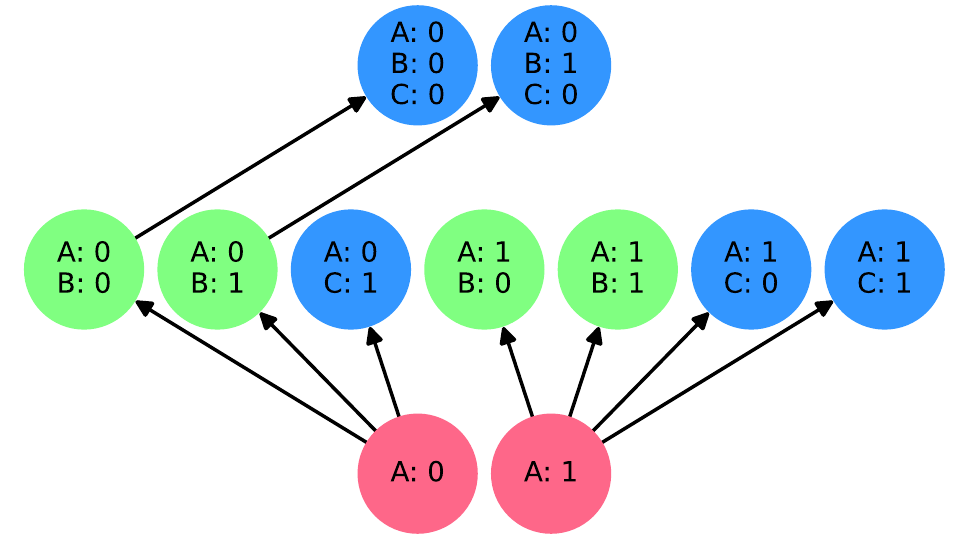}
    &
    \includegraphics[height=3.5cm]{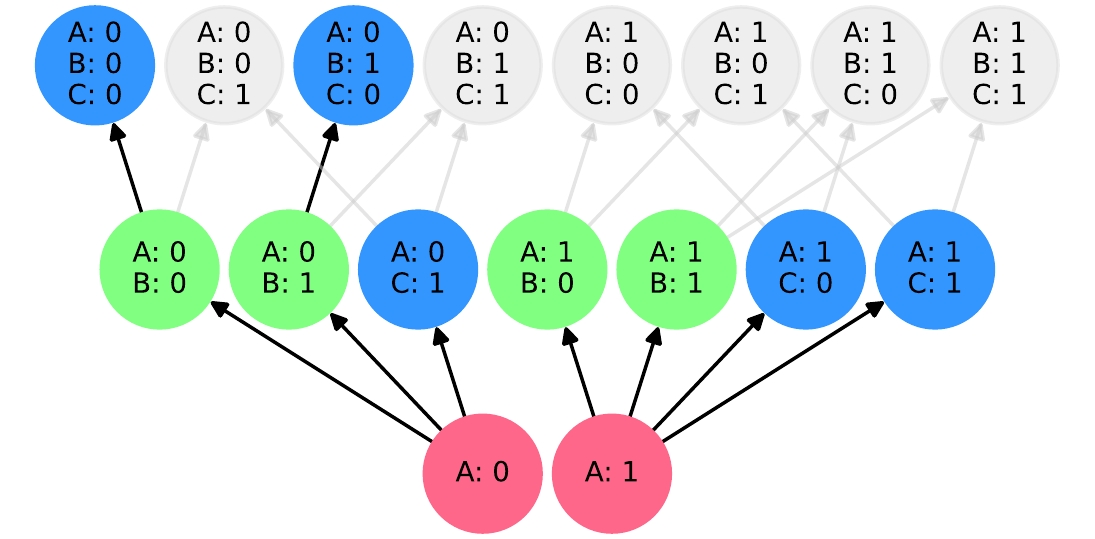}
    \\
    $\Theta_{88}$
    &
    $\Ext{\Theta_{88}}$
    \end{tabular}
\end{center}

\noindent The standard causaltope for Space 88 has dimension 36.
Below is a plot of the homogeneous linear system of causality and quasi-normalisation equations for the standard causaltope, put in reduced row echelon form:

\begin{center}
    \includegraphics[width=11cm]{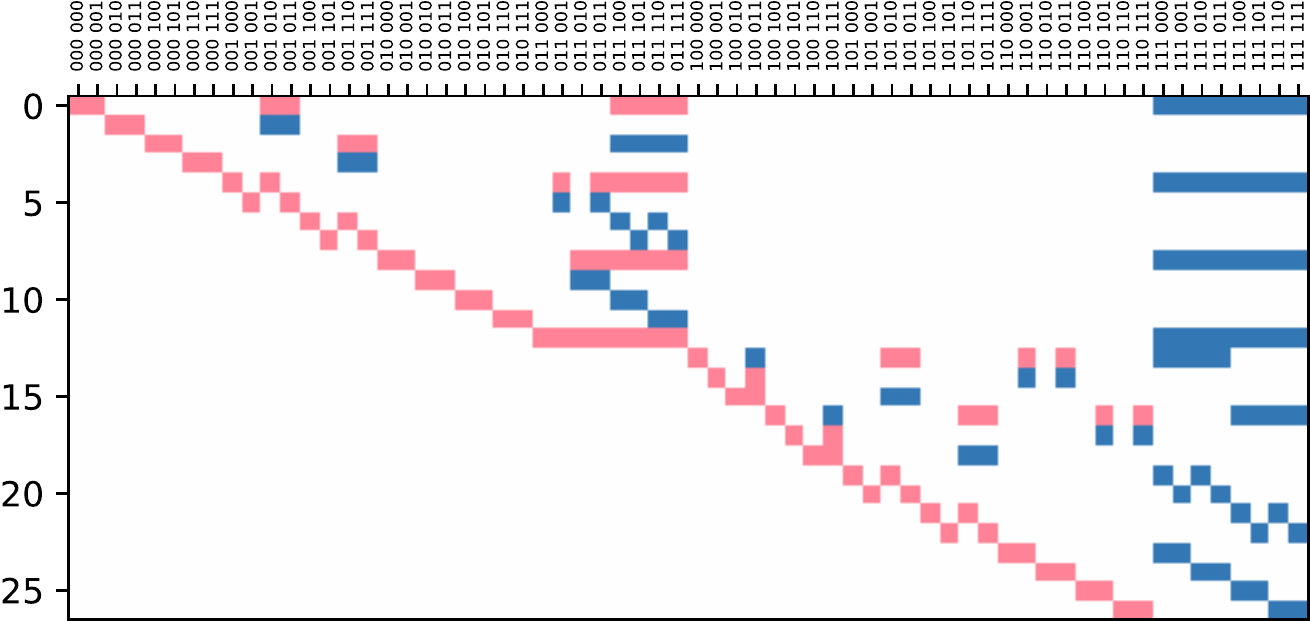}
\end{center}

\noindent Rows correspond to the 27 independent linear equations.
Columns in the plot correspond to entries of empirical models, indexed as $i_A i_B i_C$ $o_A o_B o_C$.
Coefficients in the equations are color-coded as white=0, red=+1 and blue=-1.

Space 88 has closest refinements in equivalence classes 64, 74 and 77; 
it is the join of its (closest) refinements.
It has closest coarsenings in equivalence classes 91, 94, 95 and 96; 
it is the meet of its (closest) coarsenings.
It has 2048 causal functions, 256 of which are not causal for any of its refinements.
It is a tight space.

The standard causaltope for Space 88 has 1 more dimension than those of its 3 subspaces in equivalence classes 64 and 74.
The standard causaltope for Space 88 is the meet of the standard causaltopes for its closest coarsenings.
For completeness, below is a plot of the full homogeneous linear system of causality and quasi-normalisation equations for the standard causaltope:

\begin{center}
    \includegraphics[width=12cm]{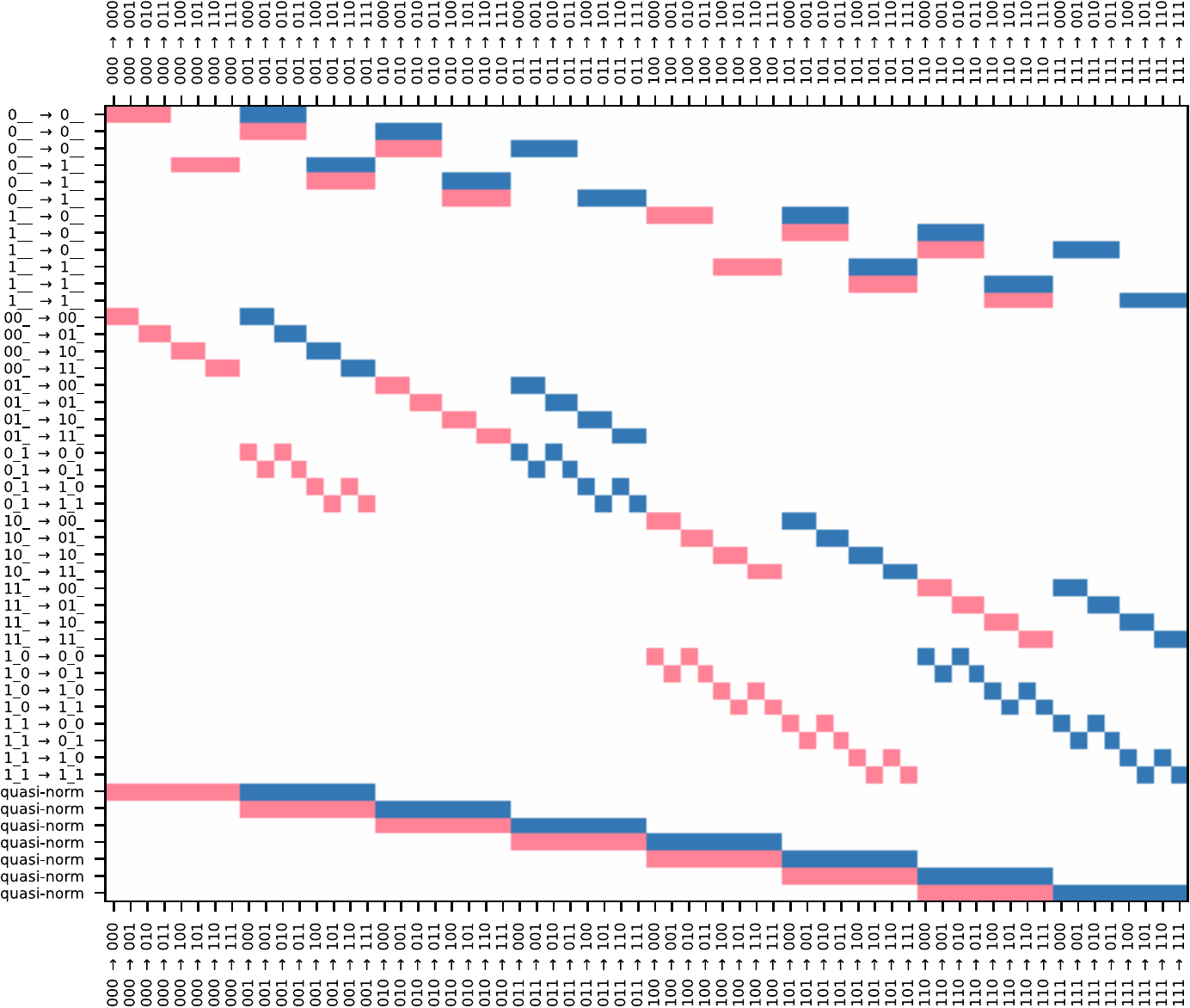}
\end{center}

\noindent Rows correspond to the 47 linear equations, of which 27 are independent.

\newpage
\subsection*{Space 89}

Space 89 is not induced by a causal order, but it is a refinement of the space 100 induced by the definite causal order $\total{\ev{A},\ev{B},\ev{C}}$.
Its equivalence class under event-input permutation symmetry contains 24 spaces.
Space 89 differs as follows from the space induced by causal order $\total{\ev{A},\ev{B},\ev{C}}$:
\begin{itemize}
  \item The outputs at events \evset{\ev{B}, \ev{C}} are independent of the input at event \ev{A} when the inputs at events \evset{B, C} are given by \hist{B/1,C/1}.
  \item The output at event \ev{B} is independent of the input at event \ev{A} when the input at event B is given by \hist{B/1}.
\end{itemize}

\noindent Below are the histories and extended histories for space 89: 
\begin{center}
    \begin{tabular}{cc}
    \includegraphics[height=3.5cm]{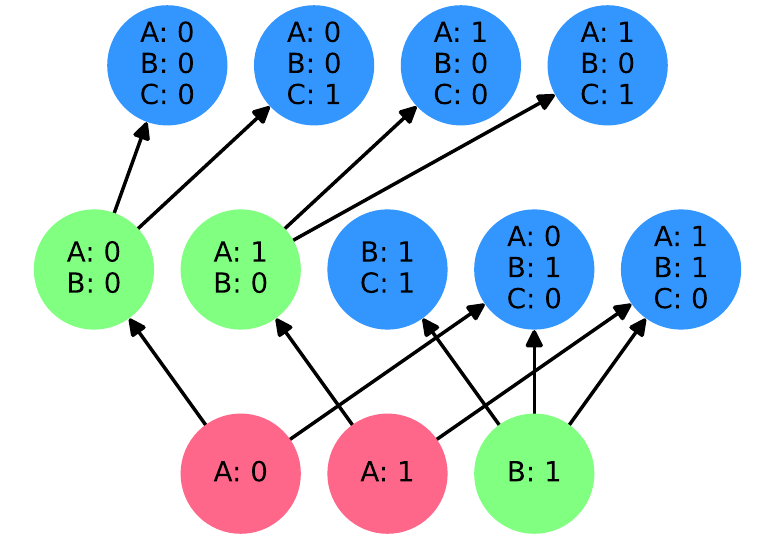}
    &
    \includegraphics[height=3.5cm]{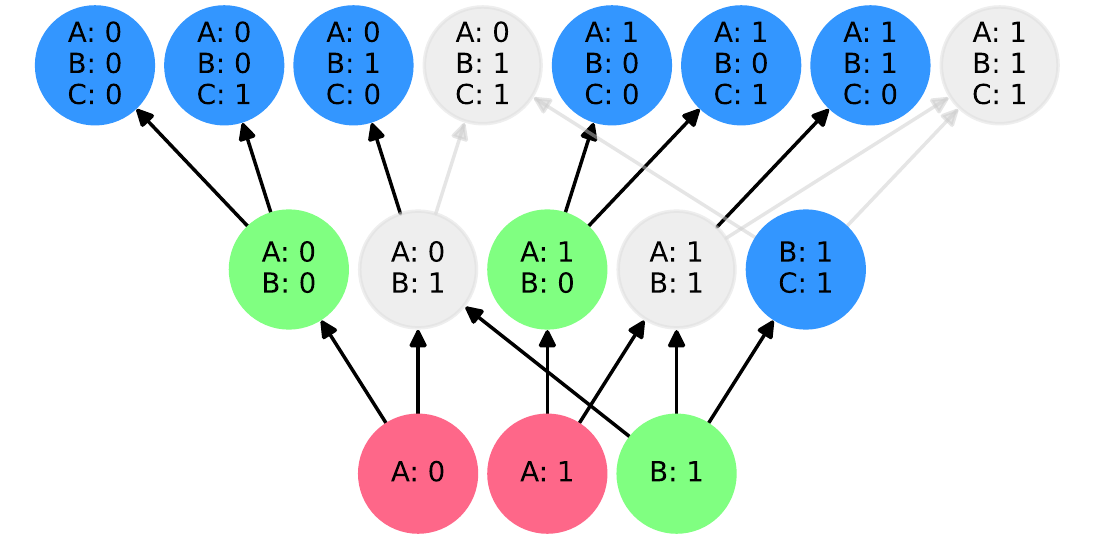}
    \\
    $\Theta_{89}$
    &
    $\Ext{\Theta_{89}}$
    \end{tabular}
\end{center}

\noindent The standard causaltope for Space 89 has dimension 39.
Below is a plot of the homogeneous linear system of causality and quasi-normalisation equations for the standard causaltope, put in reduced row echelon form:

\begin{center}
    \includegraphics[width=11cm]{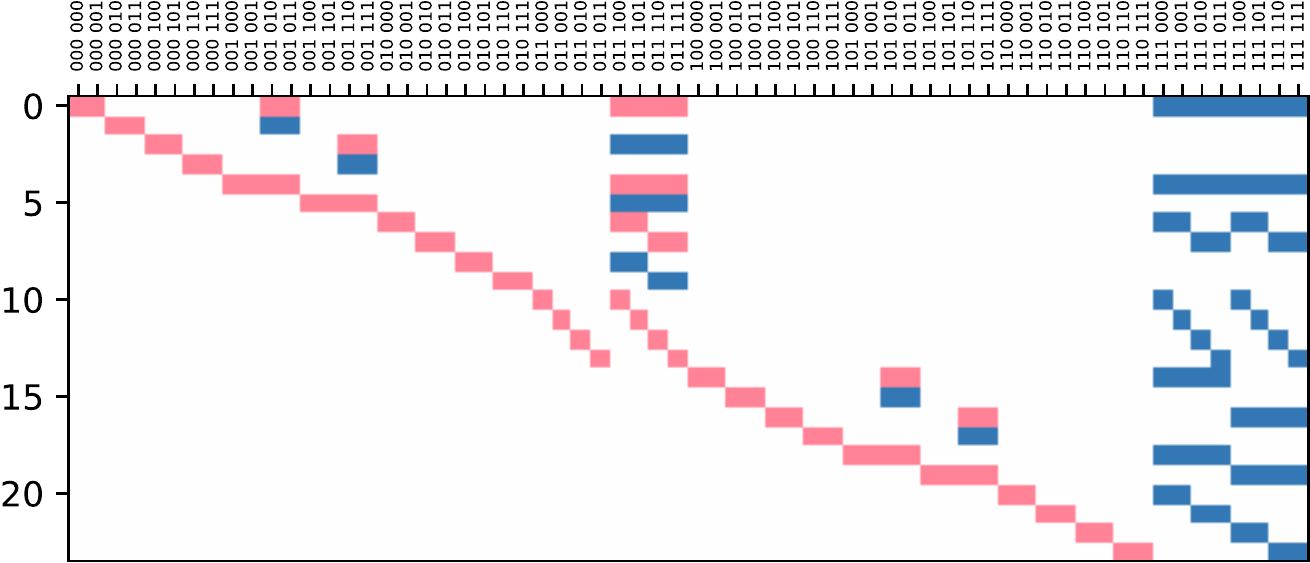}
\end{center}

\noindent Rows correspond to the 24 independent linear equations.
Columns in the plot correspond to entries of empirical models, indexed as $i_A i_B i_C$ $o_A o_B o_C$.
Coefficients in the equations are color-coded as white=0, red=+1 and blue=-1.

Space 89 has closest refinements in equivalence classes 80, 81, 82 and 84; 
it is the join of its (closest) refinements.
It has closest coarsenings in equivalence class 98; 
it does not arise as a nontrivial meet in the hierarchy.
It has 4096 causal functions, 256 of which are not causal for any of its refinements.
It is a tight space.

The standard causaltope for Space 89 has 1 more dimension than that of its subspace in equivalence class 84.
The standard causaltope for Space 89 has 2 dimensions fewer than the meet of the standard causaltopes for its closest coarsenings.
For completeness, below is a plot of the full homogeneous linear system of causality and quasi-normalisation equations for the standard causaltope:

\begin{center}
    \includegraphics[width=12cm]{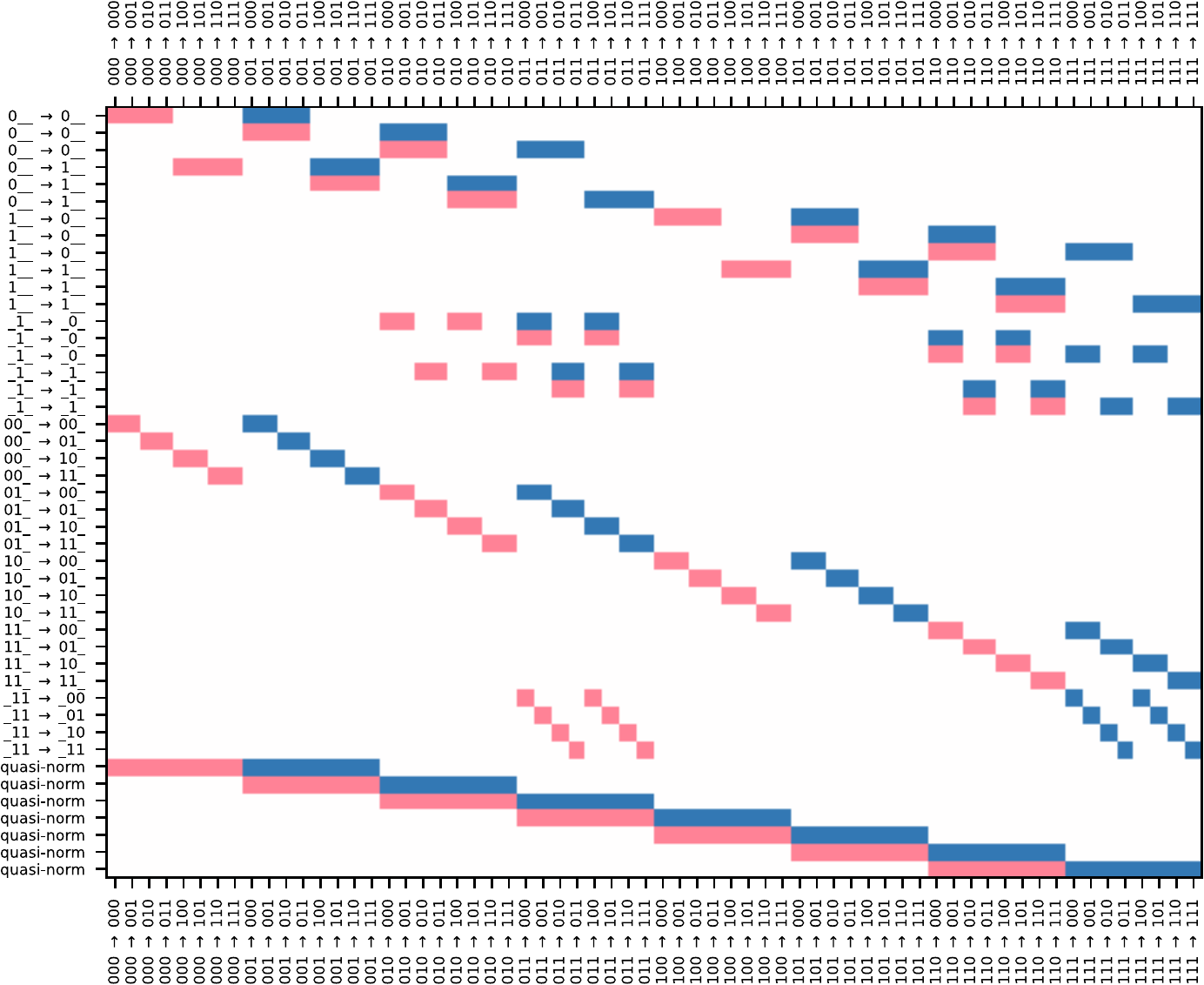}
\end{center}

\noindent Rows correspond to the 45 linear equations, of which 24 are independent.

\newpage
\subsection*{Space 90}

Space 90 is not induced by a causal order, but it is a refinement of the space 100 induced by the definite causal order $\total{\ev{A},\ev{B},\ev{C}}$.
Its equivalence class under event-input permutation symmetry contains 48 spaces.
Space 90 differs as follows from the space induced by causal order $\total{\ev{A},\ev{B},\ev{C}}$:
\begin{itemize}
  \item The outputs at events \evset{\ev{A}, \ev{C}} are independent of the input at event \ev{B} when the inputs at events \evset{A, C} are given by \hist{A/1,C/0}.
  \item The output at event \ev{B} is independent of the input at event \ev{A} when the input at event B is given by \hist{B/1}.
\end{itemize}

\noindent Below are the histories and extended histories for space 90: 
\begin{center}
    \begin{tabular}{cc}
    \includegraphics[height=3.5cm]{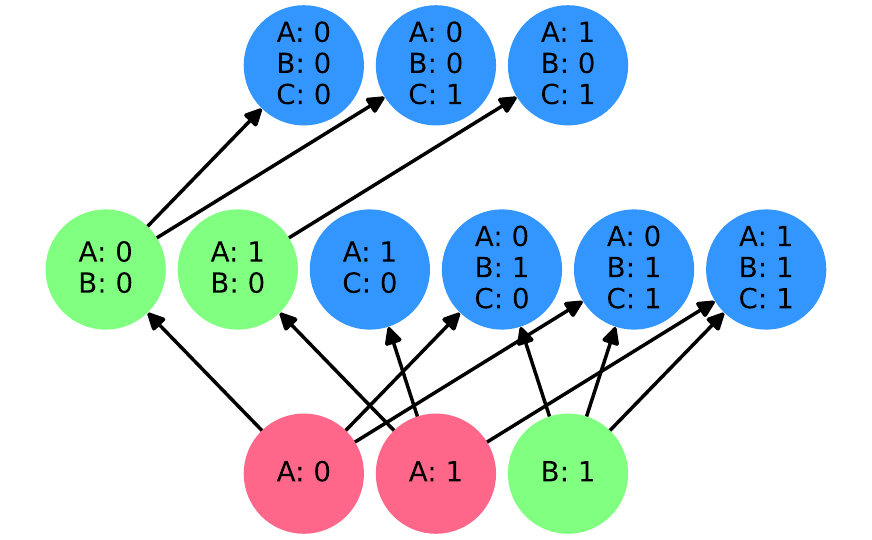}
    &
    \includegraphics[height=3.5cm]{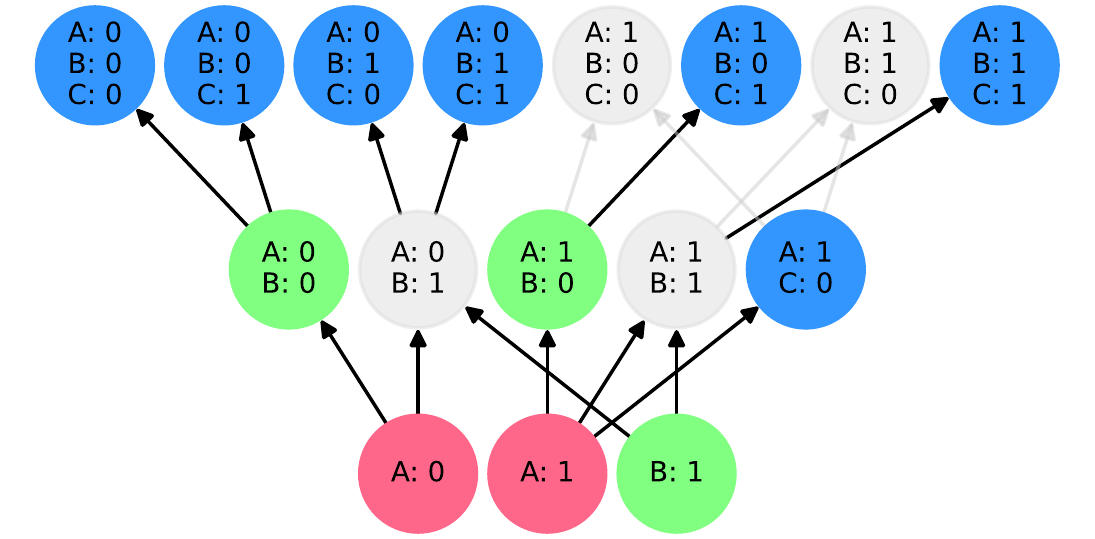}
    \\
    $\Theta_{90}$
    &
    $\Ext{\Theta_{90}}$
    \end{tabular}
\end{center}

\noindent The standard causaltope for Space 90 has dimension 39.
Below is a plot of the homogeneous linear system of causality and quasi-normalisation equations for the standard causaltope, put in reduced row echelon form:

\begin{center}
    \includegraphics[width=11cm]{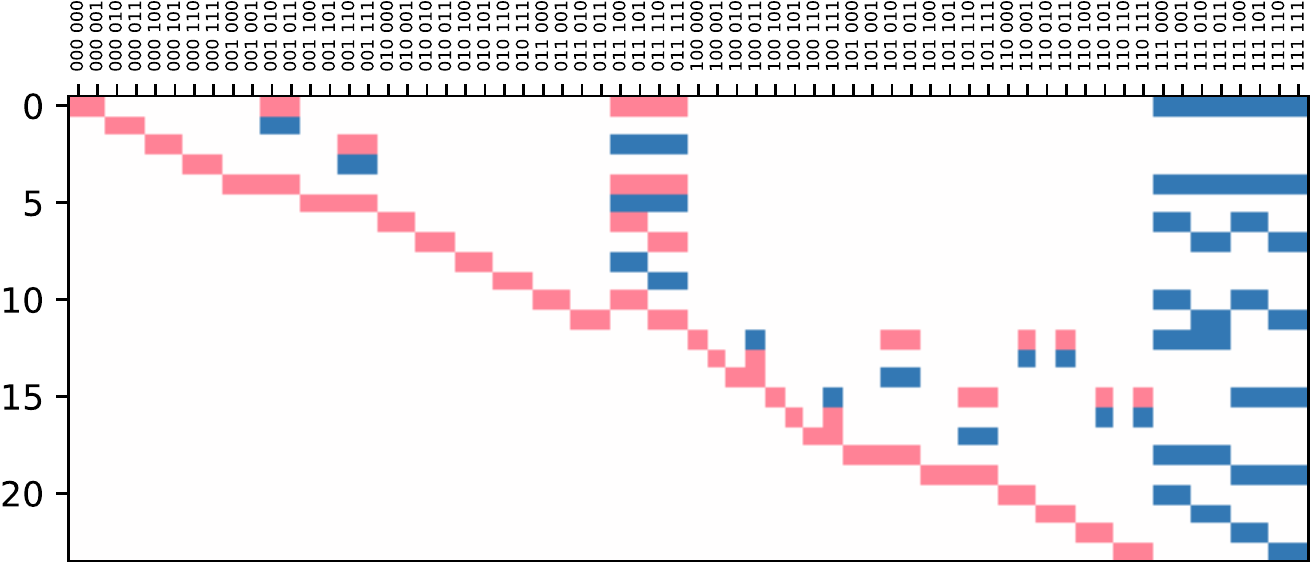}
\end{center}

\noindent Rows correspond to the 24 independent linear equations.
Columns in the plot correspond to entries of empirical models, indexed as $i_A i_B i_C$ $o_A o_B o_C$.
Coefficients in the equations are color-coded as white=0, red=+1 and blue=-1.

Space 90 has closest refinements in equivalence classes 80, 82, 83, 84, 86 and 87; 
it is the join of its (closest) refinements.
It has closest coarsenings in equivalence classes 97 and 98; 
it is the meet of its (closest) coarsenings.
It has 4096 causal functions, 896 of which are not causal for any of its refinements.
It is a tight space.

The standard causaltope for Space 90 has 1 more dimension than that of its subspace in equivalence class 84.
The standard causaltope for Space 90 is the meet of the standard causaltopes for its closest coarsenings.
For completeness, below is a plot of the full homogeneous linear system of causality and quasi-normalisation equations for the standard causaltope:

\begin{center}
    \includegraphics[width=12cm]{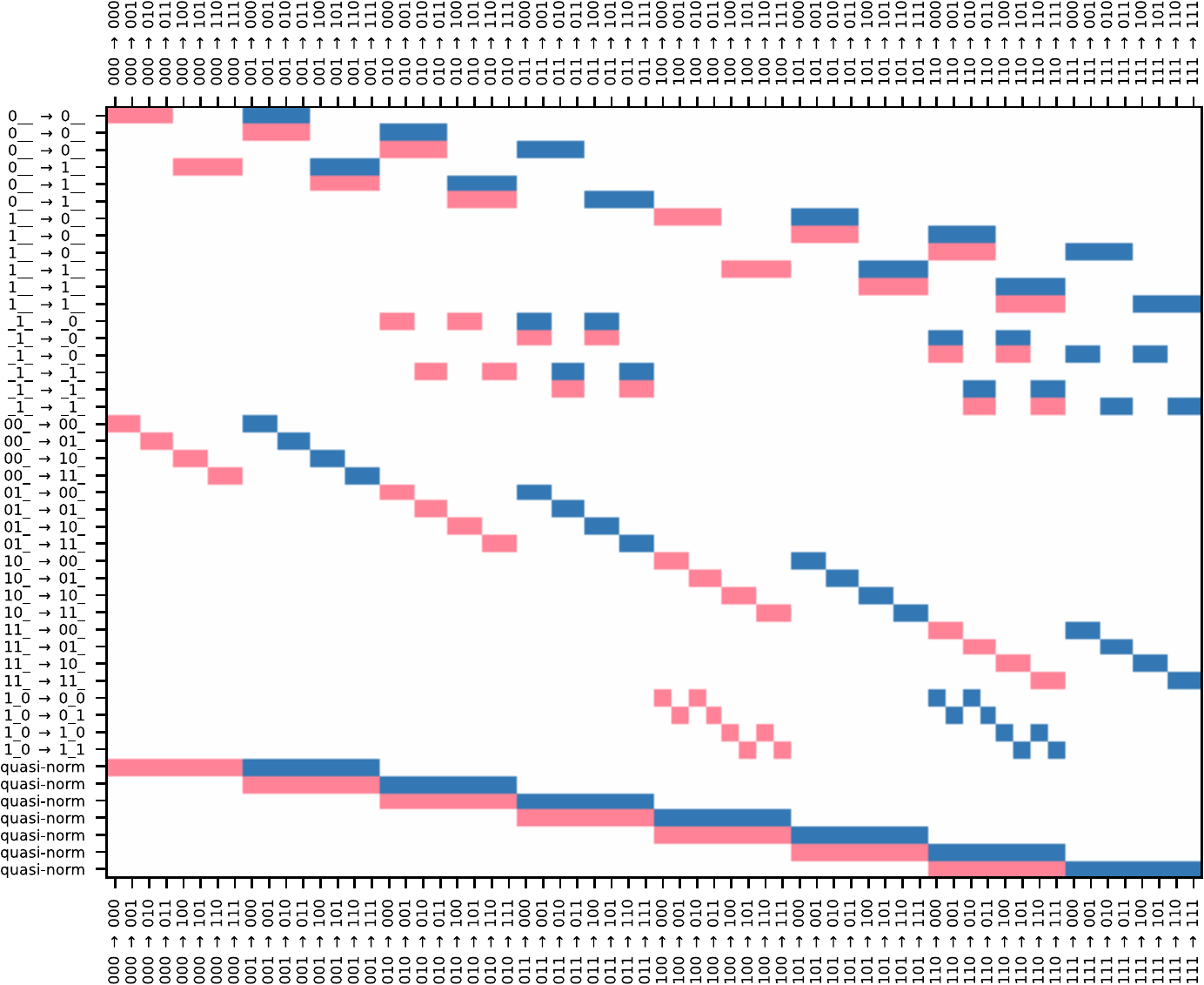}
\end{center}

\noindent Rows correspond to the 45 linear equations, of which 24 are independent.

\newpage
\subsection*{Space 91}

Space 91 is not induced by a causal order, but it is a refinement of the space 100 induced by the definite causal order $\total{\ev{A},\ev{B},\ev{C}}$.
Its equivalence class under event-input permutation symmetry contains 12 spaces.
Space 91 differs as follows from the space induced by causal order $\total{\ev{A},\ev{B},\ev{C}}$:
\begin{itemize}
  \item The outputs at events \evset{\ev{A}, \ev{C}} are independent of the input at event \ev{B} when the inputs at events \evset{A, C} are given by \hist{A/1,C/0} and \hist{A/1,C/1}.
\end{itemize}

\noindent Below are the histories and extended histories for space 91: 
\begin{center}
    \begin{tabular}{cc}
    \includegraphics[height=3.5cm]{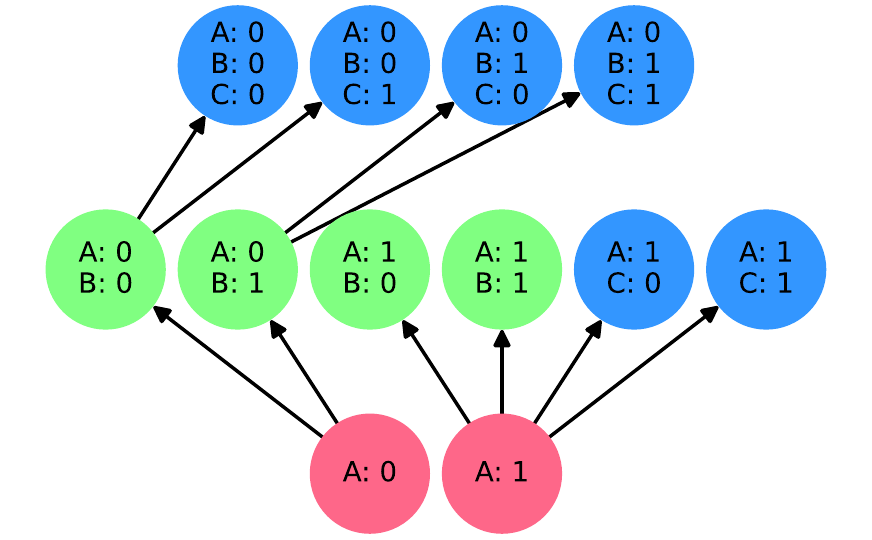}
    &
    \includegraphics[height=3.5cm]{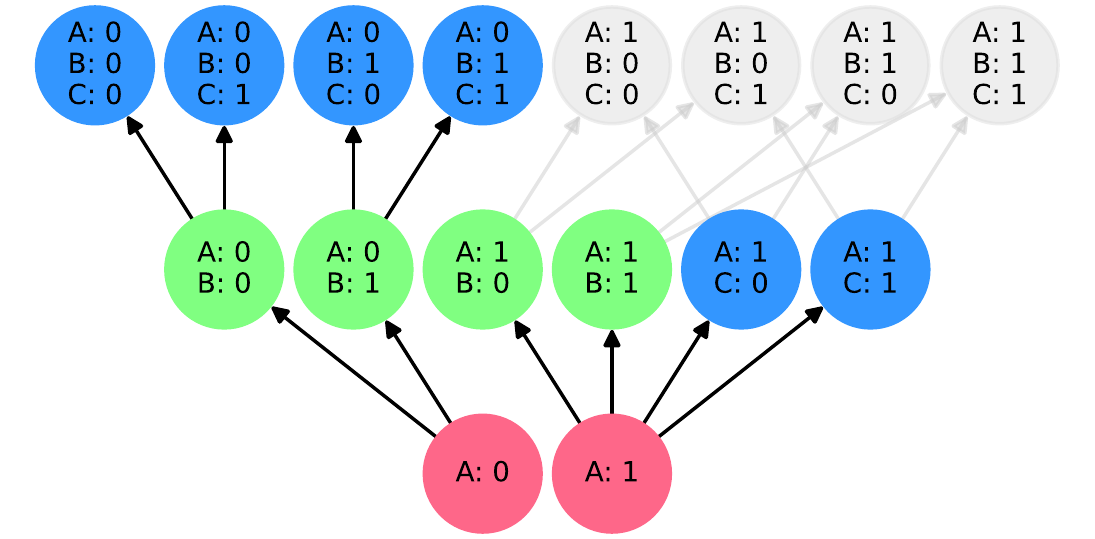}
    \\
    $\Theta_{91}$
    &
    $\Ext{\Theta_{91}}$
    \end{tabular}
\end{center}

\noindent The standard causaltope for Space 91 has dimension 38.
Below is a plot of the homogeneous linear system of causality and quasi-normalisation equations for the standard causaltope, put in reduced row echelon form:

\begin{center}
    \includegraphics[width=11cm]{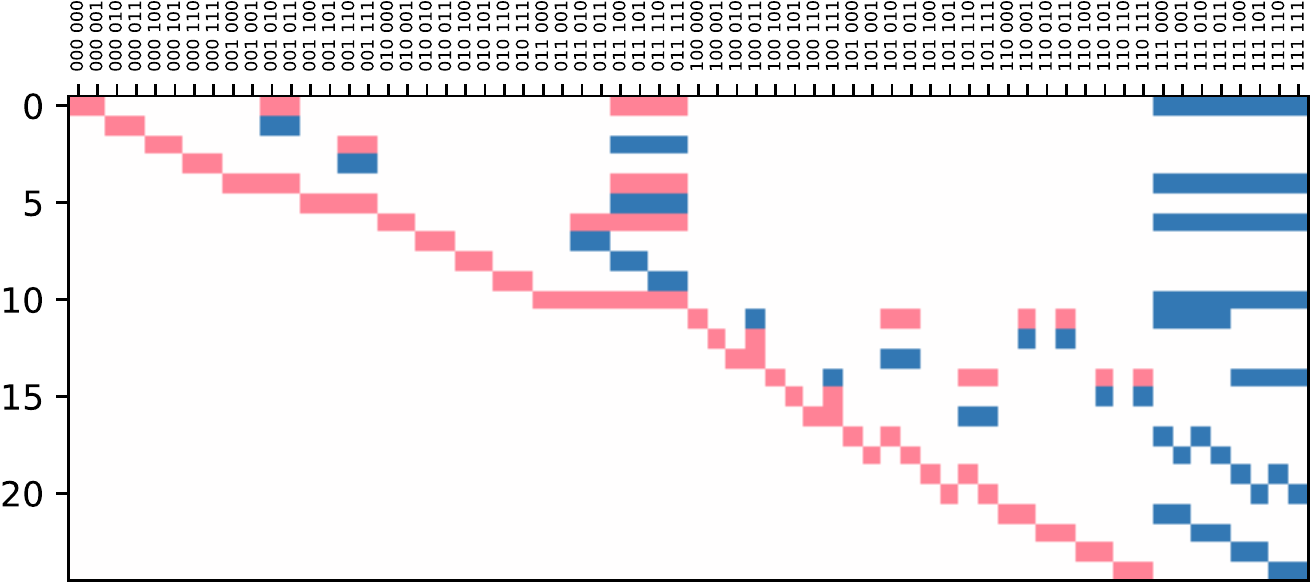}
\end{center}

\noindent Rows correspond to the 25 independent linear equations.
Columns in the plot correspond to entries of empirical models, indexed as $i_A i_B i_C$ $o_A o_B o_C$.
Coefficients in the equations are color-coded as white=0, red=+1 and blue=-1.

Space 91 has closest refinements in equivalence classes 83 and 88; 
it is the join of its (closest) refinements.
It has closest coarsenings in equivalence classes 97 and 99; 
it is the meet of its (closest) coarsenings.
It has 4096 causal functions, 576 of which are not causal for any of its refinements.
It is a tight space.

The standard causaltope for Space 91 has 1 more dimension than those of its 2 subspaces in equivalence class 83.
The standard causaltope for Space 91 is the meet of the standard causaltopes for its closest coarsenings.
For completeness, below is a plot of the full homogeneous linear system of causality and quasi-normalisation equations for the standard causaltope:

\begin{center}
    \includegraphics[width=12cm]{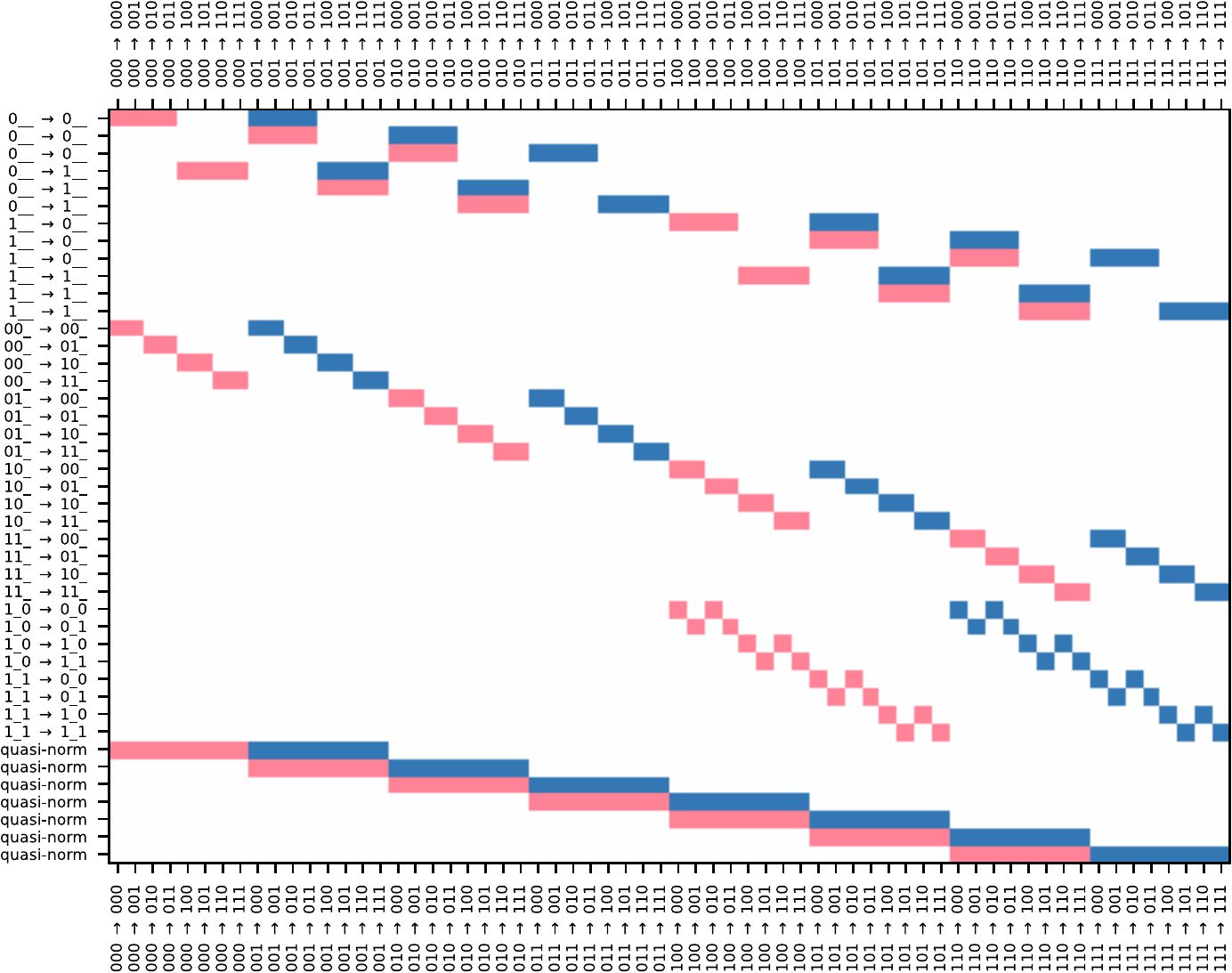}
\end{center}

\noindent Rows correspond to the 43 linear equations, of which 25 are independent.

\newpage
\subsection*{Space 92}

Space 92 is induced by the definite causal order $\total{\ev{A},\ev{C}}\vee\total{\ev{B},\ev{C}}$.
Its equivalence class under event-input permutation symmetry contains 3 spaces.

\noindent Below are the histories and extended histories for space 92: 
\begin{center}
    \begin{tabular}{cc}
    \includegraphics[height=3.5cm]{svg-inkscape/space-ABC-unique-tight-92-highlighted_svg-tex.pdf}
    &
    \includegraphics[height=3.5cm]{svg-inkscape/space-ABC-unique-tight-92-ext-highlighted_svg-tex.pdf}
    \\
    $\Theta_{92}$
    &
    $\Ext{\Theta_{92}}$
    \end{tabular}
\end{center}

\noindent The standard causaltope for Space 92 has dimension 40.
Below is a plot of the homogeneous linear system of causality and quasi-normalisation equations for the standard causaltope, put in reduced row echelon form:

\begin{center}
    \includegraphics[width=11cm]{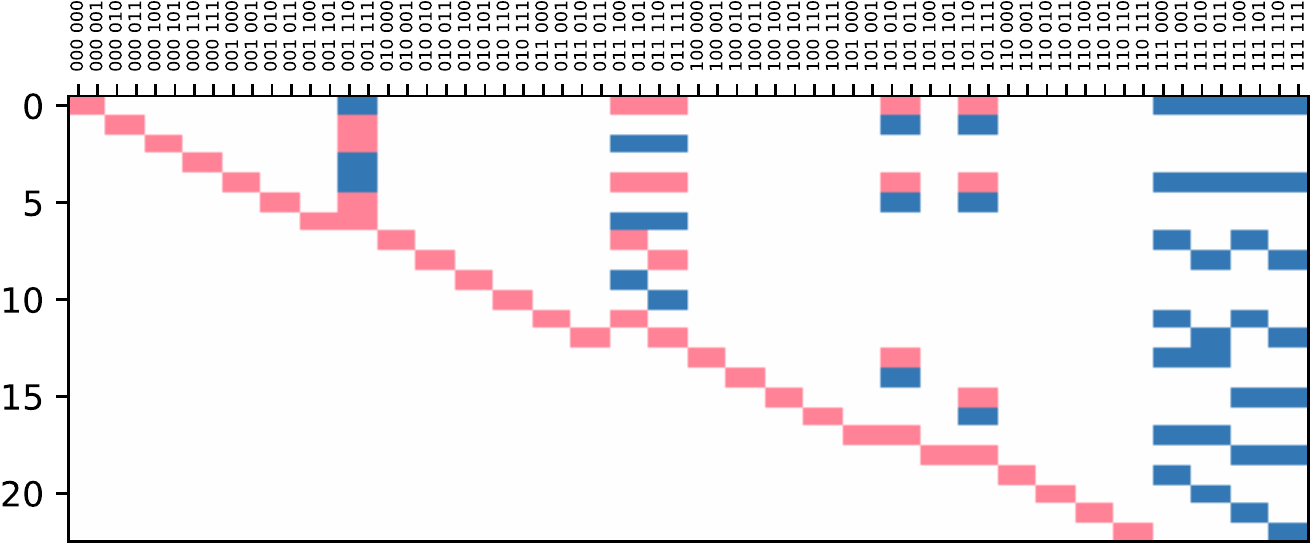}
\end{center}

\noindent Rows correspond to the 23 independent linear equations.
Columns in the plot correspond to entries of empirical models, indexed as $i_A i_B i_C$ $o_A o_B o_C$.
Coefficients in the equations are color-coded as white=0, red=+1 and blue=-1.

Space 92 has closest refinements in equivalence class 84; 
it is the join of its (closest) refinements.
It has closest coarsenings in equivalence class 98; 
it is the meet of its (closest) coarsenings.
It has 4096 causal functions, 512 of which are not causal for any of its refinements.
It is a tight space.

The standard causaltope for Space 92 has 2 more dimensions than those of its 8 subspaces in equivalence class 84.
The standard causaltope for Space 92 is the meet of the standard causaltopes for its closest coarsenings.
For completeness, below is a plot of the full homogeneous linear system of causality and quasi-normalisation equations for the standard causaltope:

\begin{center}
    \includegraphics[width=12cm]{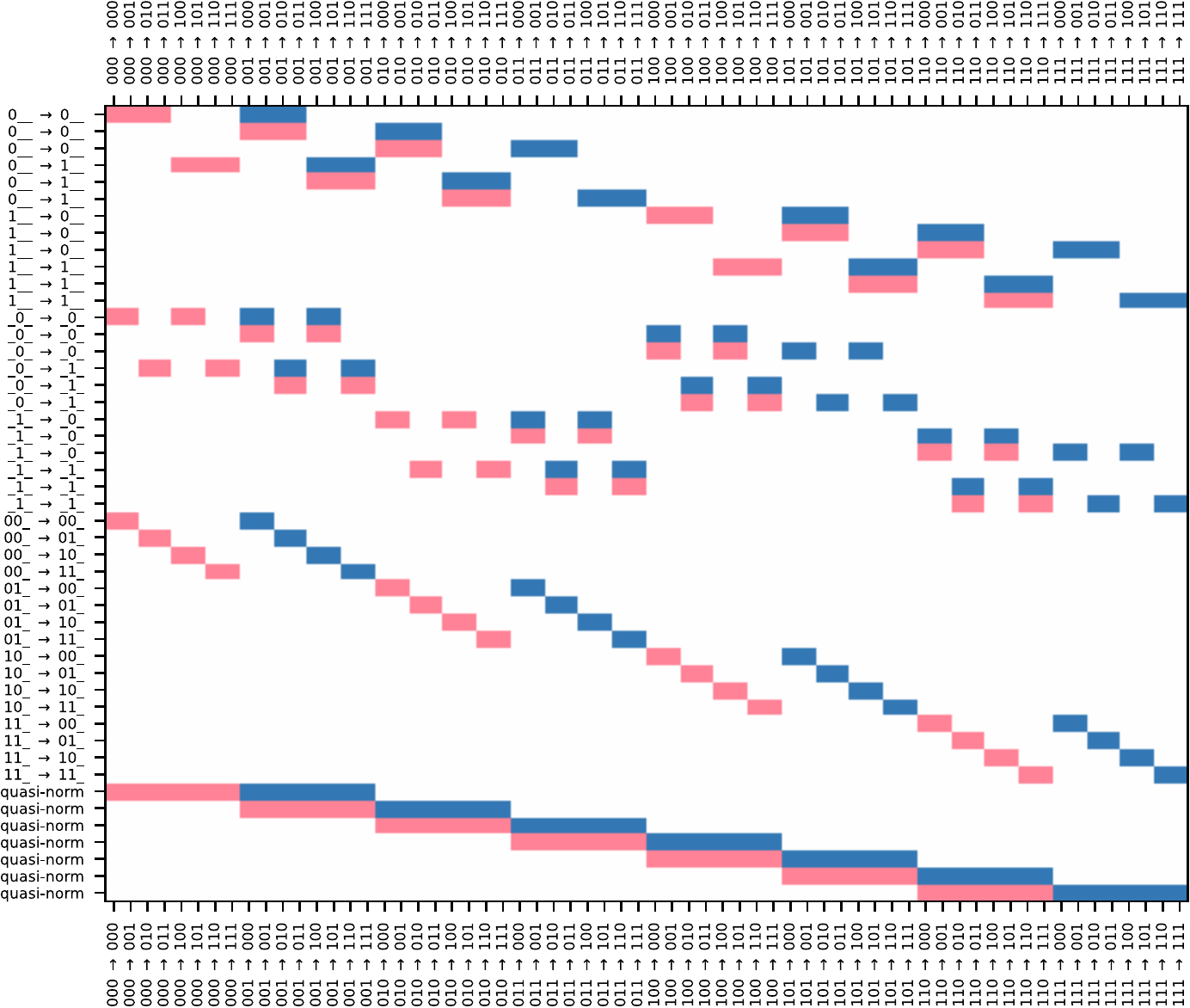}
\end{center}

\noindent Rows correspond to the 47 linear equations, of which 23 are independent.

\newpage
\subsection*{Space 93}

Space 93 is not induced by a causal order, but it is a refinement of the space induced by the indefinite causal order $\total{\ev{A},\{\ev{B},\ev{C}\}}$.
Its equivalence class under event-input permutation symmetry contains 24 spaces.
Space 93 differs as follows from the space induced by causal order $\total{\ev{A},\{\ev{B},\ev{C}\}}$:
\begin{itemize}
  \item The outputs at events \evset{\ev{A}, \ev{B}} are independent of the input at event \ev{C} when the inputs at events \evset{A, B} are given by \hist{A/0,B/0}, \hist{A/0,B/1} and \hist{A/1,B/0}.
  \item The output at event \ev{B} is independent of the inputs at events \evset{\ev{A}, \ev{C}} when the input at event B is given by \hist{B/0}.
  \item The outputs at events \evset{\ev{A}, \ev{C}} are independent of the input at event \ev{B} when the inputs at events \evset{A, C} are given by \hist{A/1,C/0} and \hist{A/1,C/1}.
\end{itemize}

\noindent Below are the histories and extended histories for space 93: 
\begin{center}
    \begin{tabular}{cc}
    \includegraphics[height=3.5cm]{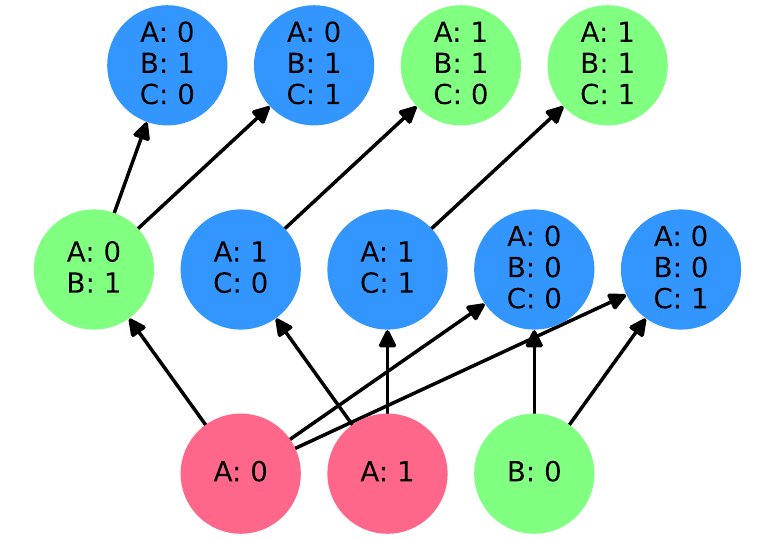}
    &
    \includegraphics[height=3.5cm]{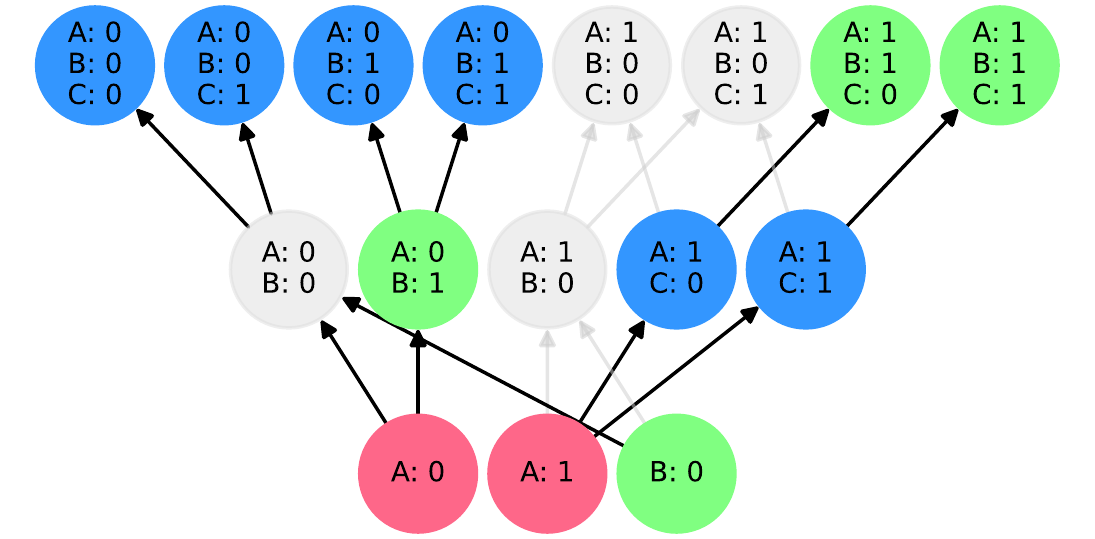}
    \\
    $\Theta_{93}$
    &
    $\Ext{\Theta_{93}}$
    \end{tabular}
\end{center}

\noindent The standard causaltope for Space 93 has dimension 39.
Below is a plot of the homogeneous linear system of causality and quasi-normalisation equations for the standard causaltope, put in reduced row echelon form:

\begin{center}
    \includegraphics[width=11cm]{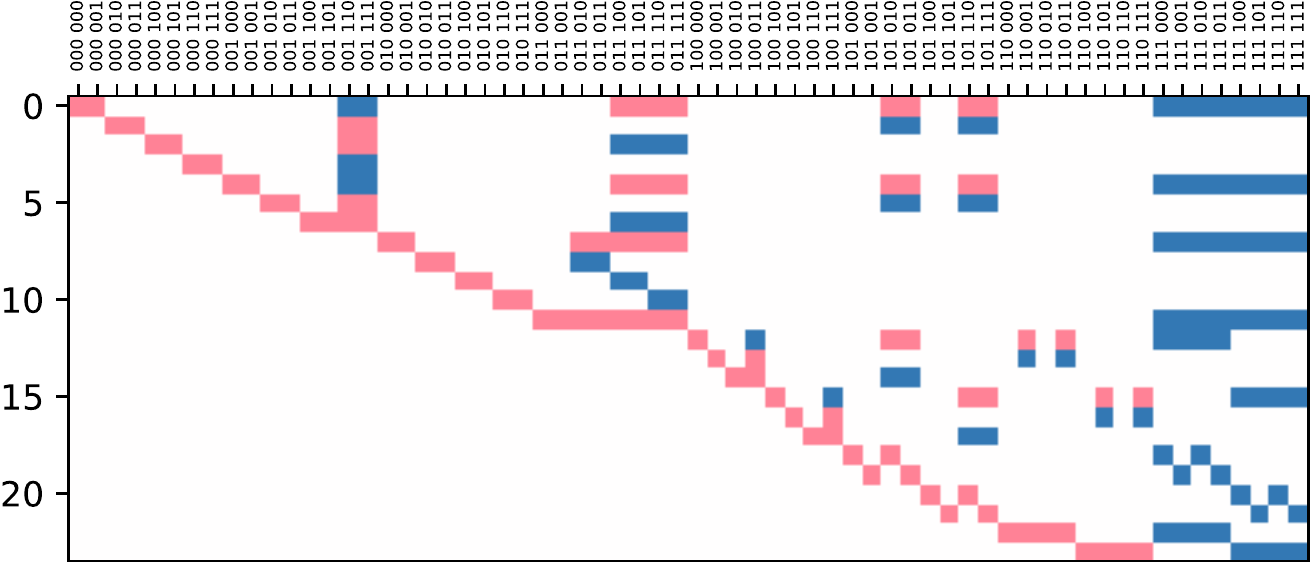}
\end{center}

\noindent Rows correspond to the 24 independent linear equations.
Columns in the plot correspond to entries of empirical models, indexed as $i_A i_B i_C$ $o_A o_B o_C$.
Coefficients in the equations are color-coded as white=0, red=+1 and blue=-1.

Space 93 has closest refinements in equivalence classes 79, 83 and 85; 
it is the join of its (closest) refinements.
It has closest coarsenings in equivalence class 99; 
it does not arise as a nontrivial meet in the hierarchy.
It has 4096 causal functions, 896 of which are not causal for any of its refinements.
It is a tight space.

The standard causaltope for Space 93 has 2 more dimensions than those of its 5 subspaces in equivalence classes 79, 83 and 85.
The standard causaltope for Space 93 has 1 dimension fewer than the meet of the standard causaltopes for its closest coarsenings.
For completeness, below is a plot of the full homogeneous linear system of causality and quasi-normalisation equations for the standard causaltope:

\begin{center}
    \includegraphics[width=12cm]{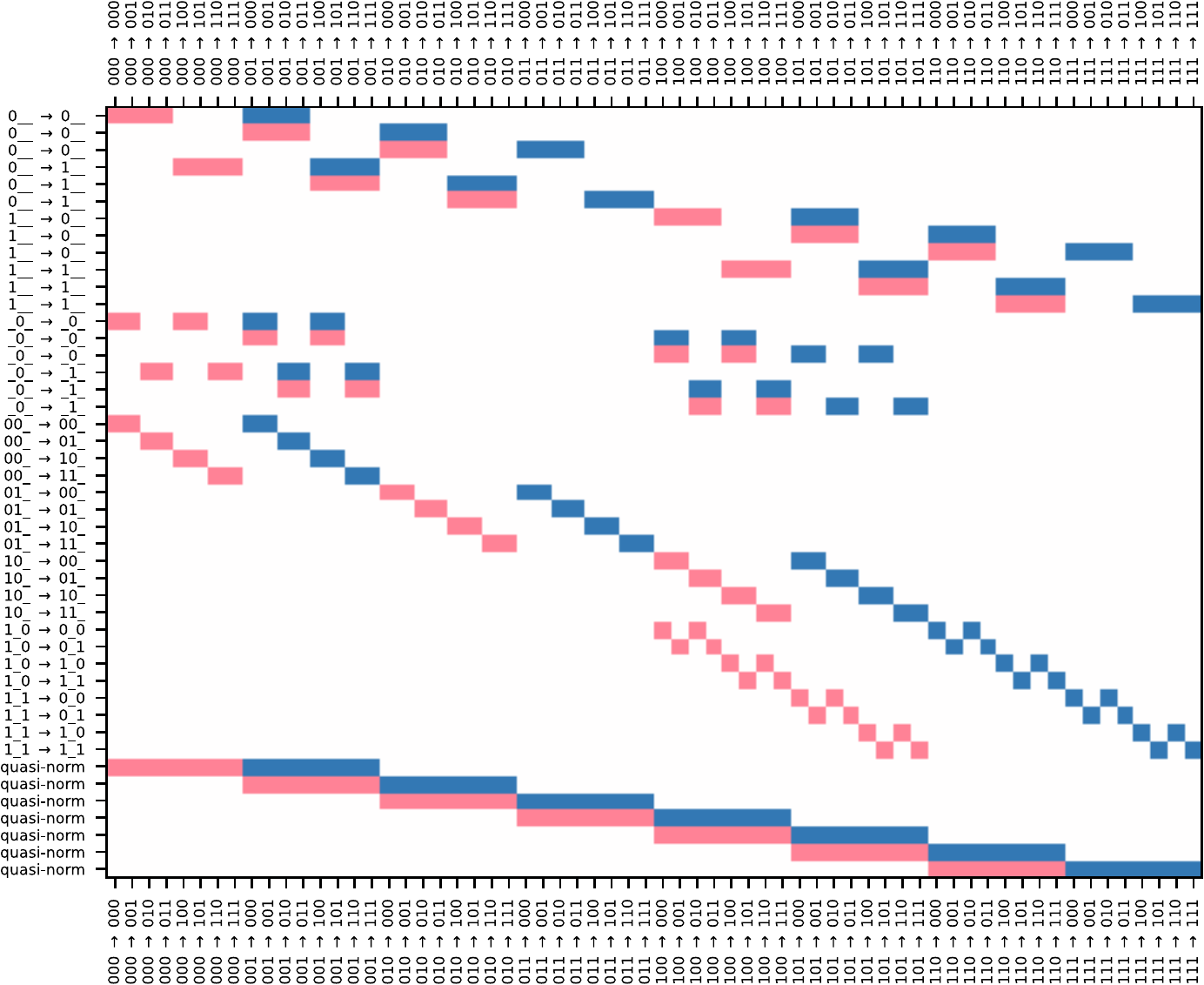}
\end{center}

\noindent Rows correspond to the 45 linear equations, of which 24 are independent.

\newpage
\subsection*{Space 94}

Space 94 is not induced by a causal order, but it is a refinement of the space 100 induced by the definite causal order $\total{\ev{A},\ev{B},\ev{C}}$.
Its equivalence class under event-input permutation symmetry contains 12 spaces.
Space 94 differs as follows from the space induced by causal order $\total{\ev{A},\ev{B},\ev{C}}$:
\begin{itemize}
  \item The outputs at events \evset{\ev{A}, \ev{C}} are independent of the input at event \ev{B} when the inputs at events \evset{A, C} are given by \hist{A/0,C/1} and \hist{A/1,C/1}.
\end{itemize}

\noindent Below are the histories and extended histories for space 94: 
\begin{center}
    \begin{tabular}{cc}
    \includegraphics[height=3.5cm]{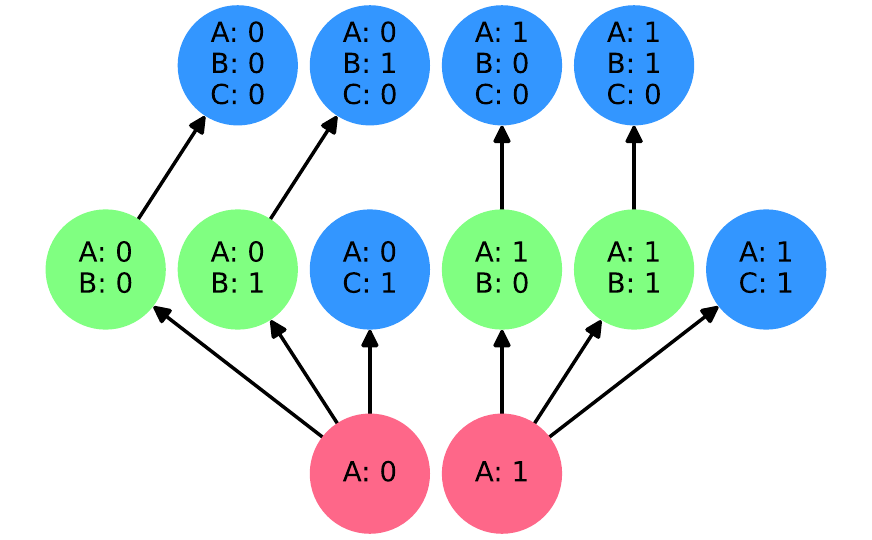}
    &
    \includegraphics[height=3.5cm]{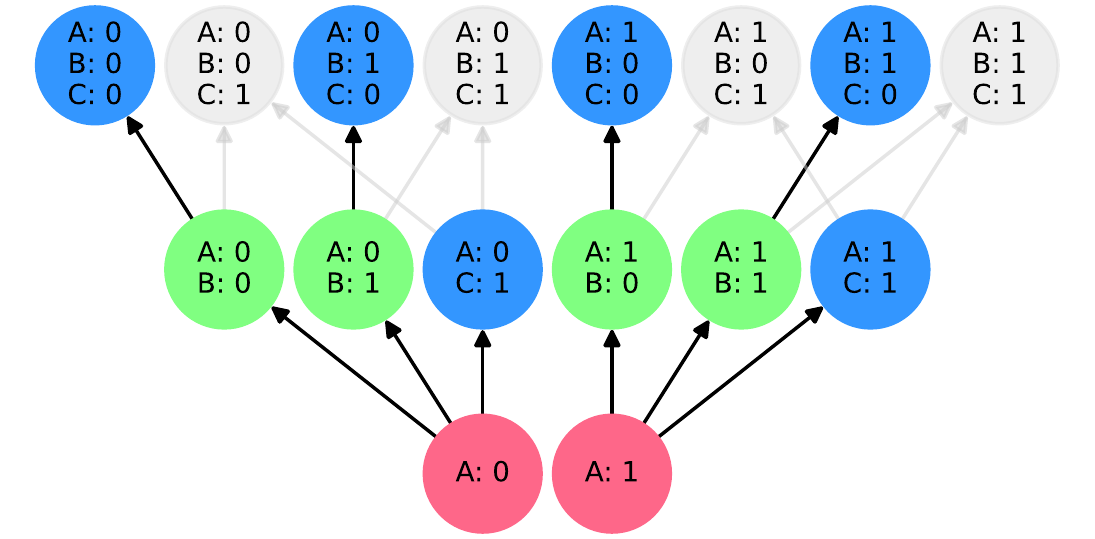}
    \\
    $\Theta_{94}$
    &
    $\Ext{\Theta_{94}}$
    \end{tabular}
\end{center}

\noindent The standard causaltope for Space 94 has dimension 38.
Below is a plot of the homogeneous linear system of causality and quasi-normalisation equations for the standard causaltope, put in reduced row echelon form:

\begin{center}
    \includegraphics[width=11cm]{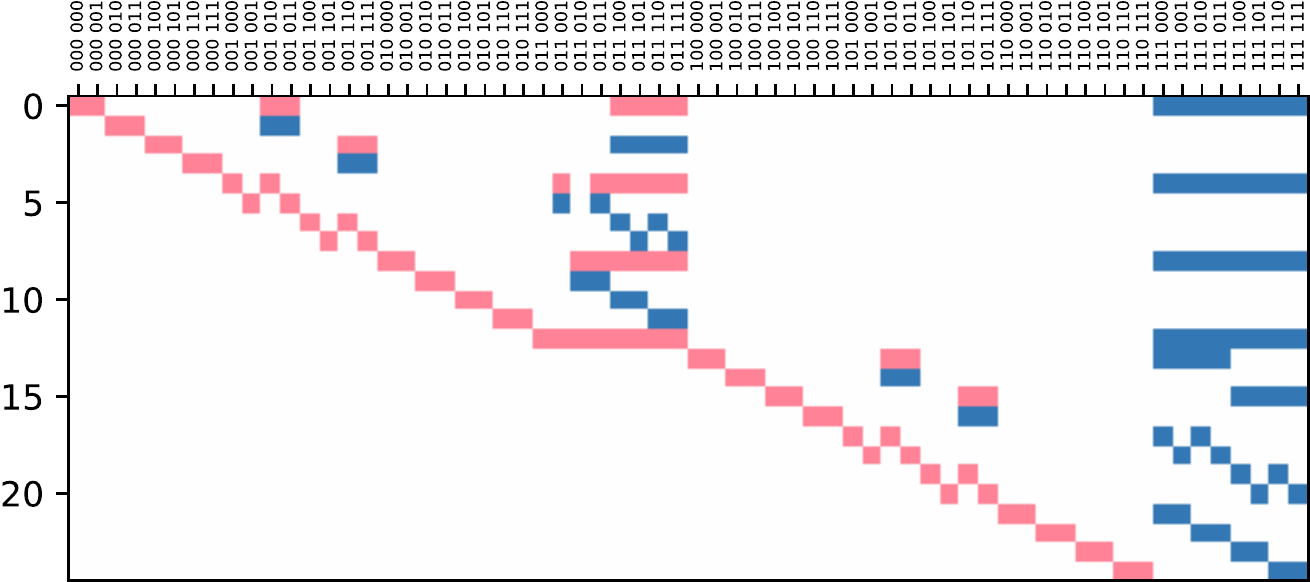}
\end{center}

\noindent Rows correspond to the 25 independent linear equations.
Columns in the plot correspond to entries of empirical models, indexed as $i_A i_B i_C$ $o_A o_B o_C$.
Coefficients in the equations are color-coded as white=0, red=+1 and blue=-1.

Space 94 has closest refinements in equivalence classes 78, 86 and 88; 
it is the join of its (closest) refinements.
It has closest coarsenings in equivalence class 97; 
it is the meet of its (closest) coarsenings.
It has 4096 causal functions, 640 of which are not causal for any of its refinements.
It is a tight space.

The standard causaltope for Space 94 has 1 more dimension than those of its 3 subspaces in equivalence classes 78 and 86.
The standard causaltope for Space 94 is the meet of the standard causaltopes for its closest coarsenings.
For completeness, below is a plot of the full homogeneous linear system of causality and quasi-normalisation equations for the standard causaltope:

\begin{center}
    \includegraphics[width=12cm]{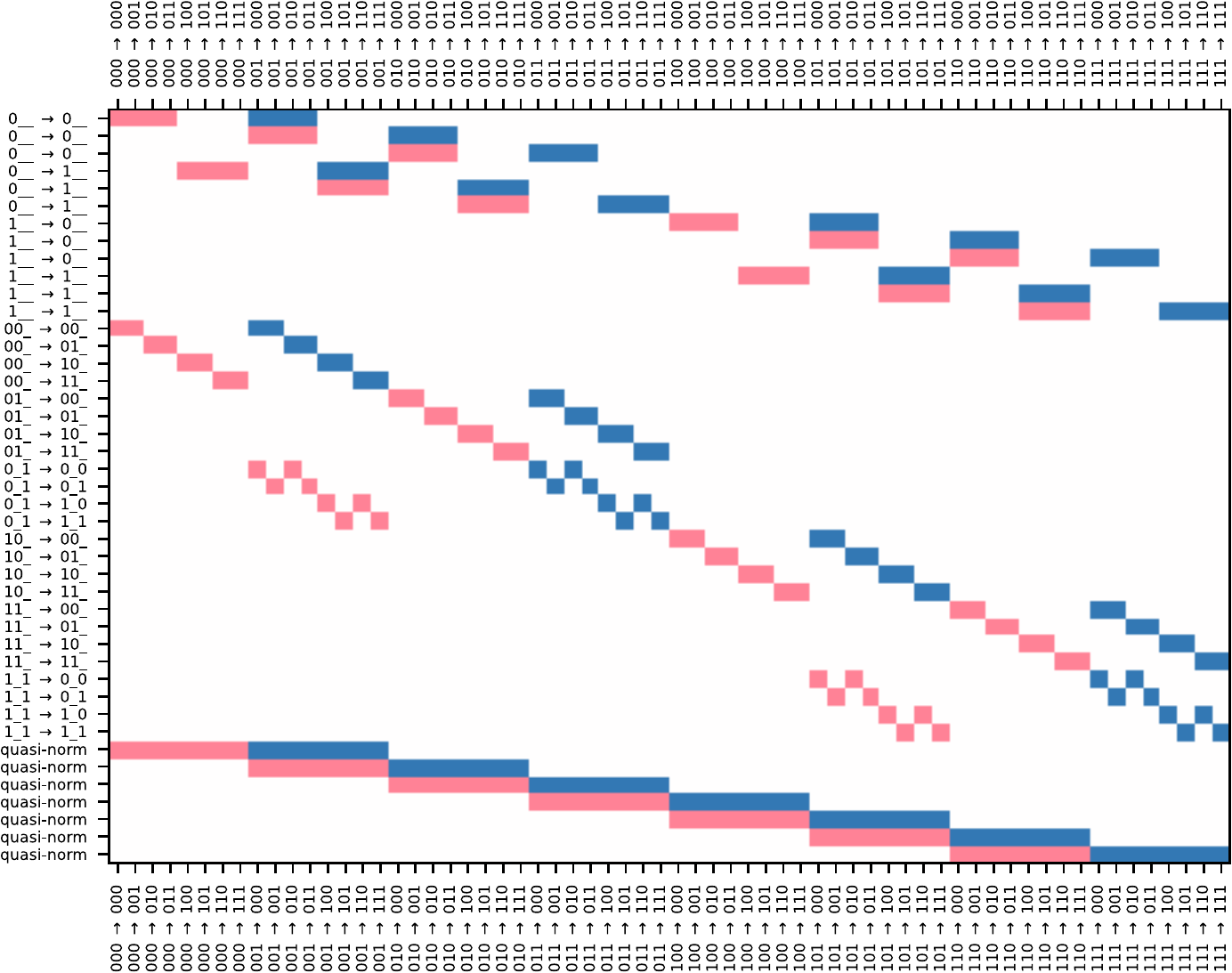}
\end{center}

\noindent Rows correspond to the 43 linear equations, of which 25 are independent.

\newpage
\subsection*{Space 95}

Space 95 is not induced by a causal order, but it is a refinement of the space 100 induced by the definite causal order $\total{\ev{A},\ev{B},\ev{C}}$.
Its equivalence class under event-input permutation symmetry contains 12 spaces.
Space 95 differs as follows from the space induced by causal order $\total{\ev{A},\ev{B},\ev{C}}$:
\begin{itemize}
  \item The outputs at events \evset{\ev{A}, \ev{C}} are independent of the input at event \ev{B} when the inputs at events \evset{A, C} are given by \hist{A/0,C/1} and \hist{A/1,C/0}.
\end{itemize}

\noindent Below are the histories and extended histories for space 95: 
\begin{center}
    \begin{tabular}{cc}
    \includegraphics[height=3.5cm]{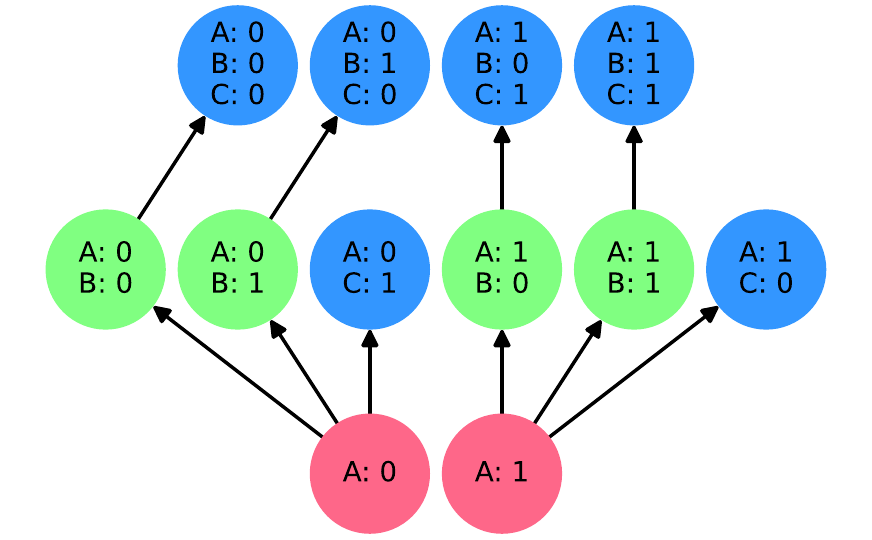}
    &
    \includegraphics[height=3.5cm]{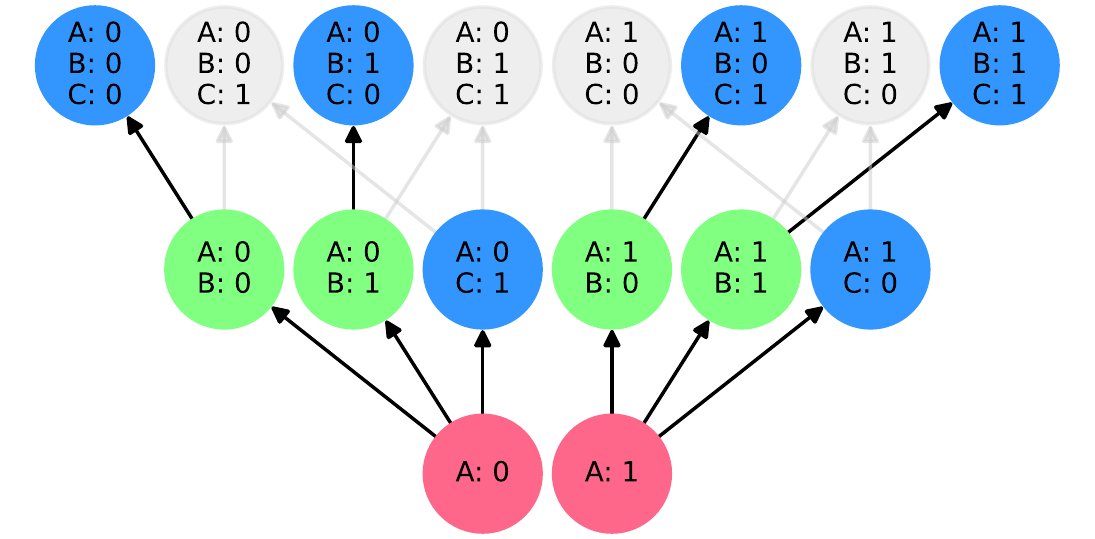}
    \\
    $\Theta_{95}$
    &
    $\Ext{\Theta_{95}}$
    \end{tabular}
\end{center}

\noindent The standard causaltope for Space 95 has dimension 38.
Below is a plot of the homogeneous linear system of causality and quasi-normalisation equations for the standard causaltope, put in reduced row echelon form:

\begin{center}
    \includegraphics[width=11cm]{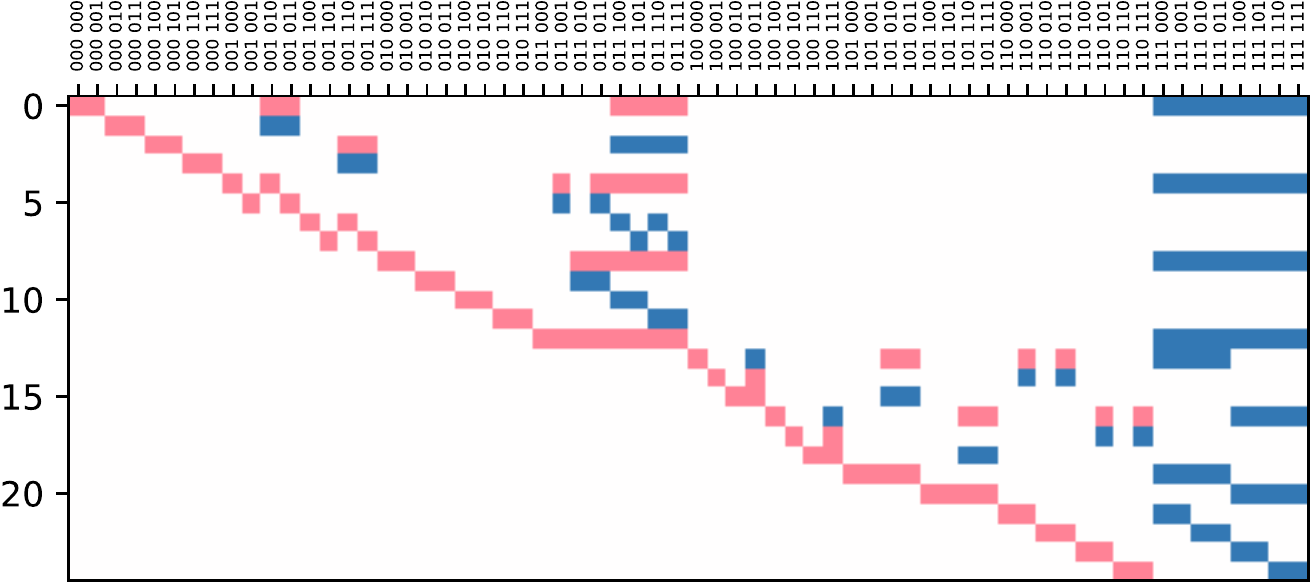}
\end{center}

\noindent Rows correspond to the 25 independent linear equations.
Columns in the plot correspond to entries of empirical models, indexed as $i_A i_B i_C$ $o_A o_B o_C$.
Coefficients in the equations are color-coded as white=0, red=+1 and blue=-1.

Space 95 has closest refinements in equivalence classes 87 and 88; 
it is the join of its (closest) refinements.
It has closest coarsenings in equivalence class 97; 
it is the meet of its (closest) coarsenings.
It has 4096 causal functions, 1024 of which are not causal for any of its refinements.
It is a tight space.

The standard causaltope for Space 95 has 1 more dimension than those of its 2 subspaces in equivalence class 87.
The standard causaltope for Space 95 is the meet of the standard causaltopes for its closest coarsenings.
For completeness, below is a plot of the full homogeneous linear system of causality and quasi-normalisation equations for the standard causaltope:

\begin{center}
    \includegraphics[width=12cm]{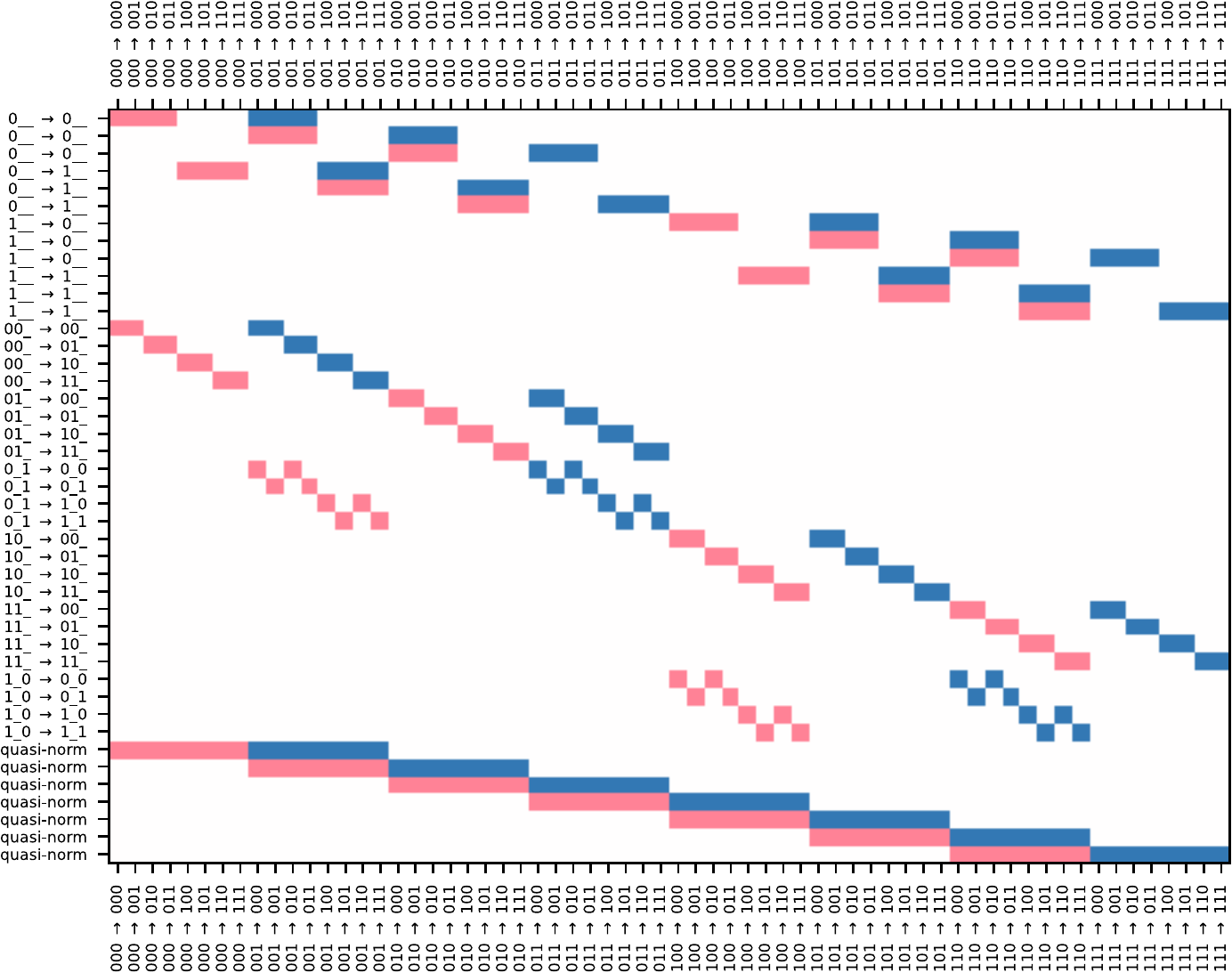}
\end{center}

\noindent Rows correspond to the 43 linear equations, of which 25 are independent.

\newpage
\subsection*{Space 96}

Space 96 is not induced by a causal order, but it is a refinement of the space induced by the indefinite causal order $\total{\ev{A},\{\ev{B},\ev{C}\}}$.
Its equivalence class under event-input permutation symmetry contains 24 spaces.
Space 96 differs as follows from the space induced by causal order $\total{\ev{A},\{\ev{B},\ev{C}\}}$:
\begin{itemize}
  \item The outputs at events \evset{\ev{A}, \ev{B}} are independent of the input at event \ev{C} when the inputs at events \evset{A, B} are given by \hist{A/0,B/0}, \hist{A/0,B/1} and \hist{A/1,B/0}.
  \item The outputs at events \evset{\ev{A}, \ev{C}} are independent of the input at event \ev{B} when the inputs at events \evset{A, C} are given by \hist{A/0,C/1}, \hist{A/1,C/0} and \hist{A/1,C/1}.
\end{itemize}

\noindent Below are the histories and extended histories for space 96: 
\begin{center}
    \begin{tabular}{cc}
    \includegraphics[height=3.5cm]{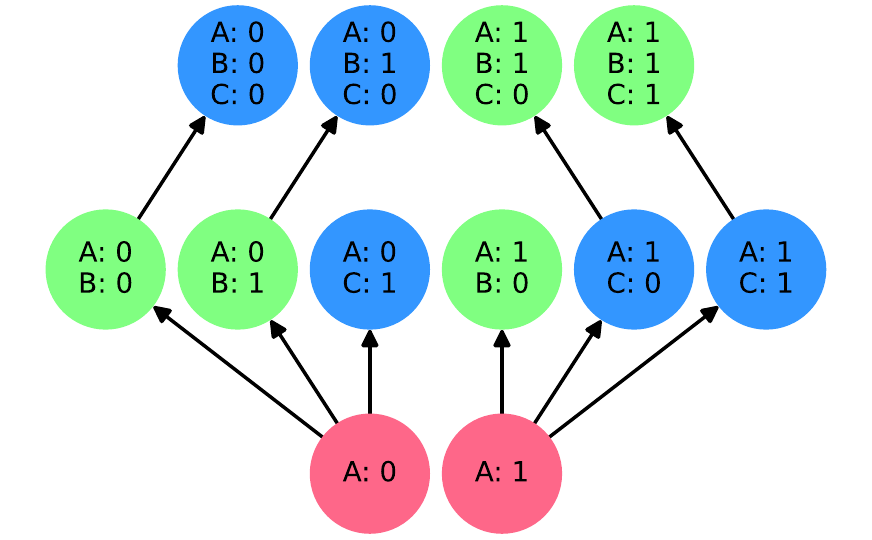}
    &
    \includegraphics[height=3.5cm]{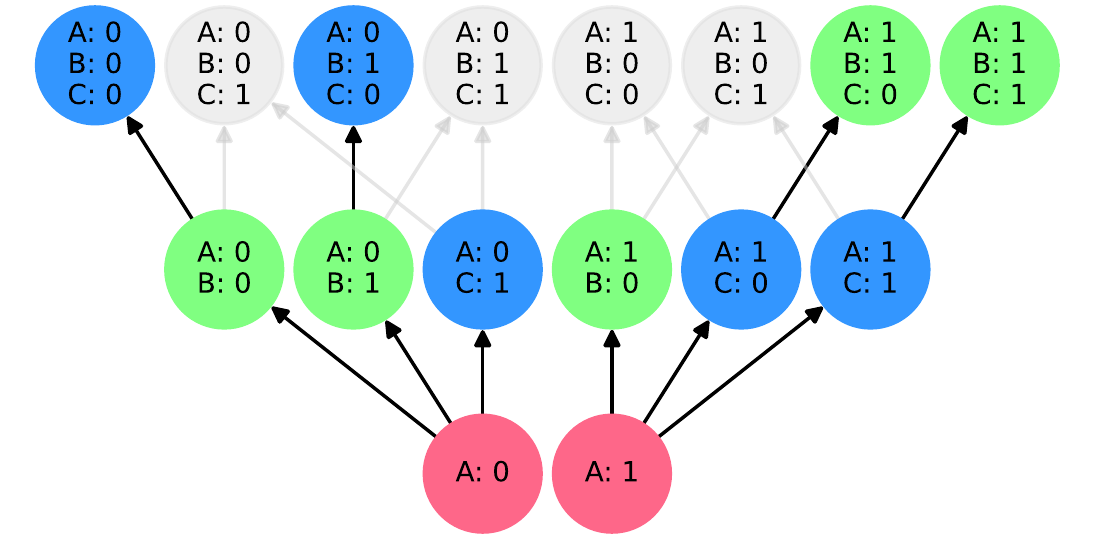}
    \\
    $\Theta_{96}$
    &
    $\Ext{\Theta_{96}}$
    \end{tabular}
\end{center}

\noindent The standard causaltope for Space 96 has dimension 38.
Below is a plot of the homogeneous linear system of causality and quasi-normalisation equations for the standard causaltope, put in reduced row echelon form:

\begin{center}
    \includegraphics[width=11cm]{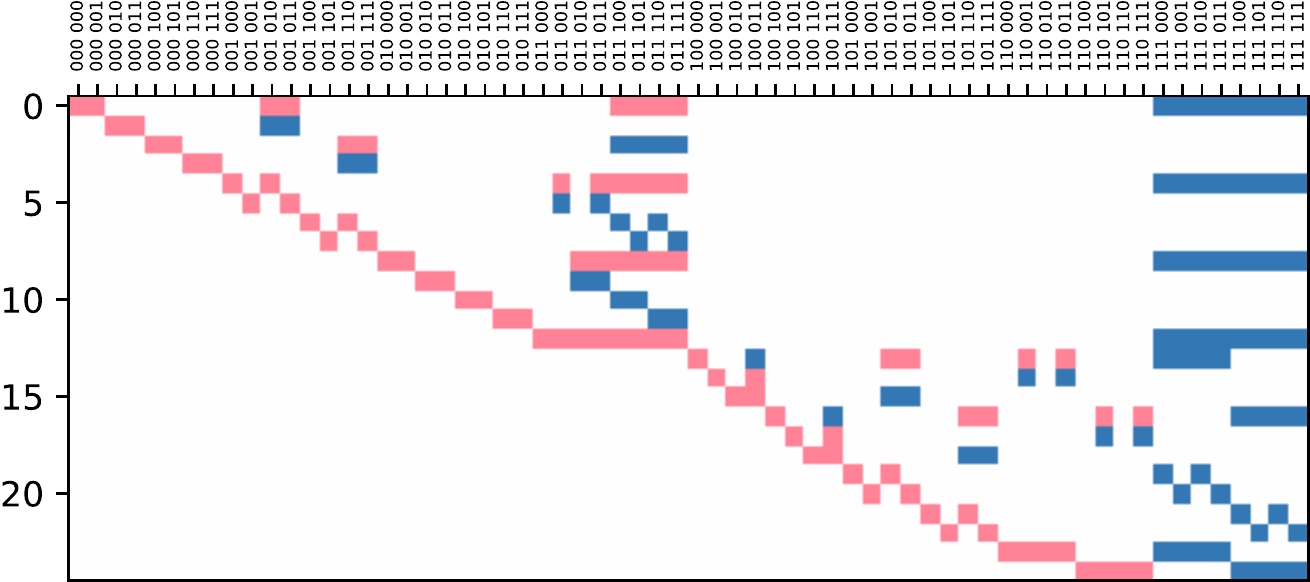}
\end{center}

\noindent Rows correspond to the 25 independent linear equations.
Columns in the plot correspond to entries of empirical models, indexed as $i_A i_B i_C$ $o_A o_B o_C$.
Coefficients in the equations are color-coded as white=0, red=+1 and blue=-1.

Space 96 has closest refinements in equivalence classes 79 and 88; 
it is the join of its (closest) refinements.
It has closest coarsenings in equivalence class 99; 
it is the meet of its (closest) coarsenings.
It has 4096 causal functions, 384 of which are not causal for any of its refinements.
It is a tight space.

The standard causaltope for Space 96 has 1 more dimension than those of its 2 subspaces in equivalence class 79.
The standard causaltope for Space 96 is the meet of the standard causaltopes for its closest coarsenings.
For completeness, below is a plot of the full homogeneous linear system of causality and quasi-normalisation equations for the standard causaltope:

\begin{center}
    \includegraphics[width=12cm]{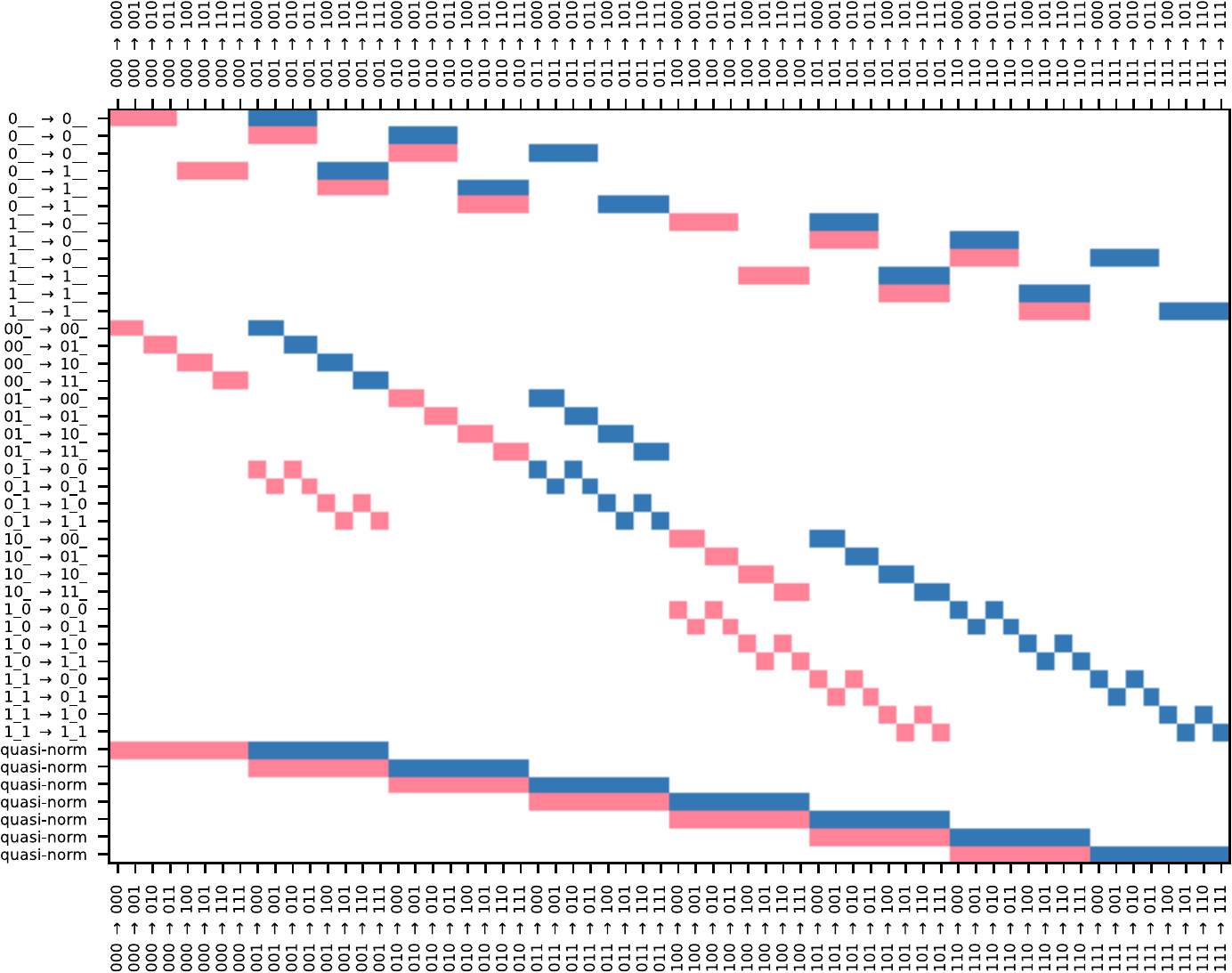}
\end{center}

\noindent Rows correspond to the 43 linear equations, of which 25 are independent.

\newpage
\subsection*{Space 97}

Space 97 is not induced by a causal order, but it is a refinement of the space 100 induced by the definite causal order $\total{\ev{A},\ev{B},\ev{C}}$.
Its equivalence class under event-input permutation symmetry contains 24 spaces.
Space 97 differs as follows from the space induced by causal order $\total{\ev{A},\ev{B},\ev{C}}$:
\begin{itemize}
  \item The outputs at events \evset{\ev{A}, \ev{C}} are independent of the input at event \ev{B} when the inputs at events \evset{A, C} are given by \hist{A/1,C/1}.
\end{itemize}

\noindent Below are the histories and extended histories for space 97: 
\begin{center}
    \begin{tabular}{cc}
    \includegraphics[height=3.5cm]{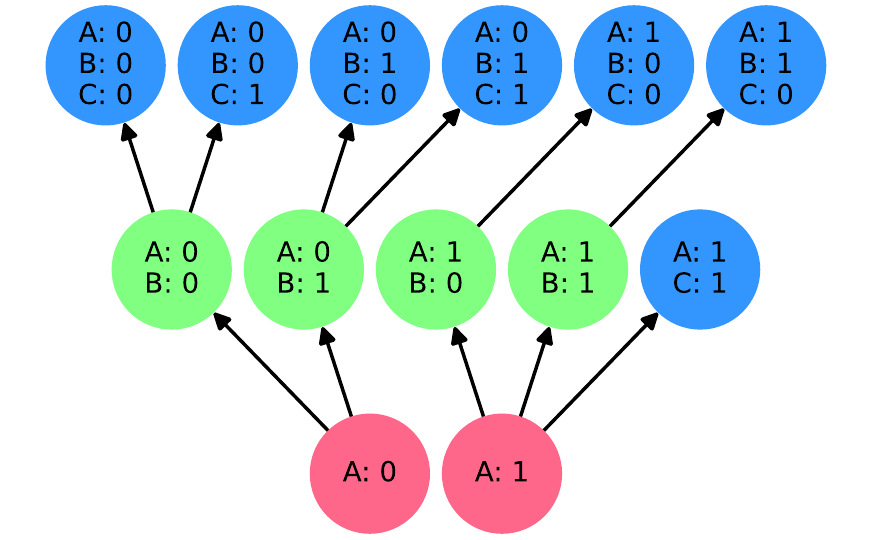}
    &
    \includegraphics[height=3.5cm]{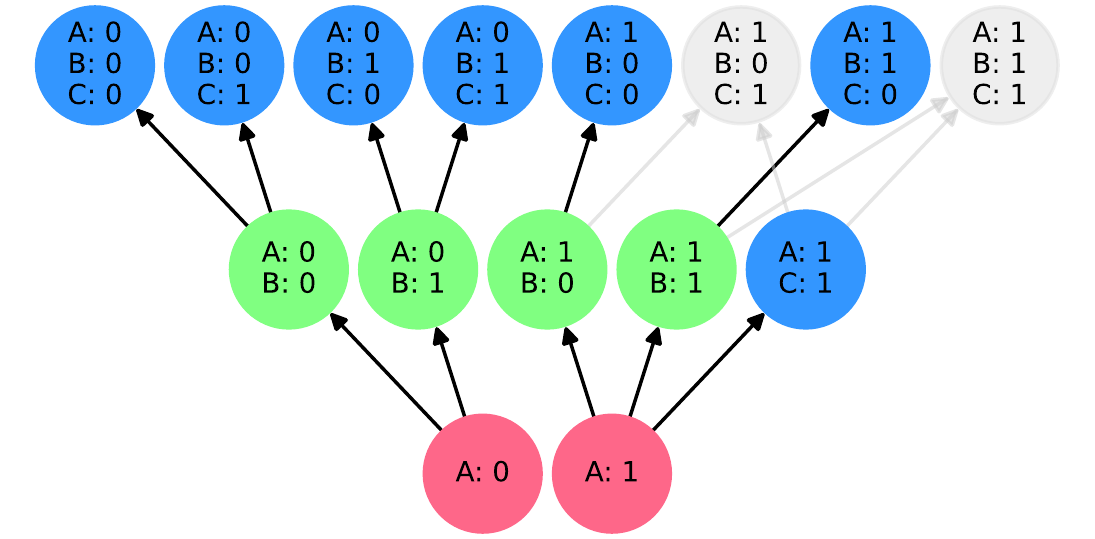}
    \\
    $\Theta_{97}$
    &
    $\Ext{\Theta_{97}}$
    \end{tabular}
\end{center}

\noindent The standard causaltope for Space 97 has dimension 40.
Below is a plot of the homogeneous linear system of causality and quasi-normalisation equations for the standard causaltope, put in reduced row echelon form:

\begin{center}
    \includegraphics[width=11cm]{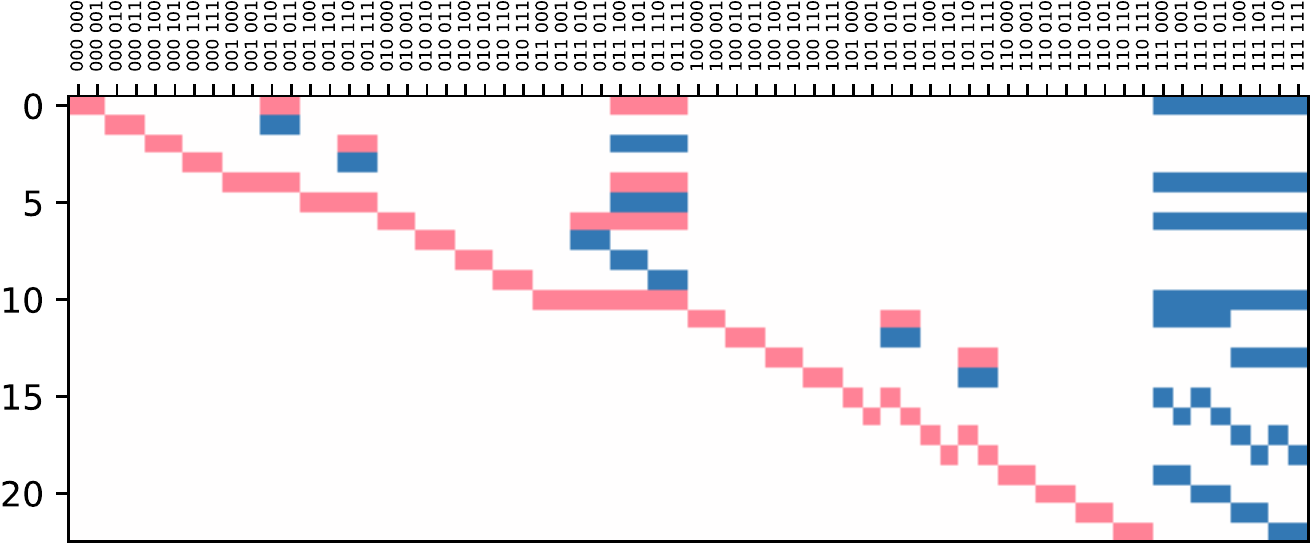}
\end{center}

\noindent Rows correspond to the 23 independent linear equations.
Columns in the plot correspond to entries of empirical models, indexed as $i_A i_B i_C$ $o_A o_B o_C$.
Coefficients in the equations are color-coded as white=0, red=+1 and blue=-1.

Space 97 has closest refinements in equivalence classes 90, 91, 94 and 95; 
it is the join of its (closest) refinements.
It has closest coarsenings in equivalence class 100; 
it does not arise as a nontrivial meet in the hierarchy.
It has 8192 causal functions, 1280 of which are not causal for any of its refinements.
It is a tight space.

The standard causaltope for Space 97 has 1 more dimension than those of its 2 subspaces in equivalence class 90.
The standard causaltope for Space 97 has 2 dimensions fewer than the meet of the standard causaltopes for its closest coarsenings.
For completeness, below is a plot of the full homogeneous linear system of causality and quasi-normalisation equations for the standard causaltope:

\begin{center}
    \includegraphics[width=12cm]{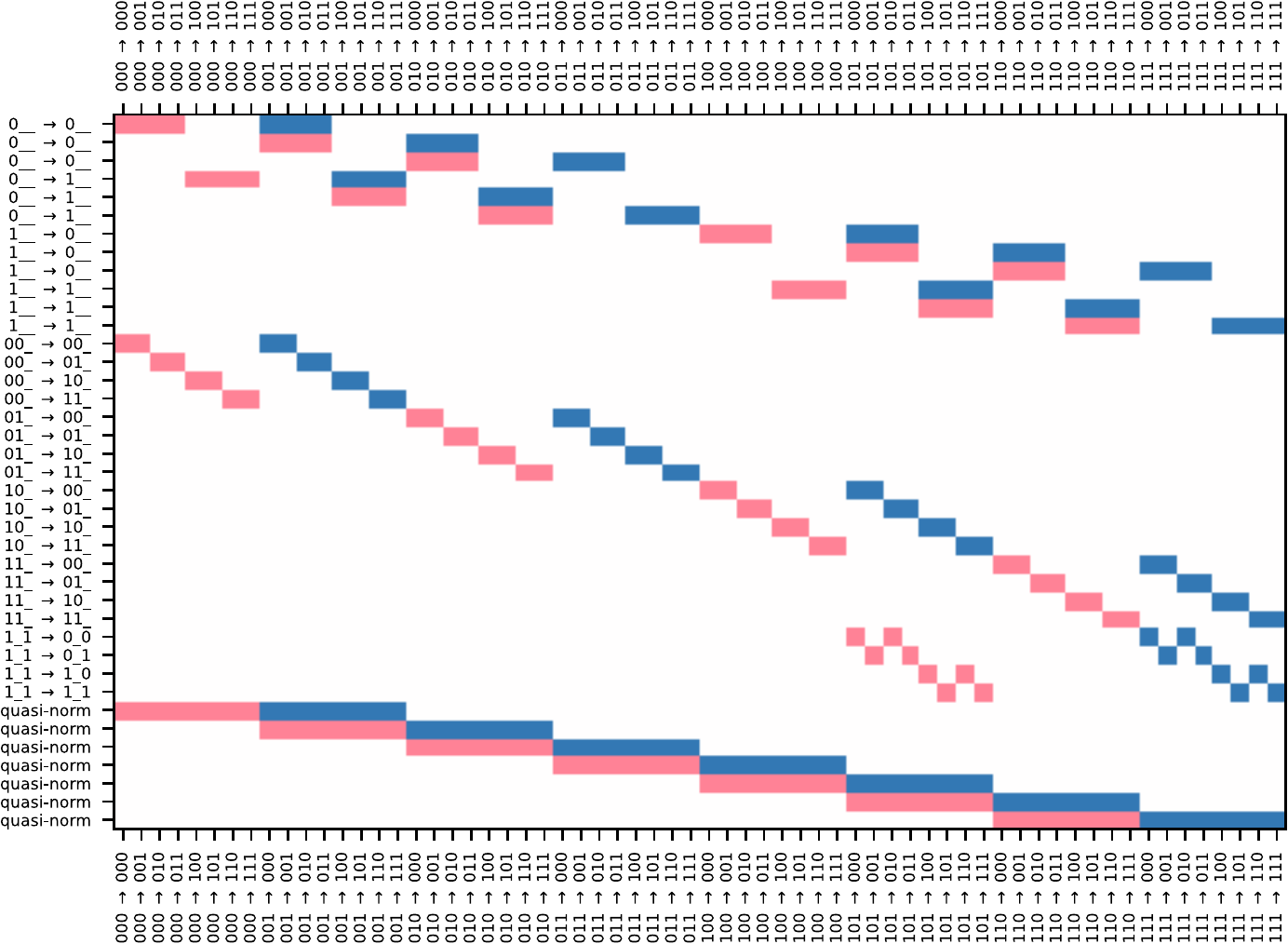}
\end{center}

\noindent Rows correspond to the 39 linear equations, of which 23 are independent.

\newpage
\subsection*{Space 98}

Space 98 is not induced by a causal order, but it is a refinement of the space 100 induced by the definite causal order $\total{\ev{A},\ev{B},\ev{C}}$.
Its equivalence class under event-input permutation symmetry contains 12 spaces.
Space 98 differs as follows from the space induced by causal order $\total{\ev{A},\ev{B},\ev{C}}$:
\begin{itemize}
  \item The output at event \ev{B} is independent of the input at event \ev{A} when the input at event B is given by \hist{B/1}.
\end{itemize}

\noindent Below are the histories and extended histories for space 98: 
\begin{center}
    \begin{tabular}{cc}
    \includegraphics[height=3.5cm]{svg-inkscape/space-ABC-unique-tight-98-highlighted_svg-tex.pdf}
    &
    \includegraphics[height=3.5cm]{svg-inkscape/space-ABC-unique-tight-98-ext-highlighted_svg-tex.pdf}
    \\
    $\Theta_{98}$
    &
    $\Ext{\Theta_{98}}$
    \end{tabular}
\end{center}

\noindent The standard causaltope for Space 98 has dimension 41.
Below is a plot of the homogeneous linear system of causality and quasi-normalisation equations for the standard causaltope, put in reduced row echelon form:

\begin{center}
    \includegraphics[width=11cm]{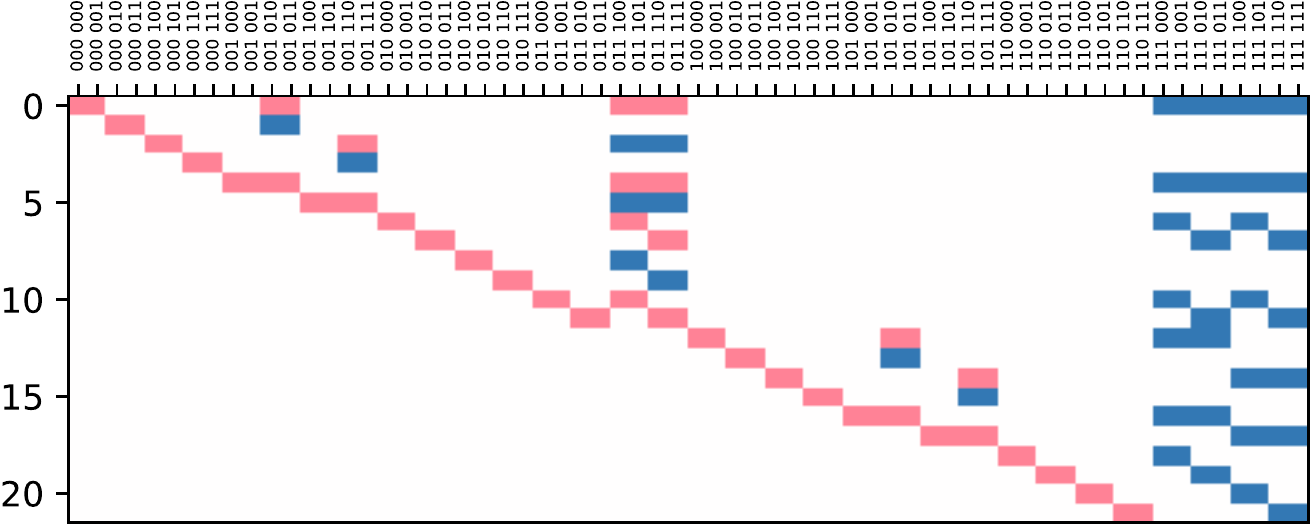}
\end{center}

\noindent Rows correspond to the 22 independent linear equations.
Columns in the plot correspond to entries of empirical models, indexed as $i_A i_B i_C$ $o_A o_B o_C$.
Coefficients in the equations are color-coded as white=0, red=+1 and blue=-1.

Space 98 has closest refinements in equivalence classes 89, 90 and 92; 
it is the join of its (closest) refinements.
It has closest coarsenings in equivalence class 100; 
it does not arise as a nontrivial meet in the hierarchy.
It has 8192 causal functions, 512 of which are not causal for any of its refinements.
It is a tight space.

The standard causaltope for Space 98 has 1 more dimension than that of its subspace in equivalence class 92.
The standard causaltope for Space 98 has 1 dimension fewer than the meet of the standard causaltopes for its closest coarsenings.
For completeness, below is a plot of the full homogeneous linear system of causality and quasi-normalisation equations for the standard causaltope:

\begin{center}
    \includegraphics[width=12cm]{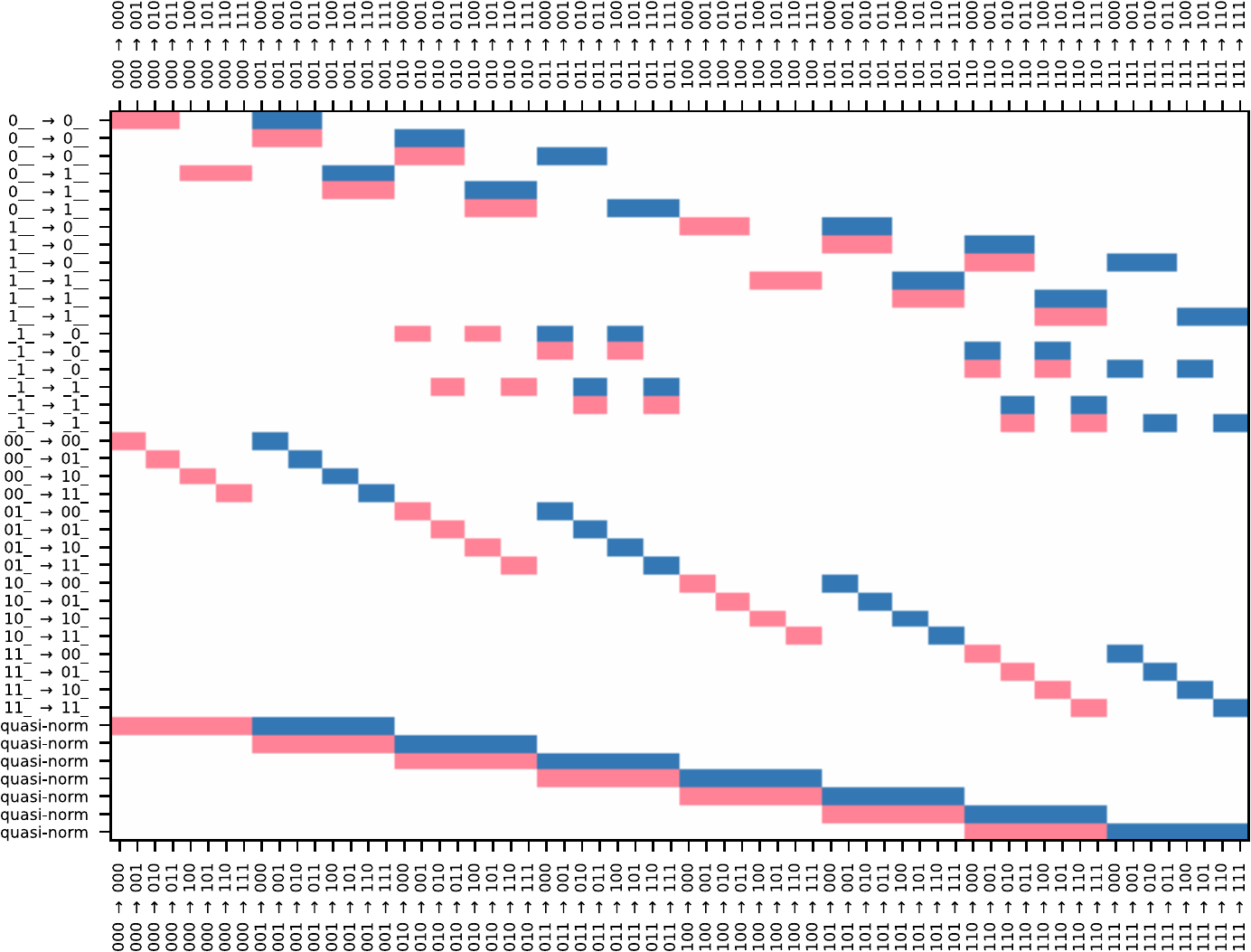}
\end{center}

\noindent Rows correspond to the 41 linear equations, of which 22 are independent.

\newpage
\subsection*{Space 99}

Space 99 is not induced by a causal order, but it is a refinement of the space induced by the indefinite causal order $\total{\ev{A},\{\ev{B},\ev{C}\}}$.
Its equivalence class under event-input permutation symmetry contains 24 spaces.
Space 99 differs as follows from the space induced by causal order $\total{\ev{A},\{\ev{B},\ev{C}\}}$:
\begin{itemize}
  \item The outputs at events \evset{\ev{A}, \ev{B}} are independent of the input at event \ev{C} when the inputs at events \evset{A, B} are given by \hist{A/0,B/0}, \hist{A/0,B/1} and \hist{A/1,B/0}.
  \item The outputs at events \evset{\ev{A}, \ev{C}} are independent of the input at event \ev{B} when the inputs at events \evset{A, C} are given by \hist{A/1,C/0} and \hist{A/1,C/1}.
\end{itemize}

\noindent Below are the histories and extended histories for space 99: 
\begin{center}
    \begin{tabular}{cc}
    \includegraphics[height=3.5cm]{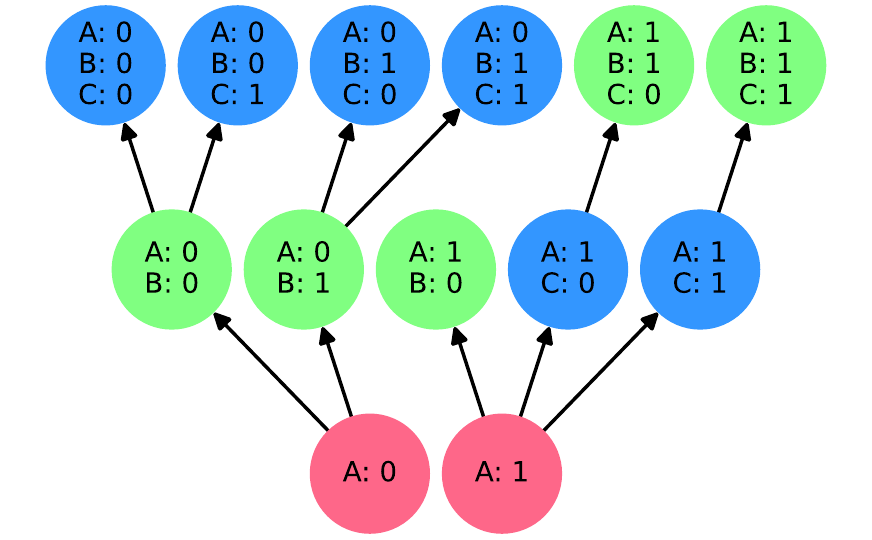}
    &
    \includegraphics[height=3.5cm]{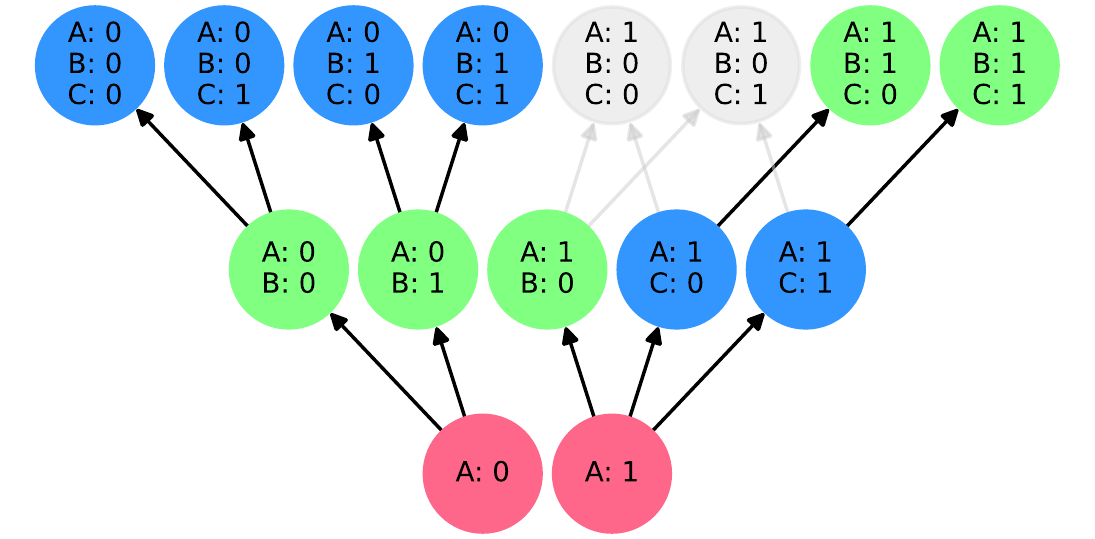}
    \\
    $\Theta_{99}$
    &
    $\Ext{\Theta_{99}}$
    \end{tabular}
\end{center}

\noindent The standard causaltope for Space 99 has dimension 40.
Below is a plot of the homogeneous linear system of causality and quasi-normalisation equations for the standard causaltope, put in reduced row echelon form:

\begin{center}
    \includegraphics[width=11cm]{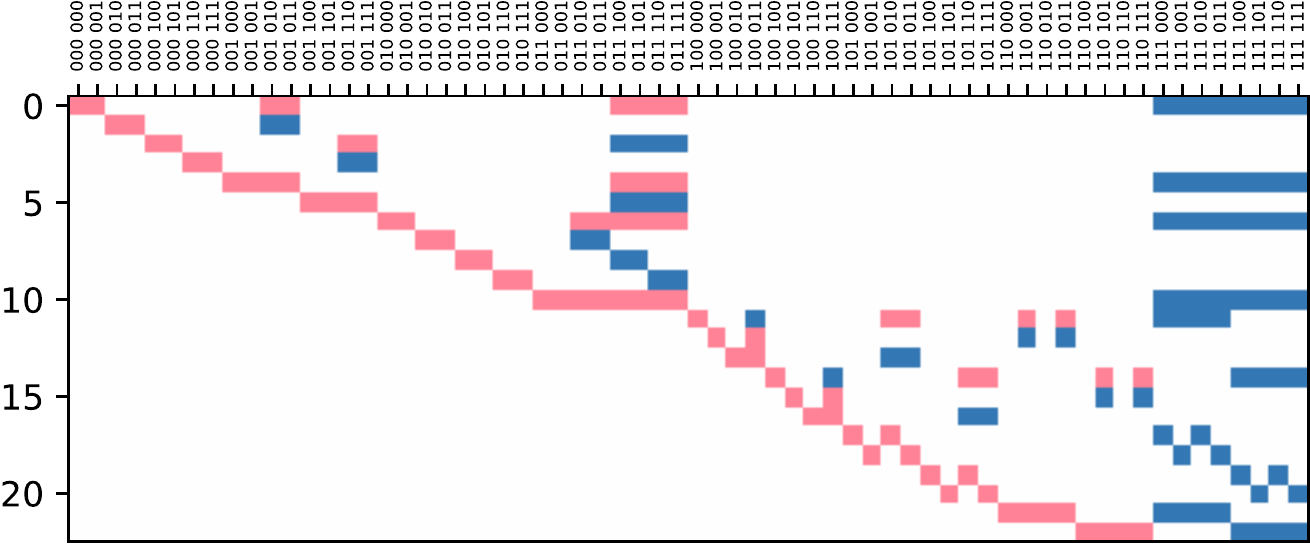}
\end{center}

\noindent Rows correspond to the 23 independent linear equations.
Columns in the plot correspond to entries of empirical models, indexed as $i_A i_B i_C$ $o_A o_B o_C$.
Coefficients in the equations are color-coded as white=0, red=+1 and blue=-1.

Space 99 has closest refinements in equivalence classes 91, 93 and 96; 
it is the join of its (closest) refinements.
It has closest coarsenings in equivalence class 101; 
it does not arise as a nontrivial meet in the hierarchy.
It has 8192 causal functions, 1024 of which are not causal for any of its refinements.
It is a tight space.

The standard causaltope for Space 99 has 1 more dimension than that of its subspace in equivalence class 93.
The standard causaltope for Space 99 has 2 dimensions fewer than the meet of the standard causaltopes for its closest coarsenings.
For completeness, below is a plot of the full homogeneous linear system of causality and quasi-normalisation equations for the standard causaltope:

\begin{center}
    \includegraphics[width=12cm]{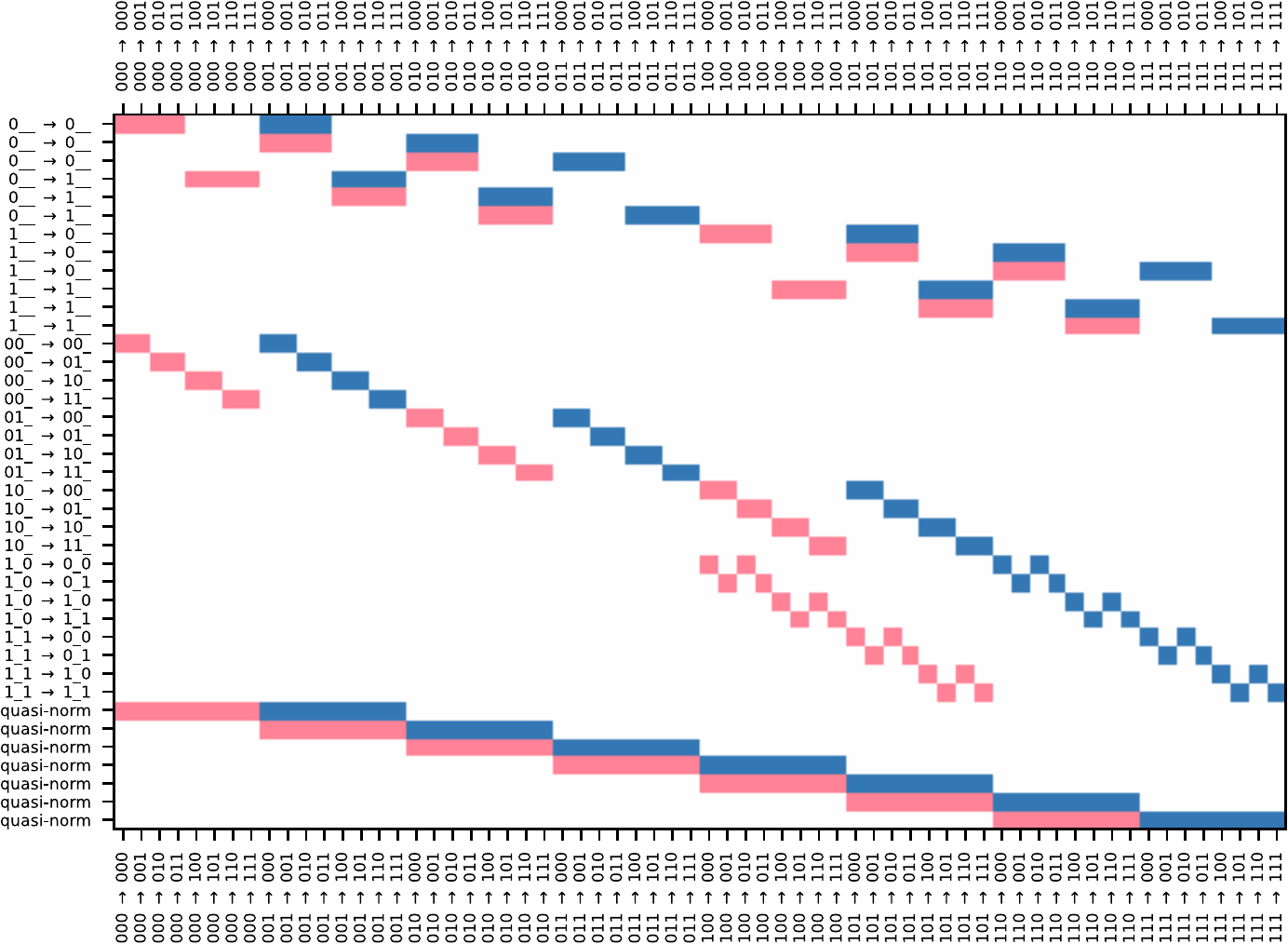}
\end{center}

\noindent Rows correspond to the 39 linear equations, of which 23 are independent.

\newpage
\subsection*{Space 100}

Space 100 is induced by the definite causal order $\total{\ev{A},\ev{B},\ev{C}}$.
Its equivalence class under event-input permutation symmetry contains 6 spaces.

\noindent Below are the histories and extended histories for space 100: 
\begin{center}
    \begin{tabular}{cc}
    \includegraphics[height=3.5cm]{svg-inkscape/space-ABC-unique-tight-100-highlighted_svg-tex.pdf}
    &
    \includegraphics[height=3.5cm]{svg-inkscape/space-ABC-unique-tight-100-ext-highlighted_svg-tex.pdf}
    \\
    $\Theta_{100}$
    &
    $\Ext{\Theta_{100}}$
    \end{tabular}
\end{center}

\noindent The standard causaltope for Space 100 has dimension 42.
Below is a plot of the homogeneous linear system of causality and quasi-normalisation equations for the standard causaltope, put in reduced row echelon form:

\begin{center}
    \includegraphics[width=11cm]{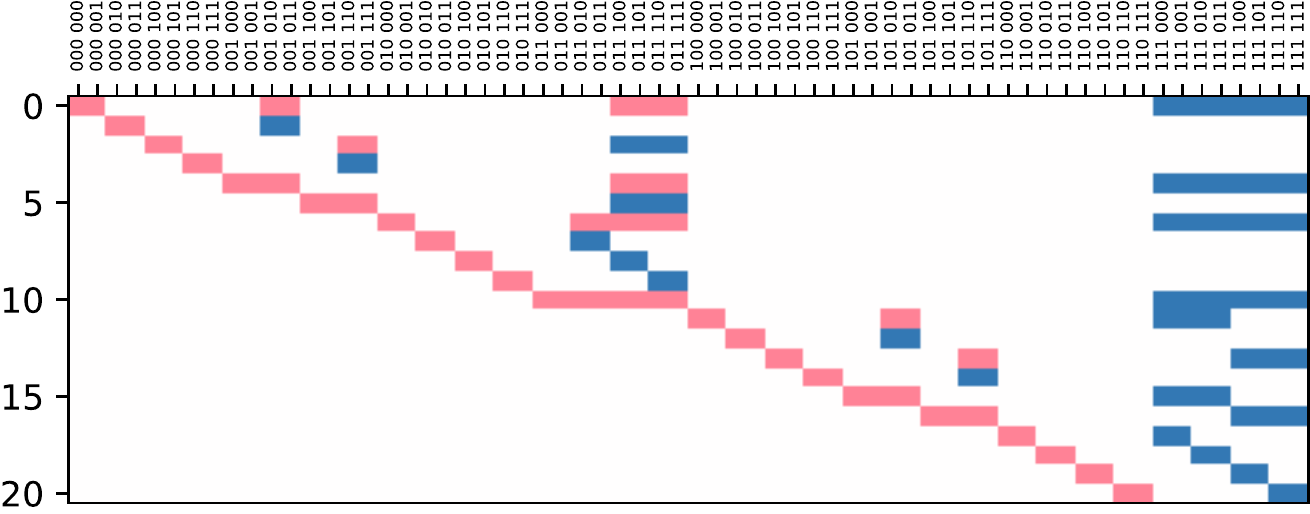}
\end{center}

\noindent Rows correspond to the 21 independent linear equations.
Columns in the plot correspond to entries of empirical models, indexed as $i_A i_B i_C$ $o_A o_B o_C$.
Coefficients in the equations are color-coded as white=0, red=+1 and blue=-1.

Space 100 has closest refinements in equivalence classes 97 and 98; 
it is the join of its (closest) refinements.
It is a global maximum of the hierarchy, with no coarsenings.
It has 16384 causal functions, 3072 of which are not causal for any of its refinements.
It is a tight space.

The standard causaltope for Space 100 has 1 more dimension than those of its 2 subspaces in equivalence class 98.
For completeness, below is a plot of the full homogeneous linear system of causality and quasi-normalisation equations for the standard causaltope:

\begin{center}
    \includegraphics[width=12cm]{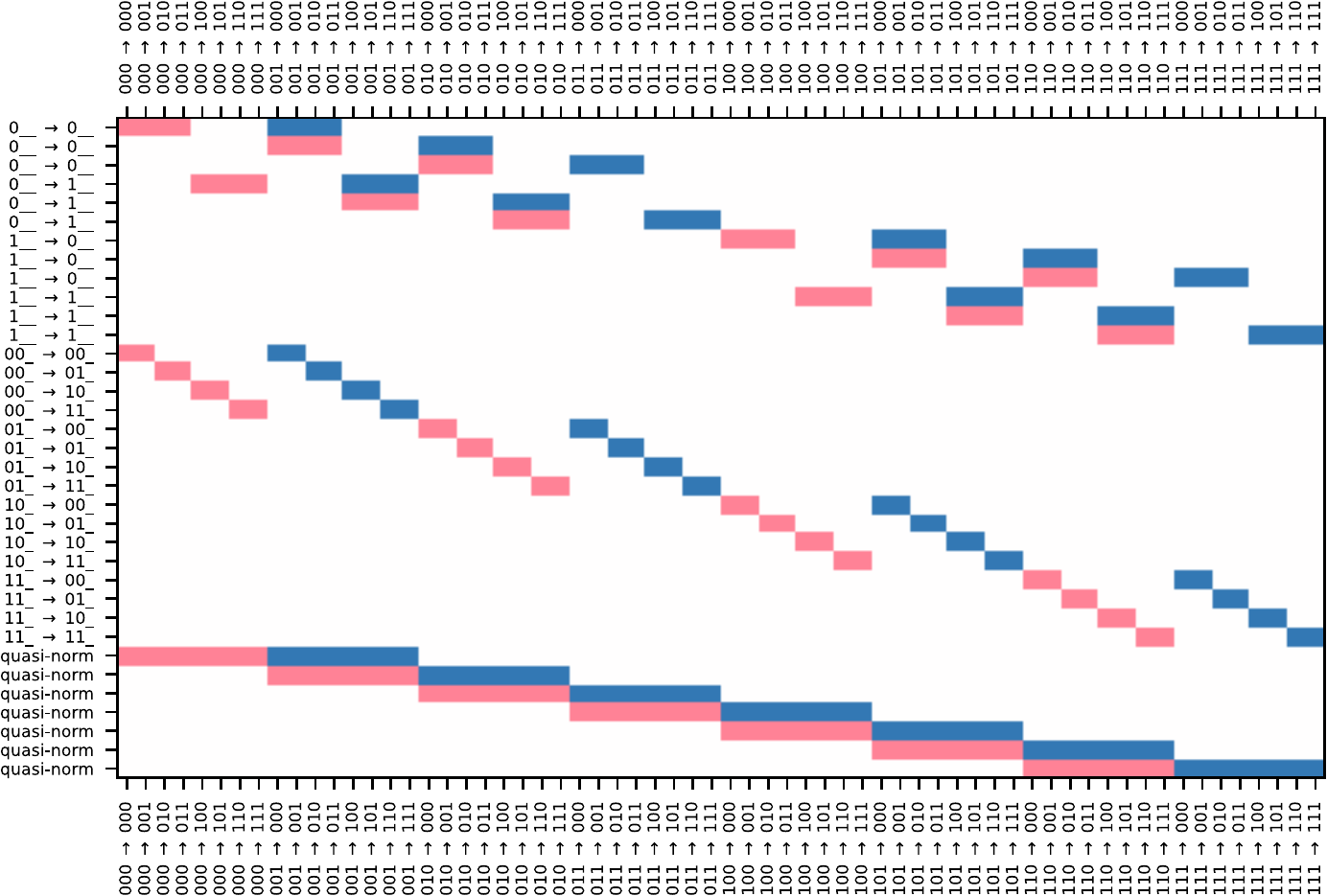}
\end{center}

\noindent Rows correspond to the 35 linear equations, of which 21 are independent.

\newpage
\subsection*{Space 101}

Space 101 is not induced by a causal order, but it is a refinement of the space induced by the indefinite causal order $\total{\ev{A},\{\ev{B},\ev{C}\}}$.
Its equivalence class under event-input permutation symmetry contains 6 spaces.
Space 101 differs as follows from the space induced by causal order $\total{\ev{A},\{\ev{B},\ev{C}\}}$:
\begin{itemize}
  \item The outputs at events \evset{\ev{A}, \ev{B}} are independent of the input at event \ev{C} when the inputs at events \evset{A, B} are given by \hist{A/0,B/0} and \hist{A/0,B/1}.
  \item The outputs at events \evset{\ev{A}, \ev{C}} are independent of the input at event \ev{B} when the inputs at events \evset{A, C} are given by \hist{A/1,C/0} and \hist{A/1,C/1}.
\end{itemize}

\noindent Below are the histories and extended histories for space 101: 
\begin{center}
    \begin{tabular}{cc}
    \includegraphics[height=3.5cm]{svg-inkscape/space-ABC-unique-tight-101-highlighted_svg-tex.pdf}
    &
    \includegraphics[height=3.5cm]{svg-inkscape/space-ABC-unique-tight-101-ext-highlighted_svg-tex.pdf}
    \\
    $\Theta_{101}$
    &
    $\Ext{\Theta_{101}}$
    \end{tabular}
\end{center}

\noindent The standard causaltope for Space 101 has dimension 42.
Below is a plot of the homogeneous linear system of causality and quasi-normalisation equations for the standard causaltope, put in reduced row echelon form:

\begin{center}
    \includegraphics[width=11cm]{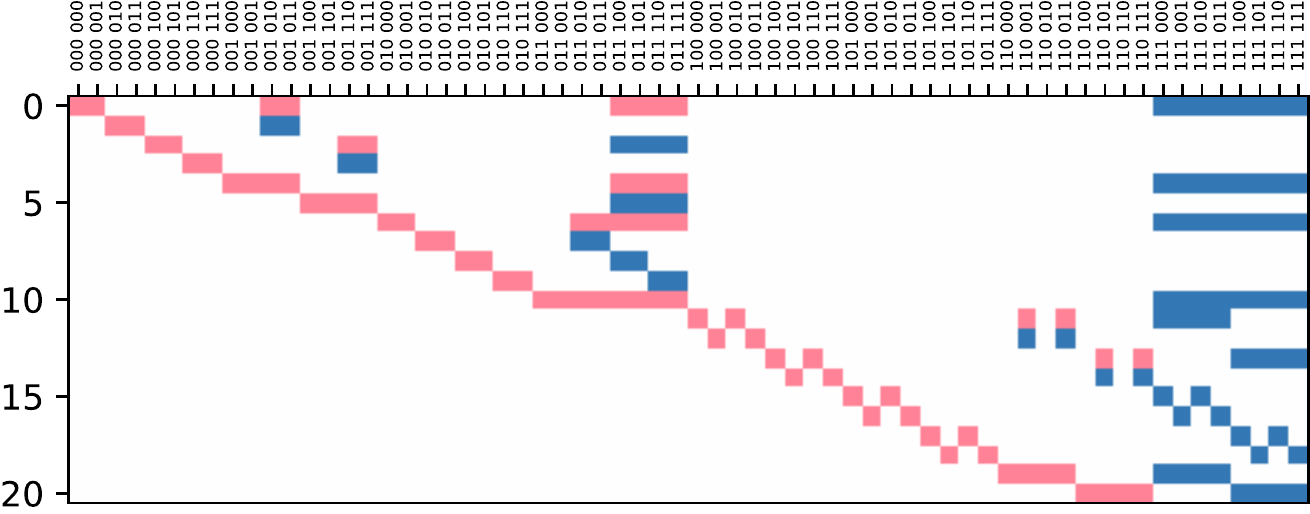}
\end{center}

\noindent Rows correspond to the 21 independent linear equations.
Columns in the plot correspond to entries of empirical models, indexed as $i_A i_B i_C$ $o_A o_B o_C$.
Coefficients in the equations are color-coded as white=0, red=+1 and blue=-1.

Space 101 has closest refinements in equivalence class 99; 
it is the join of its (closest) refinements.
It is a global maximum of the hierarchy, with no coarsenings.
It has 16384 causal functions, 7296 of which are not causal for any of its refinements.
It is a tight space.

The standard causaltope for Space 101 has 2 more dimensions than those of its 4 subspaces in equivalence class 99.
For completeness, below is a plot of the full homogeneous linear system of causality and quasi-normalisation equations for the standard causaltope:

\begin{center}
    \includegraphics[width=12cm]{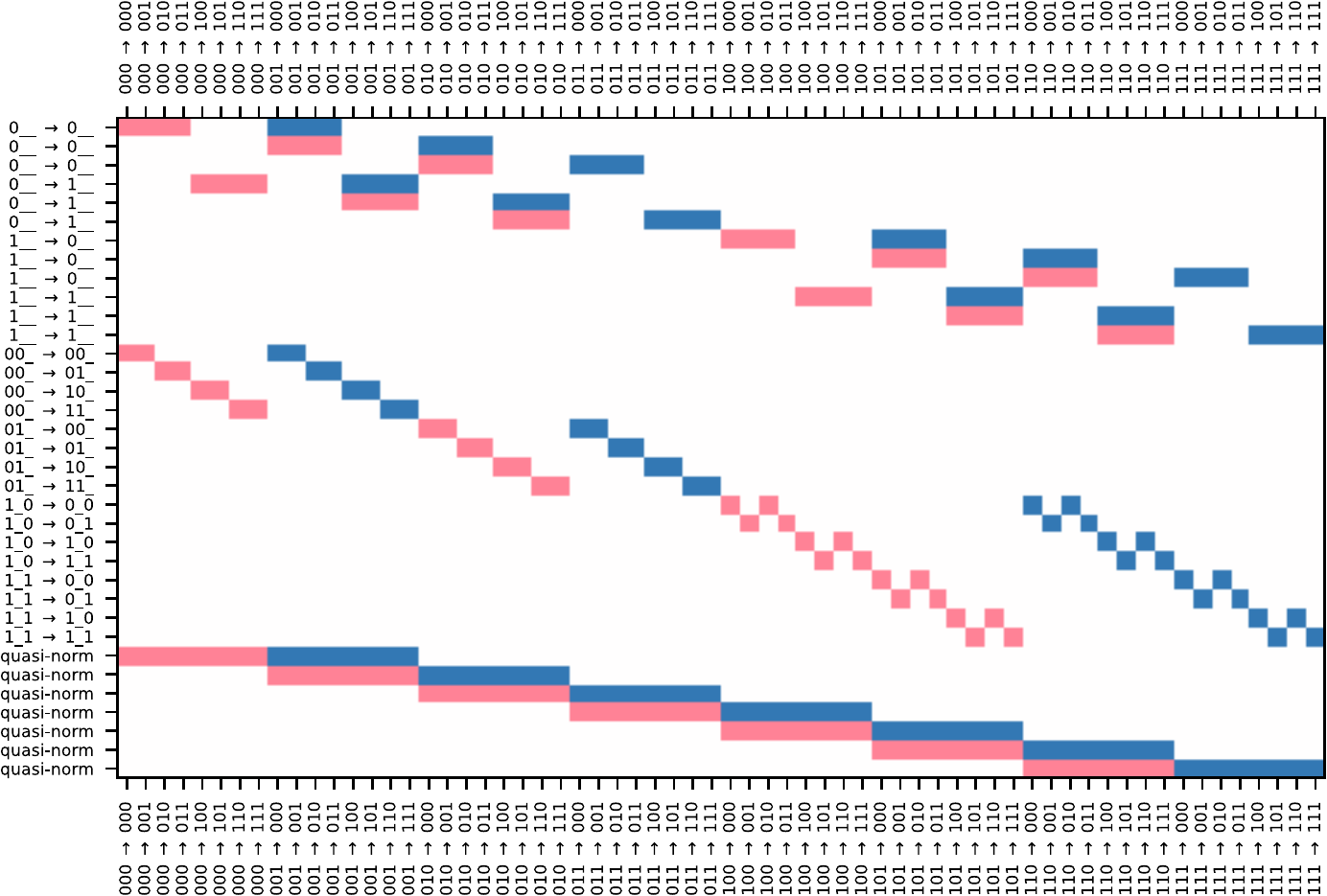}
\end{center}

\noindent Rows correspond to the 35 linear equations, of which 21 are independent.

\section{Algorithm to Find Causally Complete Spaces}
\label{appendix:space-finder-algo}

In the sub-sections below, we break down the algorithm that was used to enumerate the hierarchies of causally complete spaces on binary inputs that were presented in this work.
Each sub-section start by presenting a full listing for a block of related code, typically a single function or class method, and then proceeds to explain how the various parts work.

The algorithm is limited to the binary case, for added efficiency, but can be easily extended to arbitrary input combinations.
The algorithm can also be extended to find causally complete sub-spaces of a given space, but this requires non-trivial modification (because of the layered approach, which presumes that all histories in a layer of recursion have same-sized domains).

All code is written for Python 3.10, using PEP484 type hints.
It has been checked with the Mypy static type-checker and it has been linted with with Pylint.

\subsection{Imports and utilities}

Imports for the whole file, all from standard libraries.

\begin{minted}[firstnumber=1]{python}
from collections import deque
from collections.abc import Collection, Iterable, Iterator, Mapping, Set
from functools import cache
from itertools import chain, combinations, islice, permutations, product
from logging import Logger
from math import ceil
from numbers import Number
import sys
from time import perf_counter
from typing import (Any, BinaryIO, cast, Literal, Optional,
                    Sequence, TypeVar, Union)
\end{minted}

\noindent Below is a snippet for recursive memory size calculation of Python objects, courtesy of a StackOverflow answer.

\begin{minted}[firstnumber=last]{python}
def getsize(obj_0: Any) -> int:
    """
    Recursively iterate to sum size of object & members.
    Courtesy: https://stackoverflow.com/questions/449560/
              how-do-i-determine-the-size-of-an-object-in-python#answer-30316760
    """
    _seen_ids = set()
    def inner(obj: Any) -> int:
        obj_id = id(obj)
        if obj_id in _seen_ids:
            return 0
        _seen_ids.add(obj_id)
        size = sys.getsizeof(obj)
        if isinstance(obj, (str, bytes, Number, range, bytearray)):
            pass # bypass remaining control flow and return
        elif isinstance(obj, (tuple, list, Set, deque)):
            size += sum(inner(i) for i in obj)
        elif isinstance(obj, Mapping) or hasattr(obj, 'items'):
            size += sum(inner(k) + inner(v) for k, v in getattr(obj, 'items')())
        # Check for custom object instances - may subclass above too
        if hasattr(obj, '__dict__'):
            size += inner(vars(obj))
        if hasattr(obj, '__slots__'): # can have __slots__ with __dict__
            size += sum(inner(getattr(obj, s))
                        for s in obj.__slots__ if hasattr(obj, s))
        return size
    return inner(obj_0)
\end{minted}

\noindent The \mintinline{python}{powerset} function takes an iterable of elements some generic type \mintinline{python}{_T_co} and iterates through all possible subsets of those elements, yielded as \mintinline{python}{tuple}s of elements (rather than sets).

\begin{minted}[firstnumber=last]{python}
_T_co = TypeVar("_T_co")
def powerset(iterable: Iterable[_T_co]) -> Iterable[tuple[_T_co, ...]]:
    """
    Courtesy: https://docs.python.org/3/library/itertools.html#itertools-recipes
    powerset([1,2,3]) --> () (1,) (2,) (3,) (1,2) (1,3) (2,3) (1,2,3)
    """
    s = list(iterable)
    return chain.from_iterable(combinations(s, r) for r in range(len(s)+1))
\end{minted}

\subsection{Bitvectors}

We encode sets of non-negative in a memory-efficient way, as bitvectors.
Python natively supports bitwise operations on \mintinline{python}{int} objects, so we use those for our bitvector implementation.
The \mintinline{python}{bitvec} function takes an iterable of non-negative integers and returns the bitvector corresponding to the set of its elements, where $x_1,...,x_n \geq 0$ are encoded as $\sum_{j=1}^n 2^{x_j} \geq 0$.

\begin{minted}[firstnumber=last]{python}
Bitvec = int
""" Type alias for a bitvector, as a non-negative integer. """

def bitvec(elements: Iterable[int]) -> Bitvec:
    """
    Creates a bitvector representing the set of given
    non-negative integers.
    """
    v = 0
    for el in elements:
        assert el >= 0, f"Found negative integer {el}."
        v |= 2**el
    return v
\end{minted}

\noindent As an example, below is the encoding of the set $\{1, 3, 5, 8\}$ as a bitvector, coinciding with the number $128 = 2^8+2^5+2^3+2^1$ (bits read right-to-left).

\begin{minted}[linenos=false]{python}
bv = bitvec({1, 3, 5, 8})
print(f"{bv = } = {bv:b}")
# bv = 298 = 100101010
\end{minted}

\noindent The \mintinline{python}{sub} function takes the bitvector representations of two sets and returns the bitvector corresponding to their set difference.

\begin{minted}[firstnumber=60]{python}
def sub(u: Bitvec, v: Bitvec) -> Bitvec:
    """
    Returns the bitvector containing the elements that are
    in ``u`` but not in ``v``.
    """
    assert u >= 0, "Invalid bitvector."
    assert v >= 0, "Invalid bitvector."
    return u^(u&v)
\end{minted}

\noindent The \mintinline{python}{is_subset} function takes the bitvector representations of two sets are returns a \mintinline{python}{bool} value (\mintinline{python}{True} or \mintinline{python}{False}) stating whether the first set is a subset of the second set.

\begin{minted}[firstnumber=last]{python}
def is_subset(u: Bitvec, v: Bitvec) -> bool:
    """
    Returns whether the bitvector ``u`` is a subset of
    the bitvector ``v``.
    """
    return u == u&v
\end{minted}

\noindent The \mintinline{python}{iter_bitvec} function takes the bitvector representation of a set and iterates through the elements in the set, in strictly increasing order.

\begin{minted}[firstnumber=last]{python}
def iter_bitvec(u: Bitvec) -> Iterator[int]:
    """
    Iterates over the elements in a bitvector.
    """
    assert u >= 0, "Invalid bitvector."
    el = 0
    while u > 0:
        u, b = divmod(u, 2)
        if b == 1:
            yield el
        el += 1
\end{minted}

\noindent The \mintinline{python}{bitvec2set} function takes the bitvector representation of a set and returns the corresponding \mintinline{python}{set} instance.

\begin{minted}[firstnumber=last]{python}
def bitvec2set(u: Bitvec) -> set[int]:
    """
    Turns a bitvector into the corresponding set of non-negative integers.
    """
    return set(iter_bitvec(u))
\end{minted}

\subsection{Histories}

The code below supports histories up to 26 events, labelled by uppercase Latin alphabet letters (\mintinline{python}{"A"} to \mintinline{python}{"Z"}), with binary input values (\mintinline{python}{0} or \mintinline{python}{1}).
A ``history item'' is an event-input pair and a history is a set of such pairs, encoded as a bitvector.

\begin{minted}[firstnumber=last]{python}
Event = Literal["A", "B", "C", "D", "E", "F", "G", "H", "I", "J", "K", "L", "M",
                "N", "O", "P", "Q", "R", "S", "T", "U", "V", "W", "X", "Y", "Z"]
""" Type alias for events (single char "A" to "Z"). """

InputValue = Literal[0, 1]
""" Type alias for input/output values (binary 0 or 1). """

HistoryItem = tuple[Event, InputValue]
""" Type alias for an item (event-value pair) in an (ext) input history. """

HistoryItems = tuple[HistoryItem, ...]
""" Type alias for a variable length tuple of history items. """

History = Bitvec
"""
    Type alias for an history, a bitvector of integers indicating its items.
"""
\end{minted}

\noindent The functions \mintinline{python}{event2idx} and \mintinline{python}{idx2event} convert between events and non-negative integers in the range $0-25$:
\begin{itemize}
    \item \mintinline{python}{0} $\leftrightarrow$ \mintinline{python}{"A"}
    \item \mintinline{python}{1} $\leftrightarrow$ \mintinline{python}{"B"}
    \item ...
    \item \mintinline{python}{25} $\leftrightarrow$ \mintinline{python}{"Z"}
\end{itemize}

\begin{minted}[firstnumber=last]{python}
@cache
def event2idx(e: Event) -> int:
    """ Turns an event into the corresponding index. """
    idx = ord(e)-ord("A")
    return idx

@cache
def idx2event(idx: int) -> Event:
    """ Turns a index into the corresponding event. """
    assert 0 <= idx <= 25, "Events indexes be 0-25."
    return cast(Event, chr(ord("A")+idx))
\end{minted}

\noindent The functions \mintinline{python}{item2idx} and \mintinline{python}{idx2item} convert between history items (event-input pairs) and non-negative integers in the range $0-51$:
\begin{itemize}
    \item \mintinline{python}{0} $\leftrightarrow$ \mintinline{python}{("A", 0)}
    \item \mintinline{python}{1} $\leftrightarrow$ \mintinline{python}{("A", 1)}
    \item \mintinline{python}{2} $\leftrightarrow$ \mintinline{python}{("B", 0)}
    \item \mintinline{python}{3} $\leftrightarrow$ \mintinline{python}{("B", 1)}
    \item ...
    \item \mintinline{python}{51} $\leftrightarrow$ \mintinline{python}{("Z", 1)}
\end{itemize}

\begin{minted}[firstnumber=last]{python}
@cache
def item2idx(item: HistoryItem) -> int:
    """ Turns an history item into the corresponding index. """
    e, v = item
    return 2*event2idx(e)+v

@cache
def idx2item(idx: int) -> HistoryItem:
    """ Turns a index into the corresponding history item. """
    assert 0 <= idx <= 51, "Item indexes must be 0-51."
    return (idx2event(idx//2), cast(InputValue, idx%2))
\end{minted}

\noindent The \mintinline{python}{history} function creates a history bitvector from an event-input mapping or an iterable of history items, by first encoding the event-input pairs to non-negative integers using \mintinline{python}{item2idx}.

\begin{minted}[firstnumber=last]{python}
def history(h_items: Union[Mapping[Event, InputValue],
                           Iterable[HistoryItem]]) -> History:
    """
    Creates a history from an event-value mapping
    or an iterable of event-value pairs.
    """
    if isinstance(h_items, Mapping):
        h_items = h_items.items()
    return bitvec(item2idx(item) for item in h_items)
\end{minted}

\noindent For example, the history $\hist{A/0, B/1, D/1}$ on 4 events $\ev{A}, \ev{B}, \ev{C}, \ev{D}$ is first turned into the set \mintinline{python}{{("A", 0), ("B", 1), ("D", 1)}} of event-input, then turned into the set \mintinline{python}{{0, 3, 7}} of corresponding indices, and finally encoded into a bitvector, as the number $137 = 2^7+2^3+2^0$.

\begin{minted}[linenos=false]{python}
h = history({"A": 0, "B": 1, "D": 1})
print(f"{h = } = {h:b}")
# h = 137 = 10001001
\end{minted}

\noindent Given a history bitvector, the \mintinline{python}{history2items} function returns the corresponding immutable sequence (a \mintinline{python}{tuple}) of history items, in increasing order.

\begin{minted}[firstnumber=138]{python}
def history2items(h: History) -> HistoryItems:
    """
    Turns a history into the tuple of its items, in order of increasing index.
    """
    assert h >= 0, "Invalid history."
    return tuple(idx2item(idx) for idx in iter_bitvec(h))
\end{minted}

\noindent Given a history bitvector, the \mintinline{python}{history2items} function returns the corresponding event-input dictionary.

\begin{minted}[firstnumber=last]{python}
def history2dict(h: History) -> dict[Event, InputValue]:
    """
    Turns a history into the corresponding event-value mapping,
    with events in order of increasing index.
    """
    d: dict[Event, InputValue] = {}
    for idx in iter_bitvec(h):
        e, v = idx2item(idx)
        assert e not in d, f"History has multiple values for event {e}."
        d[e] = v
    return d
\end{minted}

\noindent Given a history bitvector, the \mintinline{python}{dom} function returns the \mintinline{python}{frozenset} of events in the history's domain.

\begin{minted}[firstnumber=last]{python}
def dom(h: History) -> frozenset[Event]:
    """
    Returns the domain of a history, as a frozenset of events.
    """
    assert h >= 0, "Invalid history."
    d = set()
    idx = 0
    while h > 0:
        h, b = divmod(h, 2)
        if b != 0:
            d.add(idx2item(idx)[0])
        idx += 1
    return frozenset(d)
\end{minted}

\noindent Given a history bitvector, the \mintinline{python}{domsize} function returns the size of the history's domain.

\begin{minted}[firstnumber=last]{python}
def domsize(h: History) -> int:
    """
    Returns the size of the domain of a history.
    """
    assert h >= 0, "Invalid history."
    size = 0
    idx = 0
    while h > 0:
        h, b = divmod(h, 2)
        if b != 0:
            size += 1
        idx += 1
    return size
\end{minted}

\noindent Given a history bitvector, the \mintinline{python}{history_sort_key} function returns a corresponding sorting key, for use with the builtin \mintinline{python}{sorted} function.
Using this key, histories are first sorted by length, and then by the items they contain (according to the sorting rules for \mintinline{python}{tuple}).

\begin{minted}[firstnumber=last]{python}
def history_sort_key(h: History) -> tuple[int, HistoryItems]:
    """
    Returns a sorting key for histories,
    sorted first by length and then by content.
    """
    h_items = history2items(h)
    return (len(h_items), tuple(sorted(h_items)))
\end{minted}

\noindent The \mintinline{python}{max_histories} function returns the sequence of maximal extended input histories on a given number of events.
The events are selected starting from \mintinline{python}{"A"}, in order.

\begin{minted}[firstnumber=last]{python}
def max_histories(num_events: int) -> tuple[History, ...]:
    """
    Returns the sequence of maximal extended input histories on a given number
    of events. Events are selected in order, starting from ``"A"``.
    """
    dom_events = tuple(idx2event(idx) for idx in range(num_events))
    return tuple(
        history(zip(dom_events, input_choices))
        for input_choices in product(cast(Sequence[InputValue], [0,1]),
                                     repeat=num_events)
    )
\end{minted}

\noindent Given a history bitvector, the \mintinline{python}{child_histories} function returns the immutable sequence of bitvectors for the history's children, the histories obtained by removing each individual event in the history's domain in turn.

\begin{minted}[firstnumber=last]{python}
def child_histories(h: History) -> tuple[History, ...]:
    """
    Given a history, returns the sequence of child histories,
    obtained by removing a single event (in turn).
    """
    h_dict = history2dict(h)
    if len(h_dict) <= 1:
        return tuple()
    return tuple(
        history((e, v) for e, v in h_dict.items() if e != removed_e)
        for removed_e in reversed(h_dict)
    )
\end{minted}

\noindent Given a sequence of history bitvectors, the \mintinline{python}{sub_histories} function returns the immutable sequence of bitvectors for all sub-histories for the given histories, in the order discovered by breadth-first search of history children.

\begin{minted}[firstnumber=last]{python}
def sub_histories(hs: Sequence[History]) -> tuple[History, ...]:
    """
    Given an ordered set of histories, returns the sequence of sub-histories,
    in order of breadth-first discovery and including the initial histories.
    """
    visited = set()
    q = deque(hs)
    sub_hs = []
    while q:
        h = q.popleft()
        if h not in visited:
            sub_hs.append(h)
            visited.add(h)
        for k in child_histories(h):
            if k not in visited:
                q.append(k)
    return tuple(sub_hs)
\end{minted}

\subsection{Event-input Permutation}

Given a sequence of events $\underline{e} = (e_1,...,e_n)$ without repetition, the elements
\[
\underline{g} = (g_E,g_{I_{e_1}},...,g_{I_{e_n}}) \in S(E) \times \prod_{j=1}^n S(\{0,1\})
\]
of the event-input permutation group on $E = \{e_1,...,e_n\}$ with binary inputs are uniquely represented by the element-wise action of $g_E$ on the sequence of events $\underline{e}$ and the action of each $g_{I_{e_j}}$ on the input value $0$:
\[
\Bigg(
    \Big(g_E(e_1),...,g_E(e_n)\Big),\,
    \Big(g_{I_{e_1}}(0),...,g_{I_{e_n}}(0)\Big)
\Bigg)
\]

\begin{minted}[firstnumber=last]{python}
PermGroupEl = tuple[Sequence[Event], Sequence[InputValue]]
"""
Type alias for elements of the event-value permutation group on a given ordered
set of events (as a unique sequence, or sequence without repetitions).
Group elements are represented by pairs:

1. an ordered set of permuted events (represented by a unique sequence),
   defining the permutation action on events
2. a sequence of values, one for each event,
   defining the permutation action on values (by bitwise XOR)
"""
\end{minted}

\noindent The \mintinline{python}{iter_perm_group} function iterates through the permutation group elements for a given sequence of unique events.

\begin{minted}[firstnumber=last]{python}
def iter_perm_group(events_list: Sequence[Event]) -> Iterator[PermGroupEl]:
    """
    Iterates through the elements of the event-value permutation group on the
    given events to permute.
    """
    num_events = len(events_list)
    assert len(set(events_list)) == num_events, "Events must not be repeated."
    for events_perm in permutations(events_list):
        for value_perm in product(cast(Sequence[InputValue], [0,1]),
                                  repeat=num_events):
            yield (events_perm, value_perm)
\end{minted}

\noindent Given a history bitvector and an event-input permutation group element, the \mintinline{python}{permute_history} function returns the bitvector for the permuted history.

\begin{minted}[firstnumber=last]{python}
def permute_history(h: History, g: PermGroupEl) -> History:
    """
    Given a history, returns the history obtained by applying the given
    element of the event-value permutation group.
    """
    h_dict = history2dict(h)
    events_perm, value_perm = g
    return history({
        e_perm: cast(InputValue, h_dict[e]^value_perm[e_idx])
        for e_idx, (e, e_perm) in enumerate(zip(sorted(events_perm),
                                                events_perm))
        if e in h_dict
    })
\end{minted}

\noindent Given a history bitvector and an iterable of event-input permutation group elements, the \mintinline{python}{history_perms} function efficiently iterates through all group elements and bitvectors of the corresponding permuted histories.

\begin{minted}[firstnumber=last]{python}
def history_perms(h: History,
                  perm_group: Iterable[PermGroupEl]
                  ) -> Iterator[tuple[PermGroupEl, History]]:
    """
    Given a history and an iterable of event-value permutation group elements,
    iterates through all pairs of permutation group element and corresponding
    permuted history.
    """
    h_dict = history2dict(h)
    for events_perm, value_perm in perm_group:
        yield (events_perm, value_perm), history({
            e_perm: cast(InputValue, h_dict[e]^value_perm[e_idx])
            for e_idx, (e, e_perm) in enumerate(zip(sorted(events_perm),
                                                    events_perm))
            if e in h_dict
        })
\end{minted}

\noindent Given a history bitvector and an iterable of event-input permutation group elements, the \mintinline{python}{history_perms} function returns the sequence of group elements stabilizing the history, in the order they are encountered and without repetition.

\begin{minted}[firstnumber=last]{python}
def stabiliser(h: History,
               perm_group: Iterable[PermGroupEl]) -> Sequence[PermGroupEl]:
    """
    Given a history and an iterable of event-value permutation group elements,
    returns the stabiliser subgroup for the history,
    as sequence of group elements.
    """
    orbit = list(history_perms(h, perm_group))
    return tuple(perm for perm, h_perm in orbit if h_perm == h)
\end{minted}

\subsection{Utilities for the \mintinline{python}{SpaceFinder} Class}
\label{appendix:space-finder-algo:section:utilities}

The \mintinline{python}{SpaceFinder} class encapsulates the data and code necessary to search for causally complete spaces on a given number of events.
Some of the required functionality is independent of the search state, and is implemented by the utilities functions below.
Given a collection of history bitvectors (as a sequence, without repetition), the \mintinline{python}{parents} function computes the mapping of histories to their sets of parents in the collection.

\begin{minted}[firstnumber=288]{python}
def parents(hs: Sequence[History]) -> dict[History, frozenset[History]]:
    """
    Given an ordered set of histories, returns the mapping of
    each history to the set of its parent histories (those of
    which they are a child history) in the set.
    """
    ps: dict[History, set[History]] = {h: set() for h in hs}
    for h in hs:
        for k in sorted(child_histories(h), key=history_sort_key):
            if k not in ps:
                ps[k] = set()
            ps[k].add(h)
    return {h: frozenset(ks) for h, ks in ps.items()}
\end{minted}

\noindent The \mintinline{python}{time_str} function provides a friendly representation of a time quantity, for use in search status updates.

\begin{minted}[firstnumber=last]{python}
def time_str(time: Union[int, float]) -> str:
    """ Formats a given time value in seconds, for printing. """
    # pylint: disable = too-many-return-statements
    assert time >= 0
    if time == 0:
        return "0s"
    if time < 1e-6:
        return f"{time*1e9:.2f}ns"
    if time < 1e-3:
        return f"{time*1e6:.2f}us"
    if time < 1:
        return f"{time*1e3:.2f}ms"
    if time < 60:
        return f"{time:.2f}s"
    time = int(time)
    if time < 60*60:
        return f"{time//60}m{time%60}s"
    if time < 24*3600:
        return f"{time//3600}h{(time%3600)//60}m"
    return f"{time//86400}d{(time%86400)//3600}h{(time%3600)//60}m"
\end{minted}

\noindent The \mintinline{python}{memory_str} function provides a friendly representation of a memory quantity, for use in search status updates.

\begin{minted}[firstnumber=last]{python}
def memory_str(mem: int) -> str:
    """ Formats a given memory value in bytes, for printing. """
    assert mem >= 0
    if mem < 1024:
        return f"{mem}B"
    if mem < 1024**2:
        return f"{mem/1024:.2f}KiB"
    if mem < 1024**3:
        return f"{mem/(1024**2):.2f}MiB"
    return f"{mem/(1024**3):.2f}GiB"
\end{minted}

\subsection{The \mintinline{python}{SpaceFinder} Class}

The \mintinline{python}{SpaceFinder} class encapsulates the data and code necessary to search for causally complete spaces on a given number of events.
This Appendix presents a simplified version of the class, with no state serialisation and no symmetry optimisaton of the top-level children subsets: for the advanced version, used in the search for causally complete spaces on 4 events, see \ref{appendix:space-finder-algo-adv} (p.\pageref{appendix:space-finder-algo-adv}).
To keep line numberings consistent with the advanced version, code omitted in the simplified version has been replaced by blank comment lines.

An object of the class is instantiated by passing a number \mintinline{python}{num_events} of events---exposed by the homonymous property---and the constructor pre-computes certain data for use during the search.
The optional \mintinline{python}{verbose} keyword argument (kwarg) determines whether status updates will be printed/logged;
the optional \mintinline{python}{update_period} kwarg determines the frequency of the updates (as a minimum number of equivalence classes discovered in between updates);

\begin{minted}[linenos=false]{python}
class SpaceFinder:
    def __init__(self, num_events: int, *,
                 verbose: bool = True,
                 update_period: Optional[int] = None) -> None:
        ...
    @property
    def num_events(self) -> int:
        ...
    ...
\end{minted}

\noindent Calling the \mintinline{python}{blank_state} method initialises the finder for a new search.

\begin{minted}[linenos=false]{python}
class SpaceFinder:
    ...
    def blank_state(self) -> None:
        ...
    ...
\end{minted}

\noindent After the search state has been set---to a blank state or to the state loaded from a file---the \mintinline{python}{find_eq_classes} method can be called to start the search, which will run until completion.

\begin{minted}[linenos=false]{python}
class SpaceFinder:
    ...
    def find_eq_classes(self) -> None:
        ...
    ...
\end{minted}

\noindent Below is a minimal example searching for causally complete spaces on 2 events: 

\begin{minted}[linenos=false]{python}
finder2 = SpaceFinder(2)
finder2.blank_state()
finder2.find_eq_classes()
\end{minted}

\noindent Basic benchmarking quantities are made available through public properties:
the \mintinline{python}{time_elapsed} property exposes the time elapsed since \mintinline{python}{find_eq_classes} was called;
the \mintinline{python}{memsize} property exposes an estimate of the amount of memory currently occupied by temporary data structures and discovered spaces;
the \mintinline{python}{perc_completed} property exposes an estimate of the percentage of search space explored so far.

\begin{minted}[linenos=false]{python}
class SpaceFinder:
    ...
    @property
    def time_elapsed(self) -> float:
        ...
    @property
    def memsize(self) -> int:
        ...
    @property
    def perc_completed(self) -> float:
        ...
    ...
\end{minted}

\noindent Below is the final portion of the output for the \mintinline{python}{finder2} 2-event example presented above, showing the time, memory and completion metrics as the algorithm progressed.

\begin{minted}{text}
[...]
      time       spaces    eq. cls     memory  completed
    2.75ms            4          1   14.20KiB   16.6667%
    2.87ms            5          2   14.22KiB   33.3333%
    2.96ms            5          2   14.22KiB   66.6667%
    3.04ms            5          2   14.22KiB   83.3333%
    3.13ms            7          3   14.25KiB  100.0000%
Found 7 spaces in 3 equivalence classes.
\end{minted}

\noindent The search process progressively uncovers equivalence classes of causally complete spaces under event-input permutation symmetry.
For reasons of memory efficiency, a single representative---the bitvector for the set of input histories in the space---is stored for each equivalence class: these representatives can be iterated using the \mintinline{python}{iter_eq_classes} property (with no guarantees on order), while their number is efficiently exposed by the \mintinline{python}{num_eq_classes} property.
The search process also efficiently computes the total number of causally complete spaces in the equivalence classes discovered so far, which is exposed by the \mintinline{python}{num_spaces} property.
The \mintinline{python}{iter_spaces} property, finally, iterates through all causally complete spaces discovered so far, by iterating through all equivalence classes and iterating over all spaces in each class by applying the event-input permutation group elements to the representative (without repetition of spaces); for each space, a pair is yielded, with the equivalence class representative as the first element and the permuted space as the second element.

\begin{minted}[linenos=false]{python}
class SpaceFinder:
    ...
    @property
    def num_eq_classes(self) -> int:
        ...
    @property
    def iter_eq_classes(self) -> Iterator[HistorySet]:
        ...
    @property
    def num_spaces(self) -> int:
        ...
    @property
    def iter_spaces(self) -> Iterator[tuple[HistorySet, HistorySet]]:
        ...
    ...
\end{minted}

\noindent In the \mintinline{python}{finder2} 2-event example presented above, there are 7 causally complete spaces in 3 equivalence classes.
Below we list both the equivalence class representatives and all spaces, using colours to highlight spaces in the same equivalence class.

\begin{minted}[linenos=false]{python}
print(f"Eq Classes = {list(finder2.iter_eq_classes)}")
print(f"All Spaces = {[space for _, space in finder2.iter_spaces]}")
\end{minted}

\begin{minted}[escapeinside=||]{text}
Eq Classes = [|\colorbox{evred}{1362}|, |\colorbox{evgreen}{1638}|, |\colorbox{evblue}{278}|]
All Spaces = [|\colorbox{evred}{1362, 820, 1558, 358}|, |\colorbox{evgreen}{1638, 1904}|, |\colorbox{evblue}{278}|]
\end{minted}

\noindent To understand which spaces were discovered, we look at the bitvector for each space (from \mintinline{python}{finder2.iter_spaces}), we extract the \mintinline{python}{int} representation of the bitvectors for the histories within (using \mintinline{python}{iter_bitvec}), and we then extract the event-input pairs for each history from the corresponding bitvector (using \mintinline{python}{history2dict}).

\begin{minted}[linenos=false]{python}
curr_eq_class_repr = None
for eq_class_repr, space in finder2.iter_spaces:
    if eq_class_repr != curr_eq_class_repr:
        curr_eq_class_repr = eq_class_repr
        print(f"Eq. class of space {eq_class_repr}:")
    print(f"  Input histories for space {space}:")
    for h in iter_bitvec(space):
        print(f"    {history2dict(h)}")
\end{minted}

\noindent The first equivalence class discovered contains the 4 spaces in the middle layer of the hierarchy of causally complete spaces on 2 events with binary inputs.

\begin{minted}[escapeinside=||]{text}
Eq. class of space |\colorbox{evred}{1362}|=10101010010:
  Input histories for space |\colorbox{evred}{1362}|=10101010010:
     1=0001 -> {'A': 0}
     4=0100 -> {'B': 0}
     6=0110 -> {'A': 1, 'B': 0}
     8=1000 -> {'B': 1}
    10=1010 -> {'A': 1, 'B': 1}
  Input histories for space |\colorbox{evred}{820}|=01100110100:
     2=0010 -> {'A': 1}
     4=0100 -> {'B': 0}
     5=0101 -> {'A': 0, 'B': 0}
     8=1000 -> {'B': 1}
     9=1001 -> {'A': 0, 'B': 1}
  Input histories for space |\colorbox{evred}{1558}|=11000010110:
     1=0001 -> {'A': 0}
     2=0010 -> {'A': 1}
     4=0100 -> {'B': 0}
     9=1001 -> {'A': 0, 'B': 1}
    10=1010 -> {'A': 1, 'B': 1}
  Input histories for space |\colorbox{evred}{358}|=00101100110:
     1=0001 -> {'A': 0}
     2=0010 -> {'A': 1}
     5=0101 -> {'A': 0, 'B': 0}
     6=0110 -> {'A': 1, 'B': 0}
     8=1000 -> {'B': 1}
\end{minted}

\noindent The second equivalence class discovered contains the 2 spaces in the top layer of the hierarchy of causally complete spaces on 2 events with binary inputs, i.e. the spaces for the 2 total orders.

\begin{minted}[escapeinside=||]{text}
Eq. class of space |\colorbox{evgreen}{1638}|=11001100110:
  Input histories for space |\colorbox{evgreen}{1638}|=11001100110:
     1=0001 -> {'A': 0}
     2=0010 -> {'A': 1}
     5=0101 -> {'A': 0, 'B': 0}
     6=0110 -> {'A': 1, 'B': 0}
     9=1001 -> {'A': 0, 'B': 1}
    10=1010 -> {'A': 1, 'B': 1}
  Input histories for space |\colorbox{evgreen}{1904}|=11101110000:
     4=0100 -> {'B': 0}
     5=0101 -> {'A': 0, 'B': 0}
     6=0110 -> {'A': 1, 'B': 0}
     8=1000 -> {'B': 1}
     9=1001 -> {'A': 0, 'B': 1}
    10=1010 -> {'A': 1, 'B': 1}
\end{minted}

\noindent The third equivalence class discovered contains the space in the bottom layer of the hierarchy of causally complete spaces on 2 events with binary inputs, i.e. the discrete space.

\begin{minted}[escapeinside=||]{text}
Eq. class of space |\colorbox{evblue}{278}|=00100010110:
  Input histories for space |\colorbox{evblue}{278}|=00100010110:
     1=0001 -> {'A': 0}
     2=0010 -> {'A': 1}
     4=0100 -> {'B': 0}
     8=1000 -> {'B': 1}
\end{minted}

\noindent In the coming sub-subsections, we will review the entire code for the \mintinline{python}{SpaceFinder} class.

\subsection{\mintinline{python}{SpaceFinder} Class -- Constructor}

The constructor accepts some search options (already discussed above) and pre-computes some data, to improve the time performance of the search process.
Data pre-computed includes:
the sequence \mintinline{python}{self._events} of events;
the sequence \mintinline{python}{self._max_histories} of maximal extended input histories;
the sequence \mintinline{python}{self._perm_group} of event-input permutation group elements;
a dictionary \mintinline{python}{self._histories_perm_dict} mapping permutation group elements to their action on histories (as a history-history mapping);
a dictionary \mintinline{python}{self._children_set} mapping histories to the \mintinline{python}{frozenset} of their children;
a dictionary \mintinline{python}{self._children} mapping histories to the list of their children;
a dictionary \mintinline{python}{self._parents} mapping histories to the \mintinline{python}{frozenset} of their parents;
a dictionary \mintinline{python}{self._domsize} mapping histories to the size of their domain;
a boolean \mintinline{python}{self._initialised} indicating whether the search state has been initialised (by calling the \mintinline{python}{blank_state} or \mintinline{python}{load_state} method);
the maximum number \mintinline{python}{self._max_space_size} of bytes occupied by the bitvector for a space of input histories;
the number \mintinline{python}{self._fixed_memsize} of bytes occupied by the data structures computed above.

\begin{minted}[firstnumber=371]{python}

class SpaceFinder:
    # pylint: disable = too-many-instance-attributes
    """
    An instance of this class encapsulates all the data used in the
    search for causally complete spaces on a given number of events.
    """

    # Options:
    _num_events: int
    _verbose: bool
    _update_period: Optional[int]
#
#
#

    # Pre-computed data:
    _events: tuple[Event, ...]
    _max_histories: Sequence[History]
    _perm_group: tuple[PermGroupEl, ...]
    _histories_perm_dict: dict[PermGroupEl, dict[History, History]]
    _children_set: dict[History, frozenset[History]]
    _children: dict[History, tuple[History, ...]]
    _parents: dict[History, frozenset[History]]
    _domsize: dict[History, int]
    _max_space_size: int
    _initialised: bool
    _fixed_memsize: int

    def __init__(self, num_events: int,
                 *, verbose: bool = True,
                 update_period: Optional[int] = None) -> None:
#
#
#
        """
        Creates a new space finder instance, which will search for causally
        complete spaces on the given number of events.

        :param num_events: the number of events
        :param verbose: whether to print status updates, defaults to True
        :param update_period: how frequently to print status updates
                              (min number of equivalence classes discovered
                              between updates), defaults to None (print update
                              after every each top-level cycle)
        :param filename: internal state is regularly saved to a binary file with
                         this name, defaults to None (don't save internal state)
        :param save_period: how frequently to save internal state (min number of
                            equivalence classes discovered between saves),
                            defaults to None (save only once, at the end)
        """
        assert update_period is None or update_period > 0
        self._num_events = num_events
        self._verbose = verbose
        self._update_period = update_period
#
#
#
#
#
#
#
        self._max_histories = max_histories(num_events)
        self._events = tuple(idx2event(idx) for idx in range(num_events))
        self._perm_group = tuple(iter_perm_group(self._events))
        hs = sorted(sub_histories(self._max_histories),
                    key=history_sort_key)
        _histories_perm_dict = {
            h: dict(history_perms(h, self._perm_group))
            for h in hs
        }
        self._histories_perm_dict = {
            p: {
                h: _histories_perm_dict[h][p]
                for h in hs
            }
            for p in self._perm_group
        }
        self._children_set = {
            h: frozenset(child_histories(h))
            for h in hs
        }
        self._children = {
            h: tuple(sorted(self._children_set[h]))
            for h in hs
        }
        self._parents = parents(hs)
        self._domsize = {h: domsize(h) for h in hs}
        self._max_space_size = sys.getsizeof(2**(2**(2*len(self._events)))-1)
        self._initialised = False
        self._fixed_memsize = getsize(self)
\end{minted}

\subsection{\mintinline{python}{SpaceFinder} Class -- Spaces}

As new equivalence classes of causally complete spaces are discovered during the search, a single representative for the equivalence class is added to \mintinline{python}{self._eq_classes}, a mutable \mintinline{python}{set}, and the number of spaces in the equivalence class is added to \mintinline{python}{self._num_spaces}.
The \mintinline{python}{num_events}, \mintinline{python}{num_eq_classes} and \mintinline{python}{num_spaces} properties expose read-only access to the number of events, of equivalence classes discovered so far, and of spaces discovered so far.
The \mintinline{python}{iter_eq_classes} property exposes an iterator over the set of representatives for equivalence classes discovered so far, while the \mintinline{python}{iter_spaces} property exposes an iterator over all causally complete spaces discovered so far (computed on the fly by permuting each equivalence class representative in turn).

\begin{minted}[firstnumber=last]{python}
    _eq_classes: set[HistorySet]
    _num_spaces: int

    @property
    def num_events(self) -> int:
        """
        Number of events.
        """
        return self._num_events

    @property
    def num_eq_classes(self) -> int:
        """
        Number of equivalence classes of causally complete spaces,
        discovered so far.
        """
        return len(self._eq_classes)

    @property
    def iter_eq_classes(self) -> Iterator[HistorySet]:
        """
        Iterate over the equivalence classes of causally complete spaces
        discovered so far, yielding a representative space for
        each equivalence class.
        """
        return iter(self._eq_classes)

    @property
    def num_spaces(self) -> int:
        """
        Number of causally complete spaces discovered so far.
        """
        return self._num_spaces

    @property
    def iter_spaces(self) -> Iterator[tuple[HistorySet, HistorySet]]:
        """
        Iterate over the causally complete spaces discovered so far.
        """
        for eq_class_repr in self.iter_eq_classes:
            space_set = bitvec2set(eq_class_repr)
            eq_class = set()
            for g in self._perm_group:
                g_action = self._histories_perm_dict[g]
                space_set_perm = {g_action[h] for h in space_set}
                space_perm = bitvec(space_set_perm)
                if space_perm not in eq_class:
                    eq_class.add(space_perm)
                    yield eq_class_repr, space_perm
\end{minted}

\subsection{\mintinline{python}{SpaceFinder} Class -- Search Metrics}

The search algorithm is recursive and operates by levels, corresponding to histories with progressively decreasing number of events in their domain.
At the top level, the algorithm iterates through sets of children for the maximal extended input histories.
After an initial step, the number of sets of children to be iterated over is stored in \mintinline{python}{self._num_todo}, while \mintinline{python}{self._num_done} is increased by one after each top-level search step is completed: the ratio \mintinline{python}{self._num_done/self._num_todo} is used as a rough estimate of completion, exposed by the \mintinline{python}{perc_completed} property.
Note that this estimate is very rough: most spaces are discovered at the beginning, so that later iterations are likely to proceed significantly faster than early ones.
We'll explore precisely how this subset selection is done later on, when talking about the search logic.

\begin{minted}[firstnumber=last]{python}
    _num_done: int
    _num_todo: int

    @property
    def perc_completed(self) -> float:
        """
        Percent estimate of search completion.
        """
        return self._num_done/self._num_todo
\end{minted}

\noindent The search status printouts also include time and memory columns.
The memory size of fixed data structures is computed exactly by the class constructor, stored in \mintinline{python}{self._fixed_memsize}, while the size of variable data structures is given as an upper-bound, computed from their length and the maximum size of a history space bitvector for the given number of events.

\begin{minted}[firstnumber=542]{python}
    _start_time: float
    _partial_spaces_visited: set[HistorySet]

    @property
    def time_elapsed(self) -> float:
        """
        Time elapsed since search started.
        """
        return perf_counter()-self._start_time

    @property
    def memsize(self) -> int:
        """
        An estimate of memory currently occupied by the auxiliary data (bytes).
        """
        partial_visited = self._partial_spaces_visited
        full_visited = self._eq_classes
        child_choices_list = self._child_choices_list
        remaining_children_list = self._remaining_children_list
        max_space_size = self._max_space_size
        memsize = self._fixed_memsize
        memsize += (sys.getsizeof(partial_visited)
                    +len(partial_visited)*max_space_size)
        memsize += (sys.getsizeof(full_visited)
                    +len(full_visited)*max_space_size)
        memsize += (sys.getsizeof(child_choices_list)
                    +len(child_choices_list)*max_space_size)
        memsize += (sys.getsizeof(remaining_children_list)
                    +len(remaining_children_list)*max_space_size)
        return memsize
\end{minted}

\noindent Finally, some utility functions are implemented to print and log all status data.

\begin{minted}[firstnumber=last]{python}
    def _print(self, *args: str, **kwargs: Any) -> None:
        """ Prints to console. """
        print(*args, **kwargs)
#
#

    def _print_status_header(self) -> None:
        if self._verbose:
            header = (f"{'time': >10} {'spaces': >12} {'eq. cls': >10}"
                      f" {'memory': >10} {'completed': >10}")
#
            self._print(header, flush=True)

    def _print_status_line(self) -> None:
        if self._verbose:
            time = time_str(self.time_elapsed)
            spaces = str(self.num_spaces)
            eq_classes = str(self.num_eq_classes)
            memory = memory_str(self.memsize)
            completed = f"{self.perc_completed:.4%}"
#
#
            line = (f"{time: >10} {spaces: >12} {eq_classes: >10}"
                    f" {memory: >10} {completed: >10}")
#
            self._print(line, flush=True)

    def _describe(self) -> None:
        """
            If in verbose mode,
            prints a summary description of the search results (so far).
        """
        if self._verbose:
            self._print(f"Found {self.num_spaces} spaces in "
                       f"{self.num_eq_classes} equivalence classes.")
\end{minted}

\subsection{\mintinline{python}{SpaceFinder._find_eq_classes}}

The private \mintinline{python}{_find_eq_classes} method contains the main space-finding logic.
It operates recursively by levels, each level corresponding to a set of histories \mintinline{python}{set(new)_hs} with a fixed number of events: 
\begin{enumerate}
    \item The \mintinline{python}{new_hs} sequence contains the candidate input histories for this level: in the search on $n$ events, level $l$ corresponds to input histories with $n-l$ events (where $l=0, ..., n-2$).
    \item The \mintinline{python}{hs} sequence contains all candidate input histories thus far, excluding those in \mintinline{python}{new_hs}.
    \item The \mintinline{python}{hs_rest} sequence has the same length as \mintinline{python}{hs} and it contains the difference between each history in \mintinline{python}{h in hs} and all the sub-histories \mintinline{python}{k in sub_histories([h])} which have been discovered so far. If this difference becomes empty, the history \mintinline{python}{h} is guaranteed to not be $\vee$-prime, and hence is removed from the candidate input histories (later on, we refer to this process as ``winnowing'').
\end{enumerate}
In the top-level (\mintinline{python}{level=0}) call to \mintinline{python}{_find_eq_classes} that starts the search, \mintinline{python}{new_hs} is set to the maximal extended input histories on the given number of events, while \mintinline{python}{hs} and \mintinline{python}{hs_rest} are set to empty.

\begin{minted}[firstnumber=691]{python}
    def _find_eq_classes(self, new_hs: Sequence[History],
                         hs: Sequence[History] = tuple(),
                         hs_rest: Sequence[History] = tuple(),
                         level: int = 0) -> Iterator[HistorySet]:
        # pylint: disable = too-many-locals
        hs_so_far = tuple(chain(new_hs, hs))
        hs_perm_dict = self._histories_perm_dict
        partial_spaces_visited = self._partial_spaces_visited
        spaces_visited = self._eq_classes
#
#
#
#
        for child_subset in self._iter_child_subsets(new_hs):
            child_subset_sorted = sorted(child_subset, key=history_sort_key)
            hs_so_far_rest = list(chain(new_hs, hs_rest))
            for k in child_subset_sorted:
                for j, h in enumerate(hs_so_far):
                    if is_subset(k, h):
                        hs_so_far_rest[j] = sub(hs_so_far_rest[j], k)
            winnowed_hs = tuple(h for j, h in enumerate(hs_so_far)
                                       if hs_so_far_rest[j])
            winnowed_hs_rest = tuple(
                h for h in hs_so_far_rest if h)
            partial_space = set(chain(child_subset_sorted, winnowed_hs))
            partial_space_bitvec = bitvec(partial_space)
            already_seen = (partial_space_bitvec in partial_spaces_visited
                            or partial_space_bitvec in spaces_visited)
            if not already_seen:
                eq_class = set()
                for p in self._perm_group:
                    p_action = hs_perm_dict[p]
                    perm_partial_space_bitvec = bitvec(p_action[h]
                                                       for h in partial_space)
                    eq_class.add(perm_partial_space_bitvec)
                    if (perm_partial_space_bitvec in partial_spaces_visited
                        or perm_partial_space_bitvec in spaces_visited):
                        already_seen = True
                        break
            if not already_seen:
                if all(self._domsize[h] == 1 for h in child_subset_sorted):
                    self._num_spaces += len(eq_class)
                    yield partial_space_bitvec
                else:
                    partial_spaces_visited.add(partial_space_bitvec)
                    yield from self._find_eq_classes(child_subset_sorted,
                                                     winnowed_hs,
                                                     winnowed_hs_rest,
                                                     level+1)
\end{minted}

\noindent At the start of the method, the sequence \mintinline{python}{hs_so_far} of all current candidate input histories is computed, and some persistent data structures are accessed and given shorter names.

\begin{minted}[firstnumber=696]{python}
        hs_so_far = tuple(chain(new_hs, hs))
        hs_perm_dict = self._histories_perm_dict
        partial_spaces_visited = self._partial_spaces_visited
        spaces_visited = self._eq_classes
\end{minted}

\noindent The rest of the method iterates over all subsets of child histories for the histories in \mintinline{python}{new_hs}.
Two different methods are used: the symmetry-optimised \mintinline{python}{self._iter_child_subsets_toplevel} is used at top-level (\mintinline{python}{level=0}), while the general purpose \mintinline{python}{self._iter_child_subsets} is used for all other levels.
In principle, the \mintinline{python}{self._iter_child_subsets} method could be used to iterate over child subsets at all levels, but the added efficiency of the \mintinline{python}{self._iter_child_subsets_toplevel} is helpful in the search on 4 events (plus, the symmetry argument is interesting in itself).
Each child subset yielded is sorted to make the search deterministic.

\begin{minted}[firstnumber=700]{python}
#
#
#
#
        for child_subset in self._iter_child_subsets(new_hs):
            child_subset_sorted = sorted(child_subset, key=history_sort_key)
\end{minted}

\noindent For each chosen subset \mintinline{python}{child_subset} of child histories, we wish to compute the part of each candidate input history in \mintinline{python}{hs_so_far} that is not covered by any sub-history discovered so far.
Excluding any sub-histories in \mintinline{python}{child_subset}, the entirety of each history in \mintinline{python}{new_hs} is uncovered---because we haven't seen any of their sub-histories yet---while the ``uncovered'' part for the histories in \mintinline{python}{hs} is given by \mintinline{python}{hs_rest}.
Hence, we initialise the sequence of ``uncovered'' parts for \mintinline{python}{hs_so_far} to \mintinline{python}{hs_so_far_rest}.
For each child history \mintinline{python}{k in child_subset}, we then proceed to subtract \mintinline{python}{k} from all its super-histories in \mintinline{python}{hs_so_far}, updating the corresponding ``uncovered'' part in \mintinline{python}{hs_so_far_rest}.
At the end, \mintinline{python}{winnowed_hs} contains the histories that have non-empty ``uncovered'' part (i.e. those which might still be $\vee$-prime), and \mintinline{python}{winnowed_hs_rest} contains the corresponding ``uncovered'' parts, to be used by recursive calls to \mintinline{python}{_find_eq_classes}.

\begin{minted}[firstnumber=706]{python}
            hs_so_far_rest = list(chain(new_hs, hs_rest))
            for k in child_subset_sorted:
                for j, h in enumerate(hs_so_far):
                    if is_subset(k, h):
                        hs_so_far_rest[j] = sub(hs_so_far_rest[j], k)
            winnowed_hs = tuple(h for j, h in enumerate(hs_so_far)
                                       if hs_so_far_rest[j])
            winnowed_hs_rest = tuple(
                h for h in hs_so_far_rest if h)
\end{minted}

\noindent The new set of candidate input histories contains the currently selected subset of child histories together with the histories that survived the winnowing step above: taken together, they form \mintinline{python}{partial_space}, a ``partial'' space of input histories so far (which is not causally complete unless we are at the bottom level $l=n-2$).
Before proceeding, we check whether we have already encountered this ``partial'' space, or any of its permutations under event-input permutation symmetry; in the process, we also compute its equivalence class.

\begin{minted}[firstnumber=715]{python}
            partial_space = set(chain(child_subset_sorted, winnowed_hs))
            partial_space_bitvec = bitvec(partial_space)
            already_seen = (partial_space_bitvec in partial_spaces_visited
                            or partial_space_bitvec in spaces_visited)
            if not already_seen:
                eq_class = set()
                for p in self._perm_group:
                    p_action = hs_perm_dict[p]
                    perm_partial_space_bitvec = bitvec(p_action[h]
                                                       for h in partial_space)
                    eq_class.add(perm_partial_space_bitvec)
                    if (perm_partial_space_bitvec in partial_spaces_visited
                        or perm_partial_space_bitvec in spaces_visited):
                        already_seen = True
                        break
\end{minted}

\noindent If \mintinline{python}{partial_space} was not encountered before, we proceed.
If we reached the bottom level---i.e. if all child histories have exactly 1 event---then \mintinline{python}{partial_space} is a representative in a newly discovered equivalence class of causally complete spaces, so we update the total number of spaces and we yield the representative.
If we instead haven't reached the bottom level, then we register the ``partial'' space \mintinline{python}{partial_space} as visited and recurse to one level below.

\begin{minted}[firstnumber=730]{python}
            if not already_seen:
                if all(self._domsize[h] == 1 for h in child_subset_sorted):
                    self._num_spaces += len(eq_class)
                    yield partial_space_bitvec
                else:
                    partial_spaces_visited.add(partial_space_bitvec)
                    yield from self._find_eq_classes(child_subset_sorted,
                                                     winnowed_hs,
                                                     winnowed_hs_rest,
                                                     level+1)
\end{minted}

\subsection{\mintinline{python}{SpaceFinder.find_eq_classes}}

The \mintinline{python}{find_eq_classes} method checks that the finder instance has been initialised, sets start time and the \mintinline{python}{self._num_eq_classes_since_last_save} counter, and then starts the search by calling \mintinline{python}{_find_eq_classes} on the maximal extended input histories for the chosen number of events.
For each equivalence class discovered, the representative yielded by \mintinline{python}{_find_eq_classes} is added to the set of equivalence classes, the \mintinline{python}{self._num_eq_classes_since_last_save} counter is increased, and a new line in the search status table is printed (if required).
When all equivalence classes have been discovered, a final line in the search status table is printed (if required), the state is saved (if required) and the number of discovered spaces and equivalence classes is printed (if required).

\begin{minted}[firstnumber=740]{python}
    def find_eq_classes(self) -> None:
        """
            Start the search for equivalence classes.
        """
        if not self._initialised:
            raise Exception("Must initialise using blank_state() "
                            "or load_state(file).")
        update_period = self._update_period
        self._start_time = perf_counter()
        self._num_eq_classes_since_last_save = 0
        init_histories = max_histories(self.num_events)
        for eq_class_repr in self._find_eq_classes(init_histories):
            self._eq_classes.add(eq_class_repr)
            self._num_eq_classes_since_last_save += 1
            if (update_period is not None
                and self.num_eq_classes % update_period == 0):
                self._print_status_line()
        if update_period is not None:
            self._print_status_line()
        self._partial_spaces_visited.clear()
#
        self._describe()
\end{minted}

\subsection{\mintinline{python}{SpaceFinder._iter_child_subsets}}

Given a sequence \mintinline{python}{hs} of histories, the private \mintinline{python}{_iter_child_subsets} method iterates through all subsets \mintinline{python}{child_subset} of \mintinline{python}{child_histories(hs)} such that each history in \mintinline{python}{h in hs} has at least one of its child histories in \mintinline{python}{child_subset}.
This is done by sorting all child histories and then iterating through all bitvectors \mintinline{python}{child_subset_bitvec} for their non-empty subsets: for each bitvector, the call to the \mintinline{python}{_child_subset} returns the corresponding subset if it satisfies the condition above, or \mintinline{python}{None} if it doesn't.

\begin{minted}[firstnumber=last]{python}
    def _iter_child_subsets(self,
                            hs: Sequence[History]
                            ) -> Iterator[set[History]]:
        child_hists: list[History] = sorted(
            {k for h in hs for k in self._children[h]},
            key=history_sort_key)
        num_child_subsets = 2**len(child_hists)
        for child_subset_bitvec in range(1, num_child_subsets):
            child_subset = self._child_subset(hs, child_hists,
                                              child_subset_bitvec)
            if child_subset is not None:
                yield child_subset
\end{minted}

\noindent In principle, this method could be used to iterate over child subsets at all levels.
However, the simplicity of its logic is paid for by the unnecessary inefficiency at the top level, where event-input permutation symmetry can be used to reduce the number of subsets that need to be explored.
For 2 events, symmetry reduces the number of subsets by about 3 times, from 16 to 6;
for 3 events, it reduces the number by about 4 times, from 4096 to 922;
for 4 events, it reduces the number by about 13 times, from 4294967296 to 315981136.
For 5 events, symmetry reduces the number of subsets by 58 times or more; however, this doesn't really matter, since the number of top-level subsets to iterate over has in excess of 20 decimal digits, making a search on 5 events impossible with our algorithm anyway.

\subsection{\mintinline{python}{SpaceFinder._child_subset}}

The private \mintinline{python}{_child_subsets} method takes as its arguments a sequence \mintinline{python}{hs} of histories, a sequence \mintinline{python}{child_hists} of their children and the bitvector for a non-empty subset of children.

\begin{minted}[firstnumber=last]{python}
    def _child_subset(self,
                      hs: Sequence[History],
                      child_hists: Sequence[History],
                      child_subset_bitvec: int,
#
#
                      ) -> Optional[set[History]]:
        # pylint: disable = too-many-arguments
        hs_still_to_cover = set(hs)
        child_subset = set()
        idx = 0
        while child_subset_bitvec > 0:
            child_subset_bitvec, b = divmod(child_subset_bitvec, 2)
            if b:
                k = child_hists[idx]
                child_subset.add(k)
                if hs_still_to_cover:
                    hs_still_to_cover -= self._parents[k]
            idx += 1
        if not hs_still_to_cover:
            return child_subset
        return None
\end{minted}

\noindent The method starts by creating a set \mintinline{python}{hs_to_cover} of histories in \mintinline{python}{hs} still to be ``covered''---i.e. where at least one child has to yet appear in the subets---and a set \mintinline{python}{child_subset} of children extracted from the bitvector.

\begin{minted}[firstnumber=782]{python}
        hs_still_to_cover = set(hs)
        child_subset = set()
\end{minted}

\noindent The method then proceeds to iterate through the bits in the subset bitvector, including into \mintinline{python}{child_subset} any child history \mintinline{python}{k} whose bit is set to 1.
The parents of \mintinline{python}{k} are then subtracted from \mintinline{python}{hs_still_to_cover} (unnecessary if it's already empty).

\begin{minted}[firstnumber=784]{python}
        idx = 0
        while child_subset_bitvec > 0:
            child_subset_bitvec, b = divmod(child_subset_bitvec, 2)
            if b:
                k = child_hists[idx]
                child_subset.add(k)
                if hs_still_to_cover:
                    hs_still_to_cover -= self._parents[k]
            idx += 1
\end{minted}

\noindent After all children indicated by the bitvector have been added to \mintinline{python}{child_subset}, we check whether there are any histories in \mintinline{python}{hs_still_to_cover}: if not, the subset is yielded; if so, \mintinline{python}{None} is yielded instead, to indicate that the subsets should not be explored.

\begin{minted}[firstnumber=793]{python}
        if not hs_still_to_cover:
            return child_subset
        return None
\end{minted}

\section{Advanced Algorithm to Find Causally Complete Spaces}
\label{appendix:space-finder-algo-adv}

This Appendix presents an advanced version of the \mintinline{python}{SpaceFinder} class from \ref{appendix:space-finder-algo}, including state serialisation features and symmetry-based reduction of the children history subsets at top-level.
We resume our story from \ref{appendix:space-finder-algo:section:utilities}, which in the advanced version includes further utility functions.

\subsection{More Utilities for the \mintinline{python}{SpaceFinder} Class}

\noindent Given a file handler (as a \mintinline{python}{BinaryIO} instance) and a collection of bitvectors for sets of histories, the \mintinline{python}{write_hsets} function serializes the bitvectors in a compact form, using the minimum number of bytes for each one.

\begin{minted}[firstnumber=331]{python}
HistorySet = Bitvec
""" Type alias for a set of histories, as a bitvector. """

def write_hsets(f: BinaryIO, hsets: Collection[HistorySet]) -> int:
    """ Writes a collection of history sets (as bitvectors) to a given file. """
    num_hsets = len(hsets)
    f.write(num_hsets.to_bytes(8, byteorder="big"))
    num_bytes_written = 8
    for hs in hsets:
        num_bytes = max(int(ceil(hs.bit_length()/8)), 1)
        f.write(num_bytes.to_bytes(2, byteorder="big"))
        hs_bytes = hs.to_bytes(num_bytes, byteorder="big")
        f.write(hs_bytes)
        num_bytes_written += num_bytes + 2
    return num_bytes_written
\end{minted}

\noindent Given a file handler (as a \mintinline{python}{BinaryIO} instance), the \mintinline{python}{_read_hsets} function iterates through all history set bitvectors serialised in the file by the \mintinline{python}{write_hsets} function.

\begin{minted}[firstnumber=last]{python}
def _read_hsets(f: BinaryIO) -> Iterator[HistorySet]:
    num_spaces = int.from_bytes(f.read(8), byteorder="big")
    for _ in range(num_spaces):
        num_space_bytes = int.from_bytes(f.read(2), byteorder="big")
        space = int.from_bytes(f.read(num_space_bytes), byteorder="big")
        assert max(int(ceil(space.bit_length()/8)), 1) == num_space_bytes
        yield space
\end{minted}

\noindent Given a file handler (as a \mintinline{python}{BinaryIO} instance), the \mintinline{python}{read_hsets_set} function returns the \mintinline{python}{set} of all history set bitvectors serialised in the file by the \mintinline{python}{write_hsets} function.

\begin{minted}[firstnumber=last]{python}
def read_hsets_set(f: BinaryIO) -> set[HistorySet]:
    """
    
    Reads a set of history sets (as bitvectors) from a given file.
    Returns a :obj:`list` or :obj:`set` depending on the value of ``mode``.
    
    """
    spaces_set: set[Bitvec] = set(_read_hsets(f))
    return spaces_set
\end{minted}

\noindent Given a file handler (as a \mintinline{python}{BinaryIO} instance), the \mintinline{python}{read_hsets_set} function returns the \mintinline{python}{list} of all history set bitvectors serialised in the file by the \mintinline{python}{write_hsets} function.

\begin{minted}[firstnumber=last]{python}
def read_hsets_list(f: BinaryIO) -> list[HistorySet]:
    """
    
    Reads a list of history sets (as bitvectors) from a given file.
    Returns a :obj:`list` or :obj:`set` depending on the value of ``mode``.
    
    """
    spaces_list: list[Bitvec] = list(_read_hsets(f))
    return spaces_list
\end{minted}

\subsection{More on the \mintinline{python}{SpaceFinder} Class}

The \mintinline{python}{SpaceFinder} class encapsulates the data and code necessary to search for causally complete spaces on a given number of events.
An object of the class is instantiated by passing a number \mintinline{python}{num_events} of events---exposed by the homonymous property---and the constructor pre-computes certain data for use during the search.
The optional \mintinline{python}{verbose} keyword argument (kwarg) determines whether status updates will be printed/logged;
the optional \mintinline{python}{update_period} kwarg determines the frequency of the updates (as a minimum number of equivalence classes discovered in between updates);
the optional \mintinline{python}{filename} kwarg determines whether the finder state will be saved to a file (at least once, at the end);
the optional \mintinline{python}{save_period} kwarg determines whether how frequently the finder state will be saved (as a minimum number of equivalence classes discovered in between updates);
the optional \mintinline{python}{logger} kwarg can be used to supply a logger for use when printing updates (which are then both printed to console and logger using the logger).

\begin{minted}[linenos=false]{python}
class SpaceFinder:
    def __init__(self, num_events: int, *,
                 verbose: bool = True,
                 update_period: Optional[int] = None,
                 filename: Optional[str] = None,
                 save_period: Optional[int] = None,
                 logger: Optional[Logger] = None) -> None:
        ...
    @property
    def num_events(self) -> int:
        ...
    ...
\end{minted}

\noindent Calling the \mintinline{python}{blank_state} method initialises the finder for a new search, while calling the \mintinline{python}{load_state} method loads previously saved search state from a file.

\begin{minted}[linenos=false]{python}
class SpaceFinder:
    ...
    def blank_state(self) -> None:
        ...
    def load_state(self, filename: str) -> None:
        ...
    ...
\end{minted}

\subsection{More on the \mintinline{python}{SpaceFinder} Class -- Constructor}

\begin{minted}[firstnumber=371]{python}

class SpaceFinder:
    # pylint: disable = too-many-instance-attributes
    """
    An instance of this class encapsulates all the data used in the
    search for causally complete spaces on a given number of events.
    """

    # Options:
    _num_events: int
    _verbose: bool
    _update_period: Optional[int]
    _toplevel_opt_depth: Optional[int]
    _save_options: Optional[tuple[str, Optional[int], bool]]
    _logger: Optional[Logger]

    # Pre-computed data:
    _events: tuple[Event, ...]
    _max_histories: Sequence[History]
    _perm_group: tuple[PermGroupEl, ...]
    _histories_perm_dict: dict[PermGroupEl, dict[History, History]]
    _children_set: dict[History, frozenset[History]]
    _children: dict[History, tuple[History, ...]]
    _parents: dict[History, frozenset[History]]
    _domsize: dict[History, int]
    _max_space_size: int
    _initialised: bool
    _fixed_memsize: int

    def __init__(self, num_events: int,
                 *, verbose: bool = True,
                 update_period: Optional[int] = None,
                 filename: Optional[str] = None,
                 save_period: Optional[int] = None,
                 logger: Optional[Logger] = None) -> None:
        """
        Creates a new space finder instance, which will search for causally
        complete spaces on the given number of events.

        :param num_events: the number of events
        :param verbose: whether to print status updates, defaults to True
        :param update_period: how frequently to print status updates
                              (min number of equivalence classes discovered
                              between updates), defaults to None (print update
                              after every each top-level cycle)
        :param filename: internal state is regularly saved to a binary file with
                         this name, defaults to None (don't save internal state)
        :param save_period: how frequently to save internal state (min number of
                            equivalence classes discovered between saves),
                            defaults to None (save only once, at the end)
        """
        assert update_period is None or update_period > 0
        self._num_events = num_events
        self._verbose = verbose
        self._update_period = update_period
        self._toplevel_opt_depth = None # search to arbitrary depth
        if filename is not None:
            save_backup: bool = True
            self._save_options = (filename, save_period, save_backup)
        else:
            self._save_options = None
        self._logger = logger
        self._max_histories = max_histories(num_events)
        self._events = tuple(idx2event(idx) for idx in range(num_events))
        self._perm_group = tuple(iter_perm_group(self._events))
        hs = sorted(sub_histories(self._max_histories),
                    key=history_sort_key)
        _histories_perm_dict = {
            h: dict(history_perms(h, self._perm_group))
            for h in hs
        }
        self._histories_perm_dict = {
            p: {
                h: _histories_perm_dict[h][p]
                for h in hs
            }
            for p in self._perm_group
        }
        self._children_set = {
            h: frozenset(child_histories(h))
            for h in hs
        }
        self._children = {
            h: tuple(sorted(self._children_set[h]))
            for h in hs
        }
        self._parents = parents(hs)
        self._domsize = {h: domsize(h) for h in hs}
        self._max_space_size = sys.getsizeof(2**(2**(2*len(self._events)))-1)
        self._initialised = False
        self._fixed_memsize = getsize(self)
\end{minted}

\subsection{More on the \mintinline{python}{SpaceFinder} Class -- Search Metrics}

In the advanced algorithm, the subsets of children being iterated over at the top level are obtained in two steps: we iterate through ``fixed'' subsets of children, and for each such ``fixed'' subset we iterate through all possible subsets of the corresponding ``variable'' children.
The list of ``fixed'' subsets is stored in \mintinline{python}{self._child_choices_list}, and the index of the ``fixed'' subset currently being explored is stored in \mintinline{python}{self._fix_child_choice_idx}.
A coarser estimate of search completion can be obtained by considering the fraction of ``fixed'' subsets that have been completely explored.

\begin{minted}[firstnumber=last]{python}
    _fix_child_choice_idx: int
    _child_choices_list: list[HistorySet]

    @property
    def fixed_toplevel_subsets_perc_completed(self) -> float:
        """
        Percent of top-level fixed child history subsets explored.
        """
        return self._fix_child_choice_idx/len(self._child_choices_list)
\end{minted}

\noindent The list of ``variable'' children subsets corresponding to the ``fixed'' subsets above is stored in \mintinline{python}{self._remaining_children_list}, the number of children in each subset is stored in \mintinline{python}{self._num_remaining_children} and the index of the ``variable'' subset currently being explored is stored in \mintinline{python}{self._var_child_subset_bitvec}.
An estimate of completion for each ``fixed'' child subset is given by the the fraction of corresponding ``variable'' child subsets that have been completely explored (out of \mintinline{python}{2**self._num_remaining_children[self._fix_child_choice_idx]} total).

\begin{minted}[firstnumber=last]{python}
    _var_child_subset_bitvec: int
    _remaining_children_list: list[HistorySet]
    _num_remaining_children: list[int]

    @property
    def var_toplevel_subsets_perc_completed(self) -> float:
        """
        Percent of top-level variable child history subsets explored
        for the current fixed child history subset.
        """
        if self._fix_child_choice_idx == len(self._child_choices_list):
            return 1.0
        curr_rem_ch = self._num_remaining_children[self._fix_child_choice_idx]
        return self._var_child_subset_bitvec/(2.0**curr_rem_ch)
\end{minted}

\noindent In the search for causally complete spaces on 2-event, there are 4 ``fixed'' subsets: the first and second with 2 ``variable'' subsets each, the third and fourth with 1 ``variable'' subset each, for a total of $2+2+1+1 = 6$ top level subsets to explore.
For concision, we print the history bitvectors rather than their full form: bitvector $1=2^0$ is $\hist{A/0}$, bitvector $2=2^1$ is $\hist{A/1}$, bitvector $4=2^2$ is $\hist{B/0}$ and bitvector $8=2^3$ is $\hist{B/1}$.

\begin{minted}[linenos=false]{python}
print(f"       Fixed | Variable  ->  Top-level subset")
for mand_subs_bv, rem_child_bv in zip(finder2._child_choices_list,
                                      finder2._remaining_children_list):
    mandatory_subset = bitvec2set(mand_subs_bv)
    remaining_children = list(iter_bitvec(rem_child_bv))
    num_opt_subsets = 2**len(remaining_children)
    for subs_indicator in range(num_opt_subsets):
        optional_subset = {
            h for i, h in enumerate(remaining_children)
            if 2**i&subs_indicator > 0
        }
        toplevel_subset = mandatory_subset|optional_subset
        optional_subset_str = ('{}' if not optional_subset
                               else str(optional_subset))
        print(f"{str(mandatory_subset): >12}"
              f" | {optional_subset_str: <8}"
              f"  ->  {str(toplevel_subset): <16}")
\end{minted}

\begin{minted}{text}
       Fixed | Variable  ->  Top-level subset
   {8, 1, 4} | {}        ->  {8, 1, 4}       
   {8, 1, 4} | {2}       ->  {8, 1, 2, 4}    
      {1, 4} | {}        ->  {1, 4}          
      {1, 4} | {2}       ->  {1, 2, 4}       
   {8, 1, 2} | {}        ->  {8, 1, 2}       
      {1, 2} | {}        ->  {1, 2}          
\end{minted}

\noindent Below is part of the printout from the search of causally complete spaces on 2 events: setting \mintinline{python}{update_period=None} forces the finder to print the current search status when it has finished exploring each ``variable'' subset at top level.

\begin{minted}{text}
Iterating over 6 top-level child history subsets.
      time       spaces    eq. cls     memory  completed fts compl. vts compl.
    2.91ms            4          1   10.35KiB   16.6667%    0.0000%   50.0000%
    3.41ms            5          2   10.38KiB   33.3333%   25.0000%  100.0000%
    3.83ms            5          2   10.38KiB   50.0000%   25.0000%   50.0000%
    5.76ms            5          2   10.38KiB   66.6667%   50.0000%  100.0000%
    6.58ms            5          2   10.38KiB   83.3333%   75.0000%  100.0000%
    7.29ms            7          3   10.41KiB  100.0000%  100.0000%  100.0000%
Found 7 spaces in 3 equivalence classes.
\end{minted}

\noindent The printout describes the following search progression:

\begin{enumerate}
    \item The 1st equivalence class is discovered while exploring $\{8, 1, 4\}$ at top-level: that's $50\%$ (1 out of 2) of the top-level subsets for mandatory subset $\{8, 1, 4\}$, and $16.666\%$ (1 out of 6) of all top-level subsets.
    \item The 2nd equivalence class is discovered while exploring $\{8, 1, 2, 4\}$ at top-level: that's $100\%$ (2 out of 2) of the top-level subsets for mandatory subset $\{8, 1, 4\}$, and $33.333\%$ (2 out of 6) of all top-level subsets. It also completes the search for $25\%$ (1 out of 4) of all mandatory subsets.
    \item No equivalence class is discovered while exploring $\{1, 4\}$ at top-level: that's $50\%$ (1 out of 2) of the top-level subsets for mandatory subset $\{1, 4\}$, and $50\%$ (3 out of 6) of all top-level subsets.
    \item No equivalence class is discovered while exploring $\{1, 2, 4\}$ at top-level: that's $100\%$ (2 out of 2) of the top-level subsets for mandatory subset $\{1, 4\}$, and $66.666\%$ (4 out of 6) of all top-level subsets. It also completes the search for $50\%$ (2 out of 4) of all mandatory subsets.
    \item No equivalence class is discovered while exploring $\{8, 1, 2\}$ at top-level: that's $100\%$ (1 out of 1) of the top-level subsets for mandatory subset $\{8, 1, 2\}$, and $83.333\%$ (5 out of 6) of all top-level subsets. It also completes the search for $75\%$ (3 out of 4) of all mandatory subsets.
    \item The 3rd equivalence class is discovered while exploring $\{1, 2\}$ at top-level: that's $100\%$ (1 out of 1) of the top-level subsets for mandatory subset $\{1, 2\}$, and $100\%$ (6 out of 6) of all top-level subsets. It also completes the search for $100\%$ (4 out of 4) of all mandatory subsets.
\end{enumerate}

\noindent This irregular progression, where many spaces are discovered while exploring the first few ``variable'' subsets and then few or none are discovered while exploring later ones, is typical of the algorithm and becomes significantly more marked for higher number of events.
For example, below is a selection of the initial part of the printout for the search on 3 events, showing all equivalence classes discovered by exploring top-level subsets originating from the first ``fixed'' subset (out of 922 total).

\begin{minted}{text}
Iterating over 922 top-level child history subsets.
      time       spaces    eq. cls     memory  completed fts compl. vts compl.
   38.23ms           48          1   82.88KiB    0.2169%    0.0000%    6.2500%
   57.58ms          439         14   83.87KiB    0.3254%    0.0000%    9.3750%
   67.41ms          583         20   85.62KiB    0.6508%    0.0000%   18.7500%
   73.10ms          619         22   86.26KiB    1.0846%    0.0000%   31.2500%
   78.23ms          622         23   86.37KiB    1.6269%    0.0000%   46.8750%
   80.07ms          634         24   86.44KiB    1.7354%    0.0000%   50.0000%
   88.90ms          694         28   86.61KiB    1.8438%    0.0000%   53.1250%
   92.54ms          718         30   86.72KiB    2.0607%    0.0000%   59.3750%
   ...
\end{minted}

\noindent The progression over ``fixed'' subsets is similarly irregular, with $96\%$ of the equivalence classes (98 out of 102) being discovered while exploring the first $50.1\%$ of subsets (462 out of 922).

\begin{minted}{text}
      time       spaces    eq. cls     memory  completed fts compl. vts compl.
   ...
  346.47ms         2617         98   98.40KiB   54.3384%   50.0000%   16.4062%
  396.66ms         2623         99   98.61KiB   95.9870%   87.5000%    0.0000%
  401.19ms         2644        102   98.75KiB   96.3124%   91.6667%   50.0000%
  407.92ms         2644        102   98.75KiB  100.0000%  104.1667%  100.0000%
Found 2644 spaces in 102 equivalence classes.
\end{minted}

\noindent Finally, here are the utility functions to print and log all status data.

\begin{minted}[firstnumber=last]{python}
    def _print(self, *args: str, **kwargs: Any) -> None:
        """ Prints to console, or logs to logger. """
        print(*args, **kwargs)
        if self._logger is not None:
            self._logger.info(*args)

    def _print_status_header(self) -> None:
        if self._verbose:
            header = (f"{'time': >10} {'spaces': >12} {'eq. cls': >10}"
                      f" {'memory': >10} {'completed': >10}"
                      f" {'fts compl.' :>10} {'vts compl.' :>10}")
            self._print(header, flush=True)

    def _print_status_line(self) -> None:
        if self._verbose:
            time = time_str(self.time_elapsed)
            spaces = str(self.num_spaces)
            eq_classes = str(self.num_eq_classes)
            memory = memory_str(self.memsize)
            completed = f"{self.perc_completed:.4%}"
            fts_completed = f"{self.fixed_toplevel_subsets_perc_completed:.4%}"
            vts_completed = f"{self.var_toplevel_subsets_perc_completed:.4%}"
            line = (f"{time: >10} {spaces: >12} {eq_classes: >10}"
                    f" {memory: >10} {completed: >10}"
                    f" {fts_completed: >10} {vts_completed: >10}")
            self._print(line, flush=True)

    def _describe(self) -> None:
        """
            If in verbose mode,
            prints a summary description of the search results (so far).
        """
        if self._verbose:
            self._print(f"Found {self.num_spaces} spaces in "
                       f"{self.num_eq_classes} equivalence classes.")
\end{minted}

\subsection{More on the \mintinline{python}{SpaceFinder} Class -- State Serialisation}

All data structures altered by the search constitute the ``search state'': this can be either initialised to blank (for new searches) or loaded from file (when continuing an existing search).
Persistent serialisation of search state was made necessary by the large amount of time and computational resources required by the search for causally complete spaces on 4 events, where it was deemed likely---at the time when the code was written---that the search might be stopped (intentionally or accidentally) and then have to resume without significant loss of progress.
The \mintinline{python}{blank_state} method sets the state for a new search: number of spaces found set to 0, no spaces visited, no equivalence classes found; it clears data about top-level subsets, if any.

\begin{minted}[firstnumber=last]{python}
    def blank_state(self) -> None:
        """
            Initialise the space finder in a blank state.
        """
        self._num_spaces = 0
        self._partial_spaces_visited = set()
        self._eq_classes = set()
        if hasattr(self, "_num_done"):
            del self._num_done
        if hasattr(self, "_num_todo"):
            del self._num_todo
        if hasattr(self, "_fix_child_choice_idx"):
            del self._fix_child_choice_idx
        if hasattr(self, "_var_child_subset_bitvec"):
            del self._var_child_subset_bitvec
        if hasattr(self, "_child_choices_list"):
            del self._child_choices_list
        if hasattr(self, "_remaining_children_list"):
            del self._remaining_children_list
        self._initialised = True
\end{minted}

\noindent The private \mintinline{python}{write} method actually performs the state writing to binary IO, dictating the binary encoding scheme used by \mintinline{python}{load_state} above.

\begin{minted}[firstnumber=last]{python}
    def _write_state(self, f: BinaryIO) -> int:
        num_bytes_written = 0
        f.write(self._num_spaces.to_bytes(8, byteorder="big"))
        f.write(self._num_done.to_bytes(8, byteorder="big"))
        f.write(self._num_todo.to_bytes(8, byteorder="big"))
        f.write(self._fix_child_choice_idx.to_bytes(8, byteorder="big"))
        f.write(self._var_child_subset_bitvec.to_bytes(8, byteorder="big"))
        num_bytes_written += 40
        num_bytes_written += write_hsets(f, self._partial_spaces_visited)
        num_bytes_written += write_hsets(f, self._eq_classes)
        num_bytes_written += write_hsets(f, self._child_choices_list)
        num_bytes_written += write_hsets(f, self._remaining_children_list)
        return num_bytes_written
\end{minted}

\noindent The \mintinline{python}{load_state} method loads the state information from a binary file by given name.

\begin{minted}[firstnumber=last]{python}
    def load_state(self, filename: str) -> None:
        """
            Load space finder state from a binary file.
        """
        with open(filename, "rb") as f:
            self._num_spaces = int.from_bytes(f.read(8), byteorder="big")
            self._num_done = int.from_bytes(f.read(8), byteorder="big")
            self._num_todo = int.from_bytes(f.read(8), byteorder="big")
            self._fix_child_choice_idx = int.from_bytes(f.read(8),
                                                         byteorder="big")
            self._var_child_subset_bitvec = int.from_bytes(f.read(8),
                                                         byteorder="big")
            self._partial_spaces_visited = read_hsets_set(f)
            self._eq_classes = read_hsets_set(f)
            self._child_choices_list = read_hsets_list(f)
            self._remaining_children_list = read_hsets_list(f)
        self._initialised = True
\end{minted}

\noindent The \mintinline{python}{save_state} method saves the state information to a binary file by given name (as well as an identical backup file).

\begin{minted}[firstnumber=last]{python}
    def save_state(self, filename: str, *, save_backup: bool = True) -> None:
        """
        Saves the state to a given file, with optional backup.
        """
        if self._save_options is not None:
            filename, _, save_backup = self._save_options
            num_bytes_written = 0
            if self._verbose:
                self._print(f"Saving to '{filename}'...", end=" ", flush=True)
            with open(filename, "wb") as f:
                num_bytes_written += self._write_state(f)
            if save_backup:
                backup_filename = filename+".bak"
                if self._verbose:
                    self._print(f"saving to '{backup_filename}'...", end=" ",
                               flush=True)
                with open(backup_filename, "wb") as f:
                    num_bytes_written += self._write_state(f)
            if self._verbose:
                self._print(f"done ({memory_str(num_bytes_written)} written).",
                           flush=True)
\end{minted}

\noindent The logic determining whether to save state at a given point of the search is encapsulated by the \mintinline{python}{_consider_saving_state} and \mintinline{python}{_save_state} private methods.

\begin{minted}[firstnumber=last]{python}
    _num_eq_classes_since_last_save: int
    def _consider_saving_state(self) -> None:
        if self._save_options is not None:
            _, save_period, _ = self._save_options
            if (save_period is not None
                and self._num_eq_classes_since_last_save >= save_period):
                self._num_eq_classes_since_last_save = 0
                self._save_state()
                self._print_status_line()
    def _save_state(self) -> None:
        if self._save_options is not None:
            filename, _, save_backup = self._save_options
            self.save_state(filename, save_backup=save_backup)
\end{minted}

\subsection{More on \mintinline{python}{SpaceFinder._find_eq_classes}}

\begin{minted}[firstnumber=691]{python}
    def _find_eq_classes(self, new_hs: Sequence[History],
                         hs: Sequence[History] = tuple(),
                         hs_rest: Sequence[History] = tuple(),
                         level: int = 0) -> Iterator[HistorySet]:
        # pylint: disable = too-many-locals
        hs_so_far = tuple(chain(new_hs, hs))
        hs_perm_dict = self._histories_perm_dict
        partial_spaces_visited = self._partial_spaces_visited
        spaces_visited = self._eq_classes
        if level == 0:
            iter_child_subsets = self._iter_child_subsets_toplevel
        else:
            iter_child_subsets = self._iter_child_subsets
        for child_subset in iter_child_subsets(new_hs):
            child_subset_sorted = sorted(child_subset, key=history_sort_key)
            hs_so_far_rest = list(chain(new_hs, hs_rest))
            for k in child_subset_sorted:
                for j, h in enumerate(hs_so_far):
                    if is_subset(k, h):
                        hs_so_far_rest[j] = sub(hs_so_far_rest[j], k)
            winnowed_hs = tuple(h for j, h in enumerate(hs_so_far)
                                       if hs_so_far_rest[j])
            winnowed_hs_rest = tuple(
                h for h in hs_so_far_rest if h)
            partial_space = set(chain(child_subset_sorted, winnowed_hs))
            partial_space_bitvec = bitvec(partial_space)
            already_seen = (partial_space_bitvec in partial_spaces_visited
                            or partial_space_bitvec in spaces_visited)
            if not already_seen:
                eq_class = set()
                for p in self._perm_group:
                    p_action = hs_perm_dict[p]
                    perm_partial_space_bitvec = bitvec(p_action[h]
                                                       for h in partial_space)
                    eq_class.add(perm_partial_space_bitvec)
                    if (perm_partial_space_bitvec in partial_spaces_visited
                        or perm_partial_space_bitvec in spaces_visited):
                        already_seen = True
                        break
            if not already_seen:
                if all(self._domsize[h] == 1 for h in child_subset_sorted):
                    self._num_spaces += len(eq_class)
                    yield partial_space_bitvec
                else:
                    partial_spaces_visited.add(partial_space_bitvec)
                    yield from self._find_eq_classes(child_subset_sorted,
                                                     winnowed_hs,
                                                     winnowed_hs_rest,
                                                     level+1)
\end{minted}

\noindent At the start of the method, the sequence \mintinline{python}{hs_so_far} of all current candidate input histories is computed, and some persistent data structures are accessed and given shorter names.

\begin{minted}[firstnumber=696]{python}
        hs_so_far = tuple(chain(new_hs, hs))
        hs_perm_dict = self._histories_perm_dict
        partial_spaces_visited = self._partial_spaces_visited
        spaces_visited = self._eq_classes
\end{minted}

\noindent The rest of the method iterates over all subsets of child histories for the histories in \mintinline{python}{new_hs}.
Two different methods are used: the symmetry-optimised \mintinline{python}{self._iter_child_subsets_toplevel} is used at top-level (\mintinline{python}{level=0}), while the general purpose \mintinline{python}{self._iter_child_subsets} is used for all other levels.
In principle, the \mintinline{python}{self._iter_child_subsets} method could be used to iterate over child subsets at all levels, but the added efficiency of the \mintinline{python}{self._iter_child_subsets_toplevel} is helpful in the search on 4 events (plus, the symmetry argument is interesting in itself).
Each child subset yielded is sorted to make the search deterministic.

\begin{minted}[firstnumber=700]{python}
        if level == 0:
            iter_child_subsets = self._iter_child_subsets_toplevel
        else:
            iter_child_subsets = self._iter_child_subsets
        for child_subset in iter_child_subsets(new_hs, level):
            child_subset_sorted = sorted(child_subset, key=history_sort_key)
\end{minted}

\subsection{More on \mintinline{python}{SpaceFinder.find_eq_classes}}

\begin{minted}[firstnumber=740]{python}
    def find_eq_classes(self) -> None:
        """
            Start the search for equivalence classes.
        """
        if not self._initialised:
            raise Exception("Must initialise using blank_state() "
                            "or load_state(file).")
        update_period = self._update_period
        self._start_time = perf_counter()
        self._num_eq_classes_since_last_save = 0
        init_histories = max_histories(self.num_events)
        for eq_class_repr in self._find_eq_classes(init_histories):
            self._eq_classes.add(eq_class_repr)
            self._num_eq_classes_since_last_save += 1
            if (update_period is not None
                and self.num_eq_classes % update_period == 0):
                self._print_status_line()
        if update_period is not None:
            self._print_status_line()
        self._partial_spaces_visited.clear()
        self._save_state()
        self._describe()
\end{minted}

\subsection{More on \mintinline{python}{SpaceFinder._child_subset}}

\begin{minted}[firstnumber=last]{python}
    def _child_subset(self,
                      hs: Sequence[History],
                      child_hists: Sequence[History],
                      child_subset_bitvec: int,
                      hs_already_covered: frozenset[History] = frozenset(),
                      children_already_chosen: frozenset[History] = frozenset()
                      ) -> Optional[set[History]]:
        # pylint: disable = too-many-arguments
        hs_still_to_cover = set(hs)-hs_already_covered
        child_subset = set(children_already_chosen)
        idx = 0
        while child_subset_bitvec > 0:
            child_subset_bitvec, b = divmod(child_subset_bitvec, 2)
            if b:
                k = child_hists[idx]
                child_subset.add(k)
                if hs_still_to_cover:
                    hs_still_to_cover -= self._parents[k]
            idx += 1
        if not hs_still_to_cover:
            return child_subset
        return None
\end{minted}

\noindent The method starts by creating a set \mintinline{python}{hs_to_cover} of histories in \mintinline{python}{hs} still to be ``covered''---i.e. where at least one child has to yet appear in the subets---and a set \mintinline{python}{child_subset} of children extracted from the bitvector; the optional arguments \mintinline{python}{hs_already_covered} and \mintinline{python}{children_already_chosen} (empty by default) are used by the symmetry-optimized algorithm of method \mintinline{python}{_iter_child_subsets_toplevel} to modify these initial sets (because, in its logic, some histories are already known to be covered and some children are already known to be included in the subset to be yielded).

\subsection{\mintinline{python}{SpaceFinder._iter_child_subsets_toplevel}}

The private \mintinline{python}{_iter_child_subsets_toplevel} method implements a symmetry-optimized version of \mintinline{python}{_iter_child_subsets}, for use by the 4-event search.
Since we are only interested in finding equivalence classes of causally complete spaces under event-input permutation symmetry, the symmetry itself can be used to fix some redundant degrees of freedom in the child history subsets at the top level (\mintinline{python}{level=0}) of the search, because the set of parent histories \mintinline{python}{hs} at top level---the maximal extended input histories---is stabilised by the whole event-input permutation symmetry group.

\begin{minted}[firstnumber=796]{python}
    def _iter_child_subsets_toplevel(self,
                                     hs: Sequence[History]
                                     ) -> Iterator[set[History]]:
        if hasattr(self, "_fix_child_choice_idx"): # loaded previous state
            assert hasattr(self, "_num_todo")
            assert hasattr(self, "_num_done")
            assert hasattr(self, "_child_choices_list")
            assert hasattr(self, "_remaining_children_list")
            assert hasattr(self, "_fix_child_choice_idx")
            assert hasattr(self, "_var_child_subset_bitvec")
            ch_choices_list = tuple(
                frozenset(iter_bitvec(child_choice_bitvec))
                for child_choice_bitvec in self._child_choices_list)
            remaining_ch_list = tuple(
                set(iter_bitvec(rem_children_bv))
                for rem_children_bv in self._remaining_children_list)
        else:
            opt_fix_ch_choices = self._opt_fix_child_choices(hs,
                                                             self._perm_group)
            ch_choices_list, num_todo, remaining_ch_list = opt_fix_ch_choices
            self._num_todo = num_todo
            self._num_done = 0
            self._child_choices_list = [bitvec(child_choice)
                                        for child_choice in ch_choices_list]
            self._remaining_children_list = [bitvec(remaining_children)
                                             for remaining_children
                                             in remaining_ch_list]
            self._fix_child_choice_idx = 0
            self._var_child_subset_bitvec = 0
        self._num_remaining_children = [len(remaining_children)
                                        for remaining_children
                                        in remaining_ch_list]
        if self._verbose:
            self._print(f"Iterating over {self._num_todo} "
                        "top-level child history subsets.")
        self._print_status_header()
        fix_child_choice_idx = self._fix_child_choice_idx
        for child_choice, remaining_children in islice(zip(ch_choices_list,
                                                           remaining_ch_list),
                                                       fix_child_choice_idx,
                                                       None):
            rem_children_sorted = sorted(remaining_children, key=history_sort_key)
            hs_already_covered = frozenset(
                h for h in hs if child_choice&self._children_set[h])
            num_child_subsets = 2**len(rem_children_sorted)
            for child_subset_bitvec in range(self._var_child_subset_bitvec,
                                                num_child_subsets):
                child_subset = self._child_subset(hs, rem_children_sorted,
                                                  child_subset_bitvec,
                                                  hs_already_covered,
                                                  child_choice)
                self._num_done += 1
                if child_subset is not None:
                    yield child_subset
                if child_subset is not None and self._update_period is None:
                    self._print_status_line()
                self._var_child_subset_bitvec += 1
                if child_subset is not None:
                    self._consider_saving_state()
            self._var_child_subset_bitvec = 0
            self._fix_child_choice_idx += 1
\end{minted}

\noindent At the start of the method, we check whether the data structures associated to symmetry-optimised search already exist: if they do, it means that we are resuming a previous search, and we don't need to create them.

\begin{minted}[firstnumber=799]{python}
        if hasattr(self, "_fix_child_choice_idx"): # loaded previous state
            assert hasattr(self, "_num_todo")
            assert hasattr(self, "_num_done")
            assert hasattr(self, "_child_choices_list")
            assert hasattr(self, "_remaining_children_list")
            assert hasattr(self, "_fix_child_choice_idx")
            assert hasattr(self, "_var_child_subset_bitvec")
            ch_choices_list = tuple(
                frozenset(iter_bitvec(child_choice_bitvec))
                for child_choice_bitvec in self._child_choices_list)
            remaining_ch_list = tuple(
                set(iter_bitvec(rem_children_bv))
                for rem_children_bv in self._remaining_children_list)
\end{minted}

\noindent If the necessary data structures don't exist, we create them instead.
The \mintinline{python}{_opt_fix_child_choices} returns a triple:
\begin{itemize}
    \item A list \mintinline{python}{ch_choices_list} of ``fixed'' children subsets, where event-input permutation symmetry has been used to remove unnecessary choices.
    \item The number \mintinline{python}{num_todo} of top-level children subsets that will need to be iterated over.
    \item A list \mintinline{python}{remaining_ch_list} of largest ``variable'' children subsets corresponding to each ``fixed'' children subset in \mintinline{python}{ch_choices_list} (which has the same length). For each ``fixed'' children subset, all subsets of the corresponding largest ``variable'' children subset will be iterated over.
\end{itemize}
We store the bitvectors for the two lists above for search state serialisation.
We initialise two counters: the \mintinline{python}{_fix_child_choice_idx} index, tracking the current ``fixed'' child subset, and the \mintinline{python}{_var_child_subset_bitvec} bitvector, tracking the current ``variable'' children subset.

\begin{minted}[firstnumber=812]{python}
        else:
            opt_fix_ch_choices = self._opt_fix_child_choices(hs,
                                                             self._perm_group)
            ch_choices_list, num_todo, remaining_ch_list = opt_fix_ch_choices
            self._num_todo = num_todo
            self._num_done = 0
            self._child_choices_list = [bitvec(child_choice)
                                        for child_choice in ch_choices_list]
            self._remaining_children_list = [bitvec(remaining_children)
                                             for remaining_children
                                             in remaining_ch_list]
            self._fix_child_choice_idx = 0
            self._var_child_subset_bitvec = 0
\end{minted}

\noindent Regardless of whether this is a new search or a resumed search, we store a list of sizes of the maximal ``variable'' child subsets, for use when computing completion percentages in search status printouts.
If required, we then print the number of top-level child subsets that will be iterated over, as well as the header for the search status metric table.

\begin{minted}[firstnumber=825]{python}
        self._num_remaining_children = [len(remaining_children)
                                        for remaining_children
                                        in remaining_ch_list]
        if self._verbose:
            self._print(f"Iterating over {self._num_todo} "
                        "top-level child history subsets.")
        self._print_status_header()
\end{minted}

\noindent After the setup, we proceed to iterate through all possible ``fixed'' children subsets \mintinline{python}{child_choice}, and the corresponding maximal ``variable'' children subsets \mintinline{python}{remaining_children}; we start from index \mintinline{python}{fix_child_choice_idx}, which might be non-zero if our search was resumed from loaded state.
For each pair, we sort the remaining children (to make the search deterministic), compute the set \mintinline{python}{hs_already_covered} of histories in \mintinline{python}{hs} that have a child in the ``fixed'' children subset, and proceed to iterate through the bitvectors of all ``variable'' children subsets, i.e. all possible subsets of \mintinline{python}{remaining_children}.

\begin{minted}[firstnumber=832]{python}
        fix_child_choice_idx = self._fix_child_choice_idx
        for child_choice, remaining_children in islice(zip(ch_choices_list,
                                                           remaining_ch_list),
                                                       fix_child_choice_idx,
                                                       None):
            rem_children_sorted = sorted(remaining_children, key=history_sort_key)
            hs_already_covered = frozenset(
                h for h in hs if child_choice&self._children_set[h])
            num_child_subsets = 2**len(rem_children_sorted)
            for child_subset_bitvec in range(self._var_child_subset_bitvec,
                                                num_child_subsets):
\end{minted}

\noindent For each bitvector, the \mintinline{python}{_child_subset} method is called to obtain the corresponding ``variable'' children subset \mintinline{python}{child_subset}, or \mintinline{python}{None} if some history in \mintinline{python}{hs} doesn't have a child in that subset.
If the subset is not \mintinline{python}{None}, we yield it; we also consider adding a line to the search status table (if required) and saving the search state (if required).
Before state serialisation, we increase the ``variable'' children subset bitvector counter \mintinline{python}{_var_child_subset_bitvec} by 1, thus moving to the next bitvector.

\begin{minted}[firstnumber=843]{python}
                child_subset = self._child_subset(hs, rem_children_sorted,
                                                  child_subset_bitvec,
                                                  hs_already_covered,
                                                  child_choice)
                self._num_done += 1
                if child_subset is not None:
                    yield child_subset
                if child_subset is not None and self._update_period is None:
                    self._print_status_line()
                self._var_child_subset_bitvec += 1
                if child_subset is not None:
                    self._consider_saving_state()
\end{minted}

\noindent When all ``fixed'' children subsets have been iterated over, we reset the corresponding bitvector counter and we increase the ``fixed'' children subset index counter by 1, thus moving to the next ``fixed'' subset.

\begin{minted}[firstnumber=855]{python}
            self._var_child_subset_bitvec = 0
            self._fix_child_choice_idx += 1
\end{minted}

\subsection{\mintinline{python}{SpaceFinder._fix_child_choices}}

The private \mintinline{python}{_fix_child_choices} method implements the core logic for the symmetry optimisation.
It takes a sequence \mintinline{python}{hs} of histories and a subgroup \mintinline{python}{stab} of the event-input permutation group.
It returns a list of pairs \mintinline{python}{(children_to_include, children_to_avoid)}, where \mintinline{python}{children_to_include} is a ``fixed'' children subset and \mintinline{python}{children_to_avoid} is used by \mintinline{python}{_opt_fix_child_choices} to compute the corresponding maximal ``variable'' children subset.
The method's logic is recusive, with a \mintinline{python}{depth} kwarg to keep track of recursion depth and a \mintinline{python}{max_depth} kwarg specifying a recursion cutoff (with \mintinline{python}{None} indicating no maximum depth).
The \mintinline{python}{children_to_include} and \mintinline{python}{children_to_avoid} optional arguments are used by the recursive calls (see later).
At the first call (\mintinline{python}{depth=0}), \mintinline{python}{hs} is set to the maximal extended input histories and \mintinline{python}{stab} is set to the entirety of the event-input permutation group.

\begin{minted}[firstnumber=857]{python}
    def _fix_child_choices(self,
                           hs: Sequence[History],
                           perm_group: Sequence[PermGroupEl],
                           children_to_include: frozenset[History]=frozenset(),
                           children_to_avoid: frozenset[History]=frozenset(), *,
                           depth: int = 0,
                           max_depth: Optional[int] = None
                           ) -> list[tuple[frozenset[History],
                                           frozenset[History]]]:
        # pylint: disable = too-many-locals, too-many-branches
        if (len(perm_group) == 1
            or not hs
            or (max_depth is not None and depth > max_depth)
            ):
            return [(frozenset(), frozenset())]
        def select_h_child_subset(s: Set[History],
                                  h_children_to_include: Set[History]) -> bool:
            if s&children_to_avoid:
                return False
            if h_children_to_include <= s:
                return True
            return False
        best: Optional[tuple[History,
                             Mapping[frozenset[History],
                                     list[PermGroupEl]]]] = None
        hs_new_fixed = []
        for h in hs:
            h_children = self._children[h]
            h_children_to_include = children_to_include&self._children_set[h]
            h_children_sel_subsets = sorted(
                (s for t in powerset(h_children)
                 if select_h_child_subset(s:=frozenset(t),
                                          h_children_to_include)),
                key=lambda s: (-len(s), sorted(s, key=history_sort_key)))
            h_children_subsets = {}
            h_children_subsets_seen = set()
            for ks in h_children_sel_subsets:
                if (not ks) or ks in h_children_subsets_seen:
                    continue
                ks_stab = []
                for p in perm_group:
                    p_action = self._histories_perm_dict[p]
                    ks_perm = frozenset(p_action[k] for k in ks)
                    if ks == ks_perm:
                        ks_stab.append(p)
                    h_children_subsets_seen.add(ks_perm)
                h_children_subsets[ks] = ks_stab
            if len(h_children_subsets) >= 1:
                if best is None or len(h_children_subsets) < len(best[1]):
                    best = (h, h_children_subsets)
            else:
                hs_new_fixed.append(h)
        if best is None:
            return [(frozenset(), frozenset())]
        best_h, best_h_children_subsets = best
        hs_new_fixed.append(best_h)
        new_hs = tuple(h for h in hs if h not in hs_new_fixed)
        seen = set()
        child_choices = []
        for ks, ks_stab in best_h_children_subsets.items():
            new_children_to_include = children_to_include|ks
            new_children_to_avoid = (children_to_avoid
                                     |frozenset(self._children[best_h])-ks)
            rec_child_choices = self._fix_child_choices(new_hs, ks_stab,
                                                        new_children_to_include,
                                                        new_children_to_avoid,
                                                        depth=depth+1,
                                                        max_depth=max_depth)
            for rec_child_choice in rec_child_choices:
                rec_children_to_include, rec_children_to_avoid = rec_child_choice
                child_choice = (rec_children_to_include|ks,
                                new_children_to_avoid|rec_children_to_avoid)
                if child_choice not in seen:
                    seen.add(child_choice)
                    child_choices.append(child_choice)
        return child_choices
\end{minted}

\noindent At the start of the method, we perform some termination checks:
\begin{enumerate}
    \item if \mintinline{python}{hs} is empty, we have no further histories to consider
    \item if \mintinline{python}{perm_group} is the trivial subgroup, we have no further symmetries available to reduce our subsets
    \item if we exceeded the max recursion depth, we terminate our search
\end{enumerate}
Whatever the case may be, we return an empty set of children to include and an empty subset of children to avoid.

\begin{minted}[firstnumber=867]{python}
        if (len(perm_group) == 1
            or not hs
            or (max_depth is not None and depth > max_depth)
            ):
            return [(frozenset(), frozenset())]
\end{minted}

\noindent We then create a small utility function \mintinline{python}{select_h_child_subset}, checking whether a given set \mintinline{python}{s} of child histories includes another given set \mintinline{python}{h_children_to_include} of child histories and is disjoint from the set \mintinline{python}{children_to_avoid}.

\begin{minted}[firstnumber=872]{python}
        def select_h_child_subset(s: Set[History],
                                  h_children_to_include: Set[History]) -> bool:
            if s&children_to_avoid:
                return False
            if h_children_to_include <= s:
                return True
            return False
\end{minted}

\noindent We will shortly be iterating over all \mintinline{python}{h in hs}, looking for a best \mintinline{python}{h} to select in this iteration.
If not \mintinline{python}{None}, \mintinline{python}{best} will contain the pair \mintinline{python}{(h, h_children_subsets)} of the current best \mintinline{python}{h in hs} and a mapping of each selected subset of the children of \mintinline{python}{h} to the corresponding stabiliser subgroup of \mintinline{python}{perm_group}.
Along the way, we will also grow a list \mintinline{python}{hs_new_fixed} of histories \mintinline{python}{h in hs} for which no selected subsets are available (i.e. the choice of subset of children histories is uniquely fixed by previous choices).

\begin{minted}[firstnumber=879]{python}
        best: Optional[tuple[History,
                             Mapping[frozenset[History],
                                     list[PermGroupEl]]]] = None
        hs_new_fixed = []
\end{minted}

\noindent For each \mintinline{python}{h in hs}, we compute the list of selected subsets.
These are subsets of the hildren of \mintinline{python}{h} which include at least all child histories of \mintinline{python}{h} that are in \mintinline{python}{children_to_include}, and include no child histories of \mintinline{python}{h} that are in \mintinline{python}{children_to_avoid}.
To make the search deterministic, the selected subsets are sorted first by decreasing length, and then by content.

\begin{minted}[firstnumber=883]{python}
        for h in hs:
            h_children = self._children[h]
            h_children_to_include = children_to_include&self._children_set[h]
            h_children_sel_subsets = sorted(
                (s for t in powerset(h_children)
                 if select_h_child_subset(s:=frozenset(t),
                                          h_children_to_include)),
                key=lambda s: (-len(s), sorted(s, key=history_sort_key)))
\end{minted}

\noindent We then proceed to look at the orbits of the selected subsets in \mintinline{python}{h_children_sel_subsets} under the permutation group \mintinline{python}{perm_group}: we select a representative from each orbit and store it in the dictionary \mintinline{python}{h_children_subsets}, together with the associated stabiliser subgroup of \mintinline{python}{perm_group}.

\begin{minted}[firstnumber=891]{python}
            h_children_subsets = {}
            h_children_subsets_seen = set()
            for ks in h_children_sel_subsets:
                if (not ks) or ks in h_children_subsets_seen:
                    continue
                ks_stab = []
                for p in perm_group:
                    p_action = self._histories_perm_dict[p]
                    ks_perm = frozenset(p_action[k] for k in ks)
                    if ks == ks_perm:
                        ks_stab.append(p)
                    h_children_subsets_seen.add(ks_perm)
                h_children_subsets[ks] = ks_stab
\end{minted}

\noindent If there were no non-empty selected subsets of children, it means that the choice of children subset for history \mintinline{python}{h} was fixed by previous choices, so we add \mintinline{python}{h} to \mintinline{python}{hs_new_fixed}.
Otherwise, we check whether the children subsets for \mintinline{python}{h} have fewer orbits under the permutation group than the previous best choice (if any): if this is the case, we store \mintinline{python}{h}, the orbit representatives and the associated stabiliser subgroups as our new best choice.

\begin{minted}[firstnumber=904]{python}
            if len(h_children_subsets) >= 1:
                if best is None or len(h_children_subsets) < len(best[1]):
                    best = (h, h_children_subsets)
            else:
                hs_new_fixed.append(h)
\end{minted}

\noindent If \mintinline{python}{best=None} at the end of the previous loop, it means that all histories are fixed, so we terminate the same way as we would at the start of the method for the case where \mintinline{python}{hs} is empty.
Otherwise, we add the best history \mintinline{python}{best_h} to the list of fixed histories---it's the one that we're fixing at this particular call.
The histories \mintinline{python}{new_hs} that will be passed to the recursive calls to \mintinline{python}{_fix_child_choices} are those histories in \mintinline{python}{hs} the subsets of which have not yet been completely fixed (i.e. those in \mintinline{python}{hs_new_fixed}).

\begin{minted}[firstnumber=909]{python}
        if best is None:
            return [(frozenset(), frozenset())]
        best_h, best_h_children_subsets = best
        hs_new_fixed.append(best_h)
        new_hs = tuple(h for h in hs if h not in hs_new_fixed)
\end{minted}

\noindent Next, we iterate through each orbit representative \mintinline{python}{ks} for the subsets of child histories of \mintinline{python}{best_h}, with associated stabiliser subgroup \mintinline{python}{ks_stab}.
The representative \mintinline{python}{ks} is exactly the set of children of \mintinline{python}{best_h} that will be included by all top-level child subsets added in this step of the iteration, while its complement \mintinline{python}{frozenset(self._children[best_h])-ks} is exactly the set of children of \mintinline{python}{best_h} that will be excluded by all top-level child subsets added in this step of the iteration.

We make a recursive call to \mintinline{python}{_fix_child_choices} to obtain a list of children to include/avoid for the histories in \mintinline{python}{hs}, where the subset \mintinline{python}{ks} is added to the previous children to be included and its complement \mintinline{python}{frozenset(self._children[best_h])-ks} is added to the previous children to be avoided.

\begin{minted}[firstnumber=914]{python}
        for ks, ks_stab in best_h_children_subsets.items():
            new_children_to_include = children_to_include|ks
            new_children_to_avoid = (children_to_avoid
                                     |frozenset(self._children[best_h])-ks)
            rec_child_choices = self._fix_child_choices(new_hs, ks_stab,
                                                        new_children_to_include,
                                                        new_children_to_avoid,
                                                        depth=depth+1,
                                                        max_depth=max_depth)
\end{minted}

\noindent We take each recursively computed pair of children to include \mintinline{python}{rec_children_to_include} and children to avoid \mintinline{python}{rec_children_to_avoid}, we add \mintinline{python}{ks} to the former and \mintinline{python}{new_children_to_avoid} to obtain a new pair \mintinline{python}{child_choice} of children to include/avoid, and we add it to \mintinline{python}{child_choices} if it doesn't yet appear in it.

\begin{minted}[firstnumber=925]{python}
            for rec_child_choice in rec_child_choices:
                rec_children_to_include, rec_children_to_avoid = rec_child_choice
                child_choice = (rec_children_to_include|ks,
                                rec_children_to_avoid|new_children_to_avoid)
                if child_choice not in seen:
                    seen.add(child_choice)
                    child_choices.append(child_choice)
\end{minted}

\noindent At the end, we return the list \mintinline{python}{child_choices}.
As an example, below is the first branch of recursion for the call to \mintinline{python}{_fix_child_choices} performed by the search on 3 events.
We start at depth 0, with the maximal extended input histories as \mintinline{python}{hs} and with the full event-input permutation symmetry group as \mintinline{python}{perm_group}.
There are no children to include and no children to avoid at depth 0.

The algorithm selects the history \hist{A/0,B/0,C/0}: it has 7 selected children subsets, falling into 3 orbits (in fact, all choices at depth 0 give the same result, because of symmetry).
A representative for each orbit is displayed below, along with the size of its stabiliser within the 48-element group \mintinline{python}{perm_group}.

\begin{minted}[linenos=false]{text}
->initial call at depth 0
  perm_group (size 48):
    whole event-input symmetry group
  children to include:
    none
  children to avoid:
    none
  best_h: {'A': 0, 'B': 0, 'C': 0}
  children: [{'A': 0, 'B': 0}, {'A': 0, 'C': 0}, {'B': 0, 'C': 0}]
  number of selected children subsets: 7
  representatives for children subset orbits: 
    0: {{'A': 0, 'C': 0}, {'B': 0, 'C': 0}, {'A': 0, 'B': 0}}, stab size 6
    1: {{'A': 0, 'C': 0}, {'A': 0, 'B': 0}}, stab size 2
    2: {{'A': 0, 'B': 0}}, stab size 4
\end{minted}

\noindent We consider representative $\{\hist{A/0,C/0}, \hist{B/0, C/0}, \hist{A/0, B/0}\}$ from depth 0 above and look at the corresponding recursive call.
The permutation group \mintinline{python}{perm_group} used by the recursive call is the 6-element stabiliser subgroup for the chosen representative, within the 48-element permutation group used by the previous call.
The child histories in the chosen representatives appear now as children to include for the recursive call, while their complement (the empty subset) appears as children to avoid for the recursive call.

At depth 1, the algorithm selects the history \hist{A/0,B/0,C/1}: it has 4 selected children subsets, falling into 3 orbits.
A representative for each orbit is displayed below, along with the size of its stabiliser within the 6-element group \mintinline{python}{perm_group}.

\begin{minted}[linenos=false]{text}
  ->recursive call at depth 1 (path=0)
    perm_group (size 6):
      (('A', 'B', 'C'), (0, 0, 0))
      (('A', 'C', 'B'), (0, 0, 0))
      (('B', 'A', 'C'), (0, 0, 0))
      (('B', 'C', 'A'), (0, 0, 0))
      (('C', 'A', 'B'), (0, 0, 0))
      (('C', 'B', 'A'), (0, 0, 0))
    children to include:
      {'A': 0, 'C': 0}
      {'B': 0, 'C': 0}
      {'A': 0, 'B': 0}
    children to avoid:
      none
    best_h: {'A': 0, 'B': 0, 'C': 1}
    children: [{'A': 0, 'B': 0}, {'A': 0, 'C': 1}, {'B': 0, 'C': 1}]
    number of selected children subsets: 4
    representatives for children subset orbits: 
      00: {{'A': 0, 'C': 1}, {'B': 0, 'C': 1}, {'A': 0, 'B': 0}}, stab size 2
      01: {{'A': 0, 'C': 1}, {'A': 0, 'B': 0}}, stab size 1
      02: {{'A': 0, 'B': 0}}, stab size 2
\end{minted}

\noindent We consider representative $\{\hist{A/0,C/1}, \hist{B/0, C/1}, \hist{A/0, B/0}\}$ from depth 1 above and look at the corresponding recursive call.
The permutation group \mintinline{python}{perm_group} used by the recursive call is the 2-element stabiliser subgroup for the chosen representative, within the 6-element permutation group used by the previous call.
The child histories in the chosen representatives are added to the children to include for the recursive call, while their complement (the empty subset) is added to the children to avoid for the recursive call.

At depth 2, the algorithm selects the history \hist{A/0,B/1,C/0}: it has 4 selected children subsets, falling into 4 orbits.
A representative for each orbit is displayed below, along with the size of its stabiliser within the 6-element group \mintinline{python}{perm_group}.
Because all stabilisers have size 1, no further steps are taken in this recursive branch.

\begin{minted}[linenos=false]{text}
    ->recursive call at depth 2 (path=00)
      perm_group (size 2):
        (('A', 'B', 'C'), (0, 0, 0))
        (('B', 'A', 'C'), (0, 0, 0))
      children to include:
        {'A': 0, 'C': 0}
        {'A': 0, 'C': 1}
        {'B': 0, 'C': 0}
        {'A': 0, 'B': 0}
        {'B': 0, 'C': 1}
      children to avoid:
        none
      best_h: {'A': 0, 'B': 1, 'C': 0}
      children: [{'A': 0, 'B': 1}, {'A': 0, 'C': 0}, {'B': 1, 'C': 0}]
      number of selected children subsets: 4
      representatives for children subset orbits: 
        000: {{'B': 1, 'C': 0}, {'A': 0, 'B': 1}, {'A': 0, 'C': 0}}, stab size 1
        001: {{'A': 0, 'B': 1}, {'A': 0, 'C': 0}}, stab size 1
        002: {{'B': 1, 'C': 0}, {'A': 0, 'C': 0}}, stab size 1
        003: {{'A': 0, 'C': 0}}, stab size 1
\end{minted}

\noindent As an alternative example, we consider representative $\{\hist{A/0,C/0}, \hist{A/0, B/0}\}$ from depth 0 instead, and look again at the corresponding recursive call.
The permutation group \mintinline{python}{perm_group} used by the recursive call is the 2-element stabiliser subgroup for the chosen representative, within the 48-element permutation group used by the previous call.
The child histories in the chosen representatives appear now as children to include for the recursive call, while their complement appears as children to avoid for the recursive call.

At depth 1, the algorithm selects the history \hist{A/1,B/0,C/0}: it has 3 selected children subsets, falling into 2 orbits.
A representative for each orbit is displayed below, along with the size of its stabiliser within the 2-element group \mintinline{python}{perm_group}.

\begin{minted}[linenos=false]{text}
  ->recursive call at depth 1 (path=1)
    perm_group (size 2):
      (('A', 'B', 'C'), (0, 0, 0))
      (('A', 'C', 'B'), (0, 0, 0))
    children to include:
      {'A': 0, 'C': 0}
      {'A': 0, 'B': 0}
    children to avoid:
      {'B': 0, 'C': 0}
    best_h: {'A': 1, 'B': 0, 'C': 0}
    children: [{'A': 1, 'B': 0}, {'A': 1, 'C': 0}, {'B': 0, 'C': 0}]
    number of selected children subsets: 3
    representatives for children subset orbits: 
      10: {{'A': 1, 'C': 0}, {'A': 1, 'B': 0}}, stab size 2
      11: {{'A': 1, 'B': 0}}, stab size 1
\end{minted}

\noindent We consider representative $\{\hist{A/1,C/0}, \hist{A/1, B/0}\}$ from depth 1 above and look at the corresponding recursive call.
The permutation group \mintinline{python}{perm_group} used by the recursive call is the full 2-element permutation group used by the previous call.
The child histories in the chosen representatives are added to the children to include for the recursive call, while their complement (the empty subset) is added to the children to avoid for the recursive call.

At depth 2, the algorithm selects the history \hist{A/0,B/0,C/1}: it has 4 selected children subsets, falling into 4 orbits.
A representative for each orbit is displayed below, along with the size of its stabiliser within the 2-element group \mintinline{python}{perm_group}.
Because all stabilisers have size 1, no further steps are taken in this recursive branch either.

\begin{minted}[linenos=false]{text}
    ->recursive call at depth 2 (path=10)
      perm_group (size 2):
        (('A', 'B', 'C'), (0, 0, 0))
        (('A', 'C', 'B'), (0, 0, 0))
      children to include:
        {'A': 0, 'C': 0}
        {'A': 1, 'C': 0}
        {'A': 0, 'B': 0}
        {'A': 1, 'B': 0}
      children to avoid:
        {'B': 0, 'C': 0}
      best_h: {'A': 0, 'B': 0, 'C': 1}
      children: [{'A': 0, 'B': 0}, {'A': 0, 'C': 1}, {'B': 0, 'C': 1}]
      number of selected children subsets: 4
      representatives for children subset orbits: 
        100: {{'A': 0, 'C': 1}, {'B': 0, 'C': 1}, {'A': 0, 'B': 0}}, stab size 1
        101: {{'A': 0, 'C': 1}, {'A': 0, 'B': 0}}, stab size 1
        102: {{'B': 0, 'C': 1}, {'A': 0, 'B': 0}}, stab size 1
        103: {{'A': 0, 'B': 0}}, stab size 1
\end{minted}

\noindent Finally, below is the full printout for a recursive path reaching down to depth 7.

\begin{minted}[linenos=false]{text}
->initial call at depth 0
  perm_group (size 48):
    whole event-input symmetry group
  children to include:
    none
  children to avoid:
    none
  best_h: {'A': 0, 'B': 0, 'C': 0}
  children: [{'A': 0, 'B': 0}, {'A': 0, 'C': 0}, {'B': 0, 'C': 0}]
  number of selected children subsets: 7
  representatives for children subset orbits: 
    0: {{'A': 0, 'C': 0}, {'B': 0, 'C': 0}, {'A': 0, 'B': 0}}, stab size 6
    1: {{'A': 0, 'C': 0}, {'A': 0, 'B': 0}}, stab size 2
    2: {{'A': 0, 'B': 0}}, stab size 4
  ->recursive call at depth 1 (path=2)
    perm_group (size 4):
      (('A', 'B', 'C'), (0, 0, 0))
      (('A', 'B', 'C'), (0, 0, 1))
      (('B', 'A', 'C'), (0, 0, 0))
      (('B', 'A', 'C'), (0, 0, 1))
    children to include:
      {'A': 0, 'B': 0}
    children to avoid:
      {'A': 0, 'C': 0}
      {'B': 0, 'C': 0}
    best_h: {'A': 0, 'B': 0, 'C': 1}
    children: [{'A': 0, 'B': 0}, {'A': 0, 'C': 1}, {'B': 0, 'C': 1}]
    number of selected children subsets: 4
    representatives for children subset orbits: 
      20: {{'A': 0, 'C': 1}, {'B': 0, 'C': 1}, {'A': 0, 'B': 0}}, stab size 2
      21: {{'A': 0, 'C': 1}, {'A': 0, 'B': 0}}, stab size 1
      22: {{'A': 0, 'B': 0}}, stab size 4
    ->recursive call at depth 2 (path=22)
      perm_group (size 4):
        (('A', 'B', 'C'), (0, 0, 0))
        (('A', 'B', 'C'), (0, 0, 1))
        (('B', 'A', 'C'), (0, 0, 0))
        (('B', 'A', 'C'), (0, 0, 1))
      children to include:
        {'A': 0, 'B': 0}
      children to avoid:
        {'A': 0, 'C': 0}
        {'B': 0, 'C': 0}
        {'B': 0, 'C': 1}
        {'A': 0, 'C': 1}
      best_h: {'A': 0, 'B': 1, 'C': 0}
      children: [{'A': 0, 'B': 1}, {'A': 0, 'C': 0}, {'B': 1, 'C': 0}]
      number of selected children subsets: 3
      representatives for children subset orbits: 
        220: {{'B': 1, 'C': 0}, {'A': 0, 'B': 1}}, stab size 1
        221: {{'A': 0, 'B': 1}}, stab size 2
        222: {{'B': 1, 'C': 0}}, stab size 1
      ->recursive call at depth 3 (path=221)
        perm_group (size 2):
          (('A', 'B', 'C'), (0, 0, 0))
          (('A', 'B', 'C'), (0, 0, 1))
        children to include:
          {'A': 0, 'B': 1}
          {'A': 0, 'B': 0}
        children to avoid:
          {'A': 0, 'C': 0}
          {'A': 0, 'C': 1}
          {'B': 0, 'C': 0}
          {'B': 0, 'C': 1}
          {'B': 1, 'C': 0}
        best_h: {'A': 0, 'B': 1, 'C': 1}
        children: [{'A': 0, 'B': 1}, {'A': 0, 'C': 1}, {'B': 1, 'C': 1}]
        number of selected children subsets: 2
        representatives for children subset orbits: 
          2210: {{'B': 1, 'C': 1}, {'A': 0, 'B': 1}}, stab size 1
          2211: {{'A': 0, 'B': 1}}, stab size 2
        ->recursive call at depth 4 (path=2211)
          perm_group (size 2):
            (('A', 'B', 'C'), (0, 0, 0))
            (('A', 'B', 'C'), (0, 0, 1))
          children to include:
            {'A': 0, 'B': 1}
            {'A': 0, 'B': 0}
          children to avoid:
            {'A': 0, 'C': 0}
            {'A': 0, 'C': 1}
            {'B': 0, 'C': 0}
            {'B': 0, 'C': 1}
            {'B': 1, 'C': 0}
            {'B': 1, 'C': 1}
          best_h: {'A': 1, 'B': 0, 'C': 0}
          children: [{'A': 1, 'B': 0}, {'A': 1, 'C': 0}, {'B': 0, 'C': 0}]
          number of selected children subsets: 3
          representatives for children subset orbits: 
            22110: {{'A': 1, 'C': 0}, {'A': 1, 'B': 0}}, stab size 1
            22111: {{'A': 1, 'B': 0}}, stab size 2
            22112: {{'A': 1, 'C': 0}}, stab size 1
          ->recursive call at depth 5 (path=22111)
            perm_group (size 2):
              (('A', 'B', 'C'), (0, 0, 0))
              (('A', 'B', 'C'), (0, 0, 1))
            children to include:
              {'A': 0, 'B': 1}
              {'A': 0, 'B': 0}
              {'A': 1, 'B': 0}
            children to avoid:
              {'A': 0, 'C': 0}
              {'A': 0, 'C': 1}
              {'A': 1, 'C': 0}
              {'B': 0, 'C': 0}
              {'B': 0, 'C': 1}
              {'B': 1, 'C': 0}
              {'B': 1, 'C': 1}
            best_h: {'A': 1, 'B': 1, 'C': 0}
            children: [{'A': 1, 'B': 1}, {'A': 1, 'C': 0}, {'B': 1, 'C': 0}]
            number of selected children subsets: 1
            representatives for children subset orbits: 
              221110: {{'A': 1, 'B': 1}}, stab size 2
            ->recursive call at depth 6 (path=221110)
              perm_group (size 2):
                (('A', 'B', 'C'), (0, 0, 0))
                (('A', 'B', 'C'), (0, 0, 1))
              children to include:
                {'A': 0, 'B': 1}
                {'A': 1, 'B': 1}
                {'A': 0, 'B': 0}
                {'A': 1, 'B': 0}
              children to avoid:
                {'A': 0, 'C': 1}
                {'B': 0, 'C': 1}
                {'B': 1, 'C': 1}
                {'A': 0, 'C': 0}
                {'A': 1, 'C': 0}
                {'B': 0, 'C': 0}
                {'B': 1, 'C': 0}
              best_h: {'A': 1, 'B': 0, 'C': 1}
              children: [{'A': 1, 'B': 0}, {'A': 1, 'C': 1}, {'B': 0, 'C': 1}]
              number of selected children subsets: 2
              representatives for children subset orbits: 
                2211100: {{'A': 1, 'C': 1}, {'A': 1, 'B': 0}}, stab size 1
                2211101: {{'A': 1, 'B': 0}}, stab size 2
              ->recursive call at depth 7 (path=2211101)
                perm_group (size 2):
                  (('A', 'B', 'C'), (0, 0, 0))
                  (('A', 'B', 'C'), (0, 0, 1))
                children to include:
                  {'A': 0, 'B': 0}
                  {'A': 1, 'B': 0}
                  {'A': 0, 'B': 1}
                  {'A': 1, 'B': 1}
                children to avoid:
                  {'A': 0, 'C': 1}
                  {'A': 1, 'C': 1}
                  {'B': 0, 'C': 1}
                  {'B': 1, 'C': 1}
                  {'A': 0, 'C': 0}
                  {'A': 1, 'C': 0}
                  {'B': 0, 'C': 0}
                  {'B': 1, 'C': 0}
                best_h: {'A': 1, 'B': 1, 'C': 1}
                children: [{'A': 1, 'B': 1}, {'A': 1, 'C': 1}, {'B': 1, 'C': 1}]
                number of selected children subsets: 1
                representatives for children subset orbits: 
                  22111010: {{'A': 1, 'B': 1}}, stab size 2
\end{minted}

\subsection{\mintinline{python}{SpaceFinder._opt_fix_child_choices}}

The private \mintinline{python}{_opt_fix_child_choices} method runs at various maximum depths to obtain a sequence of ``fixed'' children subset choices which minimises the number of top-level children subsets to be iterated over.
It does so by progressively increasing the maximum depth, stopping at the point where the overall number of top-level children subsets starts to increase (i.e. it stops at the first local minimum).

The method takes a sequence \mintinline{python}{hs} of histories and a subgroup \mintinline{python}{stab} of the event-input permutation group.
In the current implementation, where the method is called only at top-level, \mintinline{python}{hs} is always the sequence of maximum extended input histories for the given number of events, while \mintinline{python}{stab} is always the full event-input permutation group.
However, this method could also be applied at lower levels, using as \mintinline{python}{stab} the subgroup of permutations which stabilise the set of input histories \mintinline{python}{set(hs)}.

\begin{minted}[firstnumber=933]{python}
    def _opt_fix_child_choices(self, hs: Sequence[History],
                               perm_group: Sequence[PermGroupEl]
                              ) -> tuple[tuple[frozenset[History], ...],
                                         int,
                                         tuple[set[History], ...]]:
        child_hists_set = {k for h in hs for k in self._children[h]}
        toplevel_opt_depth = self._toplevel_opt_depth
        if toplevel_opt_depth is None:
            toplevel_opt_depth = len(hs)
        if self._verbose:
            self._print(f"Brute-forcing complexity: {2**len(child_hists_set)}"
                         " top-level child history subsets.")
            if toplevel_opt_depth >= 0:
                self._print("Optimising top-level child history subsets"
                            f" (max depth {toplevel_opt_depth}).")
        best: Optional[tuple[list[tuple[frozenset[History],
                                        frozenset[History]]],
                             int,
                             list[set[History]]
                            ]] = None
        for max_depth in range(-1, toplevel_opt_depth+1):
            fixed_child_choices = self._fix_child_choices(hs, perm_group,
                                                          max_depth=max_depth)
            num_todo = 0
            remaining_children_list = []
            for child_choice, children_to_avoid in fixed_child_choices:
                remaining_children = (child_hists_set
                                      -(child_choice|children_to_avoid))
                num_todo += 2**len(remaining_children)
                remaining_children_list.append(remaining_children)
            if best is None or num_todo <= best[1]:
                if self._verbose:
                    if max_depth >= 0:
                        self._print(f"  {num_todo} subsets at "
                                   f"optimisation depth {max_depth}")
                best = (fixed_child_choices, num_todo, remaining_children_list)
            else:
                break
        assert best is not None
        return (tuple(child_choice for child_choice, _ in best[0]),
                best[1], tuple(best[2]))
\end{minted}

\noindent At the start, we create the set of child histories for all histories in \mintinline{python}{hs}, set the maximum optimisation depth to the length of \mintinline{python}{hs} if not already set, and print some information about the optimisation process (if required).

\begin{minted}[firstnumber=938]{python}
        child_hists_set = {k for h in hs for k in self._children[h]}
        toplevel_opt_depth = self._toplevel_opt_depth
        if toplevel_opt_depth is None:
            toplevel_opt_depth = len(hs)
        if self._verbose:
            self._print(f"Brute-forcing complexity: {2**len(child_hists_set)}"
                         " top-level child history subsets.")
            if toplevel_opt_depth >= 0:
                self._print("Optimising top-level child history subsets"
                            f" (max depth {toplevel_opt_depth}).")
\end{minted}

\noindent

\begin{minted}[firstnumber=948]{python}
        best: Optional[tuple[list[tuple[frozenset[History],
                                        frozenset[History]]],
                             int,
                             list[set[History]]
                            ]] = None
\end{minted}

\noindent We then iterate through maximum depths from \mintinline{python}{-1} (meaning no symmetry optimisation) to \mintinline{python}{toplevel_opt_dept} (included), computing the list of children to include/avoid for each max depth.

\begin{minted}[firstnumber=953]{python}
        for max_depth in range(-1, toplevel_opt_depth+1):
            fixed_child_choices = self._fix_child_choices(hs, perm_group,
                                                          max_depth=max_depth)
\end{minted}

\noindent For each choice of children to include (the ``fixed'' children subset) and to avoid, we compute the maximum ``variable'' children subset \mintinline{python}{remaining_children}, and we increase the total number of top-level children subsets to iterate over by the number of non-empty subsets of \mintinline{python}{remaining_children}.

\begin{minted}[firstnumber=956]{python}
            num_todo = 0
            remaining_children_list = []
            for child_choice, children_to_avoid in fixed_child_choices:
                remaining_children = (child_hists_set
                                      -(child_choice|children_to_avoid))
                num_todo += 2**len(remaining_children)
                remaining_children_list.append(remaining_children)
\end{minted}

\noindent If the total number of top-level children subsets to iterate over is not greater than the current best, we store this as the new best choice; otherwise, we break (so that the best choice is the first local minimum encountered).

\begin{minted}[firstnumber=963]{python}
            if best is None or num_todo <= best[1]:
                if self._verbose:
                    if max_depth >= 0:
                        self._print(f"  {num_todo} subsets at "
                                   f"optimisation depth {max_depth}")
                best = (fixed_child_choices, num_todo, remaining_children_list)
            else:
                break
\end{minted}

\noindent We return a triple with the list of ``fixed'' children subsets, the total number of top-level children subsets, and the the list of maximum ``variable'' children subsets, obtained from the best maximum depth.

\begin{minted}[firstnumber=971]{python}
        assert best is not None
        return (tuple(child_choice for child_choice, _ in best[0]),
                best[1], tuple(best[2]))
\end{minted}

\noindent As a practical example, we look at the search for causally complete spaces on 3-events.
The following (immutable) set contains the bitvectors for all children of maximal extended input histories \mintinline{python}{hs} passed to \mintinline{python}{_fix_child_choices}:

\begin{minted}[linenos=false]{python}
frozenset({5, 6, 9, 10, 17, 18, 20, 24, 33, 34, 36, 40})
\end{minted} 

\noindent Below is the list \mintinline{python}{child_choices} obtained from the main call to \mintinline{python}{_fix_child_choices} in the 3-event search, at the optimal \mintinline{python}{max_depth=8}.

\begin{minted}[linenos=false]{python}
[(frozenset({5, 9, 17, 20, 24, 33, 36}), frozenset()),
 (frozenset({5, 9, 17, 20, 33, 36}), frozenset({24})),
 (frozenset({5, 17, 20, 24, 33, 36}), frozenset({9})),
 (frozenset({5, 17, 20, 33, 36}), frozenset({9, 24})),
 (frozenset({5, 17, 20, 33}), frozenset({36})),
 (frozenset({5, 9, 17, 20, 40}), frozenset({33, 36})),
 (frozenset({5, 9, 17, 20}), frozenset({33, 36, 40})),
 (frozenset({5, 17, 20, 40}), frozenset({9, 33, 36})),
 (frozenset({5, 6, 17, 18, 33, 36}), frozenset({20})),
 (frozenset({5, 6, 17, 18, 33}), frozenset({20, 36})),
 (frozenset({5, 6, 17, 18, 36}), frozenset({20, 33})),
 (frozenset({5, 6, 17, 18}), frozenset({20, 33, 36})),
 (frozenset({5, 6, 17}), frozenset({18, 20})),
 (frozenset({5, 9, 24, 33, 36}), frozenset({17, 20})),
 (frozenset({5, 9, 33, 36}), frozenset({17, 20, 24})),
 (frozenset({5, 24, 33, 36}), frozenset({9, 17, 20})),
 (frozenset({5, 33}), frozenset({17, 20, 36})),
 (frozenset({5, 9, 24}), frozenset({17, 20, 33, 36})),
 (frozenset({5, 9, 40}), frozenset({17, 20, 24, 33, 36})),
 (frozenset({5, 6, 9, 18}), frozenset({17, 20, 24, 33, 36, 40})),
 (frozenset({5, 6, 9, 10, 34}), frozenset({17, 18, 20, 24, 33, 36, 40})),
 (frozenset({5, 6, 9, 10}), frozenset({17, 18, 20, 24, 33, 34, 36, 40})),
 (frozenset({5, 9, 18}), frozenset({6, 17, 20, 24, 33, 36, 40})),
 (frozenset({5, 24}), frozenset({9, 17, 20, 33, 36}))]
\end{minted}

\noindent Consider, for example, the pair \mintinline{python}{(frozenset({5, 9, 17, 20}), frozenset({33, 36, 40}))} (at index 6 in the list).
The ``fixed'' children subset here is $\{5, 9, 17, 20\}$, while the children to avoid are $\{33, 36, 40\}$, so that the maximum ``variable'' children subset is:
\[
\begin{array}{rl}
&\{5, 6, 9, 10, 17, 18, 20, 24, 33, 34, 36, 40\}\\
&\backslash
\left(\{5, 9, 17, 20\}\cup\{33, 36, 40\}\right)\\
=&
\{6, 10, 18, 24, 34\}
\end{array}
\]
This gives $2^6 = 32$ of the 922 top-level children subsets, obtained by union of $\{5, 9, 17, 20\}$ with all possible subsets of $\{6, 10, 18, 24, 34\}$.



\ack
Financial support from EPSRC, the Pirie-Reid Scholarship and Hashberg Ltd is gratefully acknowledged.
This publication was made possible through the support of the ID\#62312 grant from the John Templeton Foundation, as part of the project `The Quantum Information Structure of Spacetime' (QISS), https://www.templeton.org/grant/the-quantum-information-structure-ofspacetime-qiss-second-phase.
The opinions expressed in this project/publication are those of the author(s) and do not necessarily reflect the views of the John Templeton Foundation.

\section*{Bibliography}

\bibliographystyle{unsrt}
\bibliography{biblio}

\end{document}